\documentclass[12pt,a4paper,titlepage]{book}
\pdfoutput=1
\title{\Huge \scshape TESIS}
\author{\scshape Marco Baity Jesi}
\date{\today}

\usepackage{tesis-marco} 
\newindex[myThePage]{default}{idx}{ind}{Alphabetic index}
\makenomenclature

\renewcommand{\chaptermark}[1]{\markboth{#1}{}}
\renewcommand{\sectionmark}[1]{\markright{\thesection\ --- #1}}
\fancypagestyle{plain}{%
\fancyhf{} % clear all header and footer fields
\fancyfoot[C]{\oldstylenums\thepage} % except the center

}
\makeatletter
\preto\@tabular{\fontfamily{pplx}\selectfont}
\makeatother

\titleformat{\chapter}[display]
{\bfseries\LARGE} {\filleft\MakeUppercase{\chaptertitlename}
\Huge\Roman{chapter}} {2ex} {\titlerule
\vspace{1.5ex}%
\filright}
[\vspace{1.5ex}%
]

\titleformat{\section}[block]
{\Large\normalfont}
{\bfseries\thesection}{.5em}{\titlerule\\[.8ex]\bfseries}

\begin{document}

% \fontfamily{hlhj}\selectfont
% \linenumbers

%  ************************************************************
%  PORTADA
%  ************************************************************

\thispagestyle{empty}
\vspace*{0.1cm}

\begin{center}
{\Huge \bfseries
Criticality and Energy Landscapes in Spin Glasses\\}
 \vspace*{1.3cm}
 
\begin{minipage}{0.4\textwidth}\centering
  {\large \em
Criticalidad y paisajes de energ\'{i}a en vidrios de esp\'in\\
}
 \end{minipage}
 \begin{minipage}{0.4\textwidth}\centering
{\large \em
Criticalit\`a e paesaggi di energia nei vetri di spin\\
}
 \end{minipage}
 
\vspace{1.5cm}
  {\large \textsc{PhD Thesis}}\\[2ex]
\begin{minipage}{0.45\textwidth}\centering
 {Tesis doctoral en idioma ingl\'es}
 \end{minipage}
 \begin{minipage}{0.45\textwidth}\centering
 {Tesi di dottorato in lingua inglese}
 \end{minipage}

\vspace*{1.5cm}
{\large 
Candidate:\\[2ex]
\textsc{Marco Baity Jesi}
\vspace*{1.2cm}

Supervisors:
\vspace*{0.8cm}}

\begin{minipage}{0.48\textwidth}\centering
\textsc{V\'{i}ctor Mart\'{i}n Mayor}\\[3ex]
 \includegraphics[height=3cm]{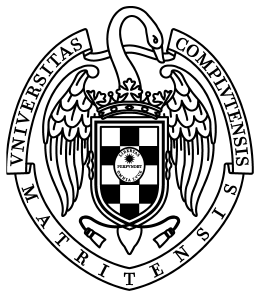}\\[3ex]
  Universidad Complutense de Madrid\\
  Facultad de Ciencias F\'{i}sicas\\
  Departamento de F\'{i}sica Te\'{o}rica I
\end{minipage}
\begin{minipage}{0.48\textwidth}\centering
\textsc{Giorgio Parisi}\\[3ex]
 \includegraphics[height=3cm]{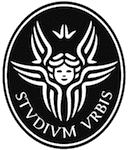}\\[3ex]
  Sapienza, Universit\`a di Roma\\
  Facolt\`a di Scienze MM.FF.NN.\\
  Dipartimento di F\`{i}sica\\[2ex]
\end{minipage}

\vspace*{1.5cm}
MMXV

\end{center}

%Pagenumbering presenta problemi con hyperref
%\pagenumbering{gobble} %No numbering
\pagenumbering{roman}
%\pagenumbering{arabic}

\newpage
\thispagestyle{empty}
$ $
\newpage

\thispagestyle{empty}
\vspace*{7cm}
\begin{flushright}
\itshape A mam\'a,\\[1ex] ~~ y a V\'ictor.
	 %A mam\'a, que me inici\'o al mundo,\\
         %a Aurora, que me inici\'o a Madrid,
         %a V\'ictor, que me inici\'o a la investigaci\'on.
\end{flushright}

\newpage
\thispagestyle{empty}
$ $

%  ************************************************************
%  TABLE OF CONTENTS
%  ************************************************************

%\renewcommand{\@pnumwidth}{5em}
\clearpage
 \tableofcontents
\phantomsection

%  ************************************************************
%  INCLUSIÓN DE CAPÍTULOS
%  ************************************************************

 \chapter*{Foreword}\label{chap:intro-general} 
 \chaptermark{Foreword}
\addcontentsline{toc}{chapter}{\protect\numberline{} Foreword}

\section*{A first acknowledgment}
\sectionmark{A first acknowledgment}
\addcontentsline{toc}{section}{\protect\numberline{} A first acknowledgment}
This dissertation is the result of a Ph.D. thesis in co-tutorship between the two universities \emph{Universidad Complutense de Madrid}, Madrid, Spain, and \emph{Sapienza, 
Universit\`a di Roma}, Rome, Italy. My supervisors were V\'ictor Mart\'in Mayor (Spanish side) and Giorgio Parisi (Italian side). I am very grateful for the 
time spent with both, and for the unquantifiable amount of things I learned from them during these years. I must acknowledge also Luis Antonio Fern\'andez P\'erez also has been at my side
helping me through with programming and showing me his complicated codes, and David Yllanes and Beatriz Seoane, who were Ph.D. students in the group before me,
were very nice receiving me in the group, and supported me when I needed it. I also wish to mention Jos\'e Manuel Sanz Gonz\'alez, who guided my steps through a large
part of my first article. 
During my thesis I also had the opportunity of a stay in the group of Matthieu Wyart, whom I desire to thank for giving me the privilege of working with him, as well
as Le Yan, my extremely valid colleague during and after my months there. I also thank Andrea Liu for having me in her research group in the upcoming last months of my Ph.D..

I also acknowledge that my thesis was funded by the FPU program of the Ministerio de Educaci\'on, Spain, that funded me with a four year fellowship,
and with extra allowances for my research stays at NYU
\footnote{New York University, New York, NY, USA. Stay from May 2$^\mathrm{nd}$ to July 31$^\mathrm{st}$, 2014.}
and at UPenn.
\footnote{University of Pennsylvania, Philadelphia, PA, USA. Stay from September 15$^\mathrm{th}$ to December 14$^\mathrm{th}$, 2015 (to be concluded).}
Further research costs such as materials and trips have been funded 
by MINECO, Spain, through the research contract No. FIS2012-35719-C02,
by the European Research Council under the European Union's Seventh Framework Programme (FP7/2007-2013, ERC grant agreement no. 247328),
by the Seventh Framework Programme (EU-FP7) through the research contract No. 287746,
and by the GDRE 224 CNRS-INdAM GREFI-MEFI.
I am also grateful to the BIFI 
\footnote{The Institute for Biocomputation and Physics of Complex Systems of the University of Zaragoza.}
for letting me use their CPU and GPU resources, and to its very professional staff.

\section*{High-performance computing in this thesis \label{sec:hpc}}
\sectionmark{High-performance computing in this thesis}
\addcontentsline{toc}{section}{\protect\numberline{} High-performance computing in this thesis}

In this thesis we present the results of several research projects on spin glasses, principally obtained through numerical simulations. 
Since this is a thesis in physics, we will mainly talk about the physical results, relegating to the background the numerical details.

Nevertheless, it is important to mention that extremely powerful numerical resources were necessary to arrive to some conclusions. Especially
\cite{janus:14c} and \cite{baityjesi:14} would have been unthinkable with normal computing resources. 

For \cite{janus:14c} I enjoyed the chance
of being part of the \index{Janus@\textsc{Janus}!collaboration}\textsc{Janus} Collaboration, a partnership of physicists and engineers that work with the \ac{FPGA}-based machine 
\index{Janus@\textsc{Janus}!computer}\textsc{Janus} \cite{janus:06, yllanes:11, janus:12} (and the recently-launched \emph{Janus II} \cite{janus:14}),
\footnote{\href{http://www.janus-computer.com/}{http://www.janus-computer.com/}}
devised expressly for Monte Carlo simulations of spin glasses.
The \textsc{Janus} computer been able to thermalize much larger lattices than conventional computers, at lower temperatures, and it can reach
times comparable with those of experiments \cite{janus:08b,janus:10,janus:10b,janus:12}.

In the case of \cite{baityjesi:14}, I was part of SCC-Computing as a member of BIFI,
\footnote{Strategic collaboration with China on super-computing based on Tianhe-1A, supported by the EU's \ac{FP7} Programme under grant agreement n°287746. \href{http://www.scc-computing.eu}{http://www.scc-computing.eu}}
a \ac{FP7} project that aimed to develop connections between European and Chinese scientists by giving European groups
the possibility to run simulations on the supercomputer 
\index{supercomputer|see{Tianhe-1A}}\index{supercomputer|see{\textsc{Janus}}}\index{Tianhe-1A}\emph{Tianhe-1A}, 
that had been the most powerful machine in the world, and at the time was ranked
number two in \emph{Top 500}.
\footnote{Top 500 is the annual ranking of the 500 most powerful computers in the world, in terms of flops. \href{http://www.top500.org}{http://www.top500.org}}
Only thanks to these extraordinary resources, added to a careful tuning of our simulations in order to get the maximum performance, it has been possible
to obtain the results shown in this dissertation.

In addition to the aforementioned facilities, I had the chance to use 
the small cluster of my group in Madrid,
the \emph{Minotauro} \ac{GPU} cluster in the Barcelona Supercomputing Center,
the \emph{Memento} and \emph{Terminus} \ac{CPU} clusters and some \acp{GPU} for benchmarking from BIFI,
and the \emph{Mercer} cluster of the New York University.\\[7ex]

\section*{Scope and organization of this dissertation}
\sectionmark{Scope and organization of this dissertation}
\addcontentsline{toc}{section}{\protect\numberline{} Scope and organization of this dissertation}

The research done during this thesis aims to make a small progress in the secular question on the nature of the glass transition.
We focused, with a mainly numerical approach, on a paradigmatic glassy system, the spin glass, and we dealt with them by seeing their 
behavior at equilibrium as well as studying the features of their rugged energy landscape.

The equilibrium properties we were interested on concerned universality in the glass transition and the fragility of the spin glass
phase under an external magnetic field. On the side of the energy landscape, it is accepted that the energy landscape plays a major
role in the slowing down of the glasses' dynamics. We tried to get a better insight by studying zero-temperature dynamics, 
by studying how the energy landscape becomes trivial when tuning certain parameters, and by analyzing lowest modes of the density of states.

The text is organized in four parts. In the following paragraphs we introduce briefly each of them.

Part I of this thesis is completely introductory on the systems we studied in this thesis, spin glasses.
Chapter \ref{chap:intro} aims to put the reader into context, by introducing spin glasses in the frame of the glass transitions in general, 
by posing a historical basis about the birth of spin glasses, mentioning and explaining the development of some major theories.
We get more technical in chapter \ref{chap:obs}, where we detail the observables that will be analyzed throughout the rest of the text. In
chapter \ref{chap:rg} we recall the reader some main concepts on scaling and renormalization group that will be useful to understand the analyses
we performed. 

Part II is dedicated to the study of critical properties of spin glasses through equilibrium simulations. We study the presence and the features
of critical lines in the presence of perturbations on paradigmatic Hamiltonians.

In chapter \ref{chap:eah3d}, that comes from \cite{janus:14c} and some unpublished results, we investigate, through Monte Carlo simulations 
with the dedicated computer \textsc{Janus}, whether the \ac{SG} phase survives the imposition of a small external magnetic field, and thus
whether there is a phase transition under the field. The two main theories on the \ac{SG} phase have different predictions, so understanding
whether there is or not a phase transition would be a strong factor for a discrimination between the two. We find very large fluctuations in the
observables we measure, and the average turns out to be a bad descriptor for our populations of measurements. Thus, we develop statistical methods and a new
finite-size scaling ansatz that let us detect very different behaviors. Some of the measurements present strong signs of criticality, while others do not. It is not possible to
determine which of the two behaviors will dominate in the thermodynamic limit, but we are able to set a temperature range where the would-be
phase transition should be searched.

The material in chapter \ref{chap:ahsg} comes from \cite{baityjesi:14}. To produce it I had the opportunity to work on large \ac{GPU} clusters in 
Spain and in China.
We do equilibrium Monte Carlo simulations on the Heisenberg spin glass with random exchange anisotropies. According to the Kawamura scenario, the chiral
and the spin glass channels couple when anisotropies are introduced. We find a phase transition for each of the order parameters, and through a
careful finite-size scaling analysis we conclude that the phase transition is unique. Moreover, the universal quantities we measure are compatible
with the Ising universality class, instead of Heisenberg, indicating that the anisotropy is a relevant perturbation in the renormalization group sense.

Part III is on spin glasses in the absence of thermal vibration. The energy landscape appears to play a fundamental role in the sluggish dynamics that
characterize a glass. It is a feature with a diverging number of dimensions, and still, it is most commonly described through a single number. This
simplification is not always suitable and it is necessary to resort to different descriptors. 

Chapter \ref{chap:hsgm}, that comes from \cite{baityjesi:15}, is a study of the energy landscape of spin glasses as a function of the number of 
spin components $m$. When $m$ is small the energy landscape is rugged and complex, with a large amount of local minima. 
An increase of $m$ involves the gradual disappearance of most of those minima, along with a growth of the correlations and a slow down of the dynamics.

In chapter \ref{chap:marginal}, that is the result of my stay at the Center for Soft Matter Research of the New York University, we show how athermal 
dynamics in spin glasses are related to crackling noise, exposing studies from \cite{yan:15, baityjesi:15c} 
and unpublished material. 
We focus on the histeresis of the \ac{SK} model, that describes spins in a fully connected graph. The dynamics along the hysteresis loop is in form
of abrupt spin avalanches. We show that these avalanches can not occur if the interactions are short-range, and that long-range interactions are a relevant
perturbation to the short-range Hamiltonian. During the avalanches, furthermore, correlations between soft spins arise spontaneously, leading naturally
the system to marginally stable states.

Chapter \ref{chap:hsgrf} describes \cite{baityjesi:15b}, where we examine soft plastic modes of Heisenberg spin glasses in a \ac{RF}, that we impose on the system
in order to get rid of the soft modes due to the rotational symmetry. At low frequencies, the density of states has a non-Debye behavior, revealing the 
presence of a \emph{boson peak}, a typical feature of structural glasses. These soft modes are localized, and they connect very near states, separated by
very low energy barriers, that we identify as classical \emph{two-level systems}. This helps to find a connection between the two main theories on the boson
peak. On one side replica theory gives a mean field description that attributes the soft modes to a fractal energy landscape, and on the other there is the 
phenomenological picture of the two-level systems, that attributes the excess of soft modes to a quantum tunneling between near states.

In part IV we give our conclusions, resuming the main results chapter by chapter.

We also include several appendices. 
Appendix \ref{app:MC} is on Monte Carlo algorithms and on parallel computing for spin glass simulations.
Appendix \ref{app:propagators} is on the measurement of connected propagators in a field.
Appendix \ref{app:hack} gives details on the creation of the \emph{quantiles} defined in chapter \ref{chap:eah3d}.
In appendix \ref{app:reglas-suma} we derive some identities that were crucial to make sure that our programs gave the correct output.
Appendix \ref{app:eah3d-errors} is about error managing.
Appendix \ref{chap:hsgrf-is-algorithm} explains the energy minimization algorithms that were used in chapters \ref{chap:hsgm} and \ref{chap:hsgrf}.

\clearpage
\section*{List of publications and presentations}
\sectionmark{List of publications and presentations}
\addcontentsline{toc}{section}{\protect\numberline{} List of publications and presentations}
To help the panel of judges we include a list of the 
%peer-reviewed 
publications and the presentations done by the candidate during his thesis.
\subsection*{Articles}
\begin{itemize}
%  \item M. Baity-Jesi, L. Yan, M. M\"uller and M. Wyart, ``Range of the interactions in self-organized criticality'', \emph{in preparation} \cite{baityjesi:15c}.
 \item M. Baity-Jesi, V. Mart\'in-Mayor, G. Parisi and S. P\'erez-Gaviro, ``Soft Modes, Localization and Two-Level Systems in Spin Glasses'', Phys. Rev. Lett. \textbf{115}, 267205 (2015) \cite{baityjesi:15b}.
 \item L. Yan, M. Baity-Jesi, M. M\"uller and M. Wyart, ``Dynamics and Correlations among Soft Excitations in Marginally Stable Glasses'',  Phys. Rev. Lett. \textbf{114}, 247208 (2015) \cite{yan:15}.
 \item M. Baity-Jesi and G. Parisi,``Inherent structures in $m$-component spin glasses'',  Phys. Rev. B \textbf{91}, 134203 (2015) \cite{baityjesi:15}.
 \item M. Baity-Jesi \emph{et al.}, ``The three dimensional Ising spin glass in an external magnetic field: the role of the silent majority'',  J. Stat. Mech. (2014) P05014 \cite{janus:14c}.
 \item M. Baity-Jesi \emph{et al.}, ``Dynamical Transition in the $D\!\!=\!\!3$ Edwards-Anderson spin glass in an external magnetic field'',  Phys. Rev. E \textbf{89}, 032140 (2014) \cite{janus:14b}.
 \item M. Baity-Jesi, L.A. Fern\'andez, V. Mart\'in-Mayor and J.M. Sanz, ``Phase transition in three-dimensional Heisenberg spin glasses with strong random anisotropies through a multi-GPU parallelization'',  Phys. Rev. B \textbf{89}, 014202 (2014) \cite{baityjesi:14}.
 \item M. Baity-Jesi \emph{et al.}, ``Janus II: a new generation application-driven computer for spin-system simulations'',  Computer Physics Communications \textbf{185} (2014) 550-559 \cite{janus:14}.
 \item M. Baity-Jesi \emph{et al.}, ``Critical parameters of the three-dimensional Ising spin glass'',  Phys. Rev. B \textbf{88}, 224416 (2013) \cite{janus:13}.
 \item M. Baity-Jesi \emph{et al.}, ``Reconfigurable computing for Monte Carlo simulations: Results and
   prospects of the \textsc{Janus} project'',  The European Physical Journal - Special Topics \textbf{210}, 33-51 (2012) \cite{janus:12b}.
\end{itemize}

\clearpage
\subsection*{Presentations}
\begin{itemize}
 \item \emph{Unifying Concepts in Glass Physics VI}, Aspen Center for Physics, Aspen (CO), USA.
	February, 1-7, 2015. Poster: ``Soft modes in $3d$ spin glasses''.
 \item Dept. of Physics and Astronomy, University of Pennsylvania, Philadelphia, USA (Visiting Andrea Liu). 
	January, $30^\mathrm{th}$, 2015. Talk: ``Random Anisotropies in Heisenberg Spin Glasses''.
 \item \emph{Transversal Seminars}. Departamento de F\'isica Te\'orica I, Universidad Complutense de Madrid, Madrid, Spain.
	January $23^\mathrm{rd}$, 2015. Talk: ``Random Anisotropies in Heisenberg Spin Glasses''.
 \item \emph{Transversal Seminars}. Departamento de F\'isica Te\'orica I, Universidad Complutense de Madrid, Madrid, Spain.
	October $31^\mathrm{st}$, 2014. Talk: ``An Introduction to spin glasses and a study on the dAT line''.
 \item \emph{Perspectives of GPU computing in Physics and Astrophysics}, Sapienza University, Rome, Italy.
	September, 15-17, 2014. Talk: ``The Effect of Random Anisotropies on Heisenberg Spin Glasses: A multi-GPU approach''.
 \item \emph{Critical Phenomena in Random and Complex Systems}, Villa Orlandi, Anacapri, Italy.
	September, 9-12, 2014. Poster: ``Phase Transition in Heisenberg Spin Glasses with Strong Random Anisotropies with a Multi-GPU Approach''.
 \item \emph{Heraeus Workshop 2014}, Institute of Materials Physics in Space, Cologne, Germany.
	September, 1-5, 2014. Poster: ``Phase Transition in Heisenberg Spin Glasses with Strong Random Anisotropies with a Multi-GPU Approach''.
 \item Department of Chemistry, Columbia University, New York, USA (Visiting David Reichman). 
	June $12^\mathrm{th}$, 2014. Talk: ``Random Anisotropies in Heisenberg Spin Glasses''.
 \item Center for Soft Matter Research, New York University, New York, USA (Visiting Matthieu Wyart).
	May $2^\mathrm{nd}$, 2014. Talk: ``Some numerical simulations on $3d$ spin glasses''.
 \item \emph{VI International Conference BIFI 2014}, Ibercaja Zentrum, Zaragoza, Spain. 
	January, 22-24, 2014. Talk: ``Phase Transition in Heisenberg Spin Glasses with Strong Random Anisotropies with a Multi-GPU Approach''.
 \item \emph{XXV IUPAP International Conference on Statistical Physics (STATPHYS 25)}, Seoul National University, Seoul, South Korea.
	July 22-26, 2013. Poster: ``Phase Transition in Heisenberg Spin Glasses with Strong Random
Anisotropies with a Multi-GPU Approach''.
 \item \emph{Partnership for supercomputing applications in science and industry}, Grand Hotel Sofia, Sofia, Bulgaria.
	April 8-10, 2013. Talk: ``Spin Glasses with a multi-GPU approach (2)''.
 \item \emph{International Workshop EU-China on Scientific Computing}, Instituto de Biocomputaci\'on y F\'isica de Sistemas Complejos (BIFI), Universidad de Zaragoza, Zaragoza, Spain.
	November 26-28, 2012. Talk: ``Spin Glasses with a multi-GPU approach (1)''.
 \item \emph{Strategic Collaboration with China - Computing Project Kick-off Meeting}, National Super Computing Center (NSCC), Tianjin, China.
	April $22^\mathrm{nd}$, 2012. Talk: ``GPU Simulations on 3d Anisotropic Heisenberg Spin Glasses''.
\end{itemize}

 \chapter*{Abstracts in other languages}
\chaptermark{Abstracts in other languages}
 \addcontentsline{toc}{chapter}{\protect\numberline{} Abstracts in other languages}
% \section*{Abstract in English \label{chap:abstract-english}}
%  \addcontentsline{toc}{section}{\protect\numberline{}Abstract in English}
% \clearpage

\section*{Resumen en castellano \label{sec:resumen}}
\sectionmark{Resumen en castellano}
 \addcontentsline{toc}{section}{\protect\numberline{}Resumen en castellano}
 \index{transici\'on v\'itrea|see{glass transition}}\index{fase v\'itrea|see{glass phase}} 
 \index{universalidad|see{universality}}\index{grupo de renormalizaci\'on|see{renormalization group}}
 \index{anisotropia|see{anisotropy}}\index{quiralidad|see{chirality}}
 \index{paisaje de energia|see{energy landscape}}
 \index{m\'inimo local|see{local minimum}}\index{estabilidad marginal|see{marginal stability}}
 \index{modos blandos|see{soft modes}}\index{campo magn\'etico aleatorio|see{random magnetic field}}
 \index{barrera energética|see{energy barrier}}
 \index{estructura inherente|see{inherent structure}}

Esta tesis tiene el objetivo de avanzar en la comprensión de la transición y la fase vítrea. Se centra en un tipo de sistema 
vítreo particular, los vidrios de espín. A pesar de la sencillez de su modelización, preguntas fundamentales, como la 
naturaleza de su fase de baja temperatura en tres dimensiones, aun siguen sin contestar.

La tesis contiene una introducción y resultados originales. La introducción se inicia con una presentación muy general 
a los sistemas vítreos. A continuación se introducen los vidrios de espín a través de una breve reseña historiográfica. 
Se pasa entonces a recordarle al lector ciertos conceptos básicos necesarios para seguir con comodidad el resto del manuscrito, 
como los observables relevantes en simulaciones Monte Carlo, la fenomenología de las transiciones del segundo órden, el \emph{scaling}, 
la universalidad y el grupo de renormalización.  

Pasemos ahora a describir los resultados originales. Se estudian aspectos diferentes de estos sistemas, con un enfoque
principalmente numérico. La disertación se divide en dos partes. En la primera parte de la tesis se 
hacen simulaciones de Monte Carlo de equilibrio, en búsqueda de propiedades críticas del vidrio de espín. Estas simulaciones 
han requerido recursos computacionales extraordinarios, como el ordenador dedicado Janus, y el supercomputador chino Tianhe-1a. 
La segunda parte de la tesis se centra en estudiar el paisaje de energía, que desempeña un papel preponderante en el crecimiento 
de los tiempos de relajación de los vidrios. La dificultad en el estudio del paisaje de energía se halla en la descripción de un 
espacio con un número divergente de dimensiones. La costumbre es describirlo a través de un único número, la energía. Esta 
simplificación no siempre es viable, por lo que es necesario recurrir a otros descriptores.

El capítulo \ref{chap:eah3d} presenta los resultados de nuestra primera campaña de Monte Carlo. La pregunta fundamental se refiere 
a la posibilidad de encontrar una transición de fase \emph{spin glass} en presencia de un campo magnético. 
Las dos principales teorías sobre la fase de baja temperatura tienen predicciones diferentes, así que 
entender el comportamiento bajo un campo magnético comportaría probablemente entender la naturaleza de la fase de baja temperatura. 
El estudio de Monte Carlo muestra que hay unas fluctuaciones tan grandes en los valores de los observables, que la media ya no es 
un buen descriptor del comportamiento colectivo. La búsqueda de buenos descriptores hace necesario desarrollar nuevos métodos estadísticos. 
Tras introducir una variable de condicionamiento adecuada (una \emph{conditioning variate}) se estudian funciones de correlación condicionadas al valor de 
dicha variable. Se hallan comportamientos muy diferentes, según el percentil de la variable de control considerado: los resultados de 
algunos percentiles sugieren la existencia de una fase vítrea (en presencia de campo), mientras en otros percentiles no se detectan señales
de una transición. 
No es posible discernir cual de los dos comportamientos dominaría en el límite termodinámico, pero se localiza el rango de temperaturas 
donde debería encontrarse la transición de fase, si la hubiese.  

El segundo trabajo de equilibrio (capítulo \ref{chap:ahsg}) se propone estudiar la transición de fase del vidrio de espín de 
Heisenberg con anisotropías aleatorias. Según el escenario de Kawamura, el canal quiral y spin glass se acoplan al introducir una 
anisotropía en el modelo. Se halla la transición de fase para cada uno de los parámetros de orden. Tras un cuidadoso análisis de los 
efectos de volúmen finito se concluye que la transición de fase es única, y que su clase de universalidad es la de Ising en lugar de 
Heisenberg, por lo cual la anisotropía es una perturbación relevante en el sentido del grupo de renormalización.

En la segunda parte de la tesis se presentan tres trabajos.
El capítulo \ref{chap:hsgm} es con un estudio de la dependencia del paisaje de energía en el número de componentes $m$ de los espines. 
Cuando $m$ es pequeño el paisaje de energía es complejo y rugoso con muchos mínimos locales, que van desapareciendo al crecer de $m$. 
También se oberva como el crecimiento de induce un crecimiento de las correlaciones. En consecuencia, la dinámica es tanto más 
lenta cuanto mayor es $m$.

En el capítulo \ref{chap:marginal} se examinan los procesos de histéresis que aparecen en el modelo de Sherrington y Kirkpatrick. 
La dinámica en los ciclos de histéresis se produce en forma de avalanchas de espines. Estas avalanchas con invariancia de escala no 
pueden ocurrir con interacciones de corto alcance. De hecho, las interacciones de largo alcance son una perturbación relevante en el 
Hamiltoniano de corto alcance. Durante estas avalanchas, además, se producen correlaciones entre espines de baja estabilidad, que 
tienden a ponerse en configuraciones frustradas entre sí, generando espontáneamente estabilidad marginal en el sistema.

Por último, en el capítulo \ref{chap:hsgrf} se presenta un estudio de los modos de baja frecuencia en el vidrio de espín de Heisenberg. En efecto, en este régimen de 
frecuencias la densidad de estados tiene un comportamiento con ley de potencia diferente del de Debye, indicando la presencia de un 
\emph{boson peak}. Ésta es una característica típica de los vidrios estructurales. Estos \emph{modos blandos}, además, son localizados, y conectan 
estados muy cercanos. La barrera energética  que separa estas parejas de estado es muy baja 
(no crece con el tamaño del sistema). Todo esto sugiere la identificación de estas parejas de estados con \emph{two-level systems} clásicos. 
Se encuentra así una conexión entre las 
dos principales teorías que explican el \emph{boson peak}. Por un lado tenemos la teoría de las réplicas que, en aproximación de campo medio, 
predice  que estos modos blandos se deben a un paisaje de energía fractal. Por otro lado, la teoría de los \emph{two-level systems} atribuye 
el boson peak al efecto tunel cuántico entre estados cercanos.

\section*{Riassunto in italiano \label{sec:riassunto}}
\sectionmark{Riassunto in italiano}
\addcontentsline{toc}{section}{\protect\numberline{}Riassunto in italiano}

\index{struttura inerente|see{inherent structure}}
\index{gruppo di rinormalizzazione|see{renormalization group}}
\index{transizione vetrosa|see{glass transition}}
\index{fase vetrosa|see{glass phase}}\index{paesaggio di energia|see{energy landscape}} 
\index{minimo locale|see{local minimum}}
\index{stabilit\`a marginale|see{marginal stability}}\index{universalit\`a|see{universality}}
\index{chiralit\`a|see{chirality}}\index{modi soffici|see{soft modes}}

L'obiettivo di questa tesi \`e di fare un passo avanti nella comprensione della fase vetrosa.
Ci si concentra in un tipo di sistema in particolare, i vetri di spin.
Nonostante la loro modellizzazione sia molto semplice, domande fondamentali, come
la natura della fase a bassa temperatura in tre dimensioni, ancora non trovano risposta.

Il testo è diviso in introduzione e risultati originali. L'introduzione comincia con una presentazione
generale dei sistemi vetrosi. In seguito si introducono i vetri di spin con
una breve rassegna storiografica sulla loro origine.
Si ricordano poi al lettore dei concetti basici necessar\^i per poter seguire comodamente
la trattazione, dalle osservabili rilevanti in una simulazione di Monte Carlo, alla fenomenologia
delle transizioni di fase di secondo ordine, allo \emph{scaling}, fino al gruppo di rinormalizzazione.

In quanto ai risultati originali, si studiano i vetri di spin sotto differenti punti di vista, con un approccio
principalmente numerico. L'esposizione è divisa in due parti.
Nella prima si fanno simulazioni Monte Carlo di equilibrio, alla ricerca di propriet\`a critiche
dei vetri di spin. Per entrambi i lavori all'equilibrio sono state necessarie risorse computazionali straordinarie,
come il computer dedicato \textsc{Janus} e il supercomputer cinese \emph{Tianhe-1a}.
La seconda parte di risultati originali 
\`e centrata nello studio del paesaggio di energia,
che sembra avere un ruolo fondamentale
nella crescita dei tempi di rilassamento dei vetri. Il paesaggio di energia \`e uno spazio con un numero divergente
di dimensioni che solitamente si descrive per mezzo di un unico numero, l'energia. Questa semplificazione \`e talvolta
eccessiva ed \`e necessario ricorrere a descrittori differenti.

Nel capitolo \ref{chap:eah3d} si espongono i risultati della prima di due campagne di Monte Carlo.
Si cerca di comprendere se in presenza di un campo magnetico esterno sussiste una transizione dalla fase paramagnetica
alla fase \emph{spin glass}.
Le due principali teorie sulla fase di bassa temperatura
hanno predizioni diverse, per cui comprendere il comportamento sotto un campo magnetico implicherebbe probabilmente
una cognizione della natura della fase a bassa temperatura.
Si trova che le fluttuazioni delle
osservabili sono cos\`i forti che la media non \`e un descrittore affidabile del comportamento collettivo.
Per questo motivo diviene necessario sviluppare dei nuovi metodi statistici in modo da avere dei buoni descrittori.
Troviamo comportamenti molto differenti: alcune delle misure suggeriscono la presenza di una transizione di fase, 
mentre altre no. Non si riesce a discernere quale dei due comportamenti dominerebbe nel limite termodinamico,
ma si localizza il rango di temperature in cui dovrebbe trovarsi la transizione di fase nel caso in cui fosse presente.

Il secondo lavoro all'equilibrio, esposto nel capitolo \ref{chap:ahsg}, si propone di studiare la transizione del vetro di spin di Heisenberg con delle anisotropie aleatorie.
Secondo lo scenario di Kawamura, l'introduzione dell'anisotropia del modello induce che il canale chirale 
 e quello spin
glass si accopp\^ino. Viene trovata una transizione di fase per ognuno dei parametri d'ordine, e in seguito a una meticolosa
analisi degli effetti di taglia finita si conclude che la transizione di fase \`e unica. 
Inoltre, le quantit\`a universali 
della transizione sono compatibili con la classe di universalit\`a di Ising invece che di Heisenberg, indicando che 
l'anisotropia \`e una perturbazione rilevante nel senso del gruppo di rinormalizzazione.

Nella seconda parte si presentano tre lavori. Il capitolo \ref{chap:hsgm}
è uno studio della dipendenza del paesaggio di energia dal numero di componenti $m$ degli spin.
Quando $m$ \`e piccolo
il paesaggio \`e complesso e rugoso, con una gran quantit\`a di minimi locali, che 
per\`o scompaiono al decrescere di $m$.

A meno di effetti di taglia finita, la crescita di $m$ induce anche un incremento delle correlazioni, e un conseguente rallentamento della dinamica.

Nel capitolo \ref{chap:marginal} si esaminano i processi di isteresi del modello di Sherrington e Kirkpatrick. 
La dinamica nel ciclo di isteresi avviene sotto forma di valanghe si spin. 
Queste valanghe non posso esserci in sistemi con interazioni a corto raggio, e le interazioni a lungo raggio sono
una perturbazione rilevante in un Hamiltoniano a corto raggio. Durante queste valanghe, inoltre, si generano autonomamente
delle correlazioni tra gli spin poco stabili, i quali tendono a mettersi in configurazioni mutuamente frustrate, portando
spontaneamente il sistema a configurazioni marginalmente stabili.

Infine, nel capitolo \ref{chap:hsgrf} si presenta uno studio
dei modi soffici  del vetro di spin di Heisenberg sotto un campo magnetico aleatorio, che viene imposto
per eliminare i modi di bassa energia dovuti alla simmetria rotazionale. 
Il comportamento a bassa frequenza della densit\`a
degli stati \`e differente da quello tipico di Debye, indicando la presenza di un \emph{boson peak}, caratteristica tipica
dei vetri strutturali. Questi modi soffici, inoltre, sono localizzati e connettono stati molto vicini separati da barriere
assai piccole, che identifichiamo come versioni classiche del \emph{two-level system}.
Questo aiuta a trovare una connessione tra le due principali teorie che spiegano il \emph{boson peak}. Da un lato c'\`e
la teoria delle repliche, che mostra in approssimazione di campo medio che questi modi soffici sono dovuti a un paesaggio
di energia frattale, e dall'altro c'\`e quello dei \emph{two-level systems}, che attribuisce il \emph{boson peak} all'effetto
tunnel quantistico tra stati vicini.

\cleardoublepage

 \pagenumbering{arabic}
%\renewcommand{\thepage}{\texorpdfstring{\oldstylenums{\myThePage}}{\myThePage}}
%\cbcolor{red}

\part{Introduction\label{part:intro}}
\chapter{Background \label{chap:intro}}
After briefly introducing the glass phase in general terms, showing how it appears in many aspects of modern society,
we make a historical presentation on birth and evolution of the \ac{SG} theory.
It is hard to propose oneself a historical approach on a research topic, since any quoted argument could need a whole treatise for
itself, so we choose the starting point that looked mostly appropriate to us, and refer to an exhaustive bibliography the interested reader.
\footnote{
In particular, in \cite{mattis:81} there is an extended historical introduction on magnetism (but not on \acsp{SG}).
Historical comments on \acsp{SG} appear in \cite{mydosh:93}; a perspective is given in \cite{sherrington:07}.
}
Moreover, since the \ac{SG} theory has by now evolved over half a century under disparate aspects, and it has fused with many
other domains of science, such as biology and computer science, it is unthinkable to use this introduction to mention all the aspects of
this stimulating branch of physics.  We will instead focus on the origin of \acp{SG} as they are known at present, and we will only touch 
on  those aspects of \ac{SG} theory that are useful to expose the results of this thesis.
\footnote{
The references herein come from an intensive bibliographic research, and are in the author's opinion the most representative of a part of 
the history of \acp{SG}. 
It may occur to the reader that some notable publication or remark, that should appear in this thesis, has been not been cited. If this were
the case, the author would thank such reader if he could inform him in order to add the missing work to further versions of this introduction.
}
Since its aim is to get into the topic and set the bases for further discussion,
the introduction on \acp{SG} is left open, and recent developments are left to the introduction of each chapter.

\section{The glass transition \label{sec:glass-transition}}
\begin{figure}[!t]
\centering
\includegraphics[width=0.7\textwidth]{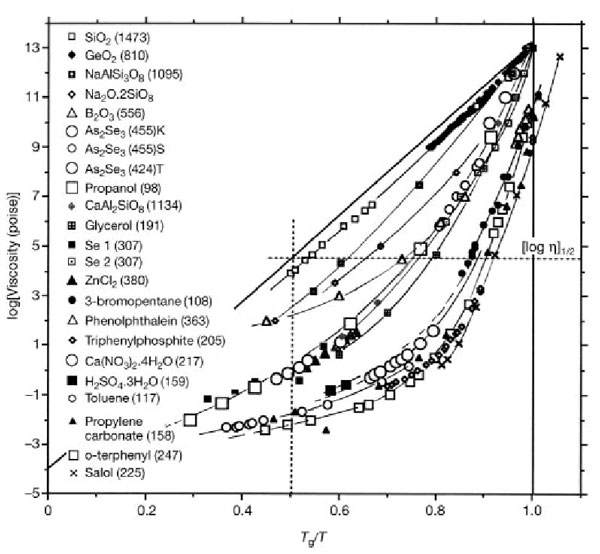}
\caption[Angell plot]{\index{Angell plot}\index{viscosity}\index{relaxation time}
Logarithm of the viscosity $\eta$ against the inverse of the temperature $T$, normalized with a constant $T_\mathrm{g}$. $T_\mathrm{g}$ is the temperature
where the viscosity is $10^{13}$ poise
(with the exception of some curves that do not meet at $T_\mathrm{g}/T=1$ because $T_\mathrm{g}$ 
was defined as the temperature where the enthalpy relaxation time is $\approx200s$).
It represents an experimental cutoff over which the relaxation times $t\sim\eta$ are too long to perform equilibrium
experiments.
On the other hand, $\eta=10^{-4}$ poise is the roughly common high-temperature limit of the viscosity.
Figure from \cite{martinez:01}.}
\label{fig:angell}
\end{figure}

\index{glass!transition|(}\index{glass!transition@transition|seealso{spin glass transition}}\index{glass!phase|(}\index{glass!phase@phase$ $|seealso{spin glass phase}}
\index{glass!structural}\index{supercooled liquid@supercooled liquid$ $|seealso{structural glass}}
If we cool a liquid quickly enough, it can happen that the sudden lack of thermal vibration arrest its dynamics before
it is able to end in the lowest-entropy configurations and crystallize. Once this happens, a glass is formed, and the material behaves
as a solid even though apparently no symmetry was broken and no phase transition took place. Simply, the viscosity and the relaxation 
times \index{viscosity}\index{relaxation time} grow so fast in a very short range of temperatures, 
that the liquid stops flowing and appears solid. In figure \ref{fig:angell}, a 
famous plot by Angell \index{Angell plot} shows this steep behavior in a set of glass formers. With a factor 2 change in temperature the viscosity grows 
8-11 orders of magnitude.
A so large growth of the relaxation times is hard to explain in the absence of a phase transition, and no
completely satisfying theory has been found. So many scenarios have been proposed to explain this phenomenon, that
is it often said that there are more glass theories than theorists.
\index{glass!transition|)}

Besides the natural interest in amorphous solid states, called structural glasses, the reason why much emphasis is put in the study of the glass transition
is the huge amount of applications that glasses have, and the immense amount of disparate systems that exhibit a glassy state.

The most commonly known glasses are silica compounds. \index{glass!structural} They are fused to a temperature where the viscosity is low and they are malleable,
and the glassy phase is obtained by quickly taking them back to room temperature. For their properties of manufacturability, low dilatancy,
uncorrosiveness and transparency they are present in many objects of our everyday life, such as windows, bottles, optical fibers, beakers 
and touchscreens.\footnote{Devices such as tablets and smartphones require high-tech glasses. The recently-developed Gorilla Glass (http://www.corninggorillaglass.com/),
for example, enjoys wide popularity.}

Still, the glassy phase presents itself in numerous different forms in technology and nature \cite{angell:95}. 
Metallic glasses \index{glass!metallic} are used for high efficiency transformers for their magnetic properties, or as an alternative to silicon 
to make molds for nanocomponents \cite{greer:95}. 
Automobile bodies and parts of boats are made of fiberglass, \index{glass!fiber} that is obtained by embedding extremely fine fibers of glass in an organic polymer plastic,
trapping the air in order to make it a good thermal insulator \cite{mayer:93,marsh:06}.
Vitrification takes places in processes related to the stabilization of labile biochemicals for commercial use \cite{crowe:98},
and in the preservation of insect life under extreme conditions of cold or dehydration \cite{crowe:98}.
Protein folding \index{protein folding} exhibits glass-like behavior \cite{weber:13},
many foods and their industrial production chain involve glassy states and dynamics \cite{blanshard:93},
and so do instances of optimization and combinatorial problems \cite{mezard:87}.

\index{spin glass|(}
Spin glasses are yet another instance of the glassy phase, characterized by an amorphous magnetic low-temperature state. Despite a very peculiar
phenomenology \cite{nagata:79,mydosh:93,vincent:97,jonason:98,herisson:02}, few or none industrial applications of \acp{SG} exist at the moment,
and it would be reasonable to query why \acp{SG} are apparently overrepresented in theoretical physics.

The main reason is due to their simplicity. Very simple Hamiltonians defined on uncomplicated graphs capture highly non-trivial behaviors, making them 
probably the most understandable models that display a glassy phase. 
Their study is useful to get an insight on the study of the glass phase in a more general sense and on complexity, since
\begin{itemize}
 \item experimental measurements are easier through the use of very sensitive magnetometers called \index{SQUID} 
	SQUIDs (Superconducting QUantum Interference Devices). See e.g. \cite{drung:07,kumar:14}.
 \item in the context of \acp{SG} it was possible to develop very advanced theoretical tools that can be reused in other contexts \cite{mezard:87,biazzo:12,charbonneau:14}.
 \item differently from structural glasses, the \ac{SG} transition is well identified in finite dimensions \cite{ballesteros:00,lee:03}.
 \item they are easier to simulate, because e.g. they are defined on graphs where the neighbors do not change with time, the degrees of freedom are binary or limited. It
 is possible to simulate far more degrees of freedom than on structural glass, making finite-sizes effects less 
 overwhelming \cite{fernandez:15}.
 \item it is possible to construct dedicated hardware for more effective numerical studies \cite{janus:06,janus:08,janus:12b,janus:14}.
\end{itemize}

Finally, \acp{SG} are often used as toy models to test the phenomenology of more complicated systems, and not seldom \ac{SG} theory was of crucial importance for relevant
advances in numerous fields. For example 
the Random First Order Transition theory \index{random!first order transition}\index{glass!structural} 
for structural glasses is inspired on the $p$-spin \ac{SG} model \index{pspin model@$p$-spin model}\cite{cavagna:09} (section \ref{sec:spin-glass-intro});
neural networks are now a branch of \ac{SG} theory, and for example the Hopfield model is known to display a \ac{SG} \index{Hopfield model}
phase and is studied with \ac{SG} tools \cite{mezard:87};
protein folding codes can be successfully obtained with \ac{SG} theory \cite{goldstein:92}, and many ideas 
from \acp{SG}es were used to understand this phenomenon \cite{wolynes:92}. 
In this dissertation we use \acp{SG} to understand marginal stability \index{marginal stability} and two-level 
systems \index{two-level system} (chapters \ref{chap:hsgrf} and \ref{chap:marginal}).

\index{glass!phase|)}
\index{spin glass|)}\index{glass!spin@spin|seealso{spin glass}}

\section{The origins of spin glass theory\label{sec:spin-glass-intro}}
During the beginning of the second half of the $20^\mathrm{th}$ century much attention has been devoted to the study of solutions
of \acf{Mn} in copper \acf{Cu}, that displayed peculiar properties that puzzled the condensed-matter community \cite{owen:56, denobel:59, zimmerman:60}.
A cusp in the susceptibility \index{susceptibility!cusp} was observed at a temperature $T_\mathrm{c}$ roughly proportional to 
the concentration of \ac{Mn} (with concentration of 0.1-10\%
$T_\mathrm{c}$ ranged between 1K and 100K), separating the paramagnetic phase \index{phase!paramagnetic} from a peculiar 
phase in which no order was identified,\index{phase!spin glass|see{spin glass!phase}}\index{spin glass!phase} though several features
discriminated it from a paramagnetic phase. It lacked
spontaneous magnetization, but after applying reasonably large fields one could observe remnant magnetization. Also, 
the susceptibility $\chi$\nomenclature[chi....]{$\chi$}{susceptibility} was practically constant instead of 
being inversely proportional to the temperature $T$, \nomenclature[T...0]{$T$}{temperature}
$\chi\propto1/T$ as the Curie law suggests for a paramagnet, \index{Curie law} and
the low-temperature specific heat was linear in $T$ instead of being proportional to $1/T^2$. %C=dQ/dT=d(U-\chi dH)/dT. U=3kT/2, \chi=1/T.

This surprising low-temperature behavior was attributed to the $s-d$ interaction \cite{marshall:60}, \index{sd interaction@$s-d$ interaction|(} 
that couples electrons of unfilled inner shells and conduction electrons. 
Depending on the involved metal, this interaction can lead both to ferromagnetism and antiferromagnetism. 
\index{ferromagnetism}\index{antiferromagnetism}
In order to explain the atypical ordered phase 
the $s-d$ interaction was supposed to be the dominant one.

This interaction was first pointed out by Zener in 1951, with a phenomenological model that did not involve the possibility of antiferromagnetism
\cite{zener:51, zener:51b, zener:51c}. Few years later the theory was further developed by Kasuya \cite{kasuya:56}, that found that 
the $s-d$ interaction can imply antiferromagnetism \index{ferromagnetism}\index{spin!wave}
and spin waves, and Yosida \cite{yosida:57}, that notices that the model from Ruderman and Kittel \cite{ruderman:54}, for the coupling between 
two magnetic moments through their hyperfine interaction with the conduction electrons, successfully describes the $s-d$ interaction.
\footnote{
A Hamiltonian for the $s-d$ interaction is also derived in \cite{mitchell:57}. More useful references on the subject: \cite{frohlich:40,bloembergen:55,vanvleck:62,mattis:81}.
}
The resulting coupling $J_{\bx\by}^\mathrm{(RKKY)}$ \index{RKKY interaction|(} between two \ac{Mn} ions 
separated by $\br$ resulting from this description is called \acf{RKKY}. It has a 
sinusoidal form that to our purposes can be represented as a pairing
\begin{equation}\label{eq:rkky}
 J_{\bx\by}^\mathrm{(RKKY)}\sim\cos{\left(\frac{\bk\cdot\br}{|\br|^3}\right)}\,,
\end{equation}
between two spins $\vec{s}_\bx$ and $\vec{s}_\by$ at distance $\br$ one from the other.
The $\bk$ is of the order of the Fermi vector, meaning that the oscillations of the cosine are very quick.
So, expression (\ref{eq:rkky}) tells us that, besides decaying as $1/r^3$, depending on the distance between 
the ions the couplings can be ferromagnetic or antiferromagnetic.

The interactions of the \ac{Cu} substrate were assumed negligible for the study of the magnetic properties of the examined CuMn alloys, and the cusp in the
susceptibility \index{susceptibility!cusp} was entirely attributed to the \ac{RKKY} interaction between the \ac{Mn} ions \cite{marshall:60}. 
Being the positions in the alloy of these ions random, both the module and the sign
of the couplings had to be treated as a random variable, and random ferromagnets \index{random!ferromagnet} 
became popular \cite{brout:59}. First modelizations involved systems
of spins under independent effective random local fields \cite{marshall:60,klein:63}, and later on disorder is assumed in the interactions \cite{montgomery:70}.
\index{sd interaction@$s-d$ interaction|)} 
\index{RKKY interaction|)}

\paragraph{The birth of spin glass theory.}\index{spin glass!theory|(}
The term \emph{spin glass} is first used in a paper by Anderson in 1970,
\footnote{Under suggestion of B.R. Coles.}
in analogy with structural glasses, to stress the presence of a low-temperature phase with unidentified order.
He defines a formally simple model where the Hamiltonian has an explicit dependence on the disorder \cite{anderson:70}.
He assumes that the dominant role is not assumed by the electrons, that have only the function of transmitting the interaction, 
but by the \ac{Mn} ions and their exchange interactions.
The interaction between the \ac{Mn} spins is given by the \ac{RKKY} interaction (\ref{eq:rkky}), \index{RKKY interaction}
whose sign depends on the distance $\br_{\bx\by}$ between two spins $\vec{s}_\bx$ and $\vec{s}_\by$ \nomenclature[s....x]{$\vec{s}_\bx$}{vector spin on site $\bx$}
and that decreases in magnitude as $\br_{\bx\by}$ increases. Since $\br_{\bx\by}$ is random and 
depends on the single realization of the alloy and of its disorder, that we will call 
\nomenclature[s.ample]{sample}{a single realization of the disorder}\emph{sample}, 
also the coupling $J_{\bx\by}$ is a random variable. So, Anderson proposed
the first \ac{SG} Hamiltonian as a Heisenberg model \index{spin glass!Heisenberg}
\begin{equation}\label{eq:H-anderson}
 \mathcal{H} = \frac{1}{2}\sum_{\bx\neq\by} J_{\bx\by}\vec{s}_\bx\cdot\vec{s}_\by\,,
\end{equation}
where the $J_{\bx\by}$ are random constants distributed through an unknown distribution that should reproduce roughly the \ac{RKKY} interaction. 
The essential novelty is thus that the ``experimental'' couplings $J_{\bx\by}^\mathrm{(RKKY)}$ are replaced by the random variables $J_{\bx\by}$.
We call 
\nomenclature[q.uenched disorder]{\small quenched disorder}{the disorder that appears in the Hamiltonian}\index{quenched!disorder} \emph{quenched disorder} 
the randomness of the $J_{\bx\by}$s, that appears directly in the Hamiltonian. Notice that being the couplings $J_{\bx\by}$ randomly negative
and positive, it is impossible to satisfy simultaneously the energy along all the bonds (we will come back to this later on). This feature is called 
\index{frustration}\emph{frustration}.
Hamiltonian (\ref{eq:H-anderson}) possesses both quenched disorder and frustration, that become the distinctive features of a \ac{SG} model \cite{young:05,kawamura:10}.
Anderson tried a mean field approach without, yet,
averaging over the disorder. He also assumed the possibility of purely nearest-neighbor interactions on a regular lattice, 
and treated the system as a set of independent clusters each with
its critical temperature, bringing back the problem of localization that in his view had been disregarded. \index{localization}\index{Anderson localization|see{localization}}
This cluster-based interpretation was well embraced by the scientific community. Experimental observations of the susceptibility cusp \index{susceptibility!cusp}
were done also in other
types of alloy such as AuFe, with similar results. The dominant interpretation was an arisal of ferro- and antiferromagnetic clusters with short-range order
that as the temperature is lowered interact at long range \cite{beck:71,cannella:72,smith:74}, 
or seeing the \ac{SG} as a sort of macroscopic antiferromagnet \cite{adkins:74}.
\footnote{This latter interpretation tried to explain the rounding of 
in the cusp of the susceptibility under an applied magnetic field. As we will discuss more 
thoroughly in chapter \ref{chap:eah3d}, it is still an open issue whether a \ac{SG} in a field undergoes
a phase transition.\index{de Almeida-Thouless}\index{transition in a field|see{de Almeida-Thouless}}}

\paragraph{The Edwards-Anderson model.}\index{spin glass!Edwards-Anderson|(}
The milestone year for the definition of \acp{SG} as a branch of theoretical physics is 1975. 
A solid basis on \ac{SG} theory was given in \cite{edwards:75,edwards:76} by
Edwards and Anderson through a very simple model that was able to describe qualitatively the experimental observations. 
Their starting idea is that in the low temperature \emph{spin glass phase} \index{spin glass!phase}
there must be some local ordering of the spins along a random preferred direction. 
Even though this direction is unknown, one can see whether an alignment is taking place by examining if after a 
time $t$ the single spins $s_\bx(t)$ have a tendency of pointing in 
the same direction. In quantitative terms, they define the 
\index{overlap}\nomenclature[q....]{$q$}{overlap}\emph{overlap}
\begin{equation}\label{eq:q-limt}
 q = \lim_{t\to\infty}\frac{1}{N} \sum_\bx^N \mean{\vec s_\bx(0)\cdot\vec s_\bx(t)}_t\,,
\end{equation}
where 
\nomenclature[...]{$\mean{\ldots}_t$ }{time average}$\mean{\obs(t)}$ is the time average of a 
generic observable $\obs$, \nomenclature[O...0]{$\obs$}{generic observable}
$\mean{\obs(t)}_t\equiv\frac{1}{t}\int_0^t dt' \obs(t')$.
Equation (\ref{eq:q-limt}) is one of several ways to define the order parameter of a \ac{SG}. 
Assuming that the equilibrium phase is ergodic, one can rewrite equation (\ref{eq:q-limt})
by replacing the time average $\mean{\ldots}_t$ with an 
\nomenclature[...]{$\mean{\ldots}$}{In the other chapters: thermal average} ensemble average $\mean{\ldots}$ to give an alternative expression for the overlap,
\begin{equation}\label{eq:q-ensemble}
q = \frac{1}{N} \sum_\bx^N \mean{\vec s_\bx}^2\,. 
\end{equation}
In the paramagnetic phase there is no favored direction, so $q=0$. On the other side, \index{phase!paramagnetic}
in the \ac{SG} phase each spin will align along a privileged direction and $q\neq0$.
In \cite{edwards:75} Hamiltonian (\ref{eq:H-anderson}) is taken into account and it is shown with a mean field 
approach that a phase transition occurs with $q$ as order parameter,
accompanied by a cusp in the susceptibility.\index{susceptibility!cusp}
Hamiltonian (\ref{eq:H-anderson}), with nearest neighbor interactions on a regular lattice, 
assumes the name of \ac{EA} model. Assuming a unitary distance between nearest neighbors,
the \ac{EA} Hamiltonian is
\begin{equation}\label{eq:H-EA}
\index{spin glass!Edwards-Anderson!Hamiltonian}
\nomenclature[H..EA]{${\cal H}_\mathrm{EA}$}{Edwards-Anderson Hamiltonian}
 \mathcal{H}_\EA = \frac{1}{2}\sum_{\norm{\bx-\by}=1} J_{\bx\by}\vec{s}_\bx\cdot\vec{s}_\by\,,
 \footnote{It is the case to make clarity on the notation for the summations.
 \nomenclature[sigma...xy]{$\sum_{\bx,\by}$}{sum over all the $\bx$ and $\by$} $\sum_{\bx,\by}$ is a sum over all the choices of $\bx$ and $\by$.
 \nomenclature[sigma...xy]{$\sum_{\bx\neq\by}$}{sum over all the $\bx$ and $\by$, except $\bx=\by$}$\sum_{\bx\neq\by}$ is a sum over all the choices of $\bx$ and $\by$, except $\bx=\by$ (in our models the positions $\bx$ are discretized).
 \nomenclature[sigma...xy]{$\sum_{\norm{\bx-\by}=1}$}{sum over all the nearest neighbor $\bx$ and $\by$} $\sum_{\norm{\bx-\by}=1}$ is a sum over all the choices of $\bx$ and $\by$ that are nearest neighbors.
 In all the previous cases each coupling is counted twice, so we put a factor 1/2 in front of the summation.
 \nomenclature[sigma...yxy]{$\sum_{\by:\norm{\bx-\by}=1}$}{sum over all the $\by$ neighbors of $\bx$}$\sum_{\by:\norm{\bx-\by}=1}$ is a sum over all the choices of $\by$ that are neighbors of $\bx$, so the summation runs 
 over a number of terms equal to the connectivity 
 \nomenclature[z....z]{$z$}{In the other chapters: connectivity}$z$. Writing $\sum_{\norm{\bx-\by}=1}$ is equivalent to $\sum_\bx\sum_{\by:\norm{\bx-\by}=1}$.
 }
\end{equation}
where for simplicity reasons the $J_{\bx\by}$ were assumed by Edwards and Anderson to come from a Gaussian \ac{pdf} $P(J)$. Different samples of an \ac{EA} spin glass will have a different realization
of the coupling, but on average they must have the same behavior, and the larger the lattice more similar the behavior will be. This assumption, that gives sense to the free energy of the \ac{SG} 
model, is called \index{self averageness}\emph{self averageness}. 
So, calling $\mathcal{F}_J$ \nomenclature[F..J]{$\mathcal{F}_J$}{sample-dependent free energy}\index{energy!free}
and $\Z_J$ \nomenclature[Z...J]{$\Z_J$}{sample-dependent partition function}\index{partition function}
the free energy and the 
partition function of a sample with a set $J$ of couplings, one is interested in the average free energy
\begin{equation} \label{eq:Fsg}\index{energy!free}
 \mathcal{F} = \int \mathcal{F}_J P(J) dJ = -k_\mathrm{B} T\int P(J) \log{\Z_J} dJ \,,
 \nomenclature[F..999]{$\mathcal{F}$}{In the other chapters: free energy}
 \end{equation}
that by writing with an over bar 
\nomenclature[...]{$\overline{(\ldots)}$}{average over the disorder}$\overline{(\ldots)}$ 
the average of the disorder assumes the form $\mathcal{F}=-k_\mathrm{B} T \,\overline{\log{\Z_J}}$.
Equation (\ref{eq:Fsg}) encloses a central difficulty in \ac{SG} theory, that is taking the average of the logarithm of $\Z_J$. This is called a 
\nomenclature[a.verage quenched]{\small average, quenched}{$\overline{\log(\Z_J)}$.} \index{quenched!average} \emph{quenched average}, 
in opposition with the easier approach, called 
\nomenclature[a.verage annealed]{\small average, annealed}{$\log(\overline{\Z_J})$.} \index{annealed average}\emph{annealed average},
of taking the logarithm of the average of $\Z_J$, resulting in the annealed free energy $\mathcal{F}_\mathrm{Ann}=\log{\overline{\Z_J}}$, 
that results incorrect at low temperatures (see e.g. \cite{mezard:87}).
To overcome the problem of this integration, Edwards and Anderson propose the 
\index{replica!theory|(}\index{replica!trick}\emph{replica trick}, that consists 
in using the identity $\displaystyle\log{(x)}=\lim_{n\to0}\frac{x^n-1}{n}$ to transform the annoying logarithm in a power law,
\footnote{The identity comes from a first order expansion of the exponential function: $x^n=\E^{n\log{(x)}}=1+n\log{(x)}+o(n^2)$.}
\begin{equation}
 \mathcal{F} = -k_\mathrm{B}T\overline{\log\Z_J} = -k_\mathrm{B}T\lim_{n\to0}\frac{\overline{\Z^n}-1}{n}\,.
\end{equation}
By artificially assuming that $n$ is an integer, one could think about $\overline{\Z^n}$ as the partition function of $n$ independent 
\index{replica}\nomenclature[r.eplicas]{replicas}{independent copies of the same sample}\emph{replicas} 
of the same system, that share the same instance of the couplings but are independent one from the other. With the help of
replicas the order parameter can be rewritten as \cite{parisi:83}
\begin{equation}\label{eq:q-replicas}\nomenclature[q....ab]{$q^{\rma\rmb}$}{overlap matrix}\index{overlap!matrix}
 q^{\rma\rmb} = \mean{\vec s_\bx^{(\rma)} \cdot \vec s_\bx^{(\rmb)}}\,.
\end{equation}
where $(\rma)$ and $(\rmb)$ indicate different replicas.\nomenclature[a....bcd]{$\rma,\rmb,\rmc,\rmd$}{replica indices}
Treating $\overline{\Z^n}$ as a set of independent replicas simplifies the calculations, although it implies
a few mathematical forcings such as taking the limit $n\to0$ with $n\in\mathbb{N}$.
Notwithstanding, although the \ac{EA} model still nowadays lacks a full analytical understanding, 
the replica trick became a very popular tool for disordered systems.

The \ac{EA}  model was promptly be extended to quantum spins \cite{sherrington:75b,fischer:75}, but we will not treat quantum \acp{SG} in this thesis, so we will
leave these models aside.
\index{spin glass!Edwards-Anderson|)}

\paragraph{The Sherrington-Kirkpatrick model.}\index{spin glass!Sherrington-Kirkpatrick|(}
Also in 1975, with the aim of giving a model for which mean field theory be valid, 
Sherrington and Kirkpatrick propose to slightly modify Hamiltonian (\ref{eq:H-EA}) by
imposing fully-connected interactions and Ising spins $s_\bx=\pm1$ \cite{sherrington:75} 
\index{spin!Ising}\nomenclature[s....x]{$s_\bx$}{Ising spin on site $\bx$}
\begin{equation}\label{eq:H-SK}\index{spin glass!Sherrington-Kirkpatrick!Hamiltonian}
  \mathcal{H}_\mathrm{SK} = \frac{1}{2}\sum_{\bx,\by} J_{\bx\by}\vec{s}_\bx\cdot\vec{s}_\by\,,
\end{equation}
where the couplings $J_{\bx\by}$\nomenclature[J...xy]{$J_{\bx\by}$}{coupling between sites $\bx$ and $\by$}
are Gaussian distributed with $\overline{J_{\bx\by}}=0$, and their variance is such that the energy is extensive, $\overline{J^2}=1/N$.
This model, for which mean field theory is valid, will be called \ac{SK} model.
Their solution, yet, has unphysical features such a negative entropy at low temperatures.
Sherrington and Kirkpatrick attributed this to an assumption they made, in their calculations, of commutativity between the limit $n\to0$ and the 
thermodynamic limit $N\to\infty$ ($N$ indicates the number of spins). Yet, it slowly became clear that the problem resided in the (yet reasonable) ansatz they made
of \emph{replica symmetry} \cite{dealmeida:78,bray:78},\index{replica!symmetry}
that the overlap (\ref{eq:q-replicas}) is the same no matter what two replicas are chosen [the \acf{RS} ansatz],
\begin{equation}\index{replica!symmetry!ansatz}
 q^{\rma\rmb} = q (1-\delta^{\rma\rmb})\,,
\end{equation}
especially after it was shown that in the \ac{SK} model the inversion of the limits is valid \cite{vanhemmen:79}.

It is worth to mention also another interesting model with disorder proposed in 1975, 
the Random Field Ising Model \cite{imry:75},\index{random!field Ising model}
\footnote{We will take inspiration from this model in chapter \ref{chap:hsgrf} to work on a system with broken rotational symmetry.
} that depicts an Ising ferromagnet in which each spin feels a random field that is not correlated with the rest of the sites.
This is not a \ac{SG} \index{spin glass} because there couplings are ferromagnetic, so there is no frustration.\index{frustration|(}
A way to define frustration quantitatively is through the Wilson loop.\index{Wilson loop} For each closed circuit in the lattice,
we can take the ordered product of all the links that form it. If this product is negative
it is not possible to find a configuration that minimizes simultaneously 
the local energy along each of the links, and the loop is said to be frustrated \cite{toulouse:77,blandin:78}.
\footnote{See the introduction of \cite{mezard:87} for a definition of frustration from every-day life examples, and \cite{parisi:95} for
an intuitive discussion on Wilson loops.}
\footnote{In this text, when we will talk about the system being more or less frustrated we will be referring to the presence of
a larger or smaller number of frustrated loops. When instead we say that two spins are mutually frustrated, we mean that the energy 
is not minimized along the bond(s) connecting the two spins.}\index{frustration|)}

\begin{figure}[!htb]
\centering
 \includegraphics[width=0.5\textwidth]{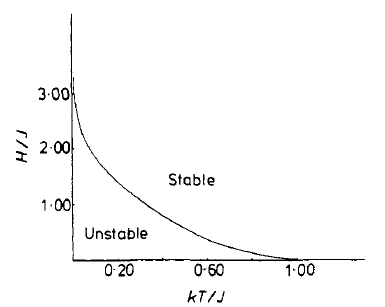}
 \caption[De Almeida-Thouless line]{Stability of the \ac{RS} solution of the \ac{SK} model in the paramagnetic phase. 
 The \ac{RS} solution is stable only at high temperatures or at high fields.
 The \acl{dat} line separates the zone of the phase diagram where the \ac{RS} phase is stable from the one where magnetic 
 ordering appears. Figure from \cite{dealmeida:78}.}
\label{fig:dat-line}
 \end{figure}

The \ac{RS} solution of the \ac{SK} model given in \cite{sherrington:75} was shown to be stable only at high temperatures \index{de Almeida-Thouless|(}
by de Almeida and Thouless \cite{dealmeida:78} (this result was promptly generalized to spins with any finite number $m$ of components \cite{dealmeida:78b}).
The paramagnetic phase is \ac{RS}, but under a certain temperature massless modes in the overlap correlation functions (replicon modes) become unstable \cite{bray:79}.
Replica symmetry, thus, becomes unstable in favor of a yet undefined \ac{SG} phase. 
Therefore all the results obtained under that
temperature, including the critical temperature, are not very useful. Also in the presence of an externally applied magnetic field 
it was shown that for low temperatures and fields
the \ac{RS} phase is not stable, so at least in the \ac{SK} modelization, there exists a \ac{SG} 
phase in a field (figure \ref{fig:dat-line}).\index{dAT|see{de Almeida-Thouless}}\index{de Almeida-Thouless!transition}\index{de Almeida-Thouless!transition}
The critical line where the \ac{RS} phase becomes unstable will be called
the \ac{dat} line. Even though the reason of this instability was suspected to be replica 
symmetry \cite{bray:78,dealmeida:78}, it was not clear how to break the symmetry between replicas in order 
to obtain a physically reasonable solution.\index{de Almeida-Thouless|)}

Perhaps with the additional stimulation of these initial failures of the replica approach, 
different approaches have been tried, such as expansions in $6-\epsilon$ dimensions of space \cite{harris:76,young:76,chen:77,southern:77} of alternative formulations
of the mean field.
In opposition with the replica method, that constructs a mean field theory after having averaged over the disorder 
with the replica trick,\index{TAP approach}\index{Thouless-Anderson-Palmer|see{TAP approach}}
Thouless, Anderson and Palmer formed a mean field theory first, including in the free energy the rebound effect of each spin on 
itself (Onsager's reaction term \cite{onsager:36,barker:73}), \index{Onsager reaction term}
and only after averaged over the disorder \cite{thouless:77}. Still, the \acf{TAP} approach was shown to be useful only at high temperatures (see e.g. \cite{mezard:87}).
Numerical simulations confirmed the validity of all the aforementioned analytical results only at high temperature \cite{kirkpatrick:78}. 

Apparently no theory was satisfactory
describing the low-temperature phase of a \ac{SG}, and no ansatz for \ac{RSB} was fully satisfactory.

\paragraph{The Parisi solution.}\index{Parisi ansatz|(}\index{replica!symmetry!breaking}\index{RSB|see{replica symmetry breaking}}
In order to find the good solution of the \ac{SK} model the replica symmetry needed to be broken, 
but $q^{\rma\rmb}$, an $n\times n$ matrix (with $n\to0$!) could be parametrized in
infinite ways, and the only \emph{modus operandi} with new ansatz for a \ac{RSB} overlap matrix was by trial and error \cite{bray:78}. It appeared also that adding new
order parameters to the model, that is giving $q^{\rma\rmb}$ the possibility to assume more than one value, shifted the negative zero-temperature entropy 
towards zero \cite{parisi:79b}. Each new order parameter is equivalent to a new breaking of the replica symmetry, so an ansatz with 2 order parameters is
called with \ac{1RSB}. It became quickly clear that the \ac{SG} phase has intriguing unseen properties when finally the good ansatz was found by Parisi in 1979,
with infinite steps of \ac{RSB}, that we call full \ac{RSB} \cite{parisi:79}.\index{replica!symmetry!breaking!full|(}

The Parisi ansatz for the matrix $q^{\rma\rmb}$ consisted in an iterative process starting from the \ac{RS} ansatz $q^{\rma\rmb} = q_0 (1-\delta^{\rma\rmb})$ 
(figure \ref{fig:SG-RSB}) \cite{parisi:80, parisi:80c,parisi:80b}.
\begin{figure}[!tbh]
\small 
\begin{eqnarray}
&\left(\begin{array}{cccccccc}
0 & & & & \multicolumn{4}{c}{\multirow{4}{*}{\Huge $q_0$}}\\
 & 0 & &  & \\
 &  & 0 &  & \\
 & &  & 0 & \\
\multicolumn{4}{c}{\multirow{4}{*}{\Huge $q_0$}} & 0 &  &  &  \\
 & & & &   & 0 &  &   \\
& & &  &  &  & 0 &   \\
 & & & &  &  &  & 0
\end{array}\right) \quad \longrightarrow \quad 
\left(\begin{array}{cccc|cccc}
0 &  &\multicolumn{2}{r|}{\multirow{2}{*}{\Large $\ \ q_1$}} & \multicolumn{4}{c}{\multirow{4}{*}{\Huge $q_0$}}\\
 & 0 & &  & \\
\multicolumn{2}{c}{\multirow{2}{*}{\Large $q_1$}}  & 0 & & \\
 &  &  & 0 & \\
\hline
\multicolumn{4}{c|}{\multirow{4}{*}{\Huge $q_0$}} & 0 & & \multicolumn{2}{c}{\multirow{2}{*}{\Large $q_1$}}\\
 & & & &  & 0 &   &  \\
& & & & \multicolumn{2}{c}{\multirow{2}{*}{\Large $q_1$}}  & 0 &   \\
 & & & &  & &  & 0
\end{array}\right) \quad \longrightarrow \quad& \\
&\left(\begin{array}{cc|cc|cc|cc}
0 & q_2 &\multicolumn{2}{c|}{\multirow{2}{*}{\Large $\ \ q_1$}} & \multicolumn{4}{c}{\multirow{4}{*}{\Huge $q_0$}}\\
q_2 & 0 & &  & \\
\cline{1-4}
\multicolumn{2}{c|}{\multirow{2}{*}{\Large $q_1$}}  & 0 &q_2 & \\
 &  & q_2 & 0 & \\
\hline
\multicolumn{4}{c|}{\multirow{4}{*}{\Huge $q_0$}} & 0 &q_2 & \multicolumn{2}{c}{\multirow{2}{*}{\Large $q_1$}}\\
\multicolumn{4}{c|}{} &q_2  & 0 &   &  \\
\cline{5-8}
\multicolumn{4}{c|}{} & \multicolumn{2}{c|}{\multirow{2}{*}{\Large $q_1$}}  & 0 & q_2  \\
\multicolumn{4}{c|}{} &  & &q_2  & 0
\end{array}\right)  \quad \longrightarrow \quad { \cdots}&
\end{eqnarray}
\caption[Replica symmetry breaking]{%
Sketch of the first two steps of replica symmetry breaking. The first $n\times n$ matrix represents the \ac{RS} ansatz, where there is total
symmetry with respect to replica exchange. The second matrix shows the first step of \ac{RSB}, the matrix is divided in blocks, and the overlap
$q^{\rma\rmb}$ can now assume two values. In the \ac{SK} model the process needs to be iterated infinite times to obtain the exact solution. The iteration
procedure is clear from the 2-step \ac{RSB}: the inner blocks are subsequently divided in smaller blocks, up to having a continuum of solutions
at the full \ac{RSB} level.
More details in the main text.
\label{fig:SG-RSB}}
\end{figure}
The $n\times n$ matrix is then parted in $n/m_1$ blocks of size $m_1\times m_1$. The off-diagonal blocks stay unchanged, but the off-diagonal
terms of the diagonal blocks now assume the value $q_1$. This is the first step of \ac{RSB}, and is called \ac{1RSB}. The second step of \ac{RSB}
is identical, and consists in iterating the symmetry breaking in each of the $n/m_1$ diagonal blocks. Each is subdivided
in $m_1/m_2$ sub-blocks of size $m_2\times m_2$. The off-diagonal sub-blocks stay the same, while the off-diagonal elements of the the diagonal
sub-blocks assume the value $q_2$. The process can be iterated infinite times, up to the full \ac{RSB} solution.
An overlap matrix constructed this way has any two rows (or columns) identical up to permutations. This property is called 
\index{replica!equivalence} replica equivalence,
and both the \ac{RS} and the \ac{RSB} matrices benefit from this property.

\begin{figure}[!htb]
 \centering
 \includegraphics[width=0.48\textwidth]{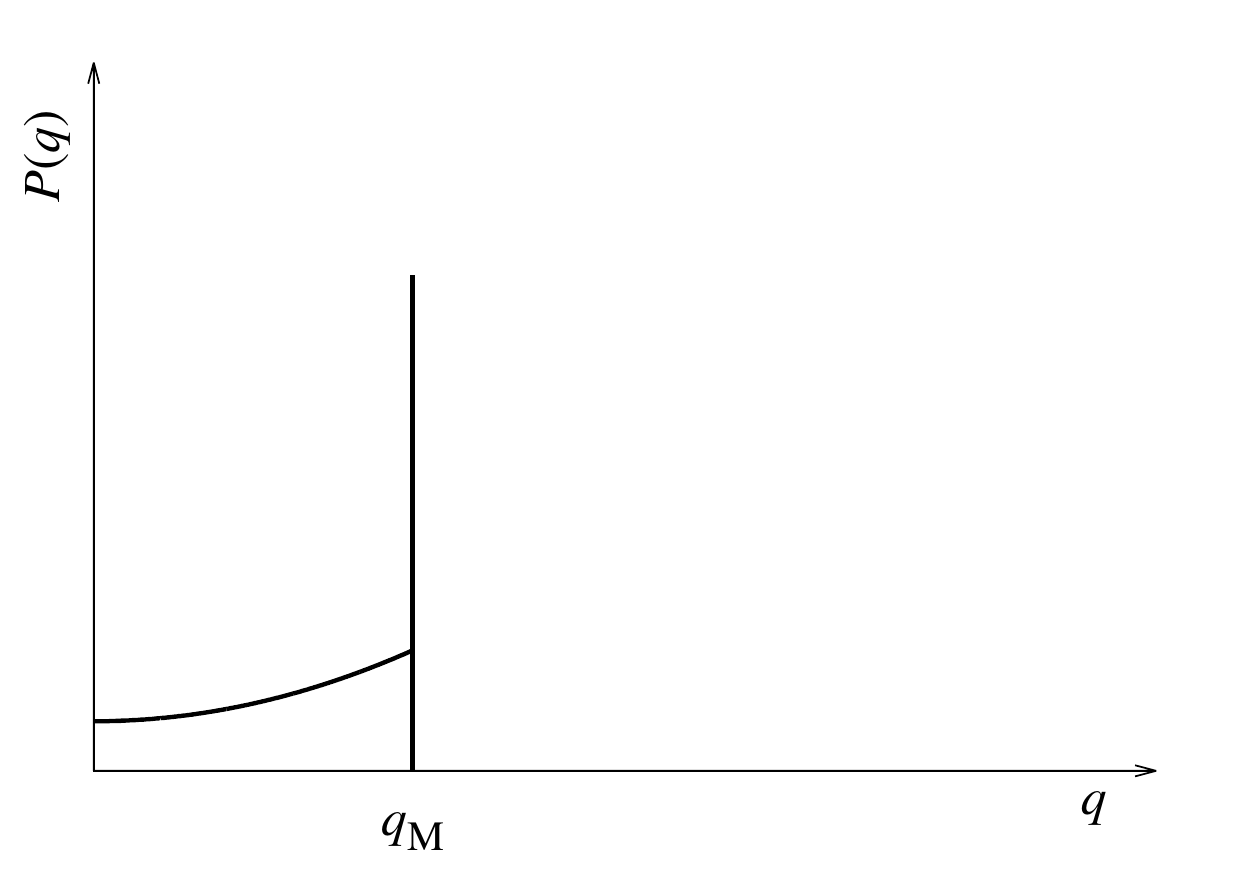}
 \includegraphics[width=0.48\textwidth]{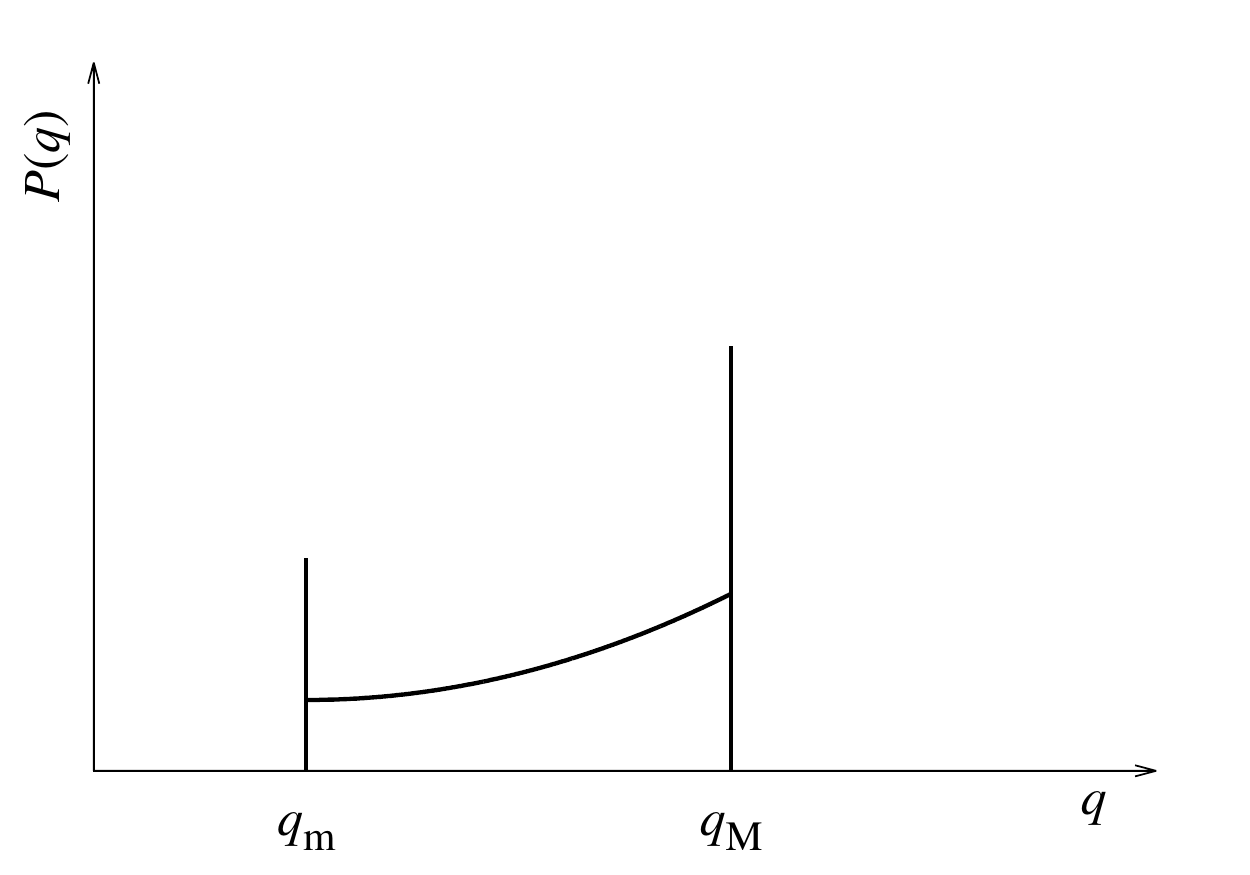}
 \includegraphics[width=0.48\textwidth]{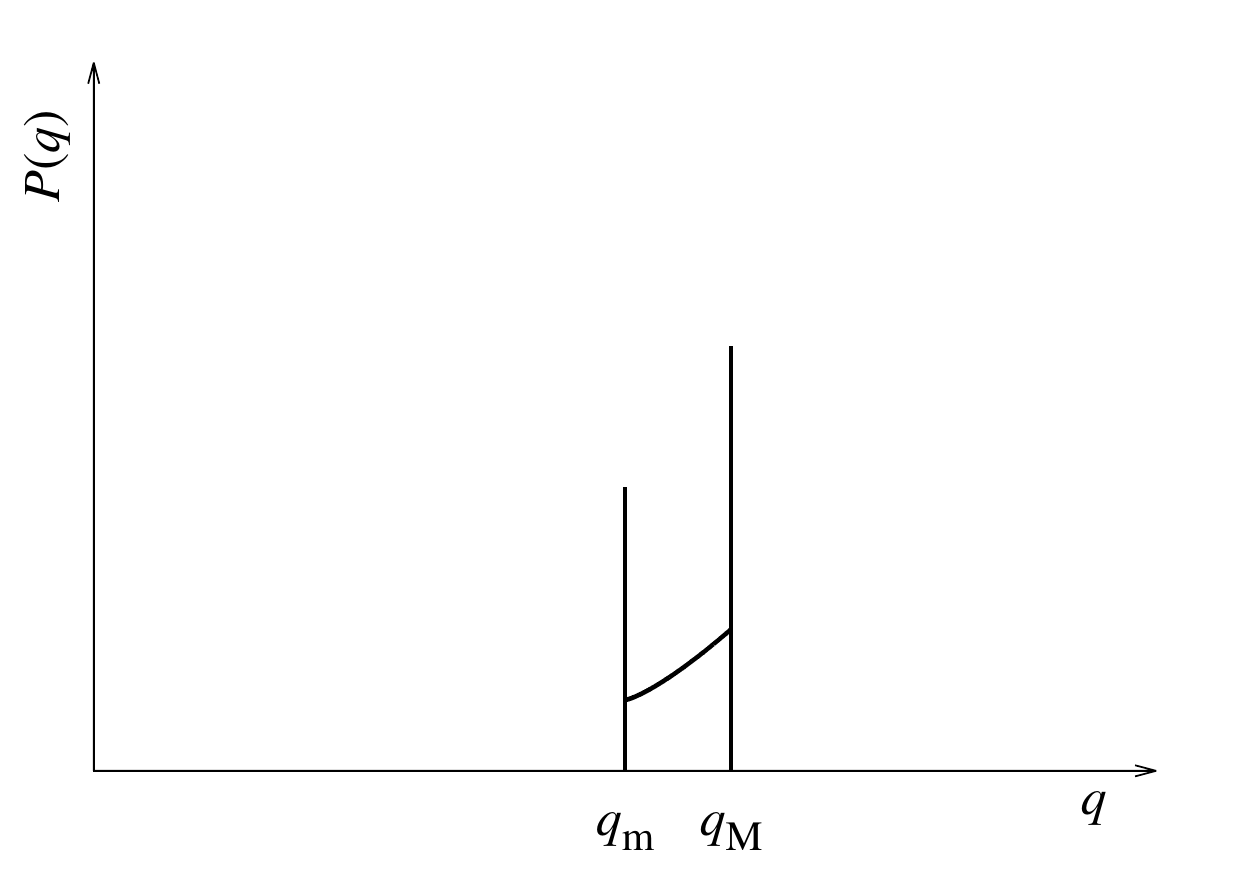}
 \includegraphics[width=0.48\textwidth]{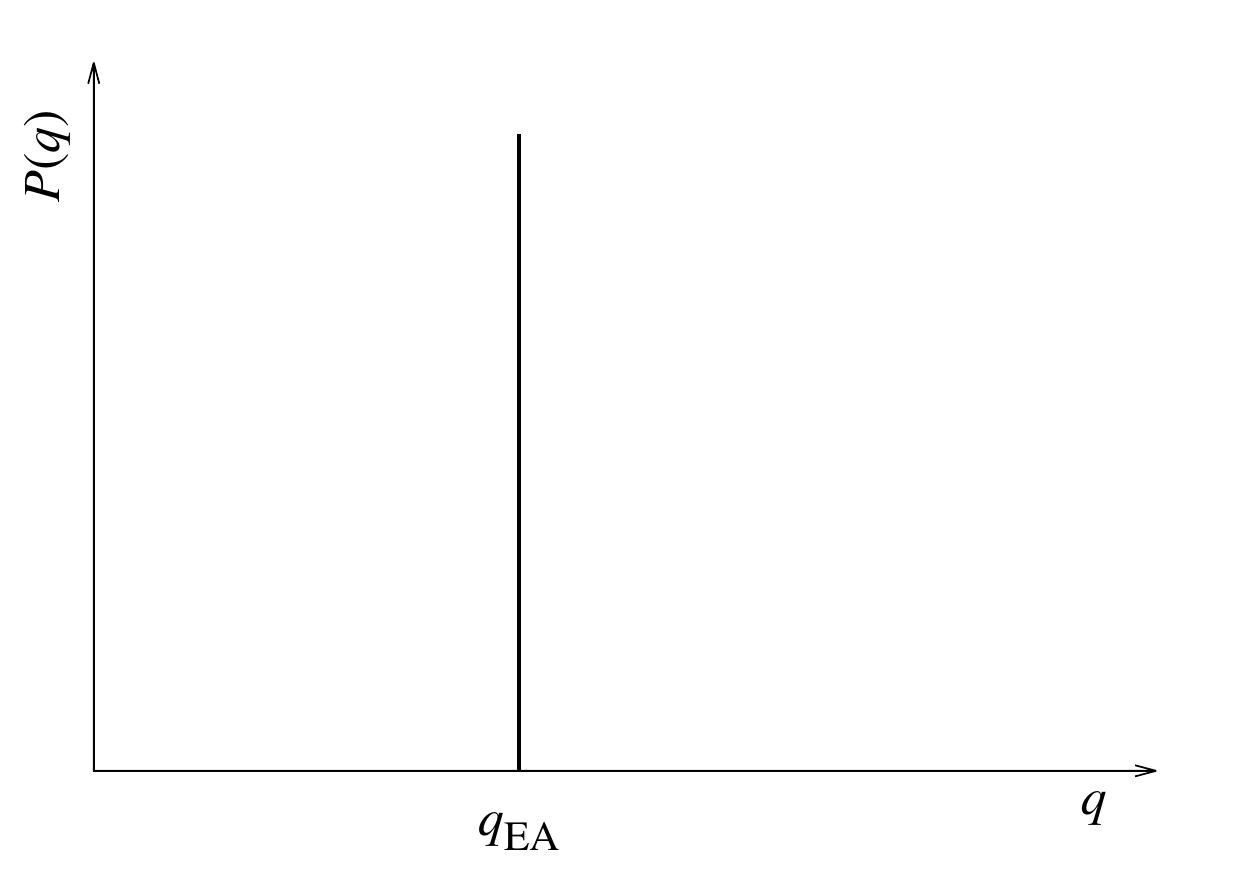}
 \caption[$P(q)$ in the \ac{SK} model]{Different instances of $P(q)$ in the \ac{SK} model. 
 \textbf{Top left}: at zero field, close to $T_\mathrm{c}$, $q_\mathrm{M}$ is proportional to $T-T_\mathrm{c}$.
 \textbf{Top right}: at small magnetic field $h$, $q_\mathrm{m}$ is proportional to $h^{2/3}$.
 \textbf{Bottom left}: at large magnetic field $h$ the \ac{dat} line is approached and the difference $q_\mathrm{M}-q_\mathrm{m}$ shrinks proportionally to the distance
 from this line.
 \textbf{Bottom right}: in the \ac{RS} phase the $P(q)$ is a delta function centered in $q_\mathrm{EA}$, that goes to zero as $h\to0$.
 }
 \label{fig:SK-pq}
\end{figure}

In the \ac{RS} phase $q^{\rma\rmb}=0~\, \forall\rma,\rmb$, so the \ac{pdf} 
of the order parameter, $P(q)$, \nomenclature[P...q]{$P(q)$}{probability distribution function of the overlap}
is a $\delta(0)$. \index{replica!symmetry}
The full \ac{RSB} ansatz
implies instead that in the \ac{SG} phase the \ac{pdf} of the order parameter is non-trivial. By simply counting the $n(n-1)$ non-diagonal values
$q^{\rma\rmb}$ can assume, one has
\begin{align}
 P(q) &= \frac{1}{n(n-1)}\sum_{\rma\neq\rmb} \delta\left(q-q^{\rma\rmb}\right) = \nonumber\\
      &= \frac{n}{n(n-1)}\left[(n-m_1)\delta(q-q_0) + (m_1-m_2)\delta(q-q_1) + \right. \\
      & \left.+ (m_2-m_3)\delta(q-q_2) + \ldots\right]\,.\nonumber
\end{align}
Once the $n\to0$ limit is taken,
\begin{equation}
 P(q) = m_1\delta(q-q_0) + (m_2-m_1)\delta(q-q_1) + (m_3-m_2)\delta(q-q_2) + \ldots~,
\end{equation}
the $P(q)$ is positive definite only if $0<m_1<m_2<\ldots<1$. One can hypothesize, as also numerical simulations suggest, 
that the $q_i$ constitute an increasing sequence, and since the sequence is infinite it is convenient to define a function $q(x)$ such that
\begin{equation}
 q(x) = q_i ~~~~\text{if}~~m_i<x<m_{i+1}\,,
\end{equation}
so after a $k$-step \ac{RSB} $q(x)$ is a piecewise function that takes at most $k+1$ different values, and when $k$ is sent to
infinity it becomes a continuous function in the interval [0,1] \cite{parisi:80}. In this representation the free energy becomes 
a functionl of $q(x)$, and has to be maximized with respect to it. It is also shown by Parisi that
\begin{align}
 q(x) &= q_\mathrm{m}~~~\text{for}~x\leq x_\mathrm{m}\,,\\[0.5ex]
 q(x) &= q_\mathrm{M}~~~\text{for}~x\geq x_\mathrm{M}\,.
\end{align}
This means that the \ac{pdf} can be rewritten as the sum of two delta functions connected by a smooth function $\tilde P(q)$ which is
non-zero only in the interval $x_\mathrm{m}<x<x_\mathrm{M}$
\begin{equation}
 P(q) = x_\mathrm{m} \delta(q-q_\mathrm{m}) + \tilde{P}(q) + x_\mathrm{M} \delta(q-q_\mathrm{M})\,.
\end{equation}
Practically, given two random states $\alpha$  and $\beta$ (chosen from $P(q)$), with mutual overlap $q^{\alpha\beta}$, with probability $x_\mathrm{M}$
$\alpha$ and $\beta$ will be the same state and they will have maximal overlap $q_\mathrm{M}$, with probability $x_\mathrm{m}$ they will be as different as it
is possible, with $q^{\alpha\beta}=q_\mathrm{m}$, and with probability $1-x_\mathrm{m}-x_\mathrm{M}$ the situation will be something in between.
The lower limits $q_\mathrm{m}$ and $x_\mathrm{m}$ depend on an external magnetic field as $h^{2/3}$. In the interval $x_\mathrm{m}<x<x_\mathrm{M}$ the function $q(x)$
depends weakly on the field, and so does $x_\mathrm{M}$. When the critical field is approached from the \ac{SG} phase the distance between the
two peaks in the $P(q)$ decreases, $x_\mathrm{m}\to x_\mathrm{M}$ and $q_\mathrm{m}\to q_\mathrm{M}$, until the $P(q)$ becomes trivial (a $\delta(q-q_\mathrm{EA})$) at
the \ac{dat} line. 
Figure \ref{fig:SK-pq} gives a better intuition on the $P(q)$.

\begin{figure}[!htb]
 \centering
 \vspace{0.5cm}
 \includegraphics[angle=270, trim=50 0 0 0, width=0.4\textwidth]{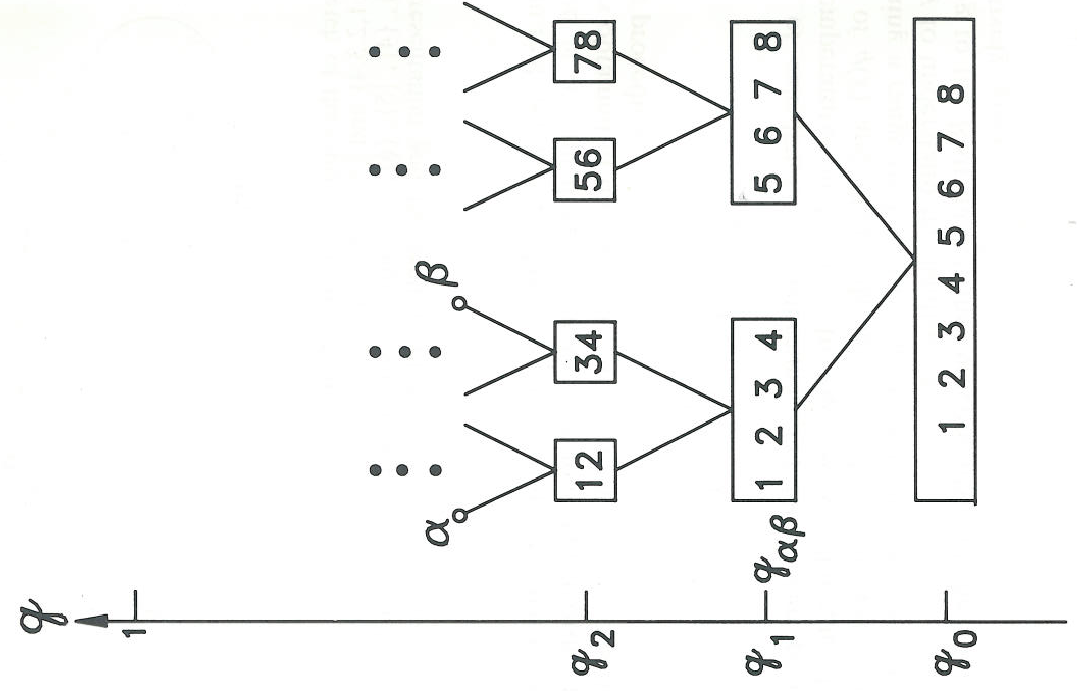}
 \includegraphics[angle=270, trim=-200 0 100 0, width=0.5\textwidth]{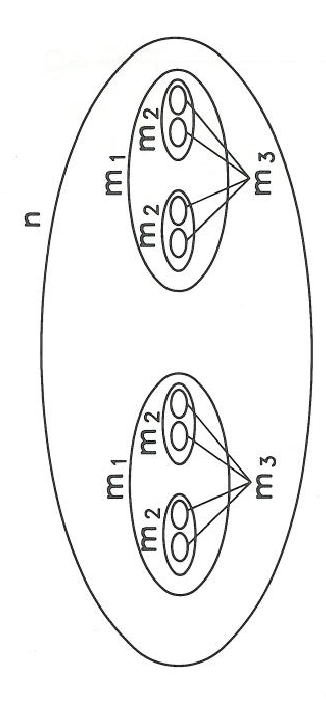}
 \caption[RSB as branching process]{\acs{RSB} as branching process. The overlap between two states $\alpha$ and $\beta$ can be seen as the first common level of 
 \ac{RSB} between $\alpha$ and $\beta$ (\textbf{left}). \index{ultrametricity}
 Another way to visualize this is to represent the \ac{RSB} process as an iterative subdivision in subsets (\textbf{right}), then the overlap between two states $\alpha$ and $\beta$
 is given by the smallest set containing both $\alpha$ and $\beta$.
 Figure from \cite{mydosh:93}.}
 \label{fig:tree}
\end{figure}

It follows from the Parisi ansatz that there is an underlying hierarchical structure in the organization of the states in the \ac{SG} phase,
that results in an ultrametric overlap space where $q^{\rma\rmc}\geq\min{(q^{\rma\rmb}q^{\rmb\rmc})}$\cite{mezard:84,mezard:85,rammal:86}. 
\index{ultrametricity} This can be seen 
by following the \ac{RSB} process as a tree (figure \ref{fig:tree}). 
At the \ac{RS} level all the states have the same overlap $q_0$, this represents the root of the tree. After one step of replica symmetry breaking
the replicas part in two groups. Replicas within the same group share have overlap $q_1$, otherwise it is $q_0<q_1$, and so on for further steps of \ac{RSB}.
The overlap between two replicas $\alpha$ and $\beta$ can be identified by returning back towards the root until the two states belong to the same group.
For example, the overlap between states $\alpha$ and $\beta$ in figure \ref{fig:tree} is $q^{\alpha\beta}=q_1$. The ultrametricity condition is easily verified by
picking three generic states.

The full \ac{RSB} $P(q)$ is sign of a \ac{SG} phase with a complex energy landscape and an infinitely large number of metastable states that are not related through evident
symmetries: ``The space of configurations consists of many 
\index{valley}\nomenclature[v.alley]{valley}{local minimum of the energy} valleys \index{local!minimum|see{valley}}
separated by high mountains (free energy barriers) whose height goes \index{energy!barrier!free}
to infinity in the infinite-volume limit'' (\cite{parisi:83}). The number of valleys is exponential in the number of spins $N$ \cite{bray:80c,dedominicis:80,young:81},
and so is the time spent in a single valley, meaning that the dynamics of a \ac{SG} are extremely slow, and 
when the system size goes to infinity ergodicity is broken \cite{mackenzie:82} in the whole \ac{SG} phase. \index{ergodicity breaking}
This was made clear at first in the infinite-range model \cite{kirkpatrick:78},\index{infinite-range model|see{$p$-spin model}}\index{pspin model@$p$-spin model}
an extension of the \ac{SK} model that mixes interactions between $p$ spins (also called the $p$-spin model). The limit $p\to\infty$ of the 
$p$-spin model yields an exactly solvable model called the \ac{REM} \cite{derrida:81}, \index{random!energy model}
where the probability of a state depends exclusively on its energy
and not on the configuration itself.
\index{spin glass!Sherrington-Kirkpatrick|)}

Despite the Parisi solution of the \ac{SK} model was physically consistent and confirmed by numerical simulations and other analytical methods 
(for example the cavity method \cite{mezard:86}),\index{cavity method} it contained some
mathematical arbitrarities, some of which we already mentioned, that made it non-rigorous. 
It took over 20 years later before it was confirmed rigorously through a mathematical proof \cite{guerra:02,guerra:03,talagrand:06}.
Nonetheless, this mean field solution of the \ac{EA} model posed a first hypothesis on the 
nature of the \ac{SG} phase in finite \index{replica!symmetry!breaking!scenario|(}
dimensions. Just as the mean field solution of the Ising model, valid in infinite dimensions, is a good qualitative descriptor of the ferromagnetic transition,
the \ac{SG} phase in a lattice of size $L\times L\times L$ would be qualitatively similar to the one detected in the \ac{SK} model. 
This means for instance that the \ac{SG} phase would resist
the application of a small magnetic field,\index{de Almeida-Thouless!transition} the $P(q)$ would be non-trivial and the overlap 
space would be ultrametric. \index{ultrametricity}
Also, in low dimensions the \ac{RSB} the domains are expected to be space-filling, i.e. with a fractal dimension $d_s=d$, and it \index{fractal dimension}
is possible to have excitations that involve a finite fraction, $O(L^d)$, of the total spins with a finite-energy cost.

This attractive 
% \footnote{\emph{It seems that if one is working from the point of view of getting beauty in one's equations, and if one has really a sound 
% insight, one is on a sure line of progress.}, Paul Dirac, The Evolution of the Physicist's Picture of Nature (1963).}
\footnote{\emph{``God used beautiful mathematics in creating the world''}, Paul Dirac, as quoted in \emph{The Cosmic Code : Quantum Physics As 
The Language Of Nature} (1982) by Heinz R. Pagels, p. 295; also in \emph{Paul Adrien Maurice Dirac : Reminiscences about a Great Physicist} (1990) edited by Behram N. Kursunoglu and Eugene Paul Wigner, p. xv.}
vision of how real \acp{SG} are is called \emph{\ac{RSB} scenario}.
\footnote{For a detailed review on the \ac{RSB} scenario see \cite{marinari:00}. See also \cite{parisi:96}.}
\index{replica!symmetry!breaking!scenario|)}
\index{replica!theory|)}\index{Parisi ansatz|)}\index{replica!symmetry!breaking!full|(}

\paragraph{The droplet picture.}\index{droplet picture|(}
Stimulated by earlier numerical domain-wall renormalization group studies of low-dimensional \acp{SG} \cite{bray:85,mcmillan:85}, 
and inspired on a schematic scaling theory of \acp{SG} proposed by Mc~Millan \cite{mcmillan:84},
Fisher and Huse proposed a new picture of the ordered phase in \acp{SG} \cite{fisher:86}, 
called \emph{droplet picture} \cite{fisher:87,huse:87,fisher:88,fisher:88b}. The theory, that derives from
a Migdal-Kadanoff approximation \cite{migdal:75,kadanoff:76} on the \ac{EA} model \cite{anderson:78}, exact 
in one dimension, describes the \ac{SG} phase of low-dimensional \acp{SG} as a ``ferromagnet in disguise'',
\footnote{Ferromagnets in disguise can be obtained, for example, by performing a random gauge 
transformation on an ordered system \cite{nishimori:01}, as it is done in the Mattis model \cite{mattis:76}.
}
with only two pure states,
with order parameter $q=\pm q_\EA$. Within a pure state, phase coexistence occurs in form of low lying
excitations (droplets) of spins in the subdominant state. The boundaries of these domains
are not fixed, but move around due to the disorder, exploiting unsatisfied links and avoiding the strongly satisfied ones.
The effect is that the droplets are non-convex, and their boundary scales as $L^{d_s}$, with $d-1\leq d_s<d$, so not space-filling. \index{fractal dimension}
The fundamental ansatz, inspired by an earlier argument from Anderson and Pond in the aforementioned Migdal-Kadanoff approach \cite{anderson:78}, is that the free-energy cost of the lowest-energy 
excitations of linear size $\ell$ is
\begin{equation}
 F_\ell\sim\gamma(T)\ell^\theta\,,
\end{equation}
where $\theta$ is the stiffness coefficient, $0<\theta<(d-1)/2$ and $\gamma$ is the stiffness modulus. A direct implication is that an infinite energy would be necessary to excite 
a finite fraction ($\ell\sim L)$ of the total number of spins, so only small excitations ($\ell\ll L$) are supported.

In the droplet picture the stiffness coefficient controls the decay of the correlations that go as
\begin{equation}
 C(|\bx-\by|) = \overline{\mean{s_\bx s_\by}^2} - \overline{\mean{s_\bx}^2 \mean{s_\by}^2} \sim \frac{1}{|\bx-\by|^\theta}\,,
\end{equation}
that entails $\overline{q^2}-\overline{q}^2\to0$, and therefore the overlap distribution is a delta function, $P(q)=\delta(q-q_\EA)$.

One last remarkable feature of the droplet theory is that the energy barrier for flipping a droplet in a field $h$ scales as $L^\theta-hL^{d/2}$. Because
of the bound $\theta<(d-1)/2$, the \ac{SG} phase is unstable to the presence of any magnetic field. This prediction in particular is in contrast with 
the \ac{RSB} theory, that predicts a \ac{dat} line for $h>0$. \index{de Almeida-Thouless!transition}
\footnote{In chapter \ref{chap:eah3d} we will try to see whether there is or not a phase transition in a field, that would discriminate the (in)correct theory.}

It is still matter of debate whether which of the two dominant theories for the \ac{SG} phase, the droplet and the \ac{RSB} scenario, 
describes well the \ac{SG} phase \cite{moore:11,parisi:12,yeo:12,yucesoy:12,billoire:13,yucesoy:13}.
It is predominantly accepted that the \ac{RSB} scenario is valid for dimensions greater than the upper critical dimension $d_u=6$, \index{critical dimension!upper}
and that the droplet picture is exact in $d=1$.
\index{droplet picture|)}

\paragraph{A different order parameter}\index{overlap!link|(}
The reason why it is hard to understand the \ac{SG} in real-life (three-dimensional) \acp{SG} could be that we are not looking
at the most useful order parameter \cite{contucci:03,contucci:05,contucci:06}. 

From a purely mathematical perspective, in the \ac{SK} model the square of overlap (\ref{eq:q-replicas})
represents the covariance of Hamiltonian (\ref{eq:H-SK}). On the other side, in a finite-dimensional \ac{EA} model, the covariance of
Hamiltonian (\ref{eq:H-EA}) is given by the square of the link overlap
\begin{equation}\label{eq:qlink-def}\nomenclature[q....link]{$q_\mathrm{link}$}{link overlap}
  q^2_\mathrm{link} = \frac{1}{Nz}\,\sum_{\bx\by}^N \sum_{\mu=1}^d q_\bx q_{\by}\,
\end{equation}
where $q_\bx=s_\bx^{(\rma)}\cdot s_\bx^{(\rmb)}$\nomenclature[q....x]{$q_\bx$}{local overlap} and $z$ is the connectivity.

Overlap and link overlap are the same in the \ac{SK} model, but in finite-dimensional lattices the two behave differently, as, for instance,
under an inversion of all the spins the change in $q$ is $O(L^d)$, while in the case of the link overlap the only changes are in the links that cross 
the domain surfaces, so the variation is $O(L^{d_s})$.\index{fractal dimension}

Droplet and \ac{RSB} theories have different predictions for the relation between $q$ and $q_\mathrm{link}$. In the droplet picture, where the surface-volume
ratio vanishes for large systems, $q_\mathrm{link}$ should be constant, with no correlation with $q$. On the other side, in the \ac{RSB} scenario 
the surfaces are space-filling, so there should be a correlation between $q_\mathrm{link}$ and $q$, implying that also $P(q_\mathrm{link})$ is non-trivial.
\index{spin glass!theory|)}
\index{overlap!link|)}
%FINITO
\chapter{Observables in simulations \label{chap:obs}}
The reason why numerical simulations became so popular in the last decades is that they are able
to give a perspective to physical phenomena orthogonal to the one coming from analytical work and experiments.
It is often not possible to validate a model, nor to make predictions that experimentalists can use
by using only analytical tools. A numerical simulation can take advantage of the knowledge of the Hamiltonian
to test it straightforwardly. As an advantage with respect to experiments, computer simulations are able to
measure a large set of observables, mostly microscopic, that are not accessible on real samples. The conjunction
of these three aspects of research makes scientific advance much more effective.
In this chapter we will discuss most of the observables that we kept track of in our simulations and analyses.

\paragraph{Some notation.}
Most of the work presented in this thesis comes from numerical simulations on systems of 
\nomenclature[N...0]{$N$}{total number of spins} $N$ spins, both in
regular 
\nomenclature[d....]{$d$}{dimensions of space}$d$-dimensional 
cubic lattices of size 
\nomenclature[L...size]{$L$}{linear lattice size}$L^d = N$ (chapters \ref{chap:eah3d},
\ref{chap:ahsg},\ref{chap:hsgm},\ref{chap:hsgrf}), and in fully connected networks (chapter \ref{chap:marginal}).
Each spin $\vec s_\bx$ occupies a position 
\nomenclature[x....bold]{$\bx$}{site}$\bx$ and has 
\nomenclature[m....99]{$m$}{In the other chapters: number of spin components}$m$ components, $\vec s_\bx = (s_{\bx,1},s_{\bx,2},\ldots,s_{\bx,m})$.
If $m=1$ we call them 
\nomenclature[Ising spin]{Ising spin}{spin with $m=1$ components}\index{spin!Ising}\emph{Ising spins}
and often remove the vector symbol, $s_\bx$.
If $m=2$ they are 
\nomenclature[X.Y spin]{XY spin}{spin with $m=2$ components}\index{spin!XY}\emph{XY spins}, while if $m=3$ we call them 
\nomenclature[H.eisenberg spin]{Heisenberg spin}{spin with $m=3$ components}\index{spin!Heisenberg} \emph{Heisenberg spins}.
The set of all the spins $\vec s_\bx$ of the system is denoted with a ket, 
\nomenclature[s....]{$\ket{\vec{s}}$}{configuration}$\ket{\vec{s}}$, and constitutes a 
\nomenclature[c.onfiguration]{configuration}{position of all the spins in the system: $\ket{s}$}\index{configuration}\emph{configuration}.

Through \ac{MC} simulations we thermalize the system at a temperature \index{Monte Carlo}\index{thermalization} $T$, 
taking them to follow the \index{Boltzmann!distribution}Boltzmann distribution 
\begin{equation}\label{eq:boltz}
 P(\ket{s}) \sim \E^{-\beta \mathcal{H}(\ket{s})}\,,
\end{equation}
where 
\nomenclature[H..a]{$\mathcal{H}$}{Hamiltonian}$\mathcal{H}$ is the model's Hamiltonian and 
\nomenclature[b..3]{$\beta$}{In chapter \ref{chap:obs}: inverse temperature}$\beta=1/k_\mathrm{B} T=1/T$ is the inverse temperature, as we set to one the 
Boltzmann constant, \index{Boltzmann!constant}
\nomenclature[k....b]{$k_\mathrm{B}$}{Boltzmann constant ($k_\mathrm{B}=1$)}$k_\mathrm{B}=1$.

Once the system is thermalized, one can take thermal averages of any measurable observable 
$\obs$, that we denote with $\mean{\obs}$. The averages over
the disorder, instead, are indicated with an over line $\overline{\obs}$. To make the notation lighter,
we use 
\nomenclature[E...o]{$E(\obs)$}{thermal and disorder average of observable $\obs$} $E(\obs)$
when both averages are performed, $E(\obs)\equiv\overline{\mean{\obs}}$.

It can be useful to define a scalar product between two configurations $\ket{s}$ and $\ket{s'}$, for which we use again Dirac's notation
\begin{equation}\label{eq:prod-conf}
\nomenclature[<>]{$\langle s\ket{\,s'}$}{scalar product between configurations $\ket{s}$ and $\ket{s'}$}
 \langle s\ket{\,s'} = \sum_\bx^N \vec s_\bx\cdot {\vec {s}}_\bx'\,.
\end{equation}
It is straightforward to define the 1- and 2-norms in this space
\begin{align}
 \normauno{s} &= \sum_\bx^N |\vec s_\bx|\,,\\
 \normados{s} &= \sum_\bx^N |\vec s_\bx|^2 = \langle s\ket{\,s}\,.
 \nomenclature[...1]{$\normauno{\ldots}$}{1-norm}
 \nomenclature[...2]{$\normados{\ldots}$}{2-norm}
\end{align}

Now that the notation is defined, we can proceed describing the set of observables $\obs$ that we measured in our simulations, that can be used
to validate theories and physical scenarios.

% \section{Local field}
% Given a generic Hamiltonian $\mathcal{H}$, the energy of a spin configuration $\ket{\sigma}$ of the system we measured was always 
% \nomenclature[E]{$E$}{Energy}
% $E = \mathcal{H}\left(\ket{\sigma}\right)/Nd$,
% so that with unitary couplings and in the absence of magnetic fields or random anisotropies it is normalized to one.
% 
% A useful concept is the local field $\vec{h}_\bx$, defined for spins with an arbitrary amount $m$ of components,
% \begin{equation}\label{eq:local-field}\index{Local field}
% \nomenclature[hx]{$\vec{h}_\bx$}{Local field}
%  \vec{h}_\bx = \frac{\partial \mathcal{H}}{\partial \vec{s}_\bx}\,.
% \end{equation}
% For Ising spins ($m=1$) we omit the vector arrow and write the local field as $h_\bx$. In this case
% we also need to define the local stability as
% \begin{equation}\index{Local stability}
% \nomenclature[lx]{$\lambda_\bx$}{Local stability}
% \lambda_\bx = h_\bx s_\bx \,,
% \end{equation}
% so the Hamiltonian can also be expressed as $\mathcal{H}=\displaystyle\sum_\bx^N \lambda_\bx$.
% 

\section{Overlaps}\label{sec:overlaps}\index{overlap|(}
We will use two replicas in order to create gauge-invariant observables \cite{mezard:87}.
To identify different replicas we use the superscripts $^{(\rma)}$,$^{(\rmb)}$,$^{(\rmc)}$ and $^{(\rmd)}$.
The definition of overlap we use depends on the model we consider and on its symmetries.

\paragraph{Ising overlap \label{sec:ising-q}}\index{spin!Ising}\index{overlap!Ising}
With Ising spins $s_\bx=\pm1$ we can define the local overlap as
\begin{equation}\label{eq:q-local-ising}
 q_\bx = s_\bx^{(\rma)}s_\bx^{(\rmb)}\,,
\end{equation}
from which we can create the global overlap
\begin{equation}\label{eq:q}
 q = \frac{1}{N}\sum_\bx^N q_\bx = \frac{1}{N}\langle s^{(\rma)}\ket{s^{(\rmb)}}\,,
\end{equation}
where we used notation \ref{eq:prod-conf}.

\paragraph{Tensorial overlap \label{sec:tens-q}}\index{overlap!tensorial}
When the spins are $m$-component vectors 
{ %Tutto st'accrocco est per mandare a capo a meta' formula
    \def\OldComma{,}
    \catcode`\,=13
    \def,{%
      \ifmmode%
        \OldComma\discretionary{}{}{}%
      \else%
        \OldComma%
      \fi%
    }%
$\vec s_\bx=(s_{\bx,1}, s_{\bx,2}, \ldots, s_{\bx,m})$
}
and $\mathcal{H}$ displays an $O(m)$ symmetry 
it is convenient to define a rotationally invariant overlap.

We define the tensorial site overlap is defined as
\begin{equation}
 \tau_{\alpha\beta}(\bx) = s_{\bx, \alpha}^{(\rma)}s_{\bx, \beta}^{(\rmb)}\,,
\end{equation}
where $\alpha,\beta=1,\ldots,m$ indicate the components of the vector.
Notice that $\tau_{\alpha\beta}(\bx)$ is not Hermitian, since
\begin{equation}
  \tau_{\alpha\beta}(\bx)^\dag = \tau_{\beta\alpha}(\bx) = s_{\bx, \beta}^{(\rma)}s_{\bx, \alpha}^{(\rmb)}\,.
\end{equation}
The order parameter is the overlap tensor \cite{fernandez:09b}:
\begin{equation}
 Q_{\alpha\beta} = \frac{1}{N}\sum_{\bx} \tau_{\alpha\beta}(\bx) \,.
\end{equation}
This quantity is not rotationally invariant, and since it is a tensor it is not easy to deal with, so we use the square overlap \cite{binder:86,coluzzi:95}
\begin{eqnarray}\label{eq:Q2-def}\nonumber
\nomenclature[Q...0]{$Q$}{tensorial overlap}
 Q^2 &=& \Tr \left[ Q Q^\dagger \right] \\
     \label{eq:Q2}
     &=&\frac{1}{N^2}\sum_{\bx,\by} \Tr\left[\tau(\bx)\tau(\by)^\dag\right] \\
     \nonumber
     &=& \frac{1}{N^2}\sum_{\bx,\by}
	(\vec{s}_{\bx}^{(\rma)}\cdot\vec{s}_{\by}^{(\rma)})
	(\vec{s}_{\bx}^{(\rmb)}\cdot\vec{s}_{\by}^{(\rmb)})\,,
\end{eqnarray}
that is $O(m)\times O(m)$ invariant (rotational invariance for replica $\rma$ and replica $\rmb$). 
Even though the $Q^2$ defined in equation \eqref{eq:Q2} is a square overlap, we will be calling it overlap when referring to it.

\subparagraph{The self overlap $Q^2_\mathrm{self}$} is defined analogously, by taking ${(\rma)}={(\rmb)}$ 
in the previous definitions.\index{overlap!tensorial!self}\index{overlap!self}
Notice that the self overlap is not identically equal to 1. It is easy to see, for example, that at
infinite temperature, in the thermodynamic limit it is equal to $Q^2_\mathrm{self}(T=\infty;L=\infty)=1/m$
(see for example the Appendix of \cite{baityjesi:11}).

\paragraph{Scalar overlap}\index{overlap!scalar}\index{overlap!spin glass}
With vector spins, if the Hamiltonian is not rotationally invariant the overlap can be expressed straightforwardly through the
scalar product between spins of different replicas. The site overlap would be
\begin{equation}\label{eq:site-scalar-overlap}
 q_{\SG,\bx} = \vec s^{\,\rma}_\bx\cdot \vec s^{\,\rmb}_\bx\,,
\end{equation}
and the global overlap
\begin{equation}\label{eq:scalar-overlap}
\nomenclature[q....scalarSG]{$q_{\SG}$}{spin glass overlap}
q_{\SG} = \frac{1}{N}\sum_\bx^N q_{\SG,\bx}\,.
\end{equation}
We will be calling $q_\SG$ the \ac{SG} overlap, to differentiate it from the \ac{CG} overlap $Q_\CG$, defined in the next paragraph.

\paragraph{Chiral overlap}\index{overlap!chiral glass}
With vector spins it is possible to define the chirality, \index{chirality}
an observable whose importance we will discuss in chapter \ref{chap:ahsg}. It
represents the amplitude and handedness of the alignment of the spins along the axis $\mu$, and is expressed with the mixed product of
three consecutive spins
\begin{equation}\label{eq:chirality}
 \zeta_{\bx,\mu} = \vec s_{\bx+\vn{e}_\mu} \cdot (\vec s_\bx \times \vec s_{\bx-\vn{e}_\mu})  ~~~,~~~\mu=1,\ldots,d,
\end{equation}
where $e_\mu$ is the unitary vector along the $\mu$ direction. We can see it as the oriented volume of the parallelepiped we can construct with the three spins.
The \ac{CG} overlap is defined similarly to the \ac{SG} one,
\begin{equation}\label{eq:site-chiral-overlap}
 \kappa_{\bx,\mu} = \zeta_{\bx,\mu}^{(\rma)}\, \zeta_{\bx,\mu}^{(\rmb)}\,,
\end{equation}
but in this case we also sum over the $d$ equivalent directions $\mu$
\begin{equation}\label{eq:chiral-overlap}
\nomenclature[q....scalarCG]{$q_{\CG}$}{chiral glass overlap}
 q_\CG = \frac{1}{Nd}\sum_{\bx,\mu}^{d,N}\kappa_{\bx,\mu}\,.
\end{equation}

\paragraph{Link overlap}\index{overlap!link}
We also measured the link overlaps,
that were shown to be equivalent to the overlaps in the description of the low temperature phase \cite{contucci:05b,contucci:06}.
In the case of Ising spins the link overlap is \index{overlap!link!Ising}
\begin{align}\label{eq:qlink-ising}
 q^2_\mathrm{link} &= \frac{1}{Nd}\,\sum_{\bx}^N \sum_{\mu=1}^d q_\bx q_{\bx+e_\mu}\\\nonumber
                   &= \frac{1}{Nd}\,\sum_{\bx}^N \sum_{\mu=1}^d s_\bx^{(\rma)}\, s_{\bx+e_\mu}^{(\rma)} s_\bx^{(\rmb)}\, s_{\bx+e_\mu}^{(\rmb)} \,,
\end{align}
while for vector spins \index{overlap!vector}
\begin{align}\label{eq:qlink-vector}
\nomenclature[Q...link]{$Q_\mathrm{link}$}{tensorial link overlap}
  Q^2_\mathrm{link}                &= \frac{1}{Nd}\sum_{\bx}^N \sum_{\mu=1}^d q^{\mu~2}_\mathrm{link} (\bx) \,, 
  \\  
  \nonumber
  q^{\mu~2}_\mathrm{link} (\bx) &= \Tr\left[ \tau(\bx)\tau(\bx + \hat e_\mu)^\dagger\right] = 
  \\[1.5ex]
			           &= 
		    (\vec{s}_{\bx}^{\,(\rma)}\cdot\vec{s}_{\bx+\hat e_\mu}^{\,(\rma)})
		    (\vec{s}_{\bx}^{\,(\rmb)}\cdot\vec{s}_{\bx+\hat e_\mu}^{\,(\rmb)})\,,
\end{align}
which is a generalization of (\ref{eq:qlink-ising}).
\index{overlap|)}

\section{Scalar correlators}\label{sec:scalar-correlators}\index{correlation!function|(}
For a given the wave vector $\vn{k}$ \nomenclature[k....]{$\vn{k}$}{wave vector} we can define the Fourier transforms of the overlap fields 
\begin{align}\label{eq:q-k}
\nomenclature[q....hsgk]{$\hat q_\mathrm{SG}(\vn{k})$}{Fourier transforms of the spin glass overlap}
\nomenclature[q....hcgk]{$\hat q_\mathrm{CG}(\vn{k})$}{Fourier transforms of the chiral glass overlap}
\hat q_\mathrm{SG}(\vn{k}) &= \frac{1}{N}\sum_\bx^N q_\bx {\mathrm e}^{\rmi \vn{k}\cdot \bx} \\
\hat q_\mathrm{CG}^\mu(\vn{k}) &= \frac{1}{N}\sum_\bx^N \kappa_\bx e^{\rmi \vn{k}\cdot \bx}\,,
\end{align}
that we use to build the wave-vector dependent susceptibilities as
\begin{align}\label{eq:chi-k}
  \chi_\mathrm{SG}(\vn{k}) &= N \overline{\langle|q_\mathrm{SG}(\vn{k})|^2\rangle}\,,\\
  \chi_\mathrm{CG}(\vn{k}) &= N \overline{\langle|q_\mathrm{CG}(\vn{k})|^2\rangle}\,.
\end{align}
Since the lattice is finite and has discrete spacings, in our simulations we measure $\hat q(\vn{k})$ for a specific set of wave vectors
that we need to compute relevant observables. Calling $k_\mathrm{min}=2\pi/L$ \nomenclature[k....min]{$k_\mathrm{min}$}{lowest wave number}
the lowest wave number allowed by periodic boundary conditions, we seek
\begin{align}
\nomenclature[k....n]{$\vn{k}_n$}{wave vector along an axis}
\nomenclature[k....11]{$\vn{k}_\mathrm{11}$}{wave vector along a diagonal of the lattice}
  \vn{k}_n &=(n k_\mathrm{min},0,0)&n=0,\ldots,L/2\,,\\[0.5ex]
  \vn{k}_\mathrm{11}&=(k_\mathrm{min},\pm k_\mathrm{min},0)\,&,
\end{align}
and the permutations of their components. 

We can then construct the susceptibilities $\chi_\SG=\chi_\SG(\vn{0})$ and 
$\chi_\CG=\chi_\CG(\vn{0})$ \nomenclature[chi....CGSG]{$\chi_\SG$}{spin glass susceptibility}\nomenclature[chi....CG]{$\chi_\CG$}{chiral glass susceptibility}
\index{susceptibility!spin glass}\index{susceptibility!chiral glass} and
the dimensionless cumulant $R_{12}$ \index{R12@$R_{12}$}
that will be useful to spot phase transitions with the finite-size scaling method (section \ref{sec:FSS}):\index{scaling!finite-size}
\begin{equation}
\nomenclature[R12]{$R_{12}$}{$\chi(\vn{k}_\mathrm{1})/\chi(\vn{k}_\mathrm{11})$}
R_{12} = \frac{\chi(\vn{k}_\mathrm{1})}{\chi(\vn{k}_\mathrm{11})}\,,
 \label{eq:def-R12}
\end{equation}
where we averaged over all the possible permutations of the components of $\vn{k}_\mathrm{1}$ and $\vn{k}_\mathrm{11}$.

We define the two-point correlation functions $C(\bx,\by)=\langle q_\bx q_\by\rangle$. \nomenclature[C...xy]{$C(\bx,\by)$}{two-point correlation function}
When the system is translationally invariant, this correlation can be expressed as a function of the separation $\br=\bx-\by$,
being called $C(\br)$ \index{correlation!function!two-point}. \nomenclature[r]{$\br$}{Euclidean distance between two points}
We compute $C(\br)$ and its Fourier transform $\hat C(\vn{k})$ as
\begin{align}\label{eq:C}
\nomenclature[C...r]{$C(\br)$}{two-point correlation function (with translational invariance)}
\nomenclature[C...rhat]{$\hat{C}(\bk)$}{Fourier transform of $C(\br)$}
      C (\br) &= \frac{1}{N} \sum_\bx^N q_\bx q_{\bx+\br}\,,\\[1ex]
      \label{eq:hatC}
 \hat{C}(\bk) &= \sum_\br C (\br)  \E^{\rmi \vn{k}\cdot (\br)}\,,
\end{align}
and consequently $C(\br)$ can be obtained back as the anti Fourier transform $C(\br) = \frac{1}{L}\displaystyle\sum_\bk \hat{C} (\bk)  \E^{-\rmi \vn{k}\cdot (\br)}$.
In appendix \ref{app:eah3d-checks} we discuss the numerical estimators of these quantities.

The wave-vector dependent susceptibilities are directly related to the correlation functions.\index{susceptibility!wave-vector dependent}
Using equations (\ref{eq:q-k},\ref{eq:chi-k}) we have
\begin{align}
\nomenclature[chi....k]{$\chi(\vn{k})$}{wave-vector dependent susceptibility}
 \chi(\vn{k}) &= N \big[\hat q_\mathrm{SG}(\vn{k}) \hat q_\mathrm{SG}(\vn{k})^*\big] =\\[1ex]
 &=\frac{1}{N}\sum_\bx^N q_\bx \E^{\rmi \vn{k}\cdot \bx}\sum_\by^N q_\by \E^{-\rmi \vn{k}\cdot \by}=\\
 &= \frac{1}{N}\sum_{\bx,\by}^N C(\bx,\by) \E^{\rmi \vn{k}\cdot (\bx-\by)} =
\end{align}
that in the presence of translational invariance and recalling equation (\ref{eq:hatC}) becomes 
\begin{align}
= \frac{1}{N}\sum_{\bx}^N\sum_{\br}^N C(\br) \E^{\rmi \vn{k}\cdot (\br)} = \hat C (\vn{k})\,.
\end{align}
This means that we can measure correlation functions both in the real and in the Fourier space, depending
on which of the procedures is more convenient numerically.

The point-to-plane correlation functions are computed from the Fourier transform of the fields,\index{correlation!function!plane}
\begin{equation}
\nomenclature[C...r]{$C(r)$}{point-to-plane correlation function}
\label{eq:Crplane-chi}
 C(r) ~=~ \frac{1}{L} \sum_{n=0}^{L-1} \E^{-\rmi \br\cdot \vn{k}_n} \chi\big(\vn{k}_n\big)     
      ~\equiv~ \sum_{y,z} C(x=r,y,z)\,,
\end{equation}
where $r$ is the modulus of the distance.\nomenclature[r]{$r$}{modulus of the distance}
Equation \eqref{eq:Crplane-chi} is equivalent if we align the wave vector along any of the three coordinate axes,
so we average over these choices.

In chapter \ref{chap:eah3d} we will use similar procedures to construct correlation functions with four replicas instead of two.

\section{Tensorial correlation functions \label{sec:tens-C}}\index{correlation!function!tensorial}
We will be measuring both point and plane correlation functions. The point correlation function is\index{correlation!function!point}
\begin{equation}
\nomenclature[C...tenspoint]{$C^\mathrm{(point)}(r)$}{point tensorial correlation function}
 C^\mathrm{(point)}(r) = \frac{1}{Nd}\sum_{\mu=1}^{d}\sum_{\bx}^N \Tr[\tau(\bx)\tau(\bx + \hat e_\mu r)^\dagger]\,,
\end{equation}
where $\mu=1$ (or $x$), 2 (or $y$), 3 (or $z$) \nomenclature[mu]{$\mu$}{generic principal axis}
is a coordinate axis, and $e_\mu$ is the unitary  vector in that direction.
We also use plane correlation functions because they decay slower and have a better signal-to-noise ratio.
If we denominate the plane-overlap tensor as the mean overlap tensor over a plane
\begin{equation}
 P^x_{\alpha\beta}(x) = \frac{1}{L^2} \sum_{y,z=0}^{L-1} \tau_{{\alpha\beta}}(x,y,z)\,,
\end{equation}
we can define the plane correlation function as
\begin{equation}\index{correlation!function!plane}
\nomenclature[C...tensplane]{$C^\mathrm{(plane)}(r)$}{plane tensorial correlation function}
 C^\mathrm{(plane)}(r) = \frac{1}{Ld}\sum_{\mu=1}^d\sum_{x=0}^{L-1} \Tr[P^\mu(x)P^\mu(x+r)^\dagger]\,.
\end{equation}
These tensorial definitions of $C(r)$ are $O(m)\times O(m)$ invariant.

\paragraph{The link-overlap} \hspace{-4mm} correlation functions are \index{correlation!function!link}
\begin{align}
\nomenclature[C...linkpoint]{$C^\mathrm{(point)}_\mathrm{link}(r)$}{link point correlation function}
\nomenclature[C...linkplane]{$C^\mathrm{(plane)}_\mathrm{link}(r)$}{link plane correlation function}
 \label{eq:Cpoint}
  C^\mathrm{(point)}_\mathrm{link}(r) &=\frac{1}{Nd^2}\sum_{\mu,\nu=1}^{d}\sum_{\bx}^N q^{\nu~2}_\mathrm{link} (\bx)q^{\nu~2}_\mathrm{link} (\bx+ r\hat e_\mu)\,,\\[1ex]
 \label{eq:Cplane}
  C^\mathrm{(plane)}_\mathrm{link}(r) &= \frac{1}{Ld}\sum_{\mu=1}^d\sum_{x=0}^{L-1} P_\mathrm{link}(x)P_\mathrm{link}(x+r)\,,
\end{align}
with
\begin{equation} 
P^x_\mathrm{link}(x) = \frac{1}{L^2d} \sum_{\nu=1}^d\sum_{y,z=0}^{L-1} q^{\nu~2}_\mathrm{link}(x,y,z)\,.
\end{equation}
One could in principle choose to subtract from those correlators the equilibrium link overlap, to obtain connected correlators, 
since the link overlap is non-zero also in the paramagnetic phase.

\section{Four-replica Correlators}
\index{correlation!function!4-replica|(}\index{propagator|see{correlation function}}
\index{correlator|see{correlation function}}\index{correlation function!connected}
\index{spin!Ising}
\label{sec:obs}
We will be working with Ising spins under an applied magnetic field $h>0$.
In this situation the order parameter $q_\EA$ \nomenclature[q....EA]{$q_\EA$}{Edwards-Anderson parameter}
is not zero even in the paramagnetic phase. 
This implies that we cannot construct connected correlation functions by means of only two replicas.
Therefore, for each sample we simulated 4 different replicas, in order to be able to
compute connected correlation functions that go to zero at infinite distance. 
In appendix \ref{app:propagators} we give more details and show that the most informative connected
correlator we can construct with 4 replicas is the replicon propagator \cite{dealmeida:78,dedominicis:06}\index{correlation function!replicon}
\begin{equation}
\label{eq:GRr}
\nomenclature[G...Rr]{$G_\mathrm{R}(\br)$}{replicon propagator}
G_\mathrm{R}(\br)=\frac{1}{N} \sum_{\bx} 
  \overline{\left( \langle s_{\bx} s_{\bx+\br} \rangle 
                 - \langle s_{\bx} \rangle \langle s_{\bx+\br} \rangle 
	    \right)^2}\,.
\end{equation}
To compute $G_\mathrm{R}$ we calculate the 4-replica field
\begin{equation}
 \nomenclature[phi...xabcd]{$\Phi_{x}^{(\rma\rmb;\rmc\rmd)}$}{connected 4-replica field}
\Phi_{x}^{(\rma\rmb;\rmc\rmd)}=\frac{1}{2}(s_\bx^{(\rma)} - s_\bx^{(\rmb)}) (s_\bx^{(\rmc)} - s_\bx^{(\rmd)})\,,
\end{equation}
where the indexes $\rma,\rmb,\rmc,\rmd$ indicate strictly different replicas.
Notice that
\begin{equation}
  \left\langle \Phi_{\bx}^{(\rma\rmb;\rmc\rmd)} \Phi_{\by}^{(\rma\rmb;\rmc\rmd)} \right\rangle
  =
  \left( \left\langle 
    s_{\bx} s_{\bx+\br} \rangle 
      - \langle s_{\bx} \rangle \langle s_{\bx+\br} \right\rangle 
	    \right)^2 \,,
\end{equation}
 so we obtain $G_\mathrm{R}$ by taking also the average over the samples
\begin{equation}
 E(\Phi_{\bx}^{(\rma\rmb;\rmc\rmd)} \Phi_{\by}^{(\rma\rmb;\rmc\rmd)}) = G_\mathrm{R}(\bx - \by)\,.
 \label{eq:GRxy}
\end{equation}
Here, and everywhere there is more than one possible permutation of the replica indices,
we average over all of them to gain statistics.

From this point on everything is formally like the two-replica construction, using $\Phi_{\bx}$ instead of $q_\bx$
to construct the susceptibilities $\chi(\bk)$.
For example correlations in the Fourier space are defined by Fourier-transforming
$\Phi_{\bx}^{(\rma\rmb;\rmc\rmd)}$, so the wave-vector dependent replicon susceptibility\index{susceptibility!replicon}\index{susceptibility!wave-vector dependent}
is expressed as
\begin{equation}
\nomenclature[chi....Rk]{$\chi_\mathrm{R}(\bk)$}{ wave-vector dependent replicon susceptibility}
\label{eq:chi}
\chi_\mathrm{R}(\bk) = \frac{1}{N} E(|\hat {\Phi}_{\bk}^{(\rma\rmb;\rmc\rmd)}|^2)
~~~,~~~
 \nomenclature[phi...kabcd]{$\hat {\Phi}_{\bk}^{(\rma\rmb;\rmc\rmd)}$}{Fourier transform of $\Phi_{x}^{(\rma\rmb;\rmc\rmd)}$}
 \hat {\Phi}_{\bk}^{(\rma\rmb;\rmc\rmd)} = \sum_\bx^N \mathrm{e}^{\rmi\bk\cdot\bx}\Phi_{\bx}^{(\rma\rmb;\rmc\rmd)}\,.
\end{equation}
Point-to-plane correlation functions are computed through equation \eqref{eq:Crplane-chi}.
\index{correlation!function|)}\index{correlation!function!4-replica|)}

\section{Correlation lengths}\index{correlation!length|(}
The correlation length is the average distance weighed with the $C(r)$. We will be constructing second-moment correlation
lengths for point and plane correlations
\begin{align}
\nomenclature[xi....point]{$\xi_2^\mathrm{(point)}$}{second moment point correlation length}
\nomenclature[xi....plane]{$\xi_2^\mathrm{(plane)}$}{second moment plane correlation length}
\label{eq:xi-punto}
 \xi_2^\mathrm{(point)} &=\sqrt{\frac{\int_0^{L/2} C^\mathrm{(point)}(r) r^4 dr}{\int_0^{L/2} C^\mathrm{(point)}(r) r^2 dr}}\,,\\[2ex]
\label{eq:xi-plano}
 \xi_2^\mathrm{(plane)} &=\sqrt{\frac{\int_0^{L/2} C^\mathrm{(plane)}(r) r^2 dr}{\int_0^{L/2} C^\mathrm{(plane)}(r) dr}}\,.
\end{align}
The difference in the definitions is due to the presence of a Jacobian term when we want to integrate the point correlation function
over the space. These two lengths would be proportional by a factor $\sqrt{6}$ if they had the same purely exponential correlation function.
Note that $\xi_2^\mathrm{(point)}$ and $\xi_2^\mathrm{(plane)}$ are proper estimators of a correlation length only when
the correlation functions $C^\mathrm{(point)}(r)$ and $C^\mathrm{(plane)}(r)$ are connected (i.e. they go to zero for large $r$). 
Otherwise, in principle they could be used to individuate if a quench penetrated in the \ac{SG} phase. In fact, depending on $m$ a quench will drive us
in a ferromagnetic or in a \ac{SG} phase. Our correlation functions are connected in the \ac{SG} phase, but they are 
not in a ferromagnetic state. Consequently, a cumulant such as $\xi_L/L$ - being $\xi_L$ the correlation length measured in a lattice of size $L$ - 
\nomenclature[xi....L]{$\xi_L$}{correlation length measured in a lattice of size $L$} will diverge as $L^{\theta/2}$ 
(see Ref.~\cite{janus:10} for a definition of $\theta$ and an explanation of this behavior) 
when $m$ is too large for a \ac{SG} phase, it will converge as $1/L$ if the quench penetrates in the \ac{SG} phase,
and it will be of order 1 right at the critical $m$, $m_\SG$, that is probably not integer, so not exactly locatable.

When the correlation function decays very quickly and the noise becomes larger than the signal, one could measure
negative values of $C(r)$, that would be amplified by the factors $r^2$ and $r^4$ in the integrals. This would imply
very large errors in $\xi$, or even the square root of a negative number. To overcome this problem, we truncated the
correlation functions when they became less than three times the error \cite{janus:09b}. This procedure introduces
a small bias, but reduces drastically the statistical error. Furthermore, the plane correlation function required 
the truncation much more rarely, therefore we compared the behaviors as a consistency check.

As shown in the appendices of \cite{baityjesi:11}, in the thermodynamic limit ($k_\mathrm{min}=0$) the second moment correlation length can be
re-expressed as
\begin{equation}
 \xi_L = \frac{1}{2 \sin{(k_\mathrm{min}/2)}} \sqrt{\frac{\chi(\bf{0})}{\chi(\vn{k}_\mathrm{min})}-1}\,.
 \label{eq:xi}
\end{equation}
being ${\vn{k}_\mathrm{min}}=(2\pi/L,0,0)$ or permutations. The sub-index $L$ stresses the dependence 
the linear size of the lattice (recall that $k_\mathrm{min}$ depends on $L$). This same definition can be used with any of the observables defined
in the previous section (\ac{SG} susceptibility, \ac{SG} susceptibility, replicon susceptibility,...).
This quantity will be used only using with plane correlations, since integrating over all the directions in the lattice to calculate $\chi(\vn{k}_\mathrm{min})$
is a cumbersome and imprecise task.
When computing $\xi_\mathrm{CG}$, one can choose $\mu$ parallel or orthogonal to the wave
vector $\vn{k}_\mathrm{min}$. As it was already observed in \cite{fernandez:09b}, there is no apparent difference between
the two options, so we averaged over all the values of $\mu$ to enhance
our statistics.

The definitions of the link correlation lengths $\xi_\mathrm{2,link}^\mathrm{(point)}$ and $\xi_\mathrm{2,link}^\mathrm{(plane)}$ 
can be obtained from equations \eqref{eq:xi-punto} and \eqref{eq:xi-plano}, by substituting the spin with link correlation functions.
\index{correlation!length|)}

\chapter{Phase transitions with a diverging length scale \label{chap:rg}}\index{phase!transition}
The topics treated in this section have been very successful in describing phase transitions and are very well
consolidated tool since the $1970$'s. Our scope here is not to give an extended treatment, that can be found elsewhere (see e.g. 
\cite{ma:76,binder:86,huang:87,cardy:96,amit:05}), but to refresh the reader's memory on some concepts that we 
will be using throughout this dissertation.

\section{Second-order-like phase transitions}\index{phase!transition!second order}\index{spin glass!transition}
The phenomenology of the spin-glass transitions we will treat is similar to that of a second-order phase transition. 
In this section we will assume Ising spins, but the description is the same with $m$-component spins.
The coherence length $\xi$, that we can define through the long-distance decay of two-point correlation function,
\begin{equation}\index{correlation!length}\index{correlation!function}
\nomenclature[xi....]{$\xi$}{correlation length}
 \langle s_{\bx+\br} s_\bx \rangle \stackrel{|\br|\rightarrow\infty}{\sim} \E^{-|\br|/\xi}\,,
\end{equation}
diverges in power law as we approach the critical point \index{exponent!critical|(}\index{exponent!critical!nu@$\nu$}
\begin{equation}\label{eq:scaling-nu}
\nomenclature[nu]{$\nu$}{critical exponent of the correlation length}
\xi\propto |t|^{-\nu}\,.
\end{equation}
In equation \eqref{eq:scaling-nu} we defined the reduced temperature 
\nomenclature[t]{$t$}{In chapters \ref{chap:rg},\ref{chap:eah3d}: the reduced temperature}$t=\frac{T-T_\mathrm{c}}{T_\mathrm{c}}$, and $T_\mathrm{c}$ is the 
critical temperature. \nomenclature[Tc]{$T_\mathrm{c}$}{critical temperature}
Mind that the symbol $t$ will represent the reduced temperature
only in this chapter, while throughout the rest of the text it will indicate the time. The exponent $\nu$ characterizes the phase
transition and sets its Universality class. \index{universality!class}
The correlation length $\xi$ \index{correlation!length} is not the only diverging observable. To fully identify the type of phase transition we can define six critical exponents 
$\alpha,\beta,\gamma,\delta,\eta,\nu$ that describe the power law behavior of the observables that are relevant in our case.

The specific heat diverges as 
\begin{equation}\label{eq:scaling-alpha}
\index{exponent!critical!alpha@$\alpha$}
\nomenclature[C...h]{$C_\mathrm{h}$}{specific heat}
\nomenclature[alpha....3]{$\alpha$}{In chapter \ref{chap:rg}: critical exponent of the specific heat}
 C_\mathrm{h}(t)\sim |t|^{\alpha}\,.
 \end{equation}
The case $\alpha=0$ can indicate a discontinuity or a logarithmic divergence.
 
The order parameter, for example the magnetization $m$ or the overlap $q$, vanishes as
 \begin{equation}\label{eq:scaling-beta}
\index{exponent!critical!beta@$\beta$}
\nomenclature[m....]{$m$}{In chapter \ref{chap:rg}: magnetization}
\nomenclature[b..3]{$\beta$}{In chapter \ref{chap:rg}: critical exponent of the order parameter}
\hat{m}(t)\sim (-t)^\beta  
 \end{equation}
when we approach the critical temperature from below.

The response to a small external field $h$,\nomenclature[h....h]{$h$}{external magnetic field}
that we call susceptibility, diverges like \index{susceptibility}
\begin{equation}\label{eq:scaling-gamma}
\index{exponent!critical!gamma@$\gamma$}
\nomenclature[gamma....3]{$\gamma$}{In chapter \ref{chap:rg}: critical exponent of the order parameter}
 \chi(t)\sim |t|^\gamma\,.
\end{equation}

If we are exactly at the critical point $t=0$, for small fields $h$ the order parameter behaves as
\begin{equation}\label{eq:scaling-delta}
\index{exponent!critical!delta@$\delta$}
\nomenclature[delta....3]{$\delta$}{In chapter \ref{chap:rg}: field critical exponent}
 m(t=0,h)\sim |h|^{1/\delta}\,,
 \end{equation}
and the correlation length decays with a power law
\begin{equation}\label{eq:scaling-eta}
\index{exponent!critical!eta@$\eta$}
\nomenclature[eta..]{$\eta$}{anomalous dimension}
 \langle s_{\bx+\br} s_\bx \rangle \stackrel{|\br|\rightarrow\infty}{\sim} |\br|^{-(d-2+\eta)}\,,
\end{equation}
and we call $\eta$ the anomalous dimension.

These critical exponents are constrained by a set of four independent \emph{scaling relations},\index{scaling!relations}
\begin{equation}\label{eq:scaling-relations}
\begin{array}{ccc}
  2\beta + \gamma  &=& 2 + \alpha \,,\\[1ex]
 2\beta\delta - \gamma  &=& 2 + \alpha \,,\\[1ex]
 \gamma  &=& \nu(2 - \eta) \,,\\[1ex]
 \nu d &=& 2-\alpha \,,
\end{array}
\end{equation}
that reduce to two the number of independent exponents. The fourth of equations (\ref{eq:scaling-relations}) relates
the exponents to the dimension of space. It is called \emph{hyperscaling} \index{scaling!relations!hyper}
relation and is valid only under the upper critical dimension $d_u$. \index{critical dimension!upper}
From the hyperscaling relation one understands directly that the universality class must depend on dimensionality,\index{universality!class}
since the critical exponents change with $d$.
 \index{exponent!critical|)}
 
\section{Real-space coarse graining}\index{coarse-graining}\index{coherence length|see{correlation length}}\index{correlation!length}
The coherence length $\xi$ represents the size of patches of highly correlated spins. One can think that patches of size $\xi$ interact one with
the other. This concept works very well in ferromagnets \cite{ma:76,huang:87,amit:05}, but though plausible it is still not fully 
developed for disordered systems \cite{harris:76,dotsenko:87,dotsenko:01,angelini:13}.
Following this idea, since $\xi$ is singular at the critical temperature, we can think to construct a block\index{block Hamiltonian}
Hamiltonian that describes the interactions between patches of spins. Let us call $b$ \nomenclature[b....4]{$b$}{In chapter \ref{chap:rg}: size of the patches in the block Hamiltonian}
the linear size of these blocks. Then there will be $L^db^{-d}$ blocks,
each including $b^d$ spins. The block variables $\sigma_\bx$ can be defined as the mean spin in the block
\begin{equation}\label{eq:coarse-graining}
\nomenclature[s..x]{$\sigma_\bx$}{block variable}
 \sigma_\bx = b^{-d} \sum_{\by\in\bx}^{b^d} s_\by\,,
\end{equation}
where the sum runs over all the spins $s_\by$ that belong to the block $\sigma_x$.
The probability distribution for the blocks of spins is 
\begin{eqnarray}
P'[\{\sigma\}] &=&  \left\langle
  \prod_\bx \delta \left( \sigma_\bx - b^{-d} \sum_{\by\in\bx}^{b^d} s_\by \right)
 \right\rangle_P \propto\nonumber\\[1ex]\nonumber
&\propto& \int \E^{-\mathcal{H}[\{s\}]/T} \prod_\bx \delta \left( \sigma_\bx - b^{-d} \sum_{\by\in\bx}^{b^d} s_\by \right) ds_1 ds_2...ds_{N} \equiv\\[2ex]
&\equiv& \E^{-\mathcal{H}_\mathrm{block}[\{\sigma\}]/T}\,,\\\nonumber
\end{eqnarray}
where with $\langle\ldots\rangle_P$ we indicate the average using the equilibrium distribution $P$ of the spins $s_\by$, 
$P=\mathcal{Z}^{-1}\E^{-\mathcal{H}[\{s\}]/T}$, being $\mathcal{Z}$ the partition function. \nomenclature[Z...]{$\Z$}{partition function}
$\mathcal{H}_\mathrm{block}$ is the block Hamiltonian deriving from the coarsening we made, and is equivalent to the original Hamiltonian as long as
we are interested in spatial resolutions larger than $b$. This is our case, since we want to use this procedure to describe diverging length scales.
Once we constructed the blocks once, we can obviously iterate the process, renormalizing each time dynamics variables and Hamiltonian.

\section{Scaling hypothesis and Widom scaling}\index{scaling!Widom}\index{scaling!hypothesis}\index{correlation!length}
The scaling hypothesis,  first conjectured by Widom \cite{widom:65}, is the reasonable assumption that if we have a phase transition with a diverging length $\xi$, then
$\xi$ is the only relevant length. It is model-independent and has been very effective in describing observations. The main
idea is that the singular behavior is completely due to the long-range correlation of spin fluctuations near $T_\mathrm{c}$. 
%We can then decompose the free energy $\mathcal{f}(t,h)$ as a slowly varying regular part $f_r$, and a singular part $f_s$.

To formalize this setting, we assume that when we coarsen the lattice in block variables the free energy remains
unchanged, $\mathcal{F}=\mathcal{F}_\mathrm{coarse}$: \nomenclature[F..coarse]{$\mathcal{F}_\mathrm{coarse}$}{coarse-grained free energy}\index{energy!free}
even though our model is short ranged, we are only interested in the long-range correlations
that arise from being at criticality. 
The renormalized temperature $\tilde{t}$ and field $\tilde{h}$ will have to be rescaled in a consonous way. This rescaling can be written
as
\begin{equation}\label{eq:scaling-rescale}
\begin{cases}
\nomenclature[t....tilde]{$\tilde{t}$}{rescaled reduced temperature}
\nomenclature[h....tilde]{$\tilde{h}$}{rescaled field}
\nomenclature[y....t]{$y_t$}{temperature scaling exponent}
\nomenclature[y....h]{$y_h$}{field scaling exponent}
 \tilde{t} = t\, b^{\,y_t}\\
 \tilde{h} = h\, b^{\,y_h}\,,
\end{cases}
\end{equation}
where $y_t$ and $y_h$ are generic exponents that describe the rescaling.
Using equation \ref{eq:scaling-rescale} and $F=F_\mathrm{coarse}$ we have that the intensive free energy scales as
\begin{equation}\label{eq:f-scale}
\nomenclature[f....t,h]{$f(t,h)$}{intensive free energy}
 f(t,h) = b^{-d} f(\tilde{t},\tilde{h}) =  b^{-d} f(t\, b^{\,y_t},h\, b^{\,y_h})\,.
\end{equation}
To obtain $y_t$ and $y_h$ as a function of the critical exponents we study the behavior of the magnetization $m$, that we can obtain by deriving 
$f$ by the magnetic field:
\begin{equation}\label{eq:scaling-m}
 m(t,h) = \frac{\partial f(t,h)}{\partial h} = b^{-d} \frac{\partial f(t\, b^{\,y_t},h\, b^{\,y_h})}{\partial h} = b^{\,y_h -d}m(t\, b^{\,y_t},h\, b^{\,y_h})\,.
\end{equation}

Since $b$ is an arbitrary scaling parameter, we can set it to grow as any diverging function of $\tilde{t}$ or $\tilde{h}$. If we place ourselves in
the zero-field limit $h=0$ it is convenient to choose $b=(-t)^{-1/y_t}$, so eq.~\ref{eq:scaling-m} becomes
\begin{equation}
 m(t,0) = (-t)^{\,(d-y_h)/y_t} m(-1,0)\,.
\end{equation}
Remembering the definition of the critical exponent $\beta$, that defines that approaching the critical point from below the magnetization goes 
to zero as $m(t)\sim (-t)^\beta$, we can determine the constraint $\beta=(d-y_h)/y_t$. 

We can also study the behavior of the system along the critical curve $t=0$. A helpful choice of $b$ is then $b=h^{-1/y_h}$, in such a way that
\begin{equation}
 m(0,h) = (h)^{\,(d-y_h)/y_h} m(0,1)\,.
\end{equation}
This time we use the definition of $\delta$, that for small $h$ sets the behavior of $m$ along the critical line as $m(0,h)\sim h^{1/\delta}$, and obtain
the constraint $\delta=y_h/(d-y_h)$.

Using equations \eqref{eq:scaling-relations} it becomes possible to reconstruct all the other critical exponents.

\section{Finite-size scaling \label{sec:FSS}}\index{scaling!finite-size|(}\index{phase!transition}
Simulations near $T_\mathrm{c}$ in a lattice of linear size $L$ are usually far from the thermodynamic, due to the extreme growth of the correlation length. 
Finite-size scaling (FSS)\acused{FSS} techniques let us measure properties of the thermodynamic limit by using $L$ as a scaling variable, just like we did with the 
parameter $b$ in the previous paragraphs. It was proposed by Nightingale \cite{nightingale:76} and developed by Binder \cite{binder:82}, and it is 
nowadays the method of choice to study this type of phase transitions (see e.g. 
\cite{binder:86,ballesteros:96,ballesteros:98,ballesteros:00,lee:03,campos:06,jorg:06,leuzzi:08,jorg:08b,hasenbusch:08,fernandez:09,banos:12,janus:13,baityjesi:14,lulli:15}
for applications of \ac{FSS} in the field of \acp{SG}).

\subsection{Spotting the transition}
If an observable $\obs$ 
diverges at the critical temperature as
$\obs\propto|t|^{x_\obs}$, then its thermal average close to the
critical point can be expressed like \index{exponent!critical!nu@$\nu$}\index{exponent!critical!omega@$\omega$}
\begin{eqnarray}
\nomenclature[omega....]{$\omega$}{largest irrelevant exponent}
  \langle \obs(L,T)\rangle &=& L^{x_\obs/\nu} \bigg [
    f_\obs(L^{1/\nu}\big(t)\big) \\\nonumber
    &+&             L^{-\omega}   g_\obs\big(L^{1/\nu} t \big)\\\nonumber
    &+&             L^{-2\omega}  h_\obs\big(L^{1/\nu} t \big) + \ldots \bigg],
  \label{eq:scalingGenerico}
\end{eqnarray}
where $f_\obs,g_\obs$ and $h_\obs$ are analytic scaling functions for observable $\obs$,\nomenclature[f....gh]{$f_\obs,g_\obs,h_\obs$}{scaling functions for an observable $\obs$}
while $\nu$ is defined in equation \eqref{eq:scaling-nu}. The exponent $\omega>0$ is the largest irrelevant exponent. It is
universal, and it expresses the corrections to the dominant scaling. The lower dots, $\ldots$, stand
for subleading corrections to scaling.\nomenclature[...]{$\ldots$}{subleading corrections}

The case $\obs=\xi_L(T)/L$ is of special interest, since $\nu$ is the critical\index{exponent!critical!nu@$\nu$}\index{xiL/L@$\xi_L/L$}
exponent for the correlation length. Then, equation \eqref{eq:scalingGenerico}
becomes in this case, up to the leading-order,
\begin{equation}
  \frac{\xi_L}{L} = f_\xi\big(L^{1/\nu}(t)\big) +\ldots.
  \label{eq:FSS-xi}
\end{equation}
Therefore, we can identify $T=T_\mathrm{c}$ ($t=0$) as the temperature where the curves $\xi_L(T)/L$ cross for all $L$ for sufficiently large $L$.
The same reasoning is valid also for $R_{12}$, defined in equation \eqref{eq:def-R12} \index{R12@$R_{12}$}
\begin{equation}
  R_{12} = f_R\big(L^{1/\nu}(t)\big) +\ldots,
  \label{eq:FSS-R12}
\end{equation}
so $R_{12}$ as well can be used to identify the phase transition, and has the feature of not depending on the susceptibility.

The cumulant $R_{12}$ (recall figure \ref{fig:xiL}) was introduced in
\cite{janus:12} to estimate the critical temperature bypassing pathologies
on $\chi({\bf 0})$ due to the fact that the overlap is non-zero in the
paramagnetic phase \cite{leuzzi:09}.

Note that the value of $\xi_L/L$ and $R_{12}$ at the crossing tends to a non-trivial universal \index{xiL/L@$\xi_L/L$}\index{R12@$R_{12}$}
quantity (see also footnote in section \ref{sec:rg-flow}):
\begin{equation}\label{eq:xi-L-cross}
  \left.\frac{\xi_L}{L}\right|_{T^{L,2L}}=
  \left.\frac{\xi}{L}\right|_{L=\infty}+A_\xi L^{-\omega}+\ldots\,,
\end{equation}
\begin{equation}\label{eq:R12-L-cross}
  \left.R_{12}\right|_{T^{L,2L}}=
  \left.R_{12}\right|_{L=\infty}+A_R L^{-\omega}+\ldots\,.
\end{equation}

If we let $T^{L,2L}$ be the temperature where $\xi_L(T)/L$ crosses 
$\xi_{2L}(T)/(2L)$, \nomenclature[T...L2L]{$T^{L,2L}$}{temperature where $\xi_L(T)/L$ crosses $\xi_{2L}(T)/(2L)$}
this regime is
reached once the $T^{L,2L}$ has converged.\index{exponent!critical!omega@$\omega$}
Yet, if $\omega$ is small, our lattice sizes may not be large enough, so we will have to take in account the aforementioned corrections to scaling.
Including the second-order corrections, the approach of the crossing temperature $T^{L,2L}$ to the asymptotic value $T_\mathrm{c}$
can be written as\index{exponent!critical!nu@$\nu$}
\begin{equation}
  T^{L,2L}-T_\mathrm{c} = A L^{-(\omega+1/\nu)} + B L^{-(2\omega+1/\nu)} + \ldots\,,
  \label{eq:scalingCorrections}
\end{equation}
where $A$ and $B$ are non-universal scaling amplitudes.

\subsubsection{Critical exponents}\index{exponent!critical}
To compute the critical exponents $\nu$ and $\eta$ we use the quotients' method, \index{scaling!finite-size!quotients' method}
taking the quotient of the same observable between different lattice 
sizes $L$ and $2L$. At the temperature $T^{L,2L}$ we get:\index{exponent!critical!nu@$\nu$}\index{exponent!critical!eta@$\eta$}
\begin{equation}
  \frac{\overline{\langle \obs_{2L}(T^{L,2L})\rangle_J}}{\overline{\langle
      \obs_{L}(T^{L,2L})\rangle_J}} = 2^{x_\obs/\nu}+A_{x_\obs} L^{-\omega}+\ldots\,.
\end{equation}
Again, $A_{x_\obs}$ is a non-universal amplitude, while the dots stand for
subleading corrections to scaling.  Therefore, if $\obs$ is the thermal
derivative of $\xi$, we can compute the $\nu$ critical exponent through the
relation
\begin{equation}
  \frac{d\xi_{2L}(T^{L,2L})/dT}{d\xi_{L}(T^{L,2L})/dT} = 2^{1+1/\nu}+A_\nu L^{-\omega}+\ldots\,.
  \label{eq:nu}
\end{equation}
To calculate $\eta$ we use the susceptibility, as $\chi\propto|T-T_\mathrm{c}|^{-\gamma}\sim L^{\gamma/\nu}$. 
Since for the scaling relations \eqref{eq:scaling-relations} 
$2-\eta = \gamma/\nu$, the susceptibility\index{susceptibility}\index{exponent!critical!gamma@$\gamma$}\index{scaling!relations}
at the critical temperature scales as
\begin{equation}
 \label{eq:chi-scaling}
 \chi_L\sim L^{2-\eta}\,,
\end{equation}
so the exponent $\eta$ can be calculated by taking the quotient between sizes $2L$ and $L$
\begin{equation}
  \frac{\chi_{2L}(T^{L,2L})}{\chi_{L}(T^{L,2L})} = 2^{2-\eta}+A_\eta L^{-\omega}+\ldots\,.
  \label{eq:eta}
\end{equation}
Due to the scaling relations \ref{eq:scaling-relations} determining the two exponents $\eta$ and $\nu$ is enough to be able
to estimate them all.
\index{scaling!finite-size|)}

\section{Universality and \acl{RG} flow \label{sec:rg-flow}}\index{renormalization group!flow|(}\index{RG|see{renormalization group}}\index{universality}
The \ac{RG} assumption is that the coarse-graining transformation (\ref{eq:coarse-graining}) will transform smoothly the free energy [equation (\ref{eq:f-scale})], that will
converge to a \index{fixed point|(}\ac{fp} in the space of the rescaled parameters [$\tilde t$ and $\tilde h$ in the case of equation (\ref{eq:f-scale})]. 
That is, when the system is looked at large enough scales, the whole behavior of the system will be given
by the \ac{fp}, that depends in a complicated way on physical parameters such as the temperature $T$, the magnetic field $h$, etc...

Now, the physical parameters can be adjusted in experiments, and can be imposed in calculations, in order to tune the regime in which the system finds itself. In the
space of the (rescaled) parameters, a \ac{fp} will attract the RG trajectories that start in a finite region around it. This region is often a hypersurface in the space 
of the scaling variables. Since all the trajectories of the hypersurface converge to the same \ac{fp}, in the infinite-size limit all these starting points will share
the same behavior. More precisely, the value to which the observables converge will be the same, 
such as $\left.\xi\right|_{L=\infty}$ and $\left.R_{12}\right|_{L=\infty}$,\index{R12@$R_{12}$}\index{xiL/L@$\xi_L/L$}
\footnote{
For systems belonging to the same universality class the correlation function scales as
\begin{equation}
C(\br,L) = \frac{1}{L^{2-\eta-d}}f_C\left(\frac{\br}{L}\right)\,.
\end{equation}
The scaling function $f_C$ is depends on the geometry of the system (ratio between the sides, type of boundary conditions, etc...),
but not on the Hamiltonian (as long as it is dominated by the same \ac{fp}).
If we take the ratio between the Fourier transforms of the correlation function $R_{12}=\frac{\chi(\bk_1)}{\chi(\bk_{11})}$,\index{R12@$R_{12}$}
the divergences even out an it tends to a constant value.
Similarly, for large $L$, $\frac{\xi_L}{L}=\frac{1}{2L\sin{(\pi/L)}}\sqrt{\frac{\chi(\bf{0})}{\chi(\bk_1)}-1}$ 
tends to a constant value, since $L\sin{(\pi/L)}\to\pi$. \index{xiL/L@$\xi_L/L$}

Even though $\left.\xi\right|_{L=\infty}$ is universal, its value is not very interesting, since it diverges. To obtain some non-trivial limit, we can
divide it by some power of $L$. The ratio $\left.\frac{\xi_L}{L^A}\right|_{L=\infty}$
has three limits, two of which are trivial. If $A>1$ we get $\left.\frac{\xi_L}{L^A}\right|_{L=\infty}=0$, while if $A<1$ then $\left.\frac{\xi_L}{L^A}\right|_{L=\infty}=\infty$, no matter the
universality class of the phase transition. Only $A=1$ gives therefore a useful indicator of the universality class, since $\frac{\xi_L}{L}$ tends to a finite value.
}
and the way they converge to this quantity also will coincide, so the critical exponents \index{exponent!critical}
will be the same. The set of all the quantities that are set by the \ac{fp}
is called \index{universality!class}\emph{universality class}. In principle, different models can fall in the same universality class as long as they are dominated by the same
\ac{fp}.

\begin{figure}[!htb]
\centering
\vspace{0.5cm}
 \includegraphics[width=1.0\textwidth]{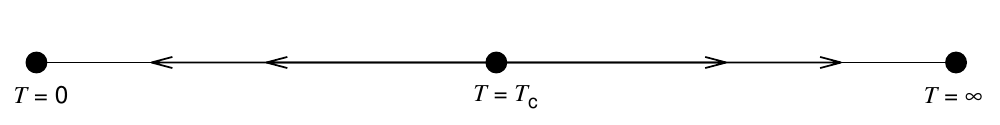}
 \caption[RG flow in the Ising model]{RG flow in the Ising model. The only control parameter is the temperature. There are two stable \acp{fp} at zero and infinite
 temperature, and one unstable \ac{fp} at the critical temperature $T_\mathrm{c}$. The arrows represent the direction of the flow.}
 \label{fig:rg-flow-ising}
\vspace{0.5cm}
\end{figure}

Let us take as an example the Ising model \cite{huang:87}\index{Ising model} with $d>1$ spatial dimensions, 
that has the temperature as only control parameter, and displays a second-order 
phase transition at a temperature $T_\mathrm{c}$. In this case the parameter space is 1 dimensional, so the critical hypersurface is a point. There is a
zero-temperature and an infinite-temperature stable \ac{fp}, respectively governing the behavior of the ferromagnetic and of the paramagnetic phases. By stable we mean 
that the \ac{fp} is attractive, and \ac{RG} trajectories starting from a neighborhood finish in those \ac{fp}s. The two are separated by a \ac{fp}
at $T_\mathrm{c}$ that represents the critical point (figure \ref{fig:rg-flow-ising}).
Any \ac{RG} trajectory starting at $T>T_\mathrm{c}$ will converge to the $T=\infty$ \ac{fp} after a large enough number of coarse-graining steps. Equivalently,
the behavior at $T<T_\mathrm{c}$ is described by the zero-temperature \ac{fp} after the system is coarse-grained enough. Moreover, the fact that the ferromagnetic phase is described
by a \ac{fp} at zero temperature means that neglecting thermal fluctuations is a fair way to treat this phase.
Being the critical \ac{fp} unstable, the only way for a trajectory to converge to it is if it starts at $T=T_\mathrm{c}$.

The ``speed'' of the rescaling is proportional to the distance from the critical temperature $t$ [recall 
the first of (\ref{eq:scaling-rescale})], so the closer we are to $T_\mathrm{c}$
the longer it will take to reach the \ac{fp}. Suppose that starting from $T>T_\mathrm{c}$ we want to reach \index{correlation!length}
a correlation length $\xi_0 = O(1)$, this will take an amount $n(T)$ of \nomenclature[n....T]{$n(T)$}{number of coarse-graining steps}
coarse-graining steps, so $\xi(T) b^{-n(T)}=\xi_0$. The smaller $|t|$, the higher $n(T)$. So, as $T\to T_\mathrm{c}$, $n(T)\to\infty$, 
meaning that $\xi(T_\mathrm{c})\to\infty$,
representing a critical point.

We stress that as long as $t\simeq0$, it will take a very large number of coarsening steps 
before the behavior of the system (for example the size of the correlation length)
start to appear more similar to that of the stable infinite-temperature \ac{fp} (to which 
it will eventually converge) than to that of the unstable critical \ac{fp}.

\subsection{Crossover behaviors \label{sec:crossover}}\index{crossover|(}\index{universality!class}
As pointed out in the previous section, when we find ourselves very close to a critical (unstable) \ac{fp}, 
the system will show for a long time (in terms of coarse-graining steps) 
echoes of that \ac{fp}'s behavior.

\begin{figure}[!htb]
\centering
\includegraphics[width=0.7\textwidth]{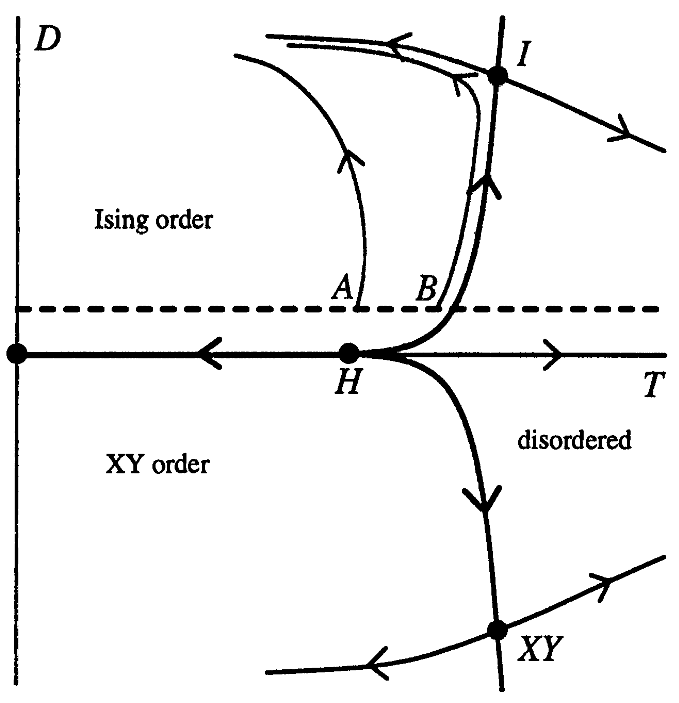}
\caption[Phase diagram of the Heisenberg model with uniaxial anisotropy]{Phase diagram of the Heisenberg model with uniaxial anisotropy. When the anisotropy parameter $D$
is positive the low-$T$ phase has Ising order, and the critical line is dominated by the Ising \ac{fp} (I). Equivalently, for $D<0$ the order is XY-like,
and the critical behavior is XY-like. Only when $D=0$ the critical behavior is Heisenberg-like. The arrows show the direction of the \ac{RG} flow.
See main text for more discussions of the figure.
Figure from \cite{cardy:96}.}
\label{fig:cross-over-cardy}
\end{figure}

To tackle the role of crossover behaviors we make an explicit example.
Let us take in account an $m=3$ Heisenberg \index{spin!Heisenberg}
magnet with single-ion uniaxial anisotropy \index{anisotropy!single-ion uniaxial}
and nearest-neighbor interactions. The Hamiltonian is
\begin{equation}
\nomenclature[H..si]{$\mathcal{H}_\mathrm{si}$}{Hamiltonian of a ferromagnet with single-ion anisotropy}
 \mathcal{H}_\mathrm{si} = -\frac{1}{2}\sum_{\norm{\bx-\by}=1} \vec s_\bx \cdot \vec s_\by - D \sum_\bx s_{\bx,z}^2\,,
\end{equation}
where $s_{\bx,z}$ is the $z$ component of spin $\vec s_\bx$.
The anisotropy term $D$ \nomenclature[D...anisotropy]{$D$}{amplitude of the anisotropy}
splits the Heisenberg $O(3)$ symmetry into a direct product of an XY $O(2)$ and an Ising symmetry.\index{spin!Ising}\index{spin!XY}
When $D=0$ the symmetry of the model is $O(3)$, and the critical behavior is governed by a Heisenberg \ac{fp}. When $D\to+\infty$ the $z$
component is infinitely favored, only configurations with $s_{\bx,z}=\pm1$ ($\forall\bx$) \nomenclature[s....xz]{$s_{\bx,z}$}{$z$ component of the spin in site $\bx$}
are eligible, and the system falls in the Ising universality class.
When $D\to-\infty$ the $z$ component is infinitely suppressed, only configurations with $s_{\bx,z}=0$ are allowed, and the critical behavior is XY.
Thus, in the RG flow diagram that we can draw in the $(T,D)$ plane, there will be three fixed points, Ising, XY and
Heisenberg. 
Figure \ref{fig:cross-over-cardy} gives a qualitative picture of what the phase diagram could look like. Two critical lines will part from
the $D=0$ fixed point.
It is reasonable that the XY and Ising universality classes for $D\neq0$ are maintained along the whole area of the phase diagram where 
the symmetries are broken, so the Ising and XY fixed points will be attractive along the critical lines.

Now, assume we are in a situation of small positive anisotropy $D$. Depending on the temperature we will find ourselves in some part of the dashed line drawn in
figure \ref{fig:cross-over-cardy}. We can start our \ac{RG} flow, for example, from point A, deep in the ferromagnetic phase, or from point B, still in
the ordered phase, but very close to the critical line. Both trajectories will eventually finish in the fixed point that describes the ordered phase. Yet, a trajectory
starting from A will head directly towards the low-temperature fixed point. The one that begins on B instead, will pass very close to the Ising \ac{fp},
and since it is a \ac{fp} it will spend a lot of time near it. As we argued in the previous section, this amount of time diverges as point B approaches the critical line.

This implies that despite the Ising \ac{fp} attracts trajectories that come away from the 
Heisenberg fixed point, when one explores the phase diagram with \index{phase!diagram}
numerical \ac{RG} methods his measurements might be biased by pure echoes of the more 
unstable Heisenberg \ac{fp}.\index{fixed point|)} On the present example there are several ways to try to
avoid this, such as (1) using a large $D$, (2) working very close to the critical 
temperature, (3) or tuning the starting point after having performed some \ac{RG} steps 
(i.e. working on very large lattices). Unfortunately these measures are seldom adoptable. In fact (1) the critical line could exist only for small $D$, so too large
anisotropies would hurl us in the paramagnetic phase, far from the critical line. This is not the case in this example (in chapter \ref{chap:ahsg} we
will succesfully use strong anisotropies to study the critical behavior of the model), but is it, by instance, the case when we deal
with \acp{SG} in a field. In that case the control parameter is the field $h$ instead of $D$, and the critical line is dominated by a fixed point at $T=0$ and finite field \cite{young:97}
(recall figure \ref{fig:dat-line}). Working at large $h$ would yes take us far from the echoes of the $h=0$ fixed point, but there is a large risk of overshootings
that would make the critical line invisible, which is a big problem especially if we are not sure whether it exists or not (see chapter \ref{chap:eah3d}).
One could then rely on working very close to the critical line (2), but this is very hard task when the position (or even the existence) of the critical line is 
unknown, or try to use extremely large lattices, that in \acp{SG} is rarely feasible because of their very sluggish dynamics \cite{janus:08b}.
The solution is to try to tune these three factors in the best possible way and to pay special attention, during the analysis, to the crossover echo effects.

For more quantitaive explanations on this and other crossover behaviors see e.g. \cite{fisher:74} and \cite{cardy:96} (where the previous example is taken from).
\index{renormalization group!flow|)}\index{crossover|)}

\subsection{A note on the distribution of the couplings \label{sec:couplings}}\index{universality!P(Jxy)@$P(J_{\bx\by})$|(}
The first \ac{SG} Hamiltonian, proposed in \cite{anderson:70,edwards:75} (see section \ref{sec:spin-glass-intro}), depended on a set of
coupling $J_{\bx\by}$ that followed a distribution of quenched couplings $P(J_{\bx\by})$. In \cite{anderson:70} $P(J_{\bx\by})$ was meant
to follow roughly the \ac{RKKY} distribution, but already in \cite{edwards:75} this idea was abandoned in favor of a Gaussian distribution,
for sake of simplicity. The \ac{EA} model described successfully the phenomenology of the spin glass, therefore it was kept. There is no
solid argument stating that the $P(J_{\bx\by})$ should be Gaussian rather than, for instance, bimodal.
The reason of this approach is often seen in theoretical physics: one simplifies the model as much as it is possible,
trying to keep track of only the most fundamental traits, so Edwards and Anderson hypothesized that it was important that $P(J_{\bx\by})$
imply frustration, but it did not have to be necessarily the real one (the one that would descend from a renormalization of the \ac{RKKY} couplings),
provided that the $P(J_{\bx\by})$ is ``decent'' enough, and for example it has a finite variance.
\footnote{Even though in this text we did not treat \acp{pdf} with non-zero mean, the mean of $P(J_{\bx\by})$is generally taken as 
a parameter \cite{mezard:87}.}

This said, a very large amount of Hamiltonians were proposed after the \ac{EA} model, and all of them tried to pick the fundamental
aspects, such as disorder, symmetries and range of the interactions, and to neglect what seemed to be unimportant, such as the exact distribution
of the couplings. 
The Gaussian \ac{pdf} has often been chosen, but depending on the context other distributions were used as well.

That these models belong to the same universality class no matter the $P(J_{\bx\by})$ is a natural hypothesis in \ac{SG} theory. 
If it were contradicted there would be no reason to choose one distribution over another, and all the results obtained by \ac{SG}
theory would have a very limited impact.

The general feeling in the \ac{SG} community has always been, indeed, that the precise distribution of the couplings is an unimportant feature
in their description, despite no proof has been given yet. Some doubts arose from numerical works in which different critical exponents
were measured (e.g. \cite{bernardi:95}), but recent careful literature suggests that it is a matter of finite-size effects, and when 
scaling corrections are taken in account the universality is confirmed.

This independence from microscopic details like the disorder distribution has been found for spin glasses \cite{vanhemmen:84,jorg:06, 
katzgraber:06,hasenbusch:08,jorg:08}, but also for other disordered systems such as
  the Random Field Ising model \cite{fytas:13}, \index{random!field Ising model}
  or disordered ferromagnets
  (either site \cite{ballesteros:98b} or bond \cite{berche:04,malakis:12}
  diluted).
\index{universality!P(Jxy)@$P(J_{\bx\by})$|)}

%FINITO
\part{Criticality\label{part:C}}
 \chapter{The Ising spin glass in a field \label{chap:eah3d}}\index{de Almeida-Thouless!transition|(}
This chapter is dedicated to the search of a would-be phase transition in 
a three-dimensional spin glass. The discussion will focus entirely
on the analysis of the data and on the results. We want to stress, in this context, that 
the equilibrium \ac{MC} simulations performed in \cite{janus:14c} required huge numerical
efforts.
On the one hand, lower temperatures drastically increased thermalization times; on the other hand, 
the significance of the results depends upon the size of the systems we are able to simulate.

The problem of enhancing the reach of our simulations is faced by resorting to advanced algorithms and techniques,
such as \ac{PT} \index{Monte Carlo!parallel tempering}
\footnote{A short discussion on \ac{PT} is given in appendix \ref{sec-app:algorithms}.}
and \ac{MSC},
\footnote{In appendix \ref{app:MSC} we describe how \ac{MSC} was implemented in the analysis stage. Multi-spin coding
\ac{MC} in the simulations \cite{seoane:13} follows roughly the same principles than in the analyses.
}
but that is still not enough.
It would not have been possible to attain the results published in \cite{janus:14c} with the mere use of ordinary \index{Janus@\textsc{Janus}!computer}
computational resources. We drew upon \ac{HPC} on one side by making use of the \textsc{Janus} dedicated computer to simulate
the largest lattices, and on the other by simulating the smaller systems on a large \ac{CPU} cluster, \emph{Memento}.

\section{The de Almeida-Thouless line in three dimensions}
\index{spin!Ising|(}\index{droplet picture}\index{replica!symmetry!breaking!scenario}\index{spin glass!Edwards-Anderson!in a field|(}
In section \ref{sec:spin-glass-intro} we explained that the nature of the \ac{SG} phase in 
three dimensions it is still matter of debate. The two dominant theories are the droplet picture
and the \ac{RSB} scenario, and they have different predictions on the presence of a \ac{SG}
phase in a field. In the droplet picture even the smallest applied magnetic field destroys the 
\ac{SG} phase, while in the \ac{RSB} scenario there is a \ac{dat} line $h_\mathrm{c}(T)$ that separates
the \ac{SG} from the paramagnetic phase.

A rather obvious way out would be the experimental study of spin
glasses in a field. Unfortunately, opposing indications have been
gleaned over the existence of a phase transition
\cite{jonsson:05,petit:99,petit:02,tabata:10}.

The \ac{RG} approach to this problem also provides
conflicting results. No \acp{fp} were found by enforcing that the
number of replicas of the replicated field theory be
zero~\cite{bray:80b}. However, \acp{fp} were found relaxing this
condition and using the most general
Hamiltonian \cite{temesvari:02}. Reasoning along this line, in
\cite{temesvari:08} (see also \cite{parisi:12}) the \ac{dat} line was computed for $d$ slightly below the upper critical dimension
$d_\mathrm{u}=6$ (the upper critical dimension remains 6 when \index{critical dimension!upper}
an external magnetic field is applied).

Equilibrium numerical simulations offer an alternative approach, which
has already been effective in establishing that a phase transition
does occur at zero field in the $d=3$ Edwards-Anderson model
\cite{palassini:99,ballesteros:00} (in agreement with
experiments~\cite{gunnarsson:91}). The same strategy has been followed
for $h>0$, with negative results \cite{young:04,jorg:08b}. Yet, this
cannot be the whole story: Recent work in $d=4$, hence below
$d_\mathrm{u}$, using a non-standard finite-size scaling method has
found clear evidence for a \ac{dat} line \cite{janus:12}. Furthermore, one
may try to interpolate between $d=3$ and $d=4$ by tuning long-range
interactions in $d=1$ chains \cite{kotliar:85, leuzzi:08}. This
approach suggests that a \ac{dat} might be present in $d=4$, but not in
$d=3$ \cite{larson:13} (yet, see the criticism in \cite{leuzzi:13}).

The problem being still open, in \cite{janus:14b} we undertook a
dynamical study of the 3-dimensional \ac{EA} spin glass with
the \textsc{Janus} dedicated computer \index{Janus@\textsc{Janus}!computer}
\cite{janus:06,janus:08,janus:09,janus:10,janus:11,janus:12b}. 
We studied very large lattices ($L^3=80^3$), in wide time scales 
(from an equivalent of $\sim1$~ps to $\sim0.01$~s), and gathered 
both equilibrium and non-equilibrium data. We focused on the increase 
of relaxation times and found a would-be dynamical transition, but at \index{relaxation time}
a suspiciously high temperature. A subsequent examination of the correlation 
length found a growth faster than predicted by the droplet theory, and 
slower than what \ac{RSB} would expect. We also examined the problem from a 
supercooled liquid point of view 
\cite{debenedetti:97,debenedetti:01,cavagna:09,castellani:05,kirkpatrick:87,kirkpatrick:89},\index{universality!class} \index{glass!structural}
motivated by the equivalence of universality classes between spin and structural glasses \cite{moore:02,fullerton:13}. At any rate, the study 
of the possible critical divergence of the correlation length allowed \index{correlation!length}
us to give upper bounds $T^\mathrm{up}(h)$ \nomenclature[T...up]{$T^\mathrm{up}(h)$}{upper bound for the critical temperature}
to the possible transition 
line for the studied fields. 

The impossibility to get concluding evidence in \cite{janus:14b},
may be due to the fact that we did not reach low enough temperatures
(our simulations fell out of equilibrium at temperatures $T$
significantly higher than $T^\mathrm{up}(h)$). In any case, a study of
the equilibrium properties of the model is mandatory if one wants to
understand the nature of the thermodynamic phases of the
three-dimensional \ac{EA} spin glass in a field.

In this dissertation we will not talk about the aforementioned out-of-equilibrium results \cite{janus:14b}.
We will instead focus on the result of equilibrium simulations performed on
\textsc{Janus}, using lattices up to $L=32$ \cite{janus:14c}.\index{Janus@\textsc{Janus}!computer}
\footnote{In \cite{janus:14b} we studied a bimodal field, while
in the work we present here $h$ is constant. Notwithstanding, we will make
comparisons with the bounds $T^\mathrm{up}(h)$ by matching
$\overline{h^2}$ in both models.} For further reference we recall
that $T^\mathrm{up}(h=0.1)=0.8$ and
$T^\mathrm{up}(h=0.2)=T^\mathrm{up}(h=0.3)=0.5$.
Analogously to what has been
already found in mean-field spin glasses on the \ac{dat} line, we
find extreme fluctuations in the model's behavior \cite{parisi:12b}. We will
propose a method to tame these fluctuations, and we will find out that,
although the average behavior does not show any sign of a phase transition,
this is not true for the medians of our observables, where we have indications
of a possible phase transition at a temperature $T_\mathrm{c}\lesssim T^\mathrm{up}(h)$.
\index{de Almeida-Thouless!transition|)}

\section{Model and simulations}\label{sec:eah3d-model-simulations}
\subsection[The \texorpdfstring{$3d$}{3d} \acl{EA} model in a field]
           {The \texorpdfstring{$\bm{3d}$}{3d} \acl{EA} model in a field}\label{sec:eah3d-model}
We consider a $3d$ cubic lattice of linear size $L$ with periodic boundary conditions. 
In each of the $N=L^3$ vertices of the lattice there is an Ising spin \index{spin!Ising}
$s_\bx = \pm1$. The spins interact uniquely with their nearest
neighbors and with an external magnetic field $h$. The Hamiltonian is
\begin{equation}
\index{spin glass!Edwards-Anderson!in a field!Hamiltonian}
\nomenclature[H..EAh]{$\mathcal{H}_{h}$}{Edwards-Anderson Hamiltonian in a field}
\label{eq:eah3d-ham}
  \mathcal{H}_{h} = - \frac{1}{2}\sum_{\norm{\bx-\by}=1}J_{\bx \by} s_\bx s_\by 
              - h\sum_\bx s_\bx\,,
\end{equation}
where the couplings $J_{\bx \by}$, which are constant during each simulation, take the values
$\pm1$ with equal probability (quenched disorder). As already stated in chapter \ref{sec:obs}, a given instance of
the bonds $J_{\bx \by}$ and of the intensity of the magnetic
field $h$ define a \emph{sample}.  We will consider real \index{replica!real}
\emph{replicas} of each sample, i.e., systems with identical couplings
$J_{\bx \by}$ and field $h$, but independent evolutions (for a
recent discussion see \cite{janus:09b} and \cite{janus:10}). In
this work we will use 4 replicas per sample.

\subsection{The simulations}\label{sec:eah3d-simulations}
For all our simulations we made use of \ac{PT}. \index{Monte Carlo!parallel tempering}\index{Monte Carlo!simulations}
\footnote{See the short note in appendix \ref{sec-app:algorithms}.} 
The whole procedure was very similar to the one in \cite{janus:12}.

\begin{table}[!htb]
\begin{tabular*}{\columnwidth}{@{\extracolsep{\fill}}ccccccccc}
$h$ & $L$ & $N_\mathrm{samples}$ & $N_\mathrm{EMCS}^{\mathrm{min}}$& $f_\mathrm{max}$ & $N^\mathrm{min}_\tau$ &$N_\mathrm{T}$  & $T_\mathrm{min}$ & $T_\mathrm{max}$\\\hline\hline
0.05  &  6  & 25600 &  $1.6\times10^6$ &   1 & 40.0 & 14 & 0.5 & 1.8 \\
0.05  &  8  & 25600 &  $3.2\times10^6$ &  16 & 40.0 & 14 & 0.5 & 1.8 \\
0.05  & 12  & 25600 &  $3.2\times10^6$ &  16 & 15.6 & 12 & 0.7 & 1.8 \\
0.05  & 16  & 12800 & $1.28\times10^7$ & 128 & 20.1 & 24 & 0.6 & 1.75 \\
0.05  & 24  &  6400 & $1.28\times10^7$ & 110 & 16.0 & 20 & 0.78 & 1.54 \\
0.05  & 32  &  2400 & $6.4\times10^7$ & 256 & 14.3 & 30 & 0.805128 & 1.54872 \\\hline
0.1   &  6  & 25600 &  $1.6\times10^6$ &   4 & 40.0 & 14 & 0.5 & 1.8 \\
0.1   &  8  & 25600 &  $3.2\times10^6$ &  16 & 40.0 & 14 & 0.5 & 1.8 \\
0.1   & 12  & 25600 &  $3.2\times10^6$ &  16 & 14.4 & 12 & 0.7 & 1.8 \\
0.1   & 16  & 12800 & $1.28\times10^7$ & 256 & 27.9 & 24 & 0.6 & 1.75 \\
0.1   & 24  &  3200 & $1.28\times10^7$ &4097 & 14.3 & 24 & 0.66 & 1.58 \\
0.1   & 32  &  1600 & $6.4\times10^7$ & 533   & 14.4    & 30 & 0.805128 & 1.54872 \\\hline
0.2   &  6  & 25600 &  $1.6\times10^6$ &   1 & 40.0 & 14 & 0.5 & 1.8 \\
0.2   &  8  & 25600 &  $3.2\times10^6$ &  16 & 40.0 & 14 & 0.5 & 1.8 \\
0.2   & 12  & 25600 &  $3.2\times10^6$ &  64 & 25.4 & 12 & 0.7 & 1.8 \\
0.2   & 16  & 12800 & $1.28\times10^7$ & 256 & 18.4 & 24 & 0.6 & 1.75 \\
0.2   & 24  &  3200 & $1.28\times10^7$ & 512 & 16.1 & 24 & 0.66 & 1.58 \\
0.2   & 32  &  1600 & $1.6\times10^7$ & 513 & 16.0 & 30 & 0.805128 & 1.54872 \\\hline
0.4   &  6  & 25600 &  $1.6\times10^6$ &   1 & 40.0 & 14 & 0.5 & 1.8 \\
0.4   &  8  & 25600 &  $3.2\times10^6$ &   4 & 30.7 & 14 & 0.5 & 1.8 \\
0.4   & 12  & 25600 &  $3.2\times10^6$ &  16 & 14.1 & 12 & 0.7 & 1.8 \\
0.4   & 16  &  3200 & $1.28\times10^7$ &  32 & 20.1 & 24 & 0.6 & 1.75 \\
0.4   & 24  &   800 & $1.28\times10^7$ &  29 & 16.1 & 24 & 0.66 & 1.58 \\
0.4   & 32  &   800 &  $3.2\times10^6$ &  16 & 16.4 & 30 & 0.805128 & 1.54872 \\\hline
\end{tabular*}
 \caption[Parameters of the simulations of \cite{janus:14c}]{Parameters of the simulations. We report the magnetic field
   $h$, the lattice linear size $L$, the number of simulated samples
   $N_\mathrm{samples}$, \nomenclature[N...samples]{$N_\mathrm{samples}$}{number of samples}
   and the basic length of a simulation in
   \ac{EMCS} $N_\mathrm{EMCS}^{\mathrm{min}}$.  In
   each simulation we measured the exponential correlation time $\tau$\index{relaxation time}\index{thermalization}
   of the \ac{PT} random walk in temperatures.  When $\tau$ was too large
   to meet our thermalization requirements, we extended the length of
   each simulation by an extension factor $f$.  We denote with
   $f_\mathrm{max}$ the greatest extension factor.  We also give the
   minimum length of a simulation $N^\mathrm{min}_\tau$ in units of
   $\tau$.  In all cases we imposed $N^\mathrm{min}_\tau>14$.
   Finally, we give the number of temperatures
   $N_T$ we used for the \ac{PT}, and the minimum and maximum temperatures\index{Monte Carlo!parallel tempering}
   $T_\mathrm{min}$ and $T_\mathrm{max}$.  }
\label{tab:nsamples-eah3d}
\end{table}

The smaller lattices ($L=6,8,12$) were simulated 
with \ac{MSC} (C code with words of 128 bits, by means of streaming \index{multi-spin coding}
extensions) \cite{newman:99,janus:12,seoane:13} on the \emph{Memento} CPU cluster 
at BIFI. See details on \ac{MSC} in appendix \ref{app:MSC}.
The larger samples ($L=16,24,32$) were
simulated on the \textsc{Janus} computer \cite{janus:06,janus:12b}. \index{Janus@\textsc{Janus}!computer}

An \ac{EMCS} consisted in 1 \ac{PT} exchange every 10 Metropolis steps for the \ac{MSC}\index{Monte Carlo!elementary step}
samples, and 1 \ac{PT} every 10 \ac{HB} for the samples simulated on \textsc{Janus}.
table \ref{tab:nsamples-eah3d} shows the relevant parameters of the simulations. The
temperatures were equally spaced between $T_\mathrm{min}$ 
and $T_\mathrm{max}$.\nomenclature[T...minmax]{$T_\mathrm{min},T_\mathrm{max}$}{lowest and highest simulated temperatures}
The intensities of the external magnetic field we chose are $h=0.05,0.1,0.2$ and $0.4$.

To check whether the samples were thermalized we measured the exponential \index{thermalization|(}
autocorrelation time of the \ac{PT} random walk in temperatures\index{relaxation time}\index{autocorrelation time|see{relaxation time}}
\index{self-correlation time|see{relaxation time}} $\tau$ \cite{fernandez:09b,janus:10,yllanes:11,janus:12}.
\nomenclature[tau....4]{$\tau$}{In chapter \ref{chap:eah3d}: exponential autocorrelation time}
 We required the simulations to last at least
$14\tau$. 
To do so without consuming computing time on already thermalized
lattices, we assigned a minimum number of \ac{EMCS}, $N_\mathrm{EMCS}^{\mathrm{min}}$,\nomenclature[N...emcs]{$N_\mathrm{EMCS}^{\mathrm{min}}$}{minimum number of EMCS}
for all the samples, and extended by a factor $f>1$ only the ones that did not \nomenclature[f....f]{$f$}{extension factor of the simulations}
meet the imposed thermalization criterion.  In table \ref{tab:nsamples-eah3d} we
report $N_\mathrm{EMCS}^{\mathrm{min}}$, the maximum extension factor
$f_\mathrm{max}$ \nomenclature[f....max]{$f_\mathrm{max}$}{maximum extension factor of the simulations}
of the simulations, and minimum number $N^\mathrm{min}_\tau$ \nomenclature[N...tau]{$N^\mathrm{min}_\tau$}{minimum number of EMCS in units of $\tau$}
of \ac{EMCS} in units of $\tau$.

Equilibrium measurements were taken offline over the second half of each
simulation. Independently of how much the simulations were extended,
we saved $N_\mathrm{m}=16$ \nomenclature[N...m]{$N_\mathrm{m}$}{number of offline measurements}
equally time-spaced configurations and performed
measurements on them. We measured four-replica observables. Therefore,
for each sample it was possible to choose quadruplets of
configurations, each from a different replica, in $N_\mathrm{m}^4$ ways.  Out of
the $N_\mathrm{m}^4$ possibilities, we chose randomly $N_\mathrm{t} = 1000$\nomenclature[N...t]{$N_\mathrm{t}$}{number of quadruplets of configurations}
combinations.  In other words, each sample participated in the
statistics with $N_\mathrm{t}=1000$ measurements.
\index{thermalization|)}

The errors were estimated with the jackknife method (appendix \ref{app:eah3d-errors}).\index{errors!jackknife}

\section{Giant fluctuations and the silent majority}
\label{sec:extended-abstract}

\subsection{No signs of a phase transition with common tools}\index{de Almeida-Thouless!transition}\index{scaling!finite-size}
A common way to locate a phase transition is to proceed as described in section \ref{sec:FSS},
by locating the temperature where the curves $\frac{\xi_L}{L}(T)$ and $R_{12}(T)$ of different lattice sizes cross.\index{xiL/L@$\xi_L/L$}\index{R12@$R_{12}$}
For sufficiently large systems, if the curves do not cross, there is no phase transition in the simulated temperature range.

\begin{figure}[!htb]
\centering
\vspace{2cm}
\includegraphics[width=\columnwidth]{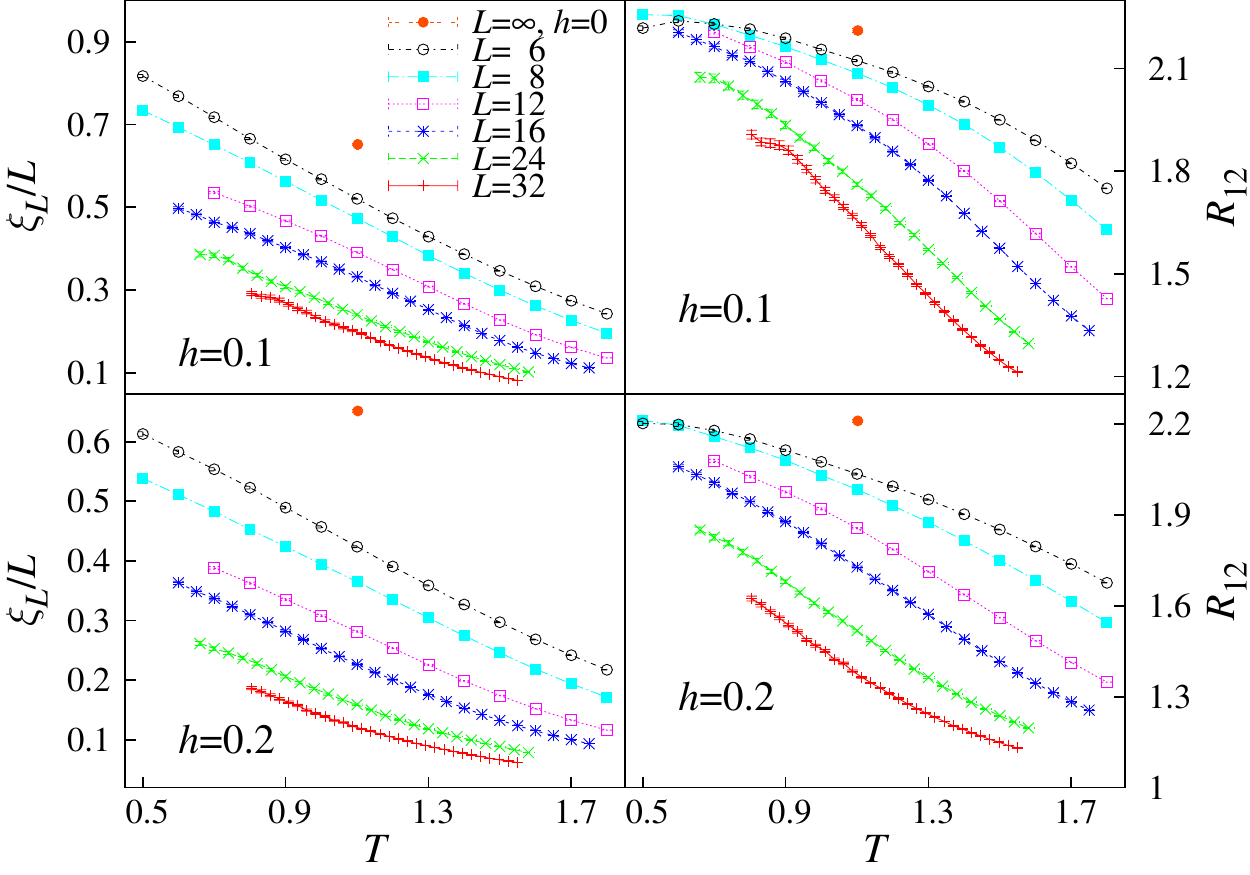}
\caption[Absence of crossings for the average $\xi_L(T)/L$ and $R_{12}(T)$]
  {The figures on the \textbf{left} show the standard correlation length $\xi_L$ in units of the\index{xiL/L@$\xi_L/L$}
  lattice size $L$ as a function of the temperature $T$, for all our lattice
  sizes.  The magnetic fields are $h=0.1$ (\textbf{top}), and $h=0.2$ (\textbf{bottom}). If
  the lattices are large enough, in the presence of a second-order phase
  transition, the curves are expected to cross at a finite temperature
  $T_\mathrm{c}(h)$. 
  The figures on the \textbf{right} show the cumulant $R_{12}$, \index{R12@$R_{12}$} which in the presence of
  a magnetic field is a better indicator of a phase transition \cite{janus:12},
  for the same magnetic fields.
  At zero field the heights of the crossings (which are universal
  quantities) are indicated
  with a	 point at $T_\mathrm{c}=1.1019(29)$.\index{universality!class}
  They are $\xi_L/L(h=0;T_\mathrm{c})=0.6516(32)$ and $R_{12}(h=0;T_\mathrm{c})=2.211(6)$ \cite{janus:13}.
  In neither case we observe signs of a crossing at the simulated
  temperatures, nor can we state that the curves will cross at lower
  temperature.
  The reader might remark that the curve for $L=32$, $h=0.1$ is not as smooth as
  one would expect from parallel tempering simulations. The reason is twofold.
  On one side the number of simulated samples is much smaller than for $L<32$, 
  and on the other side temperature chaos, which is stronger\index{temperature chaos}
  the larger the lattice, is probably present \cite{fernandez:13}.}
\label{fig:xiL}
\end{figure}

In the present case, this type of analysis yields a clear result: there is no evidence
of a crossing at the simulated temperatures, magnetic fields and sizes. This is
clearly visible from figure \ref{fig:xiL}, where the curves $\frac{\xi_L}{L}(T)$ and $R_{12}(T)$ \index{xiL/L@$\xi_L/L$}\index{R12@$R_{12}$}
should have some crossing point if we were in the presence of a phase transition. 
This is in complete qualitative agreement with earlier works on this model \cite{young:04,jorg:08b}.

\afterpage{\clearpage}

\subsection{A hidden behavior}\index{large fluctuations}
Although $\frac{\xi_L}{L}(T)$ is smaller the larger the lattice size, the\index{correlation!length}\index{correlation!length@length$ $|seealso{xiL/L}}
coherence length $\xi_L$ grows significantly even for our largest
lattice sizes. For example at $h=0.2$, $T=0.81$ we have
$\xi_{16}=6.09(4)$, $\xi_{24}=7.63(9)$ and $\xi_{32}=9.0(2)$.  The
noticeable size evolution implies that the asymptotic correlation
length $\xi_\infty$ is large compared with $L=32$.

Also, we can examine the behavior of the spin-glass order parameter,
the overlap $q$, by studying its distribution function $P(q)$.
In the absence of a phase
transition we would be in the paramagnetic phase, and $P(q)$ should be a
delta function of a positive overlap $q_\mathrm{EA}$ (so in finite systems it should
be Gaussian).

\begin{figure}[!htb]
 \includegraphics[width=\columnwidth]{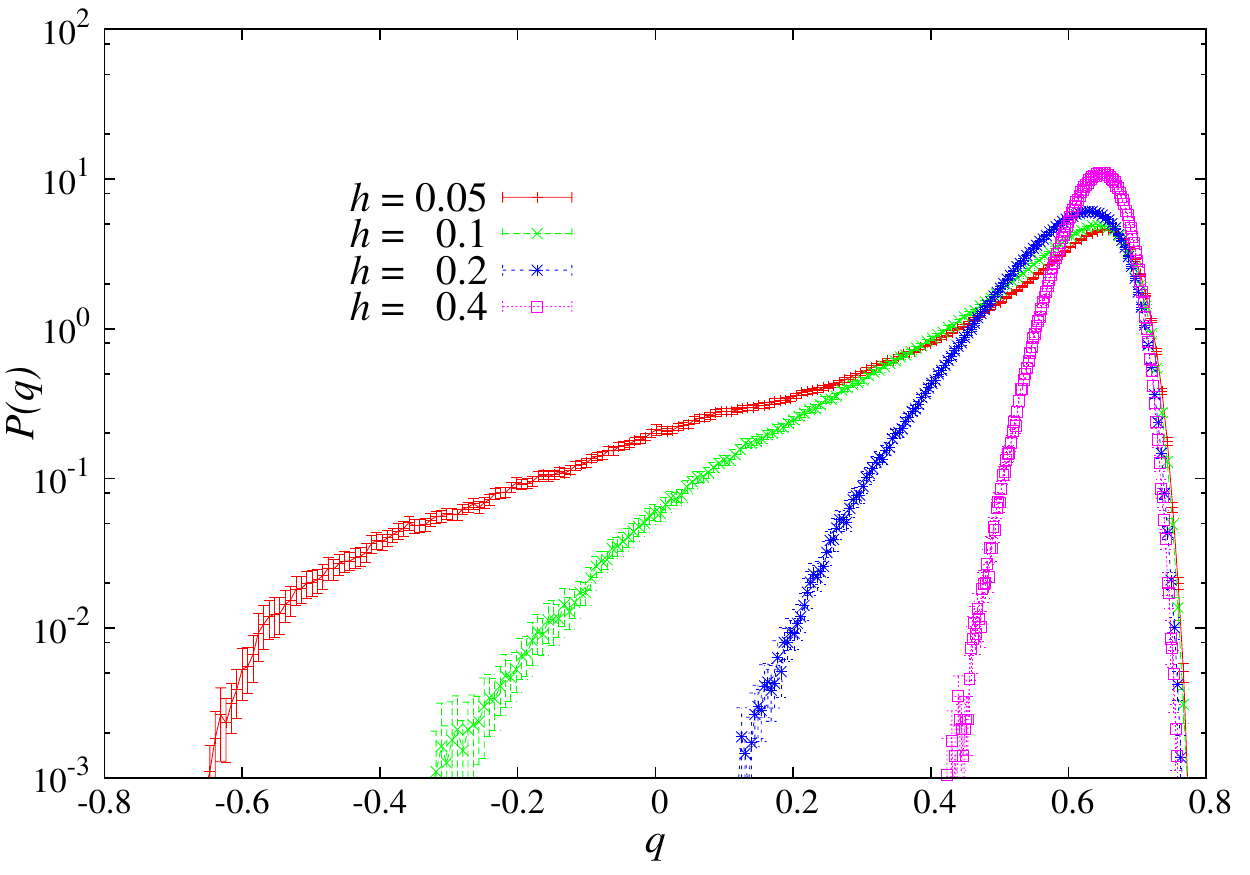}
\caption[Overlap \acs{pdf} for several fields]{The \ac{pdf} $P(q)$ of the overlap
  $q$, for our largest lattices ($L=32$) at the lowest simulated
  temperature ($T=0.805128$), for all our magnetic fields
  ($h=0.05,0.1,0.2,0.4$), see table~\ref{tab:nsamples-eah3d}.  The order
  parameter in the \ac{EA} model is the overlap $q$, and it
  is defined in the $[-1,1]$ interval (see section~\ref{sec:overlaps}).  The
  supports are wide, with exponential tails similar to those in the
  mean-field model at the \ac{dat} transition line \cite{parisi:12b}.}
\label{fig:Pq-eah3d}
\end{figure}

Instead, we can see from figure \ref{fig:Pq-eah3d} that its distribution
$P(q)$ has a very wide support, with tails that, for small enough
magnetic fields, reach even negative values of $q$. This is precisely
what was observed in the mean-field version of the model on the de
Almeida-Thouless line, and it was attributed to the contribution of
few samples \cite{parisi:12b}.

From these arguments it becomes reasonable to think that we may not be simulating large enough
lattices to observe the asymptotic nature of the system and that there may be
some hidden behavior that we are not appreciating.

\subsection{Giant fluctuations}
In fact, we find out that the average values we measure are
representative  of only a small part of the data set.  That is, the
average of relevant observables (e.g., the spatial correlation
function) only represents the small number of measurements that are
dominating it. The rest of the measurements is not appreciated by using
the average.

\begin{figure}[!b]
\includegraphics[width=\columnwidth]{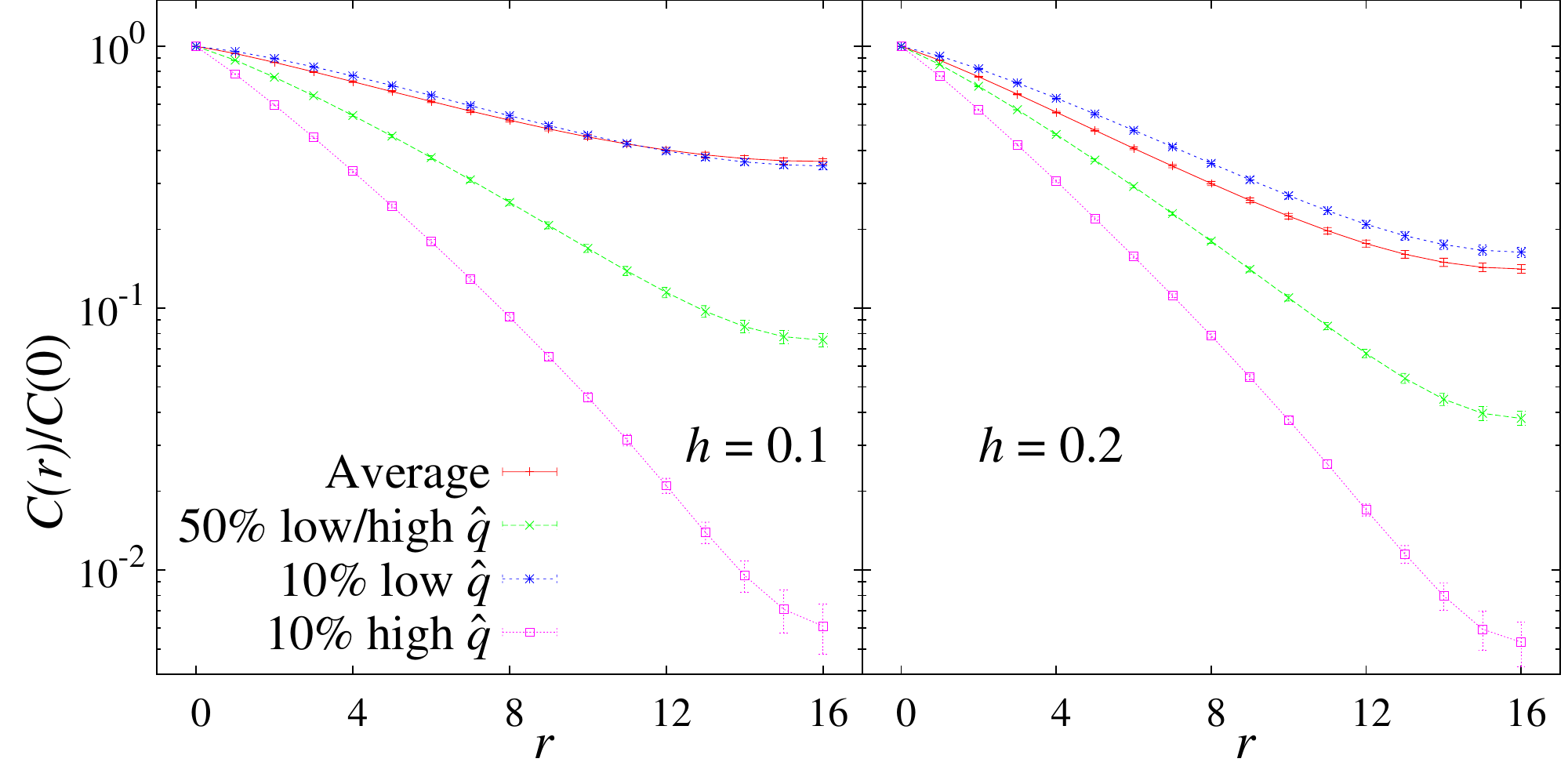}
\caption[Fluctuations in the correlation function $C(r)$]{Different instances of \index{correlation!function!plane}
the normalised plane correlation function $C(r)$ (\ref{eq:Crplane-chi}) for $L=32$, $T=0.805128$.
The field is $h=0.1$ on the \textbf{left}, and $h=0.2$ in the \textbf{right} plot.
We sort the measurements with the help of a conditioning variate\index{variate!conditioning} $\hat q$ as described in section~\ref{sec:conditional}. 
In this case $\hat{q}$ is the median overlap $q_\mathrm{med}$.
We show small sets of measurements. Namely, the ones with the $10\%$ lowest (top curve) and
highest (bottom curve) $\hat{q}$ and those whose $\hat q$ corresponds to the median
of the distribution of $\hat q$ ($50\%$ lowest/highest $\hat{q}$).
This sorting reveals extreme differences in
the \emph{fauna} of measurements.
The average and median of the correlation
functions are very different. The average is very similar to the $10\%$ lowest
ranked measures, i.e., it is only representative of a very small part of the
data. 
We normalise $C(r)$ by dividing by $C(0)$ because we measure
point-to-plane correlation functions (\ref{eq:Crplane-chi}).
The correlation functions have zero slope at $r=L/2$ due to the
periodic boundary conditions.}
\label{fig:Cr}
\end{figure}

Clearly, standard finite-size scaling methods are not adequate to these \index{scaling!finite-size}
systems, and we need to find a way to take into account \emph{all} the 
measurements. Recalling the wide distributions of figure~\ref{fig:Pq-eah3d}, 
it seems reasonable to sort our measurements according to some conditioning 
variable $\hat q$ \index{variate!conditioning}
 related to the overlaps between our replicas 
(see section \ref{sec:conditional}). This way, we find out that the
average values we measure are given by only a small part of the measurements. For
example in figure \ref{fig:Cr} we show the correlation function $C(r)$. We plot 4
estimators of $C(r)$: the average (which is the standard quantity studied in almost
all, if not all, previous work), the $C(r)$ that corresponds 
to the median of the $\hat q$ distribution, 
and the measurements with the $10\%$ highest (lowest) value
of $\hat{q}$. We see that the
average is very close to the $10\%$ lowest $\hat q$, and very far from the two other
curves. So, when we plot the average curve, we are only representing the
behavior of that small set of data.

Therefore, if we want to understand the behavior of the \emph{whole} collection of measurements, we
have to be able to find some criterion to sort them and analyse them separately.

\section{Conditional expectation values and variances}
\label{sec:conditional}

\subsection{The \acl{CV}} \index{variate!conditioning|(}\index{large fluctuations}
As we pointed out in section~\ref{sec:extended-abstract}, the behavior of the system 
is dominated by a very small number of measurements.

This means that the average over all the measurements of an
observable does not describe the typical behavior of the system. 
Furthermore, the behavior of the measurements that contribute less to the full 
averages is qualitatively different from the one of those who give the 
main contribution (see figure~\ref{fig:Cr} and later on section~\ref{sec:eah3d-results}).

We want to classify our measurements in a convenient way, in order to be able
to separate different behaviors, and analyse them separately. To
this goal, we replace normal expectation values $E({\cal{O}})$ of a generic \index{conditional expectation|(}
observable ${\cal O}$, with the expectation value $E({\cal{O}}|\hat{q})$ \nomenclature[E...Oq]{$E({\cal{O}}|\hat{q})$}{expectation value conditioned to $\hat q$}
conditioned to another random variable $\hat q$. 
Perhaps for lack of imagination $\hat q$ will be named \acf{CV}.\nomenclature[q....hat]{$\hat q$}{conditioning variate}
For each instance of
${\cal O}$ we monitor also the value of $\hat q$, and we use it to label ${\cal O}$. 
Hopefully, there will be some correlation.

The conditional expectation value is defined as the average of ${\cal O}$,
restricted to the measurements $i$ (out of the ${\cal N_\mathrm{m}}=N_\mathrm{t}N_\mathrm{samples}$ 
total \nomenclature[N..m]{${\cal N_\mathrm{m}}$}{total number of measurements}
measurements) that simultaneously yield ${\cal O}_i$ and \nomenclature[O...i]{${\cal O}_i$}{$i^\mathrm{th}$ measurement of observable $\mathcal{O}$}
$\hat q_i$ \nomenclature[q....ihat]{$\hat q_i$}{value of the CV at the $i^\mathrm{th}$ measurement} [so we are actually talking
about couples of simultaneous measurements $({\cal O}_i,\hat q_i)$]
in a small interval around $\hat q=c$, 
\begin{equation}
 E({\cal O}|\hat q = c)~=~  \frac{E\left[{\cal O}_i {\cal X}_{\hat q = c}(\hat q_i)\right]}{E\left[{\cal X}_{\hat q = c}(\hat q_i)\right]}\,.
\end{equation}
Where we have used the characteristic function
\begin{equation}\label{eq:characteristic}
\nomenclature[X..cqi]{${\cal X}_{c}(\hat q_i)$}{characteristic function}
  {\cal X}_{c}(\hat q_i) = \left\{
\begin{array}{rl}
1,& ~\mathrm{ if }~ |c - \hat q_i| < \epsilon \sim \frac{1}{\sqrt{V}} \\
0,& ~\mathrm{ otherwise}.
\nomenclature[epsilon....5]{$\epsilon$}{In chapter \ref{chap:eah3d}: width of the interval}
\end{array}
\right.
\end{equation}
In appendix~\ref{app:hack} we give technical details on the choice of
$\epsilon$. To make notation lighter, in the rest of the paper we will replace $E({\cal O}|\hat q = c)$
with $E({\cal O}|\hat q)$. \nomenclature[E...oq]{$E({\cal O}|\hat q)$}{$E({\cal O}|\hat q = c)$}

The traditional expectation value $E({\cal{O}})$ can be recovered by integrating
over all the possible values of the \ac{CV} $\hat{q}$:
\begin{equation}
 \label{eq:E-conditioned}
 E({\cal{O}}) = \int d\hat{q}\ E({\cal O}|\hat{q}) P(\hat{q})~~,~~ P(\hat q) = E[{\cal X}_{\hat q}]
~,
\end{equation}
where $P(\hat q)$ is the probability distribution function of the
\ac{CV}.\nomenclature[P...qhat]{$P(\hat q)$}{probability distribution function of the CV}
\index{conditional expectation|)}

We remark that the concept of \ac{CV} is fairly similar to the one of
control-variate. \index{variate!control}
Yet, the latter was formalised slightly differently, and with
the objective of enhancing the precision of the measures \cite{fernandez:09c}. In
\cite{janus:10,janus:10b} a procedure very similar to the present
one was followed, but the aim was constructing clustering correlation
functions, while in our case the \ac{CV} is used to analyse
\emph{separately} different behaviors outcoming from the same global data set,
so that a sensible finite-size scaling becomes possible.

\subsection{Measurements against samples}

\begin{figure}[!htb]
 \includegraphics[width=\columnwidth]{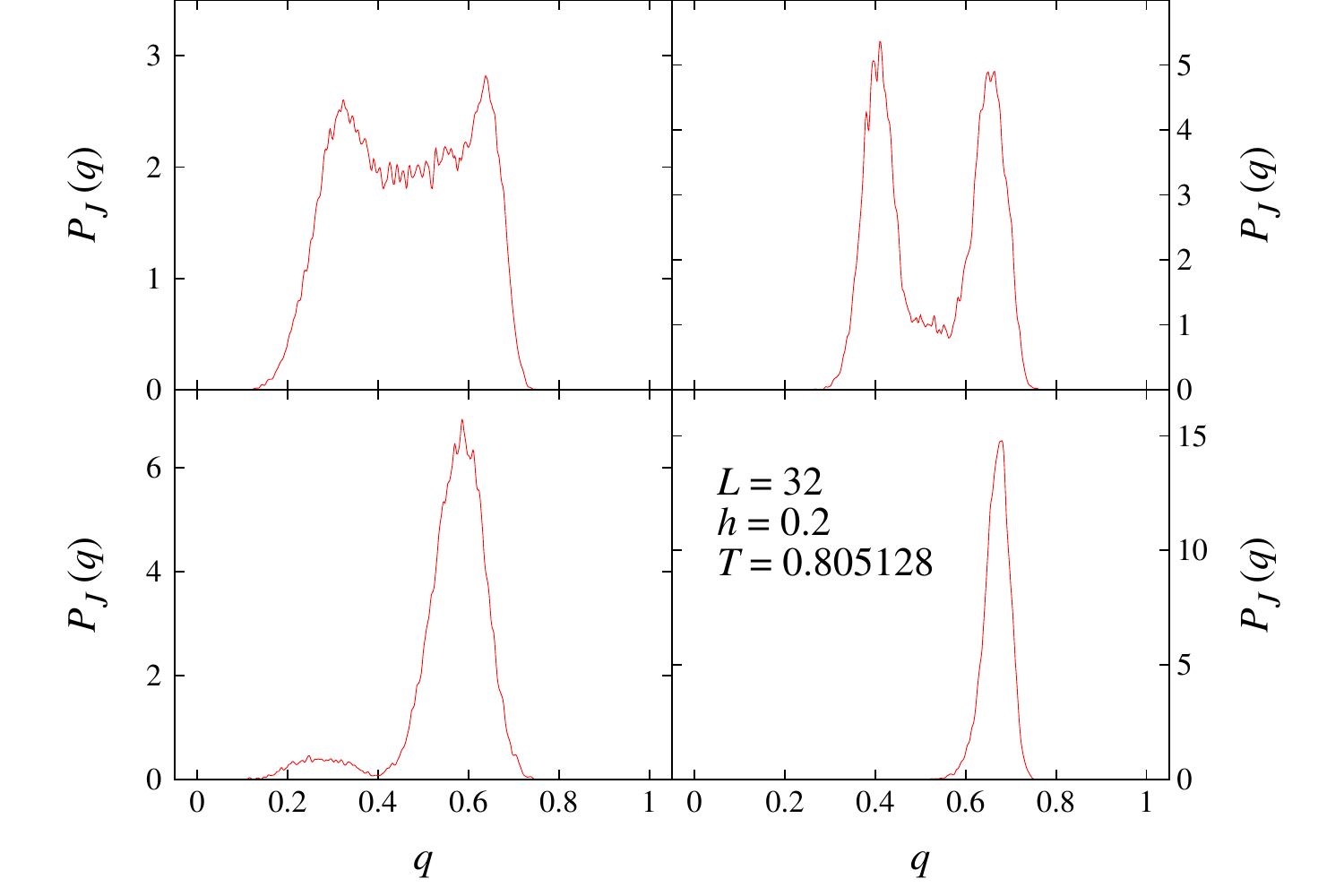}
 \caption[Sample-dependent  \acsp{pdf} $P_J(q)$]{Sample-dependent \acp{pdf} $P_J(q)$, \nomenclature[P...Jq]{$P_J(q)$}{sample-dependent distribution of the overlap}
	  for four different samples,\index{overlap!distribution!sample-dependent}
	  each representing a different type of $P_J(q)$ we encountered.
	  As well as the averaged $P(q)$, also the sample-dependent density function can be
	  wide and with a structure. The plotted data comes from samples with $L=32$, $h=0.2$ and $T=0.805128$.}
 \label{fig:PJq-eah3d}
\end{figure}

The reader may argue that a sample-to-sample distinction of the
different behaviors is more natural than a measurement-dependent one
(although intuition leads to assume that the two are related). This
was indeed our first approach to the problem (it was, in fact,
proposed in \cite{parisi:12b}). However, we found that the
approach described in the previous section is preferable,
both for practical and conceptual reasons.

On the practical side, a sample-to-sample separation implies that from
each sample we get only one data point: For any observable,
we limit ourselves to its thermal average.  In this case we would need a
limitless amount of samples to be able to construct a reasonable
$P(\hat q)$. Moreover, the simulations should last a huge number of
autocorrelation times $\tau$ if we want to have small enough errors on\index{relaxation time}
the thermal averages of each sample. Otherwise, we would introduce a
large bias that is not reduced when increasing the number of samples.

On the conceptual side, representing each sample merely with a single
number (namely the thermal expectation value), is a severe
oversimplification. As we show in figure \ref{fig:PJq-eah3d}, even though we
are in the paramagnetic phase, the behavior within each sample is far
from trivial. For a non-negligible fraction of the samples, the
overlap distribution is wide, often with a multi-peak structure. The
barriers among peaks can be deep, hence suggesting extremely slow
dynamics (which is indeed the case for physical dinamics 
\cite{janus:14b}, or for the parallel tempering dynamics \cite{hukushima:96,marinari:98b}). 

In summary, we find that using instantaneous measurements to classify
the available information is the best solution.

\subsection{The selection of the \acl{CV}}
\label{sec:select-cv}
\subsubsection{A quantitative criterion}\index{variate!conditioning!selection|(}
In appendix \ref{app:reglas-suma} we show how to decompose the moments of a generic variable $\mathcal{O}$ as sums of averages conditioned to $\hat q$.
For the variance we find that 
\begin{equation}
\label{eq:regla-de-suma-var-full}
\nomenclature[v.arO]{$\mathrm{var}({\cal O})$}{variance of $\mathcal{O}$}
\mathrm{var}({\cal O}) = 
\int_{-1}^1 d\hat{q} \, P(\hat{q})\left\{\mathrm{var}({\cal O} | \hat{q}) 
+
\big[ E({\cal O}) - E({\cal O}| \hat{q})\big]^2\right\}\,,
\end{equation}
where
\begin{equation}
 \nomenclature[v.arO]{$\mathrm{var}({\cal O} | \hat{q})$}{variance of $\mathcal{O}$ conditioned to $\hat{q}$}
\mathrm{var}({\cal O} | \hat{q}) = E\left(\big[{\cal O} - E({\cal O}|\hat{q})\big]^2 ~|~ \hat{q}\right)\,.
\end{equation}
A convenient \ac{CV} is
the one that mostly discerns the different behaviors of the model.  We can get a
quantitative criterion for the selection of a good $\hat q$ by rewriting equation (\ref{eq:regla-de-suma-var-full})
as:
\begin{equation}
\label{eq:regla-de-suma-var}
\nomenclature[c....1c2]{$c_1,c_2$}{terms of the variance}
\mathrm{var}({\cal O}) = c_1 + c_2,
\end{equation}
where 
\begin{eqnarray}
 \nonumber
c_1 &\equiv& \int_{-1}^1 d\hat{q} \, P(\hat{q})\mathrm{var}({\cal O} | \hat{q}) \,,
\\ 
\label{eq:c1-c2}
c_2 &\equiv& \int_{-1}^1 d\hat{q} \, P(\hat{q})[ E({\cal O}) - E({\cal O}|
  \hat{q})]^2\,,
\end{eqnarray}
and studying the relation between the terms $c_1$ and $c_2$.
Both are positive, and their sum is fixed independently from the used \ac{CV}.

We will show intuitively that a useful \ac{CV} has $c_2\gg c_1$.

If $c_1=0$ the fluctuations of ${\cal O}$
would be explained solely by the fluctuations of $\hat q$. 
In this case $c_2$ is large and assumes its largest possible value, meaning that
different values of ${\cal O}$ are mostly spread apart by $\hat q$.

On the other side, $c_2=0$ implies $E({\cal O}) = E({\cal O}|\hat q)$ and
signals an insensitive \ac{CV}, with null correlation 
between ${\cal O}$ and $\hat{q}$.

Equations (\ref{eq:regla-de-suma-var}) and (\ref{eq:c1-c2}) can thus be used to quantify the quality of the
\ac{CV} $\hat{q}$: We look for the highest quotient $c_2/c_1$.

\subsubsection[Candidates for \texorpdfstring{$\hat q$}{q}]{Candidates for \texorpdfstring{$\bm{\hat q}$}{q}}
\index{overlap!minimum}\index{overlap!maximum}\index{overlap!median}\index{overlap!average}

To select an appropriate \ac{CV} we need to chose ${\cal O}$
and propose some test definitions for $\hat q$.
The functions of the observables that one could use as a \ac{CV}
are infinite, but physical intuition lead us to try with simple functions of
the overlap and of the link overlap \eqref{eq:qlink-def}. \index{overlap!Ising}\index{overlap!link}
On the other side, a natural choice of ${\cal O}$ is the estimator of the replicon \index{susceptibility!replicon}
susceptibility [see (\ref{eq:chi})]. This means that
\begin{equation}
 {\cal O} \longrightarrow \frac{1}{3{\cal N}}\sum^{\cal N}_{\stackrel{\mathrm{equiv. wave}}{\mathrm{vectors}~\bk}}\left[\, |\Phi_{\bk}^{(ab;cd)}|^2 
										     + |\Phi_{\bk}^{(ac;bd)}|^2 
										     + |\Phi_{\bk}^{(ad;bc)}|^2 \,\right]\,,
\end{equation}
where ${\cal N}$ is the number of equivalent wave vectors one can construct.\nomenclature[N..0]{${\cal N}$}{number of equivalent wave vectors}
This is a 4-replica quantity [see (\ref{eq:GR})], so six instantaneous overlaps (and six link overlaps) are 
associated to each instance of the correlators. To define $\hat q$ we need to
propose a function of the six overlaps in order to get a one-to-one correspondence.

Let us reorder each 6-plet of instantaneous overlaps $\{q^{(ij)}\}$ in the form
of six sorted overlaps $\{q_{k}\}$
\begin{equation}
 \left\{q^{(\rma\rmb)},q^{(\rma\rmc)},q^{(\rma\rmd)},q^{(\rmb\rmc)},q^{(\rmb\rmd)},q^{(\rmc\rmd)}\right\} 
\longrightarrow 
\left\{q_1\leq q_2\leq q_3\leq q_4\leq q_5\leq q_6\right\}\,,
\end{equation}
and do the same thing with the link overlap
\begin{equation*}
 \left\{q_\mathrm{link}^{(\rma\rmb)},~q_\mathrm{link}^{(\rma\rmc)},~q_\mathrm{link}^{(\rma\rmd)},
       ~q_\mathrm{link}^{(\rmb\rmc)},~q_\mathrm{link}^{(\rmb\rmd)},~q_\mathrm{link}^{(\rmc\rmd)}\right\} 
 \longrightarrow ~~~~~~~~~~~~~~~~
\end{equation*}
\begin{equation}~~~~~~~~~~~~~~~~\longrightarrow 
\left\{q_\mathrm{link,1}\leq q_\mathrm{link,2}\leq q_\mathrm{link,3}\leq q_\mathrm{link,4}\leq q_\mathrm{link,5}\leq q_\mathrm{link,6}\right\}\,,
\end{equation}
The following are natural test \acp{CV}:
\begin{flalign}
\label{eq:cv}
\hat{q} =\left\{
\begin{array}{llr}
\nomenclature[q....min]{$q_\mathrm{min}$}{minimum overlap}
\nomenclature[q....minlink]{$q_\mathrm{link,min}$}{minimum link overlap}
\nomenclature[q....max]{$q_\mathrm{max}$}{maximum overlap}
\nomenclature[q....maxlink]{$q_\mathrm{link,max}$}{maximum link overlap}
\nomenclature[q....med]{$q_\mathrm{med}$}{median overlap}
\nomenclature[q....medlink]{$q_\mathrm{link,med}$}{median link overlap}
\nomenclature[q....av]{$q_\mathrm{av }$}{average overlap}
\nomenclature[q....avlink]{$q_\mathrm{link,av }$}{average link overlap}
q_\mathrm{min}     &=q_1~~~~~~~~~~~~~~~~~~~~~~~~~~~~~~~~~~~~~~~~~~~~~~~~~~~~~~~~~~~~~~~~~~~~\mathrm{(the~minimum)}\\[1ex]
q_\mathrm{link,min}&=q_\mathrm{link,1}\\[1ex]
q_\mathrm{max}     &=q_6~~~~~~~~~~~~~~~~~~~~~~~~~~~~~~~~~~~~~~~~~~~~~~~~~~~~~~~~~~~~~~~~~~~\mathrm{(the~maximum)}\\[1ex]
q_\mathrm{link,max}&=q_\mathrm{link,6}\\[1ex]
q_\mathrm{med}     &=\frac{1}{2}(q_3+q_4)~~~~~~~~~~~~~~~~~~~~~~~~~~~~~~~~~~~~~~~~~~~~~~~~~~~~~~~~~~~\mathrm{(the~median)}\\[1ex]
q_\mathrm{link,med}&=\frac{1}{2}(q_\mathrm{link,3}+q_\mathrm{link,4})\\[1ex]
q_\mathrm{av }     &=\frac{1}{6}(q_1+q_2+q_3+q_4+q_5+q_6)~~~~~~~~~~~~~~~~~~~~~~~~~~~~~\mathrm{(the~average)}\\[1ex]
q_\mathrm{link,av }&=\frac{1}{6}(q_\mathrm{link,1}+q_\mathrm{link,2}+q_\mathrm{link,3}+q_\mathrm{link,4}+q_\mathrm{link,5}+q_\mathrm{link,6})\,.\\[1ex]
\end{array}
\right.
\end{flalign}
We checked how each of the \acp{CV} sorted the overlap and link susceptibilities 
$\chi_\mathrm{R}(\bf{0})$ and $\chi_\mathrm{R}^\mathrm{link}(\bf{0})$.\index{susceptibility!replicon}
Table~\ref{tab:reglas-de-suma} depicts the $c_1$ and $c_2$ terms, and their
ratio, for all the \acp{CV}, for a single triplet $(T,L,h)$ 
and $\bk=(0,0,0)$. 
\begin{table}[!b]
\centering
\resizebox{\columnwidth}{!}{
\begin{tabular}{c||c|c|c||c|c|c}
$\hat{q}$ & $\chi_R^\mathrm{spin}$: $c_1$ & $\chi_R^\mathrm{spin}$: $c_2$ & $c_2/c_1$&  $\chi_R^\mathrm{link}$: $c_1$ & $\chi_R^\mathrm{link}$: $c_2$& $c_2/c_1$\\\hline\hline
  $q_\mathrm{min}^\mathrm{spin}$ & 399000 $\pm$ 37000 & 121000 $\pm$ 15000 & 0.30(6) & 8.35 $\pm$ 0.47 & 0.297 $\pm$ 0.023  & 0.36(5) \\[1ex] 
  $q_\mathrm{max}^\mathrm{spin}$ & 514000 $\pm$ 51000 & 6230   $\pm$ 690   & 0.012(3)& 8.54 $\pm$ 0.49 & 0.1070$\pm$ 0.0073 & 0.013(2)\\[1ex] 
  $q_\mathrm{med}^\mathrm{spin}$ & 162000 $\pm$ 10000 & 358000 $\pm$ 45000 & 2.2(4)  & 7.35 $\pm$ 0.39 & 1.30  $\pm$ 0.11   & 0.18(2) \\[1ex]
  $q_\mathrm{av}^\mathrm{spin}$  & 328000 $\pm$ 26000 & 192000 $\pm$ 28000 & 0.6(1)  & 7.51 $\pm$ 0.41 & 1.141 $\pm$ 0.094  & 0.15(2) \\[1ex] \hline
  $q_\mathrm{min}^\mathrm{link}$ & 461000 $\pm$ 46000 & 59300  $\pm$ 5800  & 0.13(3) & 8.38 $\pm$ 0.48 & 0.271 $\pm$ 0.020  & 0.032(4)\\[1ex] 
  $q_\mathrm{max}^\mathrm{link}$ & 460000 $\pm$ 46000 & 59700  $\pm$ 5900  & 0.13(3) & 8.56 $\pm$ 0.49 & 0.0838$\pm$ 0.0067 & 0.010(1)\\[1ex] 
  $q_\mathrm{med}^\mathrm{link}$ & 360000 $\pm$ 36000 & 160000 $\pm$ 18000 & 0.44(9) & 7.36 $\pm$ 0.38 & 1.29  $\pm$ 0.11   & 0.17(2)\\[1ex]
  $q_\mathrm{av}^\mathrm{link}$  & 415000 $\pm$ 42000 & 105000 $\pm$ 10000 & .25(5)  & 7.72 $\pm$ 0.42 & 0.927 $\pm$ 0.073  & 0.12(2)\\
\end{tabular}
}
\caption[Measured parameters $c_1$ and $c_2$ for the choice of the \acs{CV}]{Criterion for the choice of the
\ac{CV} $\hat q$ for $h=0.1$, $L=32$, $T=0.805128$, by looking at the 
indicators $c_1$ and $c_2$ relatively to $\chi_\mathrm{R}(\bf{0})$ and $\chi_\mathrm{R}^\mathrm{link}(\bf{0})$. 
We want the \index{susceptibility!replicon}
$\hat q$ to split as much as possible the different measured susceptibilities.
This is obtained, see (\ref{eq:c1-c2}), when the ratio $c_2/c_1$ is
maximised. From the data we see that this occurs with $\hat{q} = 
q_{\mathrm{med}}$.\index{overlap!minimum}\index{overlap!maximum}\index{overlap!median}\index{overlap!average}\index{overlap!link}}
\label{tab:reglas-de-suma}
\end{table}
The best \ac{CV} is clearly the median overlap, since it has the highest
$c_2/c_1$ ratio. The situation is similar for other choices of $(T,L,h)$.

\begin{figure}[!htb]
\centering
 \includegraphics[width=0.81\columnwidth]{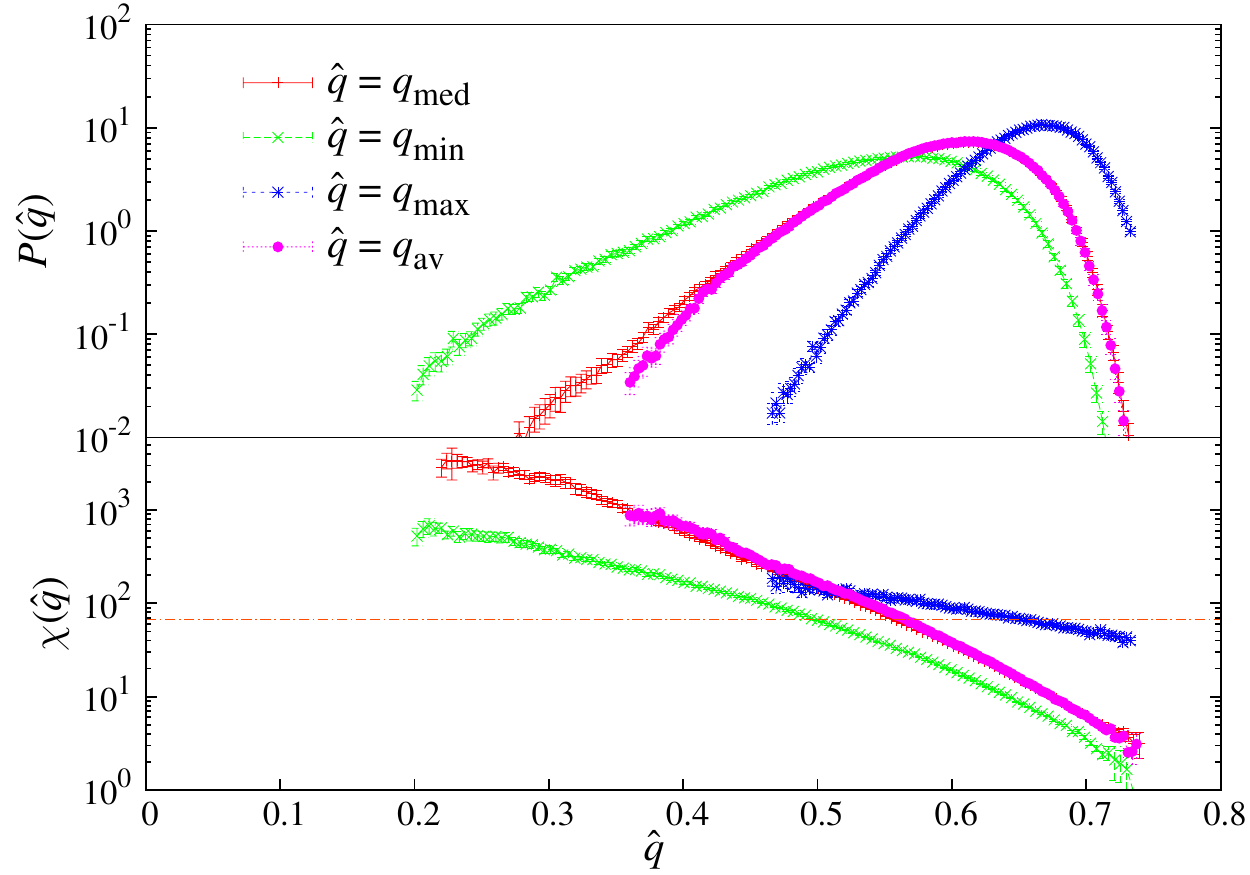}
\caption[Susceptibility $\chi(\hq)$ and \acs{pdf} of the candidate \acsp{CV}]{Features of the diverse \acp{CV} we proposed for
  $L=32$, $h=0.2$ and $T=0.805128$.  The \textbf{top} figure shows the
  histograms $P(\hat q)$ for four candidates of conditioning \index{variate!conditioning!distribution}
  variate: the minimum overlap $q_\mathrm{min}$ [of the six we can
    make with four replicas, recall equations (\ref{eq:cv})], the maximum
  $q_\mathrm{max}$, the median $q_\mathrm{med}$ and the average
  $q_\mathrm{av}$.  The histograms were constructed as explained in appendix \ref{app:hack}.
    The \textbf{bottom} figure depicts the size of the
  susceptibility $\chi$ for each value of the \ac{CV}.\index{susceptibility!replicon!conditioned}
  The horizontal line marks the value of $\chi$ when it is averaged
  over the full set of measurements.  For aesthetic reasons in both
  figures we have cut the curves at the two end points, where they
  become extremely noisy due to poor sampling.}
\label{fig:cv}
\end{figure}

For a qualitative description of the difference between the diverse
\acp{CV}, in figure~\ref{fig:cv} (top) the reader can appreciate the
probability distribution functions for each of the
\acp{CV}, while in figure~\ref{fig:cv} (bottom) we plotted the conditioned
susceptibilities. \index{susceptibility!replicon!conditioned}
From (\ref{eq:E-conditioned}) we stress that the integral of the values on
the top times the values of the bottom set yields the average susceptibility,
which is indicated with a horizontal line on the bottom plot of figure \ref{fig:cv}.
As it is also reflected by table \ref{tab:reglas-de-suma},
$q_\mathrm{max}$ is the worst \ac{CV}, as its
$\chi$ does not vary much with the fluctuations of $q_\mathrm{max}$.
The steepest slope
is obtained when the \ac{CV} is $q_\mathrm{av}$ or
$q_\mathrm{med}$, but the latter is smoother and covers a wider range of
$\chi$.

Figure~\ref{fig:cv} also displays the large deviations present in the system.
In fact one can see that the value of $q_\mathrm{med}$ at which the $P(q_\mathrm{med})$\index{variate!conditioning!distribution}
has its maximum is significantly different with respect to the value
of $q_\mathrm{med}$ at which $\chi(q_\mathrm{med})$ assumes the value of the average.

\begin{figure}[!t]
 \includegraphics[width=0.48\columnwidth]{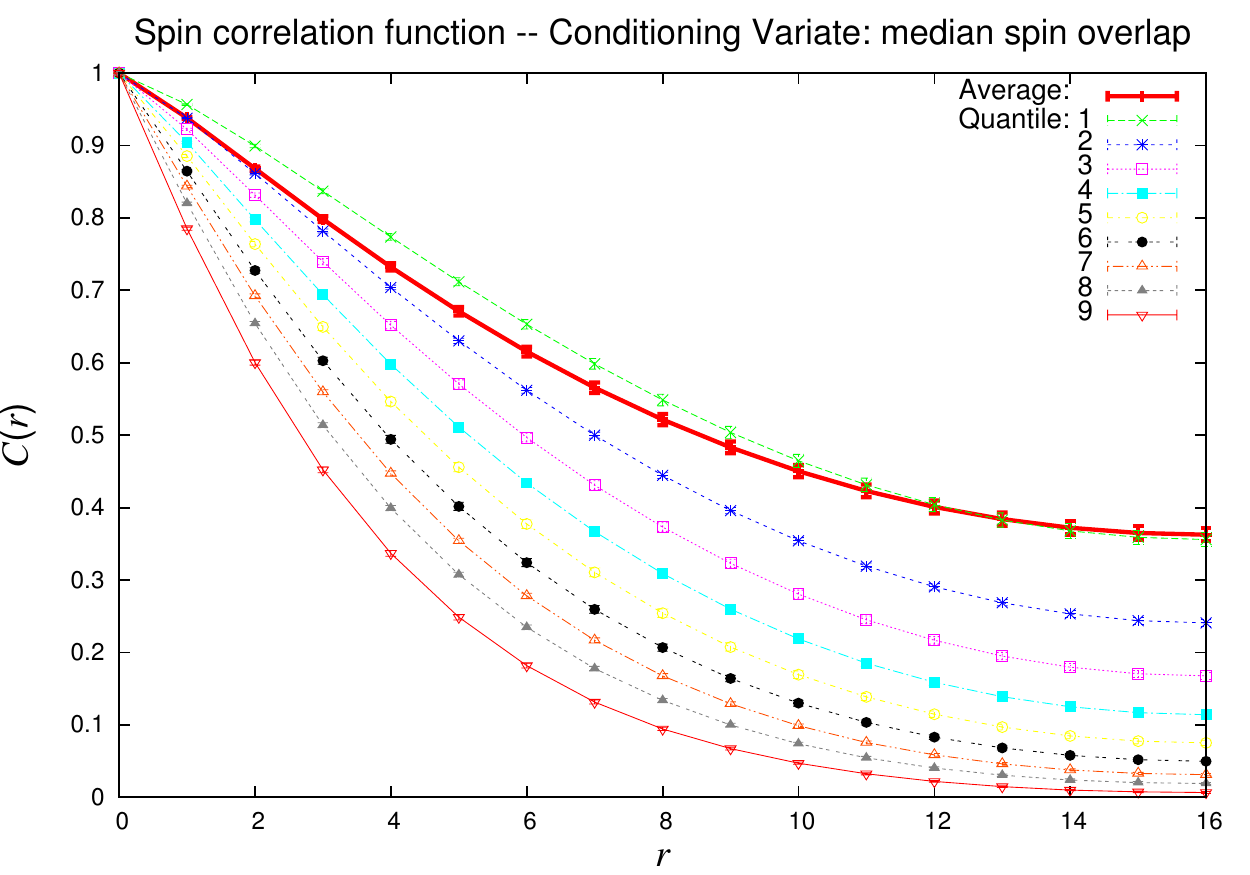}
 \includegraphics[width=0.48\columnwidth]{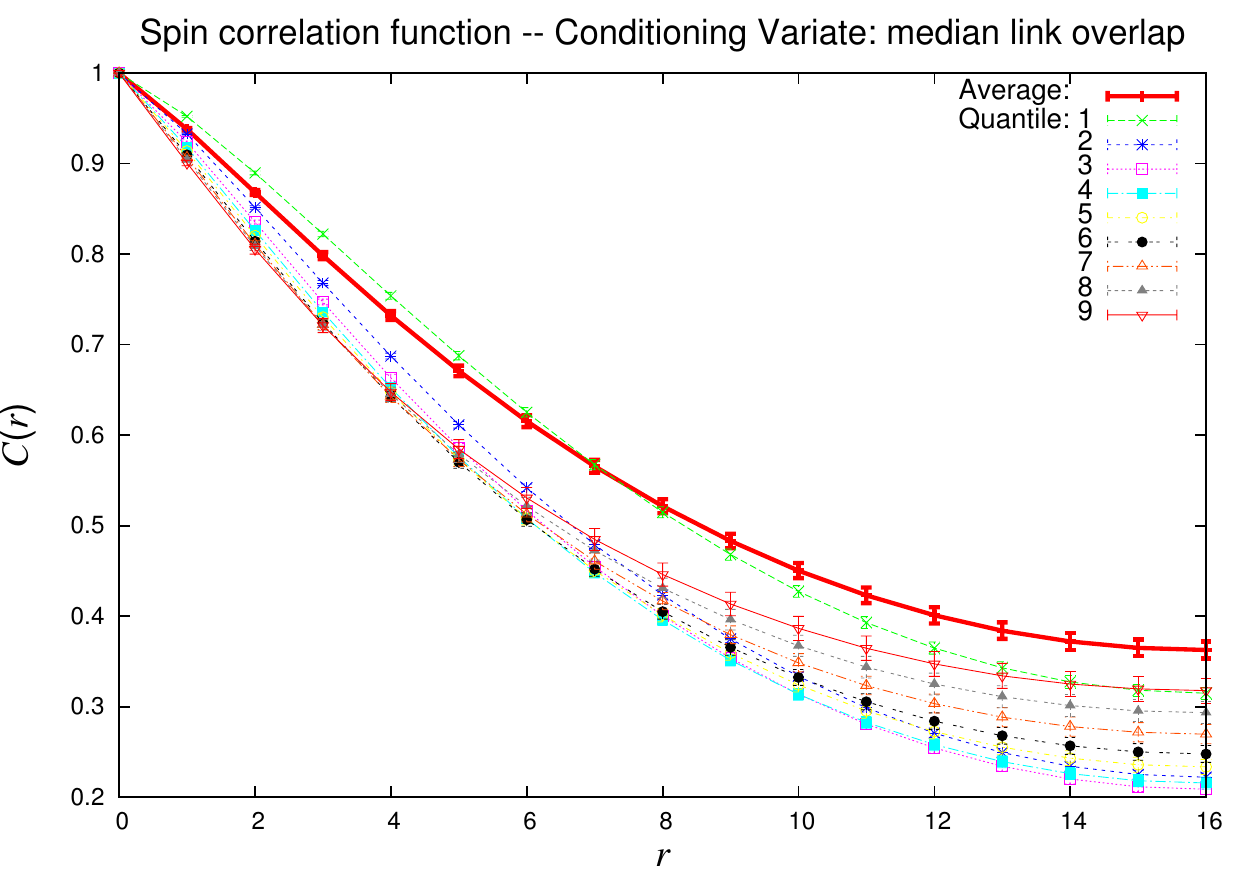}\\
 \includegraphics[width=0.48\columnwidth]{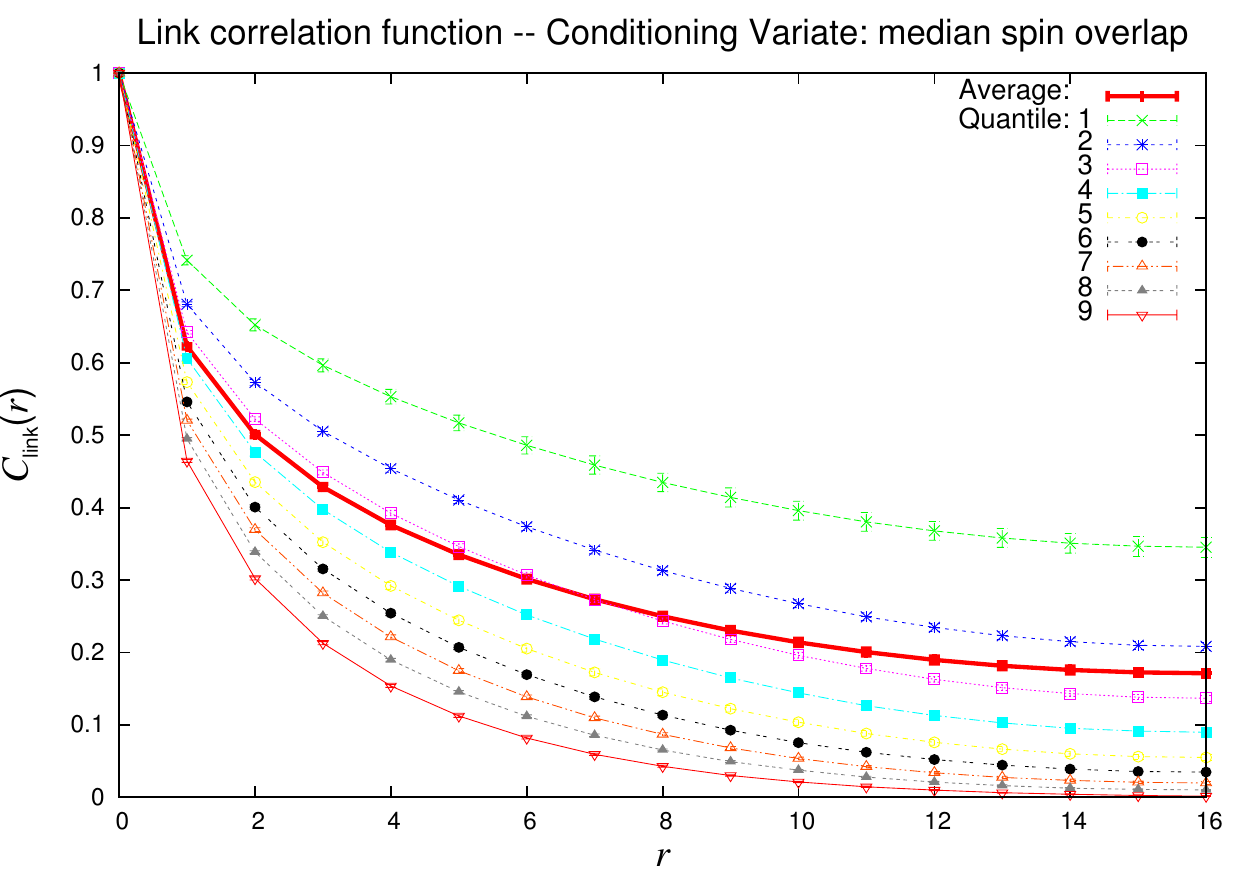}
 \includegraphics[width=0.48\columnwidth]{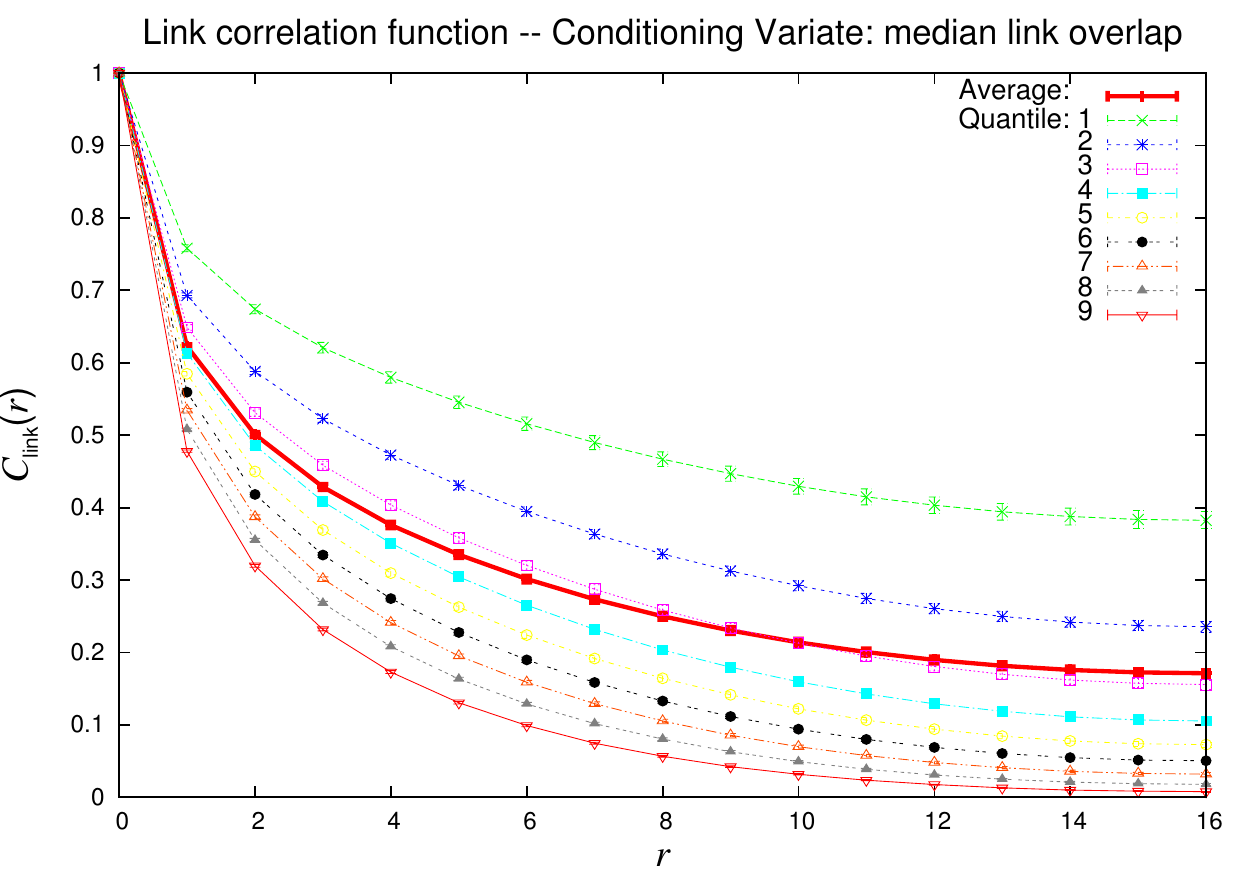}
\caption[Testing the link overlap as \acs{CV}]{Spin and link plane replicon correlation functions with $h=0.1$, $L=32$, $T=0.805128$. 
The thin lines indicate different quantiles of the conditioning variate's distribution (see section \ref{sec:median}), 
the bold lines indicate the average.\index{correlation!function!plane}\index{correlation!function!link}
\textbf{Top}: Spin correlation functions, \textbf{bottom}: link correlation functions.
\textbf{Left}: $\hat q=q_\mathrm{med}$, \textbf{right}: $\hat q=q_\mathrm{link,med}$.\index{overlap!median}\index{overlap!link}
Discussion in the main text.}
\label{fig:C-cv}
\end{figure}

Let us compare the overlap with the link-overlap signal. 
Besides the fact that the link overlaps appear to be bad \ac{CV}s, one can see from table \ref{tab:reglas-de-suma}
that on one side the fluctuations on $\chi_\mathrm{link,R}(\bf{0})$ are much smaller than $\chi_\mathrm{R}(\bf{0})$, and
on the other none of the \acp{CV} seems  to separate the behaviors (the ratio $c_2/c_1$ is much smaller). We can see this better
from figures \ref{fig:C-cv}, that depicts the results of a sorting with the median (link-)overlap on $C_\mathrm{R}(\br)$ and 
\index{correlation!function!plane}\index{correlation!function!link}
$C_\mathrm{link,R}(\br)$. The bold line stands for the average behavior, while the thin ones represent a sorting of
the data according to the quantile \index{quantile} of the distribution of the \ac{CV}.
\footnote{A quantile is the value of $\hat q$ that separates a fixed part of the \ac{pdf} (section \ref{sec:median} later on).}
If the average is in the middle of the thin lines it is a good descriptor of the data, otherwise it is a biased estimator.
Very spread thin lines indicate that $c_2\gg c_1$: the \ac{CV} separates behaviors properly.

The data illustrates that, while the average spin correlation function is not representative of the majority of outcomes for both the \ac{CV}, 
the link correlation function is well-described by its average. This suggests that the link overlap might be a more accurate indicator of 
the critical behavior of the \ac{EA} spin glass in a field. Analyses on the link-overlaps will be object of further future studies.

On another side, if we concentrate on the spin correlation function $C(r)$, we see that the link is not a suitable \ac{CV}, both because
it separates less the behaviors, and because the separation has a dependency on the distance $r$.
\index{variate!conditioning!selection|)}

\section{Quantiles and a modified finite-size scaling ansatz}
\label{sec:median}\index{quantile|(}\index{scaling!finite-size}

We stated in section~\ref{sec:extended-abstract} that the set of
measurements with low $\hat q$ has a very different behavior from the
measurements with high $\hat q$ (recall figure\ref{fig:Cr}). From now
on, we shall restrict ourselves to $\hat{q} = q_\mathrm{med}$, since\index{overlap!median}
we evinced that the median is our best \ac{CV}.  Our next
goal will be to carry out a finite-size scaling analysis based on the
$P(q_\mathrm{med})$ that lets us observe different parts of the\index{variate!conditioning!distribution}
spectrum of behaviors of the system.
\index{variate!conditioning|)}

In order to analyse separately these different sets of measures, we
divide the $P(q_\mathrm{med})$ in 10 sectors, each containing $10\%$\index{overlap!median!distribution}
of the measured $q_\mathrm{med}$.  We focus our analysis on the values
of $q_\mathrm{med}$ that separate each of these sectors. They are
called quantiles (see, e.g., \cite{hyndman:96}), and we label them
with the subscript $i=1,\ldots,9$. If we call $\tilde{q}_i(h,T,L)$ the
value of the $i^\mathrm{th}$ quantile, we can define it in the
following implicit way:\index{quantile}
\begin{equation} 
\nomenclature[q....i]{$\tilde q_i$}{$i^\mathrm{th}$ quantile}
\int_{-1}^{\tilde q_i} \mathrm{d}\hat q P(\hat q) = \frac{i}{10}\,.  
\end{equation} 
In appendix~\ref{app:hack} we explain how $\tilde{q}_i(h,T,L)$ was computed.

We can adapt to the $i^\mathrm{th}$ quantile the definitions we gave 
in section~\ref{sec:obs}:
\begin{eqnarray}
\index{xiL/L@$\xi_L/L$!quantile}\index{R12@$R_{12}$!quantile}
\index{susceptibility!wave-vector dependent!quantile}
\nomenclature[xi....Li]{$\xi_{L,i}$}{per-quantile $\xi_{L}$}
\nomenclature[chi....Ri]{$\chi_{\mathrm{R},i}(\bk)$}{per-quantile $\chi_{\mathrm{R}}(\bk)$}
\nomenclature[R...12i]{$R_{{12},i}$}{per-quantile $R_{12}$}
  \chi_{\mathrm{R},i}(\bk) &=& \frac{1}{N} E\bigg(|\hat {\Phi}_{\bk}^{(ab;cd)}|^2 ~\bigg|~ \tilde q_i\bigg)\,,\\[2ex]
 \xi_{L,i}                          &=& \frac{1}{2 \sin{(k_\mathrm{min}/2)}} \sqrt{\frac{\chi_{\mathrm{R},i}(\bf{0})}{\chi_{\mathrm{R},i}(2\pi/L,0,0)}-1}\,,\\[2ex]
 R_{{12},i}                         &=& \frac{\chi_{\mathrm{R},i}(2\pi/L,0,0)}{\chi_{\mathrm{R},i}(2\pi/L,\pm2\pi/L,0)}\,.
\label{eq:obs_sep}
\end{eqnarray}
This way we can extend the finite-size scaling methodology to the $i^\mathrm{th}$ quantile:\index{scaling!finite-size!quantile}
\begin{align}
 \left.\frac{\xi_L}{L}\right|_{T,h,L,i} &= f_{\xi_i}\left(L^{1/\nu}(T-T_\mathrm{c})\right) + \ldots\,,\\[1ex]
 \left.R_{12}\right|_{T,h,L,i} &= f_{R_i}\left(L^{1/\nu}(T-T_\mathrm{c})\right) + \ldots\,.
\end{align}
This is a new approach for \acl{FSS}. 
Although it demands a very large amount of data because it is
done over a small fraction of the measurements (in appendix \ref{app:eah3d-errors} we explain a method we used to
reduce rounding errors), it allows us to perform finite-size scaling
on selected sets of measurements.

Let us stress that no \emph{a priori} knowledge is required on
the probability distribution function $P(q_\mathrm{med})$: Quantiles are \index{variate!conditioning!distribution}
conceived in order to define a scaling that self-adapts when the volume increases.

\section{Testing the quantile approach \label{sec:eah3d-h0}}\index{quantile!test at h=0@test at $h=0$}
\begin{figure}[!tb]
\centering
\vspace{2.5cm}
 \includegraphics[width=\columnwidth]{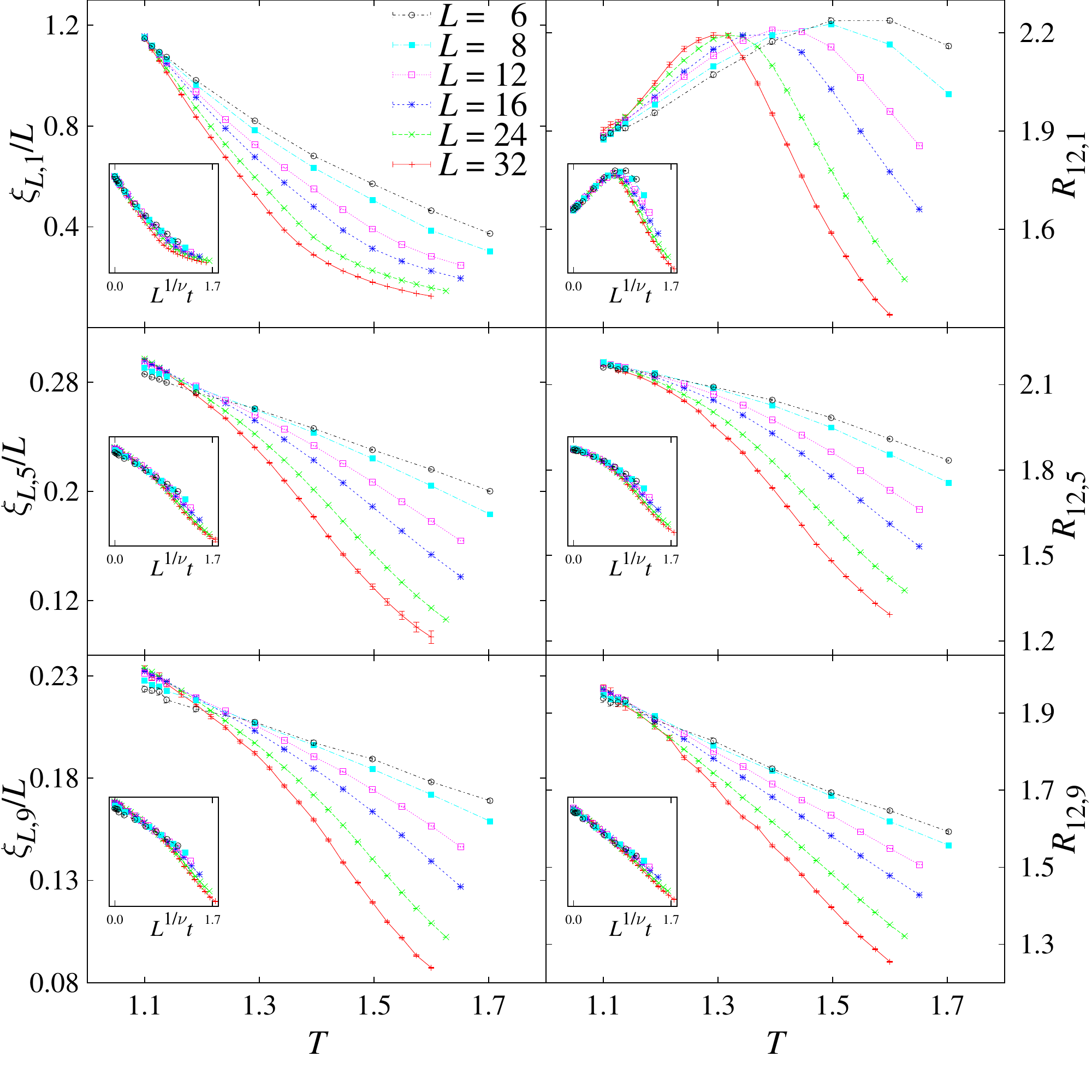}
 \caption[Cumulants $\xi_L/L$ and $R_{12}$ versus $T$, for $h=0$, quantiles 1,5,9]{
 Finite-size indicators of a phase transition, computed for
  $h=0.2$. On the \textbf{left} side we plot, for quantiles 1 (\textbf{top}), 5 (\textbf{middle})
  and 9 (\textbf{bottom}), the correlation length in units of the lattice size\index{xiL/L@$\xi_L/L$}
  $\xi_L/L$ versus the temperature, for all our lattice sizes. The \textbf{right} side is equivalent, 
  but for the $R_{12}$,\index{R12@$R_{12}$} defined in equation \eqref{eq:def-R12}.
 The curves crossings are compatible with the well-known temperature of the zero-field transition.\index{spin glass!transition}
 The data come from \cite{janus:13}.
We used 256000 samples for each lattice size.
The \textbf{insets} show the same data of the larger sets, but as a function of the scaling variable $L^{1/\nu}t$,\index{exponent!critical!nu@$\nu$}
where $t$ is the reduced temperature $t=(T-T_\mathrm{c})/T_\mathrm{c}$.}
 \label{fig:xiL-separil_h0}
\end{figure}
\afterpage{\clearpage}

We take advantage of our $h=0$ data from \cite{janus:13} to validate our new \ac{FSS} ansatz and the
quantile description, by showing its behavior in the zero-field case. 
Two replicas would be enough to construct connected correlators in $h=0$,\index{correlation!function!4-replica}
and using the 4-replica definitions proposed in section~\ref{sec:obs} only
adds noise to the results. Yet, we opted for the latter option
because the objective of the current section is the validation of the 
full procedure proposed herein.

In the absence of a magnetic field
we expect that the curves $\xi/L(T)$ and $R_{12}$ \index{xiL/L@$\xi_L/L$}\index{R12@$R_{12}$} cross no matter the quantile, since
the behavior of the system is not dominated by extreme events and crossover fluctuations.
Also, in this case the data in our hands arrive down to the critical point, so the crossings
ought to be visible.

One can see in fact from figure \ref{fig:xiL-separil_h0} that all the quantiles 
show visible signs of a crossing at $T_\mathrm{c}$ both in the case of $\xi_L/L$
and of $R_{12}$. \index{xiL/L@$\xi_L/L$}\index{R12@$R_{12}$}
Furthermore, if we plot the same data as a function of the scaling variable 
$L^{1/\nu}(T-T_\mathrm{c})/T_\mathrm{c}$ the data collapses well for all the quantiles \index{exponent!critical!nu@$\nu$}
(figure~\ref{fig:xiL-separil_h0}, insets).

\begin{figure}[!htb]
\centering
\includegraphics[width=\columnwidth]{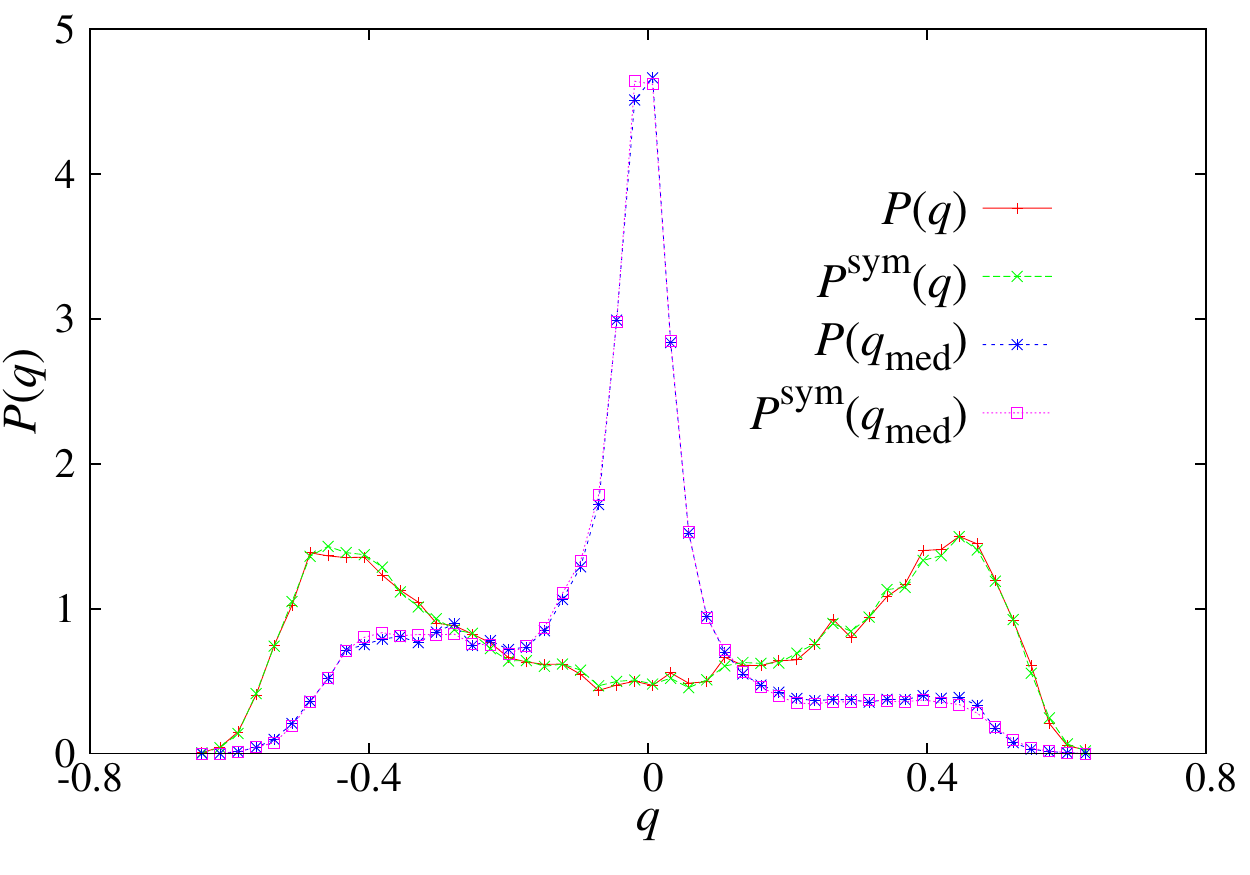}
 \caption[Overlap histograms and their symmetrized versions]{Probability distribution function for $h=0$, $L=24$, $T=1.1$. The data come from 512 samples where we
 took all the $16^4$ combinations of overlaps per sample.
 We show $P(q)$, that in null field \index{overlap!median!distribution!not symmetrical}
 is symmetric, and $P(q_\mathrm{med})$, that is not. We also plot the symmetrized histograms $P^\mathrm{(sym)}(q)$
 and $P^\mathrm{(sym)}(q_\mathrm{med})$, that overlap on the respective curves. As more extendedly explained in the main text, the symmetrized\index{symmetry!Z2@$Z_2$}
 overlap is obtained by averaging each $q_\mathrm{med}$ over the values it would acquire by imposing all the combinations of $Z_2$ symmetry (flip all
 the spins) on the configurations on which the $q_\mathrm{med}$ is calculated.}
 \label{fig:Pq-eah3d_h0}
\end{figure}

Some reader may be surprised that quantiles 1 and 9 show different
behavior, being $P(q)$ symmetrical (figure~\ref{fig:Pq-eah3d_h0}). The reason\index{variate!conditioning!distribution|(}
is that, although $P(q)$ is symmetrical, $P(q_\mathrm{med})$ is not.\index{overlap!median!distribution}\index{overlap!median!distribution!not symmetrical}
In fact, given six overlaps $q^{\rma\rmb}, q^{\rma\rmc}, q^{\rma\rmd}, q^{\rmb\rmc}, q^{\rmb\rmd},
q^{\rmc\rmd}$ coming from four configurations
$\ket{s^{(\rma)}}$,$\ket{s^{(\rmb)}}$,$\ket{s^{(\rmc)}}$,$\ket{s^{(\rmd)}}$,
each enjoying a $Z_2$ symmetry, the distribution of their median
privileges negative values.
\footnote{Let us give a simple example. Take 4 $Z_2$-symmetric single-spin systems that can assume different values 
$s_1=\pm1, s_2=\pm2, s_3=\pm3, s_4=\pm4$.
We can construct 6 overlaps $q_{ij}(s_1, s_2, s_3, s_4)$.
If we explicitate the $Z_2$ symmetry, taking all the combinations of our random variables, the histogram
of $q$ will be symmetric with zero mean. Yet, if we take the histogram of the median overlap, it will be asymmetric
with mean $\langle q_\mathrm{med}\rangle = -3$. This can easily be checked by computing all the possible combinations
of the signs of the $s_i$ and computing the median in each case: $q_\mathrm{med}(+1, +2, +3, +4) = 5$, 
$q_\mathrm{med}(+1, +2, +3, -4) = -1$, $q_\mathrm{med}(+1, +2, -3, -4) = -3.5$, and so on.
} 
We show this in figure~\ref{fig:Pq-eah3d_h0}, where we give both the $P(q)$
and the $P(q_\mathrm{med})$ for $h=0$, $L=32$, $T=1.1$. The first is
symmetrical and the second is not.  To convince the reader that the
starting configurations do enjoy $Z_2$ symmetry, we also construct the
symmetrized functions $P^\mathrm{(sym)}(q)$ and \nomenclature[P...symq]{$P^\mathrm{(sym)}(q)$}{symmetrized probability distribution function}
$P^\mathrm{(sym)}(q_\mathrm{med})$.  These two functions are obtained
by explicitly imposing the reflection symmetry $Z_2$: \nomenclature[Z...2]{$Z_2$}{reflection symmetry group}
for each measurement we\index{symmetry!Z2@$Z_2$}
construct the $2^4$ overlaps with both $\ket{s}$ and
$\ket{-s}$.  It is visible from figure~\ref{fig:Pq-eah3d_h0} that
$P^\mathrm{(sym)}(q_\mathrm{med})$ is asymmetric even though we
imposed by hand the $Z_2$ symmetry on the configurations.

\subsection[The \texorpdfstring{$P(q_\mathrm{med})$}{P(qmed)}]
           {The \texorpdfstring{$\bm{P(q_\mathrm{med})}$}{P(qmed)}}
\index{overlap!median!distribution}
To our knowledge, the median overlap
$q_\mathrm{med}$, despite its simplicity, has not been object of previous study. 
 Yet, since we base our analysis 
on this quantity, it is necessary to dedicate passing attention to its features. 

By its definition, the probability distribution $P(q_\mathrm{med})$
of the median overlap has narrower tails than $P(q)$ (recall
figure~\ref{fig:Pq-eah3d}), although from figure~\ref{fig:cv} (top) it is clear that the
strong fluctuations persist also with $q_\mathrm{med}$.

\begin{figure}[!htb]
\includegraphics[width=\columnwidth]{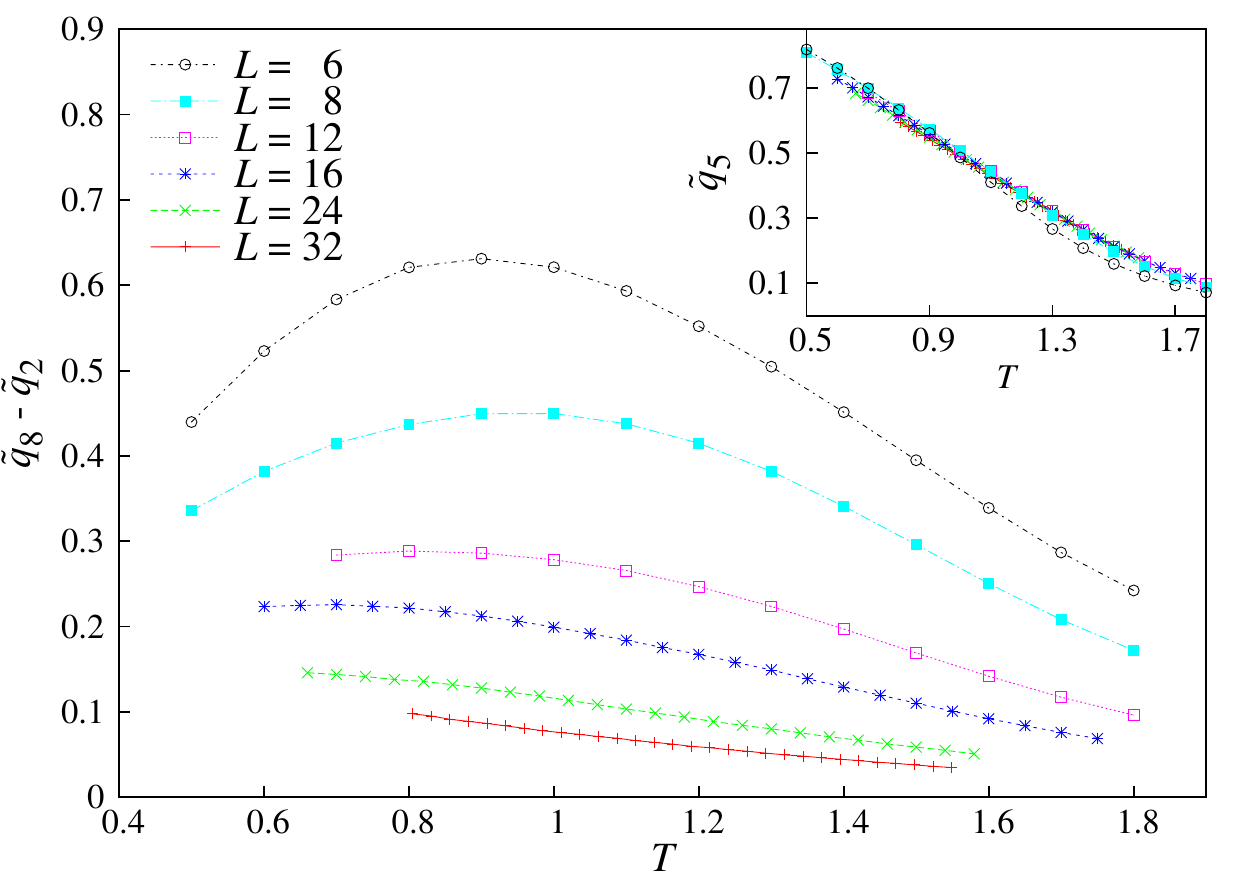}
\caption[Difference between quantiles $\tilde q_8-\tilde q_2$]{Using $q_\mathrm{med}$ as \ac{CV}, we show the\index{variate!conditioning}
temperature dependence of the difference between quantiles $\tilde q_8-\tilde
q_2$, for all our lattice sizes, in a field of intensity $h=0.2$.  This
corresponds to the width of the central $60\%$ of area of $P(q_\mathrm{med})$.
This quantity can reveal a phase transition, since in the paramagnetic phase
the $P(q_\mathrm{med})$ should be a delta function, while in the spin-glass
phase it should have a finite support. We show the central $60\%$ and not a
wider range because it is an equivalent indicator of the phase transition,
and it is safer from rare events that would vanish in the
thermodynamic limit.
In the \textbf{inset} we show the position of $5^\mathrm{th}$ quantile as a function
of temperature in all our lattice sizes. It is a very smooth curve with 
very small finite-size effects.}
\label{fig:anchura}
\end{figure}

The median of $P(q_\mathrm{med})$ corresponds to the fifth quantile. We will
prefer to call it ``$5^\mathrm{th}$ quantile'' rather than ``median of the
median overlap''.  Of the nine studied quantiles it is the smoothest and has
the least finite-size effects, as one can see from figure~\ref{fig:anchura}\index{quantile!finite-size effects}
(inset). Further analysis is given in section \ref{app:q_L}.

We remark also that the separation between the different $\tilde{q}_i$'s can be
used as order parameter, since its thermodynamic limit should be zero in the
paramagnetic phase, and greater than zero in the possible low-temperature phase
due to the (would-be) replica symmetry breaking.  Figure~\ref{fig:anchura} shows
the difference between the $8^\mathrm{th}$ and the $2^\mathrm{nd}$ quantile,
i.e., the $q_\mathrm{med}$-span of the central $60\%$ of the data. If we were
able to extrapolate a clean $L\rightarrow\infty$ limit for this curve, we would
be able to answer to whether the transition exists or not.  Unfortunately, even
for $T>T_\mathrm{c}(h=0)=1.1019(29)$, where we know that we are in the
paramagnetic phase, it is not possible to make good extrapolations since the
trend is strongly non-linear.  In section \ref{app:q_L} we will show that extrapolations to
the thermodynamic limit were only possible in the trivial case of $h=0.4$ (deep
paramagnetic phase), and that between all the quantiles, the median curve is
the one that shows less finite-size effects.

\section{A \emph{caveat} for the quantile description\label{app:Pq}}
In the absence of an applied field, the overlap probability distribution function
$P(q)$ is symmetric, with a single peak centred in $q=0$. 
In the presence of a
field, instead, we expect the $P(q)$ to be strictly positive, at least in the
thermodynamic limit. Similarly, we expect that the probability distribution
function $P(q_\mathrm{med})$ have only one peak at positive $q_\mathrm{med}$\index{overlap!median!distribution}
when a field is applied, and a peak in $q=0$ if $h=0$.

\begin{figure}[tb]
 \centering
 \includegraphics[width=0.9\columnwidth]{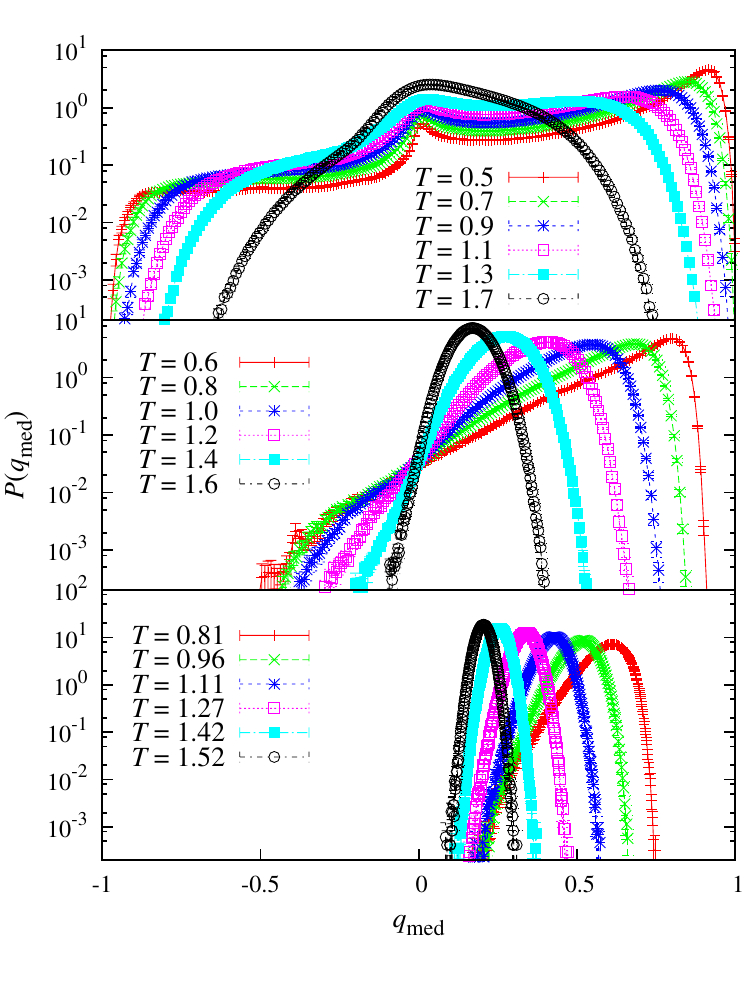}
  \caption[Median overlap probability distribution function $P(q_\mathrm{med})$]
  {Median overlap probability distribution function $P(q_\mathrm{med})$ with $h=0.2$ 
for different temperatures
(the ones from $L=32$ are an approximation to the second decimal digit).
The \textbf{top} figure shows the case of
 $L=6$, where the lowest temperature curves display a second peak around
 $q_\mathrm{med}\simeq0$, which disappears when $T$ increases.  For $L=16$ (\textbf{middle}) the
 $P(q_\mathrm{med})$ are single-peaked, but assume also negative values.  In the\index{overlap!median!distribution}
 \textbf{bottom} curve we have $L=32$, where the $P(q_\mathrm{med})$ are single-peaked
and defined only on positive $q_\mathrm{med}$, since we are closest to the
asymptotic behavior. }
\label{fig:Pq-eah3dT}
\end{figure}

If the system sizes are too small, it may occur that the $h=0$ behavior bias the
$P(q_\mathrm{med})$. This is what happens, for example, when $L=6$, $h=0.2$ and
the temperature is sufficiently low: a second peak around
$q_\mathrm{med}\simeq0$ develops upon lowering $T$ (figure~\ref{fig:Pq-eah3dT}, top). 
This second peak disappears when we increase the lattice size (figure~\ref{fig:Pq-eah3dT}, 
centre), and the $P(q_\mathrm{med})$ assumes only positive values when $L$ is large 
enough (figure~\ref{fig:Pq-eah3dT}, bottom).  The lower the field, the easier it is to find
multiple peaks, and the greater the system has to be to be able to neglect the
$h=0$ behavior. For $h=0.05$, even lattices with $L=12$ show a double peak.

A second peak in $P(q_\mathrm{med})$ is a clear signal that we are observing 
and echo of $h=0$. When we make the quantile classification, and have a
quantile on a peak, we are seeing \emph{only} non-asymptotic data. Thus,
quantile 1 for the smallest lattices gives us no relevant information. 

If we plot versus the
temperature any observable ${\cal O}$ related to the first quantile, 
the information will be biased for low temperatures, and
the bias will gradually disappear as we increase $T$. The result is
that the curve ${\cal O}(T)$ will have a strange shape and will be of
no use (see, e.g., the $h=0.05$ data in figure~\ref{fig:qL}). This is
why we did not include the $L=6$ points in the top set of 
figure \ref{fig:xiL-separil_h02} later on.

\section[Finding a privileged \texorpdfstring{$q$}{q}]{Finding a privileged \texorpdfstring{\boldmath $q$}{q}\label{app:q_L}}

\begin{figure}[!tbh]
\centering
\includegraphics[width=\columnwidth]{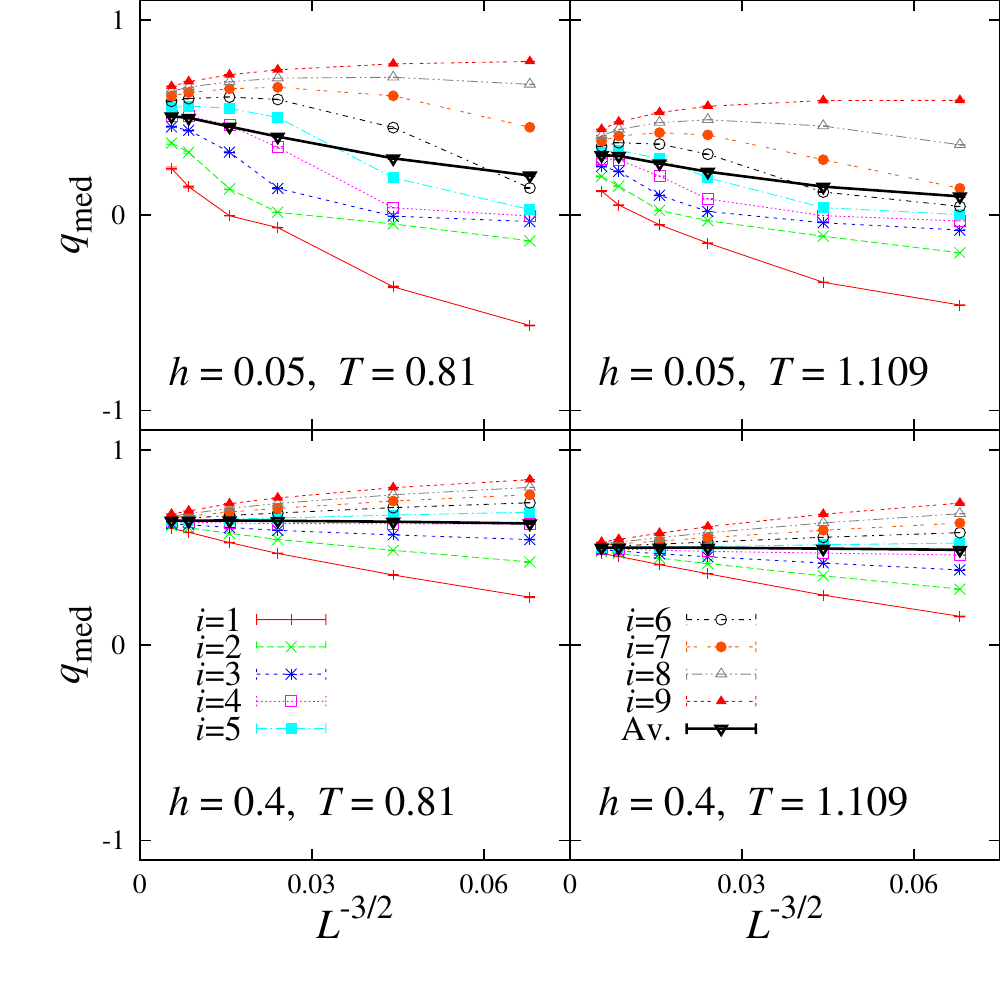}
\caption[Extrapolations to infinite size of the quantile overlap $\tilde q_i$]
{Extrapolations to infinite size of the quantile overlap $\tilde q_i$,
for $T=0.81$ (\textbf{left}) and $T=1.109$ (\textbf{right}), and fields $h=0.05$ (\textbf{top}) and $h=0.4$ (\textbf{bottom}). We show
quantiles $i=1,\ldots,9$ (thin lines), and the average behavior (bold line). 
The $h=0.4$ extrapolations to 
infinite volume were clean ($\chi^2/\mathrm{DOF} < 1$), while for $h=0.05$ (and all the other fields we simulated),
we encountered too strong and nonlinear finite-size effects to get reasonable extrapolations.\index{finite-size effects}
We choose $1/L^{D/2}$ as scaling variable because in conditions of validity of \index{central limit theorem}
the central limit theorem, the fluctuations should be of order $1/\sqrt{N}$.}
\label{fig:qL}
\end{figure}

Since all our simulations are in the paramagnetic phase the
thermodynamic limit of the $P(q)$ is a delta function, so all the
quantiles should tend to the a common $q=q_\mathrm{EA}$ in the \index{overlap!Ising}
$L\rightarrow\infty$ limit. \index{variate!conditioning!distribution|)} We tried to perform these extrapolations
at fixed (reasonably low) temperature, to see if we could look at the
problem from such a privileged position.  In figure~\ref{fig:qL} we see
this type of extrapolation for $h=0.4$ and $h=0.05$, at temperatures
$T=0.81$ and $1.109$. The first is the lowest temperature we simulated in
all our lattices, while the second is the zero-field critical
temperature \cite{janus:13}.  Since we are in the paramagnetic phase
and we are plotting $\tilde q_i$ versus the inverse lattice size, the
curves should cross at the intercept. This is indeed what appears to
happen, but although in the case of $h=0.4$, the extrapolations were
clean, for all the other simulated fields the finite-size effects were
too strong and nonlinear to make solid extrapolations. We remark, yet,
that once $L>8$ the $5^\mathrm{th}$ quantile is the one with the least
finite-size effects.\\

\afterpage{\clearpage}
\section{The silent majority}
\begin{figure}[!htb]
\centering
 \includegraphics[width=\columnwidth]{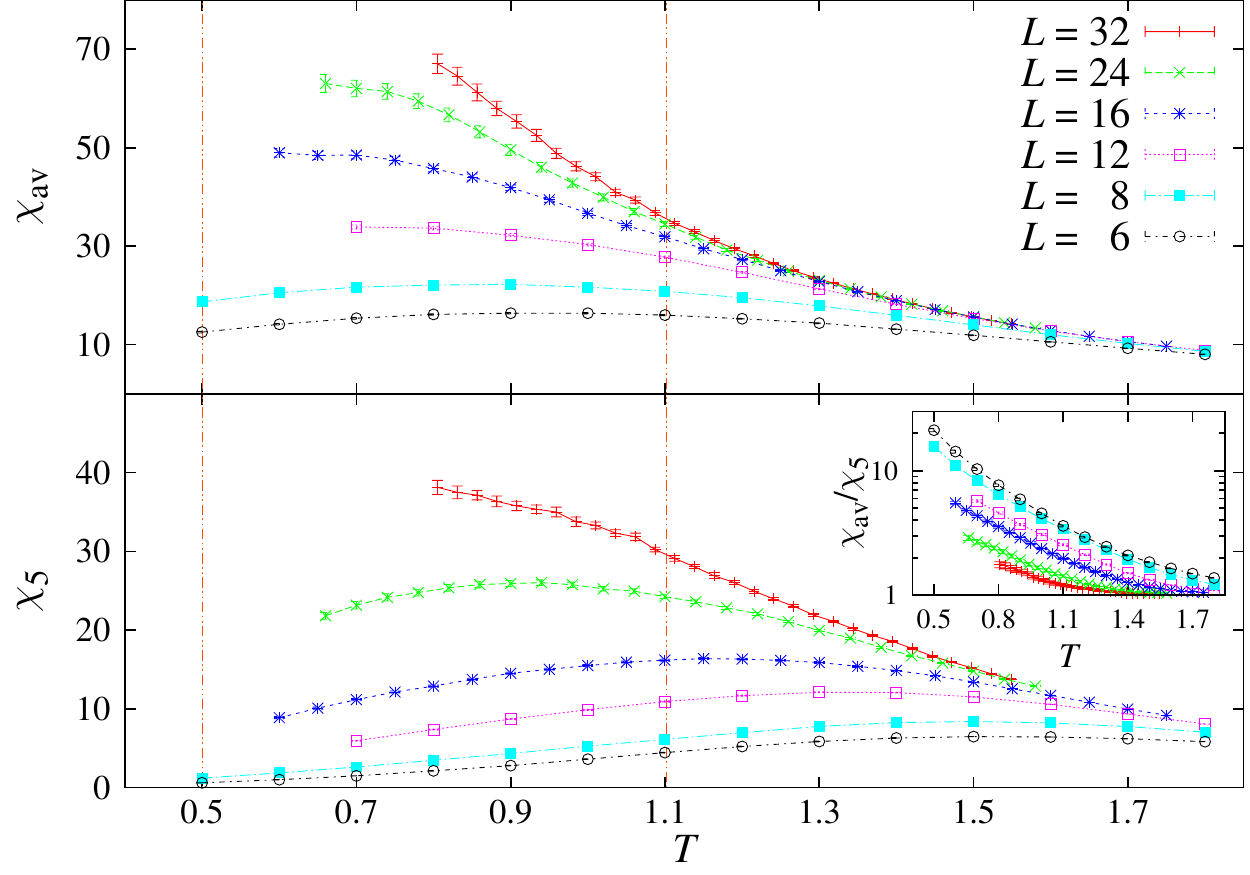}
\caption[The replicon susceptibility $\chi_\mathrm{R}$ versus $T$]
{The replicon susceptibility $\chi$ as a function of the\index{susceptibility!quantile}\index{susceptibility!replicon}
  temperature, for all the simulated lattice sizes and the field
  $h=0.2$.  We represent its average $\chi$ (\textbf{top}), and the
  $5^\mathrm{th}$ quantile $\chi_5$ with $\hat q=q_\mathrm{med}$
  (\textbf{bottom}). In both plots, the two vertical lines represent the upper
  bound of the possible phase transition $T^\mathrm{up}(h=0.2)=0.5$
  given in \cite{janus:14b}, and the zero-field critical temperature
  $T_\mathrm{c}(h=0)=1.109(29)$ \cite{janus:13}.  The amplitudes and
  the positions of the peaks of $\chi$ are strikingly different (mind
  the different scales in the $y$ axes). The \textbf{inset} shows the ratio
  between the two, which we expect to tend to an order one constant in
  the thermodynamic limit. This is actually what we see at high
  temperatures.}
\label{fig:chi} 
\end{figure}

\index{de Almeida-Thouless!transition|(}
\label{sec:eah3d-results}
As already emphasized, the behavior of the system is characterized by very strong
fluctuations, and a wide and asymmetric $P(q)$. \index{variate!conditioning!distribution} As a result, the average and
median behavior are very different. 
In figure~\ref{fig:chi}, we show the replicon susceptibility: its average $\chi$ on the left
plot, and its fifth quantile $\chi_5$. Motivated by the arguments in section \ref{sec:conditional}
all the quantiles we show in this section use the \ac{CV} $\hat q=q_\mathrm{med}$.\index{variate!conditioning}

Visibly, not only is the average susceptibility much larger than the \index{susceptibility!quantile}
$5^\mathrm{th}$ quantile, but also the two have peaks at different
temperatures. Also, finite-size effects are much stronger in the case
of $\chi_5$ (yet, recall the inset in figure \ref{fig:anchura},
finite-size effects on $\tilde q_5$ are tiny).\footnote{ We made power 
law extrapolations to $L\rightarrow\infty$ of the maxima of the susceptibility, but they 
were not satisfactory (too large $\chi^2/\mathrm{DOF}$). Only for $h=0.2,0.4$ were we able to fit the maxima's
heights and obtained $\eta(h=0.2) \approx 0.6$ and $\eta(h=0.4)\approx 0.9$.\index{exponent!critical!eta@$\eta$}}

\begin{figure}[!tbh]
\centering
\includegraphics[width=\columnwidth]{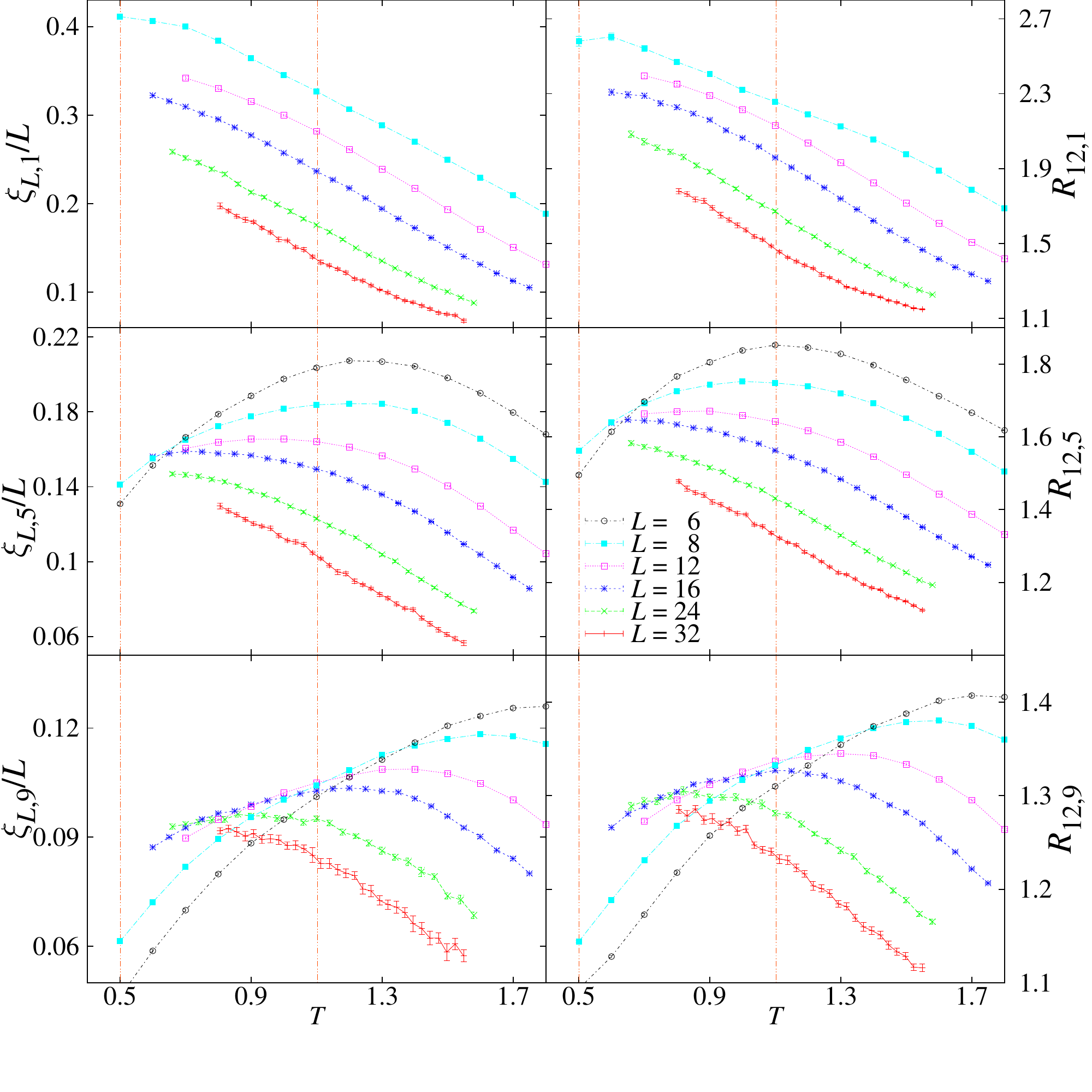}
\caption[Cumulants $\xi_L/L$ and $R_{12}$ versus $T$, for $h=0.2$, quantiles 1,5,9]
{Finite-size indicators of a phase transition, computed for
  $h=0.2$. On the \textbf{left} side we plot, for quantiles 1 (\textbf{top}), 5 (\textbf{middle})
  and 9 (\textbf{bottom}), the correlation length in units of the lattice size\index{xiL/L@$\xi_L/L$}
  $\xi_L/L$ (\textbf{left}) versus the temperature, for all our lattice sizes
  except $L=6$ (we show in section \ref{app:Pq} that the quantile description
  is not suitable for $L=6$ because there is a double peak in the\index{variate!conditioning!distribution}
  $P(q)$).  On the \textbf{right} we show analogous plots for $R_{12}$\index{xiL/L@$\xi_L/L$}\index{R12@$R_{12}$} [defined in equation \eqref{eq:def-R12}].  The
  vertical line on the left marks the upper bound $T^\mathrm{up}$ for
  a possible phase transition given in \cite{janus:14b}, while
  the one on the right marks the zero-field transition temperature
  $T_\mathrm{c}$ given in \cite{janus:13}.  Quantile 1 has the
  same qualitative behavior of the average $\xi_L/L$, shown in
  figure~\ref{fig:xiL}, while quantiles 5 and 9 suggest a scale
  invariance at some temperature $T_h<T^\mathrm{up}$.}
\label{fig:xiL-separil_h02}
\end{figure}

We show in figure \ref{fig:xiL-separil_h02} how sorting the data with the quantiles
revealed the presence of different types of behavior, by plotting the
$\xi_L/L$ and the $R_{12}$ \index{xiL/L@$\xi_L/L$}\index{R12@$R_{12}$} for quantiles 1, 5 and 9 at $h=0.2$. There are 
two vertical lines in each figure. The one
on the left represents the upper bound $T^\mathrm{up}(h)$ for the phase
transition (meaning that no phase transition can occur for $T>T^\mathrm{up}(h)$) 
given in \cite{janus:14b}, while the one on the right
indicates the zero field critical temperature
$T_\mathrm{c}=1.1019(29)$ \cite{janus:13}. 

We can see that the $1^\mathrm{st}$ quantile
has the same qualitative behavior of the average (figure \ref{fig:xiL}), but
lower values, since the main contribution to the average comes from data whose
$q_\mathrm{med}$ is even lower than $\tilde q_1$.
Moreover, one can notice that in figure~\ref{fig:xiL} the indicators $\xi_L/L$ and\index{xiL/L@$\xi_L/L$}
$R_{12}$ show a different qualitative behavior when the lattices are small ($R_{12}$\index{xiL/L@$\xi_L/L$}\index{R12@$R_{12}$}
shows a crossing). This discrepancy vanishes when we look only at the first quantile:
Separating different behaviors enhances the consistency between $\xi_L/L$ and
$R_{12}$.

The behavior of the $5^\mathrm{th}$ quantile
is quite different, since now it appears reasonable that the curves cross at
some $T\lesssim T^\mathrm{up}(h)$. The crossings become even more evident when we
consider the highest quantile.

All this is consistent with the
arguments of section~\ref{sec:extended-abstract}, where we showed how the
correlation function is dominated by a little portion of data, near the first
quantile (figure~\ref{fig:Cr}), while the behavior of the majority of the 
samples is hidden.

\begin{figure}[!htb]
\centering
 \includegraphics[width=\columnwidth]{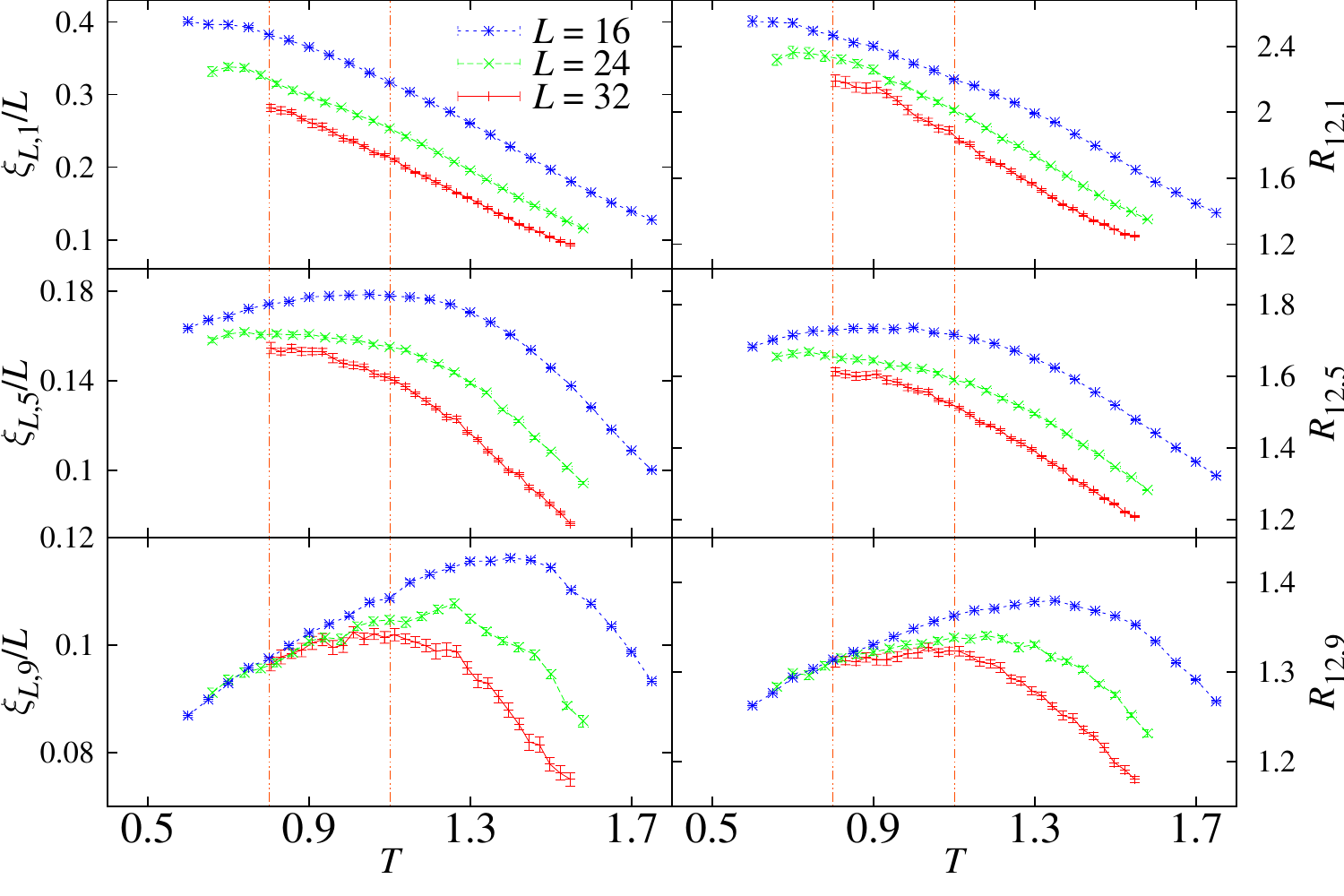}
\caption[Cumulants $\xi_L/L$ and $R_{12}$ versus $T$, for $h=0.1$, quantiles 1,5,9]{
Same as figure \ref{fig:xiL-separil_h02}, but for $h=0.1$. This time \index{xiL/L@$\xi_L/L$}\index{R12@$R_{12}$}
the effects of the zero-temperature transition are stronger, so we removed from
the plot sizes $L=6,8,12$. In section \ref{app:Pq} we show
that the quantile description is not suitable for smaller
lattices due to crossover effects from the zero-field behavior.\index{crossover}
}
\label{fig:xiL-separil_h01}
\end{figure}

Unfortunately, the high non-linearity of the curves impedes an
extrapolation of the crossing points, but they are apparently compatible with
the upper bound $T^\mathrm{up}$, and their heights apparently do not depend
on the intensity of the applied field $h$ (see also fig.~\ref{fig:xiL-separil_h01}).

The careful reader might have noticed that the upper bound $T^\mathrm{up}(h)$
for the possible phase transition given in \cite{janus:14b} is higher when
the field is lower: $T^\mathrm{up}(0.1) = 0.8 > T^\mathrm{up}(0.2) = 0.5$. It is then
justified to ask oneself how do the quantile plots look like for $h=0.1$.  We
show them in figure~\ref{fig:xiL-separil_h01}.
Since the field is lower, the effects on the double peak on the first quantile
(section \ref{app:Pq}) extend to larger lattices than for $h=0.2$. Thus, we
show only the non-biased sizes, i.e., $L>12$. 

Although the $9^\mathrm{th}$ quantile shows signs of scale invariance\index{spin glass!transition}
at $T=T^\mathrm{up}(0.1)$, the behavior of the $5^\mathrm{th}$
quantile suggests a scale invariance around $T=0.5$.  We believe that
the $5^\mathrm{th}$ quantile is a better indicator, since the position
of the fifth quantile $\tilde q_5$ has less finite-size effects (it
practically has none, figure~\ref{fig:anchura}--inset) than $\tilde
q_9$.

It is interesting to focus on the height of the crossings of each
quantile from figure \ref{fig:xiL-separil_h02}, and compare them with $h=0.2$ (figure~\ref{fig:xiL-separil_h02}). This is expected to be a universal
quantity, and in the hypothesis of a phase transition it should be the
same for both fields.  Although it is not possible to assign error
bars to the these values, it is possible to see that both for $h=0.1$
and $h=0.2$ the heights are similar
($\xi_{L,5}/L\approx0.15$, $\xi_{L,9}/L\approx0.09$,
$R_{12,5}\approx1.6$, $R_{12,9}\approx1.3$).\index{xiL/L@$\xi_L/L$}\index{R12@$R_{12}$}

\section[This is not an echo of the \texorpdfstring{$h=0$}{h=0} transition]
	{This is not an echo of the \texorpdfstring{\boldmath $h=0$}{h=0} transition}
\label{sec:zero-field}
\index{crossover}\index{universality!class}
The crossing suggested by the quantiles 5 and 9 in figure~\ref{fig:xiL-separil_h02} is unlikely
to be caused by the zero-field transition, since it appears at
$T<T_\mathrm{c}$, and shifts towards lower temperatures as the lattice size
increases. Also, the value of $\xi_L/L$ ($R_{12}$)\index{xiL/L@$\xi_L/L$}\index{R12@$R_{12}$} at the possible crossing point of the fifth quantile
is upper-bounded to $\xi_L/L\simeq0.16$ ($R_{12}\simeq1.65$), while for $h=0$ it is considerably larger 
($\xi_L(T_\mathrm{c})/L \simeq 0.28$ [$R_{12}(T_\mathrm{c})\simeq2.15$]), recall section \ref{sec:eah3d-h0}.
In this section, we will advance more arguments contradicting the notion that what was seen resulted from the effects of zero-field transition.
\index{quantile|)}

\subsection{An escaping transition}\index{crossover}
As pointed out in section~\ref{sec:extended-abstract}, there is a controversy
because we observe a wide $P(q)$, just like in the mean-field
model, but the curves $\xi_L/L(T)$ and $R_{12}(T)$\index{xiL/L@$\xi_L/L$}\index{R12@$R_{12}$} do not show any sign of a
crossing. If we were in the presence of a phase transition, a straightforward
explanation could reside in an anomalous exponent $\eta$ close to 2 \cite{baityjesi:14}, since at
the critical temperature the replicon susceptibility scales as $\chi_\mathrm{R}(L)\sim
L^{2-\eta}$ (\ref{eq:chi-scaling}). \index{susceptibility!replicon}\index{exponent!critical!eta@$\eta$}
It is possible to calculate $\eta$ with the quotients' method \cite{nightingale:76, ballesteros:96}, \index{scaling!finite-size!quotients' method}
by comparing the susceptibility $\chi_L$ \nomenclature[chi....L]{$\chi_L$}{susceptibility of a lattice of size $L$}
of different lattice sizes at the critical point $T_\mathrm{c}$:
\begin{equation}
 \label{eq:quotients}
 \frac{\chi_{2L}(T_\mathrm{c})}{\chi_L(T_\mathrm{c})} = 2^{2-\eta} + \ldots\,,
\end{equation}
where the dots stand for subleading terms.
This definition only makes sense at criticality, but we can extend 
it in an effective manner to a generic temperature. This way we can delineate 
an effective exponent\index{exponent!critical!eta@$\eta$!effective}
\begin{equation}
\label{eq:eta-eff}
\nomenclature[eta....eff]{$\eta_\mathrm{eff}$}{effective anomalous dimension}
 \eta_\mathrm{eff}(T;L,2L) = 2 - \log_2 \frac{\chi_{2L}(T)}{\chi_L(T)} \,.
\end{equation}
In case there were a phase transition at a finite temperature $T_{h}$,
we would have $\eta_\mathrm{eff}(T_{h})=\eta$. We should have
$\eta_\mathrm{eff} = 2$ in the paramagnetic phase, $\eta_\mathrm{eff}
= -1$ in the deep spin-glass phase 
\footnote{See appendix~\ref{app:eta-eff}, keeping in
mind that $\eta_\mathrm{eff} = -1$ is somewhat trivial in the limit
$h\to 0$, where $\chi$ reduces to $\chi= V E(q^2)$.} 
and signs of a crossing at $\eta_\mathrm{eff} =\eta(h=0) = -0.3900(36)$
\cite{janus:13} in the limit of a complete domination by the $h=0$
transition.

\begin{figure}[!h]
\centering
  \includegraphics[width=0.75\columnwidth]{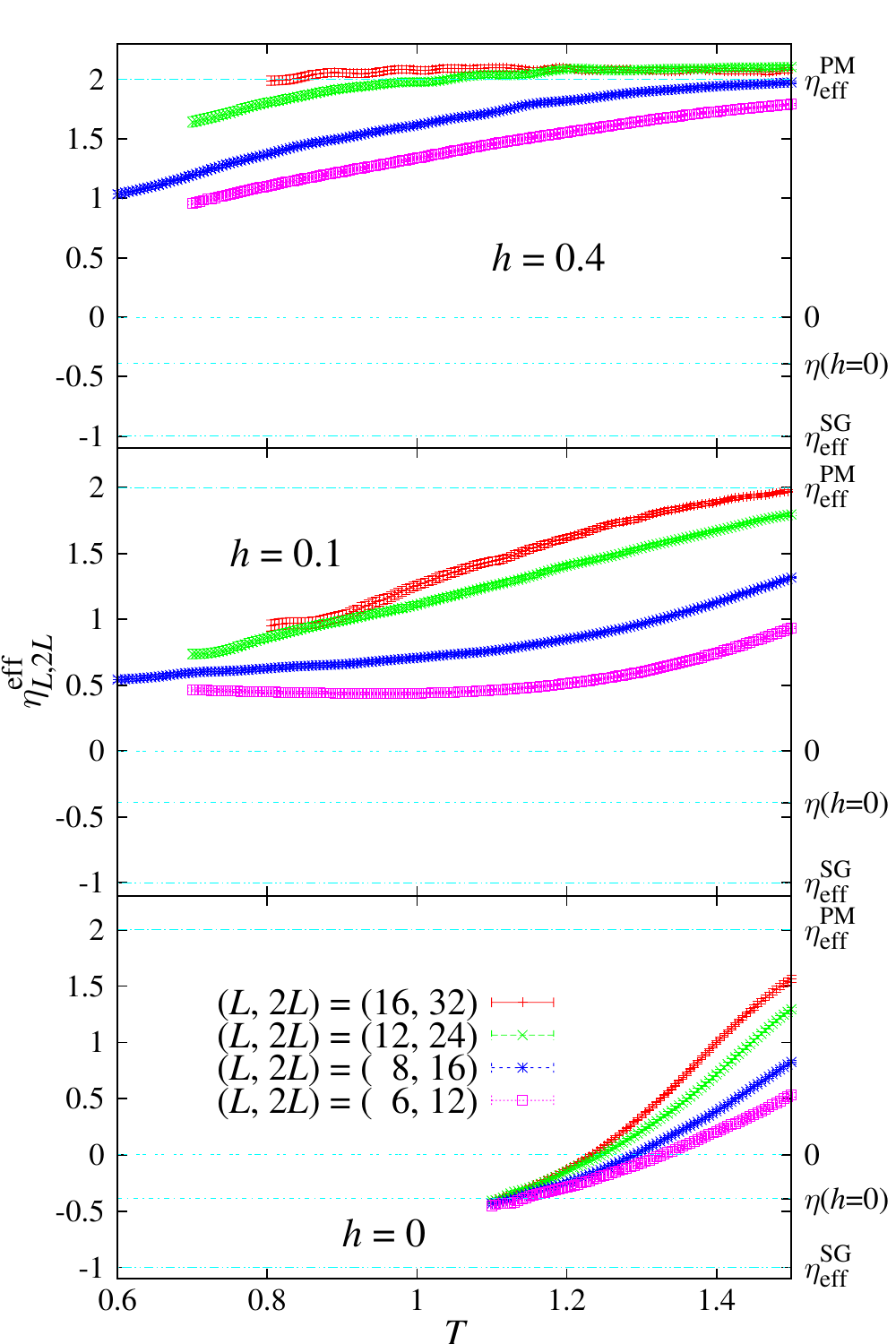}
\caption[Effective anomalous exponent $\eta_\mathrm{eff}(T)$]{We plot $\eta_\mathrm{eff}(T)$, defined in
(\ref{eq:eta-eff}), for all the pairs $(L,2L)$ we could form. 
The magnetic fields are $h=0.4$ (\textbf{top}), $h=0.1$ (\textbf{center}) and $h=0$ (\textbf{bottom}).
The $h=0$ data comes from \cite{janus:13}.\index{exponent!critical!eta@$\eta$!effective}
In each plot we uses horizontal lines to underline meaningful limits, and we
label them with a tic on the right axis.
From up to down, we depict the limit
$\eta_\mathrm{eff}^\mathrm{PM}=2$ of a system in the paramagnetic phase, the $\eta_\mathrm{eff}=0$ axis,
the zero-field value
$\eta_\mathrm{eff}(h=0,T_\mathrm{c})=-0.3900(36)$ \cite{janus:13},
and its value in a deep spin-glass phase $\eta_\mathrm{eff}^\mathrm{SG}=-1$.
Notice the difference between the case with or without a field.
For $h=0.1$ the curves appear to converge to
a positive $\eta_\mathrm{eff}\simeq0.5$, while in the latter all the curves become negative
and merge at $\eta_\mathrm{eff}(h=0,T_\mathrm{c})$.} 
\label{fig:eta-eff}
\end{figure}
\afterpage{\clearpage}%\Floatbarrier

In figure~\ref{fig:eta-eff} we show $\eta_\mathrm{eff}(T)$ for $h=0.4,
h=0.1$, and $h=0$ (the $h=0$ data come from the simulations we
performed in \cite{janus:13}).\footnote{ For each jackknife block we
  calculated $\eta_\mathrm{eff}(T)$ and made a cubic spline
  temperature interpolation.}
If a phase transition were present, but hidden by heavy finite-size effects, 
we would expect at least that the
$L$-trend of $\eta_\mathrm{eff}$ be decreasing. Contrarily, the larger our
lattices, the wider the temperature range in which $\eta_\mathrm{eff} = 2$.
The apparent phase transition shifts towards lower temperature when we
suppress finite-size effects. The data in our possession is not enough to state
whether this shift will converge to a positive temperature.  In any case, this is
compatible with the upper bounds to a possible transition given in \cite{janus:14b}.

On the other side, $\eta_\mathrm{eff}$ stays positive for all
our simulated lattices (except $h=0.05$, $L=6$), and  that even for
$T<T_\mathrm{c}(h=0)$ it tends to some value around $0.5$, so it is unlikely
that the null field transition is dominating the system's behavior.

\subsection{Scaling at \texorpdfstring{$T=T_\mathrm{c}(h=0)$}{T=Tc(h=0)}}
\begin{figure}[!b]
\centering
 \includegraphics[width=0.65\columnwidth]{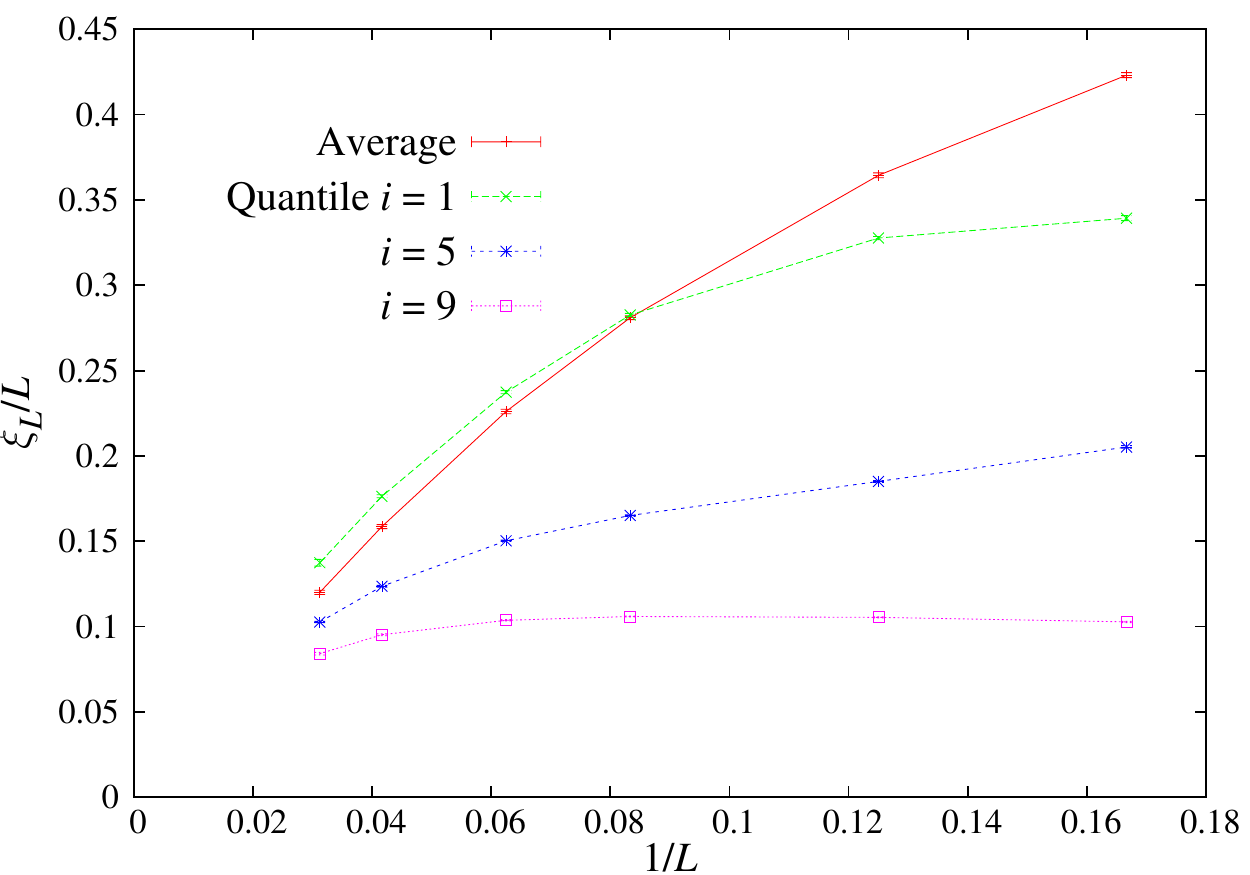}
 \caption[Scaling of $\xi_L/L$ at the null-field critical temperature]{Scaling of $\xi_L/L$ at the null-field critical temperature\index{xiL/L@$\xi_L/L$}
$T_\mathrm{c}=1.109(29)$ \cite{janus:13}, with $h=0.2$. We show the behavior of the average,
and of quantiles 1, 5 and 9.  If $L$ is large enough, $\xi_L/L$ should go as
$1/L$, while if the system is seeing purely an echo of the divergence of the $h=0$ transition
transition, then $\xi_L/L$ should be constant.}
\label{fig:xiL_Tc}
\end{figure}

\index{crossover}\index{xiL/L@$\xi_L/L$}
From the scaling with the lattice linear size of $\xi_L/L$ at
$T_\mathrm{c}=T_\mathrm{c}(h=0)$, we can get another element to discard the
hypothesis that the $h=0$ transition is biasing significantly our measures.
Assuming that there is no critical line for $h>0$, a very large correlation
length could be due to an echo of the zero-field transition or a low-temperature 
effect. 
In a theory that predicts that system is critical only at $h=0$, $T=T_\mathrm{c}$,
the effects of this echo on the $h>0$ behavior should be maximal near $T=T_\mathrm{c}$.
So, if we find a $\xi$ that is large compared to our lattice sizes
for $T<T_\mathrm{c}$, a primary check is to monitor the scaling of the coherence
length at $T_\mathrm{c}$.
figure~\ref{fig:xiL_Tc} shows the scaling of $\xi_L/L$ at
$T_\mathrm{c}$ with $h=0.2$. We plot the average, the first, the fifth and the
highest quantile. All of them show a clear decrease of $\xi_L/L$ when
increasing the lattice size, so our lattice sizes are large enough to
state that the divergence at $h=0$ is not dominating $\xi_L$'s behavior.
On the other side, we are still far from the thermodynamic limit, 
since when the lattices
are large enough, $\xi_L(T_\mathrm{c})/L$ should decay to zero linearly in\index{xiL/L@$\xi_L/L$}
$1/L$. 

\section{Overview}
\label{sec:conclusions}
We have studied the equilibrium behavior of the three-dimensional Ising
Edwards-Anderson spin glass in an external magnetic field.  Thermalizing the\index{thermalization}
system at sufficiently low temperature was a computationally hard task and
required the use of the \textsc{Janus} dedicated computer to thermalize lattice\index{Janus@\textsc{Janus}!computer}
sizes up to $L=32$, down to temperatures $T\geq0.8$. \index{Monte Carlo!simulations}

First of all, we carried out a traditional analysis of our data. We chose\index{scaling!finite-size}
observables that would be scale invariant at the critical temperature, and
compared them for different lattice sizes, looking for crossings in their
temperature curves.  With this procedure we found no traces of a phase
transition. 

Yet, the scenario is more complicated. Despite the absence of crossings, indications
that something non-trivial is going on are given by signals such as a growing\index{correlation!length}\index{susceptibility}
correlation length (even for our largest lattices), peaks in the susceptibility,
and a wide probability distribution function of the overlap.\index{overlap!distribution}

We noticed a wide variety of behaviors
within the same set of simulation parameters. Some measurements presented
signs of criticality, while others did not. So, we tried to classify them in a
meaningful way.  We sorted our observables with the help of a
\ac{CV}, \index{variate!conditioning} and came up with a quantitative criterion to select the
best \ac{CV}.  Between the ones we proposed, the function of the
instant overlaps that made the best \ac{CV} turned out to be the
median overlap $q_\mathrm{med}$.\index{overlap!median}

\index{droplet picture}\index{replica!symmetry!breaking!scenario}
As a function of the median overlap, the scenario appeared rather
non-trivial.  The averages turned out to be dominated by a very small
number of measurements.  Those with a small $q_\mathrm{med}$ behaved\index{quantile}
similarly to the average: long correlation lengths, very large
susceptibilities, and no signs of criticality.  On the other side, the
median behavior was far from the average, and the behavior of most of
the measurements was qualitatively different from the average, with
smaller correlation lengths and susceptibilities, but non-negligible
indications of scale invariance right below the upper bound
$T^\mathrm{up}(h)$ given in \cite{janus:14b}. Furthermore,
separating the different behaviors of the system we obtain mutually
consistent indications of criticality from our primary dimensionless
magnitudes $\xi_{L,i}/L$ and $R_{12,i}$.\index{xiL/L@$\xi_L/L$}\index{R12@$R_{12}$} The achievement of this
consistency is an important step forward with respect to
\cite{janus:12}, where the phase transition was revealed only
by the $R_{12}$ indicator, but it was invisible to $\xi_L/L$.

Unfortunately we were not able to make a quantitative prediction on
the critical temperatures $T_\mathrm{c}(h)$, because the observables
as a function of the lattice size and of the temperature were very
nonlinear, and the temperatures we reached were not low enough reliably to
identify the crossing points of the quantile-dependent
$\xi_{L,i}/L$ and $R_{12,i}$\index{xiL/L@$\xi_L/L$}\index{R12@$R_{12}$}.

Overall, the presence of a phase transition appears plausible from our
simulations.  Perhaps more importantly, now
the challenge is well defined: in order to be able to give,
numerically, a conclusive answer on the presence of a de
Almeida-Thouless line we need push our simulations down to
$T\simeq0.4$ (at $h=0.2$).  We believe that \emph{Janus II}, the next
generation of our dedicated computer \cite{janus:14b}, will be able to\index{Janus@\textsc{Janus}!computer}
assume this challenge.
\index{spin!Ising|)}
\index{de Almeida-Thouless!transition|)}
\index{spin glass!Edwards-Anderson!in a field|)}
%FINITO
 \chapter{Heisenberg spin glass with random exchange anisotropy \label{chap:ahsg}}
\index{spin!Heisenberg|(}\index{random!anisotropy|see{anisotropy}}\index{numerical simulations@numerical simulations$ $|seealso{Monte Carlo simulations}}
\index{spin-chirality!decoupling|see{Kawamura scenario}}\index{spin-chirality!recoupling|see{Kawamura scenario}}

In the current chapter we expose the physical results of a work that required the use of unusual computing resources
that revealed crucial for our results \cite{baityjesi:14}. We carried out a massive campaign of \ac{MC} simulations, exploiting
the \ac{GPU} clusters \emph{Minotauro},\index{GPU!cluster}\index{GPU!cluster!Minotauro}\index{Minotauro|see{GPU cluster}}\index{GPU!cluster!Tianhe-1A|see{Tianhe-1A}}
\footnote{Barcelona Supercomputing Center, Barcelona, Spain, http://www.bsc.es .}
in Barcelona, and \emph{Tianhe-1a} in Tianjin, China,\index{Tianhe-1A}
\footnote{National Super-Computing Center, Tianjin, China, http://www.nscc-tj.gov.cn/en/ .}
and developing parallel codes in C, CUDA C and MPI to run our programs on one or multiple \acp{GPU}.
\index{programming!MPI}\index{programming!CUDA}\index{programming!C}\index{programming!parallel}\index{programming!GPU}\index{computing|see{programming}}
This chapter will be dedicated to the physical results, while useful information on the computational aspects
of our campaign is supplied in appendix \ref{app:MC}.

\section{The Kawamura scenario \label{sec:ahsg:intro}} \index{Kawamura scenario|(}\index{spin glass!transition}
Already in the late '80s - early '90s there was general agreement on that experimental \acp{SG} undergo a phase transition at
sufficiently low temperature \cite{bouchiat:86,levy:88,gunnarsson:91}.

On the other hand, theoretical work in three dimensions was less advanced, even though one works
with extremely simple models. For the Ising \ac{SG} there were arguments supporting\index{spin glass!Ising}
the existence of a phase transition \cite{franz:94}, that were later confirmed
numerically \cite{palassini:99b,ballesteros:00}. In the Heisenberg case, instead,\index{spin glass!Heisenberg}
all the attempts carried out during the '80s and '90s failed in finding a
phase transition at a finite temperature
$T_\mathrm{SG}>0$ \cite{mcmillan:85,olive:86,morris:86,matsubara:91}. \nomenclature[T...SG]{$T_\mathrm{SG}$}{spin glass critical temperature}
In fact,
Matsubara et al. showed in 1991 that once a small anisotropic term is added to
the Heisenberg Hamiltonian the phase transition becomes
visible \cite{matsubara:91}. This was in agreement with a later domain-wall
computation \cite{gingras:93}. The accepted picture at the time was that the
lower critical dimension (i.e. the spatial dimension below which there is no \index{critical dimension!lower}
phase transition) lie somewhere between $3d$ and $4d$ \cite{coluzzi:95}.

However, the story was slightly more complicated. Villain and coworkers made a
provocative suggestion hypothesizing that, although maybe there was no spin 
glass transition, a different order parameter called chirality (or vorticity)\index{chirality}\index{vorticity|see{chirality}}
could be critical \cite{mauger:90}. Chirality is a scalar observable that
describes vorticity and alignment between neighboring spins [recall its definition (\ref{eq:chirality})
in chapter \ref{chap:obs}] with the idea of mapping XY and Heisenberg to \index{spin!Ising}\index{spin!XY}\index{spin!Heisenberg}
Ising \acp{SG} \cite{villain:77,villain:77b}.

Villain's idea was elaborated by
\index{Kawamura scenario} Kawamura in his \emph{spin-chirality decoupling scenario} \cite{kawamura:92,kawamura:98}. 
In the ideal
case of a purely isotropic system the spin and chiral glass order parameters \index{overlap!spin glass}\index{overlap!chiral glass}
would be decoupled, the \ac{CG} order parameter (\ref{eq:chiral-overlap}) would be critical
whereas the \ac{SG} overlap (\ref{eq:scalar-overlap}) would not display any phase transition.
The introduction of any small anisotropy would couple \index{anisotropy|(}
the two. Since real samples always have some degree of anisotropy (see the following section \ref{sec:ahsg-ani})
the \ac{SG} channel, coupled to the chiral one, would appear critical.

\index{universality!class|(}
Kawamura's scenario was apparently consistent with all the observations until
2003, when Lee and Young employed more efficient simulation algorithms and
finite-size scaling techniques to show that the spin glass channel is critical
also in the fully isotropic model (i.e. the Heisenberg limit) \cite{lee:03}.\index{spin glass!transition}
Both order parameters seemed to become positive at the same temperature.
Further simulations confirmed the existence of a \ac{SG} phase transition,
although uncertainty remains on whether the transition is unique
\cite{campos:06,fernandez:09b} or chiralities order at a slightly higher
temperature $T_\mathrm{CG}$ \cite{viet:09}.\nomenclature[T...CG]{$T_\mathrm{CG}$}{chiral glass critical temperature}
\index{Kawamura scenario|)}

A parallel issue is measuring the chiral order parameter in \index{chirality!experimental measurement}
experiments. Kawamura proposed in 2003 that the extraordinary Hall resistivity \index{extraordinary Hall resistivity}
is a simple function of the linear and non-linear \ac{CG}
susceptibilities \cite{kawamura:03}. Experiments based on this
proposal observed the chiral transition and measured, for instance, the
critical exponent $\delta$ \cite{taniguchi:07}. \index{exponent!critical!delta@$\delta$}
Interestingly enough, the value
of $\delta$ turned out to be in between spin and
chiral glass prediction. Nonetheless, it was impossible to identify a universality
class despite the critical exponents of these systems had been extensively\index{exponent!critical}
measured (at least in the \ac{SG} sector) \cite{bouchiat:86,levy:88,petit:02}: the
impression was that they change in a continuous way from the Heisenberg to the\index{crossover}
Ising limit \cite{campbell:10}, as we increased the anisotropy.

However, analogy with ferromagnetic materials suggests a different
interpretation. Anisotropy would be a relevant parameter in the sense of the
renormalization group \cite{amit:05}. There should be a new dominant \ac{fp}, \index{renormalization group}\index{fixed point}
and symmetry considerations lead to think it should belong to the
\ac{IEA} universality class. Yet, when we add a relevant \index{universality!class!Ising}
parameter to the Hamiltonian, there should be some \emph{crossover}\index{crossover}
effects (recall section \ref{sec:crossover}). In other words, one expects that while the correlation length $\xi$\index{correlation!length}
is small, the critical exponents are closer to the Heisenberg-Edwards-Anderson\index{universality!class!Heisenberg}
universality class, and that only for large enough $\xi$ the universality
class reveals its nature.

Notwithstanding, it is very hard, both numerically and experimentally, to
prepare a \ac{SG} with a large correlation length, since one should wait very long
times (it has been argued that the waiting time $t_\mathrm{w}$ \nomenclature[t....w]{$t_\mathrm{w}$}{waiting time}
required to
reach a certain coherence length is proportional to almost its seventh power,
see e.g. \cite{janus:08b,janus:09b} and \cite{joh:99}). Probably 
this explains why the largest measured correlation lengths are of
the order of only one hundred lattice spacings \cite{joh:99,bert:04}. That is a
rather small distance to reveal the true universality class, so it is
plausible that experiments will find critical exponents between the two
Universality classes.

 In fact, materials are classified
according to the degree of anisotropy in their interactions \cite{petit:02},
which turns out to be relevant in their non-equilibrium magnetic
response \cite{bert:04}. On one end of the materials' spectrum we find the
extremely anisotropic Fe$_{0.5}$Mn$_{0.5}$TiO$_3$, which is maybe the best
realization of the ideal limit of an Ising SG (Ising SGs correspond to the
idealization of uniaxial spins). On the other end, we have very isotropic
alloys such as AgMn or CuMn (whose modelization is notoriously
difficult \cite{peil:09}, due to the presence of short range spin-density wave
ordering \cite{cable:82,cable:84,lamelas:95}).

To further complicate things, in experiments one has to take in account at\index{crossover}
least two relevant crossovers. The first is the competition, that we just
pointed out, between the isotropic and the anisotropic fixed points. It is the
one we treat in this chapter.  The second crossover, that we will not address, is
about short versus long range interactions.  In fact, the Hamiltonian we treat
is short range, but some often neglected interactions, such as the \acl{DM} interaction \index{anisotropy!Dzyaloshinsky-Moriya}
(see following section \ref{sec:ahsg-ani})
have been shown to be quasi-long-range,
in the sense that the interactions are long range, but only until a cut-off
distance of the order of some tens of atomic spacings \cite{bray:82}.
\footnote{For further discussion of the crossover
between long and short range interactions see \cite{amit:05}, section 1.3.1, and \cite{cardy:96}, section 4.3.}

Recent numerical work on the Heisenberg \ac{SG} \index{anisotropy!weak}
with \emph{weak} random exchange anisotropies \cite{martin-mayor:11b}, as they would appear in nature,
found a foggy scenario over the critical properties of the model. It was observed that:
\begin{itemize}
\item The \ac{CG} critical temperature $T_{\mathrm{CG}}$ was
  significantly higher than $T_\mathrm{SG}$, in disagreement with experiments
  and expectations.
\item Apparently, the chiral susceptibility was \emph{not} divergent at \index{susceptibility!chiral glass}\index{susceptibility!spin glass}
  $T_\mathrm{CG}$. This is surprising and, apparently, in contrast with
  experiments \cite{taniguchi:07}. Technically, this lacking divergence
  appeared as a very large anomalous dimension $\eta_\mathrm{CG}\sim
  2$. \footnote{Recall that $\gamma_\mathrm{CG} = \nu (2 -\eta_\mathrm{CG})$ \index{scaling!relations}
  where $\gamma_\mathrm{CG}$ is the critical index for the \ac{CG} susceptibility, 
  while $\nu$ is the correlation-length exponent.\index{exponent!critical!nu@$\nu$}
  }
\item Introducing very weak anisotropies changed dramatically
  $T_\mathrm{SG}$. For example, the $T_\mathrm{SG}$ found by comparing
  systems of size $L=6,12$ was about twice its equivalent on the fully
  isotropic model. This is surprising, since one expects that the critical
  temperature would change very little from the isotropic case when $D$ is as
  small as in \cite{martin-mayor:11b}.
\end{itemize}

To the light of this stumble, we decided to face again the problem of the phase
transition in a model with random anisotropic exchange, but we increased drastically
two factors, the degree of anisotropy and the size of the systems, in order to collect data closer
to the attractive \ac{fp} (recall section \ref{sec:crossover}),\index{fixed point}
\footnote{The underlying assumption is that the whole critical line is dominated by the same \ac{fp}.}
that we suspected to be in the Ising
universality class for symmetry reasons that will be discussed in section \ref{sec:ahsg-model}.

In this chapter we will focus on the uniqueness of the phase transition and on
the Universality class, proposing that there is a unique transition, belonging
to the \ac{IEA} Universality class \cite{edwards:75}. We will
also give an interpretation to the results of \cite{martin-mayor:11b},
showing that the apparent inconsistencies are due to scaling corrections, that
we will try to characterize, since we believe them to be fundamental both in
the interpretation of numerical simulations and of experiments.
\index{universality!class|)}

\section{Anisotropy in spin systems \label{sec:ahsg-ani}}
Experimentally, anisotropies affect significantly the glassy response to external magnetic fields
and the behavior under cooling protocols \cite{bert:04}, and as we have mentioned in the previous section
the anisotropy is the driving element of \index{Kawamura scenario} Kawamura's spin-chirality decoupling scenario.

We quickly review here three of the principal mechanisms that lead to an anisotropy in the Hamiltonian \cite{mydosh:93}.
One one side the single-ion, and on the other the dipolar,
and the \acl{DM} anisotropies. While the first one is site-dependent and does not depend on how the
spins are coupled, the latter two are exchange anisotropies and involve the interactions between spins.

\subsection{Single-ion anisotropy}\index{anisotropy!single-ion}\index{spin!Ising}\index{spin!XY}\index{spin!Heisenberg}
Single-ion anisotropy is produced by the local crystalline electric fields of the solid. It depends on the 
spin and orbit angular moment of the modelled magnet and on the morphology of the 
crystalline structure, for example if the material is made in layers or in chains.
Certain orientations of the spins will be preferred and others will be suppressed.

The simplest form of anisotropy we can think of is a strong uniaxial anisotropy that forces the spins to point
along a single direction, that we usually identify with the $z$ axis. This is the case of the Ising spins.
Also, the system could be forced to lie on a 2$d$ plane, in that case we would talk of XY spins.
One can think Ising and XY systems as Heisenberg systems with an additional term that strongly inhibits certain
components,
\footnote{Note that the anisotropy terms in the two following Hamiltonians are equivalent, $D_\mathrm{Ising}=-D_\mathrm{XY}$, just as in section \ref{sec:crossover}.}
\begin{align}
\nomenclature[H..ising]{$\mathcal{H}_\mathrm{Ising}$}{Hamiltonian with Ising anisotropy}
\nomenclature[H..XY]{$\mathcal{H}_\mathrm{XY}$}{Hamiltonian with XY anisotropy}
\nomenclature[D...ising]{$D_\mathrm{Ising}$}{Ising anisotropy}
\nomenclature[D...XY]{$D_\mathrm{XY}$}{XY anisotropy}
 \mathcal{H}_\mathrm{Ising} &= -\displaystyle\frac{1}{2}\sum_{\norm{\bx-\by}=1} J_{\bx,\by} \vec s_{\bx} \cdot \vec s_{\by} + D_\mathrm{Ising} \displaystyle\sum_{\bx} \left((s_\bx\cdot\hat e_x)^2 + (s_\bx\cdot\hat e_y)^2\right) &,& D_\mathrm{Ising}\gg 1\,,\\[4ex]
 \mathcal{H}_\mathrm{XY} &= -\displaystyle\frac{1}{2}\sum_{\norm{\bx-\by}=1} J_{\bx,\by} \vec s_{\bx} \cdot \vec s_{\by} + D_\mathrm{XY} \displaystyle\sum_{\bx} {(s_\bx\cdot \hat e_z)}^2 &,& D_\mathrm{XY}\gg 1\,.
\end{align}
For ferromagnetic systems (not \acp{SG}), the addition of perturbations of this type to the Hamiltonian 
changes its universality class \cite{cardy:96}.\index{universality!class}
Notice that for infinite anisotropy these Hamiltonians become the usual Ising and XY Hamiltonians.

In an amorphous material this anisotropy can be random, meaning that the preferred axis along which the spins want to align varies locally.
One way to represent this effect is to choose a preferred axis, but assigning randomly how each spin couples to this axis, through a 
random term $D_\bx$ \nomenclature[D...x]{$D_\bx$}{per-site random single-ion anisotropy}
chosen from an appropriate \ac{pdf}. The resulting Hamiltonian is
\begin{equation}
\nomenclature[H..r1]{$\mathcal{H}_\mathrm{r_1}$}{Hamiltonian with random single-ion anisotropy on an established axis}
\nomenclature[H..r2]{$\mathcal{H}_\mathrm{r_2}$}{Hamiltonian with random single-ion anisotropy on a random axis}
 \mathcal{H}_\mathrm{r_1} = -\displaystyle\frac{1}{2}\sum_{\norm{\bx-\by}=1} J_{\bx,\by} \vec s_{\bx} \cdot \vec s_{\by} - \displaystyle\sum_{\bx} {D_{\bx} (\vec s_\bx\cdot \hat e_z)}^2 \,.
\end{equation}
More in general also the direction of the "easy" axis can vary, so
\begin{equation}
 \mathcal{H}_\mathrm{r_2} = -\displaystyle\frac{1}{2}\sum_{\norm{\bx-\by}=1} J_{\bx,\by} \vec s_{\bx} \cdot \vec s_{\by} - \displaystyle\sum_{\bx} {D_{\bx} (\vec s_\bx\cdot \hat n_\bx)}^2 \,,
\end{equation}
where $\hat n_\bx$ are random vectors on the sphere of radius 1.\nomenclature[n....xhat]{$\hat n_\bx$}{random vector on the sphere of radius 1}

\subsection{\acl{DM} anisotropy}\index{anisotropy!Dzyaloshinsky-Moriya}
The \acf{DM} \cite{dzyaloshinsky:58,moriya:60} interaction between two spins $\vec{s}_\bx$ and $\vec{s}_\by$ describes the scattering of a conduction electron 
by $\vec{s}_\bx$. The electron then interacts with a non-magnetic scatterer with large 
spin-orbit coupling, and ends up scattering on spin $\vec{s}_\by$.

% The DM interaction, through a spin-orbit coupling with a third spin, causes
% the interactions between spins in any SG to have a certain degree of random
% anisotropy.

This mechanism can be described with a term
\begin{equation}
\nomenclature[H..DMxy]{$\mathcal{H}^\mathrm{DM}_{\bx,\by}$}{Dzyaloshinsky-Moriya anisotropy term}
 \mathcal{H}^\mathrm{DM}_{\bx,\by} = - \vec B\cdot (\vec{s}_\bx\times\vec{s}_\by)\,,
\end{equation}
where $\vec B = \vec r_\bx\times\vec r_\by$, and $\vec r_\bx$ is the position 
of $\vec{s}_\bx$.\nomenclature[r....x]{$\vec r_\bx$}{position of $\vec{s}_\bx$}
If we write the \ac{DM} term in the form $-\vec{s}_\bx\cdot \bD^\mathrm{DM}_{\bx,\by} \vec{s}_\by$, then
\begin{equation}
\nomenclature[D...DMxy]{$\bD^\mathrm{DM}_{\bx,\by}$}{Dzyaloshinsky-Moriya anisotropy matrix}
 \bD^\mathrm{DM}_{\bx,\by} = \begin{pmatrix} 0 & B^z & -B^y \\ -B^z & 0 & B^x\\ B^y & -B^x & 0 \end{pmatrix}\,.
\end{equation}
This antisymmetric matrix has $\det \bD_\mathrm{DM} = 0$, $\Tr \bD_\mathrm{DM} = 0$, and has rank 2 (so one null eigenvalue).

\subsection{Dipolar anisotropy}\index{Kawamura scenario}\index{anisotropy!dipolar}
The dipolar anisotropy is a weak term: it is never the dominant term of the Hamiltonian. Yet, this type of anisotropy is always present in any kind of spin system, 
due to the fact that there always is a dipolar interaction between spins. This makes it a perfect candidate for the justification of the Kawamura scenario.

The dipolar interaction takes the form
\begin{equation}
 \mathcal{H}_{\bx,\by}^\mathrm{dip} = \frac{1}{r_{\bx\by}^3}\left[\vec{s}_\bx\cdot\vec{s}_\by-3(\vec{s}_\bx\cdot \hat r_{\bx\by})(\vec{s}_\by\cdot \hat r_{\bx\by})\right]\,,
\end{equation}
where $\vec r_{\bx\by}=\vec r_{\bx}-\vec r_{\by}$, and $\hat r_{\bx\by} = \vec r_{\bx\by}/|r_{\bx\by}|$.\nomenclature[r....hat]{$\hat r$}{unit vector}
We can see how the configuration that minimizes the energy actually depends on the mutual orientation of the two dipoles.
So for example, if $\vec s_\bx$ and $\vec s_\by$ are parallel to $\hat r_{\bx\by}$, the two spins will align parallel 
(the energy of the coupling is $-2/r_{\bx\by}^3$ if they are parallel, $+2/r_{\bx\by}^3$ if they are antiparallel), 
while if they are initially perpendicular to $\hat r_{\bx\by}$ they will prefer to be antiparallel 
(the energy is $1/r_{\bx\by}^3$ if they are parallel, $-1/r_{\bx\by}^3$ if they are antiparallel).
Notice that also the energy of the preferred energy minimum is different.

If we express $\mathcal{H}_{\bx,\by}^\mathrm{dip}$ in the form $\vec{s}_\bx\cdot \bD^\mathrm{dip}_{\bx,\by} \vec{s}_\by$, 
we get $D^{\alpha\beta} = \delta^{\alpha\beta}-3 r^\alpha r^\beta$.\nomenclature[delta....ab]{$\delta^{\alpha\beta}$}{Kronecker delta}
Therefore
\begin{equation}
\nomenclature[D...dipxy]{$\bD^\mathrm{dip}_{\bx,\by}$}{dipolar anisotropy matrix}
 \bD_\mathrm{dip} = \begin{pmatrix} 1-3 r^{x}r^{x} & r^{x}r^{y} & r^{x}r^{z} \\ r^{y}r^{x} & 1-3 r^{y}r^{y} & r^{y}r^{z}\\ r^{z}r^{x} & r^{z}r^{y} & 1-3 r^{z}r^{z} \end{pmatrix}
\end{equation}
is a symmetric matrix with a non-zero diagonal.
\index{anisotropy|)}

\section{The Model and its symmetries \label{sec:ahsg-model}}
We study the model introduced by Matsubara et al. \cite{matsubara:91}, which is
particularly convenient because of its simplicity. We consider $N=L^3$
three-dimensional unitary vectors $\vec s_\bx = (s_\bx^1, s_\bx^2, s_\bx^3)$ on a
cubic lattice of linear size $L$, with periodic boundary conditions. The
Hamiltonian is \index{anisotropy}
\begin{equation}\label{eq:HamiltonianDefinition}
\nomenclature[H..ANI]{$\mathcal{H}_\ANI$}{Heisenberg spin glass with a random anisotropic exchange}
  \mathcal{H}_\ANI = -\sum_{<\bx,\by>} ( J_{\bx\by} \vec s_\bx \cdot\vec s_\by +
  \sum_{\alpha\beta} s^\alpha_\bx D^{\alpha\beta}_{\bx\by} s^\beta_\by),
\end{equation}
where the indexes $\alpha,\beta$ indicate the component of the spins.
$J_{\bx\by}$ is the isotropic coupling between sites $\bx$ and
$\by$. $D_{\bx\by}$ is the anisotropy operator: a $3\times3$ symmetric
matrix, where the six matrix elements $D_{\bx\by}^{\alpha\beta}\,,
\alpha\geq\beta,$ are independent random variables, so it can be a fair descriptor of
a dipolar anisotropy.

There is quenched disorder, this means that the time scales of the
couplings $\{J_{\bx\by},D_{\bx\by}\}$ are infinitely larger than those of
our dynamic variables, so we represent them as constant in time random
variables, with $\overline{J_{\bx\by}}=\overline{D_{\bx\by}^{\alpha\beta}}=0$,
$\overline{J_{\bx\by}^2}=1$ and $\overline{(D_{\bx\by}^{\alpha\beta})^2}=D^2$.

We stress that if all the matrix elements $D^{\alpha\beta}_{\bx\by}$ are zero we
recover the fully isotropic Heisenberg model, with $O(3)$ symmetry. \index{symmetry!O(3)@$O(3)$}
However,
if the $D^{\alpha\beta}_{\bx\by}$ are non-vanishing, the only remaining
symmetry is time-reversal: \index{symmetry!Z2@$Z_2$} $\vec s_\bx \longrightarrow -\vec s_\bx$ for all
the spins in the lattice. Time reversal is an instance of the $Z_2$ \index{universality!class!Ising}
symmetry. This is the symmetry group of the \ac{IEA} model \cite{edwards:75}. Hence, we
expect that the $Z_2$ symmetry will be spontaneously broken in a unique phase
transition belonging to the \ac{IEA} Universality class (see
e.g.  \cite{gingras:93}). Of course,
underlying this expectation is the assumption that the anisotropic coupling is
a relevant perturbation in the \ac{RG} sense (as it is the case
in ferromagnets \cite{amit:05}). In fact, the infinite-anisotropy limit
can be explicitly worked out for a problem with \emph{site} anisotropy
[rather than link anisotropy as in equation \eqref{eq:HamiltonianDefinition}]:
one finds an \ac{IEA}-like behavior \cite{parisen:06,liers:07}.

As we argued in section \ref{sec:couplings} it is widely accepted that the universality class does not change with the
probability distribution of the couplings. We take
advantage of this, and choose a bimodal distribution for $J_{\bx\by}$ and\index{universality!P(Jxy)@$P(J_{\bx\by})$}
$D_{\bx\by}^{\alpha\beta}$, $J_{\bx\by}=\pm 1$ and
$D_{\bx\by}^{\alpha\beta}=\pm D$. These couplings can be stored in a single\index{programming!GPU}
bit, which is important because we are using \acp{GPU}, special hardware devices 
where memory read/write should be minimized
(appendix \ref{app:MC}).

We  chose the  two different  values $D  = 0.5,  1$.  We  want to  compare our
results with those  in \cite{martin-mayor:11b}, where simulations 
were  done  on  samples with  weak  random  anisotropies.   In that  work  the
$D_{\bx\by}^{\alpha\beta}$  did not  follow a  bimodal distribution,  but were
uniformly distributed  between $-0.05$ and $0.05$. To  make proper comparisons
we  consider   the  standard  deviation  of  the distribution.  For bimodal
distributions  it is  exactly $D$,  in \cite{martin-mayor:11b}  it is
$\overline{(D^2)^{1/2}}=1/\sqrt{1200}\simeq0.03$.

% In appendix \ref{chap:numerical-details} we treat extensively how the simulations were
% performed, giving our choices of the parameters and explaining how non-trivial issues
% related to thermalization and the parallelization of our \ac{MC} simulations on multiple GPUs.

\section{Simulation details and Equilibration \label{sec:ahsg-SimulationDetails}}
We simulated on the largest lattices to present (up to $L=64$), over a wide\index{Monte Carlo!simulations}
temperature range. \footnote{Of course the limiting factor is in the wide range of
  \emph{relaxation times,\/} rather than temperatures. However, relaxation
  times depend on a variety of implementation-dependent factors (such as the
  temperature spacing in the parallel tempering, or the number of
  \ac{OR} sweeps). Hence, comparison with other work will be easier in
  terms of temperatures.
} This has been possible thanks to an intense
use of graphic accelerators (GPUs) for the computations.  We made use of the
\emph{Tianhe-1A} \ac{GPU} cluster in Tianjin, China,\index{Tianhe-1A}
 and of the
\emph{Minotauro} \ac{GPU} cluster in Barcelona.\index{GPU!cluster!Minotauro}

We used \ac{MC} dynamics throughout all the work, mixing three different Monte Carlo
algorithms, \ac{HB}, \ac{OR} and \ac{PT} as explained in appendix \ref{app:MC}, 
\index{Monte Carlo!heatbath}\index{Monte Carlo!overrelaxation}\index{Monte Carlo!parallel tempering}
since both \ac{HB} and \ac{OR} are
directly generalized to the anisotropic exchange case, where the local field is \index{local!field}
$\vec{h_\bx} = \partial\mathcal{H}_\ANI/\partial \vec{s}_\bx= 
\sum_{\Vert{\vn x}-{\vn y}\Vert=1}[J_{\bx\by} \vec s_\by + D_{\bx\by} \vec s_{\by}]$.

All the simulations were run on NVIDIA Tesla \acp{GPU}. Except $L=64$, $D=0.5$,
where we ran on 45 parallel \acp{GPU}, each sample was simulated on a single \ac{GPU}. \index{programming!GPU!single}\index{programming!GPU!multi}
The interested reader can find in appendix \ref{app:MC} details on
how they were performed.

Table~\ref{tab:nsamples-ahsg} depicts the relevant simulation parameters. For given
$L$ and $D$, the simulations were all equally long, except for $L=64$, $D=0.5$, where\index{anisotropy}
we extended the simulation of the samples with the longest relaxation times.\index{thermalization}\index{relaxation time}
\begin{table}[!tb]
\centering
%\resizebox{\columnwidth}{!}{
\begin{tabular}{ccccccc}
      % \hline
      $D$ & $L$ & $N_\mathrm{samples}$ & $N_\mathrm{MCS}^{\mathrm{min}}$& $N_\mathrm{T}$  & $T_\mathrm{min}$ & $T_\mathrm{max}$\\\hline\hline
      0.5 & 8   & 377                  & 2.048$\times10^4$&  10 & 0.588 & 0.8\\
      0.5 & 16  & 377                  & 4.096$\times10^4$ &  28 & 0.588 & 0.8\\
      0.5 & 32  & 377                  & 3.28$\times10^5$&  45 & 0.583 & 0.8\\
      0.5 & 64  & 185                  & 4$\times10^5$ &  45 & 0.621 & 0.709\\
\hline
      1   &  8  & 1024                 & 2.048$\times10^4$ &  10 & 0.877 & 1.28 \\
      1   & 12  & 716                  & 1.68$\times10^5$ &  20 & 0.893 & 1.28\\
      1   & 16  & 1024                 & 4.096$\times10^4$ &  28 & 0.877 & 1.28 \\
      1   & 24  & 716                  & 1.68$\times10^5$&  40 & 0.900 & 1.28 \\
      1   & 32  & 1024                 & 3.28$\times10^5$&  45 & 0.917 & 1.28\\
      1   & 64  & 54                   & 3.44$\times10^5$&  45 & 1.0 & 1.16009
    \end{tabular}
%}
  \caption[Details of the simulations of \cite{baityjesi:14}]{Details of the simulations. We show the simulation
    parameters for each anisotropy $D$, and lattice size $L$.\index{anisotropy}
    $N_\mathrm{samples}$ is the number of simulated samples.
    $N_\mathrm{T}$ is the number of temperatures that were used in
    parallel tempering. The temperatures followed a geometric sequence
    between $T_\mathrm{min}$ and $T_\mathrm{max}$, and $N_\mathrm{T}$
    was chosen so that the \acs{PT}'s acceptance was around
    $15\%$.  $N_\mathrm{MCS}^{\mathrm{min}}$ is the minimum number of
    \ac{EMCS} for each simulation. The simulation for $L=64$, $D=1$ was
    intended only to locate $T_\mathrm{CG}$.
}\label{tab:nsamples-ahsg}

\end{table}

To ensure thermalization we made a \emph{logarithmic data binning}. \index{thermalization} \index{logarithmic data binning}
Each bin
had twice the length of the previous, i.e. it contained two times more \ac{EMCS}, 
and had twice the measures. More explicitly, let us call
$i_\mathrm{f}$ the last bin: $i_\mathrm{f}$ contains the last half of the
\ac{MC} time series, $i_\mathrm{f}-1$ the second quarter, $i_\mathrm{f}-2$
the second octave, and so on.  This allowed us to create a sequence of values
$\langle \obs_n(i)\rangle$, for every observable $\obs$, where $n$ indicates the
sample, and $i$ identifies the bin, that has length $2^i$ \ac{EMCS}.  A set of
samples was considered thermalized if $\overline{\langle \obs_n(i)\rangle-\langle
  \obs_n(i_\mathrm{f})\rangle}$ converged to zero.  This test is stricter than
merely requesting the convergence of the sequence of $\overline{\langle
  \obs_n(i)\rangle}$, because neighboring blocks are statistically correlated, so
the fluctuation of their difference is smaller \cite{fernandez:08b}. Physical
results were taken only from the last block.

Since the $L=64, D=0.5$ samples were the most \ac{GPU}-consuming, we were more strict with
them.  To ensure and monitor thermalization, beyond the previous criteria, we
measured the integrated autocorrelation time (mixing time) of the random walk\index{relaxation time}
in temperatures of each sample \cite{fernandez:09b,yllanes:11}. In a thermalized sample, all
the replicas stay a significant amount of time at each temperature. We made
sure that all the simulations were longer than 10 times this autocorrelation
time. The sample-to-sample fluctuations were not extreme, and the
autocorrelation times $\tau$ spanned between 10000 \ac{EMCS} to 50000 \ac{EMCS}, depending
on the sample. Finally, we decided to take measures only over the last 64000
\ac{EMCS} of each simulation.

\section{Interpolations, extrapolations and errors \label{sec:ahsg-methodology}}
\index{interpolation}\index{extrapolation}\index{errors}
We have been able to estimate the critical temperature from the crossing of
the curves $\xi/L$ at $L$ and $2L$, and the exponents $\nu$ and $\eta$ with
the method of the quotients, as described in section \ref{sec:FSS}.
\index{scaling!finite-size}\index{exponent!critical!nu@$\nu$}\index{exponent!critical!eta@$\eta$}\index{xiL/L@$\xi_L/L$}

To identify the crossing point between the pairs of curves (figures \ref{fig:xiL_SG} and \ref{fig:xiL_CG}), we used low-order
polynomial fits: for each lattice size, we took the four temperatures in the
parallel tempering nearest to the crossing point. We fitted these four data
points to a linear or quadratic function of the temperature.  The obtained
results were compatible within one standard deviation (the values reported in
this work come from the linear interpolation).  In order to calculate $\nu$ we
needed the derivative of the correlation length at the crossing point. We\index{correlation!length}
extracted it by taking the derivative of the polynomial interpolations.

However, there is a difficulty in the calculation of statistical errors: the\index{errors!jackknife}\index{errors!correlated data}
fits we had to perform came from strongly correlated data (because of the\index{Monte Carlo!parallel tempering}
parallel-tempering temperature swap). Therefore, to get a proper estimate of
the error, we made jackknife blocks, fitted separately each block, and
calculated the jackknife error \cite{amit:05}.

The whole mentioned procedure was fluid while $T_\SG^{L,2L}$ \nomenclature[T...SGL2L]{$T_\SG^{L,2L}$}{crossing temperature of $\chi_{\SG}$ with sizes $L$ and $2L$}
fell in
our simulated temperature span.  Yet, since $T_\SG^{L,2L}$ was fairly
lower than $T_\CG^{L,2L}$, \nomenclature[T...SGL2L]{$T_\CG^{L,2L}$}{crossing temperature of $\chi_{\CG}$ with sizes $L$ and $2L$}
it occurred in four cases that we did not
reach low enough temperatures in our simulations to be able to interpolate the
crossing, and we had to recur to extrapolations.  This happened with $D=1$,
$T_\mathrm{SG}^{32,64}$ and $T_\mathrm{CG}^{32,64}$, and in
the lower
anisotropy $D=0.5$, with $T_\mathrm{SG}^{16,32}$ and $T_\mathrm{SG}^{32,64}$.

The case of $T_\mathrm{SG}^{32,64}(D=1)$ and $T_\mathrm{SG}^{16,32}(D=0.5)$
was not a great issue, because the crossing point was very
near to the lowest simulated temperature, so we treated these crossings just
like the others.  

In the case of $T_\mathrm{SG}^{32,64}(D=0.5)$, instead, we had to
extrapolate at a long distance (see figure \ref{fig:xiL_SG}--top, in the next
section). Again, we performed the extrapolation through linear in temperature
fits. To make the fit of $L=64$ more stable, we took in account a progressive
number of points (i.e. we fitted to the $n$ lowest temperatures). We increased
the number of temperatures, while the crossing temperature was constant. Note
that increasing the number of temperatures in the fit results in a smaller
statistical error for the crossing-temperature. However, $\xi_L(T)/L$ is not a
linear function at high $T$ (see figure \ref{fig:xiL_SG}). Therefore a tradeoff
is needed because, when too
high temperatures were included in the fit, the crossing temperature started
to change, and we knew that curvature effects were biasing it. Our final
extrapolation was obtained from a fit performed on the 10 lowest-temperature
points. Unfortunately, this approach was not feasible for the \ac{SG}
susceptibility due to its strongly non-linear behavior. Hence, in the next
section we will not give an estimate for $\eta_\mathrm{SG}(L=64)$.

In the case of $T_\mathrm{SG}^{32,64}(D=1)$, the simulation was not
devised to reach that crossing point, and we did not extrapolate data.

% \subsection{Correlated measurements}

\section{Spin Glass Transition \label{sec:ahsg-SGtrans}}
\begin{figure}[!htb]
\centering
  \includegraphics[angle=270, width=0.8\columnwidth]{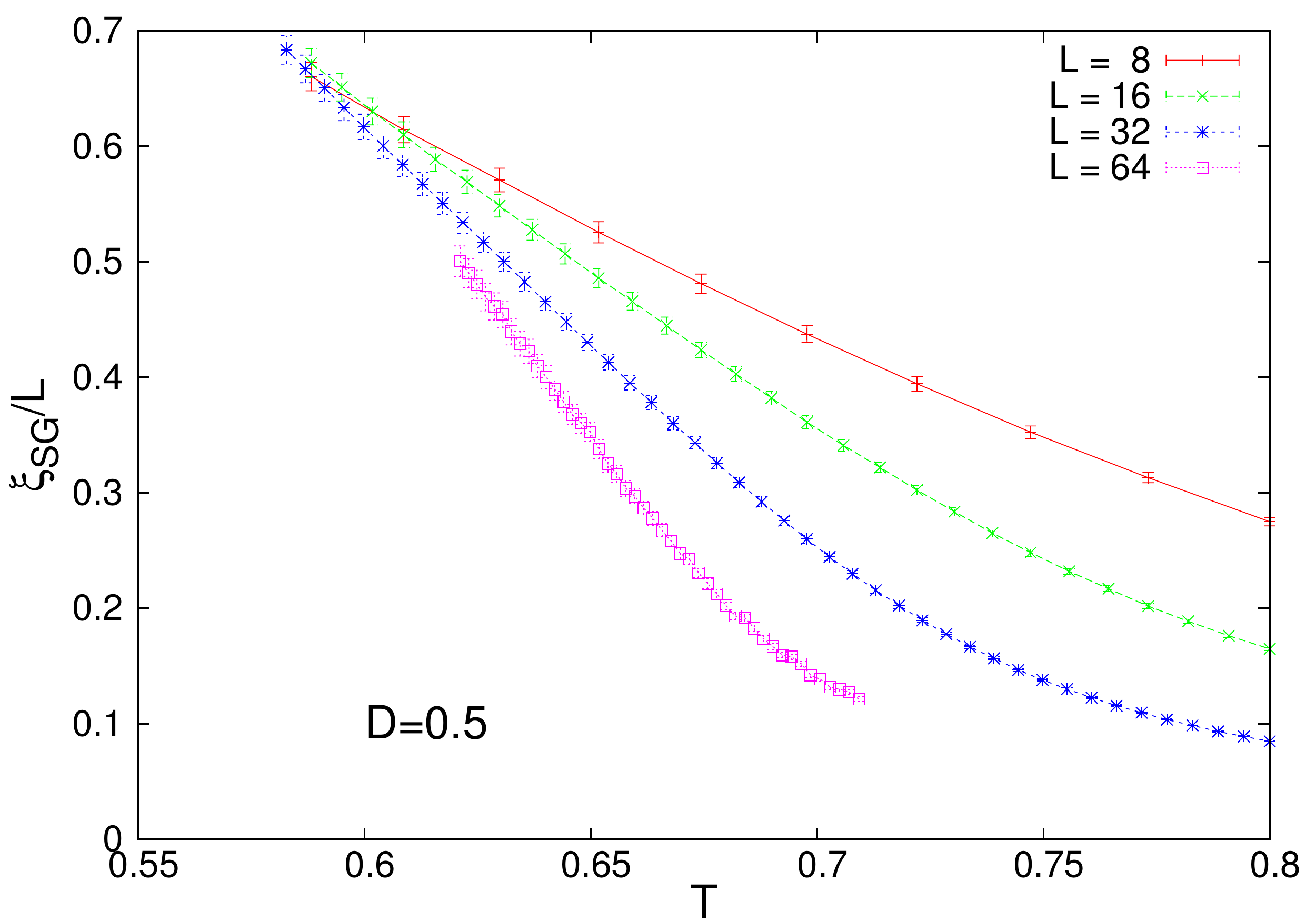}
  \includegraphics[angle=270, width=0.8\columnwidth]{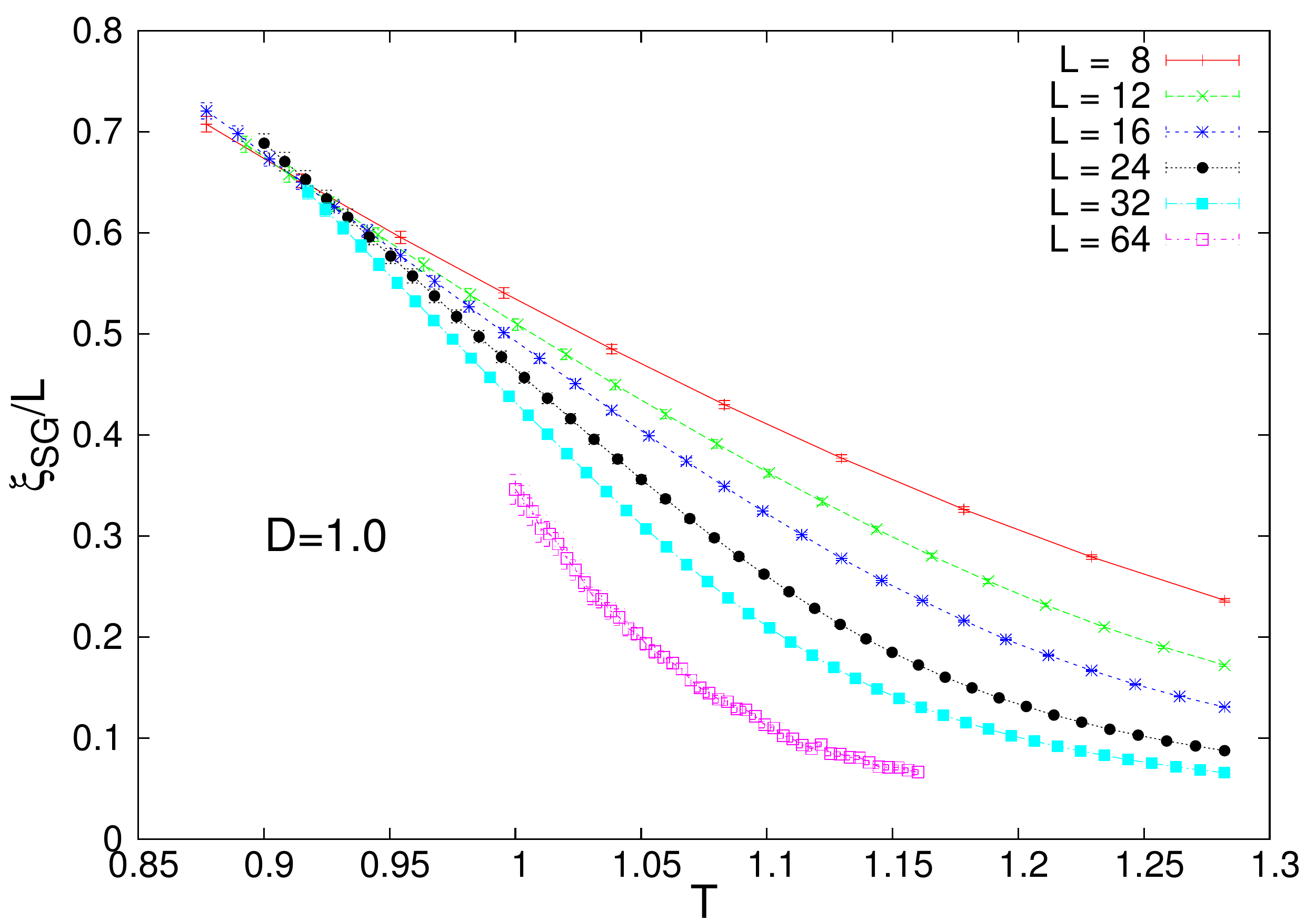}
  \caption[Crossings $\xi_\SG/L$ in the \acs{SG} sector]{Spin glass correlation length in units of
    the linear lattice size $L$ for $D=0.5$ (\textbf{top}) and $D=1$
    (\textbf{bottom}).  All the curves cross at about the same
    temperature for both anisotropies (see
    equation \eqref{eq:scalingCorrections}). The data for $D=1$, $L=64$,
    shown here for the sake of completeness, were only used for the
    chiral sector.}
  \label{fig:xiL_SG}
\end{figure}

\index{scaling!finite-size}
Figures \ref{fig:xiL_SG} show the crossings of $\xi_\mathrm{SG}(T)/L$ \index{xiL/L@$\xi_L/L$|(}
for $D=0.5, 1$. Table \ref{tab:TSG} contains the principal results on the SG
sector, providing a quantitative description of those figures.
As explained in section \ref{sec:ahsg-model}, we expect that the
transition belongs to the \ac{IEA} Universality class.\index{universality!class!Ising}
This conjecture is supported by the fact that the critical exponents
$\nu_\mathrm{SG}$ and $\eta_\mathrm{SG}$, 
and the height at which the
\nomenclature[nu....SG]{$\nu_\SG$}{exponent of the spin glass sector}
\nomenclature[nu....CG]{$\nu_\CG$}{exponent of the chiral glass sector}
\index{exponent!critical!nu@$\nu$}\index{exponent!critical!eta@$\eta$}
$\xi_\mathrm{SG}(T)/L$ cross, are compatible with those of the \ac{IEA} spin glass,
indicated in the last line of table \ref{tab:TSG}.
\begin{table}[!tb]\centering
  \begin{center} 
    %TABLE \ref{tab:TSG}. 
    \emph{Determination of the critical quantities for the \ac{SG} sector.}
  \end{center}
  \centering
%\resizebox{\columnwidth}{!}{
    \begin{tabular}{cccccc}
      $D$ & $(L,2L)$ & $T_\mathrm{SG}$   & $\nu_\mathrm{SG}$ & $\eta_\mathrm{SG}$ & $\xi_\mathrm{SG}(T_\mathrm{SG})/L$\\\hline\hline
      0.5 & (8,16)     & 0.602(18)  & 1.91(27)   & -0.388(27)  & 0.629(48)\\
      0.5 & (16,32)    & 0.577(22)  & 2.70(63)   & -0.449(67)  & 0.705(76)\\
      0.5 & (32,64)    & 0.596(14)  & 2.18(45)   & -           & 0.631(56)\\\hline
      0.5 & $\infty$ & \footnotesize 0.591(16)[0]  & \footnotesize 2.71(82)[3]   & -           & \footnotesize 0.637(87)[1]\\
      \multicolumn{2}{c}{$\chi^2/\mathrm{d.o.f.}$} & 0.55/1   &  0.47/1  & -           & 0.56/1\\\hline\hline
      1.0 & (8,16)     & 0.910(21)  & 2.38(25)   & -0.410(44)  & 0.660(34)\\
      1.0 & (12,24)    & 0.927(19)  & 2.32(28)   & -0.370(53)  & 0.629(36)\\
      1.0 & (16,32)    & 0.910(16)  & 2.37(28)   & -0.400(19)  & 0.660(35)\\\hline
      1.0 & $\infty$ & \footnotesize 0.917(32)[0] & \footnotesize 2.33(67)[0]  & \footnotesize -0.391(71)[1] & \footnotesize 0.662(83)[0]\\
      \multicolumn{2}{c}{$\chi^2/\mathrm{d.o.f.}$}  &  0.66/1 &  0.030/1   & 0.37/1         & 0.55/1\\\hline\hline
      IEA  & $\infty$ &             & 2.45(15)   & -0.375(10)  & 0.645(15)
    \end{tabular}        
%} %resizebox
  \caption[Determination of the critical quantities for the \acs{SG} sector.]
   {\label{tab:TSG} For each anisotropy $D$, and each pair of 
    \index{exponent!critical!omega@$\omega$}\index{exponent!critical!nu@$\nu$}\index{exponent!critical!eta@$\eta$}
    lattices $(L,2L)$, we obtain effective size-dependent estimates
    for $T_\mathrm{SG}$, and the universal quantities
    $\nu_\mathrm{SG}$, $\eta_\mathrm{SG}$ and
    $\xi_L(T_\mathrm{SG})/L$. The thermodynamic limit, indicated with
    $L=\infty$, is obtained by means of fits to
    equations~\eqref{eq:scalingCorrections},~\eqref{eq:nu},~\eqref{eq:eta}
    and~\eqref{eq:xi-L-cross}. Exponent $\omega$ was not a fitting
    parameter (we took $\omega_\mathrm{IEA}=1.0(1)$ from
    \cite{hasenbusch:08}).  The line immediately after the
    extrapolations displays the estimator of the $\chi^2$ figure of
    merit of each one.  $D=\mathrm{IEA}$ represents the critical
    values of the \ac{IEA} Universality class, taken
    from \cite{hasenbusch:08}.  The numbers in square
    brackets express the systematic error due to the uncertainty of $\omega_\mathrm{IEA}$. Notice that
    this systematic error is small compared to the statistical error.}
\end{table}
Hence, it is reasonable to extrapolate our results to $L\rightarrow\infty$ by
assuming the \ac{IEA} universality class. We took $\omega_\mathrm{IEA}=1.0(1)$ from\index{exponent!critical!omega@$\omega$}
\cite{hasenbusch:08}, and fitted to
equations \eqref{eq:xi-L-cross}, \eqref{eq:nu} and \eqref{eq:eta}.
\footnote{At the time these calculations were done and \cite{baityjesi:14} was submitted, 
  the most precise estimation of the critical parameters of the \ac{IEA} model was done 
  in \cite{hasenbusch:08}. At the moment of the drafting of this thesis,
  a more recent yet article from
  the \textsc{Janus} collaboration \cite{janus:13} gives a more \index{Janus@\textsc{Janus}!collaboration}
  precise determination of the critical exponents. The two estimations are compatible and \index{exponent!critical}
  using one or the other does not change qualitatively nor quantitatively our results and conclusions.
  In fact, the statistical errors on the extrapolations are much larger than those deriving from the uncertainty 
  on $\omega$ (see table \ref{tab:TSG}).} \index{exponent!critical!omega@$\omega$}
  In those fits
we took in account both the anticorrelation in the data,
\footnote{\index{errors!correlated data}
Some of the points we used for those extrapolations in chapter \ref{chap:ahsg} shared some of the data, so the measurements could not be treated as independent. 
For example, the crossing of $\xi_L/L$ for $L=8,16$, had in common the points from size $L=16$ with the pair $L=16,32$. 
This means that for the estimation of quantities deriving from the crossings, for example the thermal exponent $\nu$ [eq.(\ref{eq:nu})],
we need to take in account the non-diagonal part of the covariance matrix that gives a measure of the anticorrelation between measurements that share data.
\index{exponent!critical!nu@$\nu$}

For the described case, the typical \ac{JK} statistical error (see appendix \ref{app:JK}) coming from the diagonal part of the covariance matrix is
\begin{equation}
 \sigma_{(8,16;8,16)}^2 = (n-1)\sum_{j=0}^{n-1} \frac{(\nu_{j}^{(8,16)} - \Est(\nu^{(8,16)}) )^2}{n}\,,
\end{equation}
where $n$ is the number of \ac{JK} blocks and $\Est(\ldots)$ is the estimator of the average.
\nomenclature[E...stimator]{$\Est(\ldots)$}{estimator of the average}
The new term we need to take in account in this example is the one coupling the couple $(8,16)$ to the couple $(16,32)$
\begin{equation}
 \sigma_{(16,32;16,32)}^2 = (n-1)\sum_{j=0}^{n-1} \frac{(\nu_{j}^{(8,16)} - \Est(\nu^{(8,16)}) )(\nu_{j}^{(16,32)} - \Est(\nu^{(16,32)}) )}{n}\,.
\end{equation}
}
and
the bias arising from the uncertainty of the exponent\index{exponent!critical!omega@$\omega$}
$\omega_\mathrm{IEA}$.  Notice, from table \ref{tab:TSG}, that the dependence
on $L$ of the data is so weak, that this bias is practically negligible.
This situation is different from the one encountered in
\cite{martin-mayor:11b}, where the anisotropy fields were extremely
small ($D\simeq0.03$, see section \ref{sec:ahsg-model}). There, the finite-size effects in
the \ac{SG} sector were huge.

Overall, the strong consistency of our extrapolations to large $L$ with the
\ac{IEA} exponents shows \emph{a posteriori} that our assumption was proper.

\section{Chiral Glass Transition \label{sec:ahsg-CGtrans}}
\begin{figure}[!htb]
\centering
  \includegraphics[angle=270, width=0.8\columnwidth]{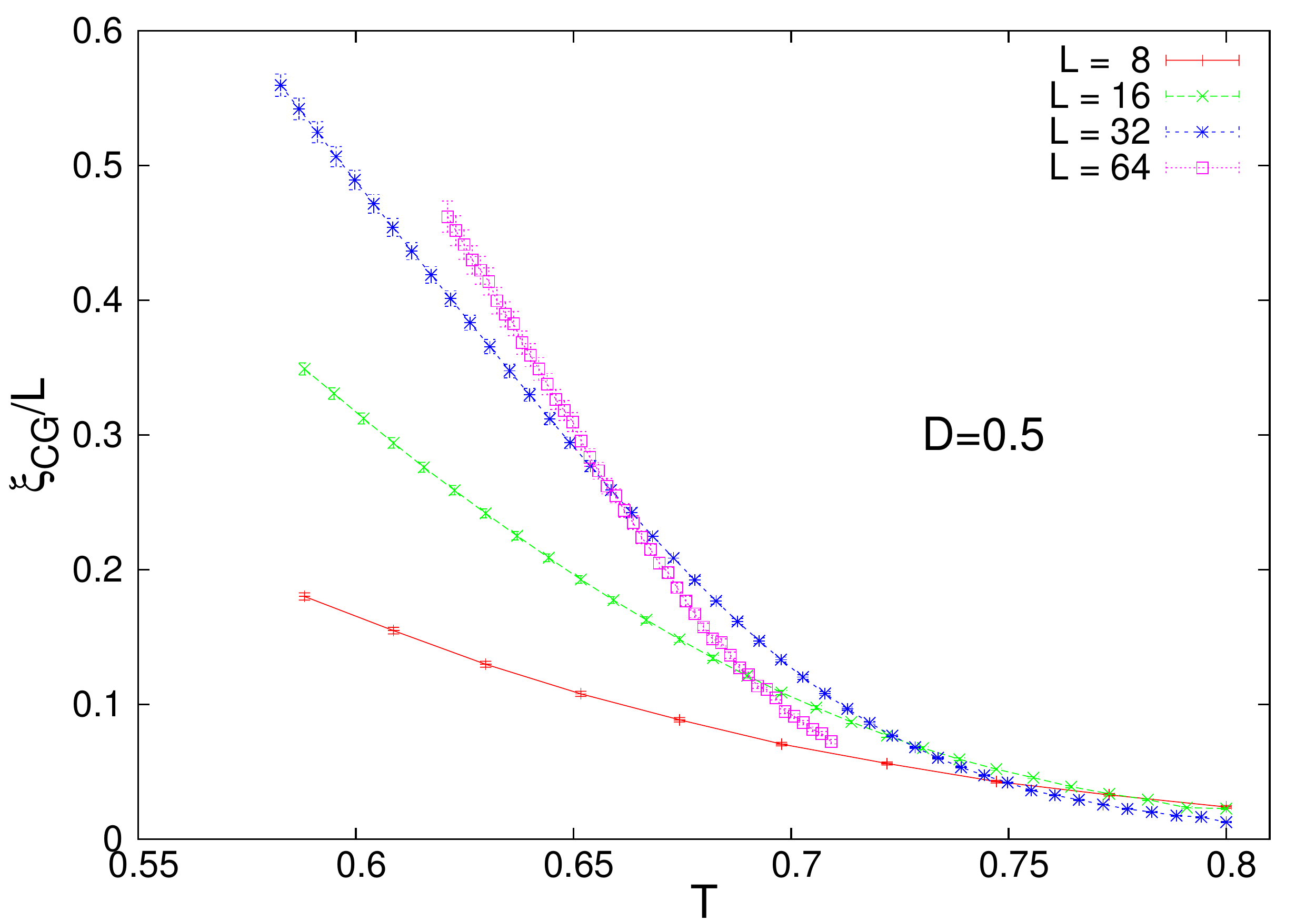}
  \includegraphics[angle=270, width=0.8\columnwidth]{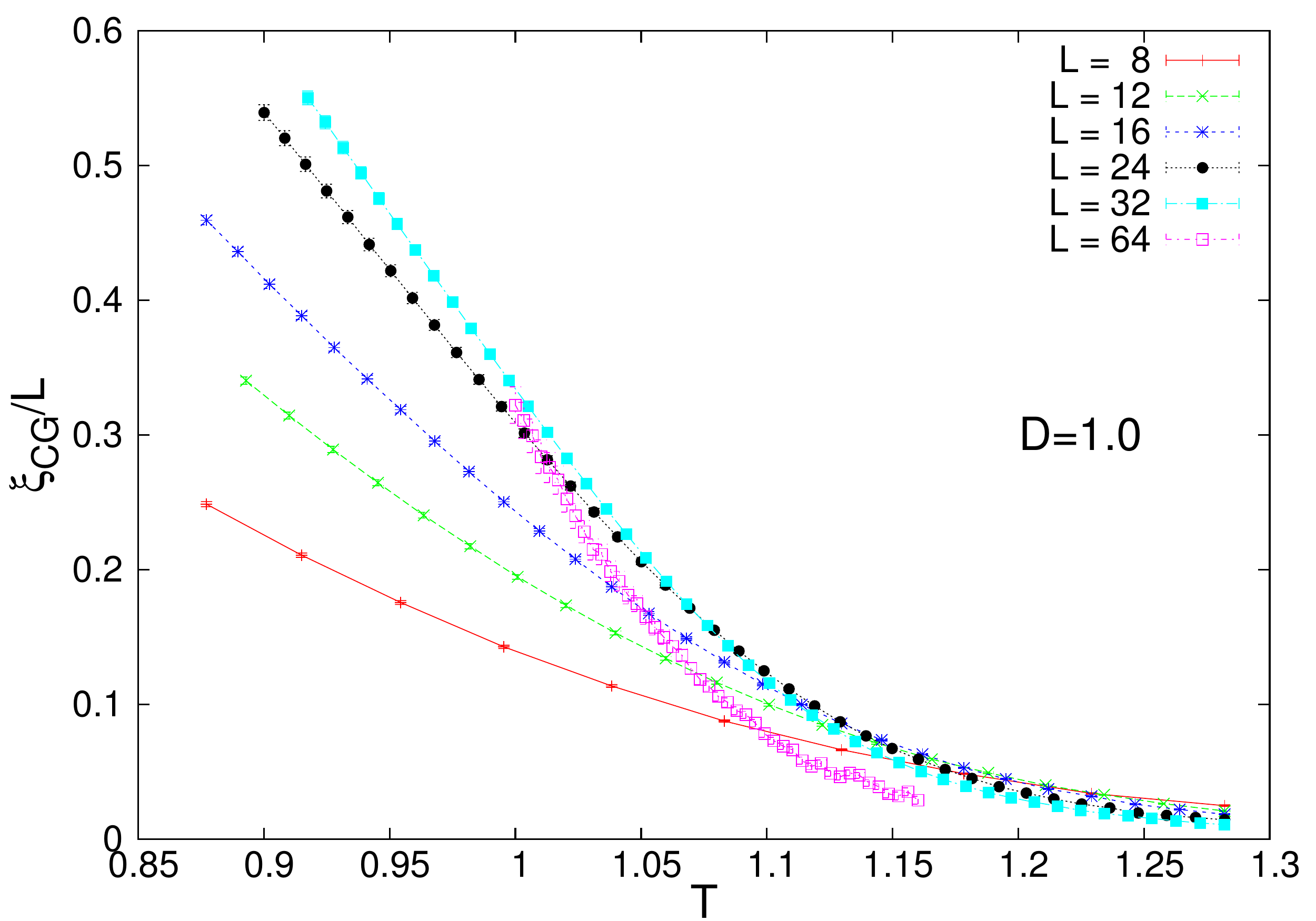}
  \caption[Crossings $\xi_\SG/L$ in the \acs{SG} sector]{Chiral Glass correlation length in units of the
    lattice size for $D=0.5$ (\textbf{top}) and $D=1$ (\textbf{bottom}). When
    $L$ grows, the crossing temperature shifts significantly towards left.}
  \label{fig:xiL_CG}
\end{figure}

In the \ac{CG} channel (figures \ref{fig:xiL_CG} and table \ref{tab:TCG}) the interpretation is slightly more controversial, since 
finite-size effects are heavy.
For the smaller lattice sizes, $T_\mathrm{CG}$ is consistently larger than $T_\mathrm{SG}$, and $\nu_\mathrm{CG}$ is incompatible
with the \ac{IEA} limit. On the other side, when $L$ is larger, $T_\mathrm{CG}$ \index{exponent!critical!nu@$\nu$}
approaches noticeably its SG counterpart, and so does $\nu_\mathrm{CG}$.
We notice that $\eta_\mathrm{CG}$ marks the distinction between these two
regimes. In fact, when $L$ is small, it is very close to 2.  This means that
the divergence of $\chi_\mathrm{CG}$ is extremely slow ($\chi\sim
L^{2-\eta}$), \footnote{Recall that $\gamma_\mathrm{CG} = \nu (2 -\eta_\mathrm{CG})$ 
where $\gamma_\mathrm{CG}$ is the critical index for the CG susceptibility, 
\index{exponent!critical!nu@$\nu$}\index{exponent!critical!eta@$\eta$}\index{scaling!relations}
while $\nu$ is the correlation-length exponent.} revealing we are still far from the
asymptotic limit. When $L$ is larger, $\eta_\mathrm{CG}$ is consistently
smaller, the divergence of $\chi_\mathrm{CG}$ is less suppressed, and we can
assume the asymptotic behavior is starting to show up. Consistently with this
observation, the value of $\xi_\mathrm{CG}/L$ at the crossing temperature
becomes sizeable [indeed, the second-moment correlation length
  \eqref{eq:xi} is well defined only if $\eta<2$, see
  e.g. \cite{amit:05}].
\begin{table}[ht]%\centering
  \begin{center} 
    \emph{Determination of the critical quantities for the \ac{CG} sector.}
  \end{center}
  \centering
%\resizebox{\columnwidth}{!}{
    \begin{tabular}{cccccc}
      $D$ & $(L,2L)$ & $T_\mathrm{CG}$ & $\nu_\mathrm{CG}$ & $\eta_\mathrm{CG}$ & $\xi_\mathrm{CG}(T_\mathrm{CG})/L$\\\hline\hline
      0.5 &  (8,16)    & 0.7762(43) & 1.45(22)   & 1.9778(23)  & 0.0321(22)\\
      0.5 &  (16,32)   & 0.7255(29) & 1.78(14)   & 1.8416(98)  & 0.0735(41)\\
      0.5 &  (32,64)   & 0.659(47)  & 2.40(47)   & 0.823(68)   & 0.258(18)\\\hline\hline
      1.0 &  (8,16)    & 1.2031(33) & 1.205(71)  & 1.9507(27)  & 0.0418(12)\\
      1.0 &  (12,24)   & 1.1472(40) & 1.72(11)   & 1.8664(51)  & 0.0691(25)\\
      1.0 &  (16,32)   & 1.1046(38) & 2.18(10)   & 1.6995(75)  & 0.1098(42)\\
      1.0 &  (32,64)   & 0.987(22)  & 2.48(84)   & 0.53(19)    & 0.368(58)
    \end{tabular}
%} %resizebox
  \caption[Determination of the critical quantities for the \acs{CG} sector.]{Same as table \ref{tab:TSG}, but for chirality. In
    this case the corrections to scaling are significant.}
  \label{tab:TCG}
\end{table}
\index{xiL/L@$\xi_L/L$|)}

\begin{figure}[!t]
\centering
  \includegraphics[angle=0, width=0.7\columnwidth]{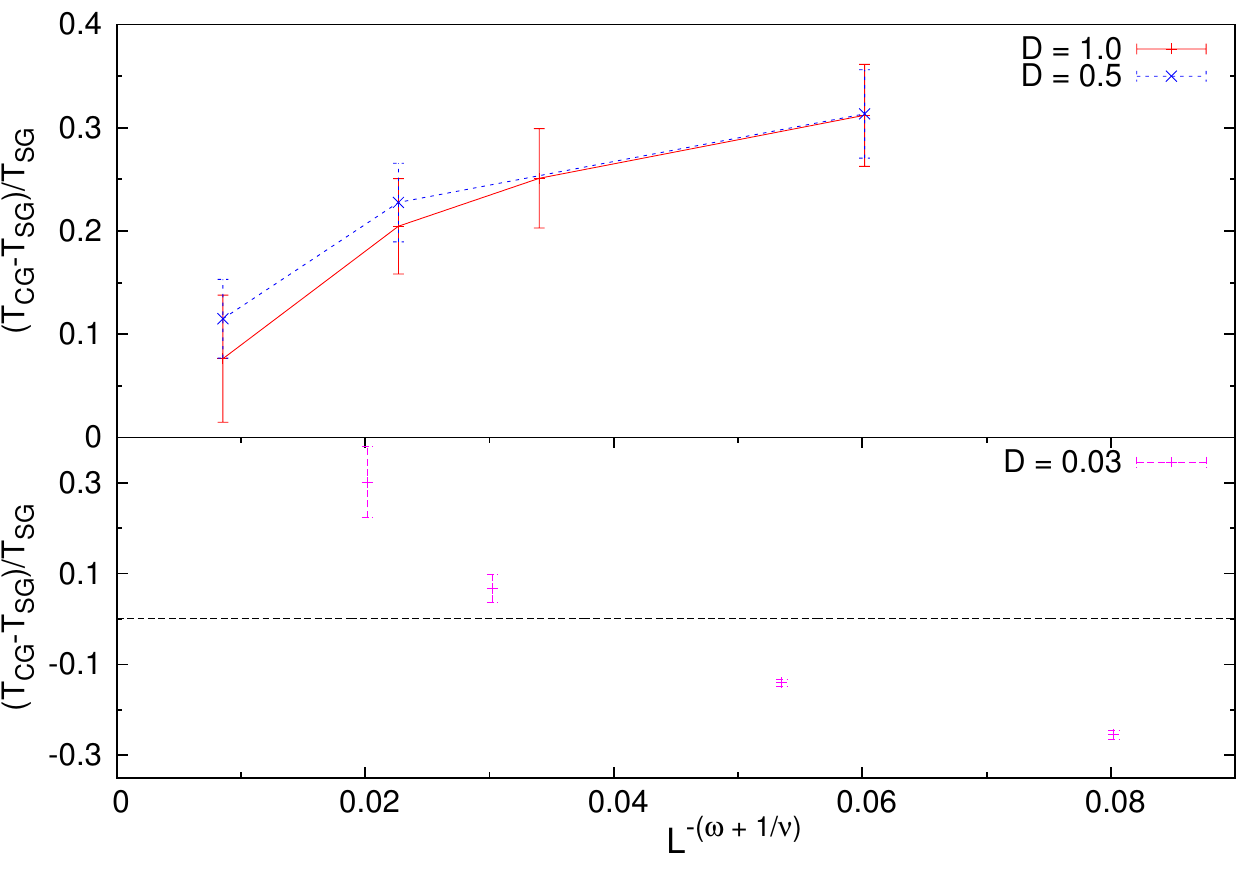}
  \caption[Difference between \acs{SG} and \acs{CG} crossings]{Difference between the chiral crossing $T_\mathrm{CG}$ 
    and the spin glass transition temperature $T_\mathrm{SG}^\infty$, 
    in units of $T_\mathrm{SG}^\infty$ (see Table~\ref{tab:TSG} 
    for the extrapolations of $T_\mathrm{SG}^\infty$).  The exponents
    $\omega_\mathrm{IEA}$ and $\nu_\mathrm{IEA}$ are taken from \cite{hasenbusch:08}. 
    In the \textbf{upper} plot we represent our\index{exponent!critical!omega@$\omega$}
    data, for $D=0.5, 1$.  The two transitions get closer when we increase $L$,
    and the approach appears faster when the lattice size increases. Notice that
    a linear interpolation between the
    two largest lattice sizes intercepts the $y$ axis compatibly with a coupling
    between the two transitions (i.e. $T_\mathrm{SG}=T_\mathrm{CG}$).  On the
    \textbf{bottom} plot we show data from \cite{martin-mayor:11b},
    where much lower anisotropies were considered. Here the scenario is\index{anisotropy}
    completely different, since the critical temperatures drift apart for large
    enough $L$. The horizontal dashed line corresponds to
    $T_\mathrm{CG}-T_\mathrm{SG}=0$.}
  \label{fig:deltaTc}
\end{figure}

\subsection{Uniqueness of the transition}\index{Kawamura scenario}
Although the \ac{SG} and \ac{CG} transitions do not coincide yet with our values of $L$ and $D$, the 
critical temperatures, as well as $\nu$,\index{exponent!critical!nu@$\nu$}\index{exponent!critical!eta@$\eta$}
become more and more similar as the linear size of the system increases. 
Moreover, the decrease of $\eta_\mathrm{CG}$ as a function of $L$ has not yet stabilized, 
so it is likely that the chiral quantities
will keep changing with bigger lattice sizes.

As explained in section \ref{sec:ahsg-model}, we expect that the \index{universality!class}\index{scaling!finite-size}
transition should belong to the \ac{IEA} Universality class.  To confirm this
expectation, we make the ansatz of a unique transition, of the \ac{IEA} Universality
class, to seek if the two critical temperatures join for $L\rightarrow\infty$.
Figure \ref{fig:deltaTc} (upper half) shows the difference between the
critical temperatures as a function of the natural scale for first order
corrections to scaling, $L^{-(\omega_\mathrm{IEA}+1/\nu_\mathrm{IEA})}$\index{exponent!critical!omega@$\omega$}
[equation \eqref{eq:scalingCorrections}]. Again, $\omega_\mathrm{IEA}$ and
$\nu_\mathrm{IEA}$ are taken from \cite{hasenbusch:08}.  Not only
figure \ref{fig:deltaTc} (top) reveals a marked increase of the speed of the
convergence for $L=64$ (to which corresponds the smallest anomalous exponent
$\eta_\mathrm{CG}$), but also, a linear interpolation to infinite
volume, taking that point and the previous, extrapolates
$T_\mathrm{SG}=T_\mathrm{CG}$ within the error.

\begin{figure}[!t]
\centering
  \includegraphics[angle=0, width=0.7\columnwidth]{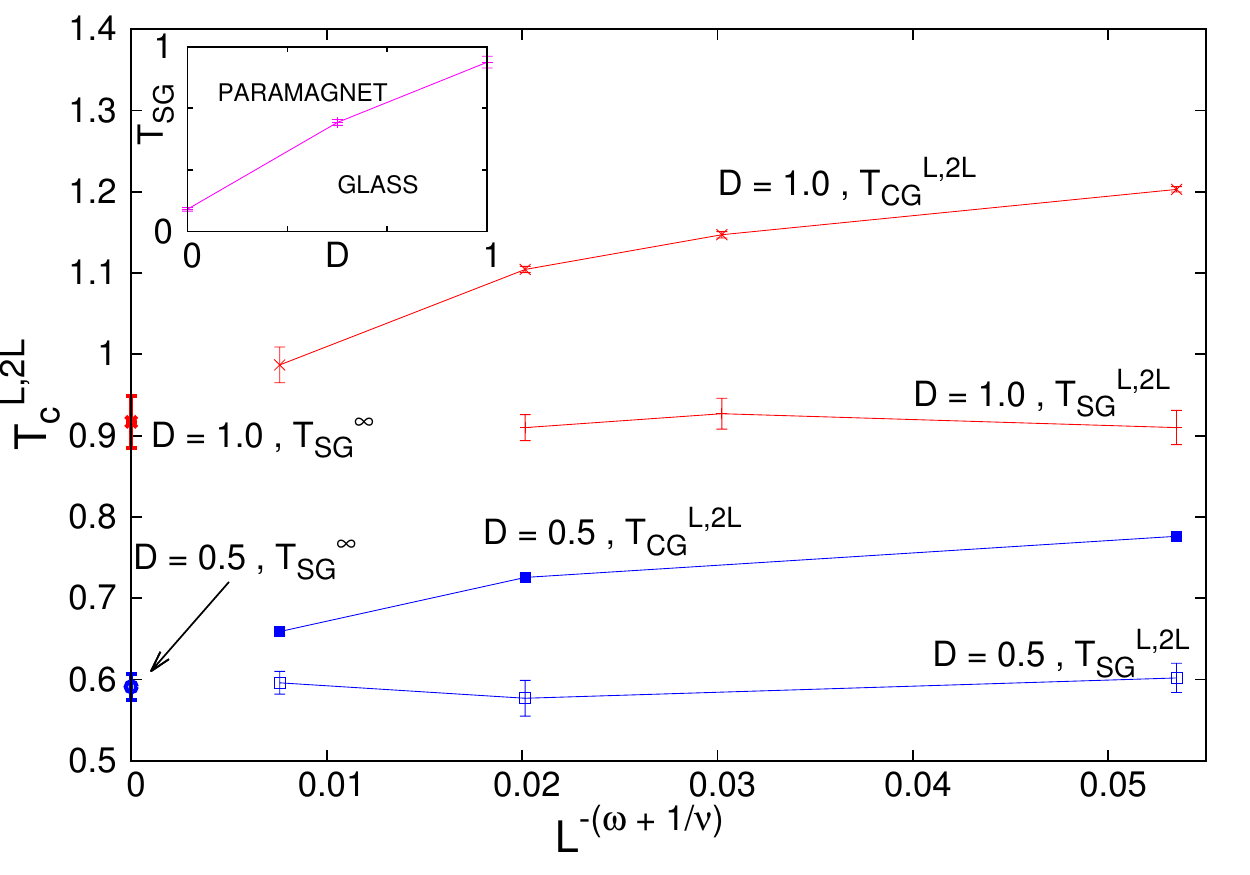}
  \caption[Crossing temperatures as a function of the scaling variable]
    {Crossing temperatures as a function of $L^{-(\omega_\mathrm{IEA}+1/\nu_\mathrm{IEA})}$ (\textbf{large} plot). 
    The points on the intercept are the $L\rightarrow\infty$ extrapolations
    from table \ref{tab:TSG}. The \textbf{inset} shows the phase diagram of the model with these same points, as 
    the most economic interpretation of our data is that\index{exponent!critical!omega@$\omega$}
    in the thermodynamic limit
    $T_\mathrm{SG}=T_\mathrm{CG}$. The $D=0$ point is borrowed from \cite{fernandez:09}.}
  \label{fig:Tc_vs_L}
\end{figure}
Figure \ref{fig:Tc_vs_L} shows how the \ac{SG} and \ac{CG} critical temperatures
approach each other with $L$. Again, $T_\mathrm{CG}$ gets closer to
$T_\mathrm{SG}$, and the speed of the approach increases with the lattice
size. 
The points in the intercept represent extrapolations to the
thermodynamic limit of the $T_\mathrm{SG}$. 
Since the observations are compatible with the ansatz of a unique phase transition,
belonging to the \ac{IEA} universality class,
we used the infinite-size limit of $T_\mathrm{SG}$ to plot the
model's phase diagram (figure \ref{fig:Tc_vs_L}, inset).
\footnote{In the phase 
diagram we show, the $D=0$ point comes
  from \cite{fernandez:09b}, where chiral and spin glass transition
  are assumed to be coupled.  There is disagreement on whether
  $T_\mathrm{SG}=T_\mathrm{CG}$ also in the isotropic case. Yet, we do plot it
  as a single transition because although $T_\mathrm{SG}$ might be lower than
  $T_\mathrm{CG}$, their best estimates are compatible (and not
  distinguishable in the plot).}

\section{Comparing with weak anisotropies}\index{anisotropy!weak|(}
Both plots of figure \ref{fig:deltaTc} show the same observable, for different
anisotropies.  The top plot depicts our data, in the case of strong
anisotropies $D=0.5,1$. The bottom one represents the case of weak
anisotropies ($D\simeq0.03$), coming from
\cite{martin-mayor:11b}.  The behavior is very different between the
two cases.  For strong anisotropies, the critical temperatures tend to meet as
we increase $L$.  That is qualitatively very different from the weak
anisotropy case, where their distance increases.  We can ask ourselves where
this qualitative difference of behavior comes from.

\begin{figure}[!t]
\centering
  \includegraphics[angle=0, width=\columnwidth,trim=0.5cm 0.5cm 5.0cm 17cm]{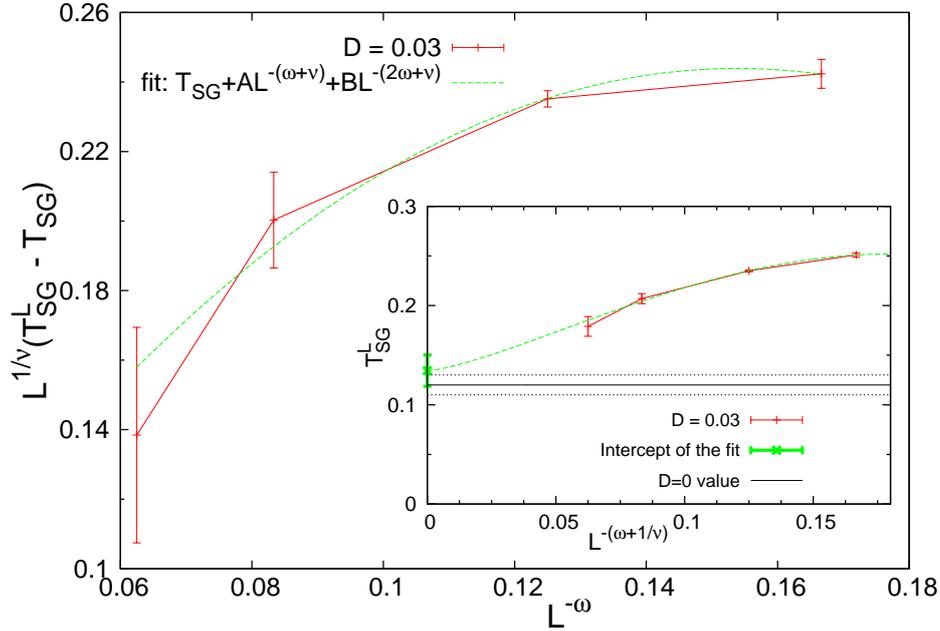}%trim=left bottom right top
   \caption[Finite-size scaling for small $D$]{Data from \cite{martin-mayor:11b}, corresponding
    to $D\simeq0.03$, with extrapolations to the
    thermodynamic limit assuming the Ising-Edwards-Anderson Universality class.
    The data is the same in both plots. The dashed line is a fit of the scaling
    in $L$, considering corrections up to the second order
    [equation \eqref{eq:scalingCorrections}].  The \textbf{large} figure displays the
    trend of the scaling variable $L^{1/\nu}(T-T_\mathrm{SG})$ as a function of
    $L^{-\omega}$. The \textbf{inset} shows the same data set, plotting
    $T_\mathrm{SG}^{L,2L}$ as a function of $L^{-\omega-1/\nu}$, see\index{exponent!critical!omega@$\omega$}\index{exponent!critical!nu@$\nu$}
    equation \eqref{eq:scalingCorrections}. The extrapolation to large-$L$
    (the point in the intercept) is compared with $T_\mathrm{SG}$ of $D=0$
    from \cite{fernandez:09b}. The full horizontal line is the central value
    of $T_\mathrm{SG}^{D=0}$, and the dashed lines define the error.  }
  \label{fig:TSG_D003}
\end{figure}

If we compare same system sizes and different $D$ in table \ref{tab:TCG}, 
we notice that finite-size effects are larger (and $\eta$ closer to two) 
the smaller the anisotropy. These differences in the finite-size effects 
are appreciable with a factor 2 change in the anisotropy (from $D=1$ to $D=0.5$), 
so it is reasonable that suppressing the anisotropy by a factor 17 or 35
will increase drastically the finite-size effects.

The most economic explanation 
is then that there is a non-asymptotic effect that disappears with much larger systems or, as we have 
seen, with larger anisotropies. In other words there is a $L^*(D)$ after which 
$T_\mathrm{SG}$ and $T_\mathrm{CG}$ start joining. For $D\simeq0.03$, $L^*$ is so large that we observe a
growing $T_\mathrm{CG}-T_\mathrm{SG}$, while for $D\geq0.5$ we find $L^*<8$.

Another peculiarity out-coming from \cite{martin-mayor:11b} arises from
the \ac{SG} transition alone.  It had been observed that a very weak perturbation
on the symmetry of the isotropic system implied huge changes in the critical
temperature, while one would expect that the transition line is smooth.

To solve this dilemma, we take advantage of having strong evidence for the Universality class of the transition.
So, we take the data from \cite{martin-mayor:11b}, and use once again the exponents $\nu_\mathrm{IEA}$ and $\omega_\mathrm{IEA}$ in 
 \cite{hasenbusch:08} to extrapolate the infinite volume limit\index{exponent!critical!omega@$\omega$}\index{exponent!critical!nu@$\nu$}
with second order corrections to scaling [equation \eqref{eq:scalingCorrections}].
The fit is good ($\chi^2/\mathrm{d.o.f.}=0.70/1$), and, as we show in figure \ref{fig:TSG_D003}, 
its $L\rightarrow\infty$ extrapolation for the critical temperature is 
compatible with $T_\mathrm{SG}(D=0)$ within one standard deviation.
Thus, taming the finite-size effects was enough to make the scenario consistent,
and the issue reduces to the fact that finite-size effects are extremely strong when 
the anisotropy is smaller.

\section{An \emph{ex post} interpretation \label{sec:ahsg-rg-flow}}
\index{renormalization group!flow}\index{crossover}\index{fixed point}\index{universality!class}
We can reinterpret the results on the Heisenberg model with
random anisotropic exchange interaction from an \ac{RG} perspective. 
It was already established that in the isotropic $D=0$ limit there is a phase transition at $T_\mathrm{SG}^{D=0}$. There are
controversies on whether $T_\mathrm{SG}^{D=0}=T_\mathrm{CG}^{D=0}$, but this is unimportant to us, because it is generally accepted
that $T_\mathrm{SG}^{D>0}=T_\mathrm{CG}^{D>0}$, though it was not verified until \cite{baityjesi:14}. Therefore in the present section we
mention the critical temperature as $T_\mathrm{SG}$.

One of our main questions was whether the universality class changes when $D>0$. Since in nature anisotropies are always present,
though weak, the problem was initially tackled by studying low random anisotropies in \cite{martin-mayor:11b}. To the light of the
remarks of section \ref{sec:crossover}, it was expectable that the numerical results be of hard interpretation.
In fact when starting the \ac{RG} flow from a small anisotropy, the system will initially feel strong effects from the $D=0$ \ac{fp}.
Furthermore, if the flow does not start close to $T_\mathrm{SG}$, the numerical simulations will only feel at first the effects of the
$D=0$ \ac{fp}, and then those of the $T=0$ or $T=\infty$ \ac{fp} (recall figure \ref{fig:cross-over-cardy} and discussion).

\begin{figure}[!t]
 \centering
 \includegraphics[width=1.0\columnwidth]{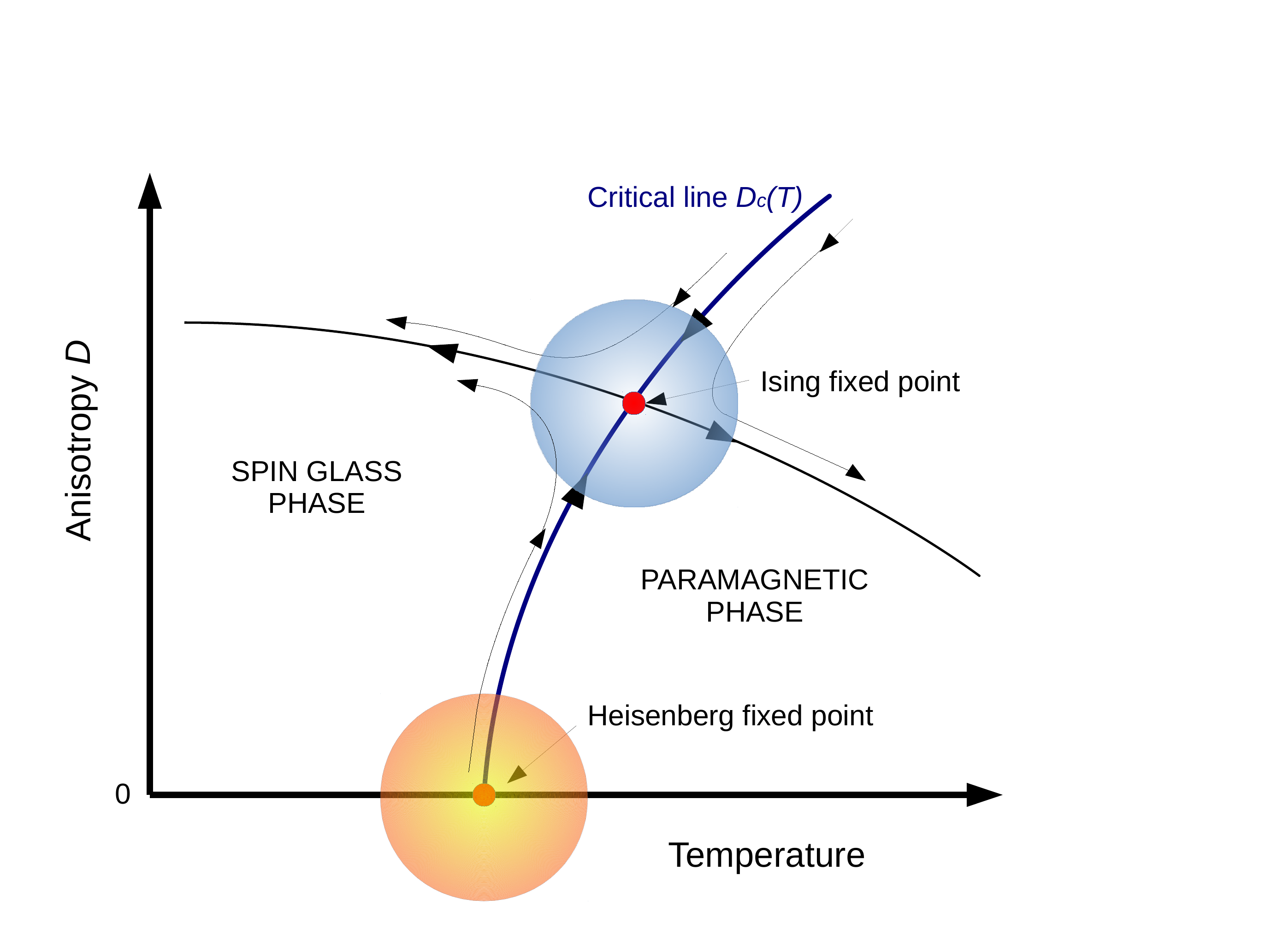}
 \caption[RG flow in the Heisenberg \acs{SG} with random anisotropies]
 {\index{renormalization group!flow}\index{crossover}\index{fixed point}
 RG flow in the Heisenberg \ac{SG} with random anisotropies. The orange zone represents the zone of the phase diagram where the
 echoes of the Heisenberg \ac{fp} are strong (even though it is not an attractive \ac{fp}). The blue area is equivalent, but for the
 Ising \ac{fp}. The Ising \ac{fp} is attractive along the critical line $D_\mathrm{c}(T)$, but it is not in the rest of the phase
 diagram, so to approach the blue from the orange zone one must follow a flow that starts very close to $D_\mathrm{c}(T)$.\index{spin glass!transition}
 Further discussions in the main text.
 }
 \label{fig:RG-flow}
\end{figure}

Of the three options that in section \ref{sec:crossover} are suggested to get away from this hard regime, we are able to adopt two,
increasing drastically both the anisotropy and the lattice sizes, and finished obtaining also a better estimate of the critical temperature. 
The result is depicted in figure \ref{fig:RG-flow}.
Starting the flow from a large anisotropy leads the system far from the zone where echoes of the $D=0$ transition are strong,
and simulating on larger lattices is equivalent to taking more \ac{RG} steps, toward the Ising \ac{fp}. Furthermore, large lattices
gave us a better estimate of the critical temperature, so our movement in the phase diagram sped towards the Ising \ac{fp} in 
an effective way. In terms of figure \ref{fig:RG-flow} we moved from the outer part of the Heisenberg fixed point influence (drawn
in orange, smaller lattices), to the zone where the Ising behavior is strong, blue zone ($L=64$), so we were able to measure an Ising behavior.
\index{anisotropy|)}

\clearpage

\section{Overview \label{sec:ahsg-conclusion}}\index{anisotropy}\index{spin glass!transition}
We performed a numerical study of the critical behavior of Heisenberg spin
glasses with strong bimodal random anisotropies. Our aim was to clarify the role of \index{scaling!corrections}
scaling-corrections, as well as the crossover effects between the Heisenberg\index{scaling!finite-size}\index{universality!class}
and Ising Universality classes, to be expected when the anisotropic
interactions are present. In fact, we show that anisotropic interactions are a
relevant perturbation in the \ac{RG} sense: no matter how small 
the anisotropy, the asymptotic critical exponents are those of the Ising-Edwards-Anderson \index{exponent!critical}
model. However, a fairly large correlation length maybe needed to reach the asymptotic regime.
This observation is relevant for the interpretation of both numerical
simulations \cite{martin-mayor:11b}, and experiments \cite{petit:02}.

It is then clear that large system sizes are needed to make progress,\index{GPU@GPU|seealso{programming GPU}}
something that calls for extraordinary simulation methods. Therefore, we
performed single-\ac{GPU} and multi-\ac{GPU} simulations to thermalize lattices up to
$L=64$ at low temperatures. As side benefit, our work provides a
proof-of-concept for \ac{GPU} and multi-\ac{GPU} massive simulation of spin-glasses with
continuous degrees of freedom. This topic is elaborated further in
Appendix~\ref{app:MC}.

We performed a finite-size scaling analysis based on phenomenological\index{xiL/L@$\xi_L/L$}
renormalization (section \ref{sec:FSS}). We imposed scale-invariance
on the second-moment correlation length in units of the system size,
$\xi_L/L$.  We followed this approach for both the chiral and spin glass order
parameters.

Our results for the spin-glass sector were crystal clear: all the indicators
of the Universality class were compatible with their counterparts in the
Ising-Edwards-Anderson model. On the other hand, in the chiral sector
scaling-corrections were annoyingly large, despite they decrease upon
increasing the magnitude of the anisotropic interactions.

Regarding the coupling of chiral and spin glass transition, our\index{Kawamura scenario}
numerical results seem to indicate that the two phase-transitions take
place at the same temperature
(i.e. $T_\mathrm{CG}=T_\mathrm{SG}$). However, it is important to
stress that we need our very largest lattices to observe this trend.
Nevertheless, what we see is in agreement with both Kawamura's
prediction and experiments, where the phase transitions are apparently
coupled, and the chiral glass susceptibility is\index{susceptibility!chiral glass}
divergent \cite{taniguchi:07}.

Moreover, we were able to rationalize the
numerical results in \cite{martin-mayor:11b} with corrections to
scaling, by assuming the Ising-Edwards-Anderson Universality class.

We remark that there are strong analogies between the interpretation of
numerical and experimental data. In both cases, there is a relevant
length scale (the correlation length for experiments, the system size for
simulations). If that length is large enough, the asymptotic
Ising-Edwards-Anderson Universality class should be observed. Otherwise,
intermediate results between Heisenberg and Ising are to be expected, and
indeed appear \cite{petit:02}.

The difficulty in reaching the asymptotic regime lies on time: the time growth\index{relaxation time}\index{correlation!length}
of the correlation length is remarkably slow ($\xi(t_\mathrm{w})\sim
t_\mathrm{w}^{1/z}$ \index{exponent!dynamical!z@$z$}\nomenclature[z....5]{$z$}{In chapter \ref{chap:ahsg}: exponent that relates space and time}
with $z\approx 7$ \cite{janus:08b,janus:09b,joh:99}, where
$t_\mathrm{w}$ is the waiting time). Indeed, the current experimental record
is around $\xi\sim$ 100 lattice spacings \cite{joh:99,bert:04}, pretty far from
the thermodynamic limit.\footnote{In a typical system \nomenclature[N...a]{$N_\mathrm{A}$}{Avogadro number}$N=L^3\sim N_\mathrm{A}\approx6\cdot10^{23} 
\Rightarrow L\simeq10^8$.} Hence attention should shift to the
study of the intermediate crossover regime. An intriguing possibility\index{crossover}
appears: one could envisage an experimental study of the crossover effects as
a function of the \emph{waiting time}. In fact, $t_w$ varies some four orders
of magnitude in current experiments \cite{rodriguez:13}, which should result in
a factor 4 variation of $\xi(t_\mathrm{w})$.
\index{spin!Heisenberg|)}

%FINITO
\part{Energy Landscapes\label{part:EL}}
 \chapter[Energy landscape of \texorpdfstring{$\bm{m}$}{m}-component spin glasses]
        {Energy landscape of \texorpdfstring{\bm{$m$}}{m}-component spin glasses \label{chap:hsgm}} \label{sec:hsgm-intro}
\chaptermark{Energy landscape of $m$-component spin glasses}\index{spin glass!theory}
Although it is established that typical spin glasses \cite{mezard:87} order at a critical 
temperature $T_\SG$ for $d\ge 3$ \cite{ballesteros:00,kawamura:01,lee:03},
the nature of the low-temperature phase of spin glasses under the upper critical dimension
$d_u=6$ is still a matter of debate (section \ref{sec:spin-glass-intro}).\index{critical dimension!upper}

Already at the dawning of spin glass theory interest had been given to the behavior
of \acp{SG} as a function of the number of spin components $m$ \cite{dealmeida:78b}.
Increasing the number of spin components $m$ reduces the number of metastable states, \index{metastable state}
and recently
renewed interest has been shown towards the properties of these models in the $m\rightarrow\infty$ limit, and their
energy landscape \cite{hastings:00}. \index{energy!landscape}
Interesting features have been pointed out in large-$m$ mean
field models, such as a Bose-Einstein condensation in which the spins condense from an $m$-dimensional 
to an $n_0$-dimensional subspace, where $n_0$ scales with the total number of spins $N$ as $n_0\sim N^{2/5}$ ~\cite{aspelmeier:04}.

It has been argued in \cite{aspelmeier:04} that the $m=\infty$ limit could be a good
starting point for the study of the low-$m$ \acp{SG},
\footnote{For example, in \cite{beyer:12} the infinite-$m$ limit
is used to derive exact relations in the one-dimensional spin glass with power law interactions.}
via $1/m$ expansions that have been used,
for instance, to try to question the presence of a \ac{dat} line \cite{moore:12b}.\index{de Almeida-Thouless!transition}
However the Hamiltonian of the $m=\infty$ model has a unique local minimum, that can be found easily 
by steepest descendent (the determination of the ground state is {\em not} an NP-complete problem). 

Explicit computations  also indicate that the $m=\infty$ model is substantially different from
any finite-$m$ model (for example there is only quasi long-range order under $T_\mathrm{SG}$, the upper critical\index{critical dimension!upper}
dimension has been shown to be $d_u=8$, and the lower critical dimension \index{critical dimension!lower} is suspected to be $d_l=8$ too
\cite{green:82,viana:88,lee:05}), and that it is more interesting to study these
models for large but finite $m$, thus reversing the order of the limits $m\rightarrow\infty$ 
and  $N\rightarrow\infty$ \cite{lee:05}.

To better understand the large (but finite) $m$ limit we undertake a numerical study in a three-dimensional 
cubic lattice. Our aim is to arrive at a quantitative comprehension of the energy landscape of systems with 
varying $m$, expecting, for example,
to observe growing correlations as $m$ increases \cite{hastings:00}.

We focus on infinite-temperature \acp{IS},\index{inherent structure|(}\nomenclature[i.nherent structure]{\small inherent structure}{the local energy minimum reached by relaxing the system}
i.e. the local energy minima that one reaches by
relaxing the system from an infinite-temperature state, that is equivalent to a random configuration. 
Examining a system from the point of view of the
\acp{IS} is a very common practice in the study of structural glasses \cite{cavagna:09}.\index{glass!structural}
Only recently the study of quenches \index{quench}
\footnote{By quench we mean the minimization of the energy throughout the best possible satisfaction
of the local constraints, i.e. a \emph{quench} is a dynamical procedure, as explained in appendix \ref{app:gs}. Be careful
not to confuse it with other uses of the same term. For example, those quenches have little to do with
the \emph{quenched approximation} used in QCD, or the \emph{quenched disorder}, that is a property of the system.}
from a high to a lower temperature has stimulated interest 
also in spin systems, both in presence and absence of quenched disorder.
\footnote{In addition to \cite{berthier:04,baityjesi:11} cited several times in this chapter, one can e.g. see \cite{blanchard:14} for systems
without quenched disorder, and \cite{burda:07} for spin glasses.}

We analyze the properties of the \acp{IS},
and we inspect the dynamics of how the system converges to those configurations.

When one performs a quench from $T=\infty$ to $0<T=T_0<T_{\SG}$, \nomenclature[T...0]{$T_0$}{final temperature of a quench}
the system is expected to show
two types of dynamics, an initial regime where thermal fluctuations are irrelevant, and
a later one where they dominate the evolution (see for example the quenches performed in \cite{berthier:04}).
We choose $T_0=0$, so we can to show that the origin of the second dynamical regime is
actually due to thermal effects. We study the quenches as a function of $m$. While on one side
in the Ising limit $m=1$ the dynamics is trivial, and correlations never become larger than a single
lattice spacing, on the other side an increasing $m$ yields a slower convergence, with the arising of low-temperature correlations 
that we can interpret as interactions between blocks of spins.

\section{Model and Simulations \label{sec:hsgm-model-sim}} 
\subsection{Model \label{sec:hsgm-model}}
The model is defined on a cubic lattice of side $L$ with periodic boundary conditions. Each of the $N=L^3$ vertices 
$\bx$ of the lattice hosts an $m$-dimensional spin $\vec{s}_{\bx}=(s_{\bx,1},\ldots,s_{\bx,m})$, 
with the constraint $\vec{s}_{\bx}\cdot\vec{s}_{\bx}=1$.
Neighboring spins $\vec{s}_{\bx}$ and $\vec{s}_{\by}$ 
are linked through a coupling constant $J_{\bx,\by}$.
The Hamiltonian is \index{spin glass!Edwards-Anderson!Hamiltonian}
\begin{equation}
\label{eq:H}
 {\cal H}_\mathrm{EA} = - \frac{1}{2}\sum_{\norm{\bx-\by}=1} J_{\bx,\by}~\vec{s}_{\bx}\cdot\vec{s}_{\by}\,,
\end{equation}
that was already defined in section \ref{sec:spin-glass-intro}.
The couplings $J_{\bx\by}$ are Gaussian-distributed,
with $\overline{J_{\bx\by}}=0$ and $\overline{J_{\bx\by}^2}=1$.
The local field $\vec h_\bx$ for (\ref{eq:H}) is $\vec h_\bx = \sum_{\by:\norm{\bx-\by}=1} J_{\bx\by} \vec s_\by$.

This Hamiltonian is invariant under the simultaneous rotation or reflection of all the spins [that belongs to the $O(m)$ \index{symmetry!o(m)@$O(m)$}
symmetry group], so the energy minimas may be found modulo a global rotation. For this reason we will use the tensorial\index{overlap!tensorial|(}
definitions of the overlap (section \ref{sec:tens-q}) and correlation functions and lengths (both point and plane, 
section \ref{sec:tens-C}), so that the observables we measure are rotationally invariant too.

When one of the defined quantities is referred to the \acp{IS}
(i.e. the final configurations of our quenches), we will stress it by putting the subscript $_\mathrm{IS}$.
\nomenclature[I.S]{$_\mathrm{IS}$}{the subscripted quantity is referred to an inherent structure}

\subsection{Simulations \label{sec:hsgm-sim}}
We are interested in the \acp{IS} from infinite temperature, hence we need to pick
random starting configurations, and directly minimize the energy.\index{numerical simulations}

The algorithm we choose is a direct quench, \index{quench}
that consists in aligning each spin to its local field $\vec{h}_{\bx}$ (appendix \ref{app:gs}).\index{local!field}
This choice was done because it allows us to compare \acp{IS} from systems with a different $m$
in a general way. For example, the \ac{SOR} (appendix \ref{app:sor})
yields \acp{IS} with different properties, depending on the value of a parameter $\Lambda$ \cite{baityjesi:11},
\nomenclature[Lambda...L]{$\Lambda$}{parameter of the SOR}
and the same $\Lambda$ is not equivalent for two different values of $m$.

\begin{table}[!b]
\centering
 \begin{tabular}{ccccc}
  $L$ & $m$ & $N_\mathrm{samples}$ & $N_\mathrm{sweeps}$ & $\mathcal{N}_\mathrm{m}$\\\hline\hline
  8   & 1  & 10000 & $10^5$ & 22\\
  8   & 2  & 10000 & $10^5$ & 22\\
  8   & 3  & 10000 & $10^5$ & 22\\
  8   & 4  & 5000  & $10^5$ & 22\\
  8   & 6  & 10000 & $10^5$ & 22\\
  8   & 8  & 10000 & $10^5$ & 22\\\hline
 16   & 1  & 1000  & $10^5$ & 22\\
 16   & 2  & 1000  & $10^5$ & 22\\
 16   & 3  & 1000  & $10^5$ & 22\\
 16   & 4  & 1000  & $10^5$ & 22\\
 16   & 8  & 1000  & $10^5$ & 22\\
 16   & 12 & 1000  & $10^5$ & 22\\
 16   & 16 & 1000  & $10^5$ & 22\\\hline
 64   & 3  & 160    & $10^5$ & 22\\
 \end{tabular}
 \caption[Parameters of the simulations of \cite{baityjesi:15}]{Parameters of our simulations. $N_\mathrm{samples}$ is the number of simulated samples, 
$N_\mathrm{sweeps}$ is the number of quench sweeps of the whole lattice, and $\mathcal{N}_\mathrm{m}$ is
the number of measures we did during the quench. We chose to follow the same roughly logarithmic
progression chosen in \cite{berthier:04}, measuring at times 2, 3, 5, 9, 16, 27, 46, 80,
139, 240, 416, 720, 1245, 2154, 3728, 6449, 11159, 19307, 33405, 57797, 100000.
}
\label{tab:param-hsgm}
\end{table}

For each sample we simulated two replicas, in order to be able to compute overlaps.
We fixed the number of full sweeps of a lattice to $N_\mathrm{sweeps}=10^5$, \nomenclature[N...sweeps]{$N_\mathrm{sweeps}$}{number of full sweeps of the lattice}
as it had already been
done in \cite{berthier:04} with quenches to finite temperature. As it can be seen in figures 
\ref{fig:evol-E}, \ref{fig:evol-q}, \ref{fig:evol-qself} and \ref{fig:evol-xi} further on, this amount
of steps was enough to guarantee the convergence to an \ac{IS} in all our simulations.
To ensure the convergence we required the last (logarithmically spaced) measurements to be equal 
within the error for each of the measured observables.

In table \ref{tab:param-hsgm} we give the parameters of our simulations.

\paragraph{Truncated correlators} When the correlation function decays very quickly and the noise 
becomes larger than the signal, one could measure \index{correlation!function!truncation}
negative values of $C(r)$, that would be amplified by the factors $r^2$ and $r^4$ in the integrals \eqref{eq:xi-punto} and \eqref{eq:xi-plano}. This would imply
very large errors in $\xi$, or even the square root of a negative number. To overcome this problem, we truncated the
correlation functions when they became less than three times the error, as it was first proposed in \cite{janus:09b}.
This procedure introduces a small bias, but reduces drastically the statistical error. Furthermore, the plane correlation function required 
the truncation much more rarely, therefore we compared the behaviors as a consistency check.

\section[Features of the inherent structures varying \texorpdfstring{$m$}{m}\label{sec:hsgm-IS-m}]
        {Features of the inherent structures varying \texorpdfstring{\bm{$m$}}{m}\label{sec:hsgm-IS-m}}
\begin{figure}[!tb]\centering
 \includegraphics[width=0.6\columnwidth]{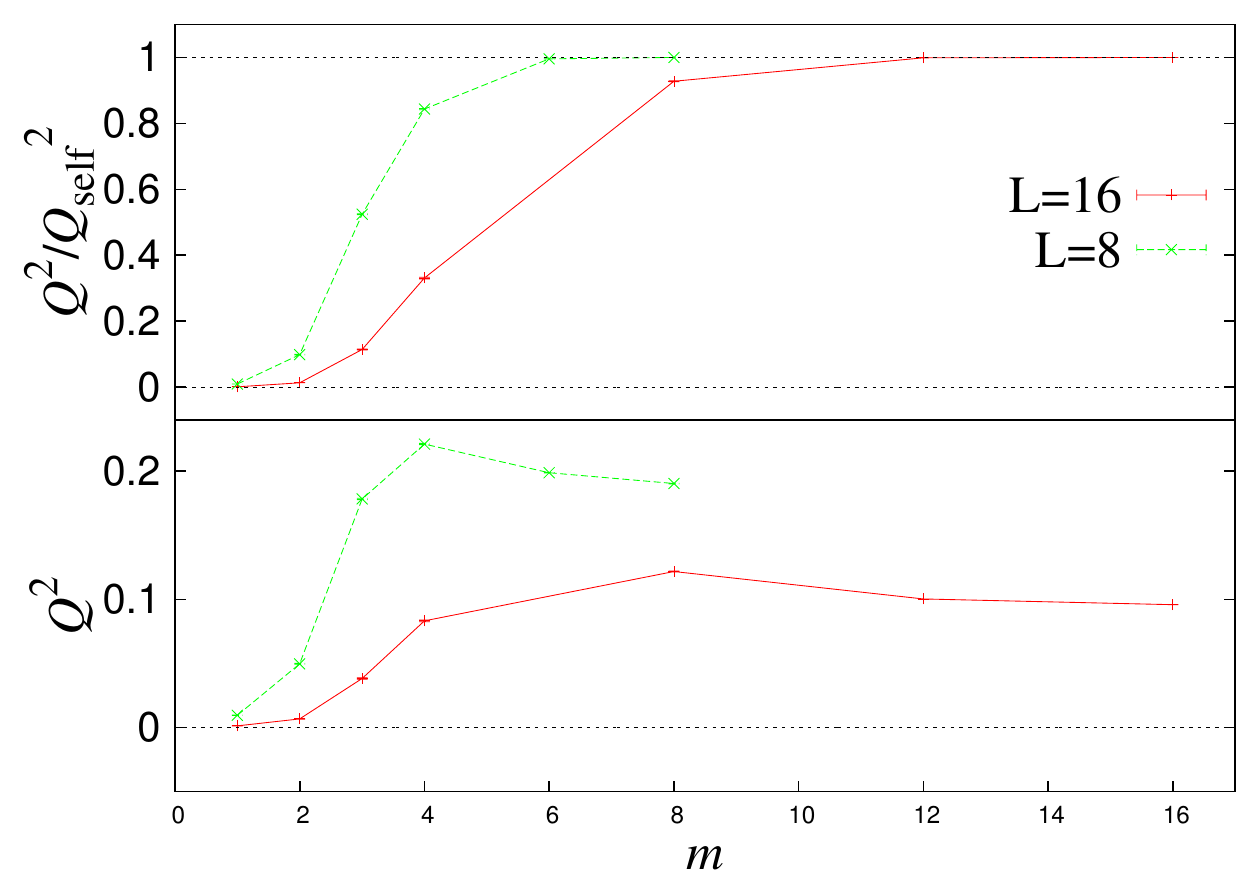}
 \caption[Dependency of the \acsp{IS}' overlaps on $m$]{Dependency of \index{overlap!tensorial}
 the \acp{IS}' overlaps from the number of components $m$
 of the spins. The \textbf{top} figure displays the overlap normalized with the self-overlap,
 showing that when $m$ is large enough the energy landscape is trivial. In the \textbf{bottom}
 we have the unnormalized overlap $Q^2$.
 The dashed horizontal lines represent the limits 0 and 1, that bound both observables.
 Error bars are present though small, so almost not visible.}
 \label{fig:q-m}
\end{figure}

We want to analyse how the model's behavior changes with $m$.
Intuitively, the more components a spin has, the easier it is to avoid frustration \cite{hastings:00}, and the\index{frustration}
simpler is the energy landscape.\index{energy!landscape}
According to this scenario, when $m$ increases, the number of available \acp{IS} decreases
down to the limit in which the energy landscape is trivial, and there is 
only one minimum. 
This should be reflected in the quantity  $Q^2/Q^2_\mathrm{self}$ (recall definition \eqref{eq:Q2-def}), that should be small when there are many 
minima of the energy, and go to 1
when there is only one inherent structure,
since all the quenches end in the same configuration. 
As shown in 
figure \ref{fig:q-m} (top), our expectation is confirmed. With Ising spins ($m=1$)
the energy landscape is so rich that \acp{IS} have practically nothing in common.
When we increase $m$ the overlaps start to grow until the limit $Q^2=Q^2_\mathrm{self}$.
By comparing the data for different $L$, we can dismiss a difference in
the behavior between discrete ($m=1$) and continuous ($m>1$) spins, since
$m=1$ for $L=8$ behaves the same as $m=2$ for $L=16$. In section \ref{sec:hsgm-dynamics} 
we will discuss aspects in which we do encounter differences. 

Since the number of available \acp{IS} depends on both $m$
and $L$, we can give an operative definition of a
ratio $(m/L)_\mathrm{SG}$ under which the number of \acp{IS} is
exponential (so $Q^2/Q^2_\mathrm{self}\simeq0$), and of a ratio $(m/L)_\mathrm{1}$
over which there is only one minimum. 
\nomenclature[m....LSG]{$(m/L)_\mathrm{SG}$}{ratio under which the number of \acp{IS} is exponential}
\nomenclature[m....L1]{$(m/L)_\mathrm{1}$}{ratio over which there is only one minimum of the energy}
This way, we can characterize finite-size effects effectively: An extremely 
small system $m/L>(m/L)_\mathrm{1}$ is trivial and has only one stable state. Increasing
the size we encounter a less trivial behavior, but to find a visible signature 
of a spin glass phase one has to have $L\geq m (L/m)_\mathrm{SG}$. From figure \ref{fig:q-m}
one can see that for $L=8$, $m_\mathrm{SG}=1$, and for $L=16$, $m_\mathrm{SG}=2$. 
Then, for example, we see that to observe a complex behavior for $m=3$
spin glasses, one should use $L>16$.

Moreover, this interpretation gives a straightforward explanation of the finite-size effects
one encounters in the energy of an inherent structure (table~\ref{tab:IS-m}). For example,
if we compare $L=8,16$ at $m=8$, we notice two incompatible energies. 
In fact, there is an intrinsic difference between the two
sizes, since $L=8$ represents single-basin systems, while $L=16$ has a variety of inherent
structures.
On the other side, finite-size effects on lower $m$ are smaller, because we are comparing
similar types of behavior.

Notice that, although the ratio  $Q^2/Q^2_\mathrm{self}(m)$ grows 
monotonously, this is not true for the pure overlap $Q^2(m)$ (figure \ref{fig:q-m}, 
bottom), that has a peak at an intermediate $m$. Moreover, the position of the peak
doubles when we double the lattice linear size, justifying the operational definitions
$(m/L)_\mathrm{SG}$ and $(m/L)_\mathrm{1}$. 
\begin{figure}[!tb]\centering
 \includegraphics[width=0.6\columnwidth]{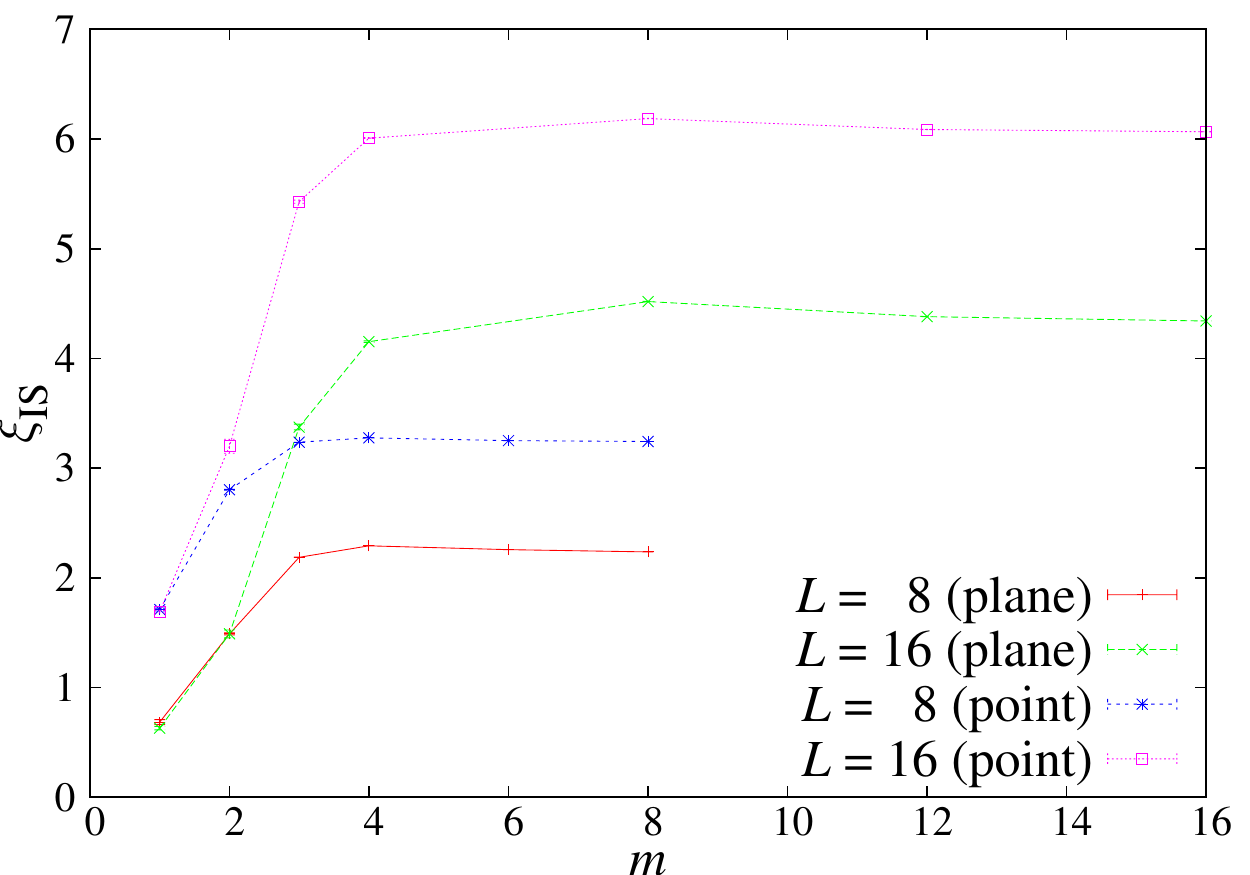}
\caption[Dependency of $\xi$ on $m$]{Dependency of the second-moment correlation length $\xi_2$ \index{correlation!length}
on the number of components of the spins $m$. We show both the plane and the point
correlation functions defined in equations \eqref{eq:Cpoint} and \eqref{eq:Cplane}, for $L=8,16$.}
\label{fig:xi-m}
\end{figure}
The same peak at intermediate $m$ is also visible in the energy and in the correlation
length (figure \ref{fig:xi-m}), indicating that there is an intrinsic difference
in the nature of the reached \acp{IS}. 
In table~\ref{tab:IS-m} we give the values
of the aforementioned observables at the \ac{IS}.
\begin{table}[b]
\centering
%\resizebox{\columnwidth}{!}{
 \begin{tabular}{ccccccc}
  \normalsize{$L$} & \normalsize{$m$} & \normalsize{$e_\mathrm{IS}$} & \normalsize{$Q^2_\mathrm{IS}$} & \normalsize{$Q^2_\mathrm{self,IS}$}  & \normalsize{$\xi^\mathrm{(plane)}_\mathrm{IS}$}& \normalsize{$\xi^\mathrm{(point)}_\mathrm{IS}$} \\\hline\hline
  8   & 1  & -0.4709(1)   & 0.0095(1)  &  1           &  0.68(2)   &  1.71(1) \\
  8   & 2  & -0.5953(1)   & 0.0497(3)  &  0.50297(2)  &  1.49(1)   &  2.802(4) \\
  8   & 3  & -0.6151(1)   & 0.1784(6)  &  0.33994(4)  &  2.188(2)  &  3.2358(7) \\
  8   & 4  & -0.6176(2)   & 0.2213(5)  &  0.26229(9)  &  2.2919(9) &  3.2760(5) \\
  8   & 6  & -0.61801(11) & 0.1989(1)  &  0.1997(1)   &  2.2567(3) &  3.2514(2) \\
  8   & 8  & -0.61797(12) & 0.1905(1)  &  0.1905(1)   &  2.2364(3) &  3.2428(2) \\\hline
 16   & 1  & -0.4721(1)   & 0.00123(6) &  1           &  0.63(2)   &  1.69(1) \\
 16   & 2  & -0.5965(1)   & 0.0067(2)  &  0.500379(8) &  1.49(4)   &  3.20(6) \\
 16   & 3  & -0.6165(1)   & 0.0382(5)  &  0.33416(1)  &  3.37(3)   &  5.43(1) \\
 16   & 4  & -0.6191(2)   & 0.0833(6)  &  0.25144(2)  &  4.153(7)  &  6.008(4)	 \\
 16   & 8  & -0.6200(1)   & 0.1218(3)  &  0.13126(5)  &  4.519(2)  &  6.187(1) \\
 16   & 12 & -0.6202(1)   & 0.10031(9) &  0.10044(9)  &  4.3814(8) &  6.087(1) \\
 16   & 16 & -0.6197(1)   & 0.0959(1)  &  0.0959(1)   &  4.3412(8) &  6.066(1)\\\hline
 64   & 3  & -0.61657(4)  & 0.00064(2) &  0.3333466(4)&  3.53(7)   &  6.74(6) \\
\end{tabular}
%}
\caption[Properties of the \acsp{IS}]{\index{correlation!length}\index{energy}
Properties of the \acp{IS}. For each choice of the parameters
we show the observables at the end of the quench: The energy \nomenclature[e....IS]{$e_\mathrm{IS}$}{energy of the IS}
$e_\mathrm{IS}$, the 
overlap $Q^2_\mathrm{IS}$, the selfoverlap $Q^2_\mathrm{self,IS}$, the point-correlation 
length $\xi^\mathrm{point}_\mathrm{IS}$ and the plane correlation length 
$\xi^\mathrm{plane}_\mathrm{IS}$.
}
\label{tab:IS-m}
\end{table}
We see in this behavior the competition between two effects.
When $m$ is small, the quench has a vast choice of valleys where to fall. 
Since, reasonably the attraction basin of the lower-energy \acp{IS} is larger,\index{attraction basin} 
the wide variety of \acp{IS} will increase the probability of
falling in a minimum with low energy and larger correlations.
When $m$ increases, the number of available valleys decreases, so it is
more likely that two different replicas fall in the same one. Yet, the \emph{quality}
of the reached \acp{IS} decreases, since the quench does not have the
possibility to choose the lowest-energy minimum.

\section{Overlap Probability Densities}
\begin{figure}[!b]\centering
 \includegraphics[width=0.8\columnwidth]{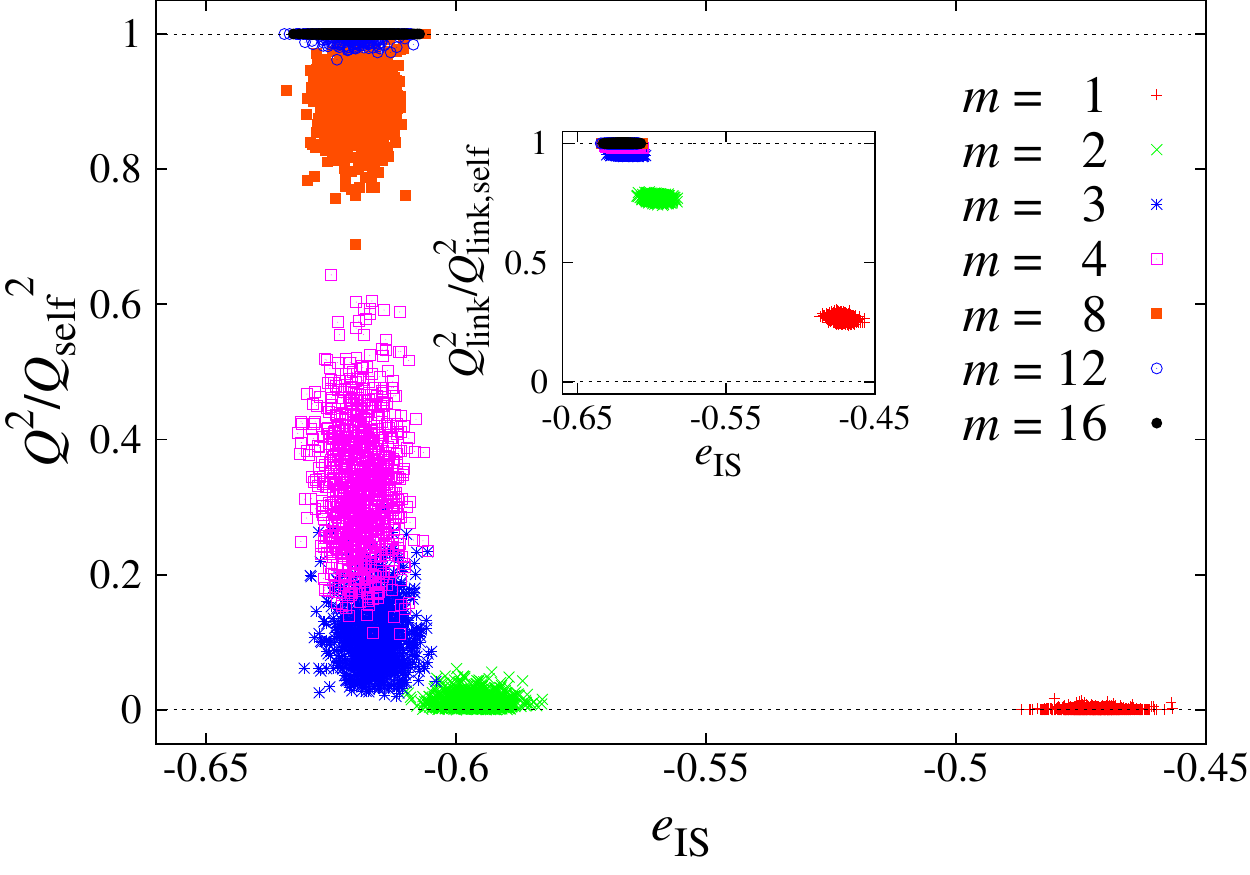}
 \caption[Scatter plots of $Q^2_\mathrm{IS}/Q^2_\mathrm{self,IS}$ vs $e_\mathrm{IS}$]
 {Scatter \index{energy}
 plots for $L=16$, at different values of $m$, of
 the overlap ratio  $Q^2_\mathrm{IS}/Q^2_\mathrm{self,IS}$ against
 mean energy between the two replicas $e_\mathrm{IS} = (e_\mathrm{IS}^{(a)}+e_\mathrm{IS}^{(b)})/2$.
 Each simulated sample contributes to the plot with a single point.
  The \textbf{inset} displays an analog plot for the link overlap.}
\label{fig:Eq}
\end{figure}

\index{overlap!distribution}
From these observations it is reasonable to think that overlap and energy of the \acp{IS}
are correlated. We looked for these correlations both on the overlap, on the selfoverlap,
and in their ratio, but with a negative result. In figure \ref{fig:Eq} we show a scatter-plot
of the ratio of the inherent structure's overlaps $Q^2_\mathrm{IS}/Q^2_\mathrm{self,IS}$
that confirms our statements. An equivalent plot for the link overlap is displayed in the inset.

\begin{figure}[!htb]\centering
 \includegraphics[width=0.95\columnwidth]{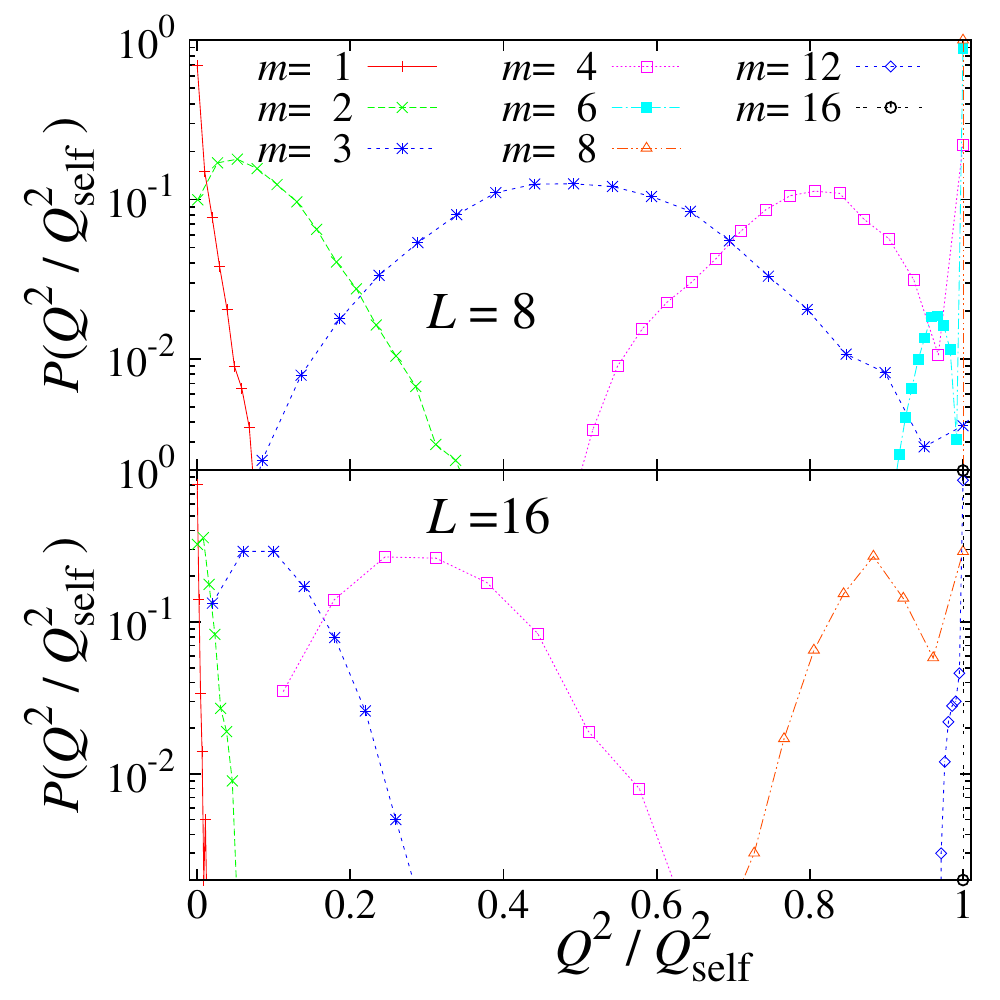}
 \caption[Overlap \acsp{pdf} of 
  the \acsp{IS} for different values of $m$.]{Overlap \acp{pdf} of 
  the \acp{IS} for different values of $m$. The \textbf{top} figure depicts
  data for $L=8$, on the \textbf{bottom} we have $L=16$. 
  The curves are normalized to plot all the curves together. The actual probability distribution
  function is obtained by dividing each point by the bin width $\Delta Q/N_\mathrm{bins}$, where 
  $\Delta Q$ is the difference between maximum and minimum $Q^2$.}
 \label{fig:Pq-hsgm}
\end{figure}

\begin{figure}[!htb]\centering
 \includegraphics[width=0.485\columnwidth]{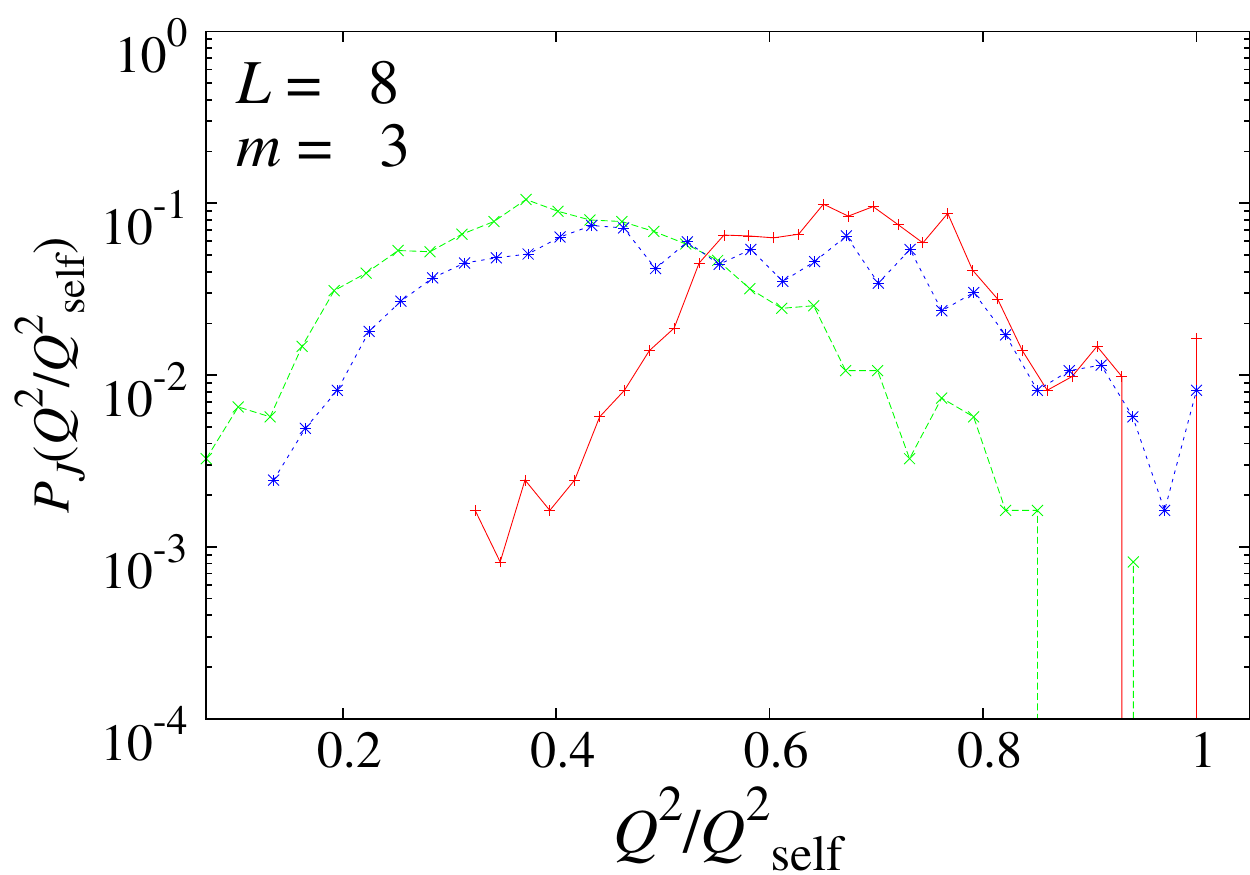}
 \includegraphics[width=0.485\columnwidth]{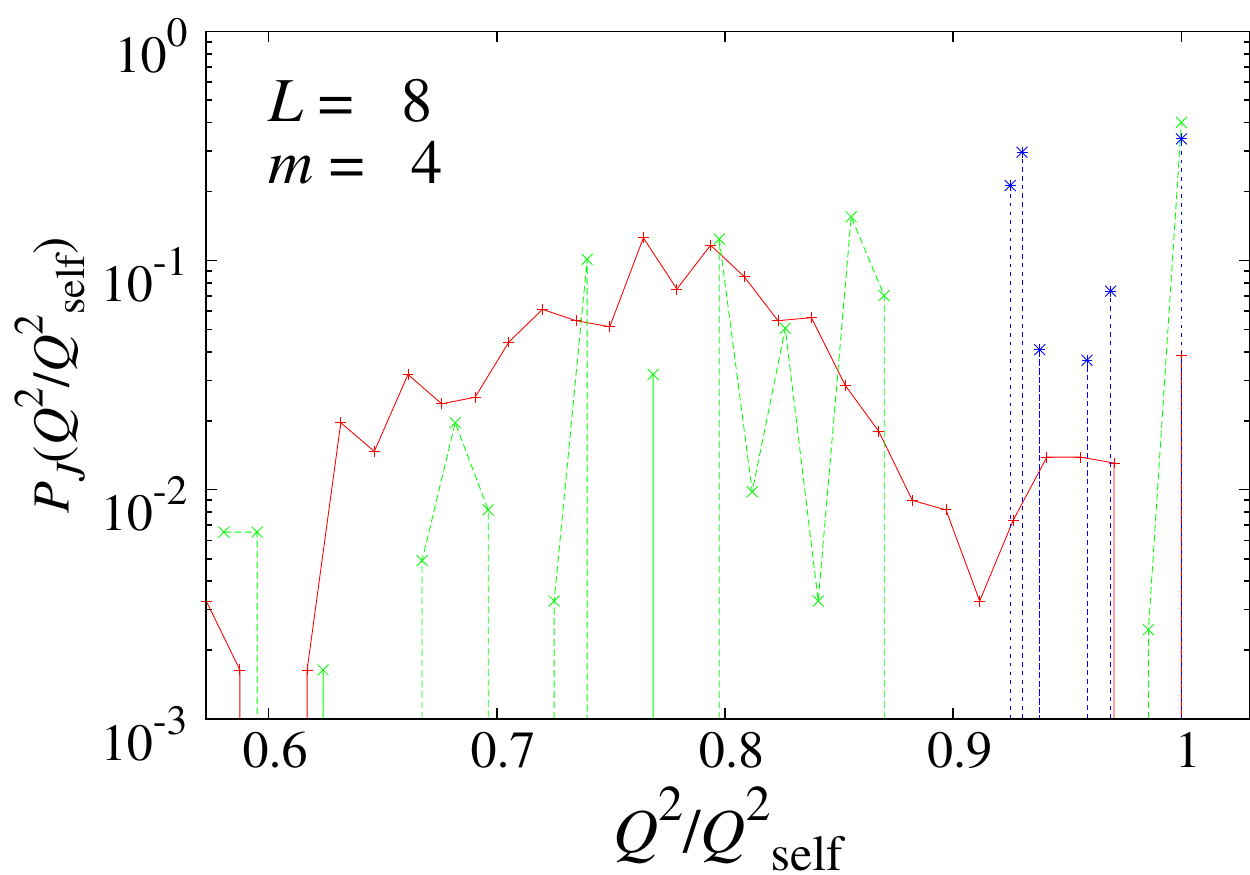}
 \includegraphics[width=0.485\columnwidth]{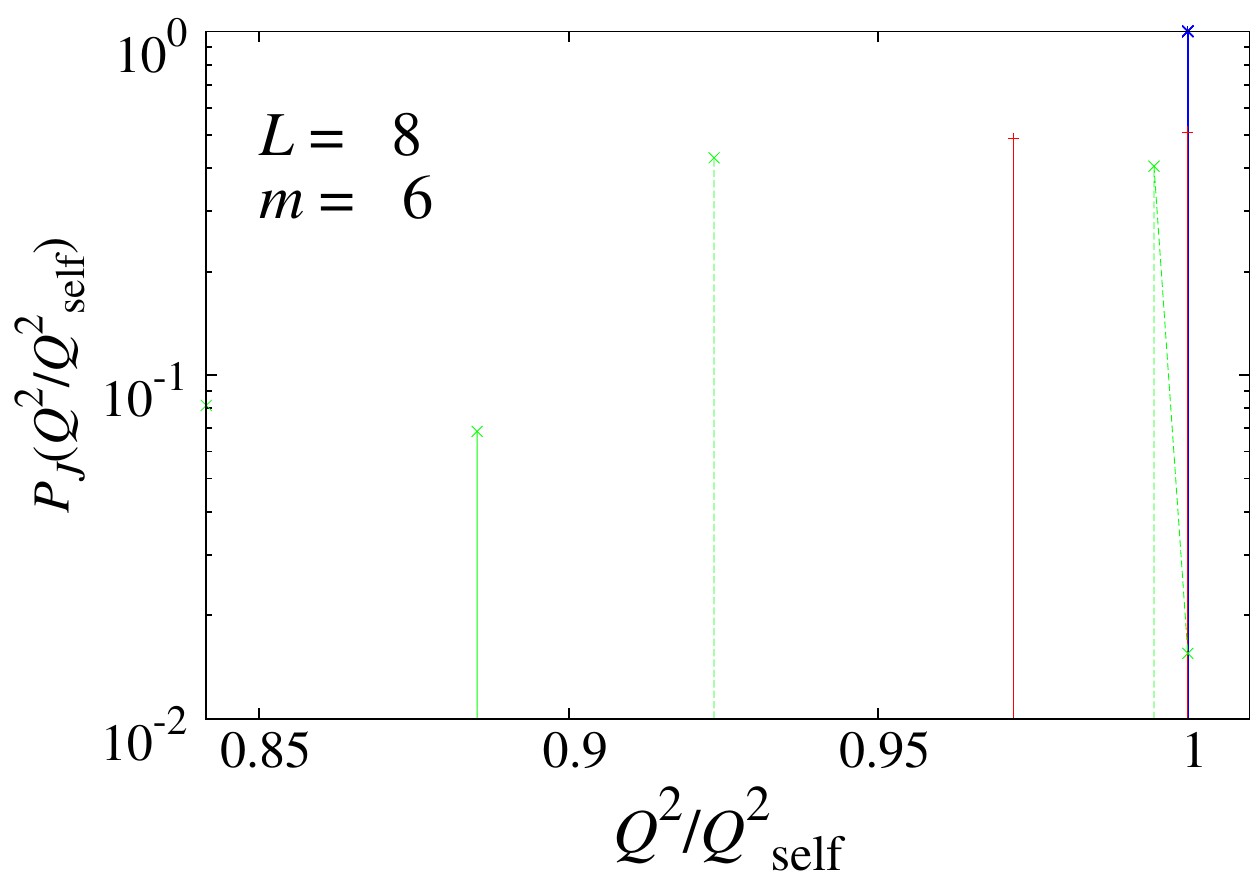}
 \includegraphics[width=0.485\columnwidth]{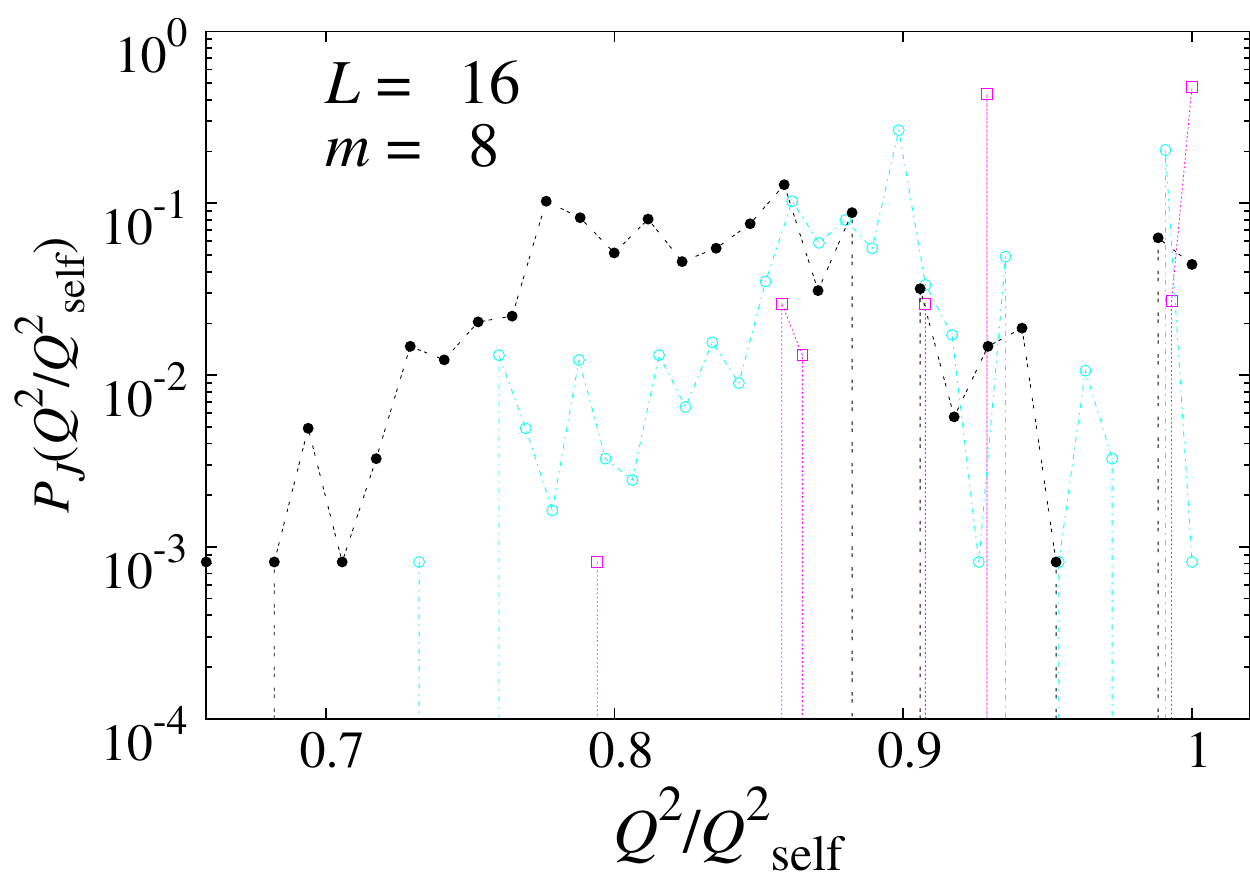}
 \caption[Sample-dependent overlap \acsp{pdf} $P_{J}(Q^2_\mathrm{IS}/Q^2_\mathrm{self,IS})$.]{
 Sample-dependent \index{overlap!distribution!sample-dependent}
 overlap \acp{pdf} $P_{J}(Q^2_\mathrm{IS}/Q^2_\mathrm{self,IS})$. 
 Each curve depicts data from a separate sample. 
 In each plot we show a selection of three samples with different shapes of the distribution. 
 The choices of the parameters are represented in the key of each plot. 
 We used two different color codes to distinguish the three plots that come from $L=8$ systems 
 (\textbf{top-left} and \textbf{right}, and \textbf{bottom left}), from the \textbf{bottom-right} plot
 that is for $L=16$. The curves are normalized as in figure \ref{fig:Pq-hsgm}.}
 \label{fig:PJq-hsgm}
\end{figure}

The cross sections of figure \ref{fig:Eq} give an idea of the energy and overlap probability
distribution functions. We show explicitly the overlap probability distribution
functions (normalized with the bin width) of the \acp{IS} in figure \ref{fig:Pq-hsgm}.
They are qualitatively different from their thermal counterparts (see, e.g., 
\cite{janus:10}). The ratio $Q^2_\mathrm{IS}/Q^2_\mathrm{self,IS}$ is bounded between 
zero and one. The distributions are extremely wide, and the phenomenology is 
quite different near the two bounds. In fact, when $m$ is large enough,
the limit $Q^2_\mathrm{IS}/Q^2_\mathrm{self,IS}=1$ changes completely the shape of the curves, 
introducing a second peak (that we could read as an echo of the Bose-Einstein condensation
remarked in \cite{aspelmeier:04}).
Around the lower bound of the $P(Q^2_\mathrm{IS}/Q^2_\mathrm{self,IS})$, instead, there is no double
peak.
We can try to give an interpretation to the presence of this second peak by looking at the overlap
distribution functions $P_J(Q^2_\mathrm{IS}/Q^2_\mathrm{self,IS})$ for a given instance of the
couplings. In figure \ref{fig:PJq-hsgm} we show that this distribution has relevant sample-to-sample fluctuations.
When we increase $m$, the number of minima of the energy, $N_\mathrm{IS}$, gradually becomes smaller. Yet,
depending on the specific choice of the couplings, $N_\mathrm{IS}$ can vary sensibly. For example in figure \ref{fig:PJq-hsgm}, top-right,
one can see that when $L=8$ and $m=4$, $N_\mathrm{IS}$ can be both large (red curve) or
of order one (blue curve). For $L=8$, $m=6$ (figure \ref{fig:PJq-hsgm}, bottom-left), the situation is similar: for the blue curve $N_\mathrm{IS}=1$,
while for others $N_\mathrm{IS}>1$.

As we similarly stated in section \ref{sec:hsgm-IS-m}, we notice that the lattice size plays
a substantial role on the properties of the reached inherent structure, since when we pass from
$L=8$ to $L=16$ histograms regarding the same $m$ cover very different ranges of $q$. We can both
see them traditionally as strong finite-size effects, or focus on $L$ as a relevant parameter (as it was
suggested, for example, in \cite{baityjesi:14}), concentrating the interest on finite $L$.

\subsection{Link Overlaps}
\begin{figure}[!t]\centering
 \includegraphics[width=0.95\columnwidth]{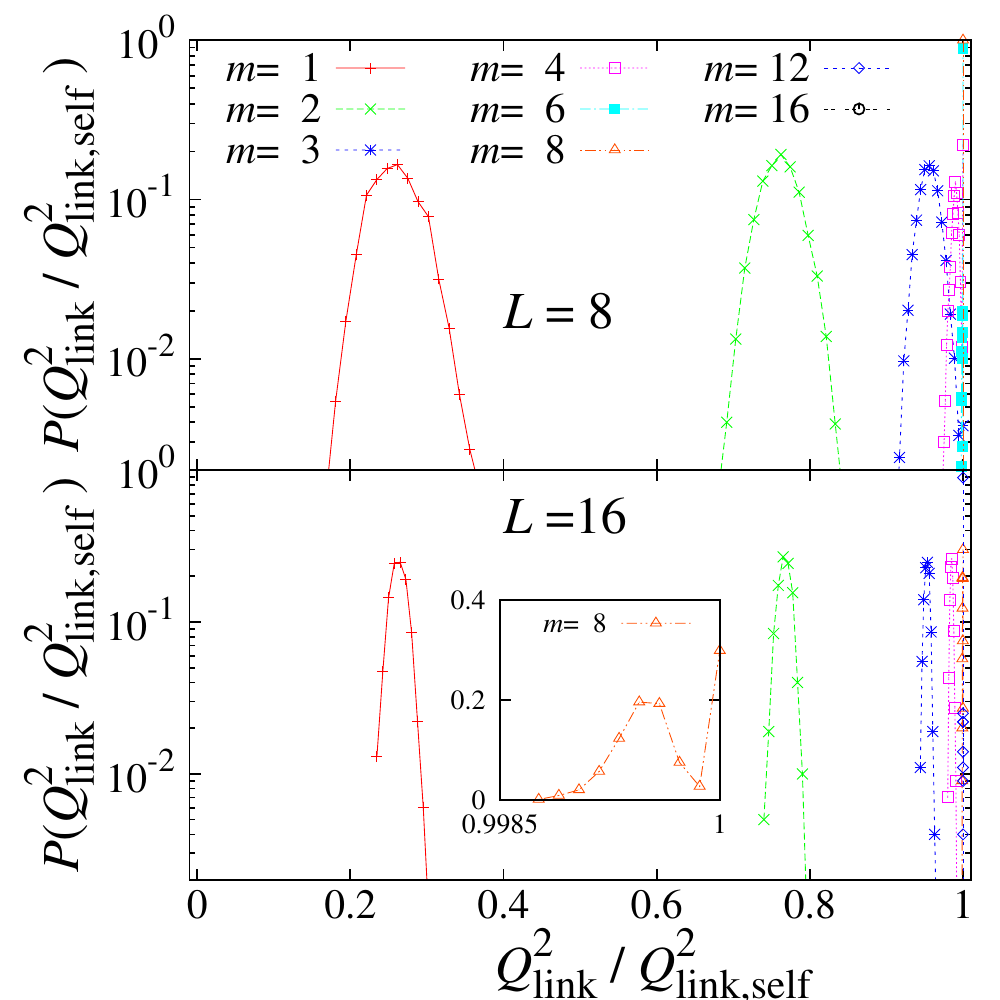}
 \caption[Link-overlap \acsp{pdf} of the \acsp{IS} for different values of $m$.]
 {\index{overlap!link!distribution!sample-dependent}
 Same as figure \ref{fig:Pq-hsgm}, but for the link overlap. 
  The \textbf{inset} shows a zoom for the $m=8$, $L=16$ data,
where we also removed the logarithmic scale on the $y$ axis.
 }
 \label{fig:Pq-hsgmlink}
\end{figure}

\index{overlap!link}
Since in the past ten years an increasing attention has been devoted to the link overlap $Q^2_\mathrm{link}$
as an alternative order parameter for the study of the low temperature region of spin
glasses \cite{krzakala:00,contucci:06,janus:10}, in figure \ref{fig:Pq-hsgmlink} we show also the link-overlap 
histograms $P(Q^2_\mathrm{link,IS}$) at the \ac{IS}.
The functions $P(Q^2_\mathrm{link,IS})$ have much smaller finite-size effects than the $P(Q^2_\mathrm{IS})$,
and are more Gaussian-like (although the Gaussian limit is impossible, since $Q^2_\mathrm{link}$ is bounded
between 0 and 1). The inset shows that the second peak on high overlaps is present also
with the link overlap.

\begin{figure}[!htb]
\vspace{2.5cm}
\centering
 \includegraphics[width=0.48\columnwidth]{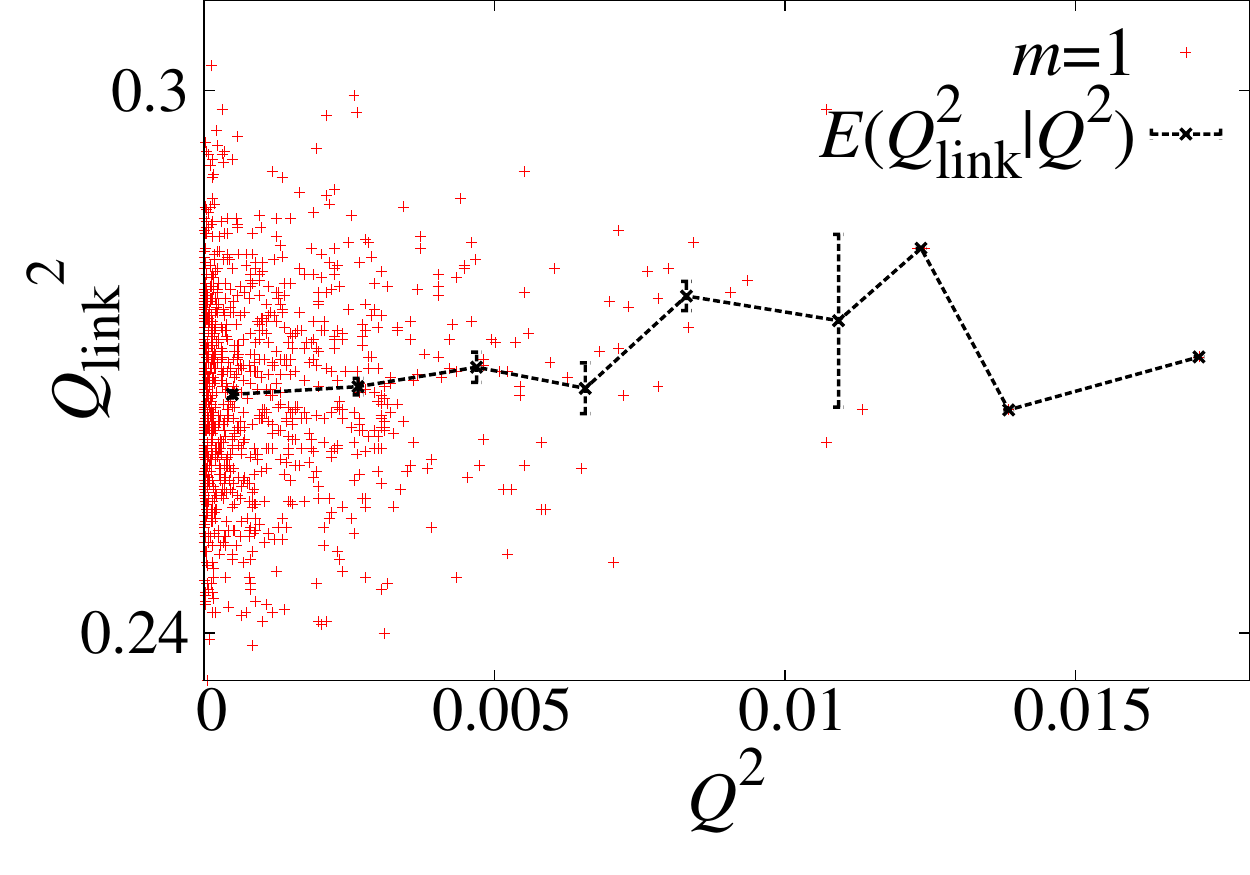}
 \includegraphics[width=0.48\columnwidth]{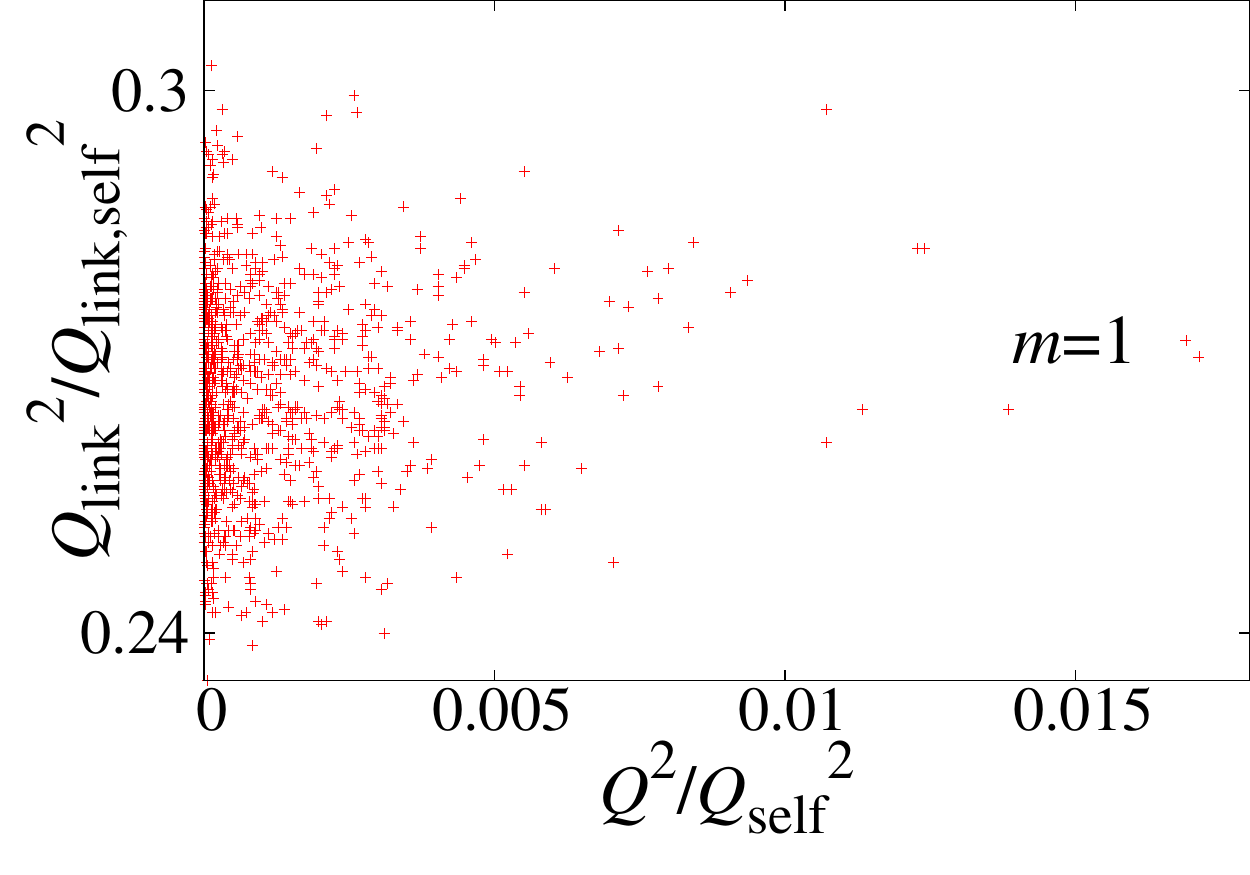}
 \includegraphics[width=0.48\columnwidth]{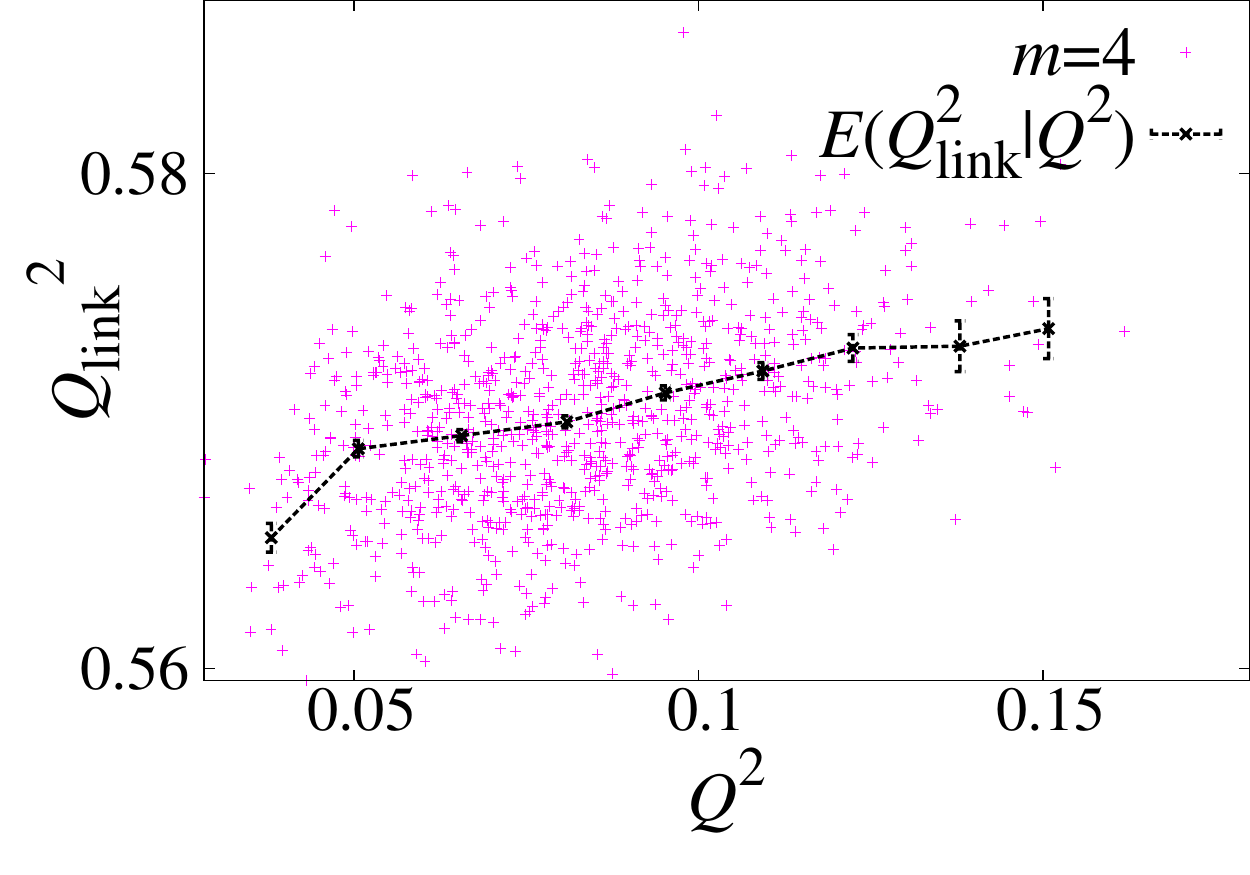}
 \includegraphics[width=0.48\columnwidth]{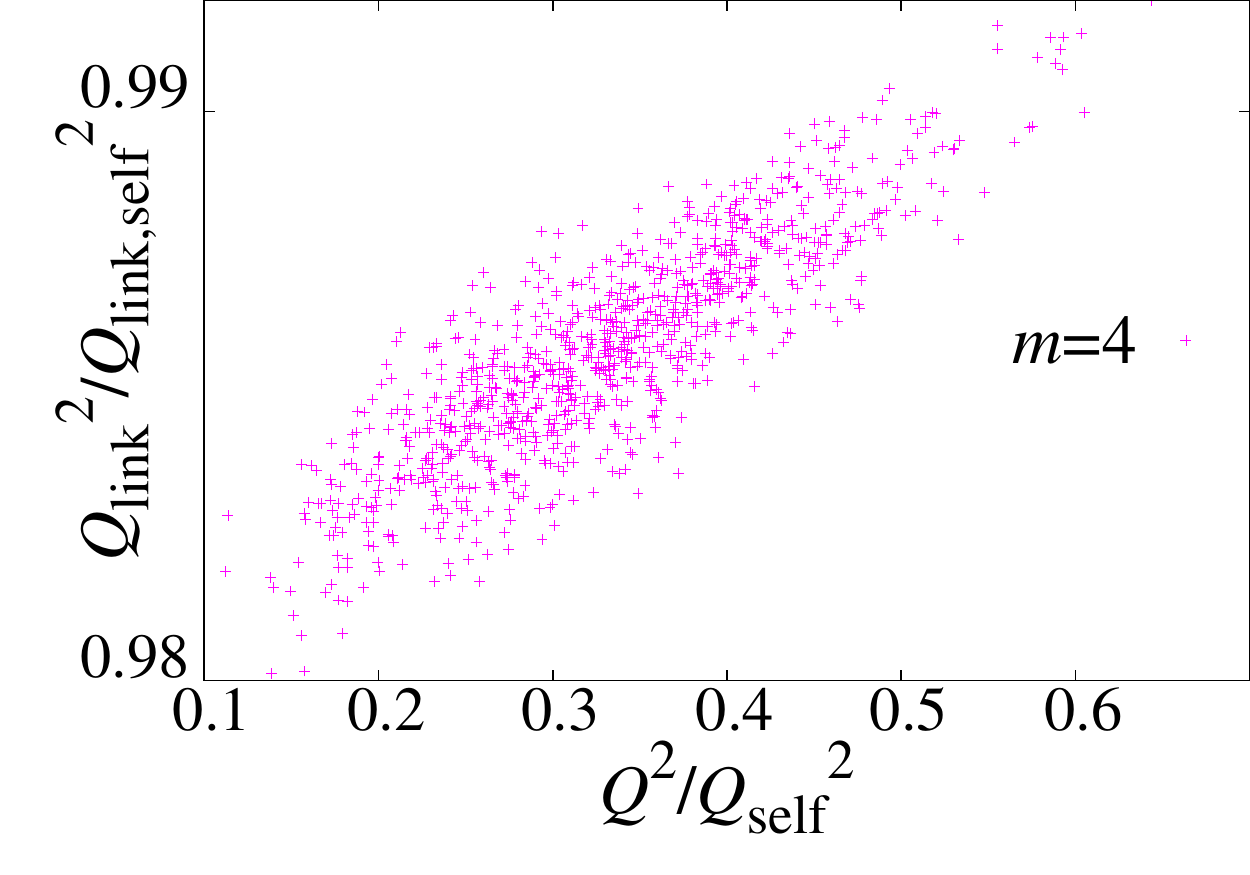}
 \includegraphics[width=0.48\columnwidth]{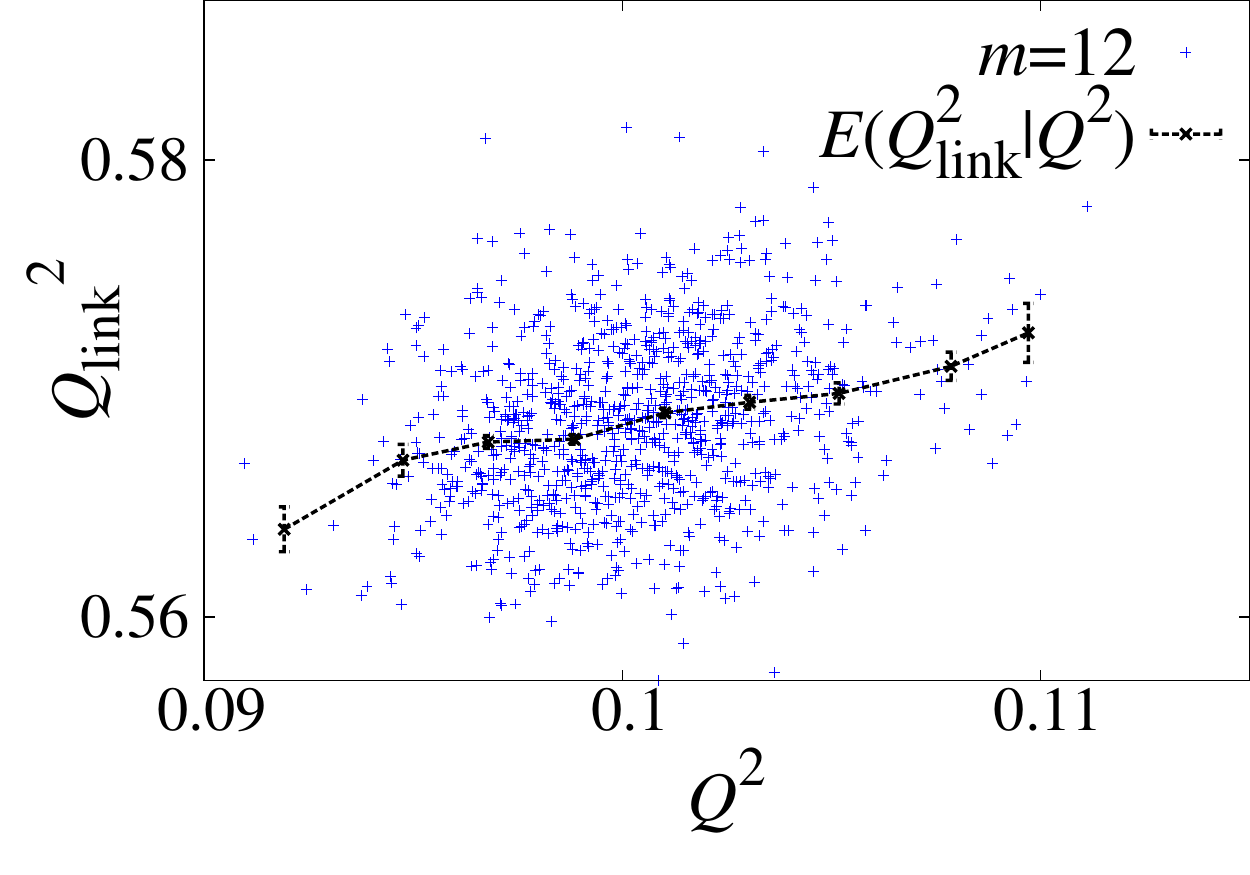}
 \includegraphics[width=0.48\columnwidth]{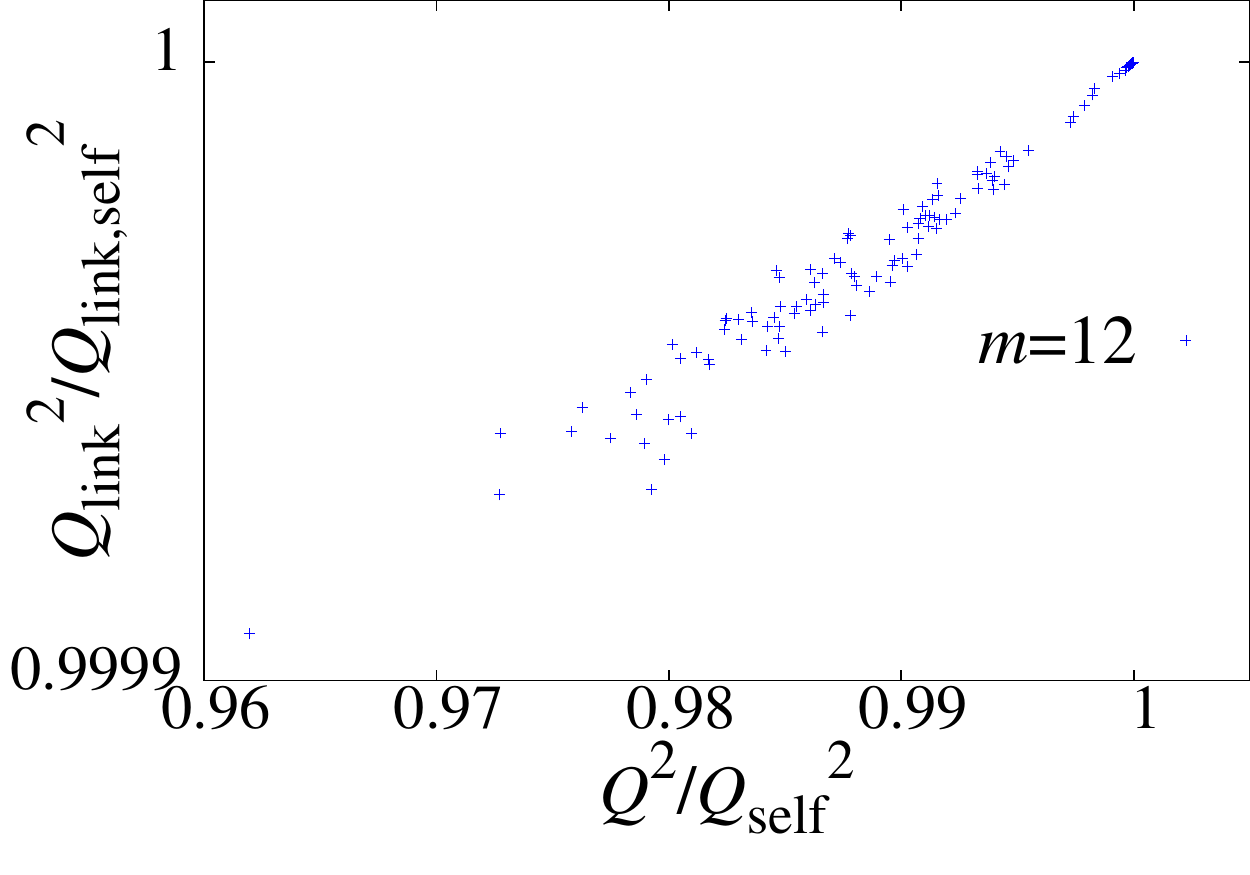}
 \caption[Correlation between the spin and the link overlap of the \acsp{IS}]
 {Correlation between the spin and the link overlap of the \acp{IS}, \index{overlap!link}
 for $L=16$ lattices, with $m=1$ (\textbf{top}), $m=4$ (\textbf{center}) and 
 $m=12$ (\textbf{bottom}). On the \textbf{left} we plot the overlaps, while on the \textbf{right} they
 are normalized with the self overlap. Normalizing with the self overlap
 increases the correlations between the two order parameters. The two top figures are the same
 because the self overlap is one when $m=1$.
 The black lines on the left plots represent $E(Q_{link}^2|Q^2)$, and they show that a correlation 
 exists also without normalization.}
 \label{fig:scatterplot_qqlink}
\end{figure}

We checked also the correlation between spin and link overlaps. At finite temperature there are
different predictions between RSB and droplet pictures. According to the \ac{RSB} picture the conditional
expectation value $E(Q^2_\mathrm{link}|Q^2)$ should to be a linear, strictly increasing function of
$Q^2$, while this should not be true in the Droplet theory (section \ref{sec:spin-glass-intro}).
When $m$ is small, this correlation is practically invisible, but it becomes extremely strong when
we increase the number of components of the spins (figure \ref{fig:scatterplot_qqlink}). 
Notice how the correlation between spin and link overlap is formidably increased when we normalize
the two with the selfoverlap.
The curves in figure \ref{fig:scatterplot_qqlink} represent $E(Q^2_\mathrm{link}|Q^2)$.
If we exclude the tails, that are dominated by rare non-Gaussian events, the trend is 
compatible with linearly increasing functions.

\clearpage

\section{Quench Dynamics}
\label{sec:hsgm-dynamics}
\begin{figure}[!b]\centering
 \includegraphics[width=0.8\columnwidth]{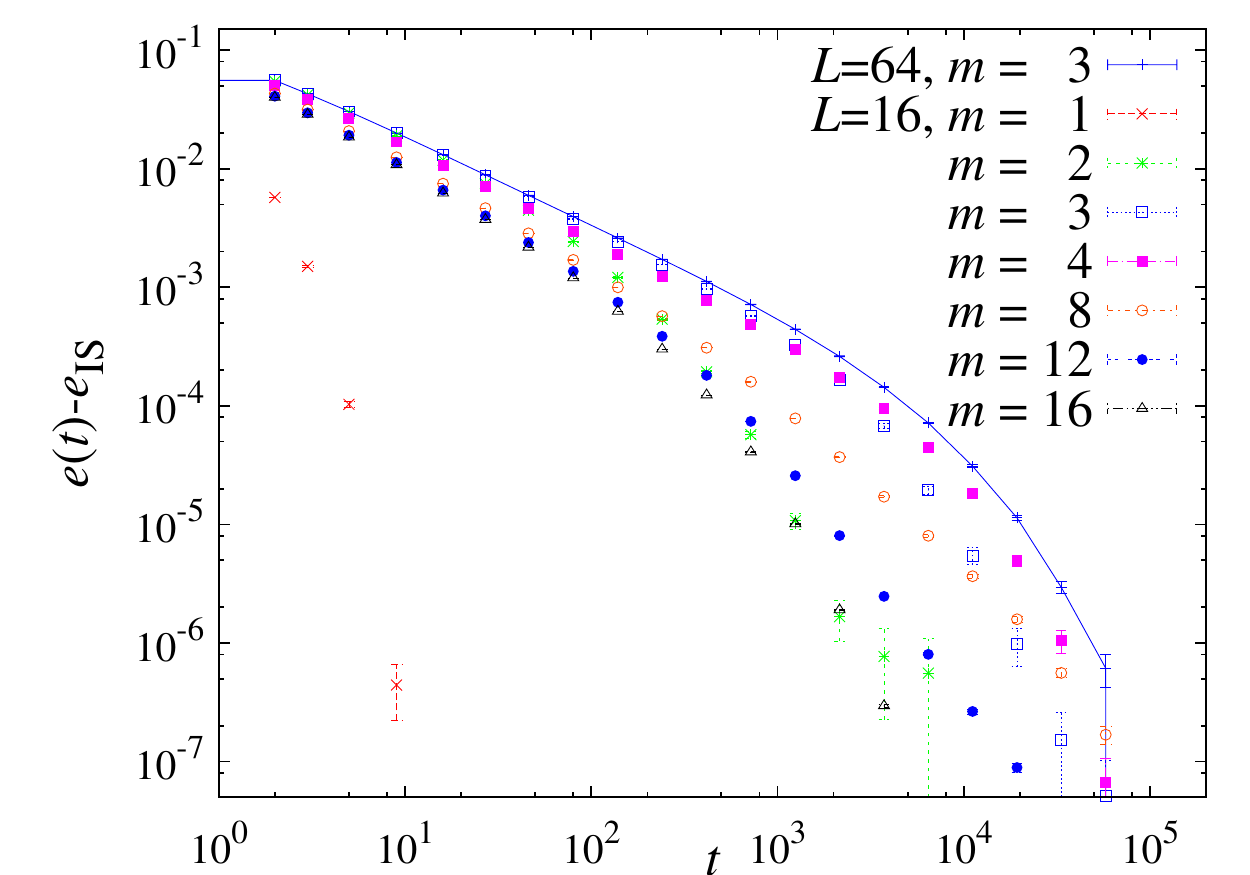}
  \caption[Evolution of the energy during a quench]{Evolution of the energy during the quench for all the simulated values of $m$,
  in $L=16$ lattices. On the $x$ axis there is the time, measured in full lattice quench sweeps. On the
  $y$ axis there is the difference between the energy at time $t$, \nomenclature[t....6]{$t$}{In chapter \ref{chap:hsgm}: time in the quench}
  $e(t) = \left(e^{(a)}(t)+e^{(b)}(t)\right)/2$, and its final value $e_\mathrm{IS} = e(t=10^5)$.
  The convergence speed is very different between continuous and discrete spins.
  To stress the finite-size effects we also show points for $L=64$, $m=3$ (points connected by segments).
  \index{energy}}
 \label{fig:evol-E}
\end{figure}

Let us get an insight on the dynamics of the quench.
For short times, the energy converges towards a minimum with a roughly power law behavior (figure \ref{fig:evol-E}). 
At longer times there is a cutoff, that grows with the system's size,
revealing a change in the dynamics after which the system converges faster to a valley.
We stress the great difference in the 
convergence rate between $m=1$ and $m>1$. We can identify two different decrease rates, depending on
whether the spins are discrete or continuous.

\begin{figure}[!htb]
\centering
\includegraphics[width=0.85\columnwidth]{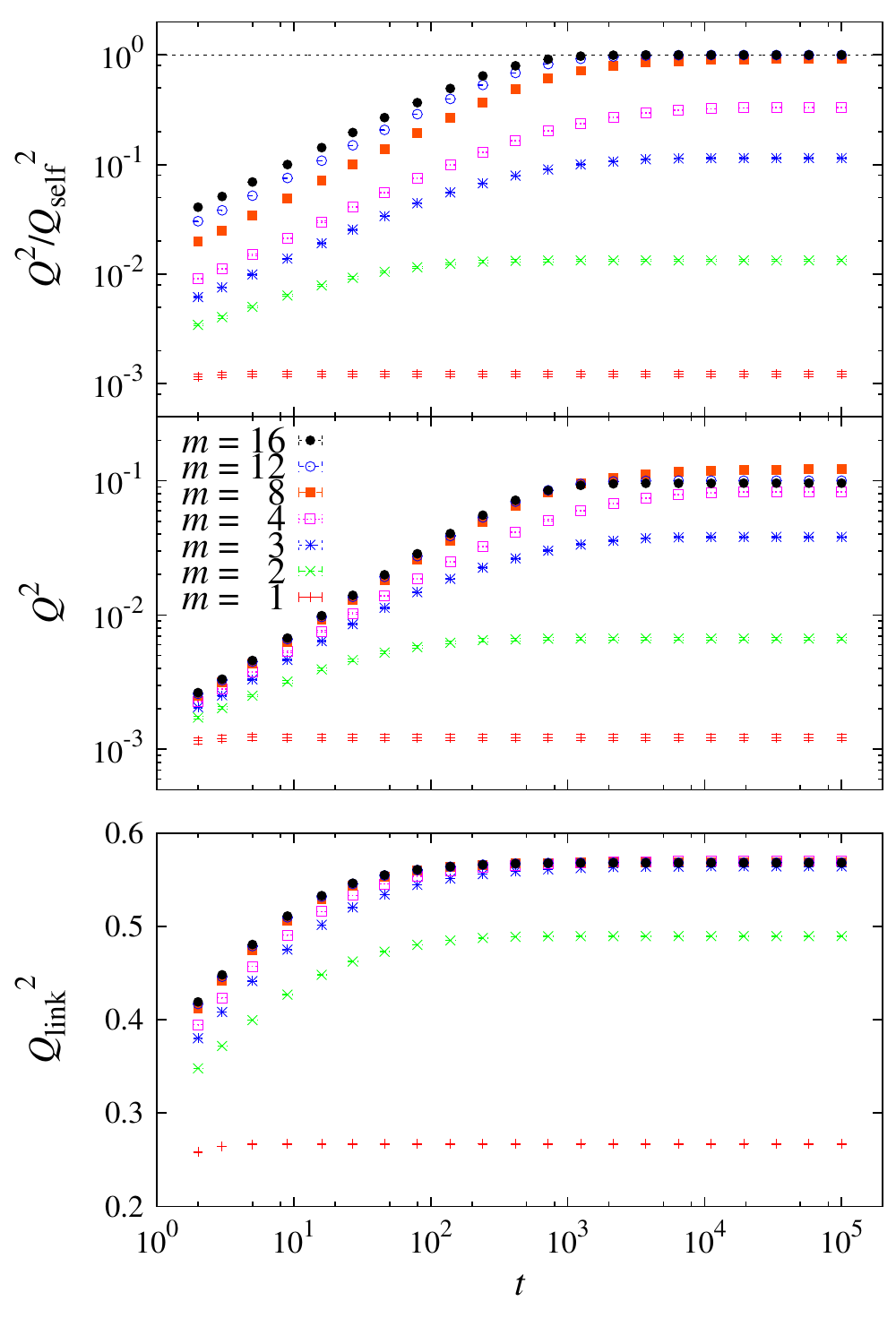}
 \caption[Time evolution of the overlaps during a quench]{\index{overlap!link}
  Time evolution of the overlaps in $L=16$ lattices. 
  In the \textbf{top} set we show the overlap $Q^2$ normalized with the selfoverlap $Q^2_\mathrm{self}$.
  On the \textbf{center} we show $Q^2$ without normalizing. Notice that differently from the top case,
  in the center plot it is the curve representing $m=8$ that reaches the highest values.
  The \textbf{bottom} plot shows that the behavior is analogous with $Q^2_\mathrm{link}$.
  }
 \label{fig:evol-q}
\end{figure}

Figure \ref{fig:evol-q} shows the evolution of the overlap for $L=16$, and gives a better understanding of why quantities
such as $Q^2_\mathrm{IS}$ are not monotonous with $m$. We show both the evolution of $Q^2/Q^2_\mathrm{self}$ (top),
and of $Q^2$ (bottom). The first one behaves as one would expect when the number of minima is decreasing to one. 
On the other side, we see from the lower plot
\emph{how} the quenches of $m=8$ reach the highest overlap. 
A possible interpretation is to ideally separate the quench in two regions. At the beginning 
there is a search of the valley with a power-law growth of $Q^2$, and later the convergence inside of the valley. 
Figure \ref{fig:evol-q} shows that the search of the valley stops earlier when $m=12,16$, i.e. when their number is
of order one.

\begin{figure}
\centering
 \includegraphics[width=0.8\columnwidth]{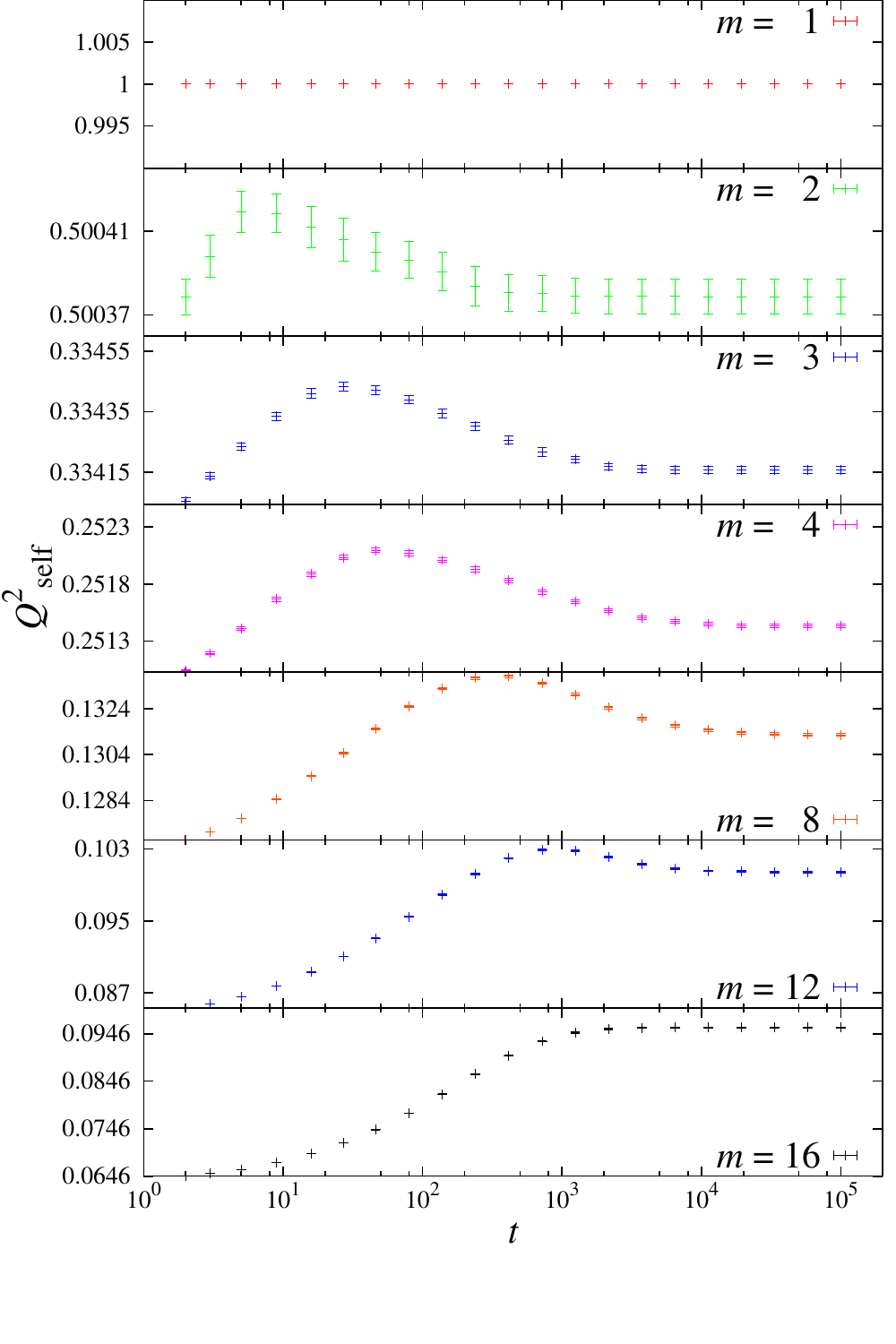}
  \caption[Time evolution of the selfoverlaps during a quench]
  {\index{overlap!self}
  Evolution of the selfoverlap $Q^2_\mathrm{self}(t)$ for lattices of size $L=16$, for different values of $m$.
  Note the differences in the $y$-scales: For small $m$ the variation of $Q^2_\mathrm{self}(t)$ is very small, while
  for the largest ones it is of the order of the self-overlap.}
   \label{fig:evol-qself}
\end{figure}

We remark on a nonlinear trend on the evolution of the selfoverlap $Q^2_\mathrm{self}(t)$. For continuous spins 
($m>1$) it has a different value at infinite and zero temperature (figure \ref{fig:evol-qself}). 
This variation is strikingly visible when $m$ is large,
but the same trends are found for $m\leq3$, though the variations are so small that it is justified that they are usually not found.
\footnote{To our knowledge, the only reference where a non-trivial behavior of the self-overlap was found is in \cite{baityjesi:11}.
Yet, in this case it was in the study of \acp{IS} from finite temperature, and in the chiral sector (they worked with $m=3$).}
Moreover $Q^2_\mathrm{self}(t)$ is highly nonlinear, and, except for the highest $m$, it overshoots before having converged.

\begin{figure}[!t]\centering
 \includegraphics[width=0.8\columnwidth]{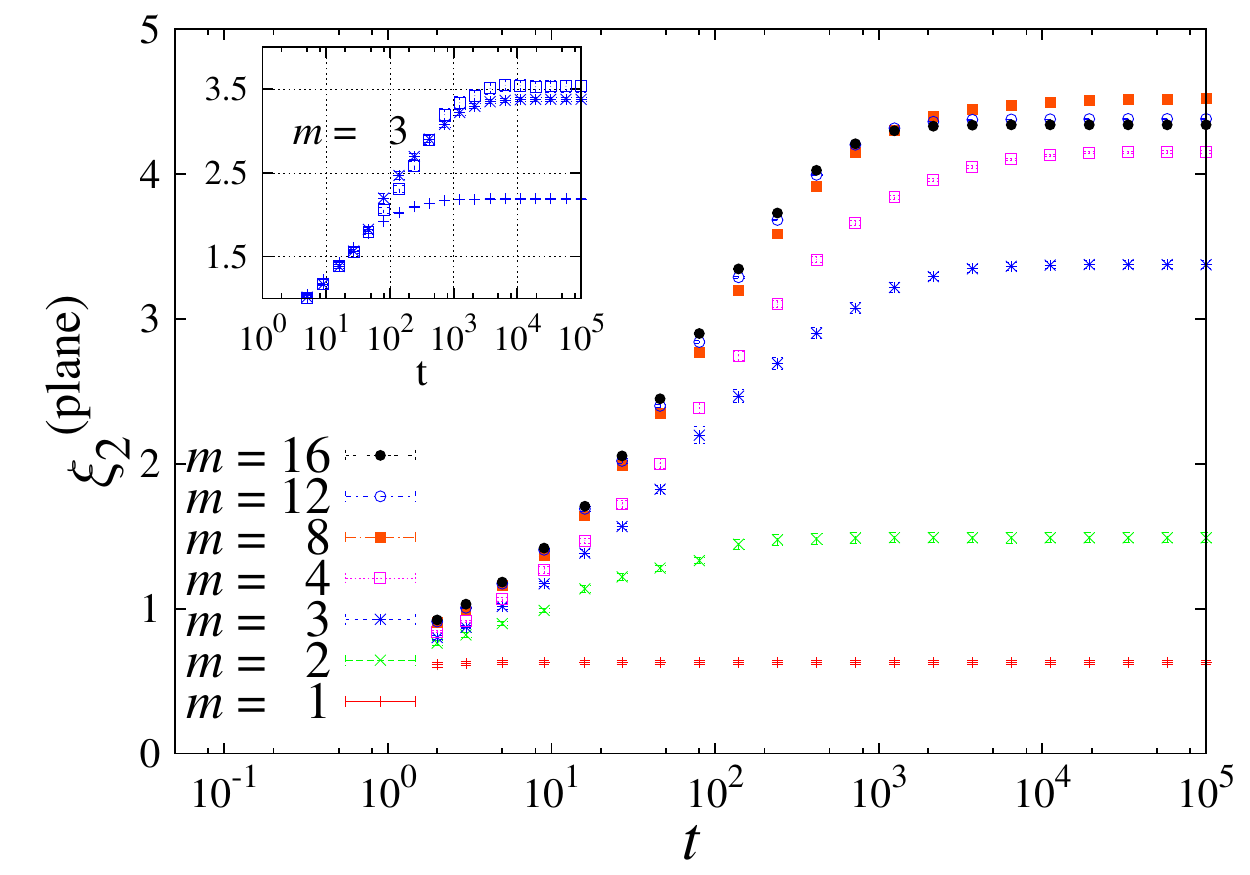}
\caption[Time evolution of the correlation length during a quench]
	{Time evolution of the plane second-moment correlation length $\xi_2^\mathrm{plane}$.\index{correlation!length}
	In the \textbf{large} figure we show every simulated $m$ for size $L$. Notice that the highest correlation
	length is reached by $m=8$. The \textbf{inset} depicts the sole case of three-dimensional spins ($m=3$) for
	sizes $L=8,16,64$.}
\label{fig:evol-xi}
\end{figure}

\index{correlation!length}
In figure \ref{fig:evol-xi} we show the evolution of the correlation lengths $\xi_2^\mathrm{plane}$ during the quenches for $L=16$
for all our values of $m$. 
We see the same variety of behaviors shown by $Q^2$ (figure \ref{fig:evol-q}), with $\xi_2^\mathrm{plane}(m=12,16)$
that abruptly stop increasing, while when $m=8$ the increase is similar but lasts longer and the change of growth is smoother.
\footnote{The point correlation length $\xi_2^\mathrm{point}$ behaves analogously.}

We can contrast our results with the ones obtained by Berthier and Young in \cite{berthier:04}
for $m=3$ Heisenberg spin glasses. In that case they measured the evolution of the coherence
length in quenches down to positive temperature $T_0>0$ ($L=60$).
They remarked two different regimes of growth of the coherence length, and attributed them to the 
passage from critical to activated dynamics.\footnote{Note that the definition of the coherence length in \cite{berthier:04} is
different from ours.}
In that case the slope of the second phase kept being positive
and $\xi$ did not appear to converge after $10^5$ lattice sweeps.
We can make a direct comparison with our quenches to zero-temperature $T_0=0$ with $L=64$ (figure \ref{fig:evol-q}, inset).
We obtain a flat second regime after $10^4$ sweeps, so we can indeed attribute the growth in the 
second regime to thermal effects.
In the inset we compare the coherence length of different lattice sizes to remark 
that although 
$\xi_2^\mathrm{plane}<4$, we are clearly far from the thermodynamic limit even for $L=16$.

\afterpage{\clearpage}

\section{Overview}
\label{sec:hsgm-conc}

We performed an extensive study of the energy landscape of three-dimensional vector spin glasses, focusing
on their dependence on the number of components $m$ of the spins. We were concerned both
with the zero-$T$ dynamics and with the properties of the \acp{IS}, remarking
various types of finite-size effects.

Increasing $m$ the number of minima in the energy landscape decreases monotonously,
down to the limit of a single state. The number of components $m_\mathrm{SG}(L)$ after which the 
number of minima becomes subexponential grows with the lattice size. Reversing the relation, 
we can operatively define $L_\mathrm{SG}(m_\mathrm{SG})$ as the smallest lattice size needed
in order to observe a complex behavior for a given $m$.

For small $m$ correlations are small and dynamics are trivial, while when $m$ becomes
larger correlations increase and the convergence to an inherent structure slows down (for a small enough $m/L$ ratio).
We remark on the competition between the $m=1$ limit, with abundance of \acp{IS},
and the large-$m$ limit where at $T=0$ there is only a single state. 

In finite systems neither the overlap, nor the correlation length, nor the energy
of the \acp{IS} is a monotonous function of $m$, as one would expect from a decreasing number
of available disordered states. They have instead a peak at an intermediate $m$.
We attribute this to the fact that when there are several minima, those of more ordered states have a larger
attraction basin, so having many \acp{IS} makes it easier to fall into a more ordered state.
If one wanted to rule out the non-monotonous behavior it could be useful to redefine the correlations
as a function of the normalized overlaps $Q^2/Q^2_\mathrm{self}$, as we have seen that the normalized overlaps
do exhibit a monotonous trend.

Also, we presented \acp{pdf}
of the spin and link order parameters $Q^2/Q^2_\mathrm{self}$ and $Q^2_\mathrm{link}/Q^2_\mathrm{link,self}$, 
noticing that the states with $Q^2/Q^2_\mathrm{self}=1$ have a major attraction basin,\index{attraction basin}
and create a second peak in the curve. Finite-size effects in the \acp{IS}' pdfs
were very heavy, as remarked also by looking at other observables, but they were minimal
if we considered the link overlap. This can suggests that perhaps the link overlap might be a better
descriptor to search a phase transition in a field (chapter \ref{chap:eah3d}).\index{de Almeida-Thouless!transition}
% Also, the dependency between $Q$ and $Q_\mathrm{link}$
% is consistent with RSB predictions.

Finally, we found a non-trivial behavior on the evolution of the self-overlap, 
that could be used as an indicator of the ``quality'' of a reached inherent structure.
\index{inherent structure|)}\index{overlap!tensorial|)}

%FINITO
 \chapter{Zero-temperature dynamics \label{chap:marginal}} \label{sec:marginal-intro}
In numerous glassy systems, such as electron \cite{efros:75,davies:82,pankov:05,palassini:12}, \index{glass!electron}
structural \cite{wyart:12,lerner:13,kallus:14} \index{glass!structural}
and spin glasses \cite{thouless:77,ledoussal:10,sharma:14}, \index{glass!spin}
it is possible to identify a set of states that exhibit a distribution of soft modes, \index{modes!soft}
unrelated to 
any symmetry, that reaches zero asymptotically. 
These states with modes infinitely close to zero constitute the manifold 
that separates stable from unstable states, \index{marginal stability}
and are said marginally stable \cite{mueller:15}.

When we relax an unstable system, it will stabilize the excitations and approach the marginally stable manifold, that we can identify as
the region of the space of states where the system becomes stable. When we treat, as we do in this chapter, discrete excitations, the marginal
manifold can be attained only in the thermodynamic limit.

Close to null temperature, when marginally stable systems are driven through an external force, the dynamics proceed through discrete 
changes in some relevant observable. The size of these rearrangements is scale-invariant, and it is usually referred to
as crackling noise \cite{sethna:01}.\index{crackling noise}

Often such scale-free bursty dynamics appears for a specific value of the force \cite{sethna:93,fisher:98}.
When the crackling noise occurs without the need to tune the
external parameters, we talk of \ac{SOC}. \index{self-organized criticality|(}
When a pseudogap is present in the density of states, and a system \index{pseudogap}
displays \ac{SOC},
\footnote{By pseudogap we mean a gap with zero width, i.e. the distribution is zero only in a point.}
then if the stability bounds are saturated 
the system is marginal \cite{mueller:15}.

The crackling responses are power-law distributed and span all the system. 
We study the arisal of crackling and of a pseudogap in \index{spin glass!Sherrington-Kirkpatrick|(}
the Sherrington-Kirkpatrick spin glass \cite{eastham:06,horner:08}, that exhibits both marginal stability and \ac{SOC}.
This is done both statically, through stability arguments, and by studying the dynamics of the crackling,
that in the \ac{SK} model appears in form of avalanches of spin flips.\index{avalanche}
At first, we focus on single- and multi-spin stability and scaling arguments. We characterize the pseudogap finding correlations \index{correlation!soft spins}
between soft spins and 
we show that an infinite number of neighbors is needed to have avalanches that span the whole system at $T=0$, confirming 
a sensation generated by numerical simulations \cite{andresen:13}.
We confirm this impression by stability arguments, indicating that an infinite number of neighbors is needed, and that
the presence of the short-range interactions irrelevant: \ac{SOC} is present in the presence of long-range interactions, and absent in their absence.
We then study what happens \emph{during} the avalanches, focusing on their dependency on the type of dynamics,
and modelizing them through different types of random walks. The same pseudogap that we find with stability arguments
arises spontaneously during the dynamics.

\section{Self-organized criticality and marginal stability in the SK model}\index{spin!Ising|(}
The \ac{SK} model, that was introduced in chapter \label{chap:spin-glass-intro} as a \ac{SG} model
for which mean field theory is valid \cite{sherrington:75}, treats 
Ising spins $s_\bx=\pm1$ at the vertices of
a fully connected graph. We are interested in its hysteresis, \index{hysteresis}
so the Hamiltonian includes a magnetic field term,
\begin{equation}
\label{eq:h-SK}
\mathcal{H}_\mathrm{SK}=-\frac{1}{2}\sum_{\bx\neq\by}J_{\bx\by}s_\bx s_\by-h\sum_{\bx}^Ns_\bx.
\end{equation}
The couplings are Gaussian-distributed, with mean $\overline{J_{\bx\by}}=0$ [the overline $\overline{(\ldots)}$ indicates an average over the instances
of the couplings]. 
The variance scales as $\overline{J_{\bx\by}^2}=J^2/N$, so the free energy is extensive and the \index{local!stability}
local stability distribution [equation \eqref{eq:plambda} later on] stays $O(1)$.

We define the local field as\index{local!field}\index{local!stability|(}
\begin{equation}
\label{eq:lambda}
\nomenclature[h....x]{$h_\bx$}{local field in the SK model}
h_\bx\equiv -\frac{\partial \mathcal{H}}{\partial s_\bx} = \sum_{\by\neq \bx}J_{\bx\by}s_\by+h\,,
\end{equation}
and the local stability of each spin as
\begin{equation}
\nomenclature[lambda....x]{$\lambda_\bx$}{local stability in the SK model}
\lambda_\bx=h_\bx s_\bx\,. 
\end{equation}
If a spin $s_\bx$ is aligned to its local field, then $\lambda_\bx>0$ and that site is stable. If $\lambda_\bx<0$ we call it unstable.
We will be interested in the distribution of local stabilities
\begin{equation}\label{eq:plambda}
\index{local!stability!distribution}
\nomenclature[rho....lambda]{$\rho(\lambda)$}{distribution of local stabilities}
 \rho(\lambda) = \frac{1}{N}\sum_\bx^N \delta(\lambda-\lambda_\bx)\,,
\end{equation}
where $\delta(\ldots)$ is a Dirac delta function.\nomenclature[delta....Dirac]{$\delta(\ldots)$}{Dirac delta function}
In a stable state, $\rho(\lambda)$ assumes only positive values, whereas if it is non-zero for negative $\lambda$ the state is unstable.
\footnote{When we say stable we mean that all the local stabilities are positive. In a thermodynamic sense those states are metastable.}
In a marginally stable state the $\rho(\lambda)$ reaches asymptotically zero, \index{pseudogap}
creating a pseudogap in the distribution of the local field. 
For small enough $\lambda$ we can expect it to scale as \cite{eastham:06}
\begin{equation}
\label{eq:pseudogap}
\index{exponent!pseudogap!theta@$\theta$}
\nomenclature[theta....pseudo]{$\theta$}{pseudogap exponent}
\rho(\lambda)\propto\lambda^{\theta}\,,
\end{equation}
for some $\theta$ that we will try to determine.\index{local!stability|)}

We work at zero temperature, focusing only on the changes that the variation of field $h$ imposes on the
energy landscape. \index{energy!landscape}
The dynamics are triggered by the variations of $h$. As soon as the field is strong enough to destabilize a spin,
that spin will flip. This flip can both stabilize the system, or destabilize some of its neighbors. When more than one spin is unstable,
the most unstable one is flipped (greedy dynamics, \cite{parisi:03}).\index{dynamics!greedy}
This dynamics is not frustrated: the flipping event decreasing the local energy of a spin also lowers the total energy, and thus stable states 
are achievable after a finite amount of steps.

\begin{figure}[!t]
 \centering
 \includegraphics[width=0.485\columnwidth]{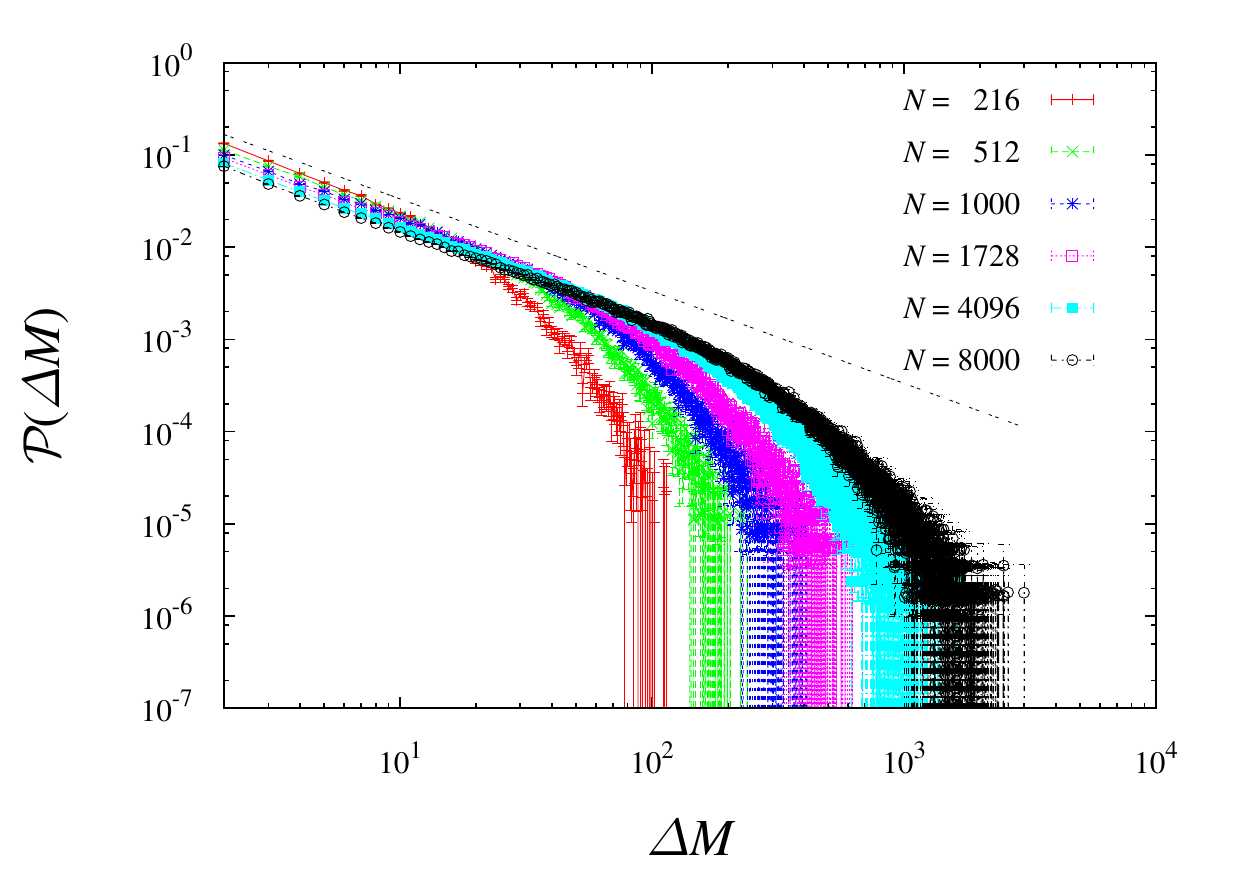}
 \includegraphics[width=0.485\columnwidth]{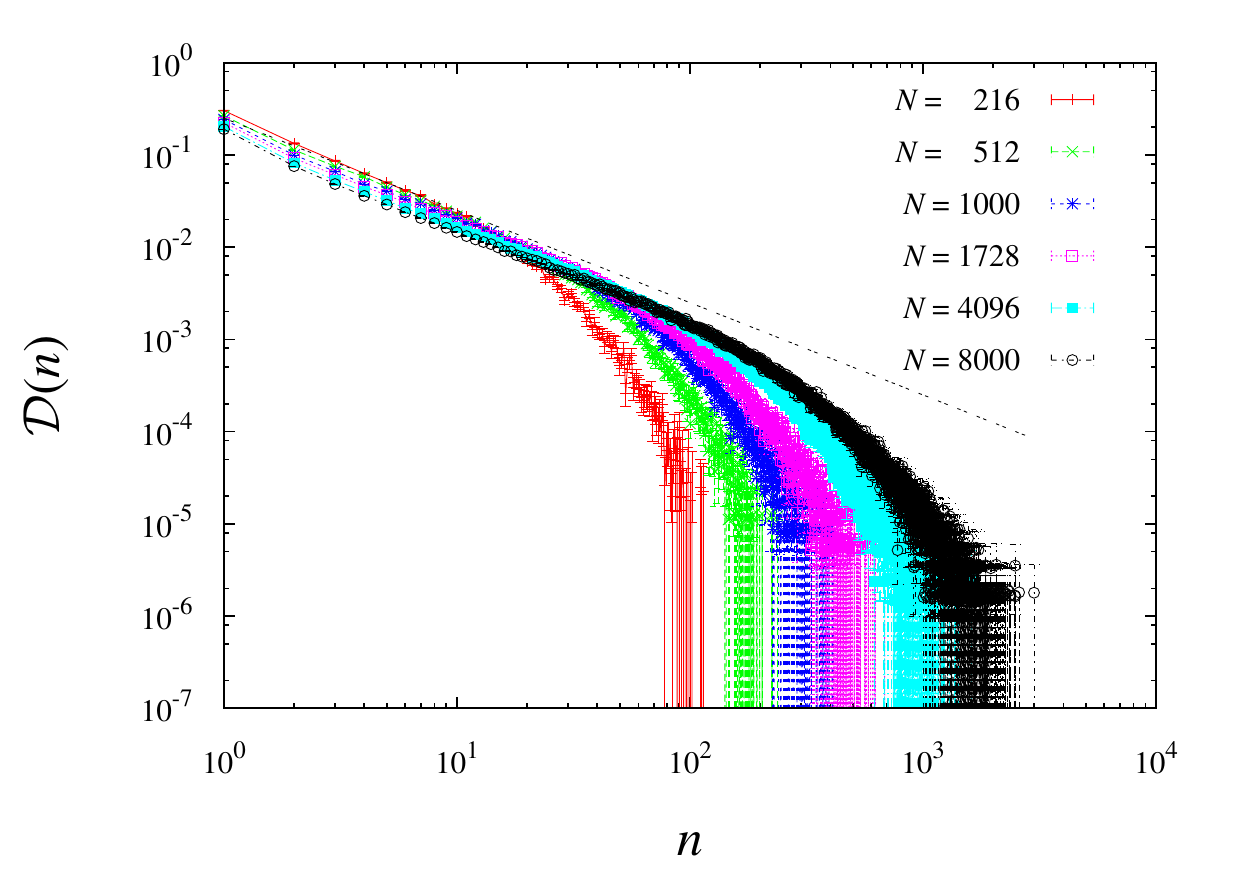}
\caption[Avalanches in the SK model]
	{Avalanches in the SK model for several system sizes. \index{avalanche}
	 \textbf{Left}: distribution of the magnetization jumps $\mathcal{P}(\Delta M)$.
	 \textbf{Right}: distribution of the avalanche sizes $\mathcal{D}(n)$.
	 The straight lines are reference curves $\propto \Delta M^{-1}$ and $\propto n^{-1}$.}
\label{fig:SK-aval}
 \end{figure}

The magnetization change $\Delta M$ \nomenclature[Delta...M]{$\Delta M$}{magnetization jump}\index{avalanche}
between the beginning and the end of the avalanche,
\footnote{The magnetization is $M=\sum_\bx^N s_\bx$.\nomenclature[M...M]{$M$}{magnetization}}
and the number of spin flips $n$, \nomenclature[n....7]{$n$}{In chapter \ref{chap:marginal}: size of the avalanche}
that we call the avalanche size,
have distributions $\mathcal{P}(\Delta M)$ \nomenclature[P..M]{$\mathcal{P}(\Delta M)$}{magnetization jump distribution}
and $\mathcal{D}(n)$ \nomenclature[D..n]{$\mathcal{D}(n)$}{avalanche size distribution}
that follow a power law
\begin{align}
\label{eq:power-laws}
 \mathcal{P}(\Delta M) &\propto \Delta M^{-\tau} \hat{p}(\Delta M/N^\beta)/\log(N)\,,\\[1ex]
 \mathcal{D}(n)        &\propto n^{-\rho} \hat{d}(n/N^\sigma)/\log(N)\,,
\end{align}
where $\hat{p}$ and $\hat{d}$ \nomenclature[p....hat]{$\hat{p}$}{scaling function}\nomenclature[d....hat]{$\hat{d}$}{scaling function}
are scaling functions and $\beta$ \nomenclature[beta....7]{$\beta$}{In chapter \ref{chap:marginal}: scaling exponent for $\Delta M$}
and $\sigma$ \nomenclature[sigma....]{$\sigma$}{scaling exponent for $n$}
are scaling exponents, with $\sigma\simeq1$ and $0.5\leq\beta\leq1$ \cite{pazmandi:99}\index{exponent!scaling!beta@$\beta$}\index{exponent!scaling!sigma@$\sigma$}.
The power law exponents are numerically found to be $\tau=\rho=1$ \cite{pazmandi:99}. 
\nomenclature[tau....7]{$\tau$}{In chapter \ref{chap:marginal}: power law exponent for $\mathcal{P}(\Delta M)$}
\nomenclature[rho....]{$\rho$}{power law exponent for $\mathcal{D}(n)$}
\index{exponent!avalanche!tau@$\tau$}\index{exponent!avalanche!rho@$\rho$}
The same values of the exponents
are found for the ground states (equilibrium avalanches) through replica calculations \cite{ledoussal:12}.\index{replica!theory}
In figure \ref{fig:SK-aval} we show both distributions $\mathcal{P}(\Delta M)$ and $\mathcal{D}(n)$.

 \clearpage

\section{Stability and correlations}

\subsection{Presence of avalanches}\index{stability!argument}
In order to have avalanches, when a spin is flipped, in average it must trigger at least another spin.
\footnote{We say at least one, and not one and only one spin, because in principle the
average number of triggered spins could be larger than one, and the avalanches stop due to the fluctuations in the number of triggered spins.}

\index{local!stability|(}
Every spin flip causes a kick $K$ \nomenclature[K...a]{$K$}{kick in the local stability of its neighbors}
in the local stability of its neighbors, that will be equal to twice the typical coupling 
$J_\mathrm{typ}$ \nomenclature[J...typ]{$J_\mathrm{typ}$}{typical coupling} between them, 
so the average kick scales as $K\sim2J_\mathrm{typ}\sim2/\sqrt{N}$. The probability that spin $s_i$ is triggered by the kick is
$P(\lambda_i<K)$, so extending it to the whole system we need
\begin{equation}\label{eq:constraint-1}
 (N-1) P(\lambda_i<K)\geq1\,.
\end{equation}\index{local!stability!distribution}
Since the kick coming from a single spin is small, we can restrict ourselves to the soft part of the $\rho(\lambda)$, so through
equation \eqref{eq:pseudogap} we get\index{exponent!pseudogap!theta@$\theta$|(}
\begin{equation}
 P(\lambda_i<K) \sim \int_0^{1/\sqrt{N}} \lambda^\theta d\lambda \sim N^{\frac{1-\theta}{2}}\,,
\end{equation}
that combined with \eqref{eq:constraint-1} implies the stability bound 
\begin{equation}
\theta\leq1\,.  
\end{equation}
If the bound is not satisfied, the avalanches fade off very quickly.

If equation \eqref{eq:constraint-1} is satisfied as an equality (we will show that this is the case), it would mean that, in a finite system, in 
average there is only one element with stability uniformly distributed in $0<\lambda_i<K$,
therefore the $\rho(\lambda)$ displays a kink for small $\lambda$ and intercepts the $y$ axis at a height $\rho(0)\sim1/\sqrt{N}$.

\paragraph{Smallest stability}
We can estimate the scaling of the least stability 
$\lambda_\MIN$ \nomenclature[lambda....min]{$\lambda_\MIN$}{smallest local stability} with a similar argument.
There has to be a fraction $\frac{1}{N}$ of spins with stability of the order of $\lambda_\MIN$ or lower, so
\begin{equation}
 \frac{1}{N} \sim \int_0^{\lambda_\MIN} \lambda^\theta d\lambda\sim \lambda_\MIN^{\theta+1}\,,
\end{equation}
that implies that the smallest stability scales as
\begin{equation}
 \lambda_{\MIN} \sim  N^{-1/(\theta+1)}\,.
\end{equation}
This also means that the minimum increase of the external field to trigger 
an avalanche scales as $h_\MIN\sim\lambda_\MIN\sim N^{-1/(\theta+1)}$.
\nomenclature[h....min]{$h_\MIN$}{smallest local field}

\subsection{Contained avalanches}\index{stability!two-spin}
Let us now consider in a stable state, a site $\bx$, with local stability of the order of $\lambda_\MIN$, and the site $\by$ that, among its neighbors,
has the lowest stability. In a finite percentage of cases, the interaction between the two sites will be unfrustrated, \index{frustration}
meaning that $s_\bx J_{\bx\by} s_\by >0$. In this situation, the energy cost of the simultaneous flip of both spins will be 
\begin{equation}\label{eq:2-spin-stability}
\nomenclature[Delta...Exy]{$\Delta E_{\bx\by}$}{energy cost of flipping $s_\bx$ and $s_\by$}
 \Delta E_{\bx\by} = 2(\lambda_\bx+\lambda_\by) - 4 |J_{\bx\by}|\,.
\end{equation}
For stability reasons, $\Delta E_{\bx\by}$ should be positive. So, to grant that the second term does not counteract the first two with
very large probability, we need $\lambda_\MIN\geq J_\mathrm{typ}$, therefore $N^{-1/(\theta+1)}\geq N^{-1/2}$, and
\begin{equation}\label{eq:constraint-2}
 \theta\geq1\,.
\end{equation}
Constraints \eqref{eq:constraint-1} and \eqref{eq:constraint-2} imply that the two bounds are saturated and the pseudogap exponent is $\theta=1$, 
confirming numerical simulations \cite{pazmandi:99}.

\index{stability!single-spin}
To extend this bound to single-flip stability, one can consider the quantity $E$, defined as the average number of spins triggered by a flip,
\begin{equation}
\nomenclature[E...E]{$E$}{mean	 number of spins triggered by a flip (see also $E(\nunst)$)}\index{local!stability!distribution}
 E = N \int_0^K\rho(\lambda)d\lambda\sim N^{(1-\theta)/2}\,.
\end{equation}
If $E\gg1$ the number of unstable spins grows exponentially, and the avalanche never stops. To avoid this possibility we must have $\theta\geq1$.
Later on we will come back to the participation of $E$ in the dynamics.

\subsection{Multi-spin stability}\index{stability!multi-spin}
We can also extend the stability criterion to a whole set $\mathcal{F}$\nomenclature[F..7]{$\mathcal{F}$}{In chapter \ref{chap:marginal}: a set of soft spins} 
of $m$ \nomenclature[m....7]{$m$}{In chapter \ref{chap:marginal}: the number of chosen softest spins}
spins that are initially stable with respect to a single spin flip.
The energy cost of such a change would be 
\begin{equation}
\label{eq:multi-spin-stability}
\nomenclature[Delta...E]{$\Delta E(\mathcal{F})$}{energy cost of flipping all the spins belonging to $\mathcal{F}$}
\Delta E(\mathcal{F})=2\sum_{\bx\in\mathcal{F}}\lambda_\bx-2\sum_{\bx,\by\in\mathcal{F}}J_{\bx\by}s_\bx s_\by\,,
\end{equation}
which is an extension of equation \eqref{eq:2-spin-stability}. To study the stability with respect to multi-spin flip excitations,
we want to compare the contribution of the two terms in \eqref{eq:multi-spin-stability}. This had been done by Palmer and Pond by taking in account
only the $m$ softest spins \cite{palmer:79}. 

\begin{figure}[!htb]
\centering
 \includegraphics[width=0.485\textwidth]{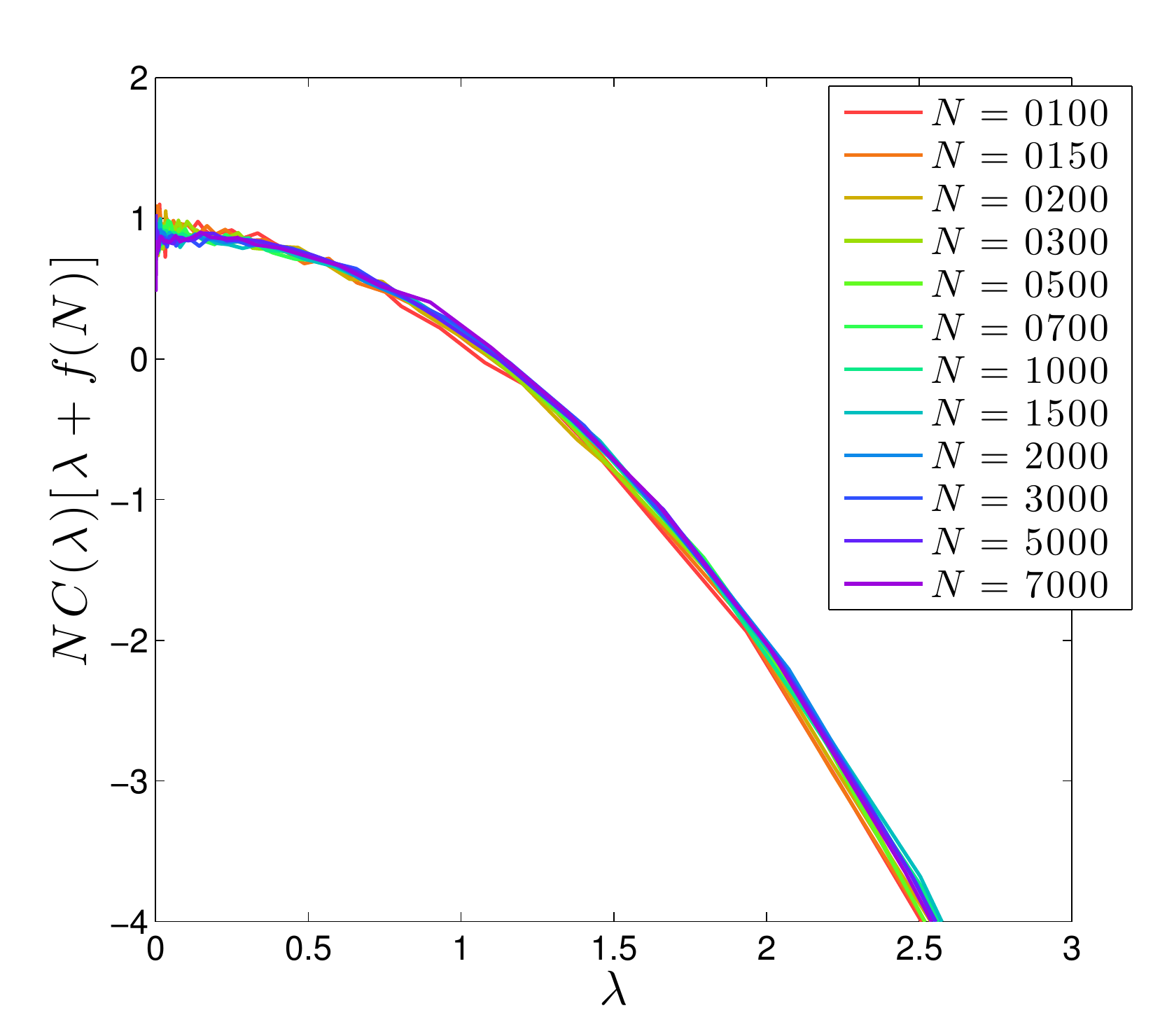}
 \caption[Frustration]{Correlation $C(\lambda)=-2\mean{s_\bx J_{\bx\by}s_\by}$ \index{correlation!soft spins}
 between the least stable spin and the spins with local
 stability $\lambda$, in locally stable states along the hysteresis loop. The function $f(N)=1.1\log(N)/N$ is added in order to obtain a collapse. 
 When $N$ goes to infinity $f(N)$ vanishes, and for $\lambda\ll1$ the curve is flat, so in this regime (small $\lambda$ and thermodynamic limit) we can expect $C(\lambda)\sim1/\lambda$.}
 \label{fig:frustrated}
\end{figure}

Calling $\lambda(m)$ \nomenclature[lambda....m]{$\lambda(m)$}{the $m^\mathrm{th}$ smallest stability}
the $m^\mathrm{th}$ smallest stability, one has that
\begin{equation}\label{eq:maxlam}\index{local!stability!distribution}
 \frac{m}{N} = \int_0^{\lambda(m)} \rho(\lambda)d\lambda \sim \lambda(m)^{\theta+1}\,,
\end{equation}
so
\begin{equation}\label{eq:lm}
\lambda(m)\sim \left(\frac{m}{N}\right)^{\frac{1}{1+\theta}}\,, 
\end{equation}
and the first term in the \ac{rhs} of equation \eqref{eq:multi-spin-stability}
scales as $m\left(\frac{m}{N}\right)^{\frac{1}{1+\theta}}$.
For the second term one has $\sum_{\bx,\by}^m J_{\bx\by}s_\bx s_\by \sim \sum_\bx^m (m/N)^{1/2}$ because of the random signs.
The contribution scales then as $m(m/N)^{1/2}$. In \cite{palmer:79} it was assumed to be positive, i.e. the softest spins are
in average unfrustrated \index{frustration} among each other, and from that a stability bound $\theta\geq1$ was recovered. We can see from figure \ref{fig:frustrated}
that for small $\lambda$ this hypothesis is not confirmed, so $\Delta E(\mathcal{F})$ is always positive in average.

The nonfulfilling of Palmer and Pond's hypothesis means that for small $\lambda$ the correlation \index{correlation!soft spins}
\begin{equation}
\nomenclature[C...flambda]{$C(\lambda)$}{correlation between the softest spin and the spins with local stability $\lambda$}
 C(\lambda) = -2\mean{s_\bx J_{\bx\by} s_\by}
\end{equation}
between the softest spin and a spin with stability $\lambda$ is positive in average, and the argument on the average scalings does not imply $\theta\geq0$.
If we postulate a behavior
\begin{equation}\label{eq:C-lambda-trend}
\index{exponent!correlation!delta@$\delta$}\index{exponent!correlation!gamma@$\gamma$}
\nomenclature[delta....7]{$\delta$}{In chapter \ref{chap:marginal}: correlation exponent (number of spins)}
\nomenclature[gamma....7]{$\gamma$}{In chapter \ref{chap:marginal}: correlation exponent (stability)}
 C(\lambda) \sim \lambda^{-\gamma} N^{-\delta}\,,
\end{equation}
we can predict the scaling 
\begin{equation}\label{eq:scaling-Jss}
\mean{-\sum_{\bx,\by}^m J_{\bx\by}s_\bx s_\by} \sim m^2 C\bigl(\lambda\left(m\right)\bigr) \sim m^{2-\frac{\gamma}{1+\theta}} N^{\frac{\gamma}{1+\theta}-\delta} 
\end{equation}
with the help of equation \eqref{eq:lm}.

\subsection{Bound due to the fluctuations}
Even though a multi-spin stability criterion on the averages does not imply a bound $\theta\geq1$ on the correlation, \index{correlation!soft spins}
it is still possible to recover that bound by studying the large fluctuations of the last term of \eqref{eq:multi-spin-stability},
that might make $\Delta E(\mathcal{F})$ negative.

Given the set of the $m'$ most unstable spins, let us consider all the subsets $\mathcal{F}$ of $m=m'/2$ spins.
\index{exponent!correlation!delta@$\delta$}\index{exponent!correlation!gamma@$\gamma$}
We can assume that the $\Delta E$ associated with each of the sets $\mathcal{F}$ are independent and Gaussian-distributed, with
\begin{align}
\label{eq:de-mean}
 \mean{\Delta E}_{m'} &= 2 m \mean{\lambda(m)}_{m'} - 2 m^2 \mean{s_\bx J_{\bx\by} s_\by}_{m'}
 &\sim \frac{m^{(2+\theta)/(1+\theta)}}{N^{1/(1+\theta)}} + m^{2-\frac{\gamma}{1+\theta}} N^{\frac{\gamma}{1+\theta}-\delta} 
 \,,\\[1ex]
\label{eq:de-var}
 \var{\Delta E} &= \langle\Delta E^2\rangle_{m'} - \mean{\Delta E}^2_{m'} = 8m^2/N\,,
\end{align}
where $\mean{\ldots}_{m'}$ is an average over the $m'$ softest sites. 
\footnote{
The second of the two terms on the \ac{rhs} of equation \ref{eq:de-mean} comes from equation \ref{eq:scaling-Jss}.
To find the first one it is necessary to calculate 
\begin{equation}\index{local!stability!distribution}
 \mean{\lambda(m)}_{m'} = \frac{\int_0^{\lambda(m')} \lambda\rho(\lambda)d\lambda}{\int_0^{\lambda(m')} \rho(\lambda)d\lambda}\,,
\end{equation}
where the maximum stability of the chosen set, $\lambda(m')$, can be evaluated through equation \eqref{eq:maxlam}.
Remembering that $m'=2m$, one obtains $\mean{\lambda(m)}_{m'}\sim \left(\frac{m}{N}\right)^{\frac{1}{1+\theta}}$, that
multiplied by $m$ gives the term that appears in equation \eqref{eq:de-mean}.
}
We neglected the non-diagonal terms in the variance.
So, from equation \eqref{eq:de-var} it descends that the fluctuations \nomenclature[X...0]{$X$}{fluctuation of $\Delta E (\mathcal{F})$}
$X=\sum_{\bx,\by}^m J_{\bx\by}s_\bx s_\by-\mean{\sum_{\bx,\by}^m J_{\bx\by}s_\bx s_\by}$ 
on $\Delta E (\mathcal{F})$ are of order $m/\sqrt{N}$.
\footnote{
We neglect the fluctuations of $\sum_\bx \lambda_\bx$, since that sum is always positive and when $m$ is large its fluctuations are small
compared to its expectation value.
}
As there are $2^{2m}$ sets $\mathcal{F}$, the number density of having fluctuation $X$ 
is $\mathcal{N}(X)\sim 2^{2m}\E^{-N X^2/m^2}$ \nomenclature[N...X]{$\mathcal{N}(X)$}{number density of having fluctuation $X$}
(if $\Delta E$ is Gaussian, $X$ has to be Gaussian with zero mean). 
We can recover the most negative fluctuation by imposing $\mathcal{N}(X_\MIN)\sim1$,
\nomenclature[X...MIN]{$X_\MIN$}{most extreme fluctuation of $\Delta E (\mathcal{F})$}
that implies straightforwardly $X_\MIN\sim-\sqrt{\frac{m^3}{N}}$.
Thus, the energy change $\Delta E(\mathcal{F})$ associated with the most negative fluctuation scales as
\index{exponent!correlation!delta@$\delta$}\index{exponent!correlation!gamma@$\gamma$}
\begin{equation}
 \label{eq:sta}
\Delta E(\mathcal{F}_{\MIN})=m^{(2+\theta)/(1+\theta)} N^{-1/(1+\theta)} + m^{2-\gamma/(1+\theta)} N^{\gamma/(1+\theta)-\delta}- m^{3/2} N^{-1/2}.
\end{equation}

The multi-spin stability condition demands that, for large $N$ and fixed $m$, the energy change $\Delta E(\mathcal{F}_{\MIN})$ stay positive. This
occurs if\index{exponent!correlation!delta@$\delta$}\index{exponent!correlation!gamma@$\gamma$}
\begin{align}
\label{eq:sca-1}
\theta&\geq 1\,, \\ 
\text{or}\notag\\
\label{eq:sca-2}
\gamma/(1+\theta)-\delta&\geq-1/2\,,
\end{align}
depending on which of the two terms in the \ac{lhs} dominates.\index{correlation!soft spins}
Nonetheless, the correlation between spins is bounded by the typical coupling, $C(\lambda)\lesssim N^{-1/2}$, so
from equation \eqref{eq:scaling-Jss} we obtain that $\gamma/(1+\theta)-\delta\leq-1/2$. Hence, if \eqref{eq:sca-1} is not
verified,$\theta<1$, then $\gamma/(1+\theta)-\delta=-1/2$ and \eqref{eq:sca-2} is saturated.

%In the condition $\theta\geq1$ ?,
The scaling with large $m$ of \eqref{eq:sta} also requires $2-\frac{\gamma}{1+\theta}\geq \frac{3}{2}$,
i.e. $\gamma\leq\frac{1+\theta}{2}\leq1$ and $\delta\leq1$.
In the relevant states all three exponents $\theta, \gamma$ and $\delta$ equal 1, and the constraints are satisfied as
exact equalities.
\index{exponent!pseudogap!theta@$\theta$|)}\index{local!stability|)}

\section{Finite-size cutoffs \label{sec:marginal-cutoffs}}\index{cutoff}
In finite systems, the avalanches are bounded by cutoffs $n_\mathrm{c}(N)$ and $\Delta M_\mathrm{c}(N)$.
The shape of the avalanche distributions gives a relation between cutoffs and average sizes of the avalanches.
In the simplest case $\rho=\tau=\sigma=\beta=1$, \index{exponent!avalanche!tau@$\tau$}\index{exponent!avalanche!rho@$\rho$}
\index{exponent!scaling!beta@$\beta$}\index{exponent!scaling!sigma@$\sigma$}
we can incorporate explicitly exponential cutoffs in the distributions of the avalanches, getting
\begin{align}
\nomenclature[n....c]{$n_\mathrm{c}$}{avalanche size cutoff}
\nomenclature[Delta...Mc]{${\Delta M}_\mathrm{c}$}{magnetization jump cutoff}
 \mathcal{D}(n)        &\propto n^{-1}\E^{-\frac{n}{n_\mathrm{c}}} \\[1ex]
 \mathcal{P}(\Delta M) &\propto \Delta M^{-1}\E^{-\frac{\Delta M}{\Delta M_\mathrm{c}}}\,,
\end{align}
so if we calculate the mean avalanche size and the mean magnetization jump,
\footnote{In this chapter the averages $\mean{\ldots}$ are averages over the avalanches.
\nomenclature[...]{$\mean{\ldots}$}{In chapter \ref{chap:marginal}: average over the avalanches}
}
they result proportional to their cutoff,
\begin{align}
\mean{n} &\propto n_\mathrm{c} \,,\\[1ex]
\mean{\Delta M} &\propto \Delta M_\mathrm{c}\,.
\end{align}
In the case that the exponents $\tau$ and $\rho$ are not equal to unity, $\mean{n}$ and $\mean{\Delta M}$ can still
be used as estimators for the cutoffs, though the relation is not linear anymore.

If the cutoffs diverge as the system size becomes infinite, the system displays \ac{SOC}, so we can search its
presence by looking at $\mean{\Delta M}$ and $\mean{n}$.

\paragraph{Scaling of $\bm{\mean{\Delta M}}$}
Let us consider an ideal driving experiment in which between the beginning and the end we vary 
the external field of $\Delta h^\mathrm{(tot)}$.
Let the driving be so slow that every time an avalanche is triggered the external field's 
variation was neglectable, so the field variation
is given only by the driving between one avalanche and the next one, $h_\MIN$, 
that as we saw scales like $N^{-1/2}$. Therefore, the number of avalanches in the experiment scales as
\begin{equation}
\nomenclature[n....av]{$n_\mathrm{av}$}{number of avalanches}
 n_\mathrm{av} = \frac{\Delta h^\mathrm{(tot)}}{h_\MIN} \sim \sqrt{N}\,.
\end{equation}

\begin{figure}[!t]
 \centering
 \includegraphics[width=0.485\columnwidth]{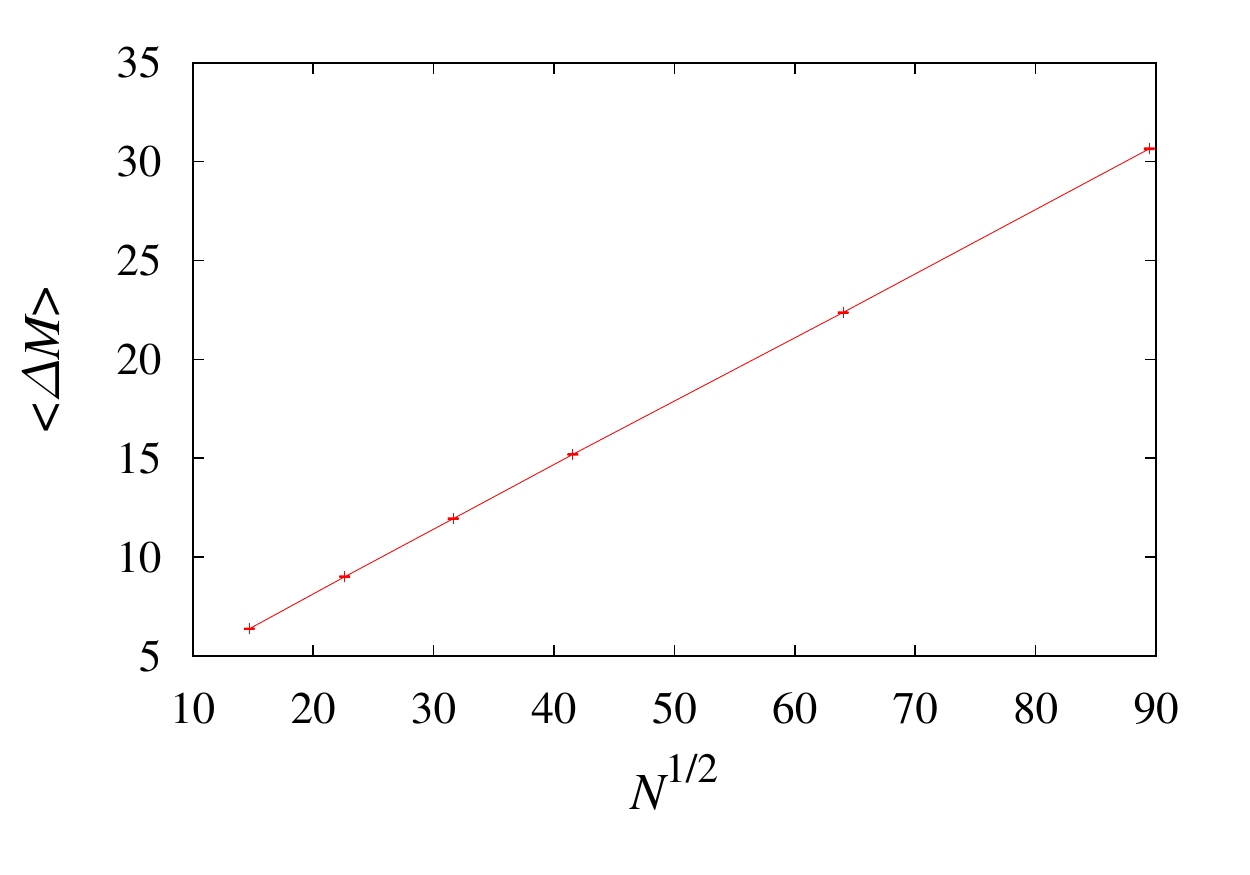}
 \caption[Scaling of $\mean{\Delta M}$ in the SK model]
	 {Scaling of the average magnetization jump $\mean{\Delta M}$ with the square root of the system size, in the \ac{SK} model.
	 }
 \label{fig:SK-mean-deltaM}
\end{figure}

Also the total magnetization, that will change extensively, 
$\Delta M^\mathrm{(tot)}\sim N $,\nomenclature[Delta...Mtot]{$\Delta M^\mathrm{(tot)}$}{total magnetization change}
is related to the number of avalanches in the experiment $n_\mathrm{av}$ by
\begin{equation}
 \Delta M^\mathrm{(tot)} \sim n_\mathrm{av} \mean{\Delta M}\,,
\end{equation}
implying 
\begin{equation}
\mean{\Delta M} \sim \sqrt{N}\,,
\end{equation}
so the cutoff goes to infinity as $N\to\infty$, and the \ac{SK} model displays \ac{SOC},
as it is confirmed in figure \ref{fig:SK-mean-deltaM}.

\paragraph{Scaling of $\bm{\mean{n}}$}
We can attempt to estimate the scaling of $\mean{n}$ by studying the energy, since, differently from the magnetization,
its evolution is monotonous in time.
In a single avalanche, the energy change is
\begin{equation}
\nomenclature[Delta...Eav]{$\Delta E_\mathrm{av}$}{energy change in a single avalanche}
\nomenclature[Delta...Eflip]{$\Delta E_\mathrm{flip}$}{energy change in a single flip}
 \mean{\Delta E_\mathrm{av}} = \mean{n} \mean{\Delta E_\mathrm{flip}}\,,
\end{equation}
where $\mean{\Delta E_\mathrm{flip}}$ is the average energy change per spin flip. Assuming that it is of the
order of the typical coupling, $\mean{\Delta E_\mathrm{flip}}\sim J_\mathrm{typ}\sim N^{-1/2}$. 

\begin{figure}[!t]
 \centering
 \includegraphics[width=0.485\columnwidth]{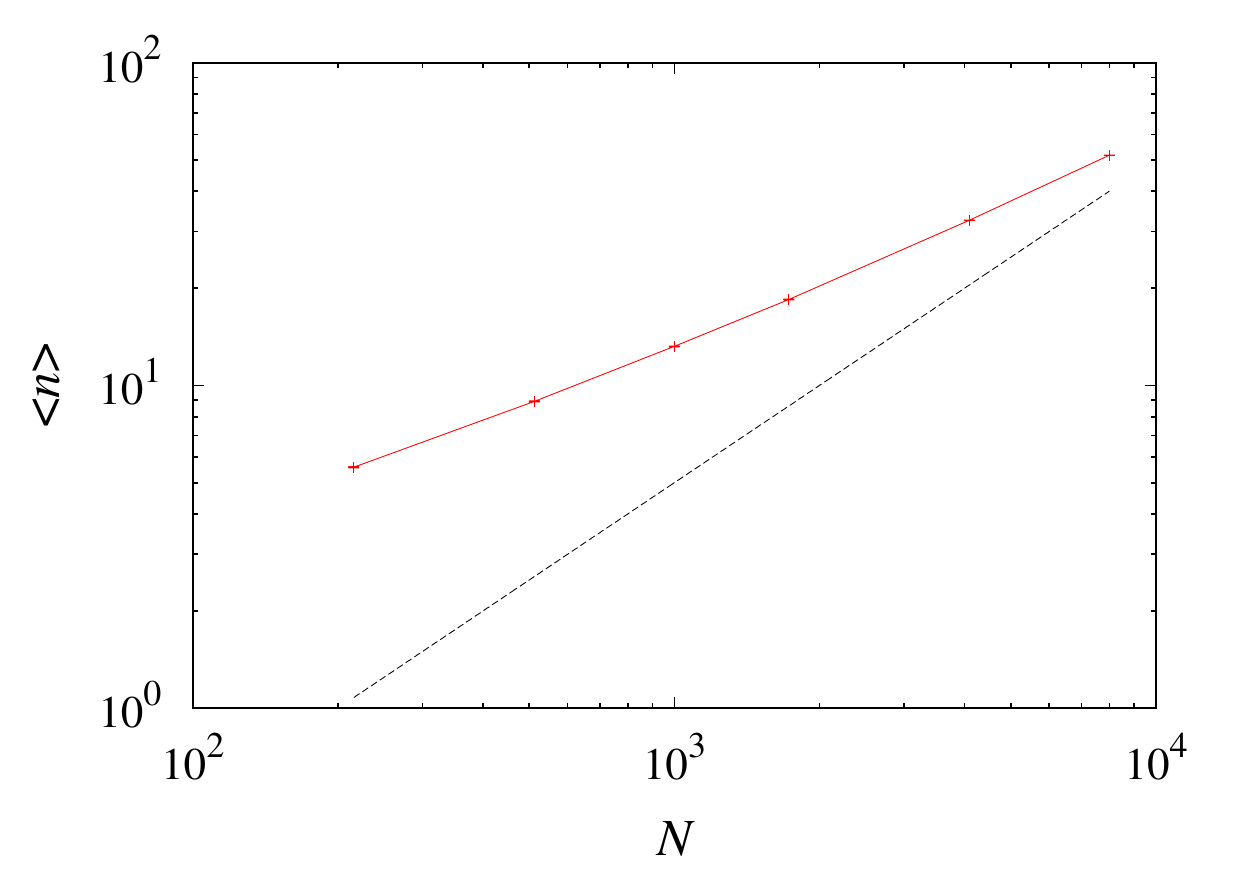}
 \caption[Scaling of $\mean{n}$ in the SK model]
	 {Scaling of the average avalanche size $\mean{n}$ with the system size $N$, in the \ac{SK} model.
	 The straight line is a reference curve $\propto N$.}
 \label{fig:SK-mean-size}
\end{figure}

For the total energy change during an avalanche, let us consider a full hysteresis loop.
Neglecting logarithmic corrections, its area $A=\sum_{i\in\text{drivings}} M dh_i \sim N$ is extensive. \nomenclature[A...big1]{$A$}{area of the hysteresis loop}
\footnote{With at most logarithmic corrections, that can be neglected in this argument.}
The total energy change, $E^\mathrm{(tot)}$, \nomenclature[E...tot]{$E^\mathrm{(tot)}$}{total energy change in a full hysteresis loop}
is zero because the experiment starts and finishes in the same point,
but it is also equal to the sum of the contributions of the avalanches and of the field drivings,
\begin{align}
 0\sim E^\mathrm{(tot)} &\sim \sum_\text{avalanches} \Delta E_\mathrm{av} + \sum_{i\in\text{drivings}} M dh_i \sim\\
                        &\sim n_\mathrm{av}\Delta E_\mathrm{av} + A \sim\\
                        &\sim N +\sqrt{N}\Delta E_\mathrm{av}\,,
\end{align}
so $\Delta E_\mathrm{av}\sim\sqrt{N}$, and as a consequence
\begin{equation}
 \mean{n}\sim N\,.
\end{equation}
In figure \ref{fig:SK-mean-size} we show that numerical data are consistent with an asymptotic behavior $\mean{n}\sim N$ (with 
possibile logarithmic corrections).\index{numerical simulations}

Both the cutoffs we recovered go to infinity with the system size, and the \ac{SK} model displays self-organized criticality.

\subsection{Short-range models}
Let us consider now models defined on a generic graph where each site has $z$ neighbors. The finite-neighbor (short-range) Hamiltonian is
\begin{equation}
\label{eq:h-SR}
\nomenclature[H..SR]{$\mathcal{H}_\mathrm{SR}$}{finite-connectivity Hamiltonian}
\mathcal{H}_\mathrm{SR}=-\frac{1}{2}\sum_{\bx}^Ns_\bx\sum_{\by\in\mathcal{V}(\bx)}^zJ_{\bx\by} s_\by-h\sum_{\bx}^Ns_\bx\,,
\end{equation}
where $\mathcal{V}(\bx)$ is the set of sites that are connected to $\bx$ through an edge of the graph.
\nomenclature[V..x]{$\mathcal{V}(\bx)$}{set of sites that are connected to $\bx$}
When the interactions are not long-range, i.e. each site has a small connectivity $z$, it has been observed numerically that 
$\theta$ stays the same,\index{exponent!pseudogap!theta@$\theta$}
but self-organized criticality vanishes \cite{andresen:13}, because the cutoffs of the power law behaviors do not diverge with the system size. Also, the pseudogap
disappears, and the intercept of the stability distribution scales as $\rho(0)\sim 1/\sqrt{z}$.\index{local!stability!distribution}

\index{local!stability}\index{stability!argument!short-range}
That $\rho(0)\sim 1/\sqrt{z}$ is expectable from the previous argument that in average there is only one element with stability uniformly distributed in $0<\lambda_i<K$.
Since now the kick $K$ is of order $1/\sqrt{z}$, the intercept is at height $\rho(0)\sim1/\sqrt{z}$, so the distribution of the stabilities
becomes
\begin{equation}
 \rho(\lambda) \sim \frac{A}{\sqrt{z}} + B\lambda\,.
\end{equation}

In these conditions the smallest stability is given by
\begin{equation}
 \frac{1}{N}\sim P(\lambda<\lambda_\MIN)\sim \frac{A'\lambda_\MIN}{z} + B'\lambda_\MIN^2\,.
\end{equation}
Since $\lambda_\MIN$ is small, we can neglect the quadratic term, so $\lambda_\MIN\sim \frac{\sqrt{z}}{N}$.
It is straightforward to see that if $z=c N$, for some finite $c$, the \ac{SK} limit is recovered.

The cutoff magnetization jump $\Delta M_\mathrm{c}$ changes consequently
\begin{equation}
 \mean{\Delta M} = \frac{\Delta M^\mathrm{(tot)}}{n_\mathrm{av}}\sim N \lambda_\MIN \sim \sqrt{z}\,.
\end{equation}
So, if the connectivity $z$ is finite the avalanches have a finite cutoff, while if it diverges we recover
the self-organized criticality of the \ac{SK} model.

\begin{figure}[!htb]
 \centering
 \includegraphics[width=0.485\columnwidth]{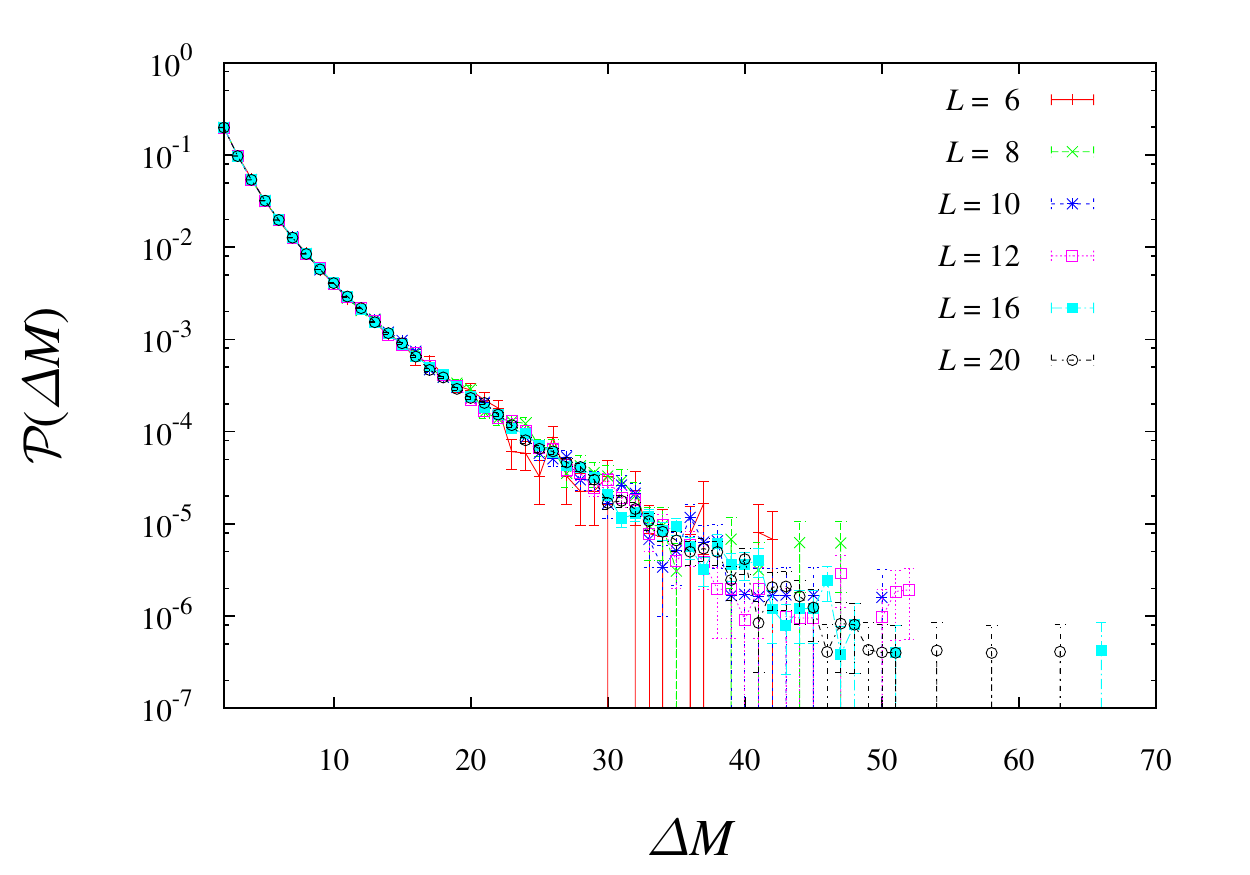}
 \includegraphics[width=0.485\columnwidth]{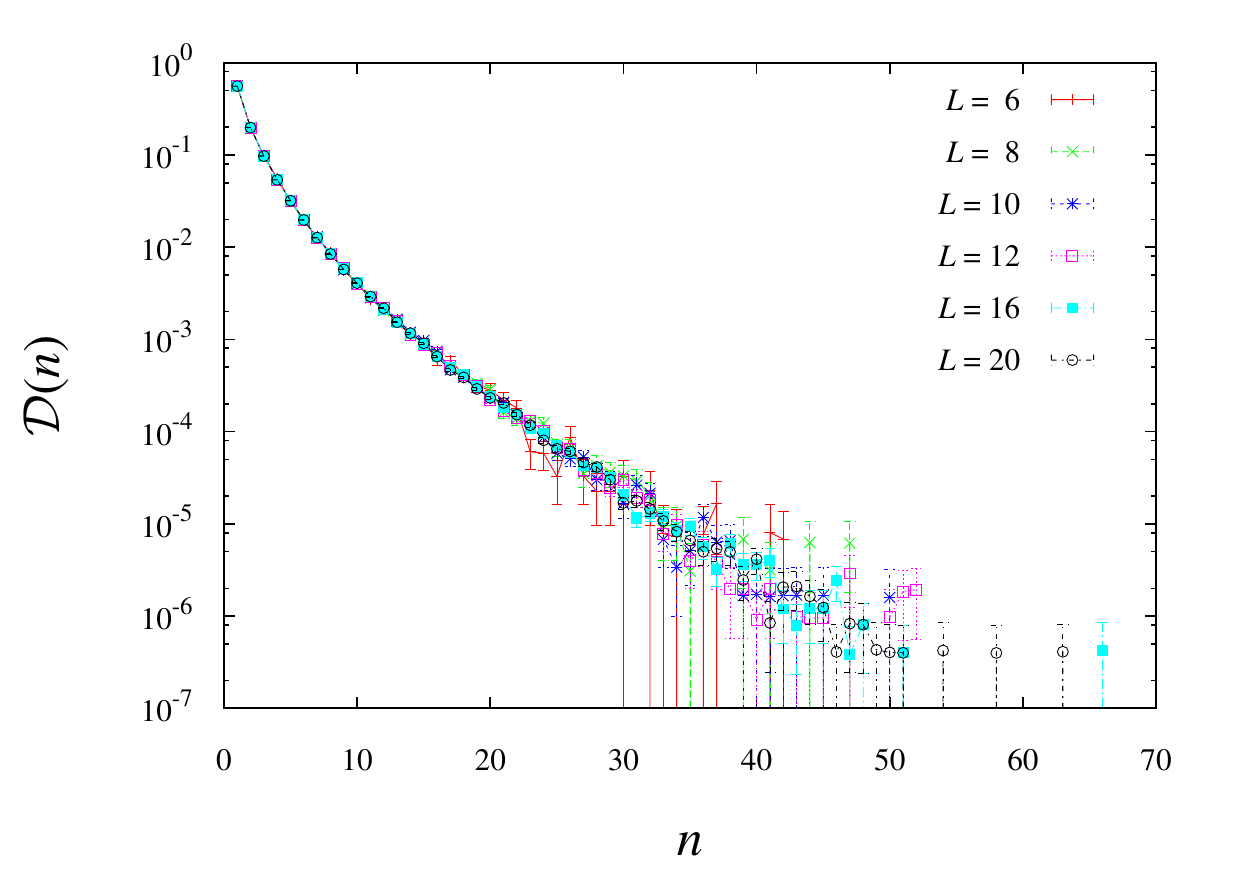}
\caption[Avalanches in the EA model]{
	 Same as figure \ref{fig:SK-aval}, but in the three-dimensional Edwards-Anderson model.
	 The system sizes are the same as figure \ref{fig:SK-aval}. In the legend we express them
	 through the linear lattice size $L$ ($L^3=N$) to stress that the interactions are between nearest 
	 neighbors of a cubic lattice.}
\label{fig:EA-aval}
 \end{figure}

\begin{figure}[!htb]
 \centering
 \includegraphics[width=0.485\columnwidth]{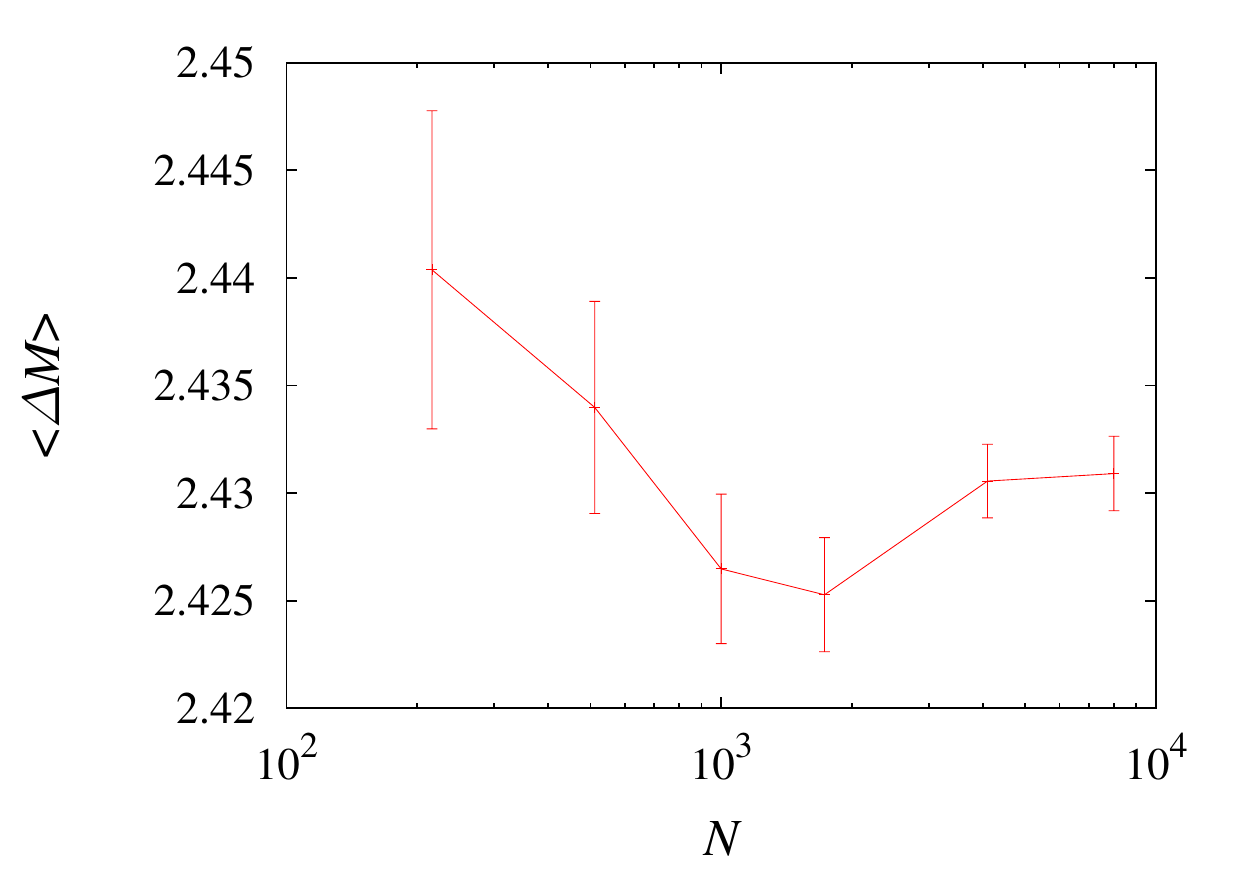}
 \includegraphics[width=0.485\columnwidth]{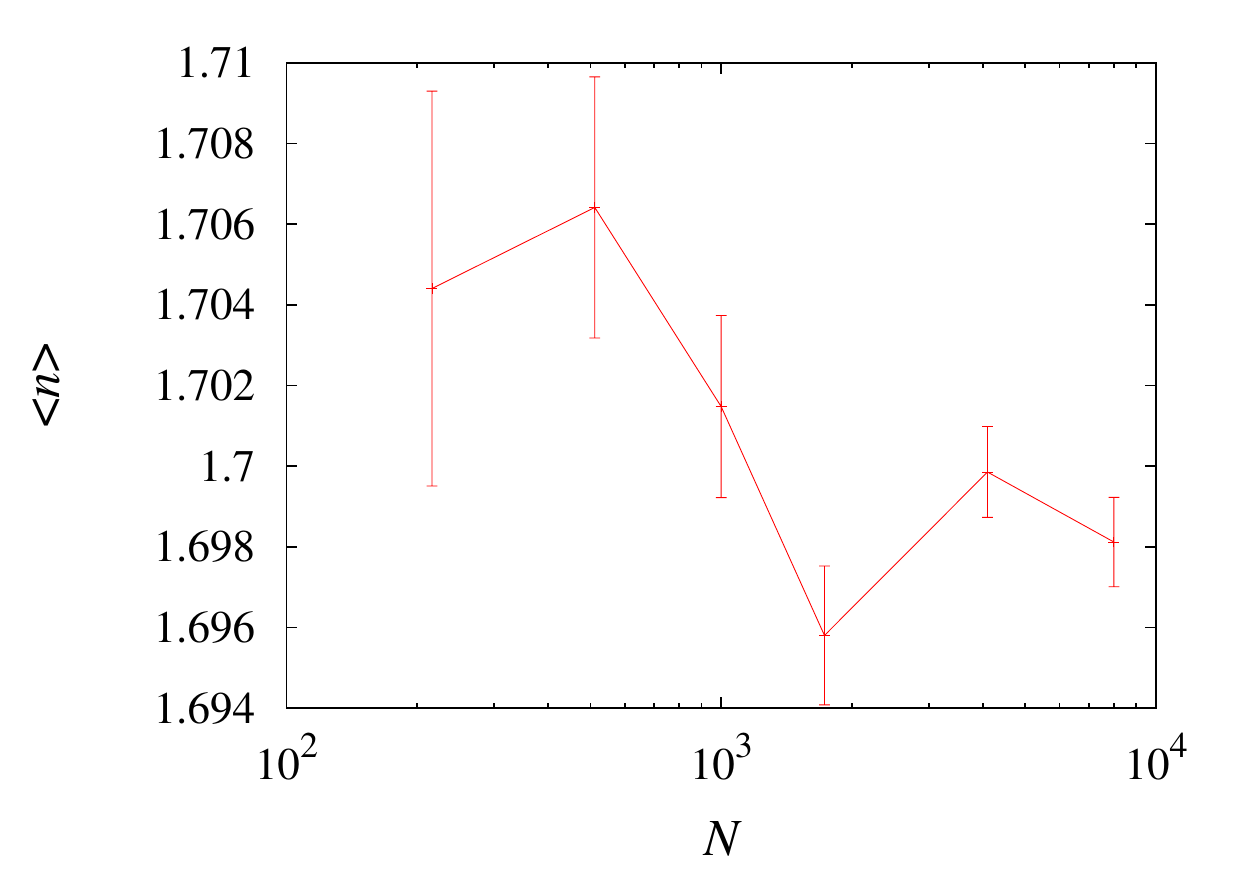}
 \caption[Scaling of $\mean{\Delta M}$ and $\mean{n}$ in the EA model]
	 {Scaling of the averages $\mean{\Delta M}$ (\textbf{left}) and $\mean{n}$ (\textbf{right}) with the system size $N$, in the EA model.
	 }
 \label{fig:EA-means}
\end{figure}

This can be seen also through the scaling of $n_\mathrm{c}$, by using the relation $\Delta E_\mathrm{av}\sim\mean{n}\mean{\Delta E_\mathrm{flip}}$.
The average energy change per flip is of the order of $\Delta E_\mathrm{flip}\sim J_\mathrm{typ}\sim \frac{1}{\sqrt{z}}$.
The hysteresis argument for $\mean{\Delta E_\mathrm{av}}$ this time yields $n_\mathrm{av} \sim \frac{1}{h_\MIN}\sim\frac{N}{\sqrt{z}}$.
Therefore 
\begin{align}
 0 \sim E^\mathrm{(tot)} &\sim  n_\mathrm{av}\Delta E_\mathrm{av} + A \\
                         &\sim N +\frac{N}{\sqrt{z}}\Delta E_\mathrm{av}\,,
\end{align}
so $\mean{\Delta E_\mathrm{av}}\sim\sqrt{z}$ and $\mean{n}\sim z$, confirming the absence of self-organized criticality in models
with finite connectivity.
One could actually expect this by looking at the distributions $\mathcal{P}(\Delta M)$ and $\mathcal{D}(n)$ in figure \ref{fig:EA-aval}. 
For all the sizes, the curves collapse to the same exponential decay, so there cannot be a scaling of the mean values (figure \ref{fig:EA-means}) 
nor of the cutoffs.

\subsection{Competition between short and long range interactions}
Long-range interaction models display \ac{SOC}, while if the interactions are short-range this is not true. 
Since the application of the concept of \ac{SOC} is related to many systems where there might be a coexistence of the two,
a question that arises spontaneously is whether it is the presence of long-range interactions that guarantees \ac{SOC}, the existence of short-range
ones that kills it, or it depends on their relative magnitude.

We define thus a model that mixes short and long-range interactions, and try to understand whether or not it displays \ac{SOC}. 
A simple way is to get an \ac{EA} model on a cubic lattice, and add to it an infinite-range interaction term.
Let the spacing between nearest neighbors in the lattic be unitary, and $L$ be the side of the full lattice.
We impose periodic boundary conditions.
Each site $\bx$ hosts a spin $s_\bx$, and interacts with the rest of the spins through a duplex network. One graph follows the geometry of the
lattice, and allows only nearest-neighbor interactions, and the other is fully connected.

The Hamiltonian is
\begin{equation}
\nomenclature[H..SL]{$\mathcal{H}_{SL}$}{Hamiltonian that mixes short- and long-range interactions}
 \mathcal{H}_{SL}= -\sum_{\langle \bx,\by\rangle} J_{\bx\by}^{(s)}s_\bx s_\by-\sum_{\bx,\by}J_{\bx\by}^{(\ell)} s_\bx s_\by - h\sum_\bx s_\bx\,,
\end{equation}
where $J_{\bx\by}^{(s)}$ \nomenclature[J...xys]{$J_{\bx\by}^{(s)}$}{amplitude of the short-range coupling}
\nomenclature[J...xyl]{$J_{\bx\by}^{(\ell)}$}{amplitude of the long-range coupling}
is the short-range coupling, and $J_{\bx\by}^{(\ell)}$ is the long-range one.
Both are gaussian random variables with zero mean $\overline{J^{(s)}_{\bx\by}} = \overline{J^{(\ell)}_{\bx\by}} = 0$, 
and variances $\overline{{J^{(s)}_{\bx\by}}^2} = J^{(s)}/z$ and $\overline{{J^{(s)}_{\bx\by}}^2} = J^{(l)}/N$. 
The limit $J^{(s)}=0$ corresponds to the \ac{SK} model, while $J^{(l)}=0$ 
is the \ac{EA} model. We work on a cubic lattice, so $z = 2d$. 

\index{stability!argument!duplex network}
We impose the stability argument separating the nearest neighbor interactions from the others\index{local!stability}
\begin{equation}\index{local!stability!distribution}
 1 \leq (N-z) \int_0^{{J}^{(\ell)}} \rho(\lambda) d\lambda + \int_0^{\tilde{J}} \rho(\lambda) d\lambda\,,
\end{equation}
with $\tilde{J}^2 = {{J}^{(\ell)}}^2 + {{J}^{(s)}}^2$.
Taken alone, the first term on the right hand side is always critical, whereas the second one is never.

To verify the presence of both terms, it is convenient to study the limit ${{J}^{(\ell)}} \ll {{J}^{(s)}}$. Since
the typical avalanches do not imply large stability jumps (figure \ref{propt} later on), the kicks on the softest modes will be dictated
by ${{J}^{(\ell)}}$, and we can assume that the stability distribution be
 $\rho(\lambda) \propto \alpha\lambda$,
where the constant $\alpha$ is to keep track of the competition between the two interactions.

\begin{figure}[!t]
 \centering
 \includegraphics[width=0.485\columnwidth]{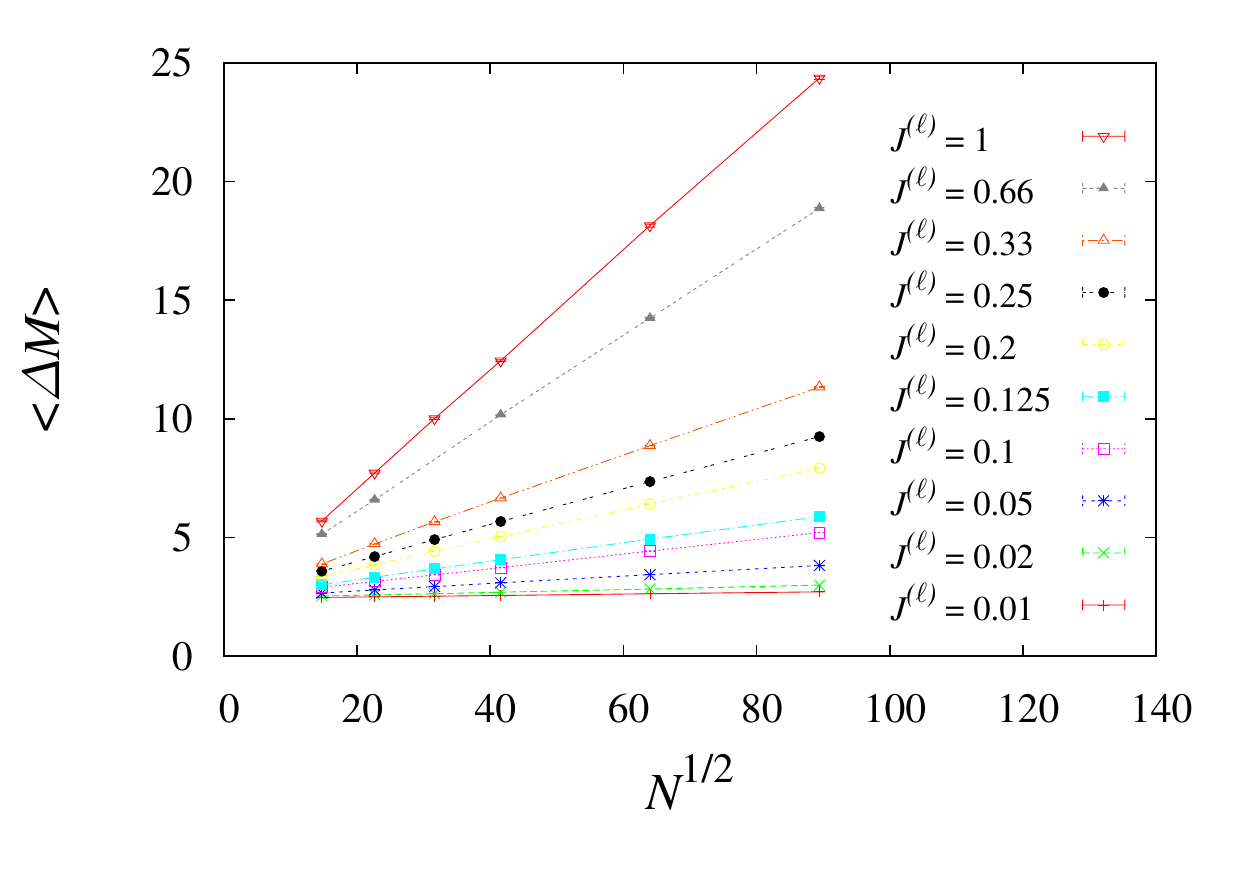}
 \includegraphics[width=0.485\columnwidth]{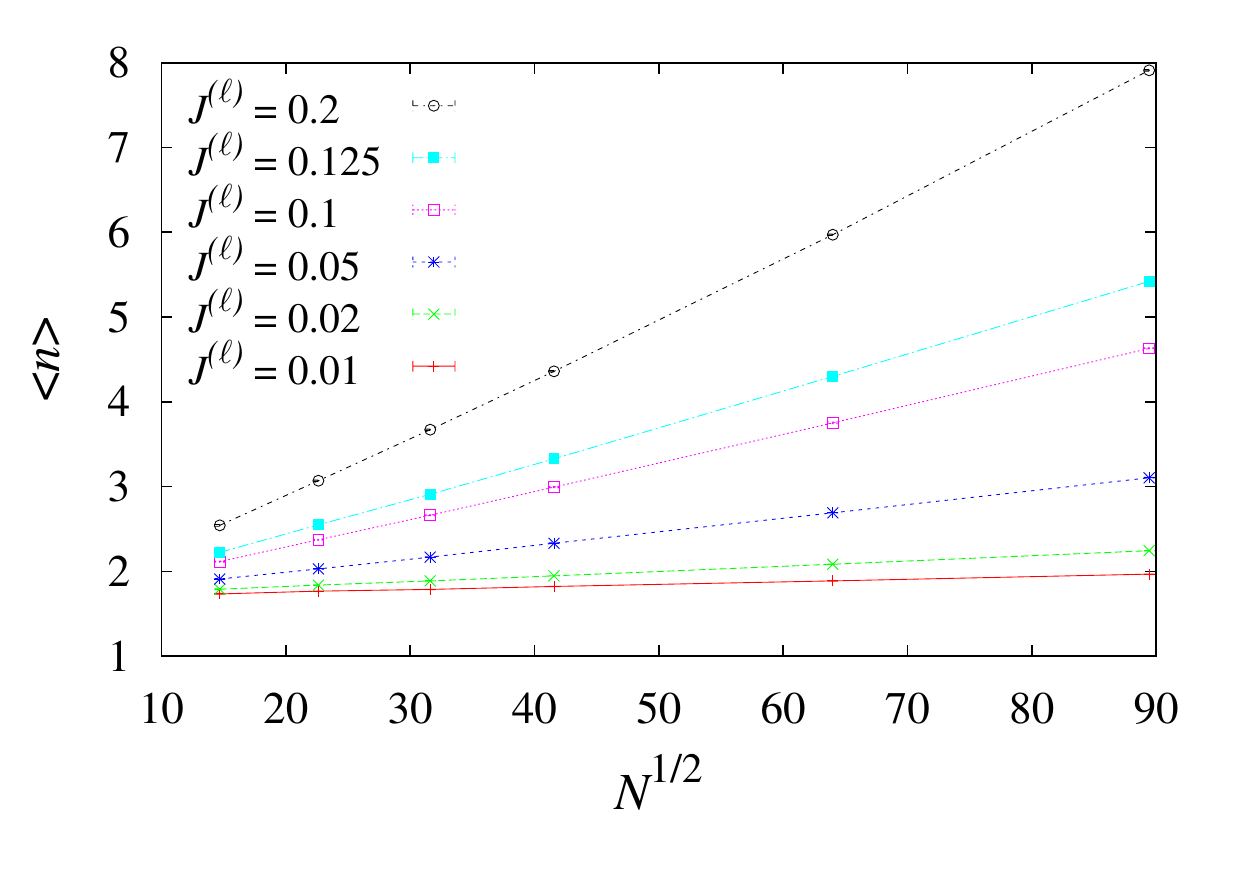}
 \caption[Scaling of $\mean{\Delta M}$ and $\mean{n}$ with mixed interactions]
	 {Scaling of the average values in the model that mixes short- and long-range interactions. 
	 The short-range coupling is kept fixed to ${{J}^{(s)}}=1$, while the amplitude ${{J}^{(\ell)}}$ is tuned in the region $0<{{J}^{(\ell)}}\leq {{J}^{(s)}}$.
	 \textbf{Left}: the average magnetization jump $\mean{\Delta M}$ scales clearly as $\sqrt{N}$.
	 \textbf{Right}: the average avalanche size follows the trend $\mean{n}\sim\sqrt{N}$ for small ${J}^{(\ell)}$.
	 }
 \label{fig:SKEA-means}
\end{figure}

The stability argument becomes then
\begin{align}
1 &\sim (N-z) \int_0^{J^{(\ell)}/\sqrt{N}}\rho(\lambda)d\lambda \sim\\
  &\sim \alpha N \int_0^{^{(\ell)}/\sqrt{N}} \lambda d\lambda\sim\\
  &\sim \alpha {{J}^{(\ell)}}^2\,,
\end{align}
so $\alpha = 1/{{J}^{(\ell)}}^2$ and 
\begin{equation}
\rho(\lambda) \propto \frac{\lambda}{{{J}^{(\ell)}}^2} \,.
\end{equation}
We can use again the argument for the scaling of the magnetization jump, $\langle\Delta M\rangle= \frac{\Delta M^{(tot)}}{\mean{n_\mathrm{av}}}$,
with this $\rho(\lambda)$. The average number of avalanches now scales as $\mean{n_\mathrm{av}}\sim\frac{1}{h_\MIN}\sim\frac{\sqrt{N}}{{J}^{(\ell)}}$,
so
\begin{equation}\label{eq:skea-scaling-dm}
\mean{\Delta M} \propto  {{J}^{(\ell)}}\sqrt{N}\,,
\end{equation}
so even in the presence a the smallest long-range interaction, as the system size grows the average magnetization jump in an avalanche diverges as $\sqrt{N}$,
as it is also confirmed numerically in figure \ref{fig:SKEA-means}, left.\index{numerical simulations}

As to the number of spins involved in the avalanche, we also find that it diverges, confirming the self-organized criticality of the model,
but this time with a different law than the SK model. In fact $\mean{n}\sim\frac{\Delta E_\mathrm{av}}{\Delta E_\mathrm{flip}}$. While $\Delta E_\mathrm{av}$
scales as ${J}^{(\ell)}\sqrt{N}$, the energy of a flip scales as $\Delta E_\mathrm{flip}\sim \sqrt{\frac{{{J}^{(\ell)}}^2}{N}+\frac{{{J}^{(s)}}^2}{z}}$.
The average number of spins taking part in an avalanche then scales like
\begin{align}
 \mean{n} &= \frac{\Delta E_\mathrm{av}}{\Delta E_\mathrm{flip}} \sim \\
	  &\sim \frac{{J}^{(\ell)}\sqrt{N}}{\frac{{J}^{(\ell)}}{\sqrt{N}}\sqrt{1+\frac{N{J^{(s)}}^2}{z{J^{(\ell)}}^2}}} \sim\\
	  &\sim \frac{{J}^{(\ell)}}{{J}^{(s)}}\sqrt{zN}\,.\label{eq:skea-scaling-n}
\end{align}
Numerical simulations, where we tune the amplitude ${{J}^{(\ell)}}$ keeping ${{J}^{(s)}}=1$ fixed, confirm this argument (figure \ref{fig:SKEA-means}, right).

\section{Dynamics}
After having given several conclusions on the self-organized criticality of the \ac{SK} based on scaling and stability arguments,
it is reasonable to ask oneself whether self-organized criticality purely a property of the visited states or the dynamics too
play an important role on the crackling.
In the following section we try to get some insight from what is happening to the system \emph{during} the avalanches.

\subsection{A non-trivial random walk}\index{random!walk!nunst@$\nunst$}
An avalanche starts when a first spin is destabilized, and it finishes when all the local stabilities are positive.
With the typical spin update, that we call greedy algorithm,\index{dynamics!greedy}
if there is more than one unstable spin, the least stable is updated first.
Calling $\nunst(t)$ \nomenclature[n....unstt]{$\nunst(t)$}{number of unstable spins after $t$ spin flips}
the number of unstable spins after $t$ spin flips, \nomenclature[t....7]{$t$}{In chapter \ref{chap:marginal}: number of elapsed spin flips in the avalanche}
this reads that the avalanche starts with $\nunst(1)=1$,
it performs a \ac{RW} in the space of $\nunst$, and it end with $\nunst(n)=0$.

The easiest guess for the dynamics is thus an unbiased \ac{RW}, \index{avalanche}
where for large avalanches $\mathcal{D}(n)$ would be the return probability\index{random!walk!return probability}
of a one-dimensional \ac{RW}.
\footnote{It would be exactly the return probability of the random walk if the avalanche started with $\nunst=0$.}
The return probability of a random walk is $P_{1d}\propto\frac{1}{\sqrt{t}}$ in $1d$ 
\nomenclature[P...1d]{$P_{1d}$}{return probability of a $1d$ RW}
\nomenclature[P...2d]{$P_{2d}$}{return probability of a $2d$ RW}
and $P_{2d}\propto\frac{1}{t\log{t}}$ in $2d$,
so the unbiased \ac{RW} scenario predicts $\rho=1/2$, \index{exponent!avalanche!rho@$\rho$}
that is different from the $\rho=1$ usually observed (recall figure \ref{fig:SK-aval}).

The \ac{RW} of $\nunst$ can be described through two equivalent auxiliary variables $E(\nunst(t))$ and $r(\nunst(t))$, 
that indicate 
\nomenclature[E...nunstt]{$E(\nunst)$}{indicator of the random walk bias}
\nomenclature[r....nunstt]{$r(\nunst)$}{indicator of the random walk bias}
the likeliness of the avalanche of shrinking or expanding:
\begin{align}
\nunst(t) &= \nunst(t-1) E(t-1)\,,\\[2ex]
\nunst(t) &= \nunst(t-1) + r(t-1)\,.
\end{align}
In an unbiased random walk $E(t) = 1 \,\forall t$ and $r(t) = 0 \,\forall t$. Random walks with constant $E<1$ ($r<0$) are attractive, meaning that there
cannot be extended avalanches, while if $E\gg1$ ($r\gg0$) the system is highly unstable and the avalanches never stop.

Since the number of triggered spins depends exclusively on the links between the flipping spin and its neighbor, which is a static property of the system, 
it is reasonable to assume - and more in a fully-connected spin glass where it makes no sense to talk of spatial domains - that $E$ and $r$ depend on 
$\nunst$ rather than on how long the avalanche lasted.

\begin{figure}[!t]
 \centering
 \includegraphics[width=0.485\columnwidth]{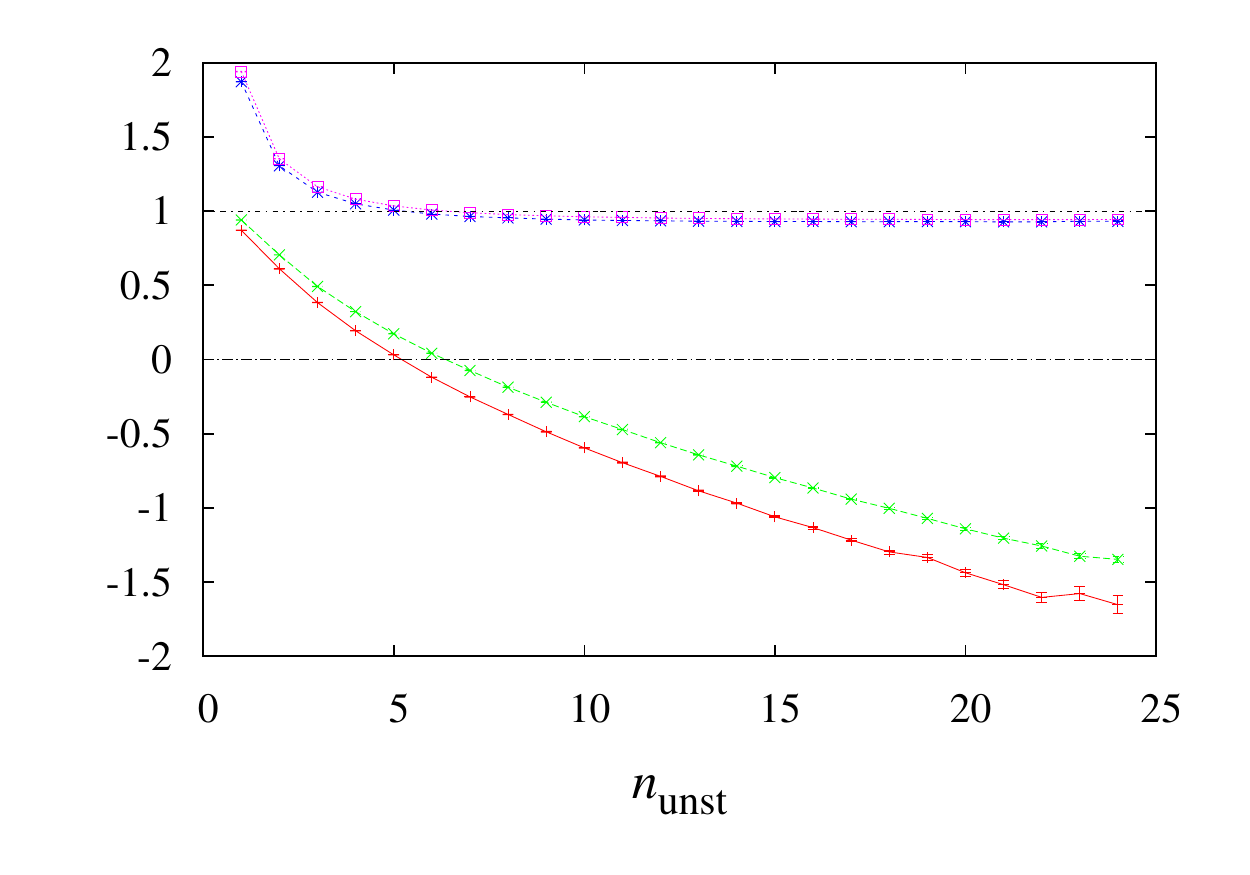}
 \includegraphics[width=0.485\columnwidth]{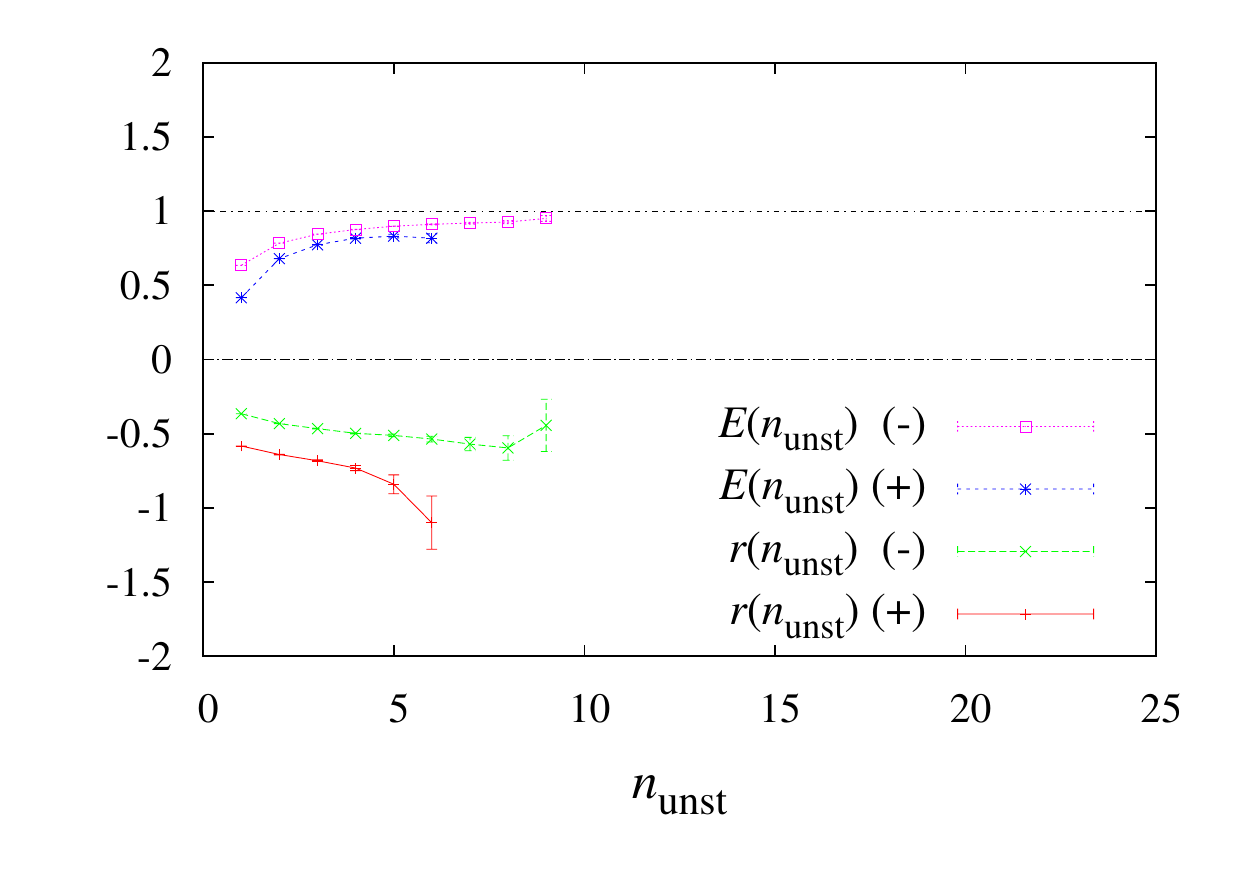}
  \caption[RW bias indicators $E$ and $r$]
	  {Indicators of the random walk bias $E$ and $r$ as a function of the number of unstable spins $\nunst$. 
	  The curves tagged with a $(+)$ indicate starting configuration with positive magnetization, those tagged
	  with $(-)$ indicate a negative magnetization. Details on the protocol are given in the main text.
	  The \textbf{left} plot shows data from the \ac{SK} model for $N=8000$. The \textbf{right} plot is from
	  the three-dimensional \ac{EA} model with $L=20$ ($N=L^3=8000$).
	  The two horizontal lines stress the values of the unbiased \ac{RW}, $E=1$ and $r=0$).}
\label{fig:rw-SKyEA}
\end{figure}

\begin{figure}[!t]
 \centering
 \includegraphics[width=0.5\columnwidth]{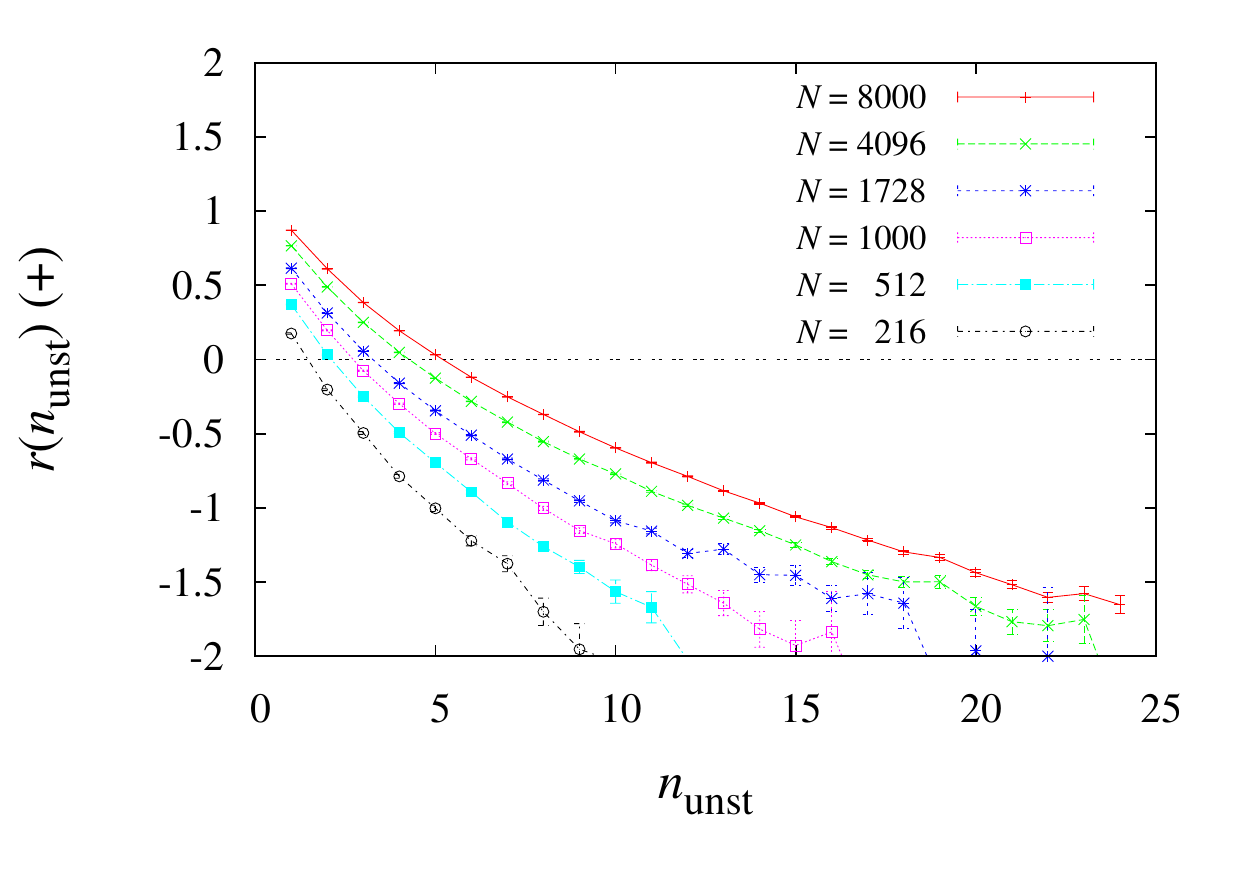}
  \caption[Finite-size effects in $r(\nunst)$]{We show $r(\nunst)$ for different lattice sizes, to stress that the point $\nunst^*$ where 
	  the curve crosses zero grows steadily with the system size. The data is from systems with positive magnetization initial conditions, $(+)$, 
	  and is qualitatively equivalent to opposite magnetization starting configurations, $(-)$.}
\label{fig:rw-L}
\end{figure}

In figure \ref{fig:rw-SKyEA} we show $E$ and $r$ for avalanches in the \ac{SK} and in the $3d$ \ac{EA} model. Both $E$ and $r$ have a marked dependency on
$\nunst$, disclosing non-trivial \acp{RW}. 
In the \ac{EA} model $E(\nunst)<1 \,\forall\nunst$, meaning that the dynamics is damped and the size of the avalanche can grow only because
of fluctuations. Mind that as $\nunst$ increases (due to ``lucky'' fluctuations), $E(\nunst)$ approaches 1, reflecting that the connectivity of the unstable
domain grows, so it becomes easier to destabilize another spin.
In the \ac{SK} model the situation is more interesting, since the dynamics is critical. Instead of 
$E(\nunst)=1 \,\forall\nunst$, that could be in principle a good ansatz for a marginal system, the avalanches
have a natural tendency to grow up to a size $\nunst^*$. \nomenclature[n....unststar]{$\nunst^*$}{$\nunst:\,E(\nunst)=1,r(\nunst)=0$}
For $\nunst>\nunst^*$, $E$ is slightly smaller than one, meaning that $\nunst^*$ is a preferred number of unstable spins. 
A size-independent $\nunst^*$ would entail that the scale invariance is only a low-resolution effect due to the fact that $E$ is smaller than one, but very close to it.
From figure \ref{fig:rw-L} we see that this is not the case: $\nunst^*$ grows as $\log(N)$.

What is clear is that the dynamics of the single spins are far from being independent, and those of the system as a whole are related on the amount of unstable spins. 
The evolutions and stabilities of the spins are correlated and there is some kind of non-trivial mechanism that keeps the system marginal during the avalanches.

\paragraph{Initial conditions}\index{avalanche!initial conditions}
It is legitimate to inquire whether different starting conditions play a pivotal role on the random walk.
In figure \ref{fig:rw-SKyEA} we compare two types of initial configuration. We start at zero field with either all spins up $(+)$ or all spins down $(-)$,
and we minimize the energy by aligning successively the most unstable spin to its local field until the system becomes stable (greedy algorithm). The two configurations
are totally equivalent, except that they have opposite remnant magnetization.
In figure \ref{fig:rw-SKyEA} we see that there is an appreciable difference between the two starting conditions,
The external field in this numerical experiment varies from 0 to 1.5, \index{numerical simulations}
that is, the data come from a large number of avalanches, $O(\sqrt{8000})$. If the information
on the initial state were lost within the first avalanche, the curves $(+)$ and $(-)$ should differ by the order of $1\%$.

\subsection{Changing the avalanche dynamics \label{sec:marginal-GRA}}
\index{greedy|see{greedy dynamics}}
\index{reluctant|see{reluctant dynamics}}\index{random!dynamics|see{aleatory dynamics}}
\index{dynamics!greedy}\index{dynamics!reluctant}\index{dynamics!aleatory}
A way to understand whether marginality is a property of the static configurations or it depends on the dynamics is to
validate it on different types of dynamics. We propose three types of single-spin-flip dynamics. The first is the one used until
now, that at each time step updates the most unstable of the spins. We call it \emph{greedy} dynamics (G)\acused{G}. The second type of dynamics
is inspired from \cite{parisi:03}, and updates the least unstable spin. This is the \emph{reluctant} algorithm (R)\acused{R}. 
It was shown in \cite{parisi:03} that minimizing the energy with R dynamics leads to inherent structures with much lower energy.
The third dynamics we test updates a random spin among those with $\lambda<0$. We call it \emph{random} dynamics (A)\acused{A}.\index{local!stability}
\footnote{We use an A, that stands for \emph{aleatory}, because the R of \emph{random} was already picked for the reluctant algorithm.}

\paragraph{Avalanche distributions}\index{local!stability!distribution}
\index{dynamics!greedy}\index{dynamics!reluctant}\index{dynamics!aleatory}
When switching to \ac{R} and \ac{A} avalanches, we remark no variation on the $\rho(\lambda)$, that for small $\lambda$ still grows linearly 
(only the amplitude changes), but we do see a difference in the exponents of the avalanche distributions. More specifically, for \ac{A} we see the
same exponents $\rho\approx1$ and $\tau\approx1$, but with \ac{R} the avalanches are significantly 
larger and have $\rho\approx1.25$, $\tau\approx1.4$.\index{exponent!avalanche!tau@$\tau$}\index{exponent!avalanche!rho@$\rho$}
In figure we show \ac{R} avalanches. 
\begin{figure}[!htb]
\centering
\includegraphics[width=0.485\columnwidth]{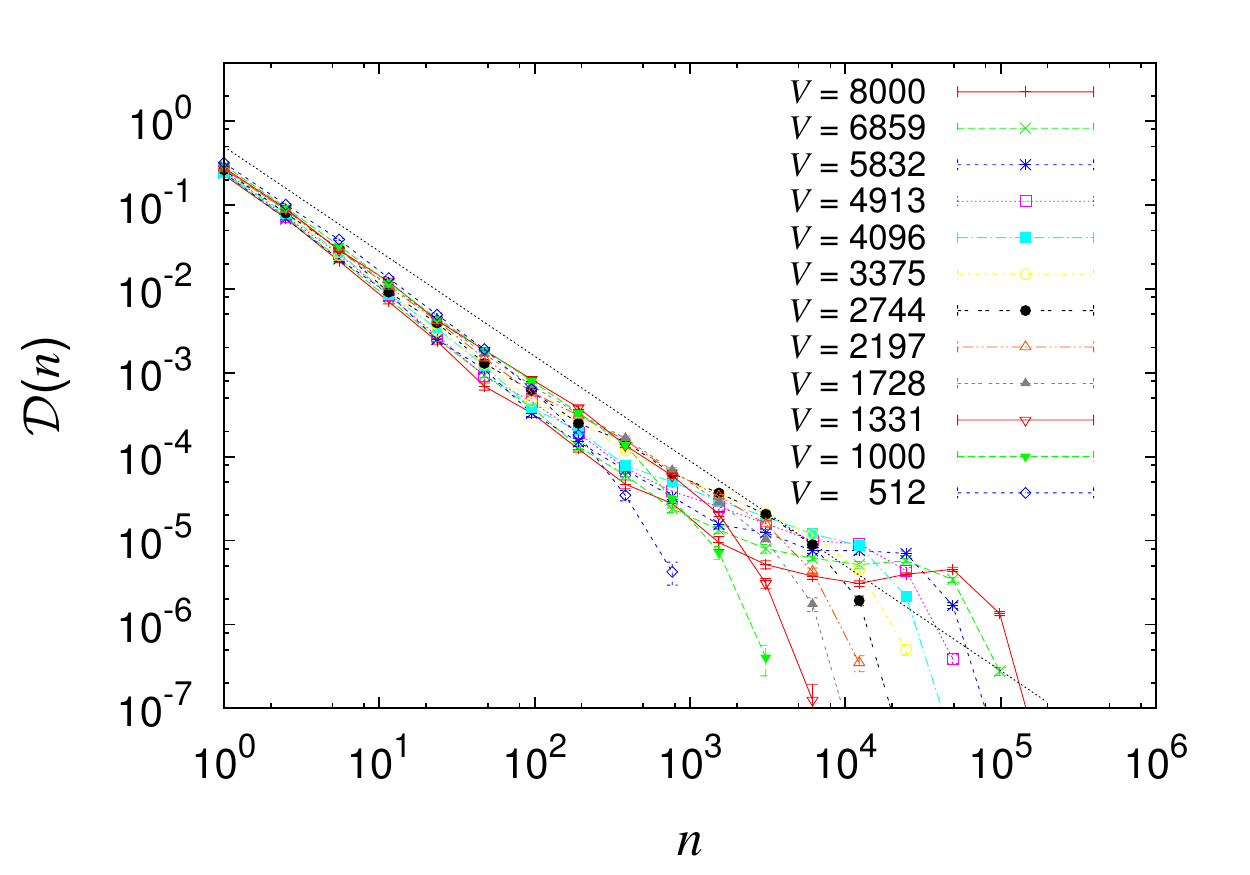}
\includegraphics[width=0.485\columnwidth]{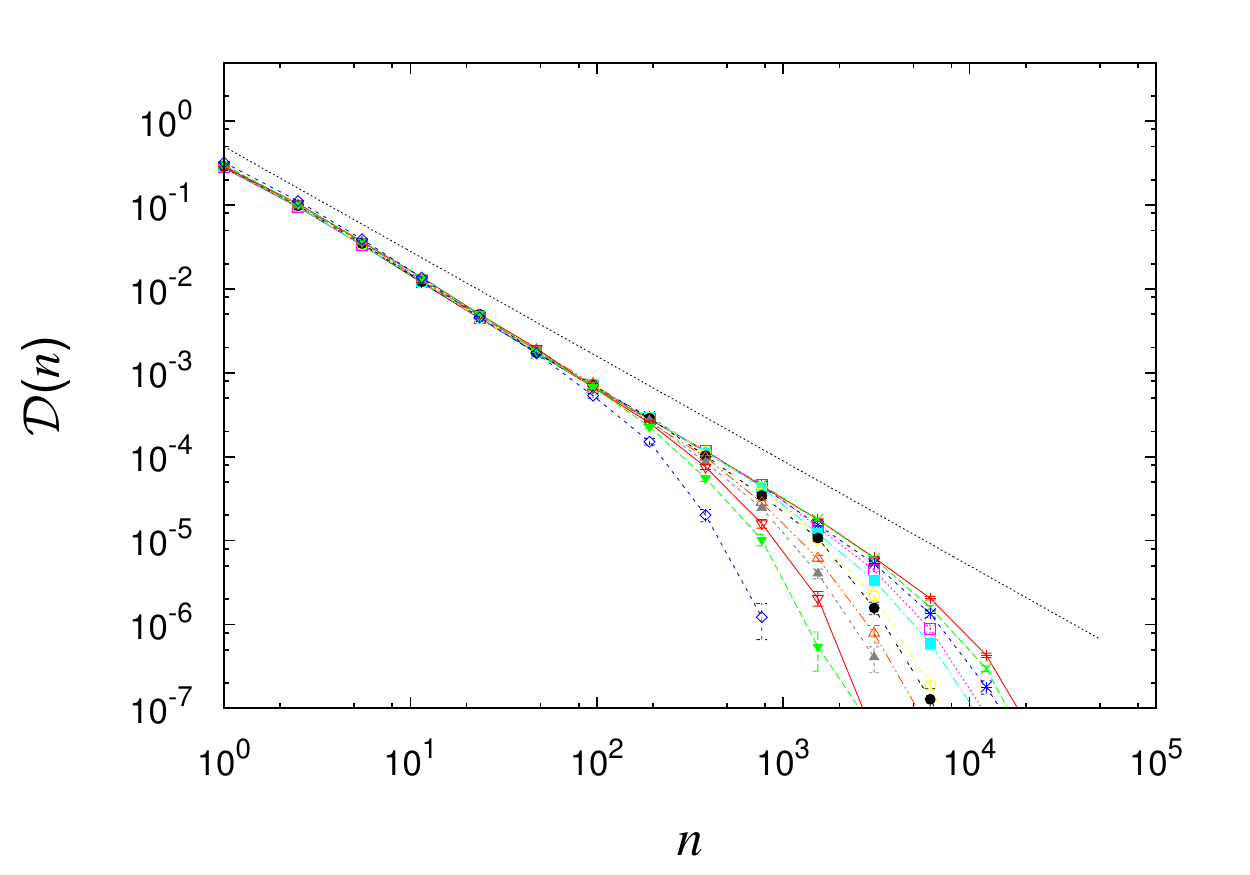}
\caption[$\mathcal{D}(n)$ in avalanches with reluctant dynamics]
	{Distribution of sizes $\mathcal{D}(n)$ in avalanches with reluctant dynamics. 
	\index{dynamics!greedy}\index{dynamics!reluctant}
	In the \textbf{left} plot the initial \ac{IS} is obtained with the \ac{G} algorithm, while on the \textbf{right} it is obtained
	with \ac{R} dynamics.\index{avalanche!initial conditions}
	The straight lines in both plots are reference curves $\propto n^{-1.25}$. Even though the finite-size behaviors are different
	depending on the starting configuration, the power law is the same, and it is different from \ac{G} dynamics.}
\label{fig:Dn-rel}
\end{figure}
The similarity between \ac{G} and \ac{A} can be attributed to the fact that the energy change in a spin flip is of the same order, 
$\Delta E_\mathrm{flip}\sim1/\sqrt{N}$, while \ac{R} dynamics implies that
the energy dissipated in a spin flip is smaller. Since the $\rho(\lambda)$ is all of order 1, the typical distance between the stabilities 
is of order $1/N$, so $\Delta E_\mathrm{flip}\sim 1/N$.
\footnote{The arguments of section \ref{sec:marginal-cutoffs} for the scaling of $\mean{\Delta M}$ and $\mean{n}$ apply also
to \ac{A} and \ac{R} dynamics. One obtains $\mean{\Delta M}\sim\sqrt{N}$ for both the dynamics, $\mean{n}\sim N$ for \ac{A}
and $\mean{n}\sim N^{3/2}$ for \ac{R} dynamics.
Numerical simulations seem compatible with these trends in the limit of very large systems.
}

\index{avalanche!initial conditions}
The data in figure \ref{fig:Dn-rel} was obtained by relaxing a totally up configuration, and once the initial \ac{IS} was found
we recorded the data of the avalanches until the overlap with the initial configuration became smaller than $Q=0.9$. This way we could grant
some dependence on the initial \ac{IS}, and compare avalanches that started with \ac{G} and \ac{R} inherent structures. We will use two letters
to identify the procedure we refer to: the first one refers to the initial \ac{IS}, the second to the avalanche dynamics, so for example
RG is a greedy avalanche starting from a reluctant \ac{IS}.
\nomenclature[G.G]{GG}{G inherent structure, G avalanche}
\nomenclature[G.R]{GR}{G inherent structure, R avalanche}
\nomenclature[R.R]{RR}{R inherent structure, R avalanche}
\nomenclature[R.G]{RG}{R inherent structure, G avalanche}
\nomenclature[G.A]{GA}{G inherent structure, A avalanche}
\nomenclature[R.A]{RA}{R inherent structure, A avalanche}

In figure  \ref{fig:Dn-rel} we compare GR and RR dynamics. Apparently, the exponent does not depend on the initial conditions, but the finite-size
effects do visibly. \index{dynamics!greedy}\index{dynamics!reluctant}
While RR avalanches display a power-law behavior with a finite-size cutoff, in GR one sees that with a probability that decreases with $N$ there
can be avalanches with a very large number of spin flips, arriving to $n>N$, that means that in average every spin flips more than once.
This suggests that \ac{G} inherent structures are in some way more unstable with respect to \ac{R} dynamics than \ac{R} inherent structures.

% \paragraph{Mean values}
% The arguments on the scaling of the mean values of section \ref{eq:marginal-cutoffs} can be applied also to \ac{R} and \ac{A} dynamics.
% For \ac{A} the arguments are the same as for \ac{G}, so the same scalings should apply. On the other side, \ac{R} dynamics implies that
% the energy dissipated in a spin flip is smaller. Since the $\rho(\lambda)$ is all of order 1, the typical distance between the stabilities 
% is of order $1/N$, so $\Delta E_\mathrm{flip}\sim 1/N$.
% Overall, these arguments imply that $\mean{\Delta M}$ scales the same with the three avalanches dynamics, while $\mean{n}$ should scale 
% as $N$ in \ac{G} and \ac{A} and as $N^{3/2}$ in \ac{R} dynamics.
% 
% \begin{figure}[!htb]
% \centering
% \includegraphics[width=0.485\columnwidth]{dm_vs_N_XX}
% \includegraphics[width=0.485\columnwidth]{n_vs_N_XX}
% \caption[]
% 	{}
% \label{fig:mean_vs_N_XX}
% \end{figure}

\begin{figure}[!htb]
\centering
\includegraphics[width=0.485\columnwidth]{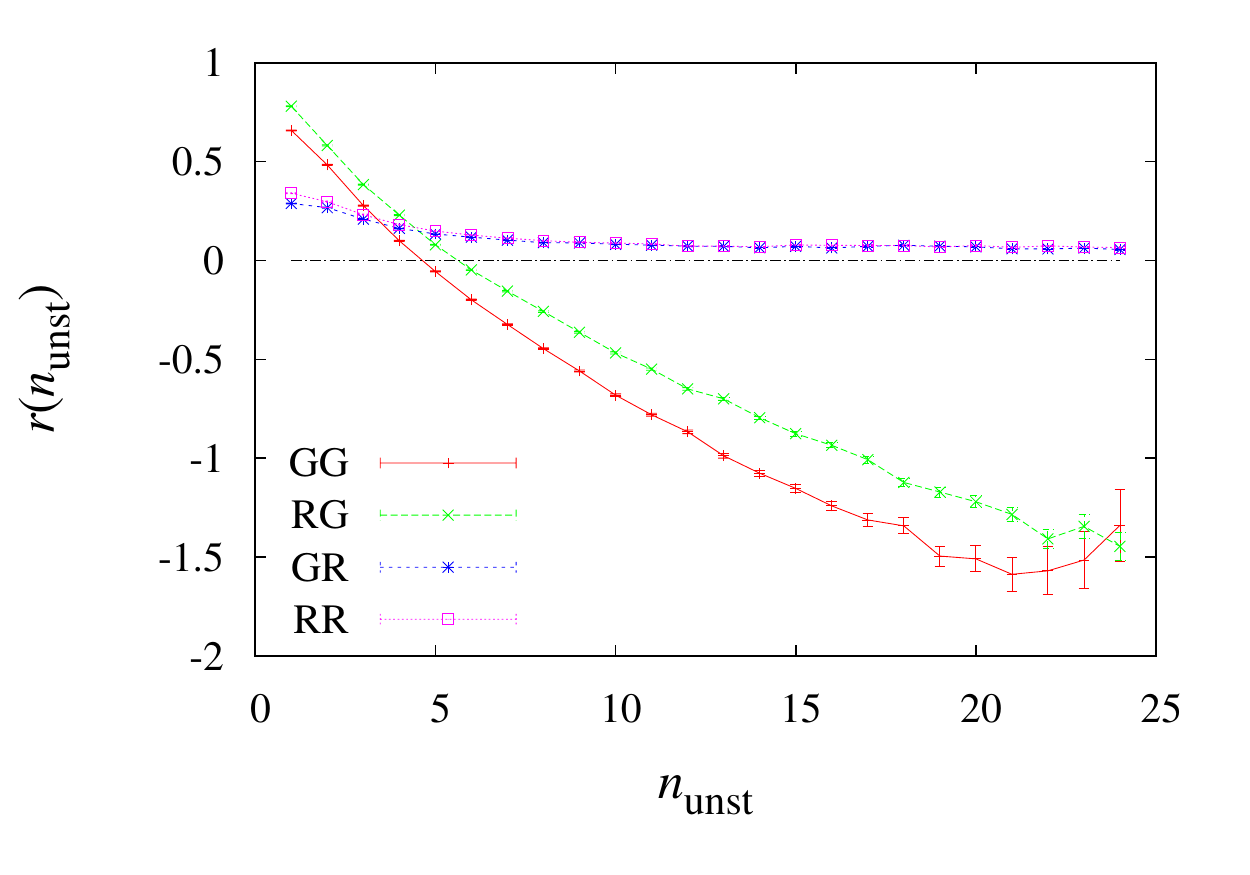}
\includegraphics[width=0.485\columnwidth]{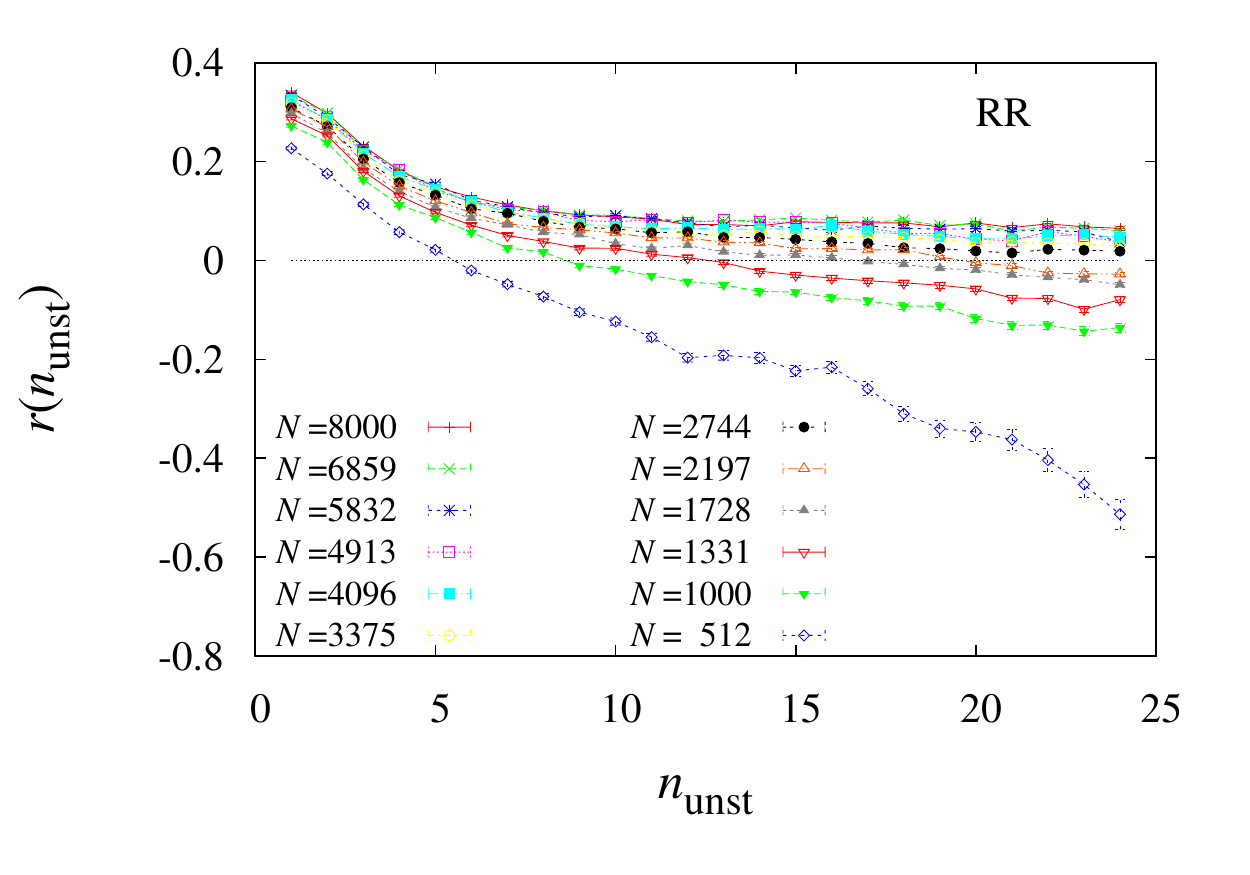}
\caption[RW bias indicator $r$ for reluctant avalanches]
	{\index{dynamics!reluctant}
	The \ac{RW} bias indicator $r$ for avalanches that start from the initial inherent structure (i.e. for $Q=0$).
	Although our data only extends to $\nunst=24$, the avalanches had also larger numbers of unstable spins.
	On the \textbf{left} we compare greedy and reluctant algorithms in $N=8000$ systems. 
	In the \textbf{right} figure we show RR data for different $N$. When the system is small, $r(\nunst)$ crosses zero at
	a finite $\nunst^*$, that grows with $N$. For $N\geq2744$ our data is not able to capture $\nunst^*$, though we still expect it to be large but finite (see also main text).
	In both plots the horizontal line stresses the unbiased value $r=0$.	
	}
\label{fig:r_V8000_Q0}
\end{figure}

\paragraph{Random walks}\index{random!walk!nunst@$\nunst$}
Seeing the avalanche as a \ac{RW} of the number of unstable spins, we see no remarkable dependency on the initial \ac{IS}, but we do notice a quite
different behavior between \ac{G} and \ac{R} avalanche dynamics (figure \ref{fig:r_V8000_Q0}, left). In the \ac{G} dynamics $r(\nunst)$
is initially positive (expansion of the avalanche preferred) becomes negative (shrinking preferred) at a finite $\nunst^*$, justifying 
avalanches of limited size. 
Differently, $r$ appears always positive in \ac{R} avalanches, indicating a tendency towards enlargement. If $r$ is always positive the avalanches can only stop due to large fluctuations 
or by saturation of the system (we have a trivial bound $\nunst<N$), that would mean that the dynamics is unstable. The power law behavior
of $\mathcal{D}(n)$ (figure \ref{fig:Dn-rel}) and the finite-size behavior of $r(\nunst)$ (figure \ref{fig:r_V8000_Q0}, right) induce to think 
that $\nunst^*$ is instead finite but large, and that its growth with the system size is significantly quicker than in \ac{G} avalanches.
\footnote{
In \ac{G} avalanches $\nunst^*$ grows logarithmically, $\nunst^*\sim \log(N)$. With \ac{R} dynamics we have little data because our measurements only go up to $\nunst=24$. 
We deduce a roughly linear scaling $\nunst^*\sim N$.
}
The different scaling of $\nunst^*$ between the \ac{G} and \ac{R} could be what leads to different exponents $\rho$ and $\tau$.
\index{exponent!avalanche!tau@$\tau$}\index{exponent!avalanche!rho@$\rho$}

\subsection{Fokker-Planck description\label{sec:fp}}\index{Fokker-Planck|(}\index{random!walk!lambda@$\lambda$}
Coming back to greedy dynamics, we will see now that the same exponents that we obtained through stability constraints arise spontaneously
from the dynamics of the avalanches in the \ac{SK} model.
Let us take in account the random walk of each local
stability in the space of the local stability space. The random walk starts when a stability becomes negative
because of an imposed external magnetic field, and it finishes when all the spins are stable again.

\index{local!stability|(}
The flipping of the spin $s_0$ changes its local stability from $\lambda_0$ to $\lambda_0'=-\lambda_0$. The stability
of all the other spins $s_\by$ in the system changes proportionally to their coupling with $s_0$,
\begin{equation}\label{eq:dinam-kick}
 \lambda_\by \to \lambda_\by' = \lambda_\by - 2 s_0 J_{0\by} s_\by\,.
\end{equation}
The stability changes have a random fluctuating part and a non-zero mean value
due to the correlations with $s_0$.  As it was similarly done in
Refs. \cite{eastham:06,horner:08}, this dynamics can be modelled with a
Fokker-Planck equation for the distribution of stabilities
$\rho(\lambda)$,\index{local!stability!distribution}
\vspace{2mm}
\begin{equation}
\label{eq:fp}
\partial_t\rho(\lambda,t)=-\partial_{\lambda}\,\left[v(\lambda,t)-\partial_{\lambda}D(\lambda,t)\right]\rho(\lambda,t)\\-\delta(\lambda-\lambda_0(t))+\delta(\lambda+\lambda_0(t)),\\
\end{equation}
\vspace{2mm}
where now the ``time'' $t$ %\nomenclature[t....7]{$t$}{In section \ref{sec:fp}: number of flips per spin that took place during the avalanche}
is the number of flips per spin that took place during 
the avalanche and the two delta functions indicate the flipping of $s_0$.\index{correlation!stability}
The drift term $v(\lambda,t)\equiv-2N\mean{s_0J_{0\by}s_\by}_{\lambda_\by=\lambda} = N C(\lambda,t)$ is the average 
positive kick that a spin with stability $\lambda$ receives [equation \eqref{eq:dinam-kick}].
\nomenclature[v....lambdat]{$v(\lambda,t)$}{drift term}\nomenclature[D...lambdat]{$D(\lambda,t)$}{diffusion constant}
The diffusion constant $D(\lambda,t) \equiv 2N\mean{J_{0\by}^2} = 2$ is the mean square of the kicks.
The dynamics have a non-trivial thermodynamic limit only if $v\sim O(1)$, meaning that $\mean{s_0J_{0\by}s_\by}\sim1/N$.
This conveys that the exponent $\delta$ from equation \eqref{eq:C-lambda-trend} must be equal to 1.

\begin{figure}[!htb]
\centering
\includegraphics[width=0.4\textwidth]{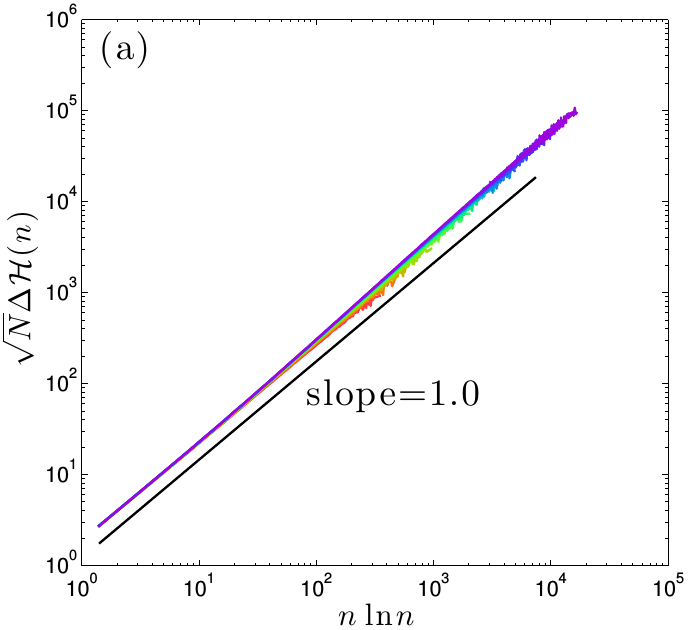}
\caption[Dissipated energy in an avalanche.]
	{The average dissipated energy $\Delta \mathcal{H}$ in avalanches of size $n$ scales as $\Delta \mathcal{H}\sim n\ln n/\sqrt{N}$. 
	$-\Delta \mathcal{H}/n$ is a measure of the typical value of the stability of most unstable spins, $\lambda_0(n)$. Thus, in the thermodynamic 
	limit,  $\lambda_0\sim \ln n/\sqrt{N} \ll 1$ even for very large avalanches.}
\label{propt}
\end{figure}

As $N\to\infty$, the lowest stability approches zero $\lambda_0(t)\to0$. We already saw, in fact, that \index{exponent!pseudogap!theta@$\theta$}
in a driving experiment with a finite field change $\Delta h$, the number of avalanches scales as $n_\mathrm{av}\sim1/h_\MIN\sim N^{1/(1+\theta)}$.
Each avalanche contains on average $\mean{n}\sim \int n\mathcal{D}(n)dn\sim N^{(2-\rho)\sigma}$ 
flip events [recall eq. \eqref{eq:power-laws}], so \index{exponent!avalanche!rho@$\rho$}
the total number of flips along the hysteresis curve is $n_\mathrm{av}\mean{n}\sim N^{(2-\rho)\sigma+1/(1+\theta)}$,\index{exponent!scaling!sigma@$\sigma$}
that is reasonably larger than $N$. A diverging number of avalanches implies that the energy dissipation in each avalanche has
to be subextensive, ruling out strongly unstable configurations with an extensive number of spins with negative 
stability $|\lambda|=O(1)$. So, as we confirm numerically \index{numerical simulations}in figure \ref{propt}, the smallest local stability
must tend to zero in the thermodynamic limit.
This observation lets us replace the delta functions in equation \ref{eq:fp} with a reflecting boundary condition at $\lambda=0$,
\begin{equation}
\label{eq:refb}
\left. \left[v(\lambda,t)-\partial_{\lambda}D(\lambda,t)\right]\rho(\lambda,t)\right|_{\lambda=0}=0\,.
\end{equation}

Since along the hysteresis loop spins flip a large amount of times, in a finite interval we have a diverging number of time steps.
At very large times a steady state must be reached. In such conditions the flux of spins must vanish everywhere, so the steady state
drift is
\begin{equation}
\label{eq:steady}
\nomenclature[v....sslambda]{$v_{\rm ss}(\lambda)$}{steady state drift term}
v_{\rm ss}(\lambda)=D\partial_{\lambda}\rho_{\rm ss}(\lambda)/\rho_{\rm ss}(\lambda)\rightarrow 2\theta/\lambda\,,%=2/\lambda 
\end{equation}
where we assumed that the steady-state stability distribution follows \eqref{eq:pseudogap}.
This implies that $\gamma=1$ in equation \eqref{eq:C-lambda-trend}.

\subsubsection{Arisal of correlations}\index{random!walk!lambda@$\lambda$}\index{correlation!arisal}
\index{local!stability!distribution}\index{exponent!correlation!delta@$\delta$}\index{exponent!correlation!gamma@$\gamma$}
We will now argue that the correlations of equation \eqref{eq:C-lambda-trend} (with $\gamma=\delta=1$)
arise naturally in the dynamics through the shifts of the local stabilities caused by the spin flips.

Let us define with $C_f(\lambda)$ and $C_f'(\lambda)$ 
\nomenclature[C...flambda]{$C_f(\lambda),C_f'(\lambda)$}{correlations between the spin $s_0$ and the spins with
local stability $\lambda$ before and after the flipping event}
the correlations between the spin $s_0$ and the spins with
local stability $\lambda$ \emph{before} and \emph{after} the flipping event. After $s_0$ flips, the stability
change is $\lambda_\bx'=\lambda_\bx+x_\bx$, where $x_\bx=-2s_0J_{0\bx}s_\bx$.
\nomenclature[lambda....primo]{$\lambda'$}{local stability after the flip}
The correlation $C_f'(\lambda)$ is an average over all the spins whose stability, after the flip, is $\lambda'$,
\begin{align}
C_f'(\lambda)&=\frac{1}{\rho'(\lambda)}\int\rho(\lambda-x)(-x)f_{\lambda-x}(x) dx,\\
\rho'(\lambda)&=\int\rho(\lambda-x)f_{\lambda-x}(x) dx.
\end{align}
$f_{\lambda}(x)$ is the Gaussian distribution of kicks $x$ given to spins of stability 
\nomenclature[f....lambdax]{$f_{\lambda}(x)$}{distribution of kicks $x$ given to spins of stability $\lambda$}
$\lambda$: $f_{\lambda}(x)=\exp\left[-\frac{(x-C_f(\lambda))^2}{4D/N}\right]/{\sqrt{4\pi D/N}}$. 
%$D=\frac{1}{2}\langle X^2\rangle=2$.
%
%To calculate the integral, notice that the Gaussian distribution $f_{\lambda}(x)$ is narrow, 
In the integrands we expand $\rho(\lambda-x)$ and $C_f(\lambda-x)$ for small $x$ and keep terms of order $1/N$, 
which yields %, $\rho(\lambda-x)=\rho(\lambda)-x\partial_{\lambda}\rho(\lambda)+o(x^2)$, 
%\[
%C_f(\lambda)=\frac{1}{\rho(\lambda)}\int\left(-x\rho(\lambda)+x^2\frac{\partial}{\partial\lambda}\rho(\lambda)\right)f_{\lambda-x}(x)\rd x
%\]
%$C_f(\lambda-x)=C_f(\lambda)-\partial_{\lambda}C_f(\lambda)x+o(x^2)$. Keep the terms to the order of $1/N$,  %but $C'\sim 1/N\lambda^2$ is negligible compare to 1 when $\lambda\gg1/\sqrt{N}$,
\begin{subequations}
\begin{align}
C_f'(\lambda)&=-C_f(\lambda)+2\frac{D}{N}\frac{\partial_{\lambda}\rho(\lambda)}{\rho(\lambda)},
\label{selfa}\\
\rho'(\lambda)&=\rho(\lambda)-\partial_{\lambda}\left[ C_f(\lambda)\rho(\lambda)-\frac{D}{N}\partial_{\lambda}\rho(\lambda)\right]. \label{selfb}
\end{align}
\end{subequations}
Thus, even if correlations are initially absent, $C_f(\lambda)=0$,
they arise spontaneously, $C_f'(\lambda)=2D\partial_\lambda\rho(\lambda)/N\rho(\lambda)$. 

\index{exponent!pseudogap!theta@$\theta$|(}
In  the steady state, $\rho'_{\rm ss}=\rho_{\rm ss}$, \nomenclature[rho....ss]{$\rho_{\rm ss}$}{steady state local stability distribution}
and equation (\ref{selfb}) implies the vanishing of the spin flux, that is, equation (\ref{eq:steady}) with $v=NC_f$.
%$C_f(\lambda)=D\partial_{\lambda}\rho(\lambda)/N\rho(\lambda)$. 
Plugged into equation (\ref{selfa}), we obtain that the correlations are steady, too, %$C'_f= C_f =v_{\rm ss}/N= 2/(N\lambda)$ 
\begin{equation}
\label{cc}
C'_f(\lambda)=C_f(\lambda)=\frac{v_{\rm ss}(\lambda)}{N}
%\frac{D}{N}\frac{\partial_{\lambda}\rho(\lambda)}{\rho(\lambda)}=\frac{D\theta}{N\lambda}
=\frac{2\theta}{N\lambda}.
\end{equation}
These correlations are expected once the quasi-statically driven dynamics reaches a statistically steady regime, and thus should be present both
during avalanches and in the locally stable states reached at their end. 
%In a stable state obtained when an avalanche stop along the hysteresis curve, the correlation $C(\lambda)$ between the softest spins must therefore take the value prescribed by Eq.(\ref{cc}).

Interestingly, equation (\ref{cc}) implies that all the bounds of equations (\ref{eq:sca-1},\ref{eq:sca-2}) are saturated if the first one is, i.e., if $\theta=1$. 
It is intriguing that the present Fokker-Planck description of the dynamics does not pin $\theta$, as according to equations (\ref{eq:steady}, \ref{cc}) 
any value of $\theta$ is acceptable for stationary states. However, additional considerations on the applicability of the Fokker-Planck 
description discard the cases $\theta>1$ and $\theta<1$. 

% { Those are related to the interesting fact that  
% that Eqs.~(\ref{fp},~\ref{refb},~\ref{steady}) with $\theta=1$ are equivalent to the Fokker-Planck equation for the radial 
% component of unbiased diffusion in $d=2$} % - the marginal dimension with respect to the return to the origin} 
% (as derived in Supplemental Material, Sec.~D), whose statistics is well known \cite{redner01,Bray13}. We can use this 
% analogy to predict $F(n)$, the number of times an {initially soft} spin flips in an avalanche of size $n$. Indeed, a 
% discrete random walker starting at the origin will visit that point $\ln(t)$ times after $t$ steps in two dimensions, and 
% thus $F(n)\sim\ln(n)$, as supported by Fig.~\ref{propt}(b). Similarly we expect times between successive flips of a given 
% spin to be distributed as $P(\delta t) \sim 1/( \delta t [\ln(\delta t)]^2)$.

\paragraph{Excluding $\theta<1$}  Our Fokker-Planck description only applies beyond the discretization scale of the kicks due to 
flipping spins, which are of order $J\sim 1/\sqrt{N}$. 
In particular, from its definition, $C(\lambda)$ must be bounded by $1/\sqrt{N}$. Taking this into account, equation (\ref{eq:steady}) 
should be modified to:
\begin{equation}
\label{003}
v_{\rm ss}(\lambda)\approx \min\{D\partial_{\lambda}\rho_{\rm ss}(\lambda)/\rho_{\rm ss}(\lambda)\sim 1/\lambda, \sqrt{N}\}.
\end{equation}
This modification has no effect when $\theta\geq 1$, since in that case $\lambda_{\min}\sim N^{-1/(1+\theta)}\geq 1/\sqrt{N}$. 
In contrast, pseudo-gaps with $\theta<1$ have $\lambda_{\min}\ll 1/\sqrt{N}$. To maintain such a pseudogap in a stationary 
state, one would require correlations much larger than what the discreteness of the model allows. Pseudogaps with $\theta<1$ 
are thus not admissible solutions of equations (\ref{eq:fp},~\ref{003}).

%Dynamically, the pseudo-gap exponent can only be the marginal one, $\theta=1$, in the SK model. Based on the Fokker 
% Planck equation and the stationarity, we have shown that the correlation, $C(\lambda)=2\theta/N\lambda$. First, 
% the correlation can not be greater than the typical coupling among spins, $J_{\rm typ}\sim1/\sqrt{N}$. However, 
% assuming a pseudo-gap profile with $\theta<1$ is achieved in stationary, there would be $N^{(1-\theta)/2}$, diverging 
% number of spins in the range $\lambda\lesssim1/\sqrt{N}$ where the drift needs to be much stronger than $J_{\rm typ}$ 
% to reinforce the pseudo-gap profile. The contradiction implies that if a stable state with $\theta<1$ is achieved, 
% a small perturbation immediately drives a diverging number of spins to be unstable, and $\theta\geq1$ when approaching 
% the steady state. 
%

\paragraph{Excluding $\theta>1$}
In this case, $\lambda_{\min}\gg 1/\sqrt{N}\sim J$. Thus when one spin flips, the second least stable spin will not flip in 
general, and avalanches are typically of size unity \cite{mueller:15}. It can easily be shown that in that case
the number of flips per spin along the loop would be small (in fact it would even vanish in the 
thermodynamic limit, which is clearly impossible). In terms of our Fokker-Planck description, the motion of 
the spin stabilities due to other flips would be small in comparison with the motion of the stabilities inbetween 
avalanches, due to changes of the magnetic field. Making the crude assumption that the magnetization 
is random for any $\lambda$, the change of external magnetic field leads to an additional diffusion term in the 
Fokker-Planck equation:\index{local!stability!distribution}
\begin{equation}
 \label{fpwithd}
\partial_t\rho(\lambda,t)=-\partial_{\lambda}(v-D\partial_{\lambda})\rho(\lambda,t)+D_h\partial_{\lambda}^2\rho(\lambda,t),
\end{equation}
where the term $D_h$ is related to the typical field increment $h_{\rm min}\sim \lambda_{\rm min}$ required 
to trigger an avalanche. Indeed $D_{h}\sim Nh_{\rm min}^2\sim N^{(\theta-1)/(\theta+1)}\gg D\sim 1$. Under these 
circumstances, equation (\ref{eq:steady}) does not hold. The dynamics would be a simple diffusion with reflecting boundary, whose 
only stationary solution corresponds to $\theta=0$, violating our hypothesis $\theta>1$.  
Thus the last term of equation (\ref{fpwithd}) provides a restoring force toward 
dominated dynamics flattens the distribution. As soon as the pseudo-gap is filled up to $\theta=1$, this diffusion 
contribution becomes sub-dominant and the dynamics is dominated by the transient dynamics concentrated in the main text. 
In stationary conditions, a typical pseudo-gap profile must thus converge to $\theta=1$.
\index{Fokker-Planck|)}

\section{Overview}\index{local!stability!distribution}
The \ac{SK} model presents \acf{SOC} in its whole hysteresis loop. That is, the external field $h$
triggers power-law distributed avalanches that span the entire system. This \ac{SOC} is strictly related to marginal stability,
since for small $\lambda$ the distribution of the local stabilities goes as $\rho(\lambda)\propto\lambda^\theta$. Through 
stability arguments we showed that to have crackling responses $\theta=1$ is needed. We extended these stability arguments to 
multiple spins, remarking that the soft spins are in average frustrated \index{frustration} with each other (the energy along their links is
not minimized): There is a correlation function $C(\lambda)$ that scales inversely with the stability $\lambda$.

We then related the averages $\mean{\Delta M}$ and $\mean{n}$ to the cutoffs of the avalanches. In order to have \ac{SOC}, the 
cutoffs need to diverge when $N\to\infty$. With scaling arguments we showed that the 
\ac{SOC} of the \ac{SK} model vanishes when one considers models with a finite
number of neighbors, as it is also confirmed by numerical simulations.
Through a model that mixes short- and long-range interactions, we showed that fully-connected interactions are a relevant 
perturbation to the short-range Hamiltonian, so
the presence of long-range interactions is strictly necessary to have \ac{SOC} in the system, independently of the presence or
not of short-range interactions, no matter their amplitude. Yet, even though the long-range interactions grant avalanches that extend
over all the system, the scaling of the avalanche sizes cutoffs is different depending on the presence of short-range interactions.

\index{random!walk!nunst@$\nunst$}
We also studied the crackling in the SK model from the point of view of the dynamics. An avalanche can be seen as a discrete \acf{RW}
of the number of unstable spins, $\nunst$. The end of the avalanche corresponds with the number of time steps that it takes the \ac{RW}
to return to zero. In critical dynamics, these \acp{RW} are non-trivial, have a preferred number of unstable spins, $\nunst^*$. For $\nunst<\nunst^*$ the 
avalanches tendentially grow, for $\nunst>\nunst^*$ they shrink, suggesting that during the avalanche there is some type of correlation between
spins that keeps the system critical. A further extensive study of the relation between $\nunst^*$, the correlations $C(\lambda)$ and the size
of the avalanches can be a key factor for the understanding of \ac{SOC}.

To figure out how much of the crackling behavior is related to the type of dynamics one chooses, and how much is more universal, we
analyzed different kinds of single-spin flip algorithm. We identified a variation in the exponents of the avalanche distributions, but more fundamental
features as the pseudogap exponent $\theta$ stay the same.

\index{random!walk!lambda@$\lambda$}
Finally, through a modelization with a \ac{RW} in the space of the spin stabilities $\lambda$, 
we found that it is the dynamics itself that, because of a strong correlation among the softest spins, 
leads the system to a marginal state with a pseudogap. \index{Fokker-Planck}
With a Fokker-Planck description of the dynamics we explained the appearance of both the 
pseudogap and the singular correlation $C(\lambda)$.
\index{spin!Ising|)}\index{spin glass!Sherrington-Kirkpatrick|)}
\index{exponent!pseudogap!theta@$\theta$|)}\index{local!stability|)}
\index{self-organized criticality|)}
%FINITO
 \chapter{Soft modes and localization in spin glasses\label{chap:hsgrf}}
\index{spin!Heisenberg|(}\index{glass!structural}\index{boson peak|(}\index{supercooled liquid}
\index{metastable state@metastable state|seealso{inherent structure}}\index{metastable state@metastable state|seealso{local stability}}
\index{modes!soft|(}\index{modes!harmonic|(}
More than 40 years ago, it became clear that supercooled liquids and amorphous
solids exhibit an excess of low-energy excitations, compared with their
crystalline counterparts \cite{phillips:81}. This excess was evinced, for
instance, from anomalies in the specific heat at low temperatures (below\index{specific heat}
10K). A number of scattering techniques such as Raman, neutron \cite{buchenau:84} and, more
recently, inelastic X-ray scattering \cite{sette:98}, have shown that these
\index{scattering!neutron}\index{scattering!Raman}\index{scattering!inelastic X-ray}
excitations are of vibrational nature, and correspond to wave vectors of a few
nm$^{-1}$ and frequencies of few mK (see e.g. \cite{monaco:09} and references
therein). The corresponding vibrational density of states $g(\omega)$ \nomenclature[g....omega]{$g(\omega)$}{density of states}
displays \index{density of states}
an excess of modes, respect to the conventional Debye behavior $g(\omega)\propto\omega^{d-1}$ ($\omega^2$ in the three-dimensional case treated herein), \index{Debye}
called \emph{boson peak}.
Despite the shape of the $g(\omega)$ depends on numerous factors, such as the
considered material, the temperature, the thermal history, etc., the presence
of the boson peak is a universal feature \cite{buchenau:84,malinovsky:91}.\index{universality}

The starting point for an analysis of vibrational excitations is 
the harmonic approximation around stable or metastable states as, 
for example, this way many low temperature properties of solids can be
calculated analytically \cite{huang:87}. 

Also in liquid systems one encounters the same phemonenology. The density of
states in liquids was extensively studied to describe their dynamics, since
for small enough times one can characterize them through independent simple
harmonic motions (instantaneous normal modes) \cite{wu:92,keyes:94,wan:94}.\index{modes!instantaneous normal}
In supercooled liquids the dynamics is so damped that it is dominated by the\index{supercooled liquid}
underlying energy landscape \cite{cavagna:09}, and it becomes natural to focus\index{energy!landscape}
the attention on the harmonic modes of the \acfp{IS}, the local minima of the\index{local!minimum|seealso{inherent structure}}
energy that can be obtained by quickly relaxing the system, to zero
temperature, obtaining metastable configurations called inherent structures\index{inherent structure}
\cite{stillinger:95,monaco:09b}. These metastable states are likely to play an
important role both in driving the sluggish dynamics of these glassy systems
\cite{grigera:03}, and in their thermodynamic properties as the temperature\index{jamming}
vanishes or the system becomes jammed \cite{xu:10}.

Two main approaches are used to explain the presence of the boson peak,
attributing it to the presence of many metastable states.  

\index{localization}\index{two-level system}
On one side, there should be a very large number of localized excitations due
to the quantum tunneling between very similar states. The system can bounce
from one state to the other with very little energy exchange. The couples of
states described through this phenomenological approach are called two-level
systems \cite{anderson:72,phillips:72,phillips:87}. Although their precise
nature has not been clarified, their presence is experimentally detectable
\cite{lisenfeld:15}.

The second cause of an excess of soft modes is motivated by the presence of
marginally stable states, \index{marginal stability} that display infinitely soft modes.  
This excess of soft modes is highly universal among strongly
disordered mean field models \cite{mezard:87}. Indeed,
by means of
replica calculations, it has been recently shown that mean field supercooled \index{replica symmetry breaking}
liquids exhibit a transition to a full \ac{RSB} phase at high enough pressure
\cite{charbonneau:14}. Full \ac{RSB} implies a complex energy landscape with a
hierarchical structure of states and a large amount of degenerate minima
separated by small energy barriers \cite{mezard:84,charbonneau:14}.  These
energy barriers can be infinitely small, along with the smallest harmonic
excitations, meaning that the system is \index{marginal stability} marginally
stable. 

Besides to the shape of the energy landscape, marginal stability is also\index{marginal stability}
caused by isostaticity \cite{wyart:12}, the condition of having as many\index{jamming}\index{isostaticity}
degrees of freedom as independent constraints, that arises at jamming
\cite{ohern:03}. The strong universality of those features in continuous\index{universality}
constraint satisfaction problems suggests that they are a key ingredient for
the understanding of the glass and the jamming transition\index{glass!transition}
\cite{franz:15,franz:15b}.

A main difference between the two scenarios is that the two-level system\index{two-level system}
picture requires the presence of strongly localized states, \index{localization}
whereas the \index{marginal stability}
marginal stability is recovered through calculations in infinite dimensions
where localization cannot play a crucial role, but a \ac{RSB} transition is \index{replica symmetry breaking}
needed.  Furthermore, the two-level system descends from a quantum description
and requires taking into account anharmonic effects, whereas the boson peak\index{anharmonicity}
predicted by \ac{RSB} theories is classical, and can be identified at the
harmonic level. Here, we somehow reconcile the two approaches by identifying
two-level systems from a purely classical and harmonic starting point.

Even though many of the tools used to explain the boson peak descend from spin\index{spin glass!theory}
glass theory, the investigation of small harmonic excitations of the
metastable states has remained relegated to the field of structural glasses.\index{glass!structural}\index{metastable state}
On one hand because in \acp{SG} no ``crystal phase'' can be reached by cooling
the system slow enough, on another, perhaps, because the two most studied
\ac{SG} models are the \ac{EA} and the \ac{SK} model, both with Ising spins,
that are discrete. In the Ising \ac{SG} the aforementioned phenomena are
difficult to study.  When the passage from paramagnetic to \ac{SG} phase is\index{spin glass!transition}
very quick, while in structural glasses there is a large range of temperatures
in the disordered phase, where the dynamics is overdamped. Furthermore, it is
not straightforward to study soft excitations in a system where the smallest
excitation is bounded by its discrete nature.

Still, as we saw in chapter \ref{chap:hsgm}, many types of \ac{SG} model with
continuous degrees of freedom are easy to define. Among those, the Heisenberg
model \eqref{eq:H-anderson}, where the spins are unitary vectors with $m=3$
components, is an epitome of the spin glass, as it is the first proposed
\ac{SG} model.  Harmonic modes can be easily studied in this model, though due
to the $O(3)$ symmetry \index{symmetry!O(3)@$O(3)$}
of the Hamiltonian, the system exhibits an excess of
trivial low-frequency modes (Goldstone modes \index{modes!Goldstone}
and spin waves) \index{spin!wave}
that make this
type of analysis less clear. We can decide, thus, to add a random magnetic\index{random!field}
field to the Heisenberg Hamiltonian to wipe out the symmetries and the soft
modes they carry, keeping only those related to marginal stability. A similar\index{marginal stability}
procedure of symmetry removal has been carried through in glass-forming
liquids, by pinning a certain fraction of particles \cite{kob:12,cammarota:13}.\index{pinning}
In those references it was shown that the glass transition survives the
pinning. Hence from the above considerations on marginal stability
\cite{mezard:87,franz:15b} we expect as well a boson peak in pinned systems.

We propose ourselves to extend these considerations to a finite-dimensional
system, the Heisenberg \ac{SG} in a random magnetic field.  This lets us
verify the extent of the universality of these phenomena. \index{universality}
On one side by
checking if the soft modes are present with a similar phenomenology on a
different type of system, and on the other by extending the ideas of marginal
stability to finite dimensions, in non-isostatic systems. As an additional
virtue, the model we study gives us the possibility of making this analysis on
unprecedentedly large systems, giving us the chance to observe scalings along
several orders of magnitude.

Here, we study the inherent structures and we do find that they are marginally\index{inherent structure}\index{marginal stability}
stable states where the distribution of eigenvalues of the Hessian matrix\index{Hessian matrix}\index{dynamical matrix|see{Hessian matrix}}
stretches down to zero as a power law. Furthermore, we find that the soft\index{localization}\index{modes!soft}
modes are localized. This cannot be revealed by computations in infinite
dimensions, though it is still possible to observe correlations in pseudo mean
field networks such as the Bethe lattice \cite{lupo:15}, and it was shown that\index{Bethe lattice}
superuniversality (the independence of the behavior on the space\index{universality!super}
dimensionality) can be recovered by removing local excitations
\cite{charbonneau:15}.

We broaden our analysis by taking in account the anharmonic effects due to the \index{anharmonicity}
complexity of the energy landscape. \index{energy!landscape} We find that the energy barriers along the\index{energy!barrier}
softest mode are extremely small, in agreement with the mean field picture,\index{modes!soft}
and that they connect very similar states with an strong relationship, that we
propose as a classical operational definition of two-level systems.\index{two-level system}

At the end of the game the scenario is consistent, with mean field theory that
does apply, but with the necessary finite-dimension corrections due to the
presence and importance of localized states.\index{localization}
\index{boson peak|)}

\section{Model and simulations}
The model we study is the three-dimensional Heisenberg spin glass in a \ac{RF}. \index{spin glass!Heisenberg}
The \ac{RF} breaks all rotational and translational symmetry, \index{symmetry!O(3)@$O(3)$}
so there should be no Goldstone bosons.\index{modes!Goldstone}
\begin{table}[!tbh]
 \centering
  \resizebox{0.75\columnwidth}{!}{
 \begin{tabular}{ccccccc}
  $H_\amp$ & $L$ & $N_\mathrm{samples}$ & $N_\mathrm{replicas}$ &  $n_\lambda$ &  $A (\ket{\vec\pi_0})$ &  $A (\ket{\vec\pi_\RAND})$ \\
\hline\hline
    50     & 192 &          10 (0)      &            2          &  35  &       -       &    -     \\
    50     &  96 &          10 (10)     &            2          &  80  &       1       &    1     \\
    50     &  48 &          70 (70)     &            2          & 500  &       1       &    1     \\
    50     &  24 &         100 (100)    &            2          & 500  &       1       &    1     \\
    50     &  12 &         100 (100)    &            2          & 500  &       1       &    1     \\    
\hline
    10     & 192 &          10 (0)      &            2          &  35  &       -       &    -     \\
    10     &  96 &          10 (10)     &            2          &  80  &       0.6     &    0.72  \\
    10     &  48 &          70 (70)     &            2          & 500  &       0.6     &    0.72  \\
    10     &  24 &         100 (100)    &            2          & 500  &       0.3     &    0.72  \\
    10     &  12 &         100 (100)    &            2          & 500  &       0.3     &    0.72  \\    
\hline
     5     & 192 &          10 (0)      &            2          &  35  &       -       &     -    \\
     5     &  96 &          10 (10)     &            2          &  80  &       0.014   &    0.3   \\
     5     &  48 &          70 (70)     &            2          & 500  &       0.014   &    0.3   \\
     5     &  24 &         100 (100)    &            2          & 500  &       0.02    &    0.3   \\
     5     &  12 &         100 (100)    &            2          & 500  &       0.024   &    0.3   \\    
\hline
     1     & 192 &          10 (0)      &            2          &  35  &       -       &    -     \\
     1     &  96 &          10 (10)     &            2          &  80  &       0.004   &    0.05  \\
     1     &  48 &          70 (70)     &            2          & 500  &       0.004   &    0.05  \\
     1     &  24 &         100 (100)    &            2          & 500  &       0.0045  &    0.05  \\
     1     &  12 &         100 (100)    &            2          & 500  &       0.0045  &    0.05  \\    
\hline
    0.5    & 192 &          10 (0)      &            2          &  35  &       -       &    -     \\
    0.5    &  96 &          10 (10)     &            2          &  80  &       0.008   &    0.022 \\
    0.5    &  48 &          70 (70)     &            2          & 500  &       0.008   &    0.02  \\
    0.5    &  24 &         100 (100)    &            2          & 500  &       0.009   &    0.022 \\
    0.5    &  12 &         100 (100)    &            2          & 500  &       0.009   &    0.022 \\    
\hline
    0.1    & 192 &          10 (0)      &            2          &  35  &       -       &     -    \\
    0.1    &  96 &          10 (10)     &            2          &  80  &       0.006   &    0.012 \\
    0.1    &  48 &         100 (70)     &            2          & 500  &       0.006   &    0.012 \\
    0.1    &  24 &         100 (100)    &            2          & 500  &       0.1     &    0.012 \\
    0.1    &  12 &         100 (100)    &            2          & 500  &       0.1     &    0.012 \\    
\hline
    0.05   & 192 &          10 (0)      &            2          &  25  &       -       &     -    \\
    0.05   &  96 &          10 (10)     &            2          &  80  &       0.06    &    0.011 \\
    0.05   &  48 &         100 (70)     &            2          & 500  &       0.06    &    0.011 \\
    0.05   &  24 &         100 (100)    &            2          & 500  &       0.42    &    0.011 \\
    0.05   &  12 &         100 (100)    &            2          & 500  &       0.36    &    0.011 \\    
\hline
    0.01   & 192 &           7 (0)      &            2          &  25  &       -       &      -   \\
    0.01   &  96 &          10 (10)     &            2          &  80  &       0.045   &    0.016 \\
    0.01   &  48 &         100 (70)     &            2          & 500  &       0.045   &    0.016 \\
    0.01   &  24 &         100 (100)    &            2          & 500  &       0.009   &    0.004 \\
    0.01   &  12 &         100 (100)    &            2          & 500  &       0.007   &    0.001 \\    
\end{tabular}
}
\caption[Simulation parameters from \cite{baityjesi:15b}]
	{Number of samples, $N_\mathrm{samples}$, and of replicas, $N_\mathrm{replicas}$, \nomenclature[N...replicas]{$N_\mathrm{replicas}$}{number of replicas}
	of our simulations. The number between parentheses is the amount of samples
	used for the forcings. \index{forcing}\index{numerical simulations}
	We indicate with $n_\lambda$ \nomenclature[n....lambda]{$n_\lambda$}{number of calculated eigenvalues}
	the number of eigenvalues \index{eigenvalue|see{Hessian matrix!spectrum}}
	we calculated from the bottom of the spectrum $\rho(\lambda)$ \index{Hessian matrix!spectrum}
	(see section \ref{sec:hsgrf-spectrum}).
	$A (\ket{\vec\pi_\RAND})$ and $A (\ket{\vec\pi_0})$ are the forcings' 
	parameters from equations 
	\nomenclature[A...big3rand]{$A (\ket{\vec\pi_\RAND})$}{amplitude parameter for forcings along $\ket{\vec\pi_\RAND}$ [see also equation \eqref{eq:Arand}]}
	\nomenclature[A...big30]{$A (\ket{\vec\pi_0    })$}{amplitude parameter for forcings along $\ket{\vec\pi_0    }$ [see also equation \eqref{eq:Arand}]}
	\eqref{eq:Arand} and \eqref{eq:A0}.}
\label{tab:hsgrf-sim}
\end{table}\afterpage{\clearpage}
The dynamic variables are spins $\vec{s}_\bx$ with $m=3$ components. They are placed at the vertices $\bx$ of a cubic lattice of linear 
size $L$ with unitary spacings. We have therefore $N=L^3$ spins, and $2N$ \ac{dof} due to the constraint $\vec{s}_\bx\cdot\vec{s}_\bx=1$.
The Hamiltonian is
\begin{equation}\label{eq:HRF}
\nomenclature[H..RF]{$\mathcal{H}_\RF$}{Hamiltonian of the Heisenberg SG with a random field}
 \mathcal{H}_\RF = -\sum_{|\bx - \by|=1} J_{\bx\by} \vec{s}_{\bx}\cdot \vec{s}_{\by} - \sum_{\bx}^{N} \vec{h}_{\bx} \cdot \vec{s}_{\bx}\,,
\end{equation}
where the fields $\vec{h}_{\bx}$ \nomenclature[h....x]{$\vec{h}_{\bx}$}{In chapter \ref{chap:hsgrf}: random field of amplitude $H_\amp$ on site $\bx$}
are random vectors chosen uniformly from the sphere of radius 
$H_\amp$.\nomenclature[H...amp]{$H_\amp$}{amplitude of the random field}
The couplings $J_{\bx\by}$ are fixed, Gaussian distributed, with $\overline{J_{\bx\by}}=0$ and $\overline{J_{\bx\by}^2}=J^2$.
% Let us call $E_\RF(\ket{s})$ the energy measured with $\mathcal{H}_\RF$ on the configuration $\ket{s}$.

The lattice sizes we simulated are $L = 12, 24, 48, 96, 192$. \index{numerical simulations}
We chose always $J=1$, and we compared it with $H_\amp = 0, 0.01, 0.05, 0.1, 0.5, 1, 5, 10, 50$.
In table \ref{tab:hsgrf-sim} we resume how many samples we simulated for each couple $(L,H)$.

\section{Calculating the \acl{dos}}
\index{density of states}\index{density of states@density of states$ $|seealso{Hessian matrix spectrum}}
\index{Hessian matrix!spectrum@spectrum$ $|seealso{density of states}}
Our goal is to study the dynamical matrix of the system. The dynamical matrix is the Hessian matrix $\M$ 
of Hamiltonian \eqref{eq:HRF}, \index{Hessian matrix}\index{inherent structure}
calculated at the local minima of the energy, that we call \acfp{IS} in analogy with structural glasses.\index{glass!structural}
Each infinite-temperature starting configuration $\ket{\vec{s}}$ can be associated to an \ac{IS} 
$\ket{\vec{s}^\mathrm{(IS)}}$ \nomenclature[s....IS]{$\vec{s}^\mathrm{\,(IS)}$}{spin of the inherent structure}
\nomenclature[s....IS]{$\ket{\vec{s}^\mathrm{\,(IS)}}$}{inherent structure} through a deterministic relaxation of the system.
\footnote{We will show in appendix \ref{chap:hsgrf-is-algorithm} that the starting temperature does not influence visibly the properties we are
studying, at least as long as we stay in the paramagnetic phase.}

\subsection{Reaching the inherent structure \label{sec:hsgrf-minimizzazione}}
As energy minimization algorithm we use the \acf{SOR} (appendix \ref{app:sor}), \index{successive overrelaxation}\index{numerical simulations}
that was successfully used in \cite{baityjesi:11} for $3d$ Heisenberg spin glasses.
This algorithm depends on a parameter $\Lambda$, and the convergence speed is maximal for $\Lambda\approx300$ \cite{baityjesi:11}.
Thus, the seek of \acp{IS} was done with $\Lambda = 300$, under the reasonable
assumption, reinforced in appendix \ref{chap:hsgrf-is-algorithm}, that a change on $\Lambda$ does not imply sensible changes 
in the observables we examine.
In fact, the concept of \ac{IS} is strictly related to the protocol one chooses to relax the system, and on the starting
configuration. From \cite{baityjesi:11} our intuition is that despite the \acp{IS}' energies do depend on these two 
elements, this dependency is small and we can neglect it (dependencies on the correlation lengths will be examined in a future work \cite{baityjesi:16}).

We validate these hypotheses in appendix \ref{app:sor-lambda-T}, where we compare the \acp{dos} both between $\Lambda=300$ and $\Lambda=1$,
and between starting configurations at different temperatures.

\begin{figure}[!t]
 \centering
 \includegraphics[width=0.48\textwidth]{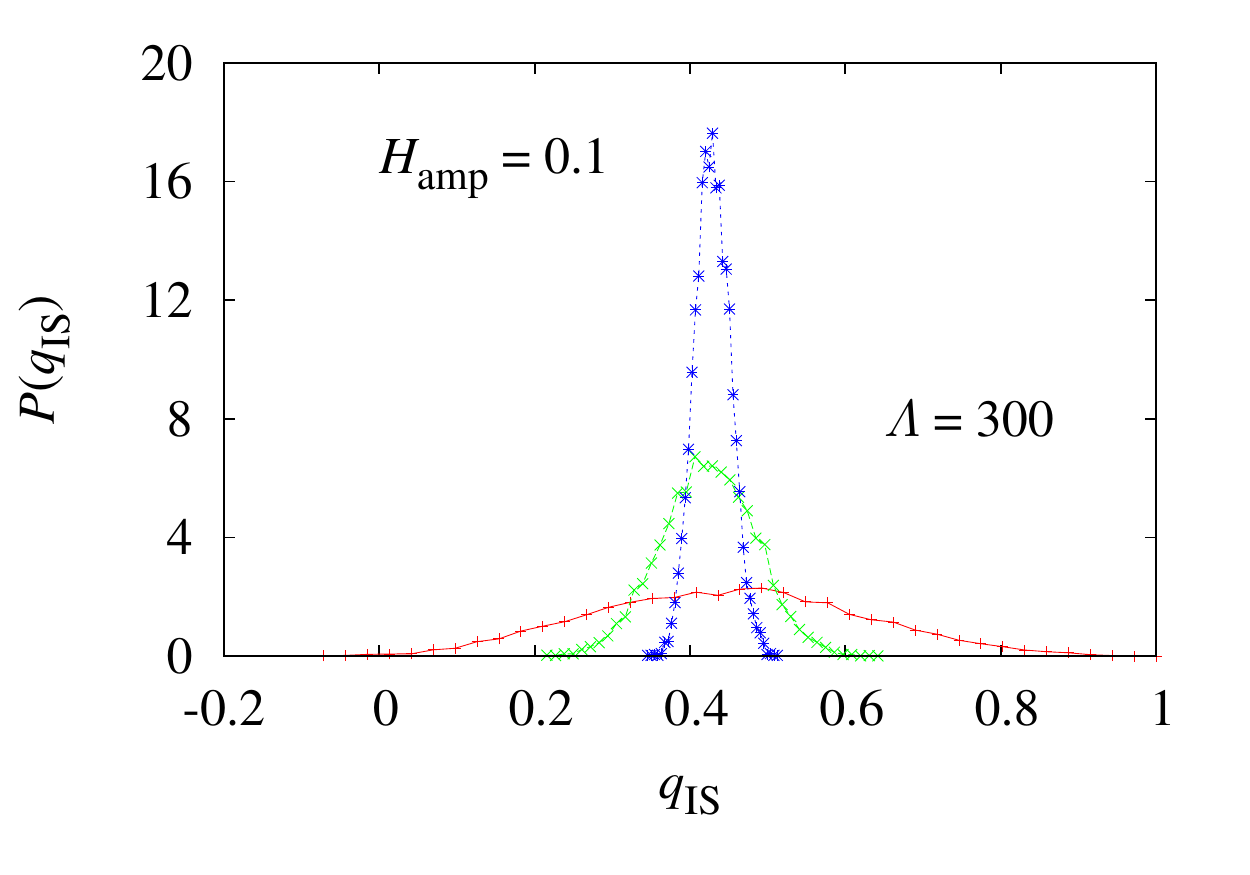}
 \includegraphics[width=0.48\textwidth]{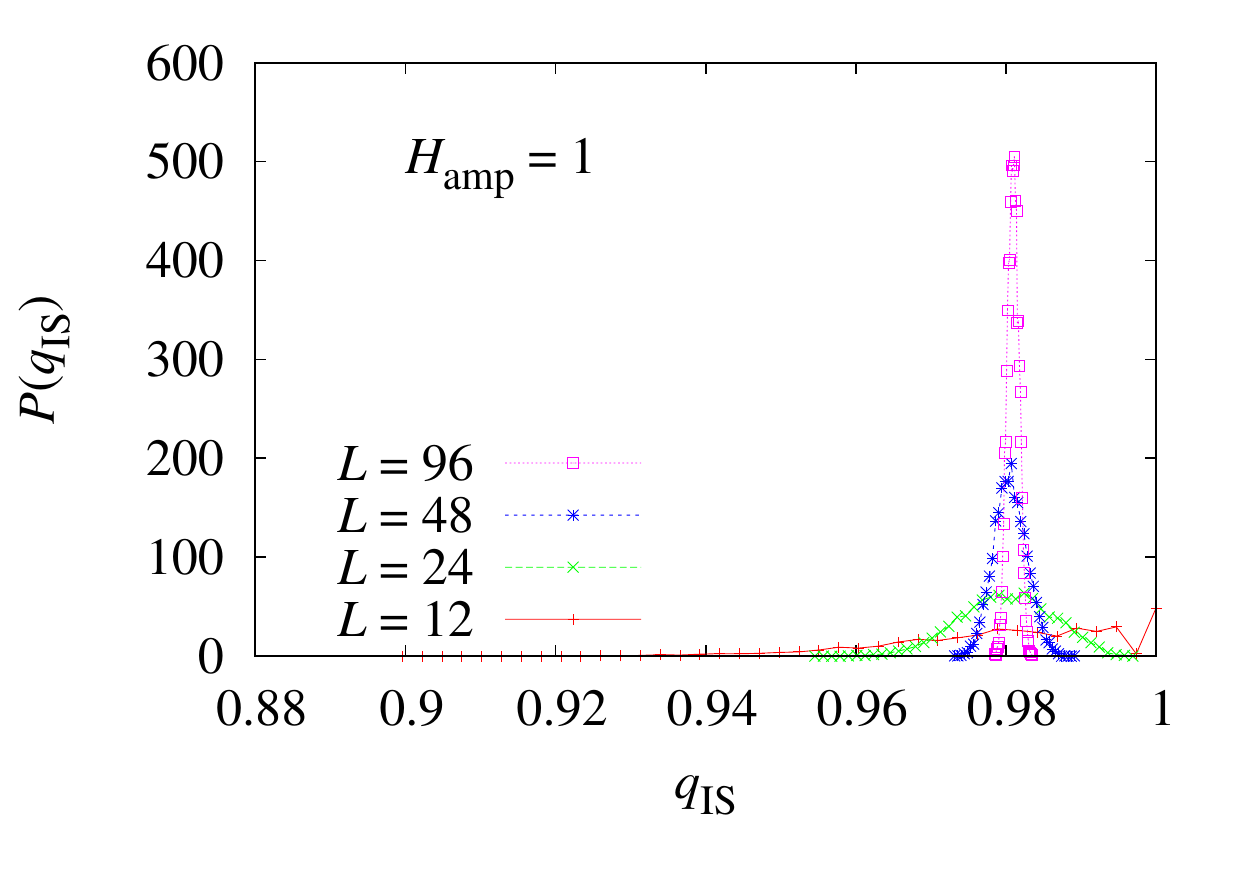}
 \caption[Distribution of the overlaps at the inherent structures.]
         {Distribution $P(q_\IS)$ of the overlaps at the inherent structures obtained with $\Lambda=300$, 
         for $H_\amp=0.1$ (\textbf{left}) and $H_\amp=1$ (\textbf{right}).\index{overlap!inherent structure!distribution}}
 \label{fig:PQ-IS}
\end{figure}

For most of the simulated fields, the \ac{pdf} of the overlap of the reached inherent structures, $P(q_\IS)$, 
\nomenclature[P...qIS]{$P(q_\IS)$}{overlap distribution of the inherent structures} is peaked around \index{inherent structure}
\index{overlap!inherent structure!distribution}
a non-zero value that is significantly far from 1 (figure \ref{fig:PQ-IS}). This means that even though all the inherent structures have 
a very large amount of spins in similar configurations, it is practically impossible with this approach (at least for $L>12$ lattices), to find
two identical inherent structures.

\subsection{The local reference frame \label{sec:hsgrf-localref}}\index{local!reference frame}
Once the \ac{IS} is found, we want to study the properties of the reached \ac{IS}. 
From the Hamiltonian at the inherent structure, $\mathcal{H}_\IS$, \nomenclature[H..IS]{$\mathcal{H}_\IS$}{Hamiltonian $\mathcal{H}_\RF$ calculated at the IS}
we \index{inherent structure}
want to compute the Hessian matrix $\M$ \index{Hessian matrix}
to study the harmonic behavior at the \ac{IS}. This is not trivial,
because it is necessary to take into account the normalization of the spins  $\vec{s}_\bx^{\,2}=1~~\forall\bx$.

To this scope we define local perturbation vectors $\vec{\pi}_\bx$, \index{perturbation|seealso{pion}}\index{perturbation|see{forcing}}
\index{local!perturbation vector|see{pion}}\index{pion|(}\nomenclature[pi....x]{$\vec{\pi}_\bx$}{pion}
and we call them pions in analogy with the nonlinear $\sigma$ model \cite{gellman:60}.\index{non-linear sigma model@non-linear $\sigma$ model}
The distinguishing feature of the pions is that they are orthogonal to the \ac{IS}, $\left(\vec{s}_\bx\cdot \vec{\pi}_\bx\right)=0$, and that their global
norm is unitary, $\langle\vec{\pi}\ket{\vec{\pi}}=1$.
\footnote{Recall the notation introduced in chapter \ref{chap:obs}, according to which $\langle \vec a\ket{\vec b}\equiv\sum_\bx \vec a_\bx\cdot\vec b_\bx$.}
We can use the pions to parametrize an order $\epsilon$ \nomenclature[epsilon....8]{$\epsilon$}{In chapter \ref{chap:hsgrf}: amplitude of the perturbation}
perturbation around the \ac{IS} as
\begin{equation}\label{eq:pions}
 \vec{s}^{\,\epsilon}_\bx = \vec{s}_\bx^\mathrm{\,(IS)} \sqrt{1-\epsilon^2\vec{\pi}_\bx^2} + \epsilon\vec{\pi}_\bx~~,~~\vec{\pi}_\bx^2\equiv \vec{\pi}_\bx \cdot\vec{\pi}_\bx\,,
\end{equation}
so the position of $\vec{s}^{\,\epsilon}_\bx$ is fully determined by $\vec{\pi}_\bx$.
As long as $\epsilon$ is small enough to grant $\epsilon^2\vec{\pi}_\bx^2<1$ $\forall\bx$, the normalization condition is naturally satisfied without the need 
to impose any external constraint.

We now build a local reference change. For each site $\bx$ we define a local basis \index{local!basis}
$\mathcal{B}=\left\{\vec{s}_\bx^\mathrm{\,(IS)}, \hat{e}_{1,\bx}, \hat{e}_{2,\bx}\right\}$,\nomenclature[B...x]{$\mathcal{B}$}{local basis}
where $\hat{e}_{1,\bx}, \hat{e}_{2,\bx}$ \nomenclature[e....1hat,e2hat]{$\hat{e}_{1,\bx}, \hat{e}_{2,\bx}$}{unitary vectors forming the basis $\mathcal{B}$}
are any two unitary vectors, orthogonal to each other and to $\vec{s}_\bx^\mathrm{\,(IS)}$, and well oriented. In our simulations they were generated
randomly.
In this basis the pions can be rewritten as
\begin{equation}
\nomenclature[a....1a2]{$a_1, a_2$}{non-trivial components of the pion}
  \vec{\pi}_\bx = (0, a_1, a_2)\,,
\end{equation}
where now they explicitly depend only on two components, with real values $a_1$ and $a_2$.
We can therefore rewrite the pions as two-component vectors $\tilde{\pi}_\bx$\nomenclature[pi....xtilde]{$\tilde{\pi}_\bx$}{pion as a function of two components}
\begin{equation}\label{eq:bidipion}
 \tilde{\pi}_\bx = (a_1, a_2)\,.
\end{equation}
At this point we completely integrated the normalization constraint with the parametrization, and we can obtain the $2N\times2N$ 
Hessian matrix $\M$, that acts on\index{Hessian matrix}
$2N$-component vectors $\ket{\tilde{\pi}}$, by a second-order development of 
$\mathcal{H}_\IS$ (the derivation of $\M$ is shown in appendix \ref{app:hsgrf-hessian}). 
The obtained matrix is sparse, with 13 non-zero elements per line (1 diagonal element, and 6 two-component vectors for the nearest-neighbors). 
The matrix element $\M^{\alpha\beta}_{\bx\by}$ is 
\begin{equation}
\nomenclature[M..xyab]{$\M^{\alpha\beta}_{\bx\by}$}{element of the Hessian matrix; $\bx,\by$ indicate the position, $\alpha\beta$ indicate the component}
 \M^{\alpha\beta}_{\bx\by} = \M_{\bx\by} (\hbe_{\alpha,\bx}\cdot\hbe_{\beta,\by})\,,\\
 \end{equation}
 with
 \begin{equation}
\nomenclature[M..xy]{$\M_{\bx\by}$}{matrix element of the Hessian matrix; $\bx$ and $\by$ stand for the position}
 \M_{\bx\by} = \delta_{\bx\by} (\vec{h}_\by^\mathrm{\,(IS)}\cdot\vec{s}_\by^\mathrm{\,(IS)})- \sum_{\mu=-D}^{D}J_{\bx\by}\delta_{\bx+\hat\mu, \by}\,,
\end{equation}
where the bold latin characters as usual indicate the site, and the greek characters indicate the component of the two-dimensional vector of equation (\ref{eq:bidipion}).

Once $\M$ is known, \index{Hessian matrix}
from each simulated $H_\amp$ we calculate the spectrum of the eigenvalues $\rho(\lambda)$ \index{Hessian matrix!spectrum}
\index{Hessian matrix!rho(lambda)@$\rho(\lambda)$|see{Hessian matrix spectrum}}
or equivalently, in analogy with plane waves \cite{huang:87}, the 
\ac{dos} $g(\omega)$, by defining $\lambda=\omega^2$.\index{density of states}
We measure the \ac{dof} both by means of a convolution
with a lorentian function with the method of the moments \cite{chihara:78,turchi:82,alonso:01}, \index{method of the moments}
and by making the explicit brute-force calculation of the lowest eigenvalues with Arpack \cite{arpack}.
\index{Arnoldi algorithm}\index{numerical simulations}\index{ArPack}
\index{pion|)}

\section{The Spectrum of the Hessian matrix \label{sec:hsgrf-spectrum}}
\begin{figure}[!thb]
 \includegraphics[width=0.49\columnwidth]{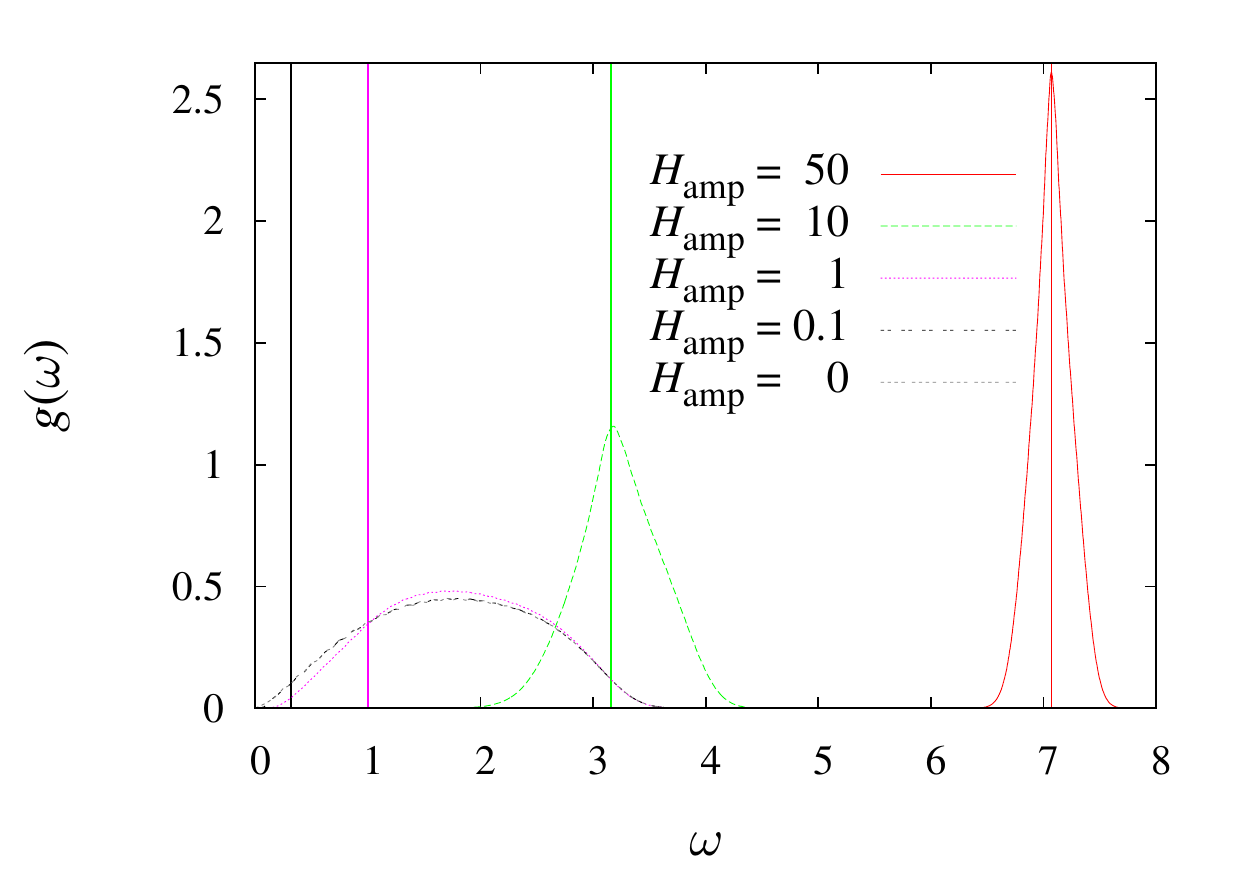}
 \includegraphics[width=0.49\columnwidth]{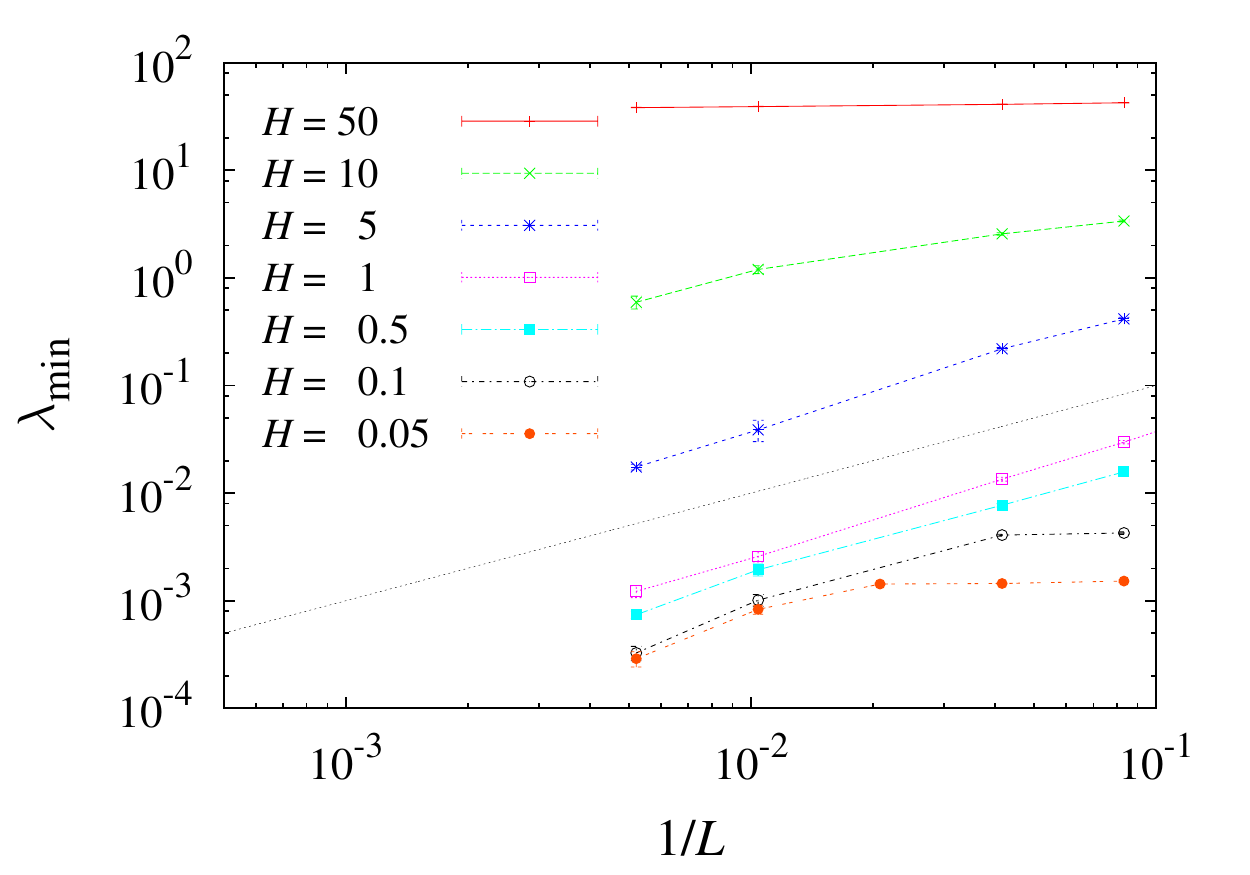}
 \caption[\acs{dos} with the method of the moments]
	 {\textbf{Left}: The \ac{dos} $g(\omega)$ calculated with the method of the moments. \index{method of the moments}
	 The vertical lines \index{density of states}\index{gap}
	  represent its face in the limit of a diagonal hamiltonian, $J=0$. 
	  The $g(\omega)$ corresponding to $H_\amp=0$ and $H_\amp=0.1$ are practically overlapped.
	  \textbf{Right}: Scaling with $1/L$ of the lowest eigenvalue $\lambda_\mathrm{min}$
	  \nomenclature[lambda....min]{$\lambda_\mathrm{min}$}{lowest eigenvalue}
	  \index{Hessian matrix!lowest eigenvalue}\index{inherent structure}
	  of the Hessian matrix $\M$ calculated at the \ac{IS}, for all the simulated fields.
	  The straight line is a reference curve $\lambda_\mathrm{min}\propto1/L$.}
 \label{fig:gomega}
\end{figure}

We find that, although for large fields there is a gap in the \ac{dos} (as one can easily
expect by calculating it exactly in the diagonal limit $H_\amp\gg J\simeq0$) when the field is small enough\index{gap}
the gap disappears and the \ac{dos} goes to zero developing soft modes (figure \ref{fig:gomega}, left).\index{modes!soft}
In the right set of figure \ref{fig:gomega} we show the scaling of the lowest eigenvalue of the Hessian. \index{Hessian matrix!lowest eigenvalue}
We see that while for very large fields it remains approximately constant, for smaller fields it approaches zero as we increase the
lattice size $L$.

It is interesting to understand the origin of these soft modes, \index{modes!soft}
so we focus on the $\rho(\lambda)$ for small $\lambda$,
or even better in its cumulative function
\begin{equation}
\index{Hessian matrix!spectrum}
\index{Hessian matrix!spectrum!cumulative}
\index{Hessian matrix!spectrum!F(l)@$F(\lambda)$|see{Hessian matrix spectrum cumulative}}
\nomenclature[F...l]{$F(\lambda)$}{cumulative of $\rho(\lambda)$}
 F(\lambda) = \int_0^\lambda \rho(\lambda')d\lambda'\,.
\end{equation}
\begin{figure}[!t]
{
    \def\OldComma{,}
    \catcode`\,=13
    \def,{%
      \ifmmode%
        \OldComma\discretionary{}{}{}%
      \else%
        \OldComma%
      \fi%
    }%
 \includegraphics[width=0.49\columnwidth]{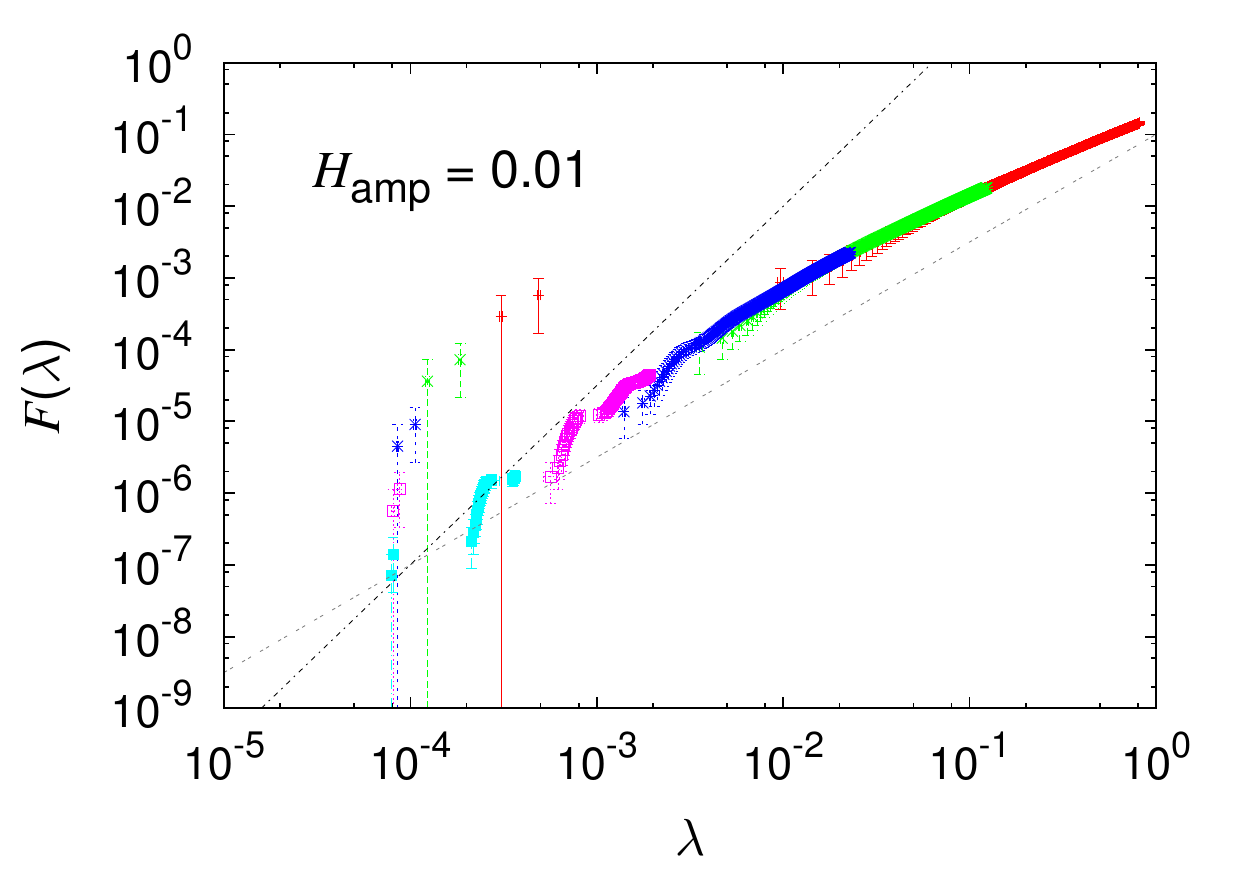}
 \includegraphics[width=0.49\columnwidth]{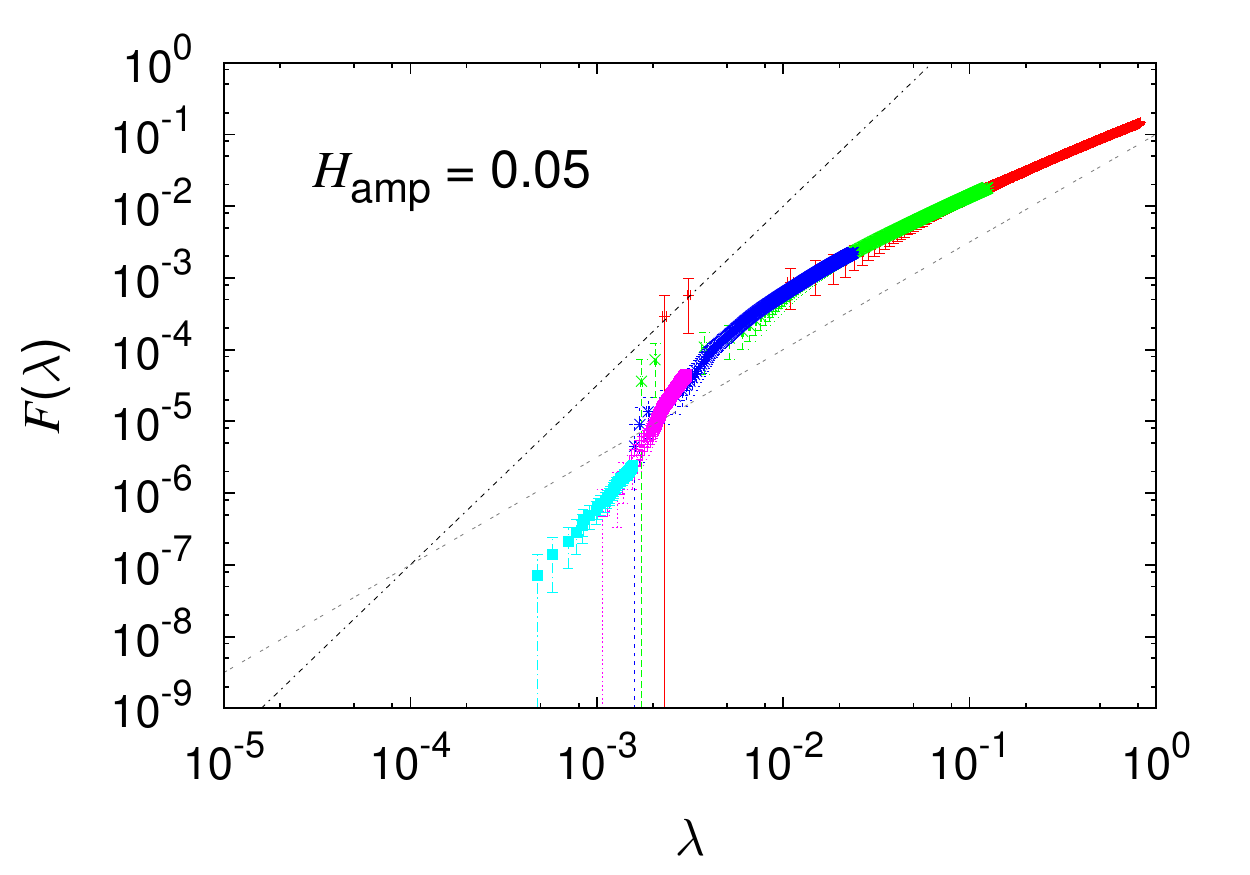}
 \includegraphics[width=0.49\columnwidth]{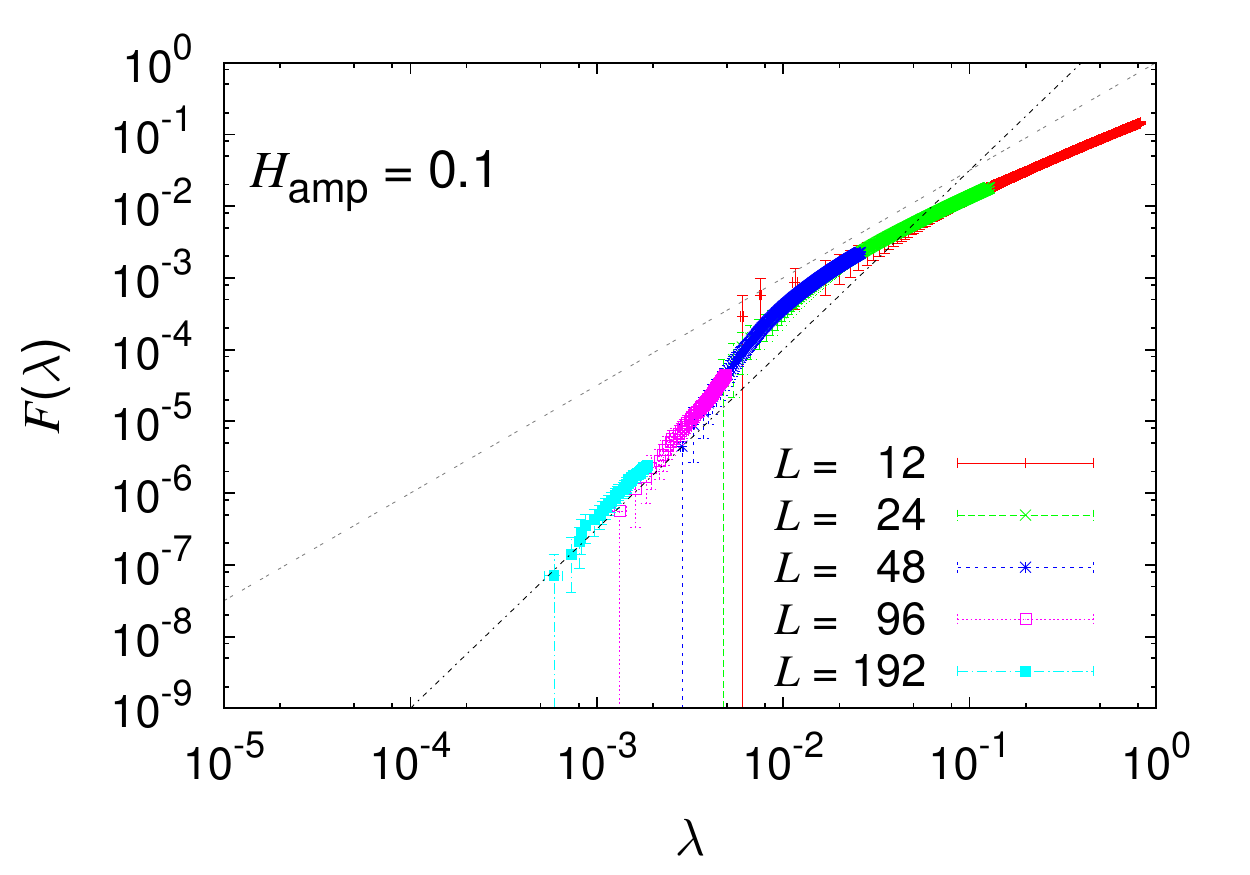}
 \includegraphics[width=0.49\columnwidth]{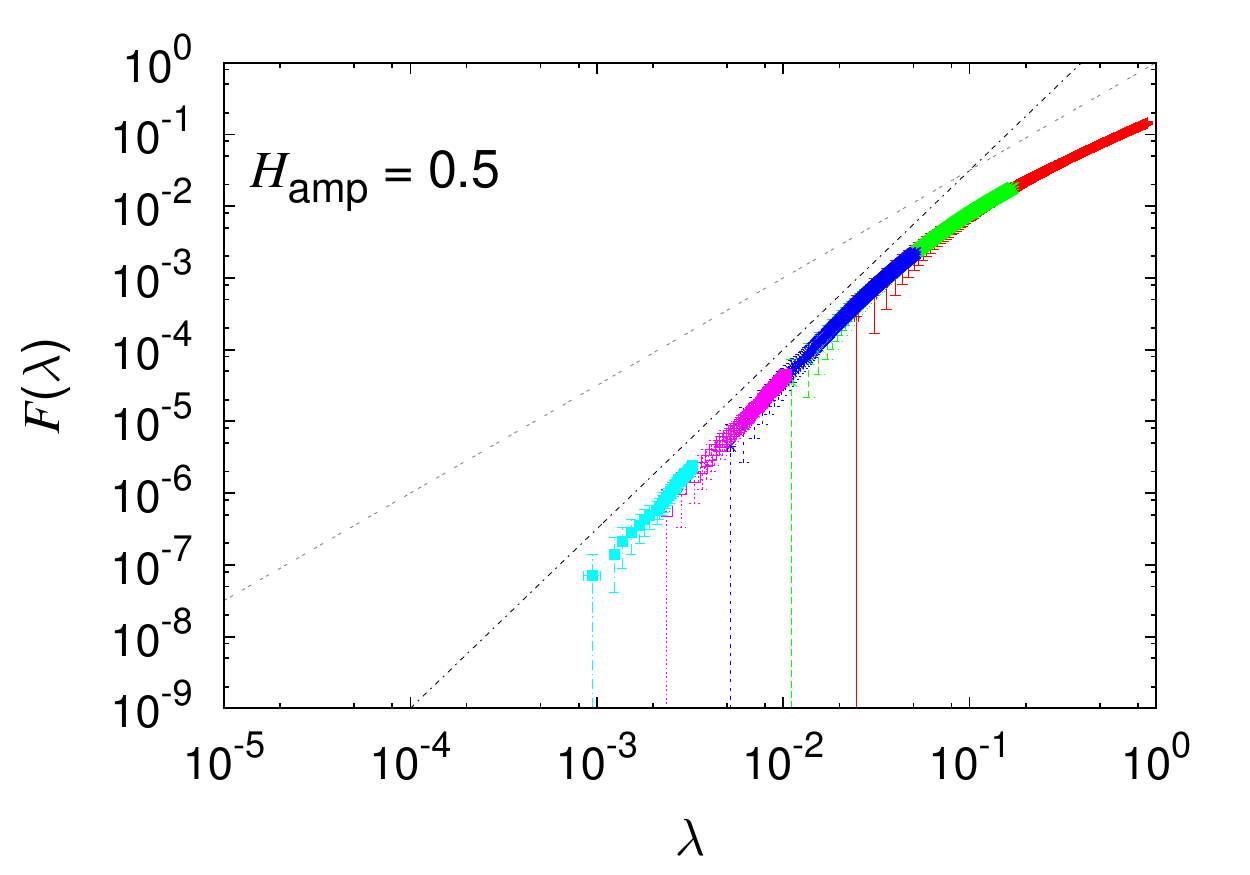}
 \caption[Cumulative distributions $F(\lambda)$ for small random fields]
	 {\index{Debye}\index{Hessian matrix!spectrum!cumulative}
	 Cumulative distributions $F(\lambda)$ for small random fields $H_\amp = 0.01, 0.05, 0.1, 0.5$. 
	 In each plot we show a black reference curve representing the power law $\lambda^{2.5}$,
	 that is our guess for a universal behavior, and a grey line indicating the Debye behavior 
	 $\lambda^{1.5}$. One could expect a Debye behavior for $\lambda>\lambda^*$, \index{crossover}
	 with $\lambda^{*}\to0$ as $H_\amp\to0$. Instead, we see an excess of eigenvalues even compared 
	 to the Debye behavior, indicating a likely \index{boson peak} boson peak.\index{boson peak}
	 Further discussions in the main text.} 
	 \label{fig:F-1}
}\end{figure}

\begin{figure}[!t]
{
    \def\OldComma{,}
    \catcode`\,=13
    \def,{%
      \ifmmode%
        \OldComma\discretionary{}{}{}%
      \else%
        \OldComma%
      \fi%
    }%
 \includegraphics[width=0.49\columnwidth]{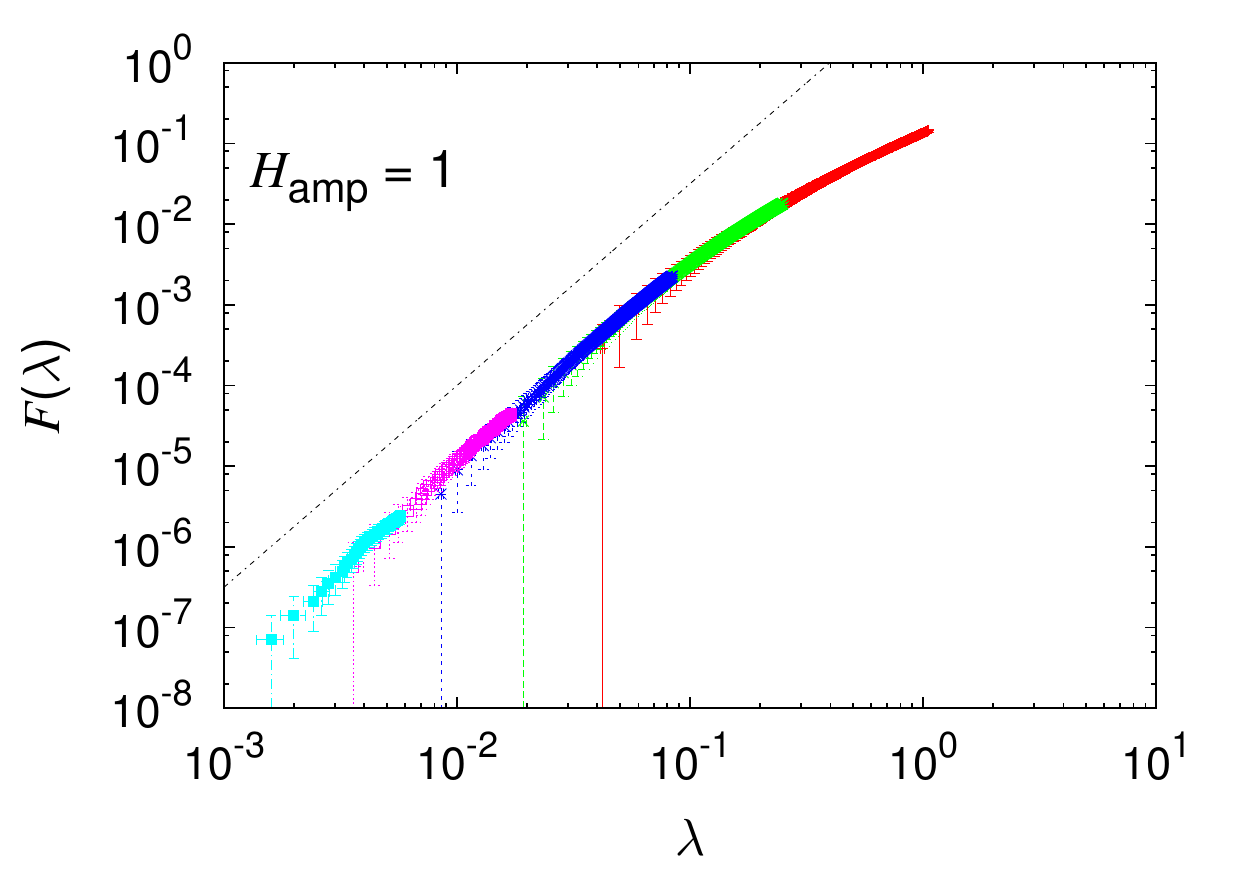}
 \includegraphics[width=0.49\columnwidth]{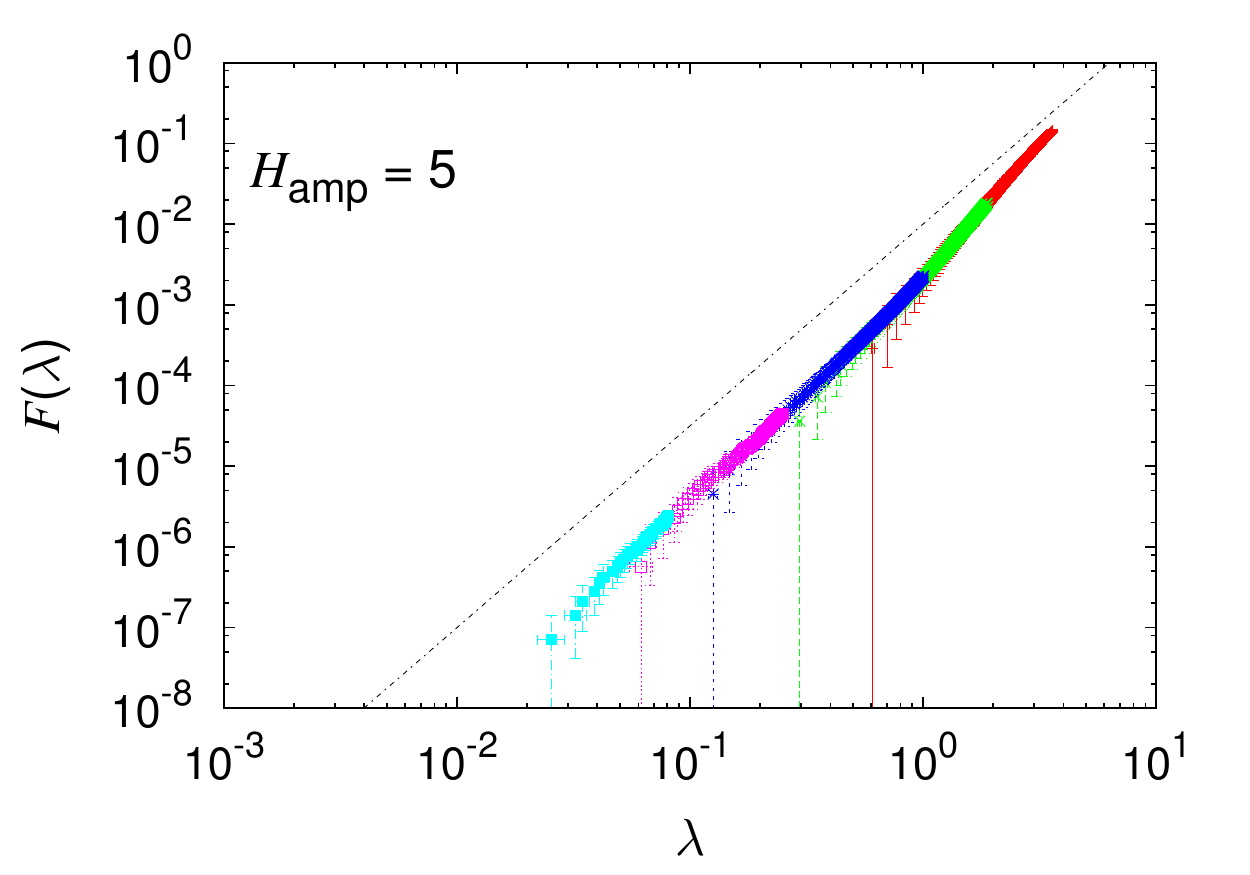}
 \includegraphics[width=0.49\columnwidth]{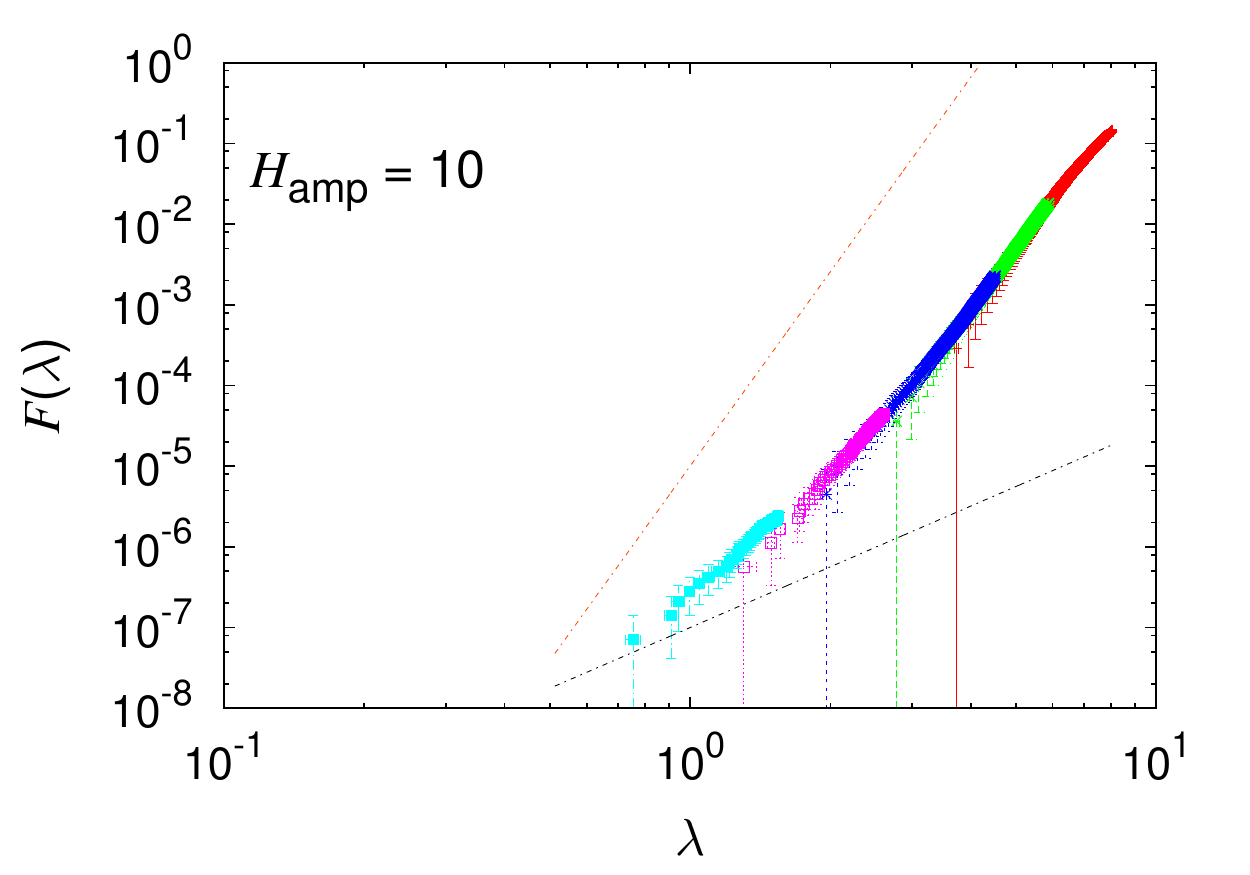}
 \includegraphics[width=0.49\columnwidth]{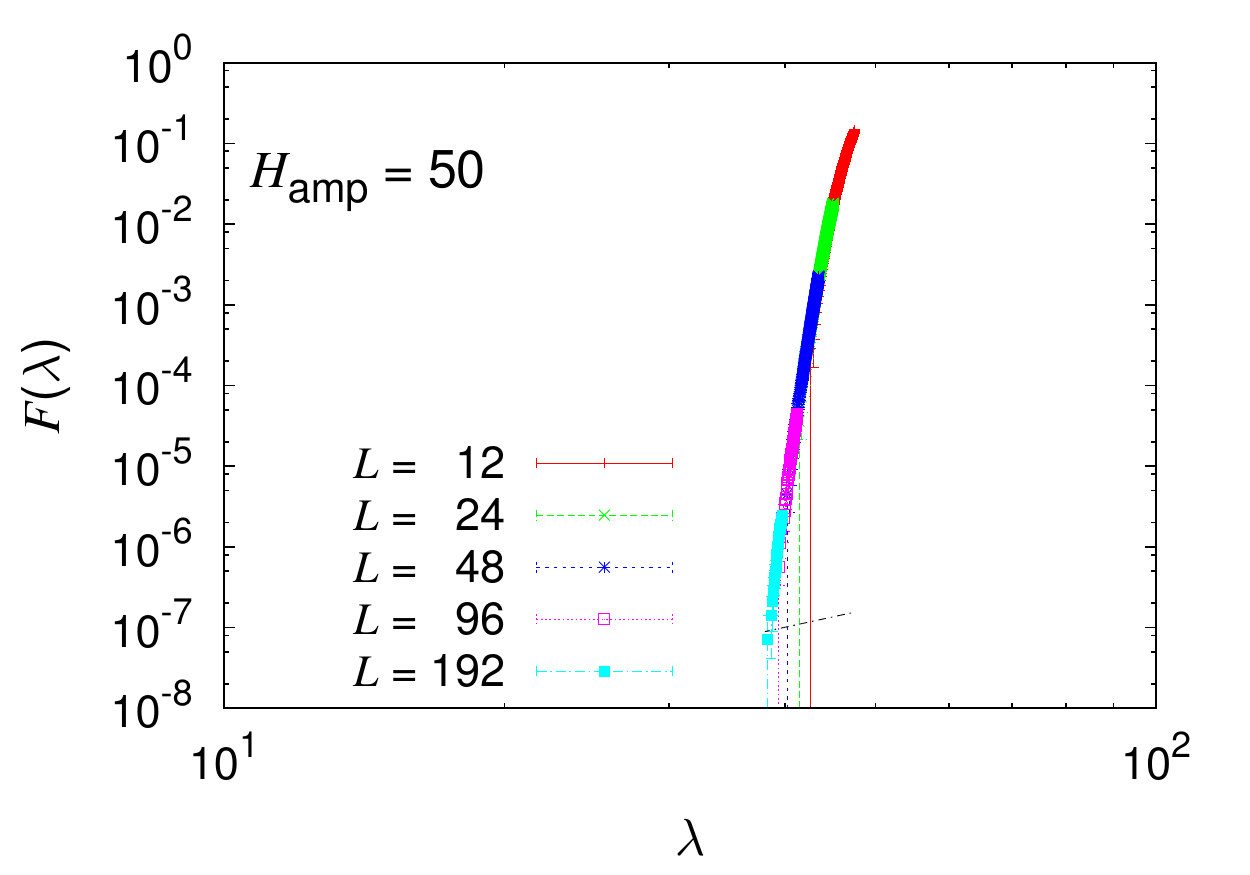}
 \caption[Cumulative distributions $F(\lambda)$ for large random fields]
         {Cumulative distributions $F(\lambda)$ \index{Hessian matrix!spectrum}
         for large random fields $H_\amp = 1, 5, 10, 50$.In each plot we show a reference curve representing the power law $\lambda^{2.5}$.
         The orange line in the \textbf{bottom left} set is proportional to $\lambda^8$.}
 \label{fig:F-2}
}
\end{figure}

In the case that there be no gap and for small $\lambda$ the function $F(\lambda)$ reach zero as a power law, we
\index{Hessian matrix!spectrum!cumulative} can define three exponents $\delta,\alpha$ and $\gamma$, 
\index{exponent!soft modes!alpha@$\alpha$|(}
\index{exponent!soft modes!gamma@$\gamma$|(}
\index{exponent!soft modes!delta@$\delta$|(}
that describe how the functions $g,\rho$ and $F$ go to zero for small $\lambda$:
\footnote{The exponents $\delta,\alpha$ and $\gamma$ have nothing to do with the critical exponents defined in chapter \ref{chap:rg}.}
\begin{equation}\label{eq:hsgrf-exponents}
\nomenclature[delta....8]{$\delta$}{In chapter \ref{chap:hsgrf}: $g(\omega)\sim\omega^\delta$      for small $\omega $}
\nomenclature[alpha....8]{$\alpha$}{In chapter \ref{chap:hsgrf}: $\rho(\lambda)\sim\lambda^\alpha$ for small $\lambda$}
\nomenclature[gamma....8]{$\gamma$}{In chapter \ref{chap:hsgrf}: $F(\lambda)\sim\lambda^\gamma$    for small $\lambda$}
g(\omega)\sim\omega^\delta\,,~~~\index{density of states}
\rho(\lambda)\sim\lambda^\alpha\,,~~~\index{Hessian matrix!spectrum}
F(\lambda)\sim\lambda^\gamma\,,\index{Hessian matrix!spectrum!cumulative}
\end{equation}
where the exponents are related by $\delta=2\alpha+1=2\gamma-1$.
In the Debye model, \index{Debye} valid for perfect crystals and based on the assumption that all the eigenvectors are plane waves, one has $\delta=d-1=2$ ($\alpha=0.5$, $\gamma=1.5$),
and this is also what one expects for our model in the absence of a field \cite{grigera:11}.
In figures \ref{fig:F-1} and \ref{fig:F-2} we show the function $F(\lambda)$ for all the fields we simulated. We were able to calculate with Arpack the lowest eigenvalues 
of the spectrum. The number of calculated eigenvalues $n_\lambda$ is shown in table \ref{tab:hsgrf-sim}.
All the plots are compared with the Debye \index{Debye} behavior $\lambda^{1.5}$ and with the power law behavior $\lambda^{2.5}$, 
because if there is some universality on the exponents $\gamma$, our data suggests it has to be around $\gamma=2.5$ (thus $\delta=4$ and $\alpha=1.5$).
%\footnote{The second decimal digit on $\gamma$ is because we chose to use conservative confidence intervals of 1/4, so our guess for the exponent is $2.75\pm0.25$.}
This is straightforward for $H_\amp = 0.1,0.5, 1, 5$, where when $\lambda$ is small
there is a clear power law behavior, with a power close to 2.5, while it can be excluded for $H_\amp=50$, where the soft modes are suppressed in favor of a gap,
as it was also clear from figure \ref{fig:gomega}.
At $H_\amp=10$ we are probably close to where the gap forms. The $F(\lambda)$ goes as a large power law $\lambda^{boh}$ when $\lambda$ is large, but at the
smallest values of $\lambda$, recovered from $L=192$, there is a slight change of power law towards something that could become 2.5.
One could also argue that a $F(\lambda)$ goes to zero as a power law for any finite $H_\amp$, as long as one looks at small enough $\lambda$. 
Numerical analysis cannot reply to questions of this type,\index{gap}
but still, even if no sharp transition is present, an empirical gap is clearly present for large $H_\amp$, since the precision of any experiment 
(numerical or real) is finite. In the case of the smallest fields $H_\amp=0.01, 0.05$, we suffer from effects from $H_\amp=0$. The spin waves do 
not hybridize with the bulk of the spectrum, and pseudo-Goldstone
modes with a very small eigenvalue appear, making it hard to extract a power law behavior.\index{Hessian matrix!spectrum}

Overall, we see good evidence for a $\gamma$ around $2.5$ at several values of $H_\amp$, and at other fields the data is not in contradiction with a hypothesis of universality
in the exponents \eqref{eq:hsgrf-exponents}. When the field is small we remark a change of trend from $\gamma\approx2.5$ to $\gamma<1.5$ 
at a value $\lambda^*$. The crossover $\lambda^*$ \index{crossover}\nomenclature[lambda....star]{$\lambda^*$}{crossover $\lambda$}\index{boson peak}
shifts towards zero as $H_\amp$ decreases. This probably indicates the presence of a \index{boson peak} boson peak, an excess of modes at low frequency. 
Signs of a boson peak in at $H_\amp=0$ can
be seen in figure \ref{fig:spectrum-dep}. In that case the mass of the spectrum is all concentrated at low $\lambda$, 
but there ought to be a Debye behavior, meaning that $\lambda^*$
is very little.
\index{exponent!soft modes!alpha@$\alpha$|)}
\index{exponent!soft modes!gamma@$\gamma$|)}
\index{exponent!soft modes!delta@$\delta$|)}

\section{Localization \label{sec:hsgrf-loc}}
\begin{figure}[!t]
\centering
 \includegraphics[width=0.48\columnwidth]{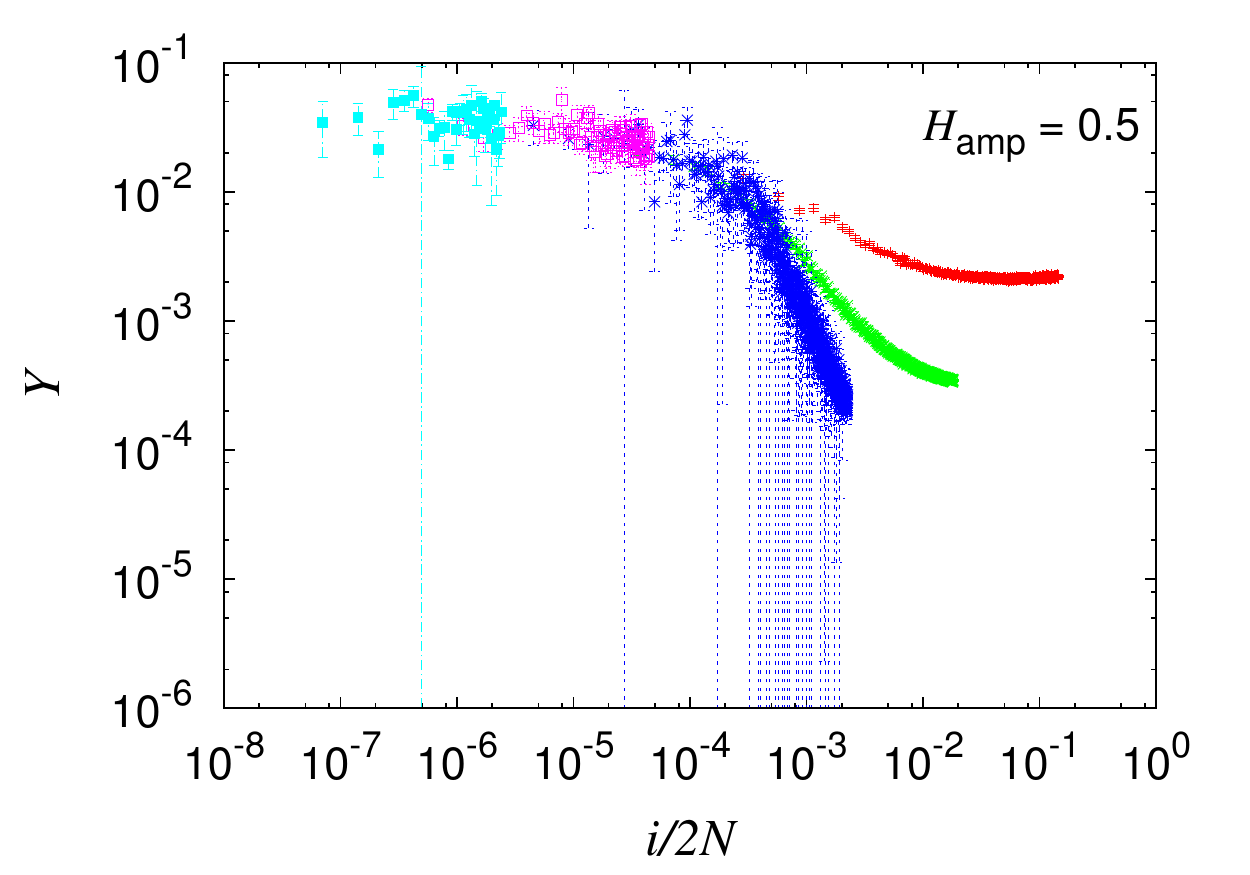}
 \includegraphics[width=0.48\columnwidth]{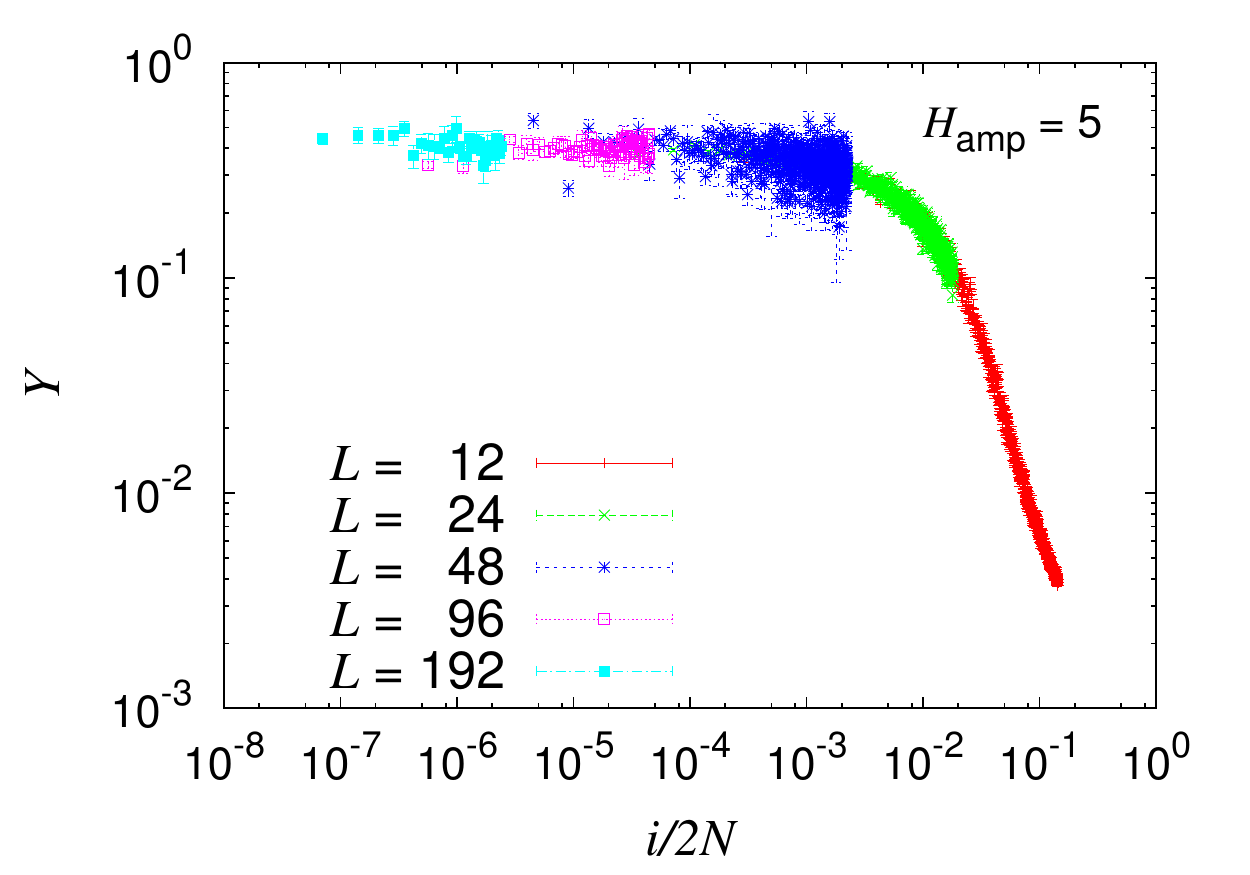}
 \caption[Participation ratio of the eigenvectors]{Participation ratio for $H_\amp=0.5$ (\textbf{left}) and $H_\amp=5$ (\textbf{right}).}
 \label{fig:loc}
\end{figure}

\index{localization|(}
We found that the application of a magnetic field does not induce a gap in the density of states.\index{gap}\index{density of states}
It goes to zero as a power law even in the presence of a not too large \ac{RF}, and it develops a gaps when the \ac{RF} is very large compared with the couplings.
What do these soft modes represent?
We want to know something more about the $2N$-dimensional eigenvectors $\ket{\tilde{\pi}_{\lambda}}$ of the matrix $\M$.\index{Hessian matrix!eigenvector}
Similarly as it happens in other types of disordered systems \cite{xu:10,degiuli:14,charbonneau:15}, the soft modes are localized,\index{modes!soft}
meaning that the eigenvectors $\ket{\pi_{\lambda}}$ are dominated by very few components.
To observe the localization we can define the inverse participation ratio 
\begin{equation}
\index{participation ratio}
\nomenclature[Y...l]{$Y_\lambda$}{inverse participation ratio}
Y_\lambda = \frac{\sum_{\bx}(|{\tilde\pi_{\lambda,\bx}|^2)^2}}{(\sum_{\bx}|\tilde\pi_{\lambda,\bx}|^2)^2} 
           = \frac{\sum_{\bx}(a_{1,i}^2+a_{2,\bx}^2)^2}{(\sum_{\bx}(a_{1,\bx}^2+a_{2,\bx}^2))^2}\,,
\end{equation}
where we coupled the two components corresponding to a single site because the local basis vectors have random directions,
so there would be no point in trying to distinguish one from the other.
If the eigenvector $\ket{\pi_{\lambda,\bx}}$ is fully localized in one site we will have $Y_\lambda = 1$. On the counterpart,
if all its components are the same (fully delocalized) we will have $Y_\lambda = 1/N$. In figure \ref{fig:loc} we show
that the softer the eigenvectors the more localized they are.
\footnote{Only in $H_\amp=0.01$ this was not clear, but we attribute it to strong echoes of the $H_\amp=0$ behavior. Due to this interference, we will basically
exclude the case of a very small field from our analysis.}
For small random fields (\ref{fig:loc}, left), we remark sizable finite-size effects, with the passage from localized to delocalized regime 
that becomes sharper as the lattice size is increased, suggesting the presence of localization threshold that separates a small fixed percentage of
localized eigenvectors from the delocalized bulk ones. For larger fields we appreciate no finite-size effects, and it appears that $\sim1\%$
of the eigenvectors is localized.

Since in a localized state the eigenvectors have a well-defined correlation length, 
\index{correlation!length!eigenvector|(}\index{correlation!length!Green|see{eigenvector correlation length}}
\index{Green's function@Green's function$ $|seealso{Green correlation length}}
\index{Green's function|(}
we can use also this criterion to probe the localization.
We can define a correlation length from Green's function $\mathcal{G}$, \nomenclature[G..G]{$\mathcal{G}$}{Green's function}
that is defined through the relation $\M\mathcal{G} = \delta_{\bx\by}$,
an is commonly used in field theory for two-point correlations.
Since $\M^{-1}$ shares eigenvectors $\psi_n$ \nomenclature[psi....n]{$\psi_n$}{eigenvector}
with $\M$ and has inverse eigenvalues $1/\lambda_n$, \index{Hessian matrix!eigenvector}\index{Hessian matrix!spectrum}
\footnote{For simplicity we use $N$-component eigenvectors $\psi_n(\bx)$ instead of the $2N$-component ones $\ket{\tilde\pi}$. The relationship
between the two can be recovered through $\psi_n^2(\bx)=\tilde\pi^2=\vec\pi^2$.}
Green's function is
\begin{equation}
 \mathcal{G}(\bx,\by) = \M^{-1}\delta_{\bx\by} = \sum_n \frac{\psi_n(\bx)\psi_n(\by)}{\lambda_n}\,,
\end{equation}
and squaring the relation
\begin{equation}
  \mathcal{G}^2(\bx,\by) = \sum_{m,n} \frac{\psi_m(\bx)\psi_m(\by)\psi_n(\bx)\psi_n(\by)}{\lambda_m\lambda_n}\,.
\end{equation}
By averaging over the disorder we gain translational invariance and $\overline{\mathcal{G}^2}$ can be written as a function of the distance $\br=\bx-\by$,
\begin{equation}
 \overline{\mathcal{G}^2(\br)} =  \overline{\sum_{m,n} \frac{1}{\lambda_m\lambda_n} \sum_\bx\left(\frac{[\psi_m(\bx)\psi_n(\bx)][\psi_m(\bx+\br)\psi_n(\bx+\br)]}{V}\right)}\,.
\end{equation}
Making the reasonable assumption that different eigenvectors do not interfere with each other, and exploiting the orthogonality condition $\sum_x\psi_m(\bx)\psi_n(\bx)=\delta_{mn}$,
we obtain the desired correlation function
\begin{equation}
\index{correlation!function!Green}
\nomenclature[C..r]{$\mathcal{C}(\br)$}{Green's correlation function}
 \mathcal{C}(\br) = \overline{\mathcal{G}^2(\br)} = \overline{\sum_{n} \frac{1}{\lambda_n^2} \psi^2_n(\bx)\psi_n^2(\bx+\br)}\,.
\end{equation}
This correlation function favors the softest modes by a factor $1/\lambda_n^2$. This is an advantage, because the bulk modes do not exhibit a finite correlation length,
so it is useful to have them suppressed.

\begin{figure}[!t]
\centering
 \includegraphics[width=0.48\columnwidth]{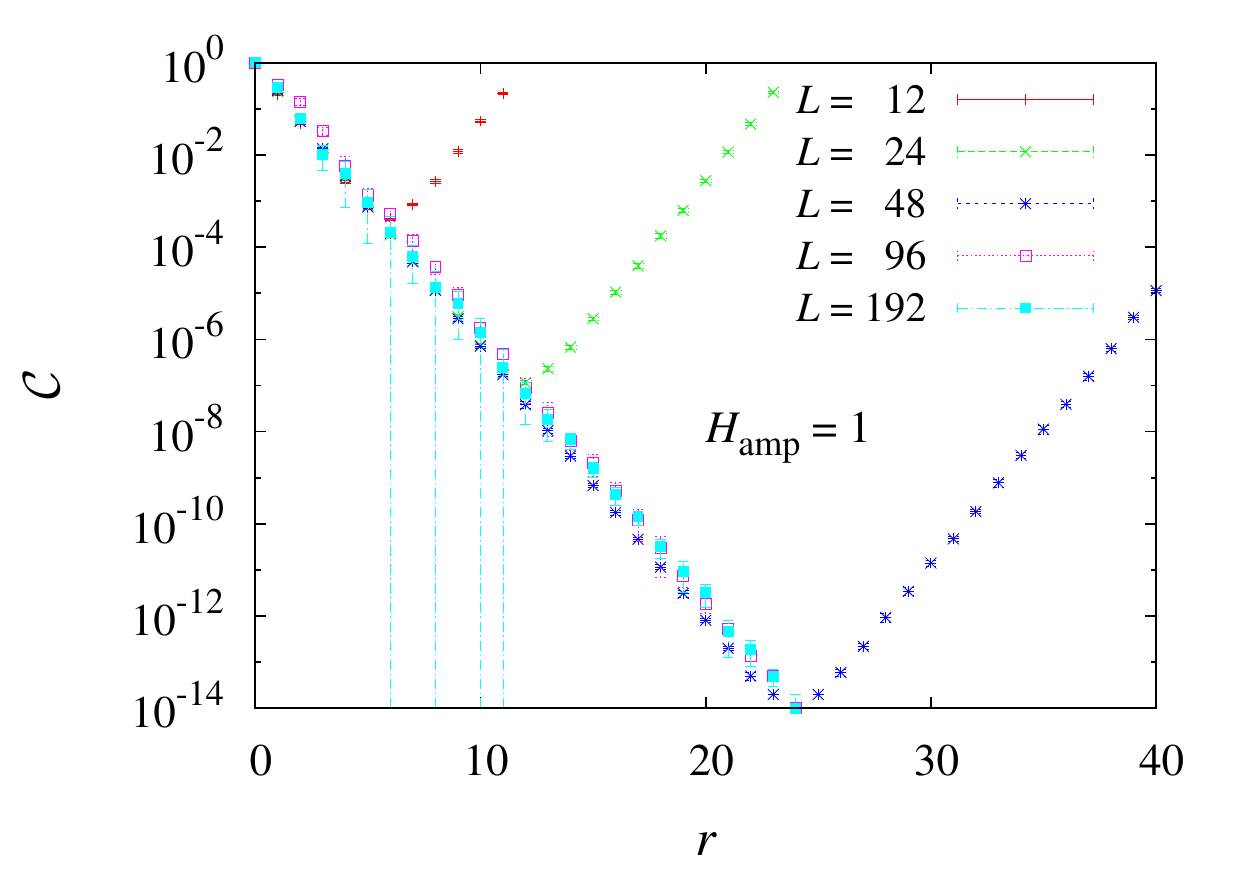}
 \includegraphics[width=0.48\columnwidth]{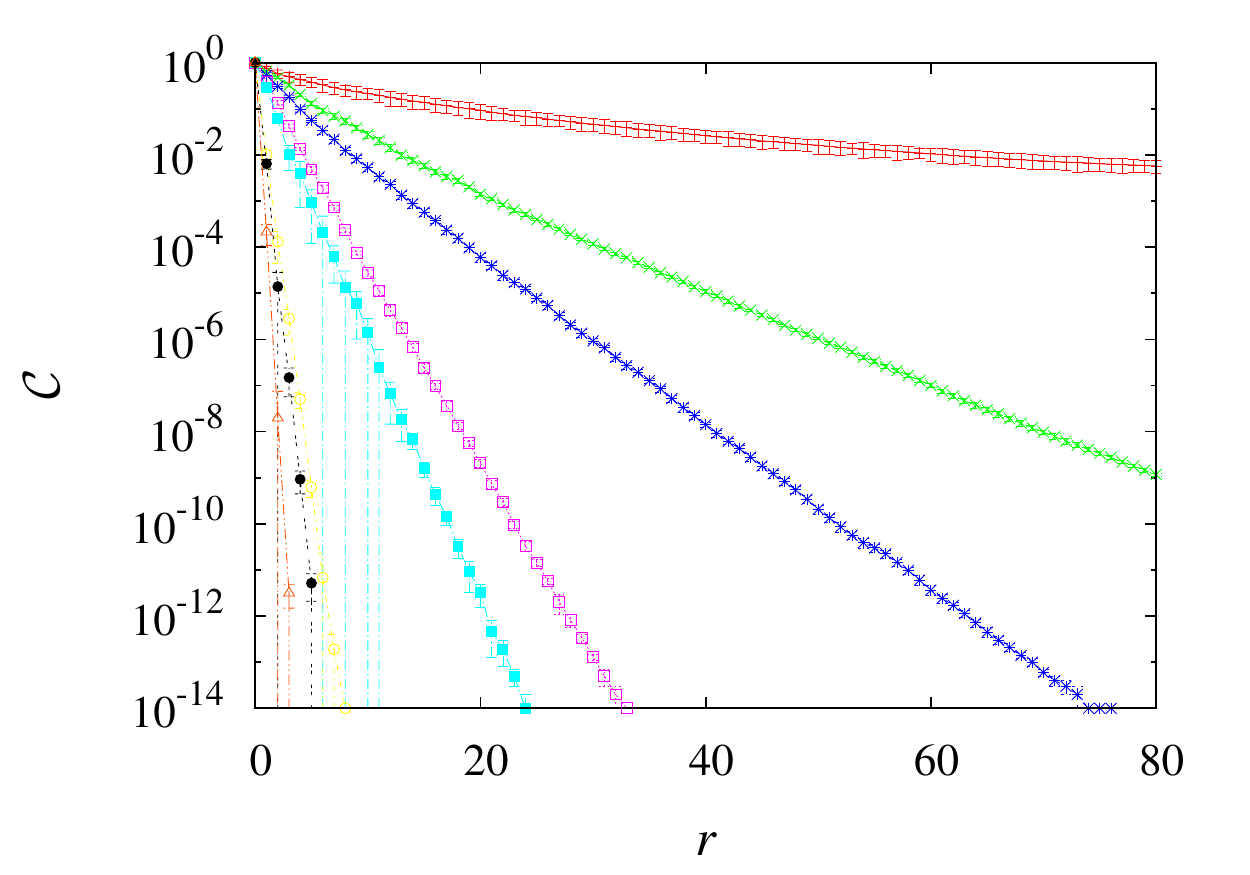}
 \caption[Correlations in the eigenvectors]
	  {
	   In the \textbf{left} set we show the correlation function $\mathcal{C}(r)$ \index{correlation!function!Green}
	   for different lattice sizes and $H_\amp=1$. Due to the periodic boundary 
	   conditions, when $r$ reaches $L/2$ the correlation function increases again.
	   On the \textbf{right} we fix the size to $L=192$ and show that the correlation length decreases with $H_\amp$.
	   The top curve, in red, is $H_\amp=0.01$, immediately under, in green, we have $H_\amp=0.05$, and so on with $H_\amp=0.1, 0.5, 1, 5, 10, 50$.
	  }
 \label{fig:hsgrf-xi}
\end{figure}

We calculated the correlations by inverting $\M$ with a conjugate gradient.
A nice exponential decay is visible (figure \ref{fig:hsgrf-xi}) to which we can associate a finite correlation length that grows as $H_\amp$ decreases.
\index{localization|)}
\index{Green's function|)}
\index{correlation!length!eigenvector|)}

\section{Anharmonicity}\index{anharmonicity|(}
The Hessian matrix $\M$ \index{Hessian matrix} is a harmonic approximation of 
the bottom of the valleys \index{valley} that carries plenty of information. Still, we can go\index{inherent structure}
beyond and take in account the effects due to the anharmonicity of the potential, and the relationship between different \acp{IS}.

The jamming point is characterized by diverging anharmonic effects (the softest modes have the smallest barriers) \cite{xu:10}. We \index{jamming}
are not able to define an equivalent of the jamming point, but we can seek for a dependency on $H_\mathrm{amp}$
of the anharmonic effects, and see for example if they diverge in null field.

\subsection{Forcings}\index{forcing|(}
\paragraph{Perturbing the Hamiltonian.}
We study the reaction of the system to an additional force along a direction $\ket{\vec\pi}$ \index{pion}
(identified with the $2N$-dimensional vector $\ket{\tilde\pi}$). 
We are interested in the softest mode, that is localized,\index{localization}
and we want to compare it with the behavior of the eigenvectors in bulk of the $\rho(\lambda)$, \index{Hessian matrix!spectrum}that are delocalized. 
Therefore we choose $\ket{\tilde\pi}=\ket{\tilde\pi_0}$ \nomenclature[pi....0]{$\ket{\tilde\pi_0}$}{softest mode}
(softest mode) and $\ket{\tilde\pi}=\ket{\tilde\pi_\RAND}$, \nomenclature[pi....RAND]{$\ket{\tilde\pi_\RAND}$}{random mode}
a vector whose components are chosen at random, with the condition $\langle \tilde\pi_\RAND\ket{\tilde\pi_\RAND}=1$.
The vector $\ket{\tilde\pi_\RAND}$ is not an eigenvector of $\M$, \index{Hessian matrix}
but it is generally a combination of all the eigenvectors of the system. Since the 
bulk eigenvectors overwhelm the soft modes by number $\ket{\tilde\pi_\RAND}$ will be representative of the bulk behavior. 
The reason why we use $\ket{\tilde\pi_\RAND}$ instead of an actual bulk eigenvector
is that with the Arnoldi algorithm we were able to compute only the \index{Arnoldi algorithm}\index{programming!ArPack|see{ArPack}}
lowest eigenvectors, so for the large lattices it was practically impossible to go beyond the localization \index{localization}threshold (recall figure \ref{fig:loc}).

With the application of a forcing along $\ket{\vec\pi}$, the Hamiltonian is modified in
\begin{equation}\label{eq:ham-forcing}
\nomenclature[H..F]{$\mathcal{H}_\F$}{forced Hamiltonian}
 \mathcal{H}_\F = -\sum_{\|\bx - \by\|=1} J_{\bx\by} \vec{s}_{\bx}\cdot \vec{s}_{\by} - \sum_{\bx}^{N} \left(\vec{h}_{\bx}+A_\F \vec{\pi}_{\bx}\right) \cdot \vec{s}_{\bx}\,,
\end{equation}
where $A_\F$ \nomenclature[A..2]{$A_\F$}{amplitude of the forcing} is the amplitude of 
the forcing along $\ket{\vec\pi}$, that will be tuned appropriately.

We stimulate the system with forcings 
of increasing amplitude, and study when this kicks the system out of the original inherent\index{inherent structure}
structure. To this scope $A_\F=A_\F(i_h)$, where $i_h\in\mathbb{N}$ \nomenclature[i....h]{$i_h$}{index of the forcing}
tunes the forcing.
\footnote{$\mathbb{N}$ is the set of the natural numbers.
\nomenclature[N...a]{$\mathbb{N}$}{set of the natural numbers}
}

The procedure is conceptually simple. Being $N_\F$ the number of forcings one wants to impose,\nomenclature[N...F]{$N_\F$}{number of forcings}
for $i_h$ in $\left\{1,\ldots,N_\F\right\}$\index{forcing!procedure}
\begin{enumerate}
 \item Start from the \ac{IS} $\ket{\vec{s}^{(\IS)}}$ of the unperturbed Hamiltonian \index{inherent structure}
$\mathcal{H}_\RF\equiv \mathcal{H}_\F(i_h=0)$.
 \item From $\ket{\vec{s}^{(\IS)}}$ minimize the energy using $\mathcal{H}_\F(i_h)$, and 
 find a new \ac{IS} for the perturbed system, $\ket{\IS(i_h)}$.\nomenclature[I.Sih]{$\ket{\IS(i_h)}$}{IS with the perturbed Hamiltonian}
 \item From $\ket{\IS(i_h)}$ minimize the energy again, using $\mathcal{H}_\F(0)=\mathcal{H}_\RF$,
 and find the \ac{IS} $\ket{\IS^{*}}$ (with elements $\vec{s}^{(\IS)*}_\bx$).
 \nomenclature[I.Sstar]{$\ket{\IS^{*}}$}{IS after the forcing}\nomenclature[s....star]{$\vec{s}^{(\IS)*}_\bx$}{spin of the IS after the forcing}
 \item If $\ket{\IS^*}=\ket{\vec{s}^{(\IS)}}$, the second minimization lead the system
 back to its original configuration, so the forcing was too weak to break through\index{energy!barrier}
 an energy barrier. On the contrary, if $\ket{\IS^*}\neq\ket{\vec{s}^{(\IS)}}$ the 
 forcing was large enough for a hop to another valley.\index{valley}
\end{enumerate}
Since this is an anharmonicity test, the same procedure for negative $i_h$ yields different results, therefore
in our simulations $i_h\in\{-N_\F,\ldots,0,\ldots,N_\F\}$.

To ensure well-defined forcings along $\ket{\vec\pi_\RAND}$, \index{forcing!normalization}
we normalized $A_\F$ with $\normauno{\ket{\vec\pi}}$, since
$\left|\sum_\bx \vec\pi_\bx\cdot\vec{s}_\bx\right| \leq \left|\sum_\bx \vec\pi_\bx\right|\leq \sum_\bx \left|\vec\pi_\bx\right|=\normauno{\ket{\vec\pi}}$.
Because $\normauno{\ket{\vec\pi}}$ scales nonlinearly with 
$N$, we multiplied back by a factor $N$, obtaining and extensive correction to the energy. For the softest
mode we analyzed the effect of intensive forcings of order $O(1)$ because larger forcings lead the system
out of the linear response regime.
The amplitudes we used can be resumed as
 \begin{align}
\label{eq:Arand}
 A_\F (i_h)&=\frac{NA i_h}{\normauno{\ket{\pi}}} ~~ \text{for} ~~\ket{\pi_\RAND}\,,\\
\label{eq:A0}
 A_\F (i_h)&= \frac{A i_h}{\normauno{\ket{\pi}}} ~~ \text{for} ~~\ket{\pi_0}\,.
 \end{align}
The amplitudes $A$ are an external parameter (of order 1), that we tried to tune in order to be in the linear
response regime for small $i_h$, and out of it for $i_h$ approaching $N_\F$. The dependency of the optimal $A$
on $L$ and $H_\amp$ was highly nonlinear. We list our choices in table \ref{tab:hsgrf-sim}.
 
\begin{figure}[!b]
 \centering
 \includegraphics[width=0.48\textwidth]{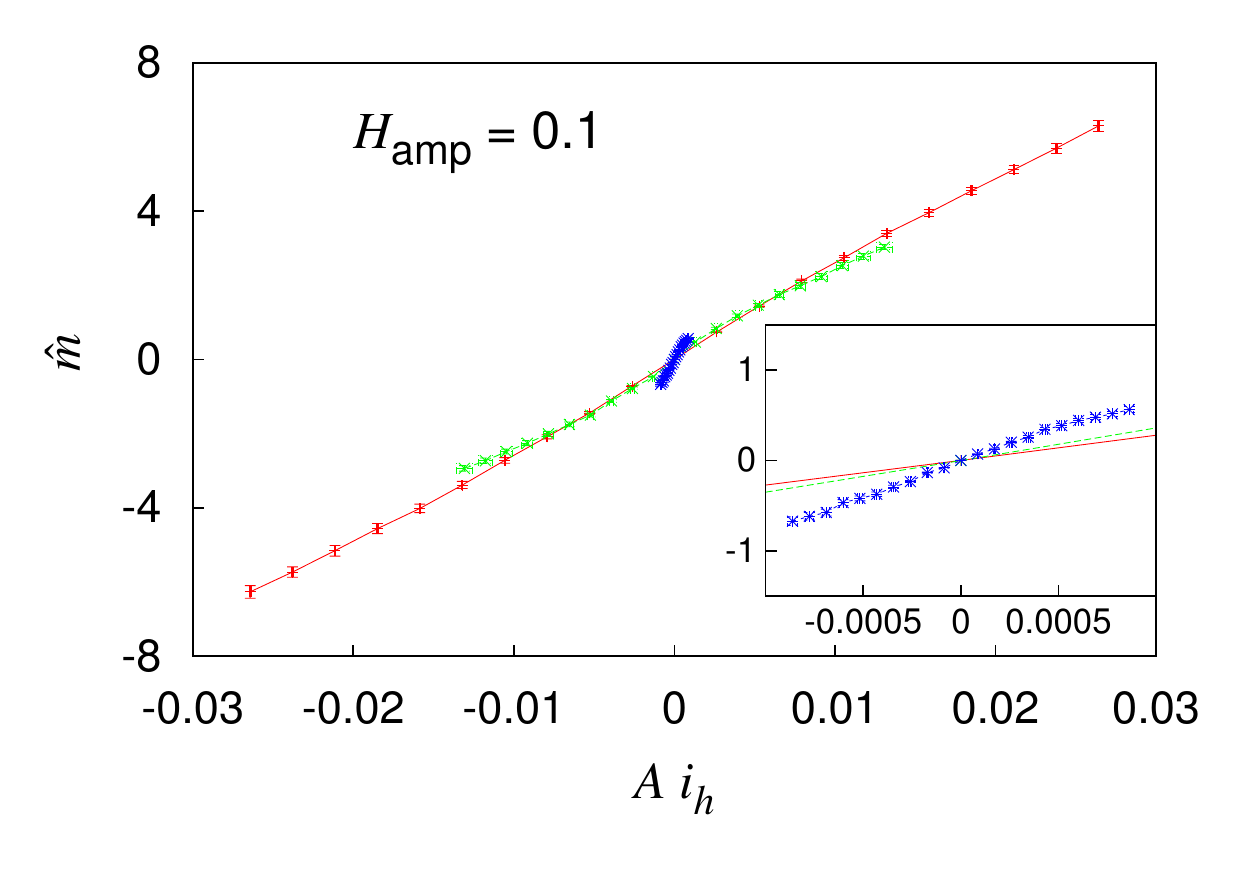}
 \includegraphics[width=0.48\textwidth]{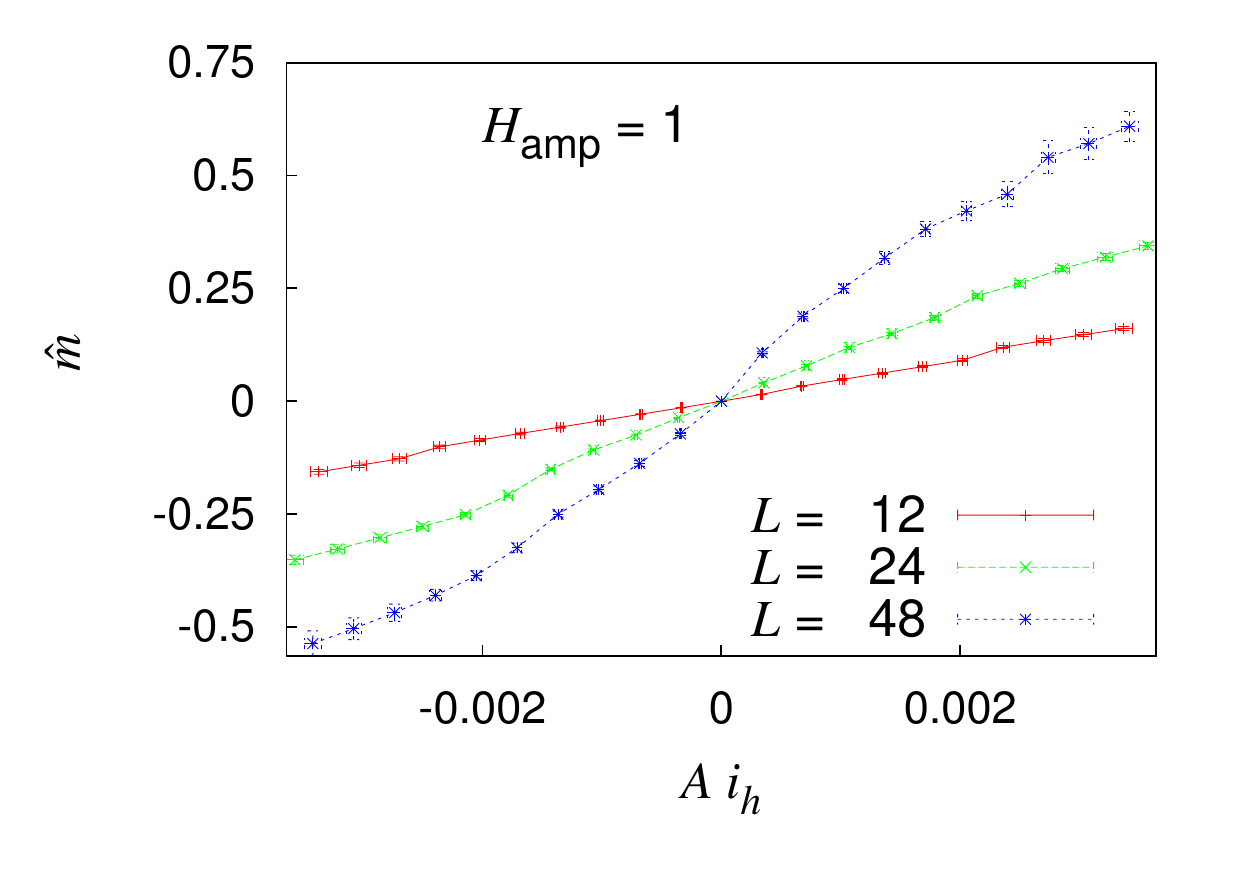}
 \caption[Polarized magnetization $\hat{m}$ of the forcings along $\ket{\vec\pi_0}$]
 {\index{polarized magnetization}
 Polarized magnetization $\hat{m}$ of the forcings along $\ket{\vec\pi_0}$,
 for $H_\amp=0.1$ (\textbf{left}) and $H_\amp=1$ (\textbf{right}). The \textbf{inset} is a zoom of the same data.}
 \label{fig:hatm-pi0}
\end{figure}

\begin{figure}[!t]
 \centering
 \includegraphics[width=0.48\textwidth]{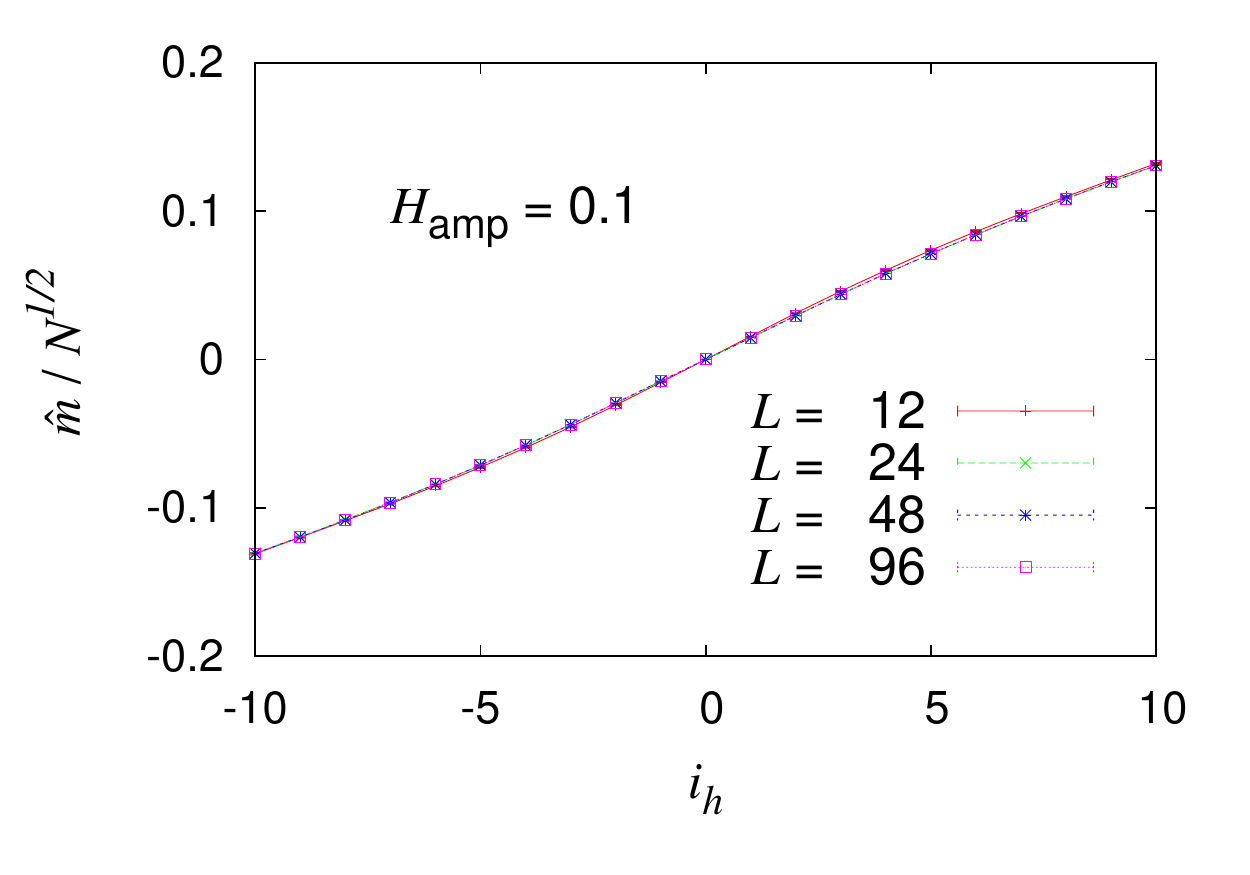}
 \includegraphics[width=0.48\textwidth]{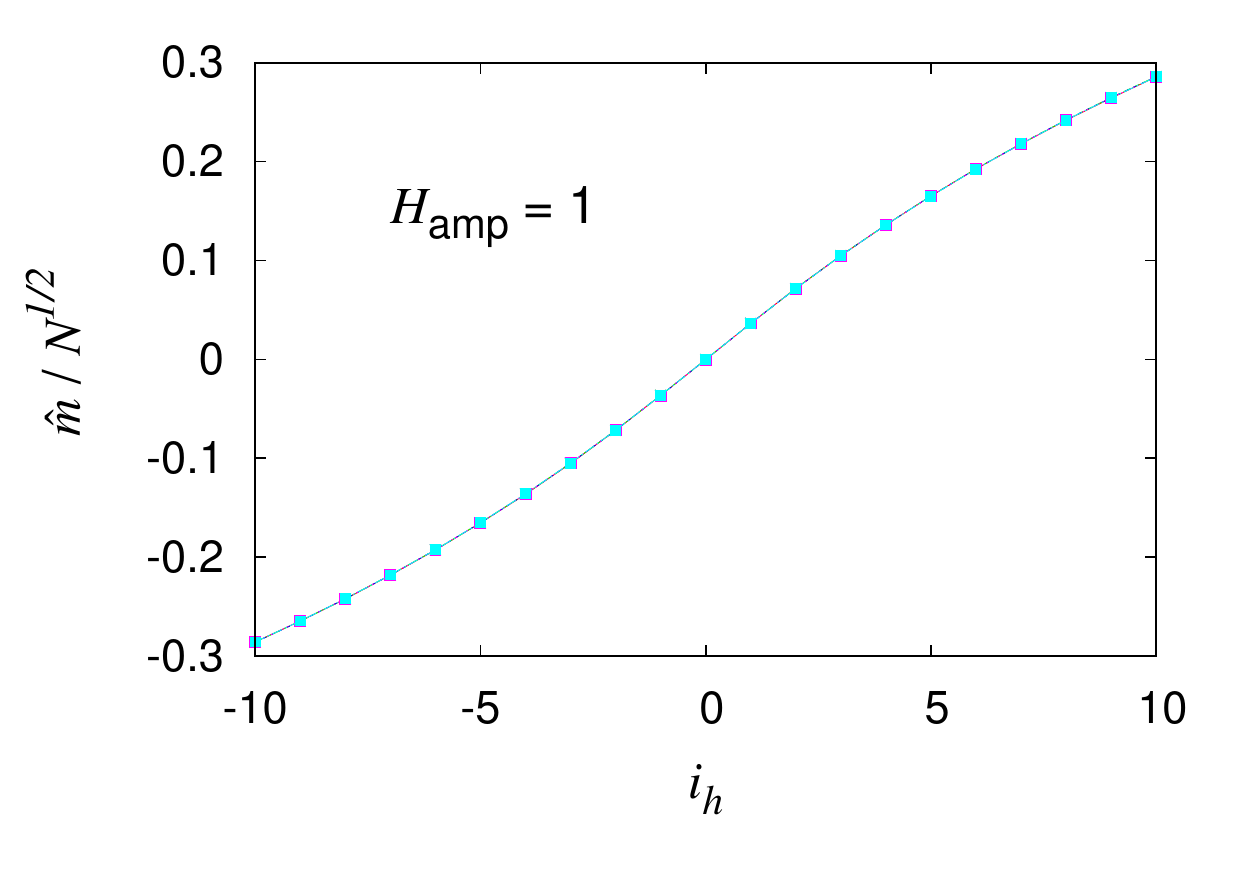}
 \caption[Recaled polarized magnetization $\hat{m}$ of the forcings along $\ket{\vec\pi_\RAND}$]
 {\index{polarized magnetization}
 Rescaled polarized magnetization $\hat{m}$ of the forcings along $\ket{\vec\pi_\RAND}$,
 for $H_\amp=0.1$ (\textbf{left}) and $H_\amp=1$ (\textbf{right}). The data are rescaled in order to collapse.}
 \label{fig:hatm-piRAND}
\end{figure}

\paragraph{Probing the linear regime}\index{forcing!linear regime}
To make sure that our forcings are not too strong, we monitor the direct reaction of the system
to the forcing.
We define a \nomenclature[m....hat]{$\hat{m}$}{polarized magnetization along the pion}\index{polarized magnetization}
``polarized magnetization'' \,$\hat{m} = \langle\IS(i_h)\ket{\vec\pi} = \sum_\bx \vec s_\bx\cdot\pi_{\bx}$, that indicates
how much the forcing pushed the alignment of the spins along the pion.
The amplitude of the forcing is tuned well if $\hat{m}(i_h)$ is close to the linear regime.
In table \ref{tab:hsgrf-sim} we show the amplitudes $A$ we used in order to be in the linear regime.
Figure \ref{fig:hatm-pi0} confirms that this was the working condition for the forcings along $\ket{\vec\pi_0}$. Figure \ref{fig:hatm-piRAND} is analogous,
but along $\ket{\vec\pi_\RAND}$. In the latter figure we rescale $\hat{m}$ by a factor $1/\sqrt{N}$ to obtain a collapse. 
In fact the normalization $\langle \vec\pi_\RAND\ket{\vec\pi_\RAND}=1$
implies that the components of $\ket{\vec\pi_\RAND}$ are of order $1/\sqrt{N}$, so the polarized magnetization is bounded by
$|\hat{m}| = |\langle IS(i_h)\ket{\vec\pi_\RAND}| \leq \sum_\bx |\vec\pi_\bx| \sim \sqrt{N}$.

The reader will notice that to be in the regime of quasi-linear response, forcings along $\ket{\pi_\RAND}$ can be extensive, 
whereas the localized forcings along $\ket{\pi_0}$ need to be of order 1. \index{localization}
%This suggests that it is practically only the soft modes that drive the modification of the configuration under an external force.

\begin{figure}[!t]
\centering
\includegraphics[width=0.48\textwidth]{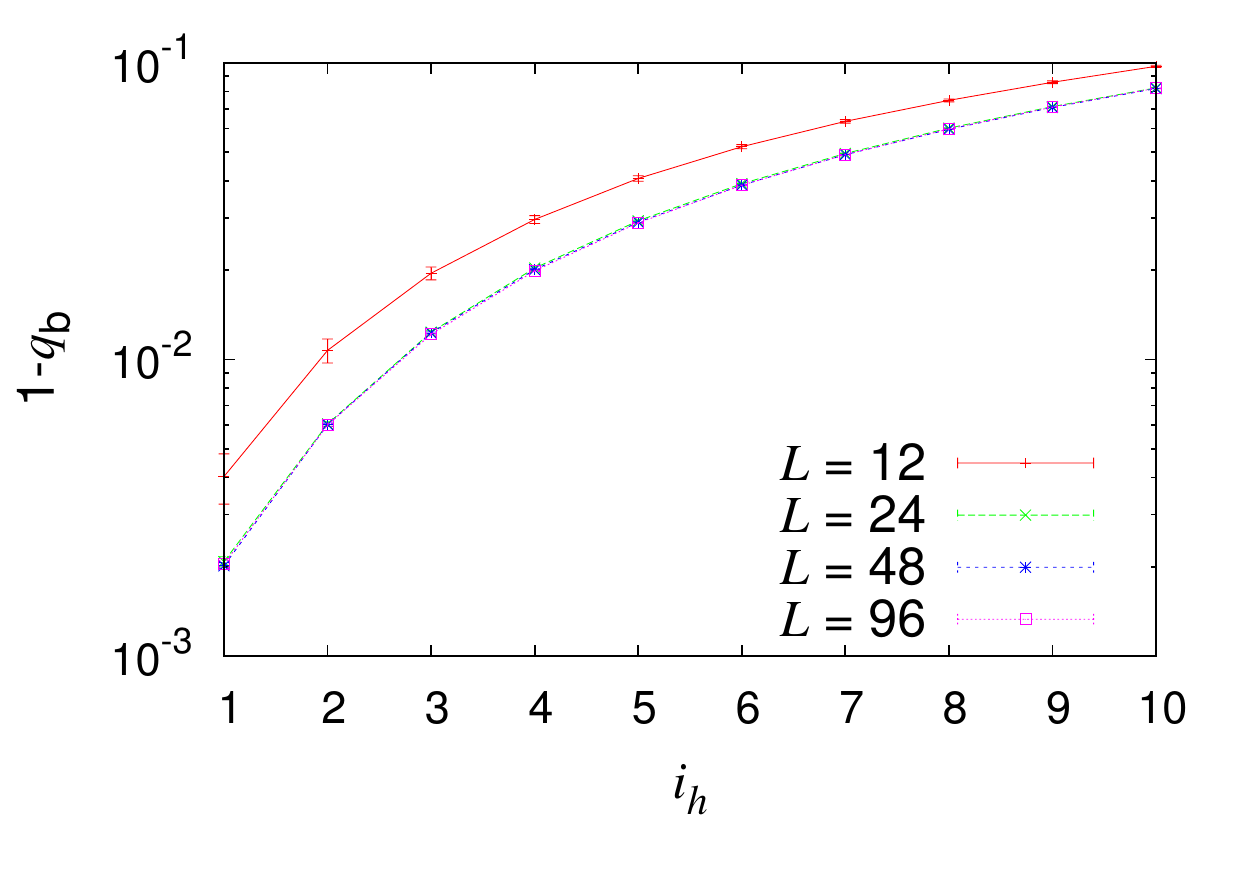}
\includegraphics[width=0.48\textwidth]{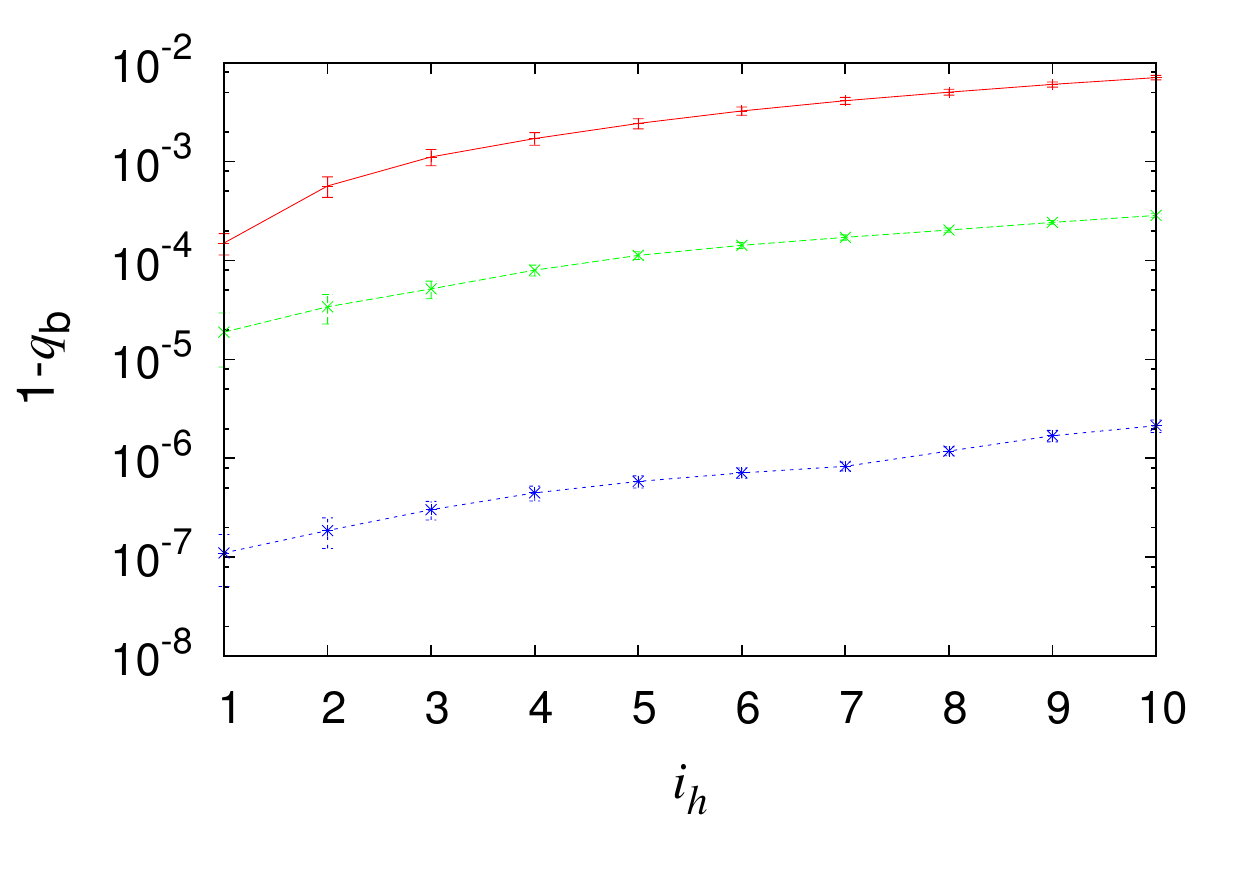}
\caption[Similarity between initial IS and forced configuration]
	{Measurement of how different the configuration is from the initial \ac{IS} is after the forcing along $\ket{\vec\pi_\RAND}$ (\textbf{left}),
	and along $\ket{\vec\pi_0}$ (\textbf{right}), for $H_\amp=0.1$. We plot $1-q_\mathrm{b}$ to make the figure clearer.
	Finite-size effects are neglectable for forcings along $\ket{\vec\pi_\RAND}$ and sizable along $\ket{\vec\pi_0}$.
	}
\label{fig:qb}
\end{figure}

\paragraph{The perturbed configuration}
After the first minimization (with the perturbed Hamiltonian) but before the second, we measure
the overlap $q_\mathrm{b}$ between $\ket{\vec{s}^{(\IS)}}$ and $\ket{\IS(i_h)}$, 
\nomenclature[q....b]{$q_\mathrm{b}$}{overlap between initial and perturbed IS}
 ~$q_\mathrm{b}=\langle\vec{s}^{(\IS)}\ket{\IS(i_h)}/N$ (figure \ref{fig:qb}),
and the energy difference $\Delta E$, in terms of $\mathcal{H}_\RF$, between $\ket{\IS(i_h)}$ and 
\nomenclature[Delta...E]{$\Delta E$}{In chapter \ref{chap:hsgrf}: energy difference in terms of $\mathcal{H}_\RF$}
 $\ket{\vec{s}^{(\IS)}}$, ~$\Delta E = E_\RF\big(\ket{\IS(i_h)}\big) - E_\RF\big(\ket{\vec{s}^{(\,\IS)}}\big)$.
The maximum value of $\Delta E$ before a hop to another valley \index{valley}should give an estimate of height of the barrier.\index{energy!barrier}
Still, it may happen that the minimum of the energy with Hamiltonian \eqref{eq:ham-forcing} have an energy 
lower than $E_\RF\big(\ket{\vec{s}^{(\,\IS)}}\big)$, so in a strict sense $\Delta E$ is not positive definite.
To overcome this issue, we resort to the energy difference $\Delta E^*$, \nomenclature[Delta...Estar]{$\Delta E^*$}{reverse energy barrier}
in terms of $\mathcal{H}_\RF$, between $\ket{\IS(i_h)}$ and 
 $\ket{(\IS^*}$, ~$\Delta E = E_\RF\big(\ket{\IS(i_h)}\big) - E_\RF\big(\ket{\vec{s}^{(\IS)}}\big)$, that 
 measures the barrier from the arriving \ac{IS} instead of the starting one. It has 
the advantage of being positive definite, and even though the eigenvector $\ket{\vec\pi_0}$ of the 
forcing is not associated to that \ac{IS}, we will see that the two \acp{IS} are so similar that it is a reasonable descriptor.

\paragraph{Ending in a new valley.}\index{valley}\index{energy!landscape|(}
For each $A_\F(i_h)$ we measure the overlap $q_\mathrm{if}$ 
\nomenclature[q....if]{$q_\mathrm{if}$}{overlap between configurations before and after the forcing}
between the two minimas of $\mathcal{H}_\RF$, the initial \ac{IS},
$\ket{\vec{s}^{(\,\IS)}}$, and the final one, $\ket{\IS^*}$. Na\"ively, checking that $q_\mathrm{if}<1$ in principle is a good
criterion to establish whether the system escaped to another valley. We proceeded similarly, in terms of the spin
variations between initial and final configuration, through the quantities
\begin{align}
\label{eq:wx}
\nomenclature[w....x]{$w_\bx$}{local spin variation}
 w_\bx       =& 1 - \left(\vec{s}^{\,(\IS)}_\bx \cdot \vec{s}^{\,(\IS)\,*}_\bx\right)\,,\\
\label{eq:W}
\nomenclature[W...W]{$W$}{global spin variation}
 W           =& \sum_\bx^N w_\bx = N - \langle\vec{s}^{\,(\IS)}\ket{\IS^*} = N (1-q_\mathrm{if}) \,,\\
\label{eq:w}
\nomenclature[w..w]{$\mathcal{W}$}{cumulant of the spin variation}
 \mathcal{W} =& \frac{\sum_\bx^N w_\bx^2}{\left(\sum_\bx^N w_\bx\right)^2}\,.
\end{align}
The local variation $w_x$ measures the change between the beginning and the end of the process. If the spin
stayed the same then $w_x=0$, while if it became uncorrelated with the initial position $w_x=1$ in average.
If one and only one spin becomes uncorrelated with its initial configuration, the variation of $W$ is $\Delta W=1/N$.
\nomenclature[Delta...W]{$\Delta W$}{variation of $W$ after the forcing}
Similar variations $\Delta W$ do not mean that one spin has decorrelated and the others have stayed the same, this
is impossible because $\ket{\vec{s}^{(\IS)}}$ and $\ket{\IS^*}$ are \acp{IS} and collective rearrangements are needed. 
A $\Delta W=1/N$ means instead that the overall change is equivalent to a single spin becoming independent of its 
initial state.
This is, for a rearrangement, the minimal change in the $W$ that we can define. Since the spins in our model
are continuous variables, we impose $\Delta W = 1/N$ as a threshold to state whether there was or not
a change of valley.

The cumulant $\mathcal{W}$ is an indicator of the type of rearrangement that took place. If the rearrangement is completely localized
(only one spin changes), $\mathcal{W}=1$, whereas if it is maximally delocalized (all the spins have the same variation), then $\mathcal{W}=1/N$. 

\subparagraph{Falling back in the same valley.}\index{valley}\index{numerical simulations}
Even though the forcing is along a definite direction, since the energy landscape is very irregular,
it may happen that stronger forcings lead the system to the originary valley. For example it may happen that
$i_h=2$ lead the system to a new valley, and $i_h=3$ lead it once again to the same valley of $i_h=1$. To exclude
these extra apparent valleys we label each visited valley with its $W$, and assume that two valleys with the same 
label are the same valley.
These events are not probable, and even less likely it is that this happen with two different but
equally-labelled valleys, so we neglect the bias due to this unlucky possibility.

\subsection{Rearrangements}\index{forcing!rearrangement}
To delineate the effect of the forcings, we want to study, for every couple $(H_\amp,L)$, 
the probability that a forcing of amplitude $A_\F$ lead the system to a new valley, 
to distinguish the behavior of soft from bulk modes.

Furthermore, once the system made its first jump to a new valley, it is not excluded that a bigger forcing
lead it to a third minimum of the energy. One can ask himself what is the probability $P_{L}^{H}(A_\F, n)$
\nomenclature[P...HLhn]{$P_{L}^{H}(A_\F, n)$}{probability of changing valley}
that $n$ \nomenclature[n....8]{$n$}{In chapter \ref{chap:hsgrf}: number of visited valleys}
new valleys are reached by forcing the system with an amplitude up to $A_\F(i_h)$, and to try to evince 
a dependency on sistem size $L$ and random field amplitude $H_\amp$. Even though $n$ is bounded by $i_h$, this does not necessarily mean that if we
made smaller and more numerous forcings $n$ could not be larger.
On another side, if for a certain parameter choice rearrangements are measured only for large $i_h$, it is reasonable to think
that these represent the smallest possible forcings to fall off the \ac{IS}.

To construct $P_{L}^{H}(A_\F, n)$, for every replica and sample
we start from $i_h=0$ and increase $|i_h|$ either in the positive or negative direction (the two are accounted for
independently). The value we assign to $P_{L}^{H}(A_\F, n)$ is the number of systems that had $n$
rearrangements after $i_h$ steps, divided by the total number of forcings, that is $2N_\mathrm{replicas}N_\mathrm{samples}$.

\begin{figure}[!t]
\centering
\includegraphics[width=0.48\textwidth]{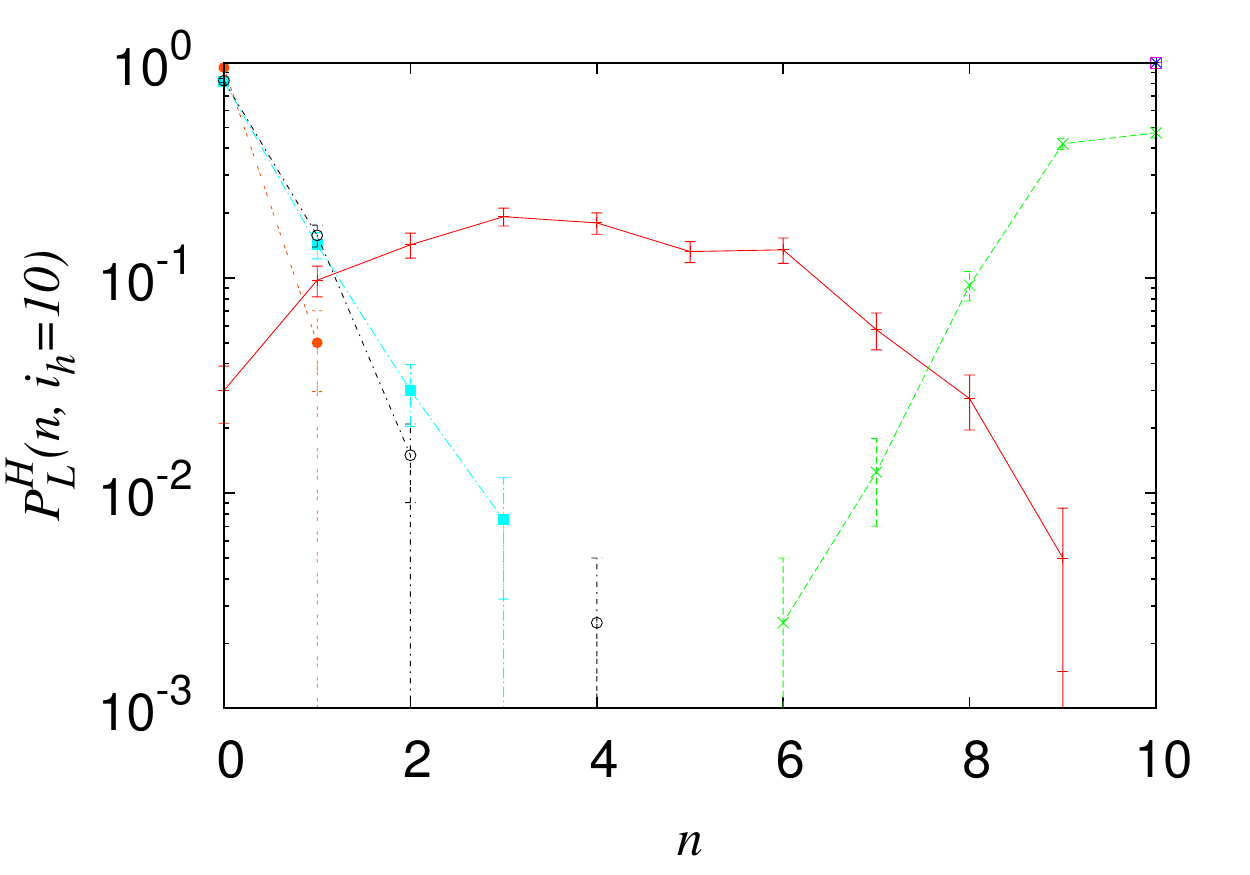}
\includegraphics[width=0.48\textwidth]{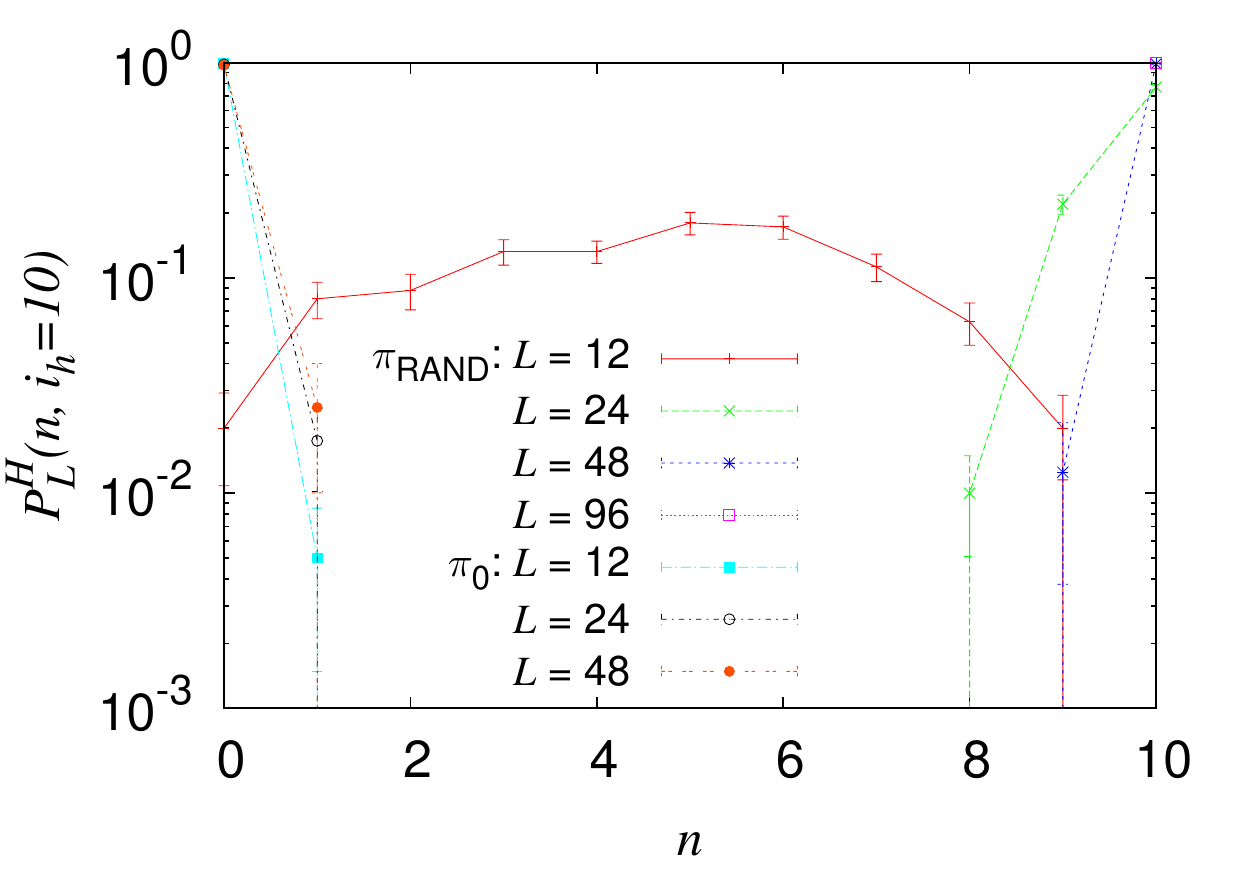}
\caption[Probability of $n$ rearrangements]
	{Probability of there being exactly $n$ changes of valley after $i_h=N_\F=10$ forcing steps. \index{valley}
	The data come from $H_\amp=0.1$ (\textbf{left}) and $H_\amp=1$ (\textbf{right}).
	If $P_{L}^{H}(A_\F, n)=1$ for $n=0$ it means that the forcings were not strong enough to ever get out of \index{inherent structure}the initial \ac{IS}. On the 
	contrary, $P_{L}^{H}(A_\F, n)=1$ for $n=10$ means that every single step lead the system to a new \ac{IS}. The latter scenario is realized
	in the case of forcings along $\ket{\vec\pi_\RAND}$, especially when the system size is large.
	On the other side, forcings along $\ket{\vec\pi_0}$ display a small but finite amount of rearrangements.}
\label{fig:first-rearrangement-1}
\end{figure}

\begin{figure}[!t]
\centering
\includegraphics[width=0.48\textwidth]{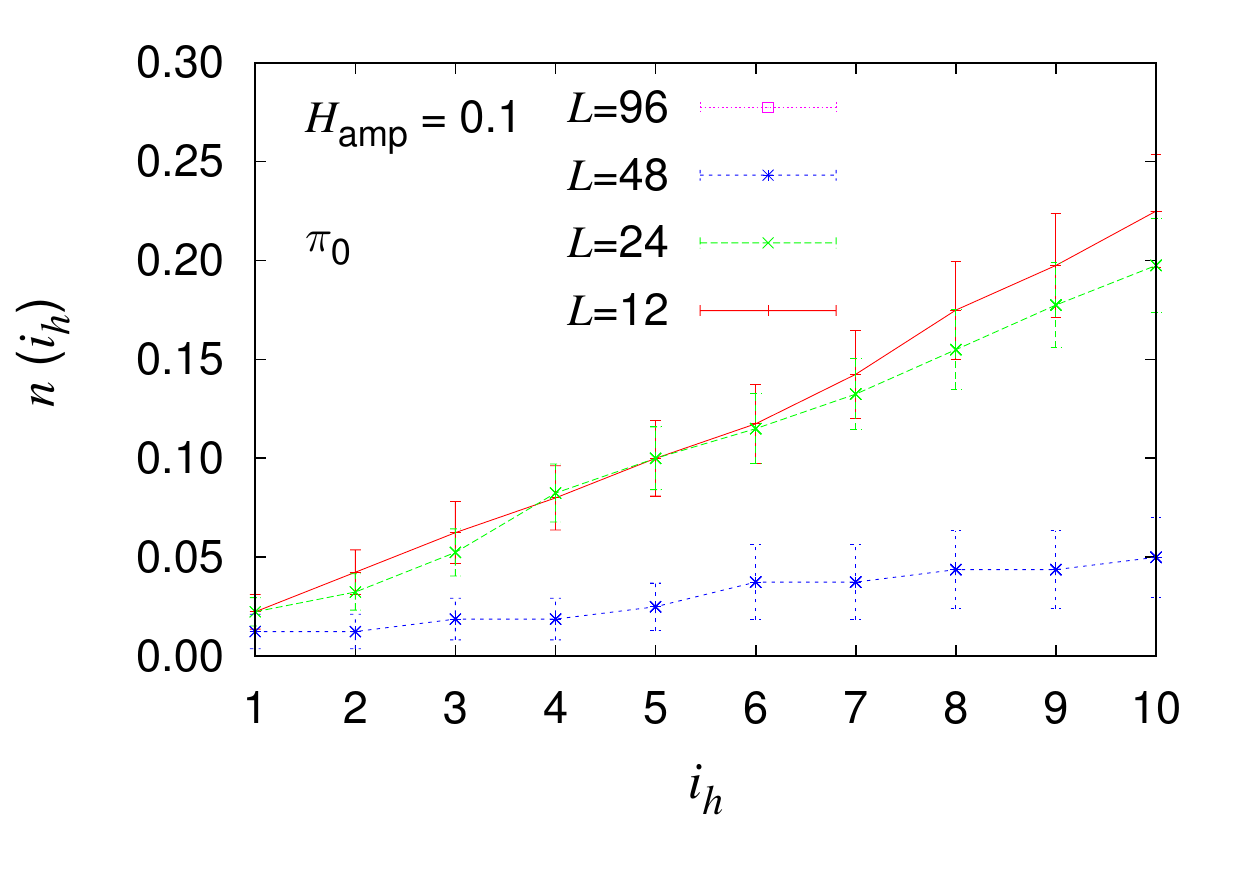}
\includegraphics[width=0.48\textwidth]{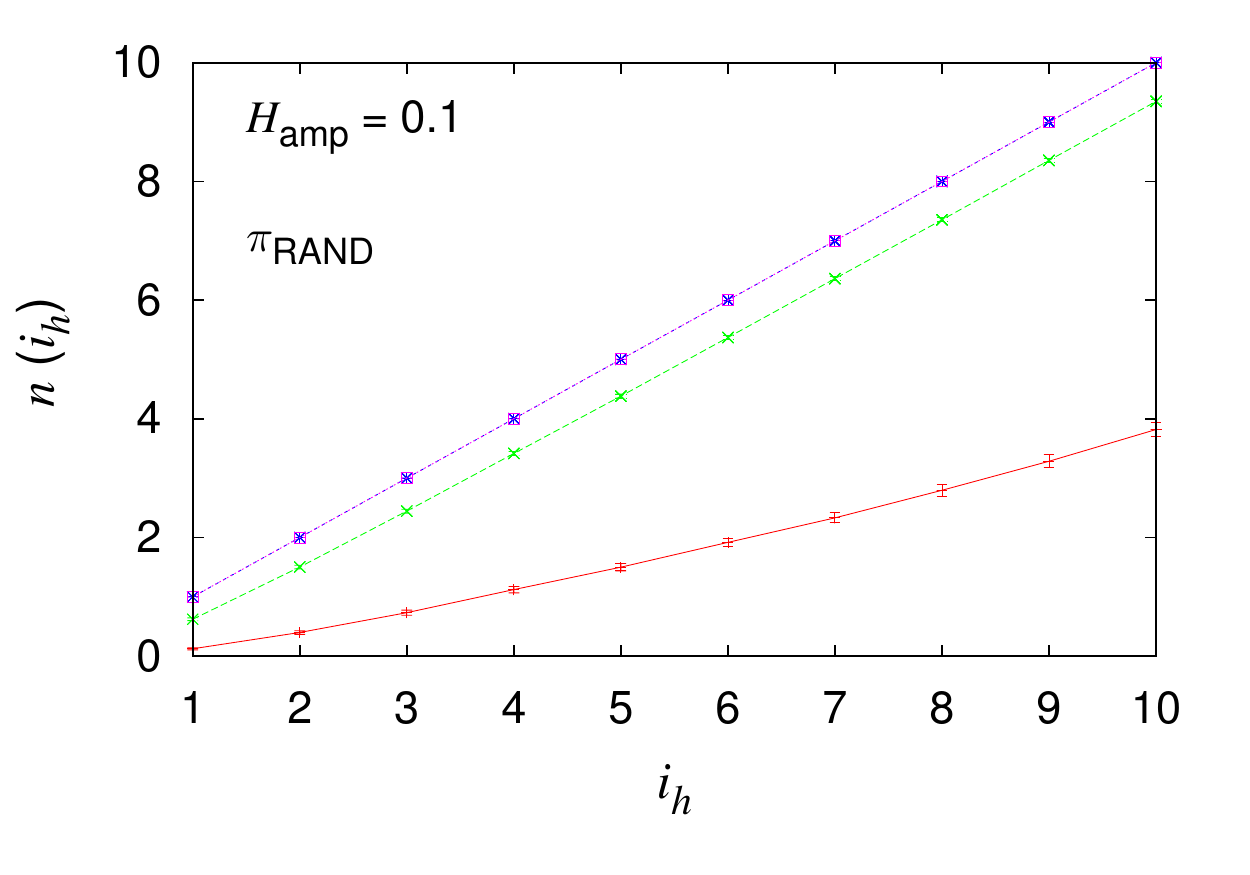}
\includegraphics[width=0.48\textwidth]{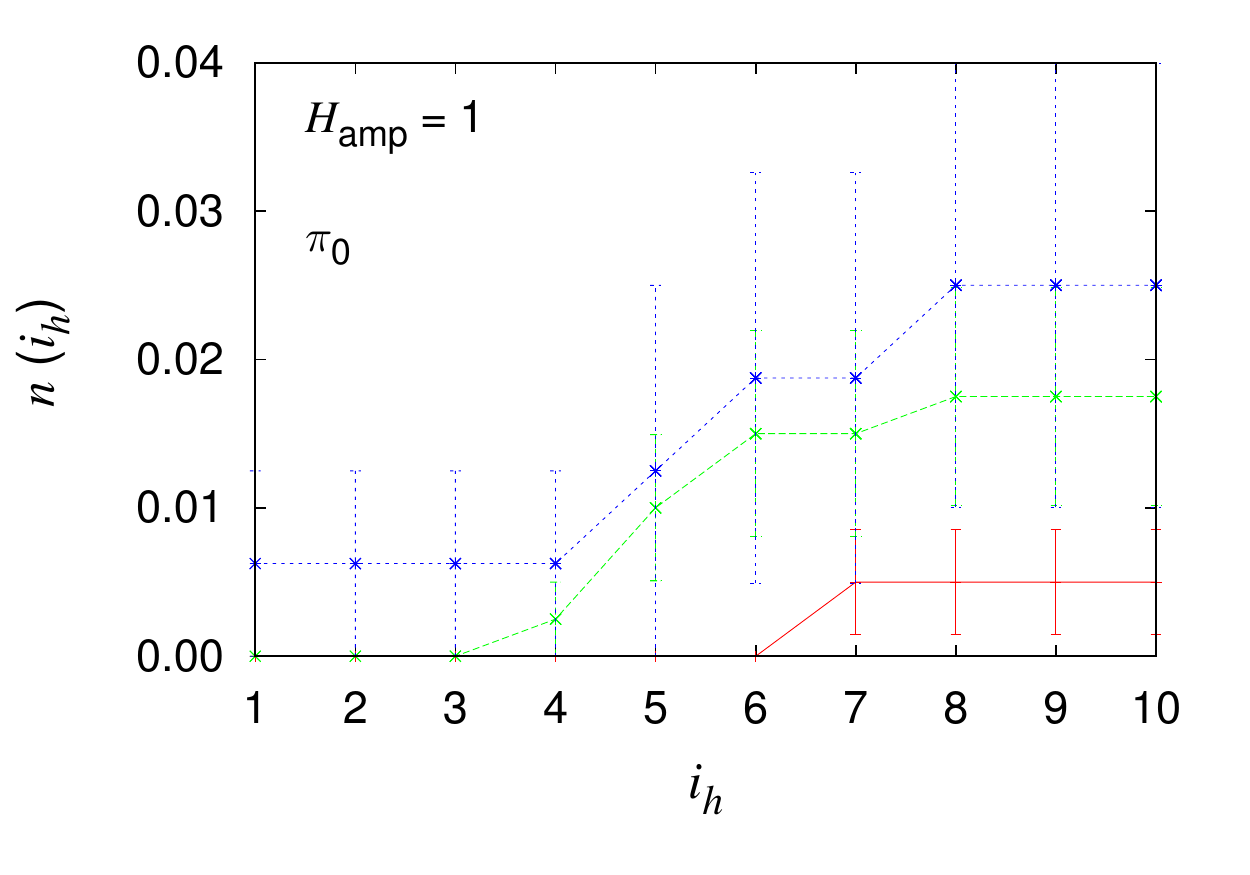}
\includegraphics[width=0.48\textwidth]{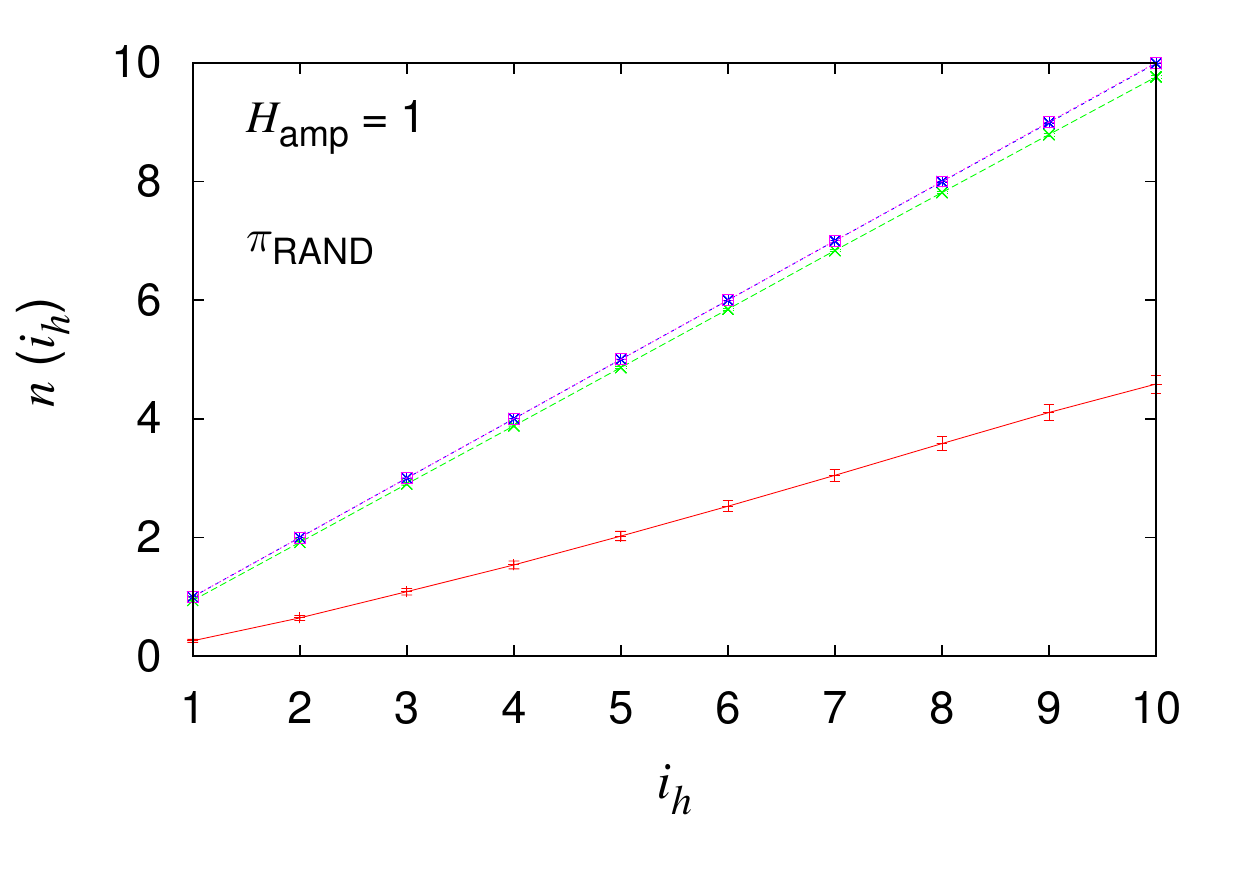}
\caption[Average number of rearrangements]
	{Average number of rearrangements $n(i_h)$ for forcings along $\ket{\vec\pi_0}$ (\textbf{left}) and along $\ket{\vec\pi_\RAND}$ (\textbf{right}). 
	The data come from $H_\amp=0.1$ (\textbf{top}) and $H_\amp=1$ (\textbf{bottom}). \index{inherent structure}
	When the lattice becomes large enough, the forcings along $\ket{\vec\pi_\RAND}$ lead to a new \ac{IS} every time $i_h$ is increased.
	The data from the $\ket{\vec\pi_0}$ and $H_\amp=1$ can be said to be in the regime
	of first rearrangement.
	}
\label{fig:first-rearrangement-2}
\end{figure}

\paragraph{First rearrangement.}\index{rearrangement!first}
In figure \ref{fig:first-rearrangement-1} we show the probability of measuring exactly $n$ rearrangements after
$i_h=N_\F=10$ forcing steps. 
\footnote{We do not show data regarding forcings for $H_\amp=10, 50$, because no arrangement takes place. 
Most likely the energy landscape is too trivial.}
Even though both for $\ket{\vec\pi_\RAND}$ and $\ket{\vec\pi_0}$ we are in the linear response regime, the behavior
is very different between the two types of forcing. In the first case every single forcing step we impose leads the system\index{valley}
to a new valley. In the second rearrangements are so uncommon that even though the probability of having exactly one rearrangement is
finite, that of having more than one becomes negligible for large samples.
It is then reasonable to think that any rearrangement we measure for $\ket{\vec\pi_0}$, it occurs for the smallest possible forcing, and
even when more than one occurs, these jumps are between \emph{neighboring valleys}, where by neighboring we mean that no smaller forcing
would lead the system to a different \ac{IS}.
To convince ourselves of this we can give a look at the average number of rearrangements after $i_h$ forcing steps, $n(i_h)$ (figure \ref{fig:first-rearrangement-2}). 
\footnote{
Because $P_{L}^{H}(A_\F, n)$ is not defined over all the samples
(it is hard to reach many different valleys and it may not happen in all the simulations),
the errors on $P_{L}^{H}(A_\F, n)$ were calculated by resampling over the reduced data sets
with the bootstrap method.
}
When $i_h$ is small no new \acp{IS} are visited and $\mean{n}=0$, while for larger $i_h$, $\mean{n}$ is positive but small, so we can call
these changes of valley ``first rearrangements'', i.e. rearrangement between \emph{neighboring valleys}.

\subsection{Two-level systems}
\begin{figure}[!t]
\centering
\includegraphics[width=0.48\textwidth]{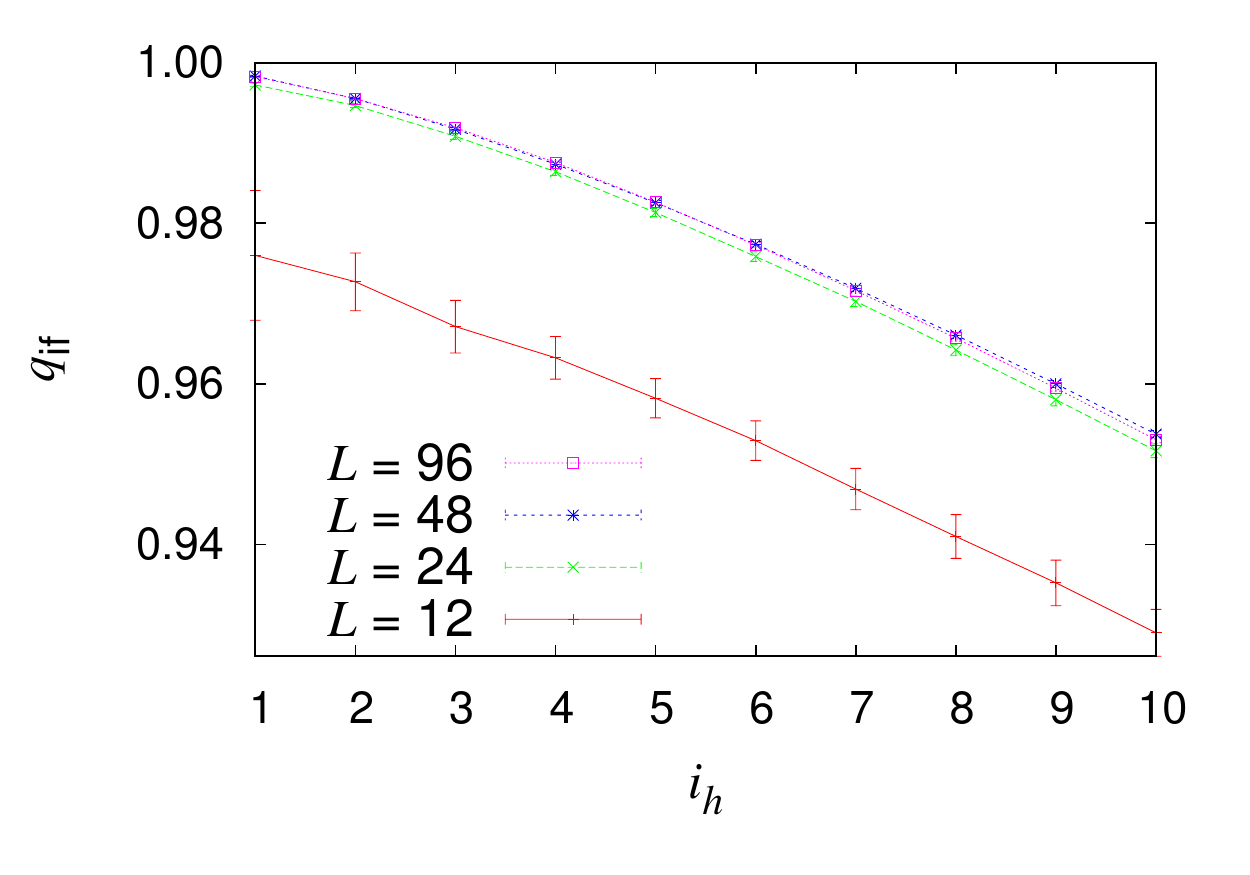}
\includegraphics[width=0.48\textwidth]{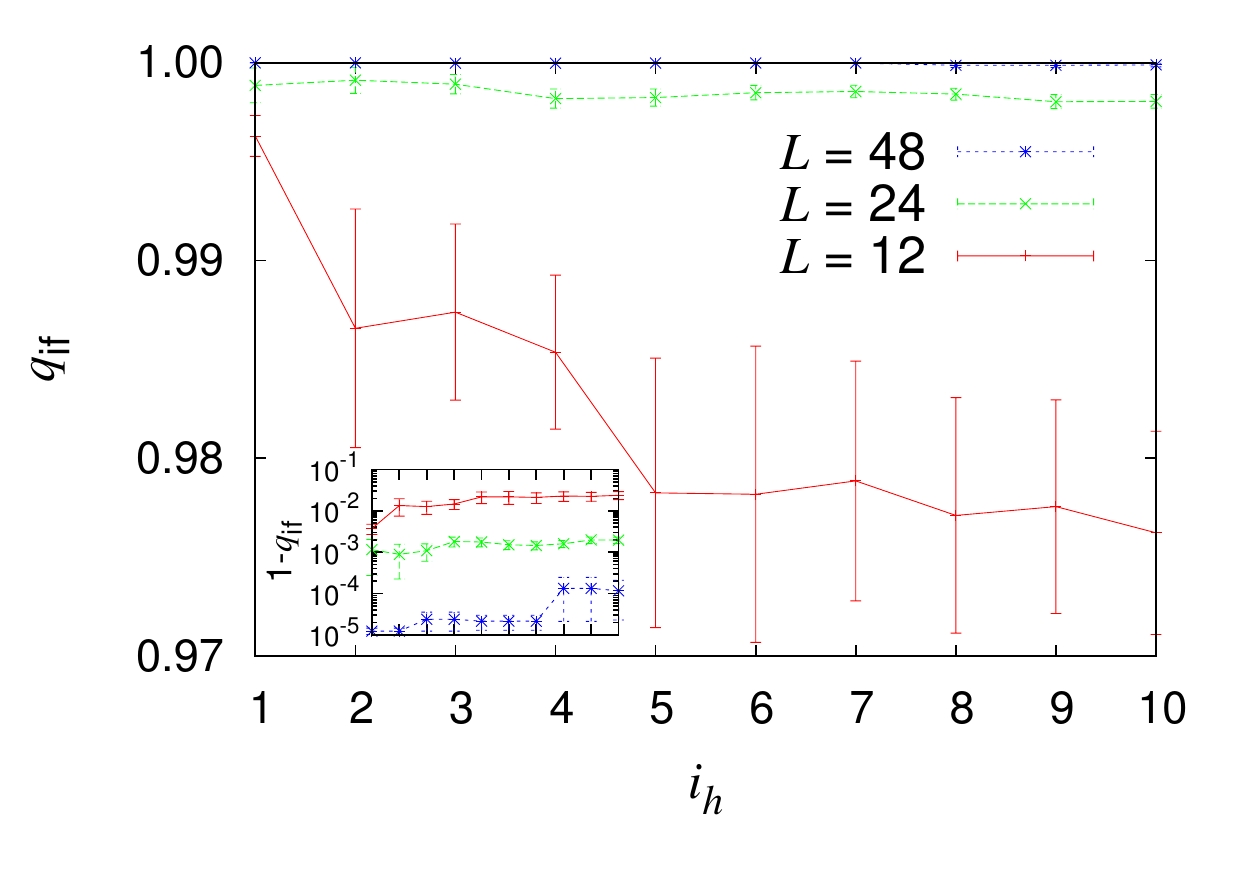}
\includegraphics[width=0.48\textwidth]{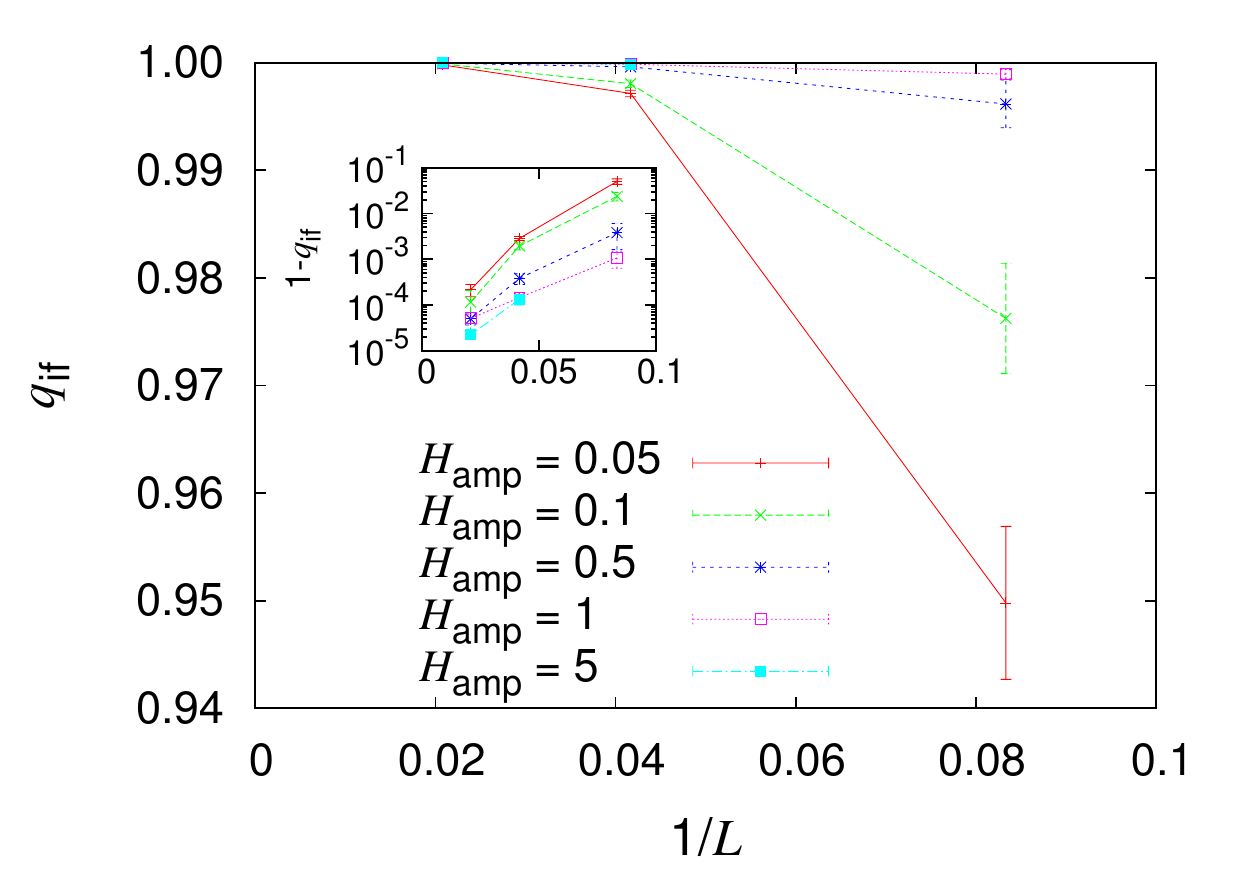}
\includegraphics[width=0.48\textwidth]{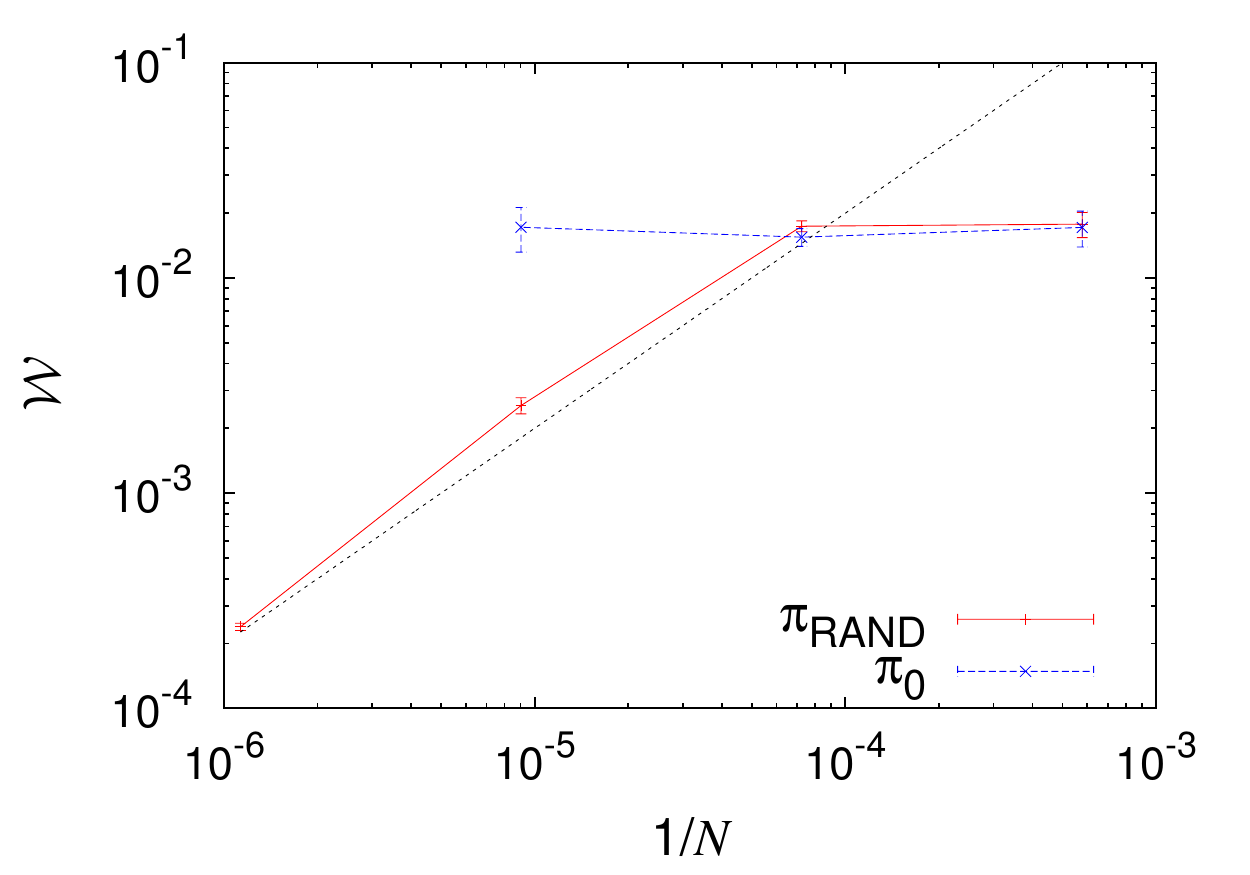}
\caption[Overlaps in the forcings]{The two top plots show the overlap \index{overlap!forcing}\index{overlap!scalar}\index{rearrangement}
				   $q_\mathrm{if}=\langle\vec s^{\,(\IS)}\ket{IS^*}/N$ between the starting and the final \ac{IS},
                                   for $H_\amp=0.1$. 
                                   \textbf{Top left}: $q_\mathrm{if}$ in forcings along $\ket{\vec\pi_\RAND}$. 
                                   \textbf{Top right}: $q_\mathrm{if}$ in forcings along $\ket{\vec\pi_0}$.
                                   The overlaps $q_\mathrm{if}$ are consistently larger than the typical overlap between two inherent structures (figure \ref{fig:PQ-IS}, left), peaked around $P(q_\IS)\simeq0.4$.
                                   The \textbf{bottom left} set shows data for forcings along $\ket{\vec\pi_0}$ for all the interesting $H_\amp$. A point is missing, for $H_\amp=5$, $L=12$, because we only registered
                                   a single rearrangement for this data set.
                                   Both \textbf{insets} display $1-q_\mathrm{if}$ from the same data of the corresponding larger plot, 
                                   to stress that the overlaps $q_\mathrm{if}$ never reach 1 (this is redundant, because $q_\mathrm{if}<1$ by definition, since it is the overlap between two 
                                   different configurations).
                                   The \textbf{bottom right} figure depicts the type of rearrangement that takes place between the initial and final \ac{IS}.
                                   The cumulant $\mathcal{W}$ is defined in \eqref{eq:w}; $\mathcal{W}=1$ indicates a completely localized rearrangement, where only a single spin changed position, 
                                   while $\mathcal{W}=1/N$ indicates a fully delocalized change of the spins. 
                                   It is visible that a random forcing leads to a completely delocalized rearrangement (the dotted line is $\sim1/N$), 
                                   whereas a localized forcing implies a localized rearrangement with no appreciable dependency on the system size.
                                   }
\label{fig:forcing-overlaps}
\end{figure}

\index{two-level system|(}
In the spectrum of $\M$, $\rho(\lambda)$, \index{Hessian matrix!spectrum}\index{modes!soft} we remarked
an extensive number of very soft modes, with a localized eigenstate (section \ref{sec:hsgrf-loc}). The eigenstates can connect
different \acp{IS} through the forcing procedure described in this section. The connection caused by such states is privileged, because the 
couples of \acp{IS} are innaturally near to each other. In figure \ref{fig:forcing-overlaps} we show the mean overlap between initial and final \ac{IS}, 
$q_\mathrm{if}=\langle\vec s^{\,(\IS)}\ket{IS^*}/N$.\index{overlap!forcing}\index{overlap!scalar}

As expectable, the rearrangements are localized when we stimulate the system along the softest mode, \index{rearrangement}
and delocalized when it is along a random direction (figure \ref{fig:forcing-overlaps}, inset).
The overlaps $q_\mathrm{if}$ are much closer to 1 than the overlaps of independent \acp{IS} shown in 
figure \ref{fig:PQ-IS}, meaning that the \acp{IS} are somewhat clustered in tiny
groups that are represented by a single \ac{IS}. This could be an operational definition of 
classical two-level system, i.e. a system in which there are two very close states
connected by a soft mode, where the transitions from one state to the other can be treated as independent 
of the rest of the system \cite{anderson:72,phillips:72,phillips:87,lisenfeld:15}.

\begin{figure}[!t]
\centering
\includegraphics[width=0.48\textwidth]{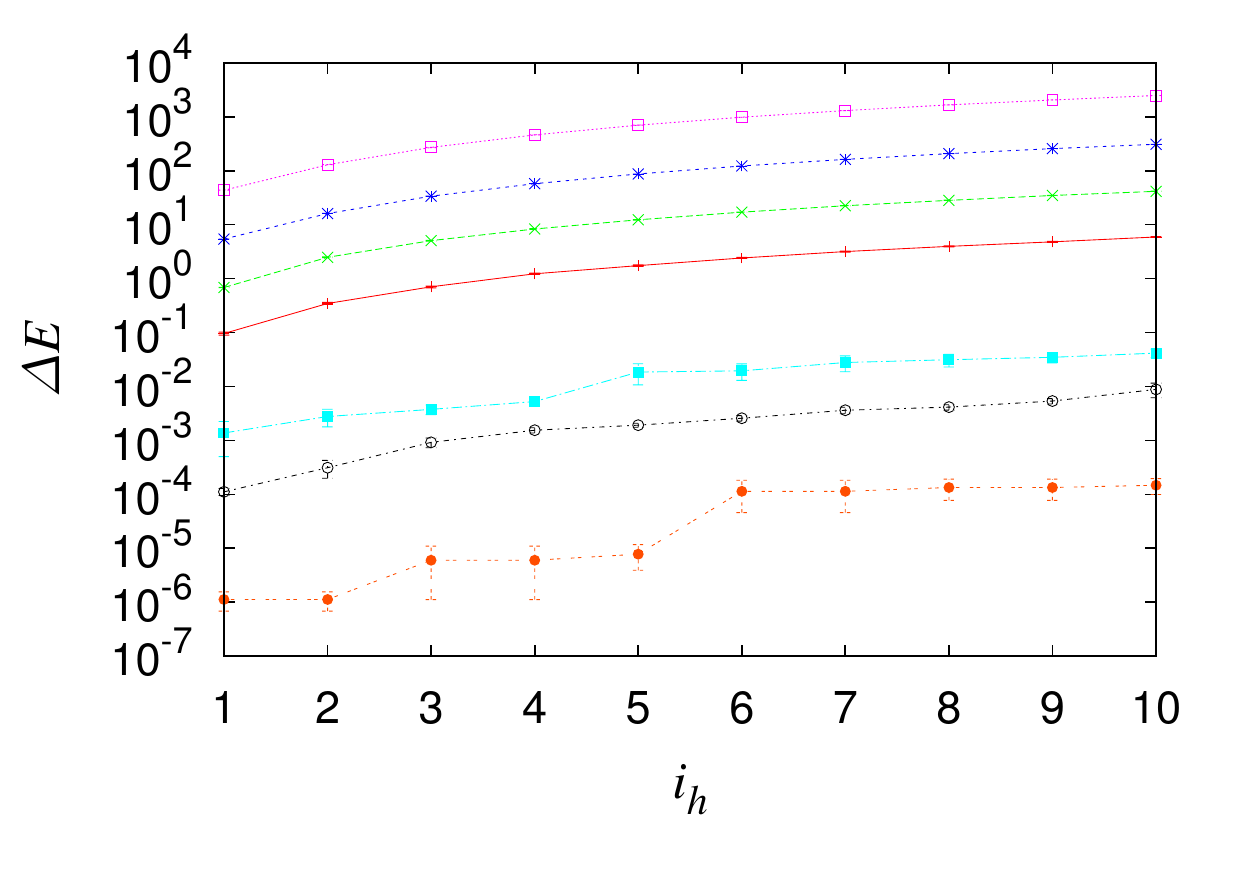}
\includegraphics[width=0.48\textwidth]{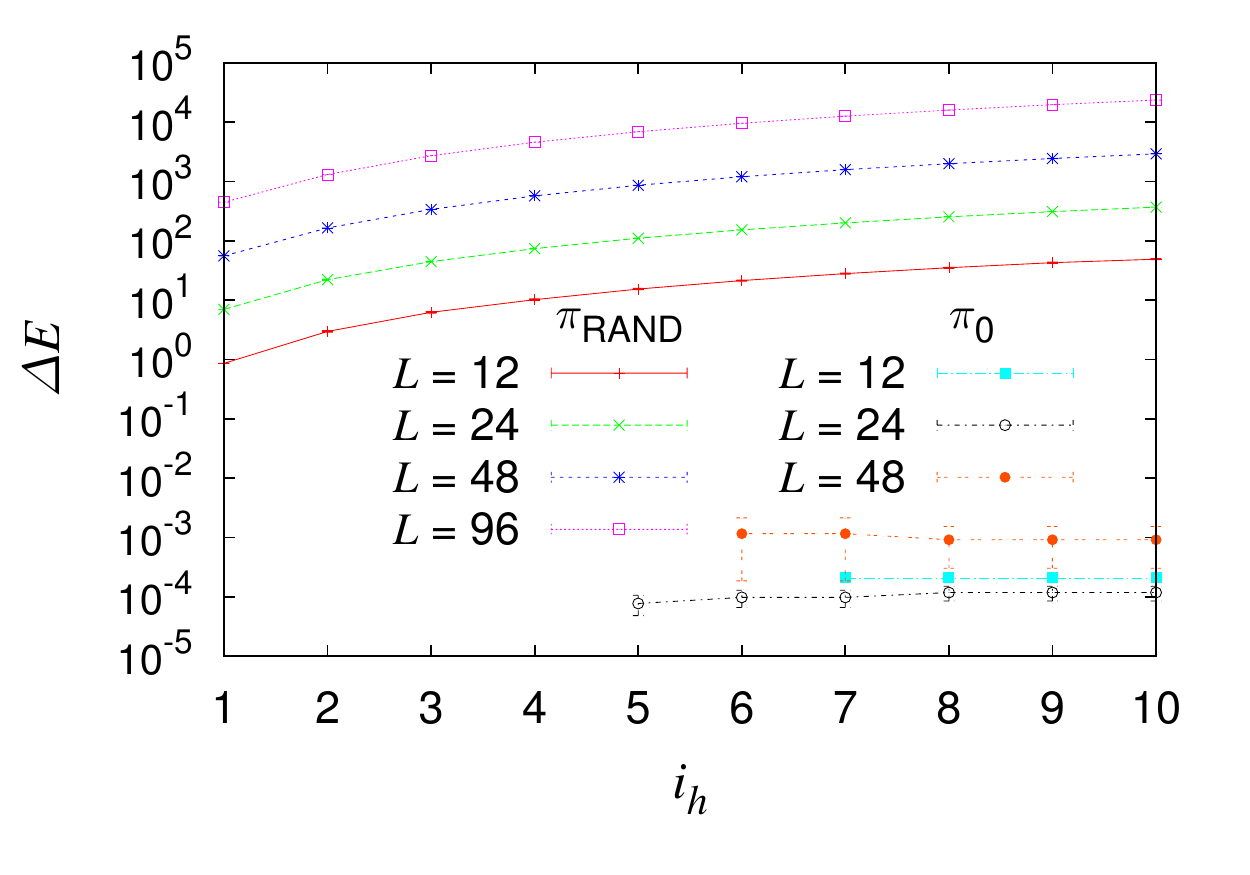}
\caption[Average energy barrier]{Average energy barrier $\Delta E^*$ for \index{energy!barrier}
				 forcings along $\ket{\vec\pi_\RAND}$ and $\ket{\vec\pi_0}$, for random fields of amplitude 
				 $H_\amp=0.1$ (\textbf{left}) and $H_\amp=1$ (\textbf{right}). In the right set, some $i_h$ are not represented
				 because for weak forcings along $\ket{\vec \pi_0}$ there were no rearrangements.
                                 }
\label{fig:forcing-barriers}
\end{figure}

To reinforce the idea of two-level system, we see that while the energy barriers from random forcings increase\index{energy!barrier}
with the system size (the growth is $O(N)$), while those within the two-level system (along the softest mode) do not (figure \ref{fig:forcing-barriers}).
\index{two-level system|)}
\index{anharmonicity|)}
\index{energy!landscape|)}

\section{Overview}

The introduction of a random field,\index{random!field} besides extinguishing the rotational symmetry, \index{symmetry!O(3)@$O(3)$}
changes the response of the Heisenberg spin glass to
soft excitations. In the absence of field the density of states is expected to 
go as $g(\omega)\propto\omega^2$ \cite{grigera:11,franz:15b}.\index{density of states}
Very strong random fields suppress the soft modes,\index{modes!soft} and a gap \index{gap}appears in 
the density of states $g(\omega)$. Still, soft modes do resist the
application of a random field when it is not too large. The data are compatible with the absence of a gap, where for small $\omega$ the density of states
grows as $g(\omega)\propto\omega^{4}$.
\index{exponent!soft modes!delta@$\delta$}

\index{localization}
It appears that a finite fraction of the modes is localized, suggesting a localization transition when the system size becomes large.

\index{anharmonicity|(}\index{forcing}
Besides the density of states, that consists in a harmonic approximation of the metastable states, \index{metastable state@metastable state|seealso{valley}}
we make an anharmonic analysis by imposing
an external force on the system.
The reaction of the spin glass has a strong dependency on the direction of application of the force. Extensive corrections to the Hamiltonian are needed
to be able to move the spins in the direction of a forcing along a random direction, while order 1 forcings are enough to obtain the same result
pushing along the softest mode, suggesting that it is the softest mode that drives the change. 

\index{energy!landscape}\index{valley}\index{inherent structure}
Even though the response appears in both cases concentrated along the softest modes, seldom the softest mode leads the system to a new inherent
structure, whereas a delocalized forcing drives it to explore many new valleys of the energy landscape.
Forcings of order one along the softest mode are the smallest we can impose in order to have a jump toward another inherent structure. The rearrangement
in the change of inherent structure is localized, and the energy barrier does not grow with the system size.

The most attractive feature of the valleys reached with a forcing along the softest mode is that their overlap with the initial inherent structure is
very high, much higher than the typical overlap expected for independent inherent structures. This means that there are couples of metastable states
with a fundamental relation between them, connected by a soft mode, with a small energy barrier. \index{energy!barrier}
This could be used as an operational definition of classical two-level system.\index{two-level system}

\index{spin!Heisenberg|)}
\index{modes!soft|)}
\index{modes!harmonic|)}
\index{anharmonicity|)}
\index{forcing|)}

\part{Conclusions and Outlook}
 \chapter{Conclusions  \label{chap:conclusiones}}
\index{spin!Ising@Ising$ $|seealso{Ising spin glass}}\index{spin glass!Ising@Ising$ $|seealso{Ising spin}}
\index{spin!XY@XY$ $|seealso{XY spin glass}}\index{spin glass!XY@XY$ $|seealso{XY spin}}
\index{spin!Heisenberg@Heisenberg$ $|seealso{Heisenberg spin glass}}\index{spin glass!Heisenberg@Heisenberg$ $|seealso{Heisenberg spin}}
\index{density of states!Debye@Debye$ $|see{Debye}}
\index{supercomputer|see{Minotauro}}
\index{thermalization@thermalization|seealso{Monte Carlo simulations}}
\index{programming!parallel@parallel|seealso{tiling}}

\section{General considerations}
% \addcontentsline{toc}{section}{\protect\numberline{} General considerations}
It is almost one century that scientists from several domains, going from\index{glass!phase}\index{glass!transition}
physics, to chemistry, engineering, mathematics and computer science, gathered to understand the 
nature of the glass transition.
In 1995, Anderson stated:
``The deepest and most interesting unsolved problem in solid state theory is probably 
the theory of the nature of glass and the glass transition'' \cite{anderson:95}.
Twenty years later, in 2015, despite  great steps forward,
the main answers on the glass transition are still unanswered.

It would be pretentious to think to make a revolutionary advance in a single Ph.D. thesis,
as the scientific advance is usually the sum of a very large series of small contributions.
It is like removing all the corns from a huge cob. Every single corn is important, even though
from the point of view of the full cob it might seem extremely small.

Removing a corn consists in advancing under any known point of view, from conceiving
new theories to developing new methodologies and instruments, to finding some new non-trivial behavior.
It is up to the researcher to decide which perspective is more suited to 
his profile and the problem he tackles, but he should always keep in mind the multidisciplinarity 
of the problem, and possibly include it in his approach.

% In this thesis we focused on spin glasses. 
% On one side we proved the usefulness of specific hardware 
% such as \ac{GPU} clusters and \ac{FPGA} machines to answer relevant questions, opening the path for new ways
% to validate theories.
% On the other side, we dealt with the glass transition under several points of view, contributing with
% a finite amuont of small corns, to what one day will hopefully be the full unraveling of the whole cob.
In this thesis we dealt with the glassy phase under several points of view, focusing on spin glasses. 
Our approach was mainly numerical, with a strong imprint due to a theoretical physics background. We worked on
simplified systems that carry only few essential features, enough to yield the phenomenology we wish
to understand.
On one hand we studied the critical behavior of canonical spin glasses, trying to understand how the spin glass phase 
and transition change under perturbations, focusing on concepts like universality and critical dimensions.
On the other hand we tried to get a better view on the energy landscape, a feature with a diverging number of degrees
of freedom that is usually described only through a single number, the energy.

We 
contributed with a finite amount of small corns, to what one day will hopefully be the full unraveling of the whole cob.

In the following sections we outline shortly the results we achieved in this thesis.
More extended conclusions are given at the end of each chapter.

\section{State of the art computing}
% \addcontentsline{toc}{section}{\protect\numberline{} State of art computing}
In this thesis we showed the usefulness of special hardware to achieve meaningful results. 
The resources we used were never used before for the problems we attacked,
so our work is a proof of concept for these approaches.

The data in chapter \ref{chap:eah3d} were obtained with the dedicated computer \textsc{Janus}, \index{Janus@\textsc{Janus}!computer}\index{Tianhe-1A}
an \acs{FPGA}-based computer designed specifically for Ising spin glass simulations.\index{spin!Ising} 
With this machine it was possible to thermalize on unprecedentedly large lattices, at uniquely low temperatures.\index{thermalization}
Even though this machine has been operating since 2008, before the beginning of my research career, 
each of the results achieved with \textsc{Janus} represents a proof of the suitability of a dedicated \ac{FPGA}-based computer.

In chapter \ref{chap:ahsg} we simulated on Heisenberg spins, \index{spin!Heisenberg}
for which \textsc{Janus} is not optimized. We resorted
then to \acp{GPU}. \index{GPU}
At the moment of the publication of \cite{baityjesi:14}, despite their popularity, 
no physical result had been obtained through \ac{GPU} simulations on a Heisenberg \ac{SG} (and to our knowledge neither on Ising spins). Even at present date,
we are only aware of benchmarks \cite{yavorskii:12,bernaschi:14} performed on Heisenberg spin glasses with \acp{GPU}.
Besides our work, only on Ising systems \acp{GPU} have been used to obtain new insight on spin glasses \cite{manssen:15,manssen:15b,lulli:15}.

Moreover, our work can provide even further guiding because we used a large \ac{GPU} cluster and ran simulations with
tens of \acp{GPU} in parallel.

\section{The Ising spin glass in a magnetic field}\index{spin glass!Ising}
% \addcontentsline{toc}{section}{\protect\numberline{} The Ising spin glass in a magnetic field}
We studied the three-dimensional Ising Edwards-Anderson spin glass in an external uniform magnetic field.\index{de Almeida-Thouless!transition}
We showed that the finite-size fluctuations are so marked, that searching signs of criticality becomes
highly challenging. Taking the averages of the observables hides the behavior of the majority of the
measurements, so we needed to develop more sophisticated statistical analysis tools. We classified our
measurements through a \emph{conditioning variate}, \index{variate conditioning} a function of the observables that helps to
distinguish different types of behavior, and proposed a new finite-size
scaling ansatz based on the quantiles of the conditioning variate's distribution.
In some cases the model appeared critical, and 
in others it did not. We were not able to extrapolate which of the two dominates in the thermodynamic limit,
but we could identify the region where the would-be phase transition should be searched.

\section{Heisenberg spin glass with a strong random exchange anisotropy}\index{anisotropy}\index{spin glass!Heisenberg}
% \addcontentsline{toc}{section}{\protect\numberline{} The Heisenberg spin glass with random exchange anisotropy}
We made equilibirum simulations on the three-dimensional Heisenberg model with the addition of a random
exchange anisotropy. 
We found both the chiral and the spin glass phase transitions.
Through a careful finite-size scaling analysis we conclude that the two critical channels are coupled,
so the phase transition is unique.\index{universality!class}
The exponents that we calculate are compatible with those of the Ising Edwards-Anderson spin glass, so
in the \ac{RG} sense the exchange anisotropy is a relevant perturbation on the Heisenberg Hamiltonian.

\section{Energy landscape of \texorpdfstring{$\bm{m}$}{}-component spin glasses}\index{energy!landscape}\index{inherent structure}\index{spin glass!vector}
% \addcontentsline{toc}{section}{\protect\numberline{} Energy landscape of $m$-component spin glasses}
We studied vector spin glasses in three dimensions, focusing on the role of the number of spin components $m$.
We performed an extensive study of the energy landscape and of the zero-temperature dynamics 
from an excited state.
An increase of $m$ implies a decrease of the amount of minima of 
the free energy, down to the trivial presence of a unique minimum.
 For little $m$ correlations are small 
and the dynamics are quickly arrested, while for larger $m$ low-temperature correlations crop up and the 
convergence is slower, to a limit that appears to be related with the system size. 

\section{Zero-temperature dynamics}
\index{correlation!stability@stability$ $|seealso{soft spins correlation}}\index{correlation!soft spins@soft spins$ $|seealso{stability correlation}}
\index{avalanche!dynamics|see{dynamics}}
% \addcontentsline{toc}{section}{\protect\numberline{} Zero-temperature dynamics}
We analyzed the \index{hysteresis}  hysteresis properties of the \ac{SK} \index{spin glass!Sherrington-Kirkpatrick}
model at zero temperature. The states along the
hysteresis loop are marginal, meaning that the density of stability goes to zero as 
$\rho(\lambda)\propto\lambda^\theta$, and exhibit \acl{SOC}. 
We analyzed the stability of these configurations, and found that previous scaling arguments on the averages,\index{correlation!stability}
granting $\theta=1$, were not exact due to the presence of correlations $C(\lambda)$ between soft spins. 
This correlation diverges as $1/\lambda$, and implies that they are mutually frustrated. 
The value $\theta=1$ was still recovered by analyzing the fluctuations of the mean values.

Also, we stated through scaling arguments that \acl{SOC} requires that each site have an infinite number of
neighbors, so the \ac{SK} model is critical, and the \ac{EA} model is not. By mixing both short- and long-range 
interactions in a duplex network, we argued, giving predictions that we confirmed numerically, that the long-range
couplings are a relevant perturbation to the short-range Hamiltonian. That is, as long as there are long-range
interactions, a sufficiently large system will display crackling over the whole system.

\index{marginal stability}
Finally, we studied the dynamics of the avalanches. We found that the type of spin update influences the power laws\index{random!walk}
of the crackling, but maintains the rest of the features. Furthermore, an avalanche can be represented as a random
walk in the number of unstable spins, and this random walk has a bias that depends on how unstable the system is in
each moment. Lastly, we described the avalanche dynamics through a random walk in the space of the local stabilities.
We found that the correlations between soft spins arise spontaneously during the avalanche, and we saw that the same
exponents found statically arise also dynamically.

\section{Soft modes and localization in spin glasses}\index{spin glass!Heisenberg}
% \addcontentsline{toc}{section}{\protect\numberline{} Soft modes and localization in spin glasses}
We examined the soft plastic modes in the three-dimensional Heisenberg spin glass under a random field that broke\index{modes!soft}
the rotational symmetry, in order to suppress the modes due to symmetries. We studied small fluctuations around the
inherent structures, both at at the harmonic and anharmonic levels.

This analysis was motivated by the phenomenology of supercooled liquids, where an excess of low-frequency modes (the boson peak) is found\index{boson peak}
over the Debye\index{Debye} behavior of the density of states,\index{density of states} $g(\omega)\propto\omega^2$. These excitations are important because they
dominate the plastic response of the sample.

We chose Heisenberg spin glasses as a platform to study those behaviors to (i) show that the boson peak appears in diverse types of disordered system, (ii) spin glasses
are better understood than structural glasses, and easier to simulate, letting us analyze much larger systems, and (iii) Heisenberg spin glasses, differently from
Ising ones, have continuous symmetries and degrees of freedom, so the analysis of the soft modes is analogous to that of sphere systems.

We found that for large random fields a gap \index{gap} appears in the density of states, while when the fields are small
the density of states departs from the Debye behavior, with $g(\omega)\propto\omega^4$.
These modes are localized, and they connect similar states connected by small energy\index{energy!barrier}
barriers, that we identify as a classical version of two-level systems.

\clearpage
\section{Future challenges and opportunities}
On the long term, the advances exposed in this thesis will represent a small step towards a full comprehension of the glass transition, a few corns of a giant cob.

On a shorter term, a whole set of new research opportunities has been brought to light thanks to this thesis.

\index{GPU}
From a computational point of view, we opened the door to \ac{GPU} computing in spin glasses, showing the feasibility and the effectiveness of \ac{GPU}
simulations. Future numerical works on this type of processing units is no more pioneering work aiming to understand whether this possibility is 
effective, but a full exploitation of the new possibilities of parallel computing.
We also proved the usefulness of the construction of machines highly optimized for the resolution of one single type of hard problem, in our case spin
glass simulations. \index{Janus@\textsc{Janus}!computer} The \textsc{Janus} computer gave us the access to regimes that were unthinkable with any other
type of resource, and its success paved the way for its next generation, \emph{Janus II}.

The statistical analyses of the large fluctuations in chapter \ref{chap:eah3d} open the way for a new approach to this types of problems. Studying
large deviation problems as a function of the quantile can reveal interesting results in topics such as temperature chaos \cite{fernandez:13}, and our
new finite-size scaling ansatz might reduce drastically the size of the systems necessary to deduce confidently the behavior in the thermodynamic limit 
with the due further comprehension \cite{billoire:14b}. Moreover, we showed two promising ways to unravel the question of the existence of a \ac{dat} 
line \index{de Almeida-Thouless!transition} in three dimensions. On one hand we showed in chapters \ref{chap:eah3d} and \ref{chap:hsgm} that the link 
overlap has far less fluctuations than the normal overlap, meaning that a traditional type of analysis on the link overlap could reveal whether or not
the spin glass phase persists under an applied magnetic field. On the other hand, we also found the region of temperature in which the phase transition would lie,
if it were present. Having a tangible temperature range where to verify the existence of a \ac{dat} line defines definitively the effort needed to answer
this question, that might be to the reach of \emph{Janus II}.

The finding of a coupling between the chiral and spin glass channel in chapter \ref{chap:ahsg} confirms a part of the Kawamura scenario. The main remaining question
is now on the coupling between the two in zero field \cite{fernandez:09b,viet:09}. 
It would be also interesting to verify the competition between chiral and spin glass sector in the two-dimensional XY spin glass, where the stiffness of chiral and spin glass
sectors appear to be different \cite{weigel:08}.
We also emphasized that the crossover\index{crossover} regime is in practical means more important than the thermodynamic behavior, since in this problem both simulations and
experiments are completely immersed in the crossover region. Therefore, quantities such as critical exponents might be more useful if accounted for as a function
of the size of the system or of the simulation time, using tools such as the time-length dictionary developed in \cite{janus:10}.

A similar type of approach could be used with the crossover of the spin glasses' behavior as a function of $m$ shown in chapter \ref{chap:hsgm}. The fact that the frustration
decreases the number of spin components can also be a starting point for models (physical and sociological) where $m$ is a function of the site, meaning that some sites are 
more susceptible to frustration than others.

The analysis of the dynamics of spin glass avalanches presented in chapter \ref{chap:marginal} still has many open points. Random walks in the space of local stabilities 
could be used to recover more relations between the exponents, and analytical computations on \acp{RW} in the number of unstable spins can explain what the values of $\tau$ 
and $\rho$ depend on. Furthermore, we expect these results obtained on the correlations in the \ac{SK} model to extend to other marginally stable systems such as sphere 
packings: opening a soft contact should imply that contacts carrying small forces should see their force increase.

Finally, the study of the soft modes in chapter \ref{chap:hsgrf} opens a whole new field of studies of the soft modes in spin glasses, and might sign a new trait of 
unification between sphere packings and spin glasses. To this objective, a search of a boson peak in zero field becomes necessary, possibly as a function of the
initial temperature of the relaxations. It would be mostly desirable to be able to search inherent structures from configuratios thermalized in the spin glass phase.
Also, a study of the correlations lengths in the inherent structures would help to unify the understanding on spin and structural glasses \cite{baityjesi:16}.

%manca tutto

\addtocontents{toc}{\protect\setcounter{tocdepth}{1}}
\appendix
\titleformat{\chapter}[display]
{\bfseries\LARGE} {\filleft\MakeUppercase{\chaptertitlename}
\Huge\Alph{chapter}} {2ex} {\titlerule
\vspace{1.5ex}%
\filright}
[\vspace{1.5ex}%
]
\renewcommand{\thesection}{\texorpdfstring{\textsc{\Alph{chapter}}.\oldstylenums{\arabic{section}}}{\Alph{chapter}.\arabic{section}}}
\renewcommand{\thesubsection}{\texorpdfstring{\textsc{\Alph{chapter}}.\oldstylenums{\arabic{section}.\arabic{subsection}}}{\Alph{chapter}.\arabic{section}.\arabic{subsection}}}
\renewcommand{\thesubsubsection}{\texorpdfstring{\textsc{\Alph{chapter}}.\oldstylenums{\arabic{section}.\arabic{subsection}.\arabic{subsubsection}}}{\Alph{chapter}.\arabic{section}.\arabic{subsection}.\arabic{subsubsection}}}
\renewcommand{\theequation}{\textsc{\alph{chapter}}.\oldstylenums{\arabic{equation}}}
\renewcommand{\thefigure}{\textsc{\alph{chapter}}.\oldstylenums{\arabic{figure}}}
\renewcommand{\thetable}{\textsc{\alph{chapter}}.\oldstylenums{\arabic{table}}}
%\renewcommand{\thechapter}{alpha{chapter}}

%  ************************************************************
%  APÉNDICES
%  ************************************************************

\part{Appendices}
\renewcommand{\sectionmark}[1]{\markright{\textit{\Alph{chapter}.\oldstylenums{\arabic{section}}}\ --- #1}}
 \chapter{Monte Carlo on Heisenberg spin glasses \label{app:MC}}
\index{Monte Carlo!simulations}\index{programming!CUDA@CUDA$ $|seealso{coalescence}}
The appendix is structured as follows. Section \ref{sec-app:algorithms} is general about all the \ac{MC}
simulations presented in chapter \ref{chap:ahsg}, though it treats the specific algorithms that we have
used with no reference to their implementation, so it is referenced also in chapter \ref{chap:eah3d}. 
However, the implementation is often crucial.
The simulations of chapter \ref{chap:ahsg} were so demanding that we have used special hardware described in
section \ref{sec-app:hardware}. 
This special hardware speeds up the simulations
thanks to a massive parallelization of the calculations, so in section \ref{sec-app:effectiveGPU} we give
some brief details about it.
Finally, we address in section \ref{sec-app:PRNG} some issues regarding the generation of pseudo-random
numbers.

\section{Simulation algorithms \label{sec-app:algorithms}}

For the thermalization of our vector \ac{SG} we used a blend\index{thermalization}
of several \ac{MC} dynamics. Specifically, our \ac{EMCS} consisted of (in sequential order):
\index{Monte Carlo!elementary step}
\begin{itemize}
\item 1 full lattice sweep with the \acf{HB} algorithm \cite{amit:05,krauth:06},\index{Monte Carlo!heatbath}
\item $L$ lattice sweeps of microcanonical \acf{OR} algorithm \cite{brown:87,amit:05},\index{Monte Carlo!overrelaxation}
\item 1 \acf{PT} sweep \cite{hukushima:96,marinari:98,yllanes:11}.\index{Monte Carlo!parallel tempering}
\end{itemize}
Heatbath by itself would provide correct (but inefficient) dynamics. It
actually mimics the natural evolution followed by real \acp{SG} (that
never reach equilibrium near or below the critical temperature). For this
reason we enhance it with two more algorithms. However, \ac{HB} does play a
crucial role, since it is irreducible (i.e. the full configuration space is
reachable, at least in principle), at variance with \ac{OR}, which
keeps the total energy constant, and \ac{PT}, which changes the
temperature but not the spin configuration.

Crucial to perform the \ac{HB} and \ac{OR} dynamics is the
factorization property of the Boltzmann weight for the
Hamiltonians \eqref{eq:eah3d-ham} and \eqref{eq:HamiltonianDefinition}. The conditional
probability-density for spin $\vec s_{\bx}$, given the rest of the spins of
the lattice is
\begin{equation}\label{eq:ProbabilidadCondicionada}
  P(\vec s_{\bx}\,|\, \{\vec s_{\by}\}_{\by\neq\bx})\propto \E^{(\vec s_{\bx}\cdot\vec h_{\bx})/T}\,,
\end{equation}
where $\vec h_{\bx}$ is the \index{local!field}\emph{local field} produced by the lattice
nearest-neighbors of spin $\vec s_{\bx}$.
\footnote{\index{local!field}
In the \ac{IEA} model in a magnetic field of chapter \ref{chap:eah3d} $h_\bx=\displaystyle\sum_{\by:\norm{\bx-\by}=1}^d J_{\bx\by}s_\by+h$,
in the Heisenberg model with random anisotropic exchange of chapter \ref{chap:ahsg}, $\vec{h}_\bx = \displaystyle\sum_{\by:\norm{{\vn x}-{\vn y}}=1}^d[J_{\bx\by} \vec s_\by + D_{\bx\by} \vec s_{\by}]$.}

In the \ac{HB} update,\index{Monte Carlo!heatbath} a new orientation for spin $\vec s_{\bx}$ is drawn
from the conditional probability \eqref{eq:ProbabilidadCondicionada}, see
\cite{amit:05} for instance.

The \ac{OR} update is deterministic. \index{Monte Carlo!overrelaxation}
Given a spin $\vec s_{\bx}$ and
its local field, we change the spin as much as possible while keeping the
energy constant:
\begin{equation}
  \vec s_{\bx}^\mathrm{\,new}=2\vec h_{\bx}\frac{\vec h_{\bx}\cdot\vec
    s_{\bx}^\mathrm{\,old}}{h_{\bx}^2} - \vec s_{\bx}^\mathrm{\;old}\,.
\end{equation}
Contrarily to \ac{HB}, the order in which the spins are updated is
important in \ac{OR}. Accessing the lattice randomly increases the
autocorrelation time in a substantial way. On the other hand, a sequential\index{microcanonical wave}
update generates a microcanonical wave that sweeps the lattice. The resulting\index{tiling}
change in the configuration space is significantly larger. A similar
microcanonical wave is generated with other types of deterministic lattice
sweeps. For instance, one could partition the lattice in a checker-board way
and first update all spins in the black sublattice, updating the white spins
only afterwards.

The combination of \ac{HB} and \ac{OR} has been shown to be effective
in the case of isotropic \acp{SG} \cite{pixley:08} and other models with
frustration \cite{alonso:96,marinari:00c}. However, if one is interested on very
low temperatures or large systems, \ac{PT} is often crucial.\index{Monte Carlo!parallel tempering}  For
each sample we simulate $N_T$ \nomenclature[N...T]{$N_T$}{number of temperatures}
different copies of the system, each of them at
one of the temperatures $T_1<T_2<\ldots<T_{N_T}$. A \ac{PT} update
consists in proposing, as configuration change, a swap between configurations
at neighboring temperatures. The exchange is accepted with the Metropolis
probability
\begin{equation}
\index{Monte Carlo!Metropolis}
 P = \min\left[1, \E^{-\beta \Delta E}\right]\,,
\end{equation}
where $\Delta E$ is the energy difference \nomenclature[Delta...E]{$\Delta E$}{In appendix \ref{app:MC}: energy difference in the PT swap}
between the two configurations and $\beta$
is the inverse temperature.
One of the two systems involved in the swap will decrease its energy, so 
that change will be automatically accepted. In order to accept the swap both the 
configuration changes need to be accepted, so the swap is generally accepted with 
probability $\E^{-\beta |\Delta E|}$.
Evidently, the acceptance is higher if the temperatures $T_i$ are
closer to each other, since the energy of the configurations will be
similar. Notice that exchanging configurations is equivalent to exchange
temperatures, so instead of swapping configurations one can swap temperatures,
reducing the data transfer to a single number.

\section{Parallel computing}\index{numerical simulations!parallel}
We discuss now part of the implementation of our codes on the specific hardware that we disposed of.

\subsection{Hardware features}\label{sec-app:hardware}\index{GPU}
The \acp{GPU} we used were of the Tesla generation, produced by NVIDIA, with a
SIMD architecture (Single Instruction, Multiple Data), optimized
for the parallel processing of large amounts of double precision data. 

We had access to Tesla M2050 GPUs in the \emph{Tianhe-1A} supercomputer in Tianjin,\index{Tianhe-1A}
 China, and Tesla M2090 GPUs on the \emph{Minotauro} cluster\index{GPU!cluster!Minotauro}
 in Barcelona, Spain. Despite the extremely high performances
claimed by NVIDIA (e.g. 665 Gflops in double precision in the case of the
M2090 \acp{GPU}), it is practically impossible to reach that limit, because the
major bottleneck does not reside in the computing speed, but in the memory
access.
Yet \acp{GPU} keep being a valid tool to simulate on \acp{SG}, as they
typically allow the same function to be launched concurrently on thousands of
threads. This is exactly what we need, since we can update simultaneously
different replicas, and also non-neighboring spins within the same replica,
because the interactions are only between nearest neighbors.

\subsection{Effective GPU coding}\label{sec-app:effectiveGPU}\index{programming GPU}
The optimization of the \ac{GPU} code required a great effort. In fact, between the first 
and the last version of the program, we gained a speed-up factor of 100. 

The complexity of the Monte Carlo algorithms, that require the definition of a very large 
number of variables, is what finally limits the speed of the program, since they exceed
the number of registers in the \ac{GPU} (this effect is called \emph{register spilling} \index{register spilling}
\cite{nvidia:15}: some of the variables have to be stored in the global memory, slowing
down their access).

To limitate the memory access, we opted to simulate the model with binary couplings $D^{\alpha\beta}_{\bx\by}=\pm D$, 
and $J_{\bx\by}=\pm 1$, in order to be able to store in a single byte the coupling 
between two sites. Since $D_{\bx\by}$ is symmetric there are 6 independent entries $D^{\alpha\beta}_{\bx\by}$, plus one for $J_{\bx\by}$.
The extra bit stayed unused.
Also, we limited the size of the lattice to powers of 2, in order to get be able to evaluate the lattice positions with 
biwise operations and to achieve a \emph{coalesced} memory access, \index{coalescence}
as explained in section \ref{sec-app:coalescence}. We also maximized the 
use of the level 1 cache memory and tiled the system in columns, \index{tiling}
updating independently two groups of non-neighboring tiles. The black tiles
are updated first, and the white are updated in a second kernel call, in order to avoid sinchronization conflicts.

Issues of this type with single-GPU coding on spin systems are extensively treated in works such as \cite{bernaschi:11, yavorskii:12, lulli:14}, 
so let us focus on the complications related to the use of multiple \acp{GPU}. We describe now in practical means the procedure of simulations for $L=64$
that mixed CUDA and \ac{MPI}.\index{programming!CUDA}\index{programming!MPI}

For each sample we simulate $2 N_T$ replicas, because we need two replicas per 
temperature to be able to calculate overlaps. We use $N_\mathrm{GPU}$ \acp{GPU}, \nomenclature[N...GPU]{$N_\mathrm{GPU}$}{number of GPUs}
and each hosts two replicas, 
not necessarily at the same temperature, hence $N_\mathrm{GPU}=N_T$. Since the interactions are only between nearest neighbors, 
we can update simultaneously up to half of the spins with\index{tiling}
two independent kernel calls (one for the black tiles and one for the white). Yet, there are only 65535 threads per \ac{GPU}
\cite{nvidia:15}, and $2 L^3=524288$ sites, so each thread has to update at least 4 spins. 
Since the major bottleneck is the memory access, we work with $N_\mathrm{threads}=2^{15}=32768$ threads, 
\nomenclature[N...threads]{$N_\mathrm{threads}$}{number of threads}
assigning a row of 8 spins to each, 
along the $x$ axis. This way we can minimize the number of reads from global memory, and we 
give a direction to the \ac{OR} spin wave. Adjacent rows are updated in different kernel calls.

\subsection{Coalescent memory reading \label{sec-app:coalescence}} \index{coalescence|(}
Changing the way we read from memory gives \ac{GPU} programs a dramatic speedup, and the only effort necessary to obtain this
is to change the indexing of the memory locations. 

When a single multiprocessor is given some thread blocks to deal with, the scheduler executes them in groups of 32 threads, 
called warps. A warp executes one instruction at a time, and the maximum performance is achieved when all the threads in the 
warp have a similar execution path. To get coalesced reading, the consecutive threads have to read from consecutive memory 
positions, in order to maximize bandwidth of the memory bus \cite{nvidia:15}. So, for 
example, if thread 1 reads from the memory position 612, thread 2 would make an effective read from position 613. In order to obtain this we have to reorganize the 
memory indexing in order to have thread 2 pointing to position 613. This is often automatically realized in simple arrays, but not when the spatial geometry comes to play with
tiling or with the indexing of the $J_{\bx\by}s$.

\index{tiling}
In the specific case of our spin indexing, we want neighboring rows to be called by neighboring threads. Yet, when we say neighboring rows, we mean neighboring rows
within the same kernel call, not in the actual lattice. It is like if we  compress  together all the white tiles and only then we worry about proximity.
The first site of the white row $i$ ($i$ runs only over the white tiles) has to be stored besides the first site of row $i+1$, 
and so on. This means that their address in memory has to differ only in the least significant bit. 
The $z$ coordinate is the same both for $i$ and $i+1$. The same happens for the $x$ coordinate, since both threads sweep the row in the same way. 
On the $y$ axis, since we update one row of every two, the least significant bit $y_0$
also is the same. Hence the least significant bit of the coalesced reading has to be $y_1$, the second least $y_2$, and so on. On table \ref{tab:bits}, line 4, we give an example
of coalesced memory access. 
Since there are $2^{15}$ threads, $i_\mathrm{th}$ has 15 significant bits.
To get the index of the starting site $i_\mathrm{row}$ of each row we need information on:
\begin{itemize}
 \item Which replica were updating. There are two replicas, so 1 bit is enough.
 \item The $z$ coordinate. It can assume $L=64$ different values, so it requires 6 bits.
 \item The $x$ coordinate is not constant. We just need the one of the first site of the row. Rows are 8 sites long, so we can only fit 8 along a side. That makes 3 bits.
 \item The $y$ coordinate. Since adjacent rows are updated in different kernel calls, $y$ has to change of 2 lattice spacings each time we change row, and half of the $y$ choices are 
 forbidden. We need 5 bits for $y$.
\end{itemize}
The mapping from $i_\mathrm{th}$, associated with the thread to the index $i_\mathrm{row}$ that indicated the initial site of the tile, 
is shown on the second line of table \ref{tab:bits}.
\begin{table}[!t]
\centering
% \begin{tabular}{cc}
% $i_\mathrm{th}$: & \fbox{$t_{14}$}\fbox{$t_{13}$}\fbox{$t_{12}$}\fbox{$t_{11}$}\fbox{$t_{10}$}\fbox{$t_{9}$}\fbox{$t_{8}$}\fbox{$t_{7}$}\fbox{$t_{6}$}\fbox{$t_{5}$}\fbox{$t_{4}$}\fbox{$t_{3}$}\fbox{$t_{2}$}\fbox{$t_{1}$}\fbox{$t_{0}$}\\[3ex]
% $i_\mathrm{row}$: & \fbox{$r_0$}\fbox{$x_5$}\fbox{$x_4$}\fbox{$x_3$}\fbox{$z_5$}\fbox{$z_4$}\fbox{$z_3$}\fbox{$z_2$}\fbox{$z_1$}\fbox{$z_0$}\fbox{$y_5$}\fbox{$y_4$}\fbox{$y_3$}\fbox{$y_2$}\fbox{$y_1$}\\[3ex]
% $i_\mathrm{site}$: & \footnotesize\fbox{$r_0$}\fbox{$z_5$}\fbox{$z_4$}\fbox{$z_3$}\fbox{$z_2$}\fbox{$z_1$}\fbox{$z_0$}\fbox{$y_5$}\fbox{$y_4$}\fbox{$y_3$}\fbox{$y_2$}\fbox{$y_1$}\fbox{$y_0$}\fbox{$x_5$}\fbox{$x_4$}\fbox{$x_3$}\fbox{$x_2$}\fbox{$x_1$}\fbox{$x_0$}\\[3ex]
% $i_\mathrm{site}^\mathrm{coalesced}$: & \footnotesize\fbox{$r_0$}\fbox{$y_0$}\fbox{$x_5$}\fbox{$x_4$}\fbox{$x_3$}\fbox{$x_2$}\fbox{$x_1$}\fbox{$x_0$}\fbox{$z_5$}\fbox{$z_4$}\fbox{$z_3$}\fbox{$z_2$}\fbox{$z_1$}\fbox{$z_0$}\fbox{$y_5$}\fbox{$y_4$}\fbox{$y_3$}\fbox{$y_2$}\fbox{$y_1$}
% \end{tabular}
\resizebox{\columnwidth}{!}{
\begin{tabular}{|r|c|c|c|c|c|c|c|c|c|c|c|c|c|c|c|c|c|c|c|}
\hline
\raisebox{6ex}[3ex][2ex]{} $i_\mathrm{th}$:  &       &          &       &          & $t_{14}$ & $t_{13}$ & $t_{12}$ & $t_{11}$ & $t_{10}$ & $t_{9}$ & $t_{8}$ & $t_{7}$ & $t_{6}$ & $t_{5}$ & $t_{4}$ & $t_{3}$ & $t_{2}$ & $t_{1}$ & $t_{0}$\\[1ex]
\hline\hline
\raisebox{6ex}[3ex][2ex]{} $i_\mathrm{row}$:  &       &          &       &          & $r_0$    & $x_5$    & $x_4$    & $x_3$    & $z_5$    & $z_4$   & $z_3$   & $z_2$   & $z_1$   & $z_0$   & $y_5$   & $y_4$   & $y_3$   & $y_2$   & $y_1$\\[1ex]
\hline\hline
\raisebox{6ex}[3ex][2ex]{} $i_\mathrm{site}$: & $r_0$ & $z_5$    & $z_4$ & $z_3$    & $z_2$    & $z_1$    & $z_0$    & $y_5$    & $y_4$    & $y_3$   & $y_2$   & $y_1$   & $y_0$   & $x_5$   & $x_4$   & $x_3$   & $x_2$   & $x_1$   & $x_0$\\[1ex]
\hline\hline
\raisebox{6ex}[3ex][2ex]{} $i_\mathrm{site}^\mathrm{coalesced}$: 
                   & $r_0$ & $y_0$    & $x_5$ & $x_4$    & $x_3$    & $x_2$    & $x_1$    & $x_0$    & $z_5$    & $z_4$   & $z_3$   & $z_2$   & $z_1$   & $z_0$   & $y_5$   & $y_4$   & $y_3$   & $y_2$   & $y_1$\\[1ex]
\hline
\end{tabular}
}
\caption[How to obtain coalescent reading]{\captionsize 
A step-by-step example of how to obtain coalescent reading for an $L=64$ lattice. 
On the \textbf{first line} we show the \emph{thread index}. It has 15 significant bits, since we use $2^{15}$ concurrent threads. We have to
use them to identify each tile with the starting point of the row (\textbf{second line}). We use the most significant bit to identify the replica. Since $L$ contains 8 rows, we need only
3 digits to identify their starting point on the $x$ axis, but we need all the information on the $z$ axis, and only 5 bits for the $y$ axis, since there is the constraint of having
to simulate non-neighboring rows.
On the \textbf{third line} we show an easy way to organize the bits to identify a site once we started moving along the row, in case of non-coalescent reading. 
It is straightforwardly deducible from $i_\mathrm{row}$. 
The \textbf{last row} shows how to organize the bits to get coalescence. The replica index stays where it is, the eleven following bits are shifted 7 positions to the right, and the final
seven are shifted 11 positions to the left. This way consecutive threads access consecutive memory positions.
More details in the main text.
}
\label{tab:bits}
\end{table}
The index $i_\mathrm{row}$ needs only 3 bits to store its $x$ position, because since the rows are of 8 sites along the $x$ axis there are only 8 tiles.
By adding the three bits (table \ref{tab:bits}, line 3) we obtain an uncoalesced memory read of site $i_\mathrm{site}$. From this one we obtain the 
coalesced read by moving the bits around in order to force the changes of indexing to the least significant bit.
Practically, it is obtained by shifting seven positions to the right the 6 $z$-bits plus the 5 $y$-bits except $y_0$, and with an 11 position shift towards left
of the remaining $y_0$ plus the 6 $x$-bits.
Notice that this type of reading is very convenient since it only implies unsophisticated bit-to-bit operations, and it is valid for any $L$ power of 2.  
This is why almost all our simulations were with $L = 8, 16, 32, 64$.

The remaining information on the actual position on the lattice is given by a binary parity parameter that the kernel gets from the input. The parity tells us whether $y$ is even or odd 
(if $y_0=1$ or $y_0=-1$), or in other words, if the kernel call regards black or white cells.
\footnote{For $x=0$ it tells us if $y$ is even or odd, but for $x=8$, it tells us if $y$ is odd or even, and so on, because in each layer of rows the parity has to change in order
to not update simulaneously neighboring rows.} The index $i_\mathrm{site}$ indicates the position of the single site once one took in account the parity and the position along the row.
\index{coalescence|)}

\subsubsection{MPI parallelization}\label{sec-app:MPI}\index{programming!MPI}\index{programming!GPU!multi}
\nomenclature[m.aster]{master}{the node that manages the instructions}
\nomenclature[s.lave]{slave}{node that receives the instructions}
To simulate $N_T$ temperatures with \ac{MPI} we used $N_T+1$ cores. $N_T$ of them, called slaves, were in charge of measurements and 
updates on two lattices, using the resources of a \ac{GPU} each. 
The remaining one, called master, did not use any \ac{GPU} and was dedicated to the 
\ac{PT} and to the management of the relationships between slaves.\index{Monte Carlo!parallel tempering}
The expedients for the simulations that we described in the previous sections are valid at the level of the slave.

\begin{figure}[ht]
\includegraphics[angle=270, width=\columnwidth]{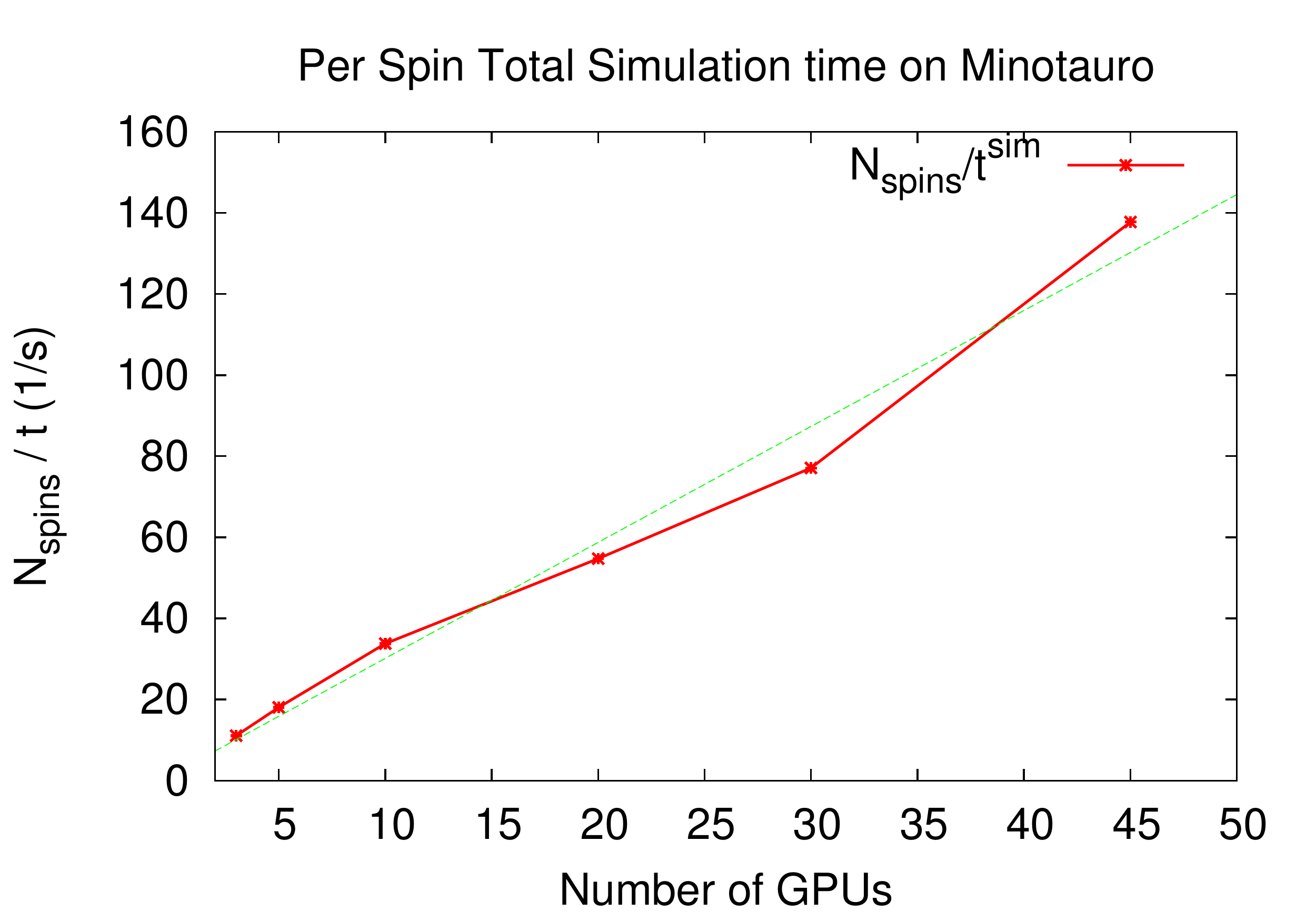}
\caption[Scaling of the computing time with $N_\mathrm{GPU}$]{Scaling of the computing time with the number of \acp{GPU} $N_\mathrm{GPU}$. 
Benchmark performed on the \emph{Minotauro} \index{GPU!cluster!Minotauro}
\ac{GPU} cluster (Barcelona Supercomputing Center, Barcelona, Spain).\index{programming!GPU!multi}}
\label{fig:scalingGPU}
\end{figure}

Each 1 \ac{HB} + $L$ \ac{OR} sweeps, we do \ac{PT}. \index{Monte Carlo!elementary step}
We measure on the device (the \ac{GPU}) the energy of each replica, and we pass this information to the master. The master makes the \ac{PT} iterations,
that require a negligible amount of time, and assigns a new temperature to each replica. 
The memory transfer overhead is minimum in this case. It becomes an issue when we have to
\begin{enumerate}
 \item Perform 2-replica measurements (e.g. overlaps)
 \item Write on disk (measurements and backup)
\end{enumerate}
since we are forced to pass the entire configuration via \ac{MPI}. 
The nature of the system we are simulating is of help, since we can dilute measures (and writes) almost as 
much as we desire, as long as we have enough measurements to perform decent 
averages. The \ac{MPI} extension turned out to be very effective, since not only the multi-GPU version 
of the algorithms was as fast as the single-GPU, but also the speed
had a linear scaling with the number of \acp{GPU} (see figure \ref{fig:scalingGPU}).

\subsection{Parallel Pseudo-Random Number Generator \label{sec-app:PRNG}}
\index{random!number generator}
Pseudo-random number generators (PRNGs) \acused{PRNG} are a critical issue in the
implementation of stochastic algorithms \cite{knuth:81}, but even more in cases
like ours, where each of the $N_\mathrm{threads}$ threads had to carry its own
\ac{PRNG}, and we had a large number of them acting in parallel on the same
lattice. This became a major problem especially in the simulations with \ac{MPI},
where a huge number of \acp{PRNG} was concentrated on only two lattices. It was
crucial to guarantee the statistical independence of the $N_\mathrm{threads}$
pseudo-random sequences. We consider three different aspects: ($a$) the \ac{PRNG}
that each thread uses, ($b$) the initialization of the generators and ($c$)
our tests on the generators.

\subsubsection{The generator}
\index{random!number generator!congruential}
\index{random!number generator!Parisi-Rapuano}
We resorted to a linear combination of Parisi-Rapuano with congruential
generators \cite{fernandez:09}.

With the Parisi-Rapuano sequence \cite{parisi:85}, the $n^\mathrm{th}$ pseudo-random number $P_n$ is generated through the following
relations:
\begin{eqnarray}
  y_n &=& (y_{n-24} + y_{n-55})\, \mathrm{mod\,2^{64}}\\\nonumber
  P_n &=& y_n ~ \text{XOR} ~ y_{n-61}\,,
\end{eqnarray}
where XOR is the exclusive OR logic operator, and $y_i$ are 64-bit unsigned
integers.  Although some pathologies have been found in the 32-bit
Parisi-Rapuano PRNG \cite{ballesteros:98c}, it looks like its 64-bit version is
solid \cite{fernandez:05}.

On the other side, we used a 64-bit congruential generator, where the
$n^\mathrm{th}$ element of the sequence, $C_n$, was given
by \cite{knuth:81,lecuyer:99}:
\begin{equation}\label{eq:congruential}
  C_n = (C_{n-1} \times 3202034522624059733 + 1 )\,\mathrm{mod\,2^{64}}\,.
\end{equation}
Also this generator is not reliable when used alone \cite{fernandez:09,ossola:04}.

The final pseudo-random number $R_n$ was obtained by summing $P_n$ and $C_n$:
\begin{equation}\label{eq:ParisiRapuanoCongruencial}
  R_n = (P_n+C_n)\,\mathrm{mod\,2^{64}}\,.
\end{equation}

\subsubsection{Initializing the generators}
\index{random!number generator!initialization}
We have found that problems arise if special care is not devoted to the
initialization of the random numbers. This is particularly important in the
case of multiple \acp{GPU} where $N_\mathrm{threads}=32768$ threads concurrently
update the spins in only two lattices.

We need one \ac{PRNG} for the master, that performs \ac{PT}, and $N_\mathrm{threads}$
independent generators for each slave. It is not trivial to avoid periodicities when not only one wants 
$N_\mathrm{threads}N_\mathrm{GPU}+1\sim 1.5\times10^6$ \ac{PRNG}, but it is crucial for them
to be reproducible, monitorizable and backupable. Starting each simulation with over a million seeds is 
not a realistic option, but any simplification can be crucial for the
simulation. 

We decide to use one seed per slave, plus one for the master, and refresh the \ac{PRNG} every time a backup is done.
\footnote{
In order to gain in speed and space in disk, we decided not to save the random wheel when we had to make backups. We limited ourselves to a refresh of the random wheels
with new seeds read from the \emph{urandom} device. In this manner, we only had to save $N_\mathrm{backups}(N_\mathrm{GPU}+1)$ long long integers per simulation.
} 
That makes 46 unsigned long long integer seeds (64 bits each).
Passing the \ac{PRNG} to the kernels is a major bottleneck in our simulations. A combination of a congruential generator with the Parisi-Rapuano wheel 
is a fair solution in terms of speed and memory passage to
the kernel, but the Parisi-Rapuano wheel contains 256 elements (passing them back to the master takes forever), and it is not trivial to initialize properly 
a very large amount of wheels starting from a single seed.

The starting point for each node is a single seed. From that we have to initialize a whole set of $N_\mathrm{threads}$ \acp{PRNG}, so it is clear that special care
is needed to obtain independent initializations. 

\paragraph{Implementation.} For the initialization of the $N_\mathrm{threads}$ generators through 
a single seed we resorted to the Luescher generator, we employed\index{random!number generator!Luescher|(}
the \emph{full luxury} version, which is fireproof but slow \cite{luescher:94}. This is how we proceeded to obtain a large set of pseudo-independent \ac{PRNG} out
of a single seed.
\begin{enumerate}
 \item Use the initial seed to initialize a 64-bit congruential \ac{PRNG} \eqref{eq:congruential}.
 \item Generate $\sim 1000$ random numbers with the congruential \ac{PRNG}, in case the initial seed was not chosen properly (e.g. it was too small).
 \item Use the congruential generator to initialize a Luescher wheel, that requires 256 24-bit elements (although we only need 24 for the initialization, 
 plus an auxiliary variable). Each 24-bit entry for the Luescher wheel is obtained through 3 subsequent call of the congruential. From each call we pick the 
 8 most significant bits, and append the three together construct the 24-bit number.
 \item Generate $\sim 1000$ random numbers with Luescher's wheel. 
 \item Use the Luescher wheel to fill up the state vector of the 64-bit \acp{PRNG} in equation \eqref{eq:ParisiRapuanoCongruencial}. Each entry is obtained through
 8 Luescher calls, and taking the 8 most significant bits from each.
\end{enumerate}
In addition to the \acp{PRNG}, also the couplings are formed by using Luescher's algorithm.
We were probably
excessively cautious, given the high quality of the full-luxury generator, but
initialization takes only a small fraction of the total computing time, and we wanted to grant the threads
sufficiently independent \acp{PRNG}.\index{random!number generator!Luescher|)}

\subsubsection{Tests}\index{random!number generator!test}
We tested with success our random sequences through the whole battery of tests proposed in \cite{marsaglia:95}.
To be sure the sequences were reliable also with concurrent threads, we also generated $N_\mathrm{threads}$
sequences and tested them \emph{horizontally}, i.e. taking first the first number of each sequence,
then the second, and so on.

Also, we made simulations with ferromagnetic couplings demanding the energies to be equal,
up to the $7^\mathrm{th}$ significant digit, to those obtained with an independent
CPU program, that had been already used to produce publications such as \cite{fernandez:09b}.

Finally, it has been pointed out that local Schwinger-Dyson relations \index{Schwinger-Dyson relations}
(see e.g. \cite{rivers:90}) can be useful to assess the quality of
\acp{PRNG} \cite{ballesteros:98c}. The relevant identity here is
\begin{equation}
  2 T \left\langle \vec s_\bx\cdot \vec h_\bx \right\rangle - \left\langle (\vec h_\bx)^2 - (\vec s_\bx\cdot \vec h_\bx)^2\right\rangle = 0 \,.
\end{equation}
We averaged it over all the sites in the lattice, in order to obtain a more stringent test for the simulations.

 \chapter{Four-Replica Correlators\label{app:propagators}}\index{correlation!function!4-replica|(}
In this appendix we give details on the 4-replica correlators used in chapter \ref{chap:eah3d}. 
In section \ref{app:need4} we motivate the need of four different replicas,
in section \ref{app:GRGL} we explain how to find the replicon and longitudinal connected \index{correlation!function!replicon}\index{correlation!function!longitudinal}
correlation functions $G_\mathrm{R}$ and $G_\mathrm{L}$, we show that the signal carried
by $G_\mathrm{L}$ is much smaller than that of $G_\mathrm{R}$, and we give an estimation of the value that 
the effective anomalous exponent $\eta_\mathrm{eff}$ \index{exponent!critical!eta@$\eta$!effective}
\nomenclature[eta....eff]{$\eta_\mathrm{eff}$}{effective anomalous exponent}
defined in section \ref{sec:zero-field} should acquire in the spin glass phase (section \ref{app:eta-eff}).\index{spin glass!phase}
Section \ref{app:MSC} is dedicated to an implementation of the \ac{MSC} technique in our analyses.

In the presence of an external field the overlap is non-zero even in the paramagnetic phase, so the correlation functions $C(\br)$ [equation (\ref{eq:C})] do not 
go to zero for large distances.
We need therefore to explicitly construct correlators that go to zero. Two natural constructions that can be measured directly are 
\begin{align} 
\index{correlation!function!connected}
\nomenclature[Gamma...1,Gamm2]{$\Gamma_1(\bx,\by),\Gamma_2(\bx,\by)$}{connected correlators}
\label{eq:gamma1}
\Gamma_1 (\bx,\by) &=\overline{\left[\langle s_\bx s_{\by}\rangle -\langle s_\bx\rangle\langle s_{\by}\rangle\right]^2}\,,\\[1ex]
\label{eq:gamma2}
\Gamma_2 (\bx,\by) &=\overline{\left[\langle s_\bx s_{\by}\rangle^2 -\langle s_\bx\rangle^2\langle s_{\by}\rangle^2\right]}\,.
\end{align}
In section \ref{app:GRGL} we will show how $\Gamma_1$ and $\Gamma_2$ relate to the correlators of the replicated field theory.\index{replica!theory}

\section{The need for four replicas \label{app:need4}}
If we use only two replicas to calculate $\Gamma_1 (\bx,\by)$ and $\Gamma_2 (\bx,\by)$, we will introduce an annoying systematic error in
our measurements. \index{errors!systematic|(}\index{bias|see{systematic errors}}
Let us examine, for example, $\Gamma_2$, reexpressing it as a function of the overlaps using equation (\ref{eq:q-local-ising}),
$\Gamma_2 (\bx,\by) = \overline{\mean{q_\bx q_\by} - \mean{q_\bx}\mean{q_\by}}$.

During a single run of $N_\mathrm{MC}$ \ac{EMCS} \nomenclature[N...MC]{$N_\mathrm{MC}$}{number of EMCS}
and samplings $q_{\bx,t} (t=1,\ldots,N_\mathrm{MC})$, \nomenclature[q....xt]{$q_{\bx,t}$}{instant measurement}
we measure an estimator $\left[q_\bx\right]$
of the overlap's thermal average $\mean{q_\bx}$,
\begin{equation}
 \nomenclature[...]{$\left[\ldots\right]$}{estimator of the thermal average}
 \left[q_\bx\right] = \frac{1}{N_\mathrm{MC}}\sum_{t=1}^{N_\mathrm{MC}} q_{\bx,t}\,.
\end{equation}
The expected value and its estimator are related by
\begin{equation}
 \left[q_\bx\right] = \mean{q_\bx} + \eta_\bx\frac{\sigma_{\bx}}{\sqrt{N_\mathrm{MC}/(2\tau)}}\,
\end{equation}
where $\tau$ \nomenclature[tau....99B]{$\tau$}{In appendix \ref{app:MC}: integrated autocorrelation time}
is the integrated time related to $q_\bx$
\footnote{See e.g. \cite{amit:05} for informations on the relation between integrated time and number of independent measurements.},
$\eta_\bx$ is gaussian \nomenclature[eta....x]{$\eta_\bx$}{Gaussian noise}
with $\overline{\eta_\bx}=0$ and $\overline{\eta_\bx}^2=1$ that stands for 
the fluctuations around the mean, and $\sigma_\bx$\nomenclature[sigma....x]{$\sigma_\bx$}{amplitude of the noise}
is the amplitude of these fluctuations.

The estimated correlation function is then\index{correlation!function!estimated}
\begin{align}
\notag
 \left[\Gamma_2 (\bx,\by)\right] &= \overline{\left[q_\bx q_\by\right] - \left[q_\bx\right]\left[q_\by\right]} = \\[1.5ex]
\notag
 &= \overline{\mean{q_\bx q_\by} - \mean{q_\bx}\mean{q_\by}} \\[2ex]
  &+ \overline{\eta_{\bx\by}\frac{\sigma_{\bx\by}}{\sqrt{N_\mathrm{MC}/(2\tau)}}}
  + \overline{\eta_\bx\frac{\sigma_{\bx}}{\sqrt{N_\mathrm{MC}/(2\tau)}}} 
 + \overline{\eta_\by\frac{\sigma_{\by}}{\sqrt{N_\mathrm{MC}/(2\tau)}}} 
  + \overline{\eta_\bx\eta_\by\frac{\sigma_{\bx}\sigma_{\by}}{\left(N_\mathrm{MC}/(2\tau)\right)}}\,.
\end{align}
When averaging over the disorder the three terms that are linear in $\eta$ are linear in $\eta$ disappear because $\overline{\eta}=0$.
On the contrary, since $\eta_\bx$ and $\eta_\by$ are correlated $\overline{\eta_\bx\eta_\by}\neq0$, therefore 
the last term represents a bias of order $o(N_\mathrm{MC}^{-1})$ that does not disappear with an average over the disorder.

Since the disorder fluctuations are $o(N_\mathrm{samples}^{-1/2})$, as long as $N_\mathrm{MC}\gg N_\mathrm{samples}$ we can neglect this bias. 
As this is not necessarily true, so we recur to four-replica measurements to have uncorrelated fluctuations. 
With an analogous procedure to the one we just presented, the reader will notice that there
is no bias in the four-replica estimators we present in the next sections.
\index{errors!systematic|)}

\section{Computing the Replicon and Longitudinal correlation functions \label{app:GRGL}}\index{correlation!function!replicon}\index{correlation!function!longitudinal}
%Once disorder averages are performed our systems are translationally invariant, so the correlators between two points $\bx,\by$ depend only on their distance.
With 4 replicas we can construct 3 different correlators \index{correlation!function!4-replica}%at distance $\br$.
\begin{align}
\label{eq:G1}
G_1(\bx,\by)&=\overline{\langle
  s_{\bx} s_{{\by}} \rangle^2} =\nonumber\\*
&= \overline{\langle
  s_{\bx}^{(a)} s_{{\by}}^{(a)} s_{\bx}^{(b)} s_{{\by}}^{(b)} \rangle}
\,,\\[1ex]
\label{eq:G2}
G_2(\bx,\by)&=\overline{\langle s_{\bx} s_{{\by}} \rangle\langle s_{\bx}\rangle \langle 
s_{{\by}}\rangle} =\nonumber\\*
&= \overline{\langle
  s_{\bx}^{(a)} s_{{\by}}^{(a)} s_{\bx}^{(b)} s_{{\by}}^{(c)} \rangle}\,,\\[1ex]
\label{eq:G3}
G_3(\bx,\by)&=\overline{\langle
  s_{\bx}  \rangle^2\langle s_{{\by}}\rangle^2}=\nonumber\\*
&= \overline{\langle
  s_{\bx}^{(a)} s_{{\by}}^{(b)} s_{\bx}^{(c)} s_{{\by}}^{(d)} \rangle}\,.
\end{align}
None of those goes to zero for large distances $\|\bx-\by\|$, but, in the
paramagnetic phase they all tend to the same
value, $q_\mathrm{EA}$, 
when $\|\bx-\by\|\longrightarrow\infty$. So, to create connected correlators, we can
make two linearly independent combinations of them, and obtain the
basic connected propagators of the replicated field theory
\cite{dedominicis:98, dedominicis:06}~\footnote{In the effective field theory the
  longitudinal ($G_\mathrm{L}$ \nomenclature[G...L]{$G_\mathrm{L}$}{longitudinal propagator}) 
  and anomalous ($G_\mathrm{A}$ \nomenclature[G...A]{$G_\mathrm{L}$}{anomalous propagator})\index{susceptibility!anomalous}\index{susceptibility!longitudinal}
  propagators are degenerated. $G_\mathrm{R}$ is the replicon propagator.} 
\begin{align}
\index{correlation!function!replicon}\index{correlation!function!longitudinal}
\label{eq:GR}
G_\mathrm{R}&=G_1-2G_2+G_3\,,\\
\label{eq:GL}
G_\mathrm{L}&=G_1-4G_2+3G_3\,.
\end{align}
$G_\mathrm{R}$ the $G_\mathrm{L}$ are easily related to $\Gamma_1$ and $\Gamma_2$ by expanding their \index{correlation!function!connected}
expressions in equations (\ref{eq:gamma1}, \ref{eq:gamma2}). The first relation is direct,
\begin{equation}
\begin{array}{rcl}
 \Gamma_1(\bx,\by) =& \overline{\left[\langle s_\bx s_\by\rangle -\langle s_\bx\rangle^2\langle s_{\by}\rangle\right]^2} &=\\[1ex]
                   =& \overline{\mean{s_\bx s_\by}^2 - 2\mean{s_\bx s_\by}\mean{s_\bx}\mean{s_\by} +\mean{s_\bx}\mean{s_\by}} &= G_\mathrm{R}(\bx,\by)\,.
\end{array}
\end{equation}
To expand $\Gamma_2$ we complete a square
\begin{align}
 \Gamma_2(\bx,\by) &= \overline{\left[\langle s_\bx s_{\by}\rangle^2 -\langle s_\bx\rangle^2\langle s_{\by}\rangle^2\right]} = \nonumber\\
                   &= \overline{\left(\mean{s_\bx s_\by}^2 - 2\mean{s_\bx s_\by}\mean{s_\bx}\mean{s_\by} +\mean{s_\bx}^2 \mean{s_\by}^2\right)} + \nonumber\\
                   &+2\left(\overline{\mean{s_\bx s_\by}\mean{s_\bx}\mean{s_\by} - \mean{s_\bx}^2 \mean{s_\by}^2}\right) = \nonumber\\[1ex]
                   &= G_\mathrm{R}(\bx,\by) + 2\left[G_2(\bx,\by) - G_3(\bx,\by) \right]\,.
                   \label{eq:GLgamma2}
\end{align}
We can rewrite equation (\ref{eq:GLgamma2}) in the more convenient form $\Gamma_2-\Gamma_1=2(G_2-G_3)$.
Notice finally from equations (\ref{eq:GR},(\ref{eq:GL})) and equation (\ref{eq:GLgamma2}) that $G_\mathrm{L}=G_\mathrm{R}-2(G_2-G_3) = 2\Gamma_1-\Gamma2$.

The relations between $G$s and $\Gamma$s can be resumed as
\begin{equation}
\label{eq:G-rel}
\begin{array}{rl}
\index{correlation!function!replicon}\index{correlation!function!longitudinal}\index{correlation!function!connected}
 G_\mathrm{R}  &= \Gamma_1 \,,\\[1ex]
G_\mathrm{L}  &= 2 \Gamma_1  - \Gamma_2 \,,\\[1ex]
2\left(G_2 - G_3\right) &=  \Gamma_2  - \Gamma_1  = G_\mathrm{R}  - G_\mathrm{L} \,.
\end{array}
\end{equation}

The definitions (\ref{eq:GR},\ref{eq:GL}), valid at equilibrium, were used in \cite{janus:14b} 
in an out-of-equilibrium context, for lattices of size $L=80$. In that work
it had been noticed that the replicon is the only correlator that carries a significant signal.\index{correlation!function!replicon}

\begin{figure}
 \includegraphics[width=\textwidth]{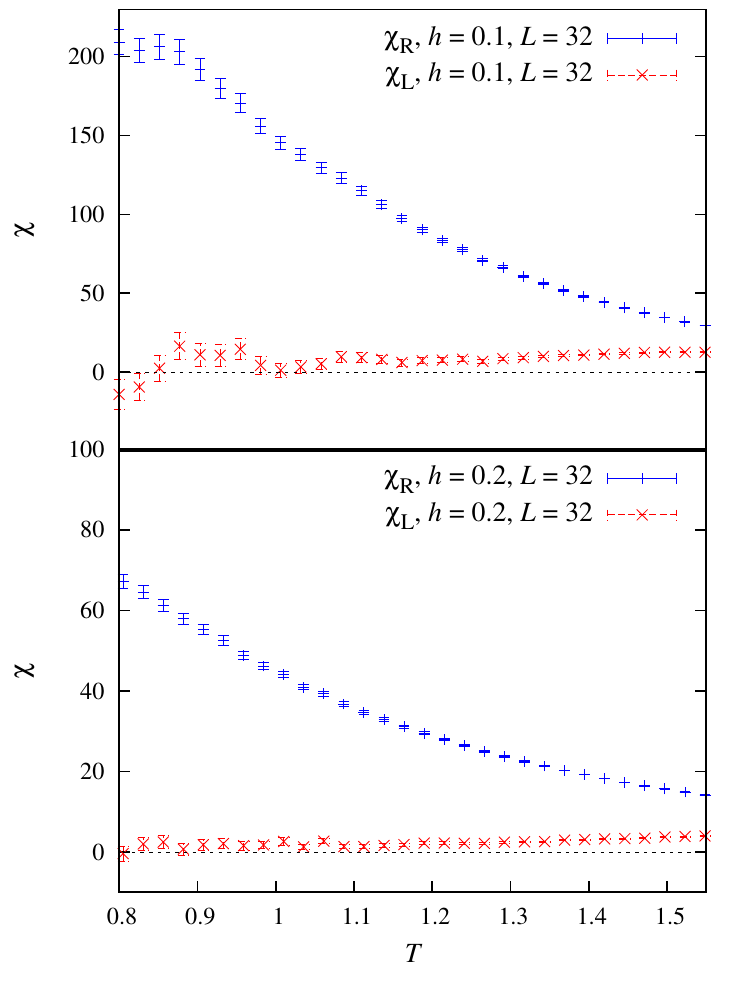}
  \caption[Replicon and longitudinal susceptibilities $\chi_\mathrm{R}$ and $\chi_\mathrm{L}$]
	  {Replicon and longitudinal susceptibilities as a function of $T$ in our equilibrium simulations, for the fields $h=0.1,0.2$ 
	  in our largest lattice sizes ($L=32$).\index{susceptibility!replicon}\index{susceptibility!longitudinal}
	  Just as in \cite{janus:14b} the signal carried by the longitudinal propagator is much smaller than that of
the replicon.
}
 \label{fig:GRGL}
\end{figure}

Also in the present work we measured both signals, and we can confirm
that the same phenomenology is observed in completely thermalised
systems. In figure \ref{fig:GRGL} we plot both the replicon \index{susceptibility!replicon}\index{susceptibility!longitudinal}
susceptibility $\chi_\mathrm{R}$ and the longitudinal susceptibility
$\chi_\mathrm{L}$, at $h=0.1, 0.2$. The figure is qualitatively very
similar to figure 13 of \cite{janus:14b}, where it is shown that
$\chi_\mathrm{R}$ carries a significant signal, while
$\chi_\mathrm{L}$ is very close to zero.\\

\index{correlation!function!4-replica|)}

\subsection{The effective anomalous dimension in the spin-glass phase\label{app:eta-eff}}\index{exponent!critical!eta@$\eta$!effective|(}
The value of the effective anomalous exponent $\eta_\mathrm{eff}$ (section~\ref{sec:zero-field}) in the deep spin-glass phase
can be predicted by using the fact that $G_\mathrm{R}$ is dominant with respect to $G_\mathrm{L}$.
\index{correlation!function!replicon}\index{correlation!function!longitudinal}

In fact, in a \ac{RSB} situation the overlap $q$ is defined over a finite range, so the
overlap's variance $\sigma^2_q = E(q^2)-E(q)^2$ is of order one:
\begin{equation}
 \mathrm{RSB}\Rightarrow \sigma^2_q\sim1\,.
\end{equation}
Now, on general grounds (see for instance \cite{fisher:91}) we can expect
\begin{equation}
\left[E(q^2)-E(q)^2\right] \sim \overline{\langle q^2\rangle-\langle q\rangle^2}\,,
\label{eq:gg}
\end{equation}
and remark that the \ac{rhs} is $\hat \Gamma_2 ({\bf0})/N$, the \index{correlation!function!4-replica}
zero-moment Fourier transform of $\Gamma_2$ [defined in (\ref{eq:gamma2})].\footnote{The correlation functions $G(\bx,\by)$ and 
$\Gamma(\bx,\by)$ are averaged over the disorder. Once this average is performed we can integrate out one of the
two spatial dependencies and write them as $G(\br)$ and $\Gamma(\br)$. There is no ambiguity in this notation: when
these function are written as depending on two parameters, it is the two positions $\bx$ and $\by$, when there is
only one parameter it is $\br=\bx-\by$.}
We have then that in \ac{RSB} conditions
\begin{equation}
 \hat\Gamma_2({\bf0})\sim N\sigma^2_q \stackrel{\mathrm{RSB}}{\sim} N\,.
\end{equation}
$\hat\Gamma_2$ can be related to the replicon and longitudinal susceptibilities through (\ref{eq:G-rel}),
\index{susceptibility!replicon}\index{susceptibility!longitudinal}
that imply that $\hat\Gamma_2({\bf0}) = 2\chi_\mathrm{R} + \chi_\mathrm{L}$.
Now, in the beginning of this section we found out empirically that the longitudinal
susceptibility is subdominant with respect to the replicon channel (figure~\ref{fig:GRGL}), so 
in the large-volume limit, in the presence of \ac{RSB}, the replicon susceptibility scales like the volume:
\begin{equation}
\index{susceptibility!replicon}
\label{eq:rsb-implies}
 \mathrm{RSB}\Rightarrow\chi_R\sim N\,.
\end{equation}
Let us recall (\ref{eq:quotients}) and impose the just-found implication.
We have then
\begin{equation}
 2^{D} \stackrel{\mathrm{RSB}}{=} \frac{\chi_{\mathrm{R},2L}}{\chi_{\mathrm{R},L}} \equiv 2^{2-\eta_\mathrm{eff}}\,,
\end{equation}
therefore in the spin-glass phase we would have $\eta_\mathrm{eff}=-1$.\\
\index{exponent!critical!eta@$\eta$!effective|)}

\section{Measuring the propagators with \acl{MSC} \label{app:MSC}}\index{correlation!function!4-replica|(}
We now write the correlators in a way that is useful for \acf{MSC}, and then we show explicitly how \ac{MSC} coding
was done on these quantities.

\subsection{Correlators as simple functions of simple fields \label{app:Xfields}}
A simple way to construct unbiased quantities is to define them as functions of fields of differences.
With four replicas we can define
\begin{equation}\label{eq:fieldX}
\begin{array}{rl}
\nomenclature[X...1xX2x]{$X_1(\bx),X_2(\bx)$}{fields of differences}
 X_1(\bx) &= (s_\bx^\rma - s_\bx^\rmb)(s_\bx^\rmc - s_\bx^\rmd)\,,\\
 X_2(\bx) &= s_\bx^\rma s_\bx^\rmb - s_\bx^\rmc s_\bx^\rmd\,.
\end{array}
\end{equation}
These are the quantities we actually measure, we want to relate them with the correlation
functions $G_\mathrm{R}$ and $G_\mathrm{L}$ (equations \ref{eq:GR}\ref{eq:GL}). \index{correlation!function!replicon}\index{correlation!function!longitudinal}

Expanding the $X_1$ field correlator we get
\begin{equation}
 \mean{X_1(\bx)X_1(\by)} = 4\mean{s_\bx^\rma s_\bx^\rmc s_\by^\rma s_\by^\rmc}
                         - 8\mean{s_\bx^\rma s_\bx^\rmc s_\by^\rma s_\by^\rmd}
                         + 4\mean{s_\bx^\rma s_\bx^\rmc s_\by^\rmb s_\by^\rmd}
 \,.
\end{equation}
On the other side rewriting the replicon propagator $G_\mathrm{R}$ \index{correlation!function!replicon}as a function of four replicas yields
\begin{equation}
 G_\mathrm{R} (\bx,\by) = \overline{\mean{s_\bx^\rma s_\by^\rma s_\bx^\rmb s_\by^\rmb}
                        -2\mean{s_\bx^\rma s_\by^\rma s_\bx^\rmb s_\by^\rmc}
                         +\mean{s_\bx^\rma s_\by^\rma s_\bx^\rmc s_\by^\rmd}}\,,
\end{equation}
so
\begin{equation}
 G_\mathrm{R} (\bx,\by) = \frac{1}{4}\overline{\mean{X_1(\bx) X_1(\by)}}\,.
\end{equation}

Equivalently, an expansion of the $X_2$ field correlator returns
\begin{align}
 \mean{X_2(\bx)X_2(\by)}&= \mean{s_\bx^\rma s_\bx^\rmb s_\by^\rma s_\by^\rmb} 
			 - \mean{s_\bx^\rma s_\bx^\rmb s_\by^\rmc s_\by^\rmd}
			 - \mean{s_\bx^\rmc s_\bx^\rmd s_\by^\rma s_\by^\rmb}  
			 + \mean{s_\bx^\rmc s_\bx^\rmd s_\by^\rmc s_\by^\rmd} =\nonumber\\[1ex]
			&= 2\left( \mean{s_\bx s_\by}^2 - \mean{s_\bx}^2 \mean{s_\by}^2\right)\,.
\end{align}
By averaging it over the disorder we can relate it to the non-connected correlators of equations (\ref{eq:G1},\ref{eq:G2},\ref{eq:G3}),
\begin{align}
 \frac{1}{2}\overline{\mean{X_2(\bx)X_2(\by)}} &= G_1(\bx,\by) - G_3(\bx,\by) =\\
  &= 2 G_\mathrm{R}(\bx,\by) - G_\mathrm{L}(\bx,\by)\,,
\end{align}
where for the second relation we used equations (\ref{eq:G-rel}).
The expression of $G_\mathrm{L}$ in terms of the fields $X_i$ becomes
\begin{equation}
 G_\mathrm{L} (\bx,\by) = \frac{1}{2}\overline{\mean{X_1(\bx) X_1(\by)}} - \frac{1}{2}\overline{\mean{X_2(\bx) X_2(\by)}}\,.
\end{equation}
Since it is possible to construct the fields $X_i$ with three independent permutations of 
the replicas ($X_i^{\rma\rmb\rmc\rmd},X_i^{\rma\rmc\rmb\rmd}$ and $X_i^{\rma\rmd\rmb\rmc}$), 
we compute correlators starting from each of those permutations and then average to reduce the fluctuations.

\subsection{Plane correlators}
Since we average over the disorder, the replicon \index{correlation!function!replicon} 
and longitudinal correlation \index{correlation!function!longitudinal} functions can be rewritten as a function of the distance vector $\br$.
We concentrate on the $G_\mathrm{R}(\br)$ because it carries the most signal and it is the one we used in our anaylses. It is expressed as
\begin{equation}
G_\mathrm{R}(\br) = \overline{\frac{1}{4}\sum_\bx\mean{X_1(\bx)X_1(\bx+\br)}}\,.
\end{equation}
For the convolution theorem, analogously as we did in equation (\ref{eq:chi-k}), we can write its Fourier transform as
\begin{equation}
\nomenclature[G...Rk]{$\hat{G}_\mathrm{R} (\bk)$}{Fourier transform of $G_\mathrm{R}(\br)$}
 \hat{G}_\mathrm{R} (\bk)=\overline{\frac{N}{4}\mean{|\hat{X}_1(\bk)\hat{X}_1(-\bk)|^2}}\,,
\end{equation}
where, for $\bk=(k,0,0)$,
\begin{equation}
\nomenclature[X...1khat]{$\hat{X}_1(\bk)$}{Fourier transform of $P(\ell)$}
\hat{X}_1(\bk) = \frac{1}{N}\sum_\ell^L \E^{\rmi k\ell} P(\ell)\,,
\end{equation}
and $P(\ell)$ is the field averaged over a plane with $x_1=\ell$
\begin{equation}
\nomenclature[P...l]{$P(\ell)$}{$X_1$ averaged over the $\ell^\mathrm{th}$ plane}
  P(\ell) = \sum_{y,z} X_1(\ell,y,z)\,.
\end{equation}
Clearly, one can choose any plane orientation, though some are easier to code than others. In our analyses we chose planes
orthogonal to the vectors of the euclidean basis and to the diagonals of the lattice [vectors of the type (1,1,0) and (1,1,1)].

The computationally demanding part of the computation of $G_\mathrm{R}(\br)$ \index{correlation!function!replicon} 
consists in creating the plane fields $P$ for all the samples and replica choices. Once we have those,
the remaining operations are of order $L$  and are quickly performed. In the next section we show how it 
was possible to speed up this problematic part of the analysis.

\subsection{Multi-spin coding \label{app:msc}}\index{multi-spin coding}\index{numerical simulations}\index{programming!parallel}
We present now \ac{MSC} \cite{jacobs:81} as an extremely fast technique to be able to calculate the elementary bricks through which we can construct our correlators.
We will show how to use \ac{MSC} to extract the plane sums $\sum_{\bx\in\text{plane}}X_1(\bx)$ from the configurations. Once they are calculated for all the planes of each direction (for example
the directions can be $x,y,z$ and the single planes are the $L$ possible plains one can construct along each direction), the core of the arithmetic operations
is done, and correlation functions are constructed quickly.

%\textbf{Creo que se introduce en \cite{jacobs:81}, pero no puedo abrirlo porque no tenemos permiso en la complu.}
\index{spin!Ising}
In a \ac{MC} simulation on Ising spins, the na\"ive approach is to store the information of each spin with an integer variable. This results in
a large waste of memory, since an integer number of $n_b$ \nomenclature[n....b]{$n_b$}{number of bits}
bits could store information for $n_b$ spins at a time. Since a bit assumes the values $b=0,1$,\nomenclature[b....99B]{$b$}{In appendix \ref{app:msc}: bit}
the spin's value is $s=1-2b$. If instead of using an integer for a
single spin we use it for $n_b$ spins, not only do we gain in memory, but also in speed. In fact, operations on the spins are highly parallelizable,
so if one performs bitwise operations on the integers storing the spins, he can ideally gain a performance factor of $n_b$.
This is the idea of multispin coding. 

Clearly this technique presents a long series of caveats and complications, since \emph{only} bitwise operations are
allowed. Storing binary magnitudes such as spins is easy, but updating them in a \ac{MC} simulations is non trivial, since the energy barriers can assume 
several values, and also it is not possible to use the same random number to update spins of the same lattice.

\paragraph{Storing the lattices}
The easiest way to parallelize is to treat groups of $n_b$ samples at a time, assigning to an $n_b$-bit integer, 
that we call a \emph{word}, \nomenclature[w.ord]{word}{$n_b$-bit integer}
the value of
the spin $s_\bx$ (or $b_\bx$, if we want to talk in terms of bits) for each of the samples. 
The bits of the word $u^{\rma}_\bx$, \nomenclature[u....ax]{$u^{\rma}_\bx$}{word}
indicating site $\bx$ and replica '$\rma$' will be
\begin{equation}
 u^\rma_\bx = [b_\bx^{\rma,1},b_\bx^{\rma,2},\ldots,b_\bx^{\rma,n_b}]\,,
\end{equation}
where we labelled with an extra superscript the different samples (i.e. bits). 
To store the full configuration of the $n_b$ samples we need $4N$ words: a word per site per replica.

The words $u^\rma_\bx$ are stored in variables of type \texttt{MYWORD}, where
\texttt{MYWORD} is usually an $n_b$-bit integer. In this work we used $n_b=128$.\index{programming!C|(}
In our C code we use triple arrays to store the configurations, so the full configurations are stored in arrays of the type \texttt{MYWORD u[NT][NR][N];} where
\texttt{NT} is the number of measurements $N_\mathrm{m}$ 
we use (recall section \ref{sec:eah3d-simulations}) and \texttt{NR} is the number of replicas,
and \texttt{N} is the number of spins $N$ in a single lattice.

If $N_\mathrm{samples}$ is a multiple of $n_b$ the method is then
fully optimized, otherwise it is enough to discard a number of bits from the last group of samples.

\subsection{Replicon correlator with \ac{MSC}}\index{correlation!function!replicon}\index{programming!multi-spin coding}
We will not face the task of explaining how to perform a \ac{MSC} simulation, that is alreadly done in
literature, for example in \cite{jacobs:81,seoane:13}. We will instead focus on how we multi-spin coded the analysis of the correlation 
function $G_\mathrm{R}$ ($G_\mathrm{L}$ is similar).

We already described in section \ref{app:Xfields} how it is possible to obtain $G_\mathrm{R}$ and $G_\mathrm{L}$ from the
fields $X_1$ and $X_2$ [equation (\ref{eq:fieldX})], that are simple enough to allow for a \ac{MSC} computation: The field $X_1$ takes only the values
$-4,0,4$, while $X_2$ takes $-2,0,2$, so they can be stored with two bytes each (per site per sample). We want to use \ac{MSC} to construct
the plane average $P(\ell)$ of $X_1$ and $X_2$, that is the most computationally demanding part of the analyses.

At the beginning of the \ac{MSC} computation we have 4 replicas $u_\bx^\rma,u_\bx^\rmb,u_\bx^\rmc,u_\bx^\rmd$ with which to construct $X_1(\bx)$ 
(for $G_\mathrm{R}$ we do not need $X_2$). \footnote{We do it with the three independent permutations of the replicas $X_1^{\rma\rmb\rmc\rmd}(\bx)$, $X_1^{\rma\rmc\rmb\rmd}(\bx)$ 
and $X_1^{\rma\rmd\rmb\rmc}(\bx)$.}

The \ac{MSC} operations have to be iterated over all the sites. Once the loop over the sites is finished the per-site analysis is over, global quantities 
are created and the \ac{MSC} part is finished. The loop over the sites is the bulk of \ac{MSC}, where we compute $n_b$ per-site observables at a time
through bitwise operations. In C the loops appears as (along with some variable declaration)
\begin{lstlisting}[name=msc,caption={C code for \ac{MSC}: Variable declarations. The // symbols indicate that the rest of the line is commented.},label={lst:msc1}]

//Where we store the final overlaps - 6 permutations, n_b samples
int q       [ 6][NUMBITS];
int temporal[12][NUMBITS];
    
//Temporary variables to store the four spins
MYWORD spinA,spinB,spinC,spinD;
MYWORD temp;

//Temporary variables to store the six overlaps
MYWORD spinAB,spinAC,spinAD,spinBC,spinBD,spinCD;

//space_N[12] is a set of buffers, used to store large vectors
//They are defined as global variables 
//MYWORD space_N[12][N];

//Buffers for large vectors
aguja1_AB_plus =space_N[0]; //Store positive values of X1_ABCD
aguja1_AB_minus=space_N[1]; //Negative values of X1_ABCD
aguja1_AC_plus =space_N[2]; //Positive values of X1_ACBD
aguja1_AC_minus=space_N[3]; //Negative values of X1_ACBD
aguja1_AD_plus =space_N[4]; //Positive values of X1_ADBC
aguja1_AD_minus=space_N[5]; //Negative values of X1_ADBC
agujaQ_AB=space_N[6];       //Overlaps qAB
agujaQ_AC=space_N[7];       //Overlaps qAC
agujaQ_AD=space_N[8];       //Overlaps qAD
agujaQ_BC=space_N[9];       //Overlaps qBC
agujaQ_BD=space_N[10];      //Overlaps qBD
agujaQ_CD=space_N[11];      //Overlaps qCD

for(site=0; site<N; site++)
{
	spinA=u[i0][0][site];
	spinB=u[i1][1][site];
	spinC=u[i2][2][site];
	spinD=u[i3][3][site];
  .
  .
  .
\end{lstlisting}

The first step is calculating the overlaps between couples of replicas. \index{multi-spin coding!overlaps}
The XOR logic gate ($\wedge$ in C) between two bits returns 1 if they are different, and 0 if they are the same.
It can be used to represent the overlap between two spins.
Calling $b^{\rma\rmb}_\bx=b^{\rma}_\bx\wedge b^{\rmb}_\bx$ the value of the bit representing the overlap $q_\bx^{(\rma\rmb)}$, will be
\begin{equation}\label{eq:bitwise-q}
q_\bx^{(\rma\rmb)} =
\begin{cases}
  +1,&\text{~if~~} b^{\rma\rmb}_\bx=0\\
  -1,&\text{~if~~} b^{\rma\rmb}_\bx=1
 \end{cases}
\,.
\end{equation}
Calling \texttt{_my_xor(out,in1,in2)} a function (or macro) that returns as \texttt{out} the bitwise XOR between \texttt{in1} and \texttt{in2},
the code continues as
\begin{lstlisting}[name=msc,caption={C code for \ac{MSC}: Computing overlaps with \ac{MSC}},label={lst:msc2}]
  .
  .
  .
//Overlaps computed with XOR gates
	_my_xor(spinAB,spinA,spinB); // AB=A^B
	_my_xor(spinAC,spinA,spinC); // AC=A^C
	_my_xor(spinAD,spinA,spinD); // AD=A^D
	_my_xor(spinBC,spinB,spinC); // BC=B^C
	_my_xor(spinBD,spinB,spinD); // BD=B^D
	_my_xor(spinCD,spinC,spinD); // CD=C^D

//Store the local overlaps
	agujaQ_AB[site]=spinAB;
	agujaQ_AC[site]=spinAC;
	agujaQ_AD[site]=spinAD;
	agujaQ_BC[site]=spinBC;
	agujaQ_BD[site]=spinBD;
	agujaQ_CD[site]=spinCD;
  .
  .
  .
\end{lstlisting}

For the fields $X_1$ the calculation is more involved, because we need to use 2 bits. Among the several possibilities,
we decide to use the two necessary bits independently. One bit stores the positive values, and the other stores the negative values. So,
if the two are the same, the value of the variable is zero, otherwise it is $+1$ or $-1$ depending on which of the two is non-zero.

The difference between two spins $s_\bx^\rma-s_\bx^\rmb$ can assume the values $-2,0,2$. It is zero if the are the same, i.e.
if their overlap is equal to $q_\bx^{\rma\rmb}=1$ [and $b^{\rma\rmb}_\bx=0$, for equation (\ref{eq:bitwise-q})]. Taking the example of
the field $X_1^{\rma\rmb\rmc\rmd}$, if either $b^{\rma\rmb}_\bx=0$ or $b^{\rmc\rmd}_\bx=0$, then the whole product is zero. For the field
$X_1^{\rma\rmb\rmc\rmd}$ to be non zero we need $q_\bx^{\rma\rmb}=q_\bx^{\rmc\rmd}=-1$ [$b^{\rma\rmb}_\bx=b^{\rmc\rmd}_\bx=1]$.

The AND gate ($\&$ in C), returns a 0 unless both input bits are 1, so $X_1^{\rma\rmb\rmc\rmd}\neq0$ if and only if $b^{\rma\rmb}_\bx\,\&\,b^{\rmc\rmd}_\bx=1$.
In that case we have to understand what sign it assumes.

Given $s_\bx^\rma -s_\bx^\rmb\neq0$, if $s_\bx^\rma=1$ then $s_\bx^\rma -s_\bx^\rmb=2$, and if $s_\bx^\rma=-1$ then $s_\bx^\rma -s_\bx^\rmb=-2$. The same
holds for $s_\bx^\rmc -s_\bx^\rmd$.
So, the product between the aforementioned differences, $X_1^{\rma\rmb\rmc\rmd}$, is inferable by comparing $s_\bx^\rma$ with $s_\bx^\rmc$
\begin{equation}
 \mathrm{sign}(X_1^{\rma\rmb\rmc\rmd})=
 \begin{cases}
  +, &\text{~if~~} q_\bx^{\rma\rmc}= +1~(b^{\rma\rmc}_\bx=0)\,\\
  -, &\text{~if~~} q_\bx^{\rma\rmc}= -1~(b^{\rma\rmc}_\bx=1)\,.
 \end{cases}
 \end{equation}
 To represent this with bitwise operations first we calculate the auxiliary value \texttt{temp}. 
 Having \texttt{temp=1} is a necessary condition for a positive $X_1^{\rma\rmb\rmc\rmd}$, so 
 (\texttt{temp} AND $b^{\rma\rmc}$) is 1 if and only if $X_1^{\rma\rmb\rmc\rmd}=1$. This means that
 we can store the bit (\texttt{temp} AND $b^{\rma\rmc}$) for the negative values of $X_1$. Equivalently,
 for the positive values we can use a NAND [NOT AND, $\sim\&$ in C (the simple not is $\sim$)] gate.
 The following commented code clarifies the procedure
 \footnote{The code contains the logic-gate macros for the AND gate, \texttt{_my_and(out,in1,in2)}, and for the NAND, \texttt{_my_andnot(out,in1,in2)}. In both
 cases the two words \texttt{in1} and \texttt{in2} are the input, and \texttt{out} is the output.}
\begin{lstlisting}[name=msc,caption={C code for \ac{MSC}: Creating the fields with \ac{MSC}},label={lst:msc3}]
  .
  .
  .
//////////////////////////////////////////////////
// We want to create the following fields       //
//                                              //
// X1[0][i]=(u[a][i]-u[b][i])*(u[c][i]-u[d][i]) //
// X1[1][i]=(u[a][i]-u[c][i])*(u[b][i]-u[d][i]) //
// X1[2][i]=(u[a][i]-u[d][i])*(u[b][i]-u[c][i]) //
//////////////////////////////////////////////////


//First field
// X1AB=(sA-sB)*(sC-sD) //
	_my_and(temp,spinAB,spinCD);               // temp=AB&CD : 
	                                           // temp=0 ==> X1=0
	                                           // temp=1 ==> X1=-2,+2
// (~AC)&(AB&CD) :
//~AC gives positive values ==> store in aguja1_AB_plus
	_my_andnot(aguja1_AB_plus[0],spinAC,temp);
//   AC &(AB&CD) : 
// AC gives negative values ==> store in aguja1_AB_minus
	_my_and(aguja1_AB_minus[0],spinAC,temp);   
	aguja1_AB_plus++;                          //Pass to the next site
	aguja1_AB_minus++;                         //Pass to the next site

	
	
	
//Second field
// X1AC    [B<>C]
	_my_and(temp,spinAC,spinBD);               // temp=AC&BD : 
						   // temp=0 ==> X1=0
						   // temp=1 ==> X1=-2,+2
// (~AB)&(AC&BD) :
//~AB gives positive values ==> store in aguja1_AC_plus 
	_my_andnot(aguja1_AC_plus[0],spinAB,temp); 
//   AB &(AC&BD) : 
// AB gives negative values ==> store in aguja1_AC_minus
	_my_and(aguja1_AC_minus[0],spinAB,temp);   
	aguja1_AC_plus++;                          //Pass to the next site
	aguja1_AC_minus++;                         //Pass to the next site

	
	
	
//Third field
// X1AD    [B<>D]
	_my_and(temp,spinAD,spinBC);               // temp=AD&BC : 
						   // temp=0 ==> X1=0
						   // temp=1 ==> X1=-2,+2
// (~AD)&(AD&BC) :
//~AD gives positive values ==> store in aguja1_AD_plus 
	_my_andnot(aguja1_AD_plus[0],spinAC,temp); // (~AD)&(AD&BC) :~AD gives positive values ==> store in aguja1_AD_plus 
//   AD &(AD&BC) :
// AD gives negative values ==> store in aguja1_AD_minus
	_my_and(aguja1_AD_minus[0],spinAC,temp);   //   AD &(AD&BC) : AD gives negative values ==> store in aguja1_AD_minus
	aguja1_AD_plus++;                          //Pass to the next site
	aguja1_AD_minus++;                         //Pass to the next site
	
}// close the loop for(site=0; site<N; site++)
  .
  .
  .
\end{lstlisting}

Once the \ac{MSC} loop is finished we have $3N$ words (one per site per permutation) each containing the site-dependent field $X_1^{\rma\rmb\rmc\rmd}(\bx)$
for the set of $n_b$ samples. 
The final step is to transform this in a sample-dependent quantity over which it is possible to perform normal arithmetic operations. Practically, we want to
transform the bits in numbers.

To this objective we call a generic function 
\texttt{suma_booleana(buffer, size, n_bits, obs)} that takes the \texttt{buffer} where the $N$ $n_b$-words are stored,
and yields an array of $n_b$ elements - one per sample - each containing information on the variable over the whole system. In other words we pass from $N$ words
each describing a site, to $n_b$ values, each describing a sample. This can be done through $O(\log(N))$ operations.

In the following listing we show how this was done with the overlap, with the array \texttt{q[6][NUMBITS]}, defined in listing \ref{lst:msc1}, that contains
the count of how many overlaps $q_\bx=-1$ there are in each system, for the six combinations of the replicas and the $n_b$ samples. In general this function
will need as extra input also the size of the lattice \texttt{size}=$N$, and the number of bits \texttt{n_bits} that are necessary to construct that number 
(usually \texttt{n_bits}$=\log_2 N$).
\begin{lstlisting}[name=msc,caption={C code for \ac{MSC}: From the multi-spin to the traditional formalism},label={lst:msc4}]
  .
  .
  .
for (k=0;k<6;k++) //Loop over the 6 overlaps (AB,AC,AD,BC,BD,CD)
{
	//q[i][ibit] counts how many local overlaps q_x=-1 there are
	suma_booleana(space_N[6+k],N,bits_of_N, q[k] );
}
  .
  .
  .
\end{lstlisting}

Regarding the correlation functions the situation is slightly more complicated. We want to average the field $X_1$ not over the whole lattice, but over specific planes,
in order to be able to compute the correlation at distance $r$.
We define \texttt{NPLANES} planes, along the directions we want to average over (privileged directions are easier to code), and loop over them.
For each direction we make a loop over the distances, and for each distance we perform the following operations:
\begin{enumerate}
 \item[(A)] The first step to average $X_1$ over the plane is to create a buffer with only the sites regarding that plane. This is done for the 
	    3 permutations of the replica indices. For each permutation we have the positive- and the negative-value buffer, that makes 6 buffers in total.
 \item[(B)] We expand each of the six buffers with \texttt{sum_booleana}, this time over an $L*L$-dimensional space. We store those data, regarding a single $r$
	    of a single direction, in 6 temporal variables \texttt{temporal} (declared in listing \ref{lst:msc1}).
 \item[(C)] We store each plane with an array \texttt{sumplane} (declared in listing \ref{lst:msc1}) that depends on the parameters of all the nested loops: plane direction \texttt{o},
	    plane position \texttt{r}, replica permutation \texttt{k}, and sample \texttt{ibit}. The storage has to be performed through the operation 
	    \path{temporal[2*k][ibit]-temporal[2*k+1][ibit]}, because \texttt{temporal[2*k][ibit]} stores the number of sites with $X_1(\bx)=1$, and \texttt{temporal[2*k+1][ibit]}
	    has information on the number of sites with $X_1(\bx)=-1$, so the full sum $\sum_{\bx\in\mathrm{plane}} X_1(\bx)$ is obtained by subtracting one from the other.
\end{enumerate}
The C code is as follows
\begin{lstlisting}[name=msc,caption={C code for \ac{MSC}: Storing the $X_1$ regarding each plane},label={lst:msc5}]
  .
  .
  .
//space_N[12] & space_S[6] are sets of buffers to store large vectors
//They are defined as global variables 
//MYWORD space_N[12][N],
//MYWORD space_S[ 6][L*L];


for (o=0;o<NPLANES;o++) // Loop in plane orientations
{ 
	for (r=0;r<L;r++)
	{
		/////////
		// (A) //
		/////////
		for (j=0;j<L*L;j++)
		{
			i=plane[o][r][j];
			for (k=0;k<6;k++)
				space_S[k][j]=space_V[k][i];
		}

		/////////
		// (B) //
		/////////
		for (k=0;k<6;k++)
		{
			suma_booleana(space_S[k],L*L,bits_de_S,temporal[k]);
			
			//In temporal[k] we have, for each sample, 
			//the sum of the X1 of type k of a plane:
			// 0 <= temporal[k][ibit] <= L*L
			//
			//k even: counts how many positive X1
			//k  odd: counts how many negative X1
			}

		/////////
		// (C) //
		/////////
		
		//Loop over the n_b samples
		for (ibit=0;ibit<NUMBITS;ibit++) 
		{
			//Loop over the 3 permutations of the replica indices
			for (k=0;k<3;k++)                 
				sumplane[k][o][r][ibit]=temporal[2*k][ibit]-temporal[2*k+1][ibit];

			//sumplane is declared as a global variable
			//int sumplane[6][NPLANES][L][NUMBITS];
			//
			//temporal[2*k][ibit] counts the number of times X1(x)=1
			//temporal[2*k+1][ibit] the number of times X1(x)=-1
			//temporal[2*k][ibit]-temporal[2*k+1][ibit]: sum_x X1(x)
	    }//ibit
	}//r
}//o

\end{lstlisting}
At this point the analysis can procede in the traditional way, by computing the plane correlators with \texttt{sumplane}.

At the end of the full procedure we will have to proceed with the correct normalization of the correlators,
taking in account for example that the $X_1$ we calculated is a factor 4 smaller than its actual value.

\index{correlation!function!4-replica|)}
\index{programming!C|)}

\chapter{Technical details on the creation of quantiles\label{app:hack}}\index{quantile|(}\index{variate!conditioning}
To grant the reproducibility of our results in chapter \ref{chap:eah3d}, we give details on how we
proceeded in the labelling of the observables with the \acf{CV}, and
over the definition of the quantiles. Section \ref{app:create-Pq} is dedicated to the construction
of the \ac{pdf} of the \ac{CV} and section \ref{app:defining-quantiles} to that of the quantiles.
In section \ref{app:2rep} we show that by using two-replica instead of four-replica correlation functions\index{correlation!function!2-replica}
the quantile description give a similar result, with the first quantiles do not show signs of scale invariance,
but the $\xi/L$ and $R_{12}$ related to the median do suggest a phase transition.\index{xiL/L@$\xi_L/L$}\index{R12@$R_{12}$}

\section{Creating the \texorpdfstring{$P(\hat q)$}{P(q)} \label{app:create-Pq}}\index{variate!conditioning|(}
As already explained in section \ref{sec:eah3d-model-simulations}
the analysis we conduct uses instantaneous realisations of the
observables, instead of the average over the equilibrium regime. 
This is because computing $P(\hat q)$ properly requires as many instances of the
overlap as possible.

Operatively, we discard the first half of each simulation from the measurements
because out of equilibrium. We divide the second half of the simulation time-series in 16
blocks, and for the 4 replicas we save the final configuration of each
block. This gives us $16^4$ configurations over which we can potentially
compute overlaps for a single sample. Since it is not feasible to make measurements over
the $16^4$ times per sample, for $N_\mathrm{t}$
times we pick 4 random numbers
between 1 and 16 to create an instant measure. This way we increase our
statistics of a factor $N_\mathrm{t}$, obtaining ${\cal N_\mathrm{m}}=N_\mathrm{samples}(L,T,h)\times N_\mathrm{t}$
measures for each triplet $(L,T,h)$. We used $N_\mathrm{t}=1000$.

With the 4 replicas it is possible to compute 6 different overlaps $q_i$
($i=1,...,6$), and one instance of most observables, for example the replicon\index{susceptibility!replicon}\index{susceptibility!longitudinal}
susceptibility $\chi_\mathrm{R}$. Our ansatz is that $\chi_\mathrm{R}$ and the overlaps have some
type of correlation, so we label $\chi_\mathrm{R}$ with some function of the overlaps
$\hat q(q_1,...,q_6)$, that we called conditioning variate.

The random variable $\hat q$ will have a probability distribution \index{variate!conditioning!distribution}
function $P(\hat q)$ that we want to calculate numerically, in order
to be able to work on the quantiles.  Since our objective is not to
individuate exactly the quantiles, but to compute observables related
to a particular quantile, we coarse grain the range of definition of
the $P(\hat q)$.  This is done by making a binning of the $P(\hat q)$ 
[equations  (\ref{eq:binning1},\ref{eq:binning2}) here below].
This way, each conditioned expectation value of a generic observable,
$E({\cal O}| \hat q)$, can be calculated over a reasonable amount of
measurements, and we have exactly one conditioned expectation value
for each bin of the $P(\hat q)$. Integrals such as those in
(\ref{eq:E-conditioned}) and (\ref{eq:regla-de-suma-var}) are
computed as sums over the histogram bins.  Furthermore, the described
histogramming procedure has the advantage that errors can be\index{errors!jackknife}
calculated in a very natural way with the \ac{JK} method.

In order to have, as $L$
increases, both a growing number of bins, and of points per bin, we choose bins
of width $\Delta \hat q = 1/\sqrt{aV}$. \nomenclature[a....a]{$a$}{parameter for the width of the bins of the $P(\hat q)$}
We add the restriction of having at
least 150 bins, in order to be able to define the quantiles properly (with large bins it
could happen that a single bin contain more than $10\%$ of the \ac{pdf},
and we want
to avoid the eventuality of two quantiles in the same bin).  We
verified that there is no appreciable difference in the results between
$a=1,2,4$. Larger $a$ implies a too large error, because the bins are too
small, while with smaller $a$ the bins are too few. The results we show
throughout this thesis have $a=2$.

To compute the conditional expectation values defined in section~\ref{sec:conditional} 
we use the following estimators:
\begin{eqnarray}
\label{eq:binning1}
 E({\cal O}|\hat q = c)&\approx& \frac{\frac{1}{{\cal N}_\mathrm{m}} \sum_i^{\cal N_\mathrm{m}} {\cal O}_i {\cal X}_{c}(\hat q_i)}
			       {\frac{1}{{\cal N}_\mathrm{m}} \sum_i^{\cal N_\mathrm{m}}{\cal X}_{c}(\hat q_i)}~,\\[2ex]
\label{eq:binning2}
 P(\hat q)             &\approx&
		\frac{1}{\cal N_\mathrm{m}}\sum_i^{\cal N_\mathrm{m}} {\cal X}_{c=\hat q}(\hat q_i)\,,
\end{eqnarray}
where with the symbol ``$\approx$'' we stress that the quantity is an estimator that converges
to the exact value only in the limit of an infinite number of measurements $\mathcal{N}_\mathrm{m}$. ${\cal X}_c$ is the characteristic function
defined in equation \eqref{eq:characteristic}.\\
\index{variate!conditioning|)}

\section{Defining the quantiles \label{app:defining-quantiles}}\index{quantile!operational definition}
As stated in section~\ref{sec:median}, the quantiles are the points that
separate definite areas under $P(\hat q)$. Therefore, the
$i^\mathrm{th}$ quantile $\tilde q_i$ is defined by means of the
cumulative distribution $X(\hat q)$ of $P(\hat q)$, via the implicit
relation
\begin{equation} 
X(\tilde q_i) = \int_{-1}^{\tilde
q_i} \mathrm{d}\hat q\ P(\hat q) = \frac{i}{10}\,.  
\end{equation} 
Since this is a
continuous relation, and our binning is discrete, it is most probable that
the quantile fall between two neighbouring bins. To evaluate the observables
right at the position of the quantile, we make linear interpolations between
the two bins.

Let us call $i^-_\mathrm{bin}$($i^+_\mathrm{bin}$) the bin just under (over)
quantile $i$. Observable ${\cal O}_i$ at quantile $i$ will be a linear
combination of the values it assumes at $i^-_\mathrm{bin}$ and
$i^+_\mathrm{bin}$:
\begin{equation}
{\cal O}_i = p\, {\cal O}_{i^-_\mathrm{bin}} + (1-p)\, {\cal O}_{i^+_\mathrm{bin}}\,,
\end{equation}
where the interpretation of the indices is straightforward, and $0\leq p\leq1$
is the interpolation weight 
\begin{equation}
  p = \frac{X(\tilde q_i) - X(\hat q_{i^+_\mathrm{bin}})}{X(\hat q_{i^-_\mathrm{bin}}) - X(\hat q_{i^+_\mathrm{bin}})}\,.
\end{equation}

\section{Quantiles with 2-replica correlators \label{app:2rep}}\index{correlation!function!2-replica}
To have well behaving (connected) correlators in the presence of a magnetic field we needed to use
4 replicas for each instance of them. As explained in sections \ref{sec:conditional} and 
\ref{sec:median}, since the overlap is a 2-replica observable, we had to choose a function 
of the 6 overlaps in order to have a one-to-one correspondence between conditioning variates
and the correlators. The functions we tried out were the minimum, the maximum, the median
and the average of the 6 overlaps.

Now, it is legitimate to ask oneself if the fluctuations we observed would also be visible 
having $q$ as conditioning variate. Although this is not possible with the replicon\index{correlation!function!replicon}
correlation function $G_\mathrm{R}$, we can renounce to have a connected correlation function,
and study the fluctuations of the 2-replica point-to-plane correlator\\
\begin{equation}\label{eq:g_2_r}
\nomenclature[G...2ncr]{$G_2^\mathrm{nc}(r)$}{2-replica non-connected plane correlation function}
 G_2^\mathrm{nc}(r) = \sum_{y,z} E(\,q_{(0,0,0)}\, q_{(r,y,z)}\,) \,,
\end{equation}
which allows us to have $q$ as a conditioning
variate. $G_2^\mathrm{nc}(r)$ is the total correlation between the
origin, $(0,0,0)$, and the plane $x=r$. Of course, one could
equivalently consider the planes $y=r$ or $z=r$. One can displace
freely the origin, as well. We average over all these $3V$ choices.

\begin{figure}[t]
\centering
 \includegraphics[width=0.6\columnwidth]{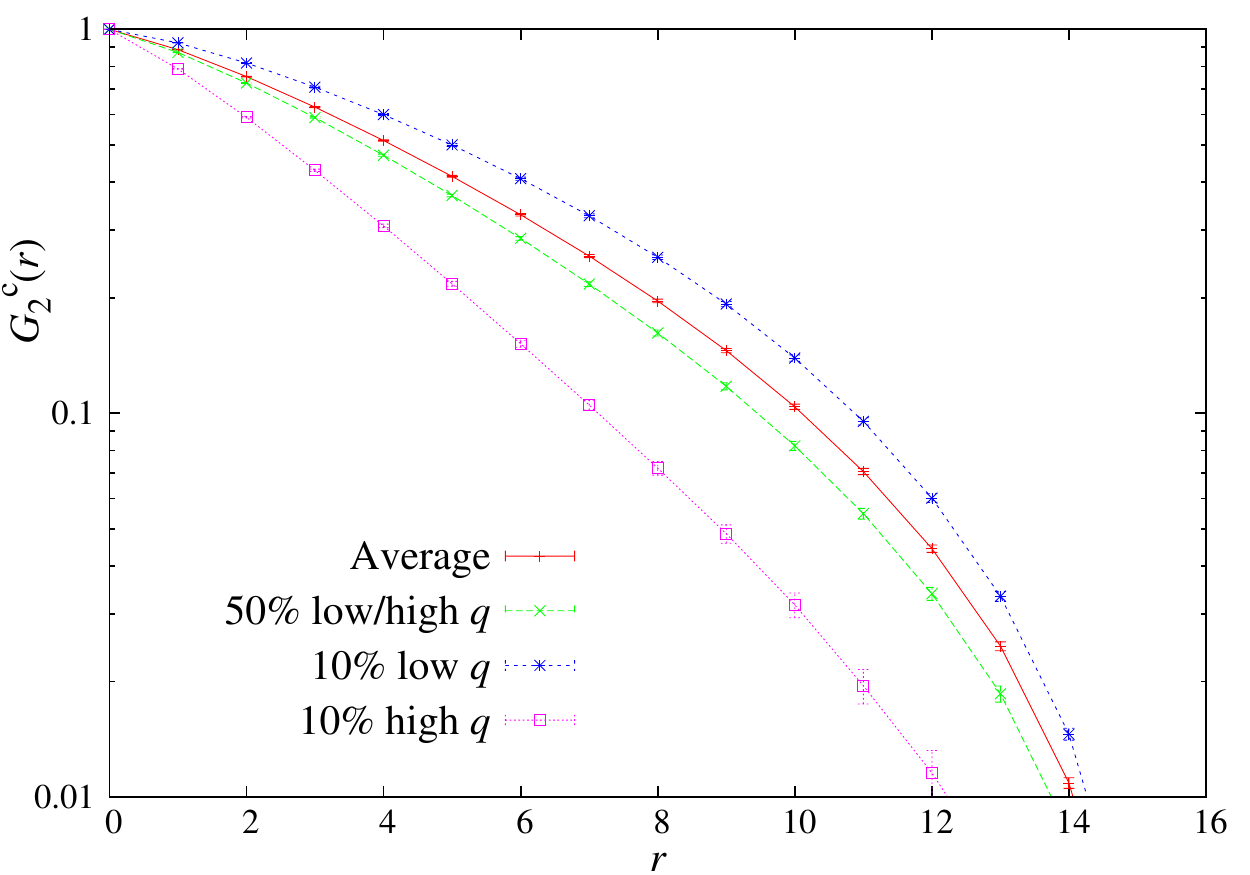}
 \caption[Two-replica correlation function $G_2^\mathrm{c}(r)$ for different quantiles]
   {Same as figure \ref{fig:Cr}, but for the 2-replica connected
   correlation function $G_2^\mathrm{c}(r)$ (\ref{eq:g_2_r_norm}). We show $L=32$ data from $h=0.2$,
   $T=0.805128$.  Note that $G_2^\mathrm{c}(r)$ is bound to be 1 at
   $r=0$, and 0 at $r=L/2$, so the fluctuations between different
   quantiles are even stronger than they may appear.}
\label{fig:G2}
\end{figure}

At this point, it is possible to compare with previous work that
studied fluctuations with 2-replica correlators \cite{parisi:12b}.
Furthermore, we can construct the pseudoconnected correlation function
\begin{equation}\label{eq:g_2_r_norm}
\nomenclature[G...2cr]{$G_2^\mathrm{c}(r)$}{2-replica connected plane correlation function}
 G_2^\mathrm{c}(r) = \frac{G_2^\mathrm{nc}(r) - G_2^\mathrm{nc}(L/2)}{G_2^\mathrm{nc}(0) - G_2^\mathrm{nc}(L/2)} \,,
\end{equation}
which forcedly is one for $r=0$, and goes to zero for $r=L/2$. In figure \ref{fig:G2} we show that
the same dramatic fluctuations encountered with $G_\mathrm{R}$ (figure \ref{fig:Cr}) are also present here.

\begin{figure}[tbh]
 \includegraphics[width=0.495\columnwidth]{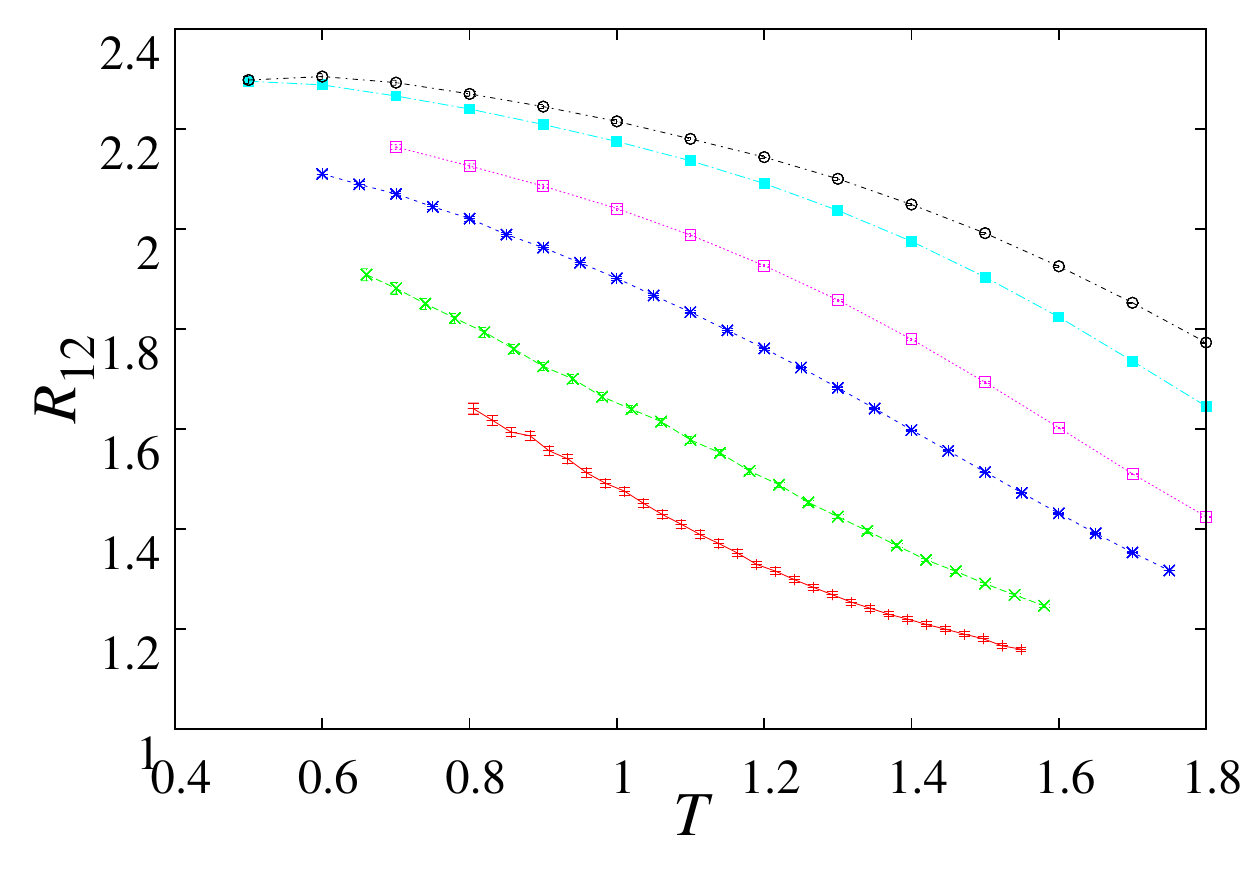}
 \includegraphics[width=0.495\columnwidth]{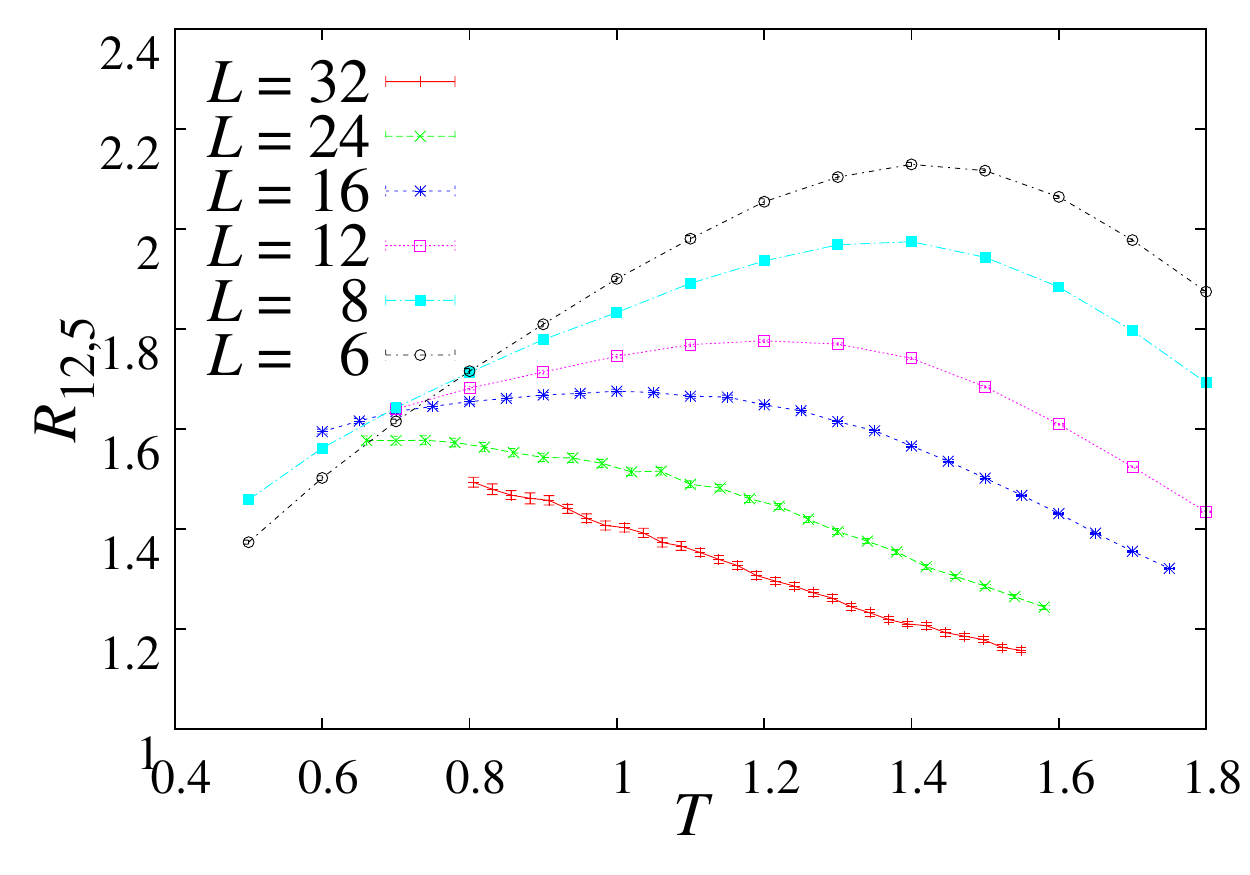}
 \caption[The $R_{12}$ cumulant computed from the two-replica
   correlation function]{The $R_{12}$ cumulant computed from the two-replica\index{R12@$R_{12}$}
   correlation function (\ref{eq:g_2_r}) rather than from four
   replicas. The field is $h=0.2$.  On the \textbf{left} side we show the
   average behavior, and on the \textbf{right}, the $5^\mathrm{th}$ quantile,
   with the plain overlap $q$ (\ref{eq:q}) as conditioning
   variate.}
\label{fig:R12-global-mediana_2rep}
\end{figure}

The overall results, figure \ref{fig:R12-global-mediana_2rep}, are\index{R12@$R_{12}$}
consistent with the picture we draw in section \ref{sec:eah3d-results}. On
the one hand, the standard data average hides all signs of a phase
transition.  On the other hand, the fifth quantile displays signs of
scale invariance.

\index{quantile|)}

\chapter{Decomposing conditional expectations \label{app:reglas-suma}}\index{regla de suma|see{sum rule}}\index{sum rule}
We want to derive here some useful relations pertinent to the conditioned expectations of chapter \ref{chap:eah3d} that can be used
to have a quantitative criterion for the conditioning variate (section \ref{sec:select-cv}) and to check that 
the statistical analysis code is reliable.

\section{Variance}
In section \ref{sec:select-cv} we used the integral rule 
\begin{equation}
\label{eq:regla-de-suma-var-full-app}
\mathrm{var}({\cal O}) = E\left(\big[\mathcal{O} - \Eo \big]^2\right) = \int_{-1}^1 d\hat{q} \, P(\hat{q})\left\{\mathrm{var}({\cal O} | \hat{q}) + \big[ E({\cal O}) - E({\cal O}| \hat{q})\big]^2\right\}\,,\\
\end{equation}
\begin{equation}
\nomenclature[v.aroq]{$\varoq$}{conditional variance}
 \varoq = E\left(\big[{\cal O} - E({\cal O}|\hat{q})\big]^2 ~|~ \hat{q}\right)\,,\\[2ex]
\end{equation}
to choose the best \ac{CV}.\index{variate!conditioning!distribution}
The $P(\hq)=E[\mathcal{X}_{\hq}]$, 
when computed numerically, is actually an empirical probability over the whole set of $\mathcal{N}_\mathrm{m}$ measurements,
\begin{equation}
\nomenclature[h....q]{$h(\hq)$}{histogram of $\hq$}
P(\hq) \approx h(\hq) 
 = \frac{\sum_{i}^{\mathcal{N}_\mathrm{m}} \mathcal{X}_{\hq_i}(\hq)}{\int_{-1}^1 d\hq\sum_{i}^{\mathcal{N}_\mathrm{m}}\mathcal{X}_{\hq_i}(\hq)}\,,\\[1ex]
\end{equation}
where the $i$ labels the measurements, 
and $\hq_i$ \nomenclature[q....hi]{$\hq_i$}{value of the \ac{CV} for measurement $i$} the value of the \ac{CV} for measurement $i$.
  
Relation (\ref{eq:regla-de-suma-var-full-app}) is easily shown to be true by applying equation 
(\ref{eq:E-conditioned}) to the variance of $\mathcal{O}$, $\mathrm{var}({\cal O})$
and summing zero to it:
\begin{align}
\nonumber  & E\left(\big[\mathcal{O} - \Eo \big]^2\right) = \\[1.5ex]
\nonumber =&\int_{-1}^1 d\hq\Pq  E\left(\big[\mathcal{O} - \Eo \big]^2\right |\hq) =\\[1.5ex]
\nonumber =& \int_{-1}^1 d\hq \Pq \left\{ E \left( \mathcal{O}^2|\hq \right) +\Eo^2 -2\Eo\Eoq + \left[ \Eoq^2-\Eoq^2\right] \right\} =  \\[1.5ex]
          =& \int_{-1}^1 d\hq \Pq \left\{ \varoq + \big[ E({\cal O}) - E({\cal O}| \hat{q})\big]^2\right\}\,.\\[1.5ex]\notag
\end{align}

\section{Higher moments}
The same procedure can be used to find a relation for higher moments.
The skewness of observable $\mathcal{O}$ is
\begin{align}
 \mathrm{S} (\mathcal{O}) = E\left[\big[\mathcal{O}-\Eo \big]^3\right] =  \int_{-1}^1 d\hq \Pq E\left(\big[\mathcal{O}-\Eo \big]^3 |\,\hq\right)\,.\\[1ex]
\end{align}
To simplify the notation let us write
\begin{align}
\nomenclature[E...Oh]{$\hEo$}{$\Eoq$}
\nomenclature[S...Oh]{$\hSo$}{$\Soq$}
\nonumber \hEo &= \Eoq\,,\\[1ex]
 \hSo &= \Soq = E\left( \big[\mathcal{O}-\Eoq \big]^3| ~\hq\right)\,,
\end{align}
so, opening the cube,
\begin{align}
 \nonumber \mathrm{S}(\mathcal{O}) =&\\[1.5ex]
 \nonumber =&\int_{-1}^1 d\hq \Pq \left\{ \hE{\mathcal{O}^3 - \Eo^3 - 3\mathcal{O}^2\Eo +3\mathcal{O}\Eo^2}\right\}= \\[1.5ex]
 \nonumber =&\int_{-1}^1 d\hq \Pq \left\{ \left[\hE{\mathcal{O}^3} +2\hEo^3 - 3 \hE{\mathcal{O}^2}\hEo\right] + \right.\\[1.5ex]
 \nonumber  &\left.- \Eo^3 + 3\hE{\mathcal{O}^2\hEo+ 3\hE{\mathcal{O}^2}\Eo- 3\hE{\mathcal{O}^2}\Eo - 2\hEo^3}  \right\}=\,,
\end{align}
 the term in square brackets is equal to $\hSo$
\begin{align}
 \nonumber =&\int_{-1}^1 d\hq \Pq \left\{ \hSo + 3\hE{\mathcal{O}^2}\left[ \hEo-\Eo\right] +\hEo^3-\hEo^3 -\Eo^3 \right.+\\[1.5ex]
 \nonumber  &~~~~~~~~~~~~~~~~~~~~~~~~~~~~~~~~\left.- 3\hEo^2\Eo+3\hEo\Eo^2-2\hEo^3+3\hEo^2\Eo  \right\} =\\
 \nonumber =&\int_{-1}^1 d\hq \Pq \left\{ \hSo + 3\hE{\mathcal{O}^2}\left[ \hEo-\Eo\right] + \left[\hEo -\Eo\right]^3\right.+\\[1.5ex]
 \nonumber  &~~~~~~~~~~~~~~~~~~~~~~~~~~~~~~~~~~~~~~~~~~~~~~~~~~~~~~~~~~~~~~~~~~~~~~~~~~~~~\left.-3\hEo^2\left[\hEo -\Eo\right]   \right\} =\\[1.5ex]
 \nonumber =&\int_{-1}^1 d\hq \Pq \left\{ \hSo + \left[\hEo -\Eo\right]\left(3\,\varoq +\left[\hEo -\Eo\right]^2\right)\right\} \,,
 \end{align}
that can also be rewritten as
\begin{equation}\label{regla-suma-skewness}
 \mathrm{S}(\mathcal{O}) = \int_{-1}^1 d\hq \Pq \left\{  \hSo + 3\,\varoq\left[\hEo -\Eo\right] +\left[\hEo -\Eo\right]^3   \right\}\,.\\[1ex]
\end{equation}

Operatively, in our spin systems we define two types of skewness of the overlap, depending on the replicas we use
\begin{align}
\nomenclature[S...2q]{$\mathrm{S}_2(q)$}{two-replica skewness}
\nomenclature[S...3q]{$\mathrm{S}_3(q)$}{three-replica skewness}
 \mathrm{S}_2(q) &= E\left[ \left(q^\mathrm{(ab)} - E(q)\right)^3\right]\,,\\[1ex]
 \mathrm{S}_3(q) &= E\left[ \left(q^\mathrm{(ab)} - E(q)\right)\left(q^\mathrm{(ac)} - E(q)\right)\left(q^\mathrm{(bc)} - E(q)\right) \right]\,.\\[1ex]
\end{align}
Applying equation (\ref{regla-suma-skewness}) to $\mathrm{S}_2(q)$ is straightforward, while for $\mathrm{S}_3(q)$ we have to apply some little modification
specifying the replica
\begin{align}\label{eq:s3}
  \nonumber\mathrm{S}_3(q) =& \int_{-1}^1 d\hq \Pq \left\{\hE{q^\mathrm{(ab)}q^\mathrm{(ac)}q^\mathrm{(bc)}} - E(q)\hE{q^\mathrm{(ab)}q^\mathrm{(bc)}+q^\mathrm{(ac)}q^\mathrm{(bc)}+q^\mathrm{(ab)}q^\mathrm{(ac)}}\right.\\[1.5ex]
                           &\left.+ E(q)^2\hE{q^\mathrm{(ab)}+q^\mathrm{(ac)}+q^\mathrm{(bc)}} - E(q)^3\right\}\,.
\end{align}
The terms in equation \ref{eq:s3} can be easily computed in our analysis out of the four simulated replicas
\begin{align}
 \hE{q^\mathrm{(ab)}q^\mathrm{(ac)}q^\mathrm{(bc)}} &= \frac{1}{4}\displaystyle\sum_{\alpha\neq\beta\neq\gamma} \hE{ q^{(\alpha\beta)}q^{(\alpha\gamma)}q^{(\beta\gamma)} } \,,\\[1ex] 
 \hE{q^\mathrm{(ab)}q^\mathrm{(bc)}} &= \frac{1}{12}\displaystyle\sum_{\alpha\neq\beta\neq\gamma}\hE{ q^{(\alpha\beta)}q^{(\beta\gamma)} + q^{(\alpha\gamma)}q^{(\beta\gamma)}+q^{(\alpha\beta)}q^{(\alpha\gamma)}} \,,\\[1ex]
 \hE{q^\mathrm{(ab)}} &= \frac{1}{6}\displaystyle\sum_{\alpha\neq\beta}\hE{q^{(\alpha\beta)}} \,,
\end{align}
where the indices $\alpha,\beta,\gamma$ in the sums indicate the different replicas.

We give the same expression for the kurtosis $\mathrm{K}=E\left[ \left(q^\mathrm{(ab)} - E(q)\right)^4\right]$\nomenclature[K...K]{$\mathrm{K}$}{kurtosis}
\begin{align}
 \notag \mathrm{K}  =&\int_{-1}^1 d\hq \Pq \left\{\hat{\mathrm{K}}(q) + \left[\hE{q}-E(q)\right]^4 +4\,\hat{\mathrm{S}}(q) \left[\hE{q}-E(q)\right] +\right.\\
          &\left.+ 6\,\var{q|\hq}\left[\hE{q}-E(q)\right]^2 \right\}\,,
\end{align}
where we introduced $\hat{\mathrm{K}}(q) = \hE{\big[q-\hE{q} \big]^4}$.\nomenclature[K...qh]{$\hat{\mathrm{K}}(q)$}{$\hE{\big[q-\hE{q} \big]^4}$}

More in general, we find that for the $n^\mathrm{th}$ moment
$K_n(\cal{O})$\nomenclature[K...nO]{$K_n(\cal{O})$}{$n^\mathrm{th}$ moment of $\mathcal{O}$}
\begin{equation}
\label{eq:regla-de-suma-generalizada}
K_n({\cal O}) = \int_{-1}^1 d\hat{q} \, h(\hat{q}) \sum_{i=0}^n
\left(\begin{array}{c} n\\i
\end{array}\right)
 K_i({\cal O} | \hat{q}) \left[ E({\cal O}) - E({\cal O}| \hat{q})\right]^{n-i}\,,
\end{equation}
where we have to notice that $ K_1({\cal O} | \hat{q})= 0$. 

\section{Consistency checks on the correlation functions \label{app:eah3d-checks}}\index{correlation!function}
Since in our analyses we often measure both the correlation function $C(\br)$ [equation \ref{eq:C})] 
and its Fourier transform $\chi(\bk)$ [equation \ref{eq:hatC})],
it is useful from a programming point of view to have some constraints that tie one to the other. 
Our programs were quite intricated,  and these constraints, despite their easy derivation, 
revealed crucial to keep the code under control.

Since $C(\br)=C(-\br)$, and because of the periodic boundary conditions,
when we calculate correlation functions along an axis, $C(r)=C(L-r)$,and $\hat{C}(k)=\hat{C}(L-k)$.
Moreover, the wave numbers restrict to $k=2\pi n/L$ ($n=0,\ldots,L-1$), so let us label them with the integer index $n$, $\hat{C}(k(n))=\hat{C}(n)$.

These symmetries give us the chance to create simple constraints on the correlators to check their consistency. 
The correlation function has to be expressable as anti Fourier transform of the $\hat{C}(k)$ through
\begin{align}
C\left(r\right) = \hat{C}\left(0\right) + 2 \sum_{n=1}^{L/2-1} \hat{C}(n) \cos{\left(\frac{2\pi n}{L}\right)}+\hat{C}\left(\frac{L}{2}\right)\,.
\end{align}
On the reverse way, we easily get basic constraints on the $\hat{C}(n)$ for some specific value of $n$:
\begin{align}
\label{eq:check-C0}
 \hat{C}\left(0\right)   &= C\left(0\right) + 2\sum_{r=1}^{L/2-1}C\left(r\right) + C\left(\frac{L}{2}\right) \,,\\[1ex]
\label{eq:check-CL2}
 \hat{C}\left(\frac{L}{2}\right) &= C\left(0\right) + 2\sum_{r=1}^{L/2-1}C\left(r\right)\left(-1\right)^r + C\left(\frac{L}{2}\right)\left(-1\right)^{L/2} \,.
\end{align}
%where in equation (\ref{eq:check-CL2}) we assumed that $L/2$ is an even number, as it was in all our simulations.
We can also get a constraint for $\hat{C}(L/4)$,
\begin{equation}
\notag
 \hat{C}\left(\frac{L}{4}\right) = C\left(0\right) + 2\sum_{r=1}^{L/2-1}C\left(r\right)\cos{\left(\frac{\pi r}{2}\right)} + C\left(\frac{L}{2}\right)\cos{\left(\frac{\pi L}{4}\right)} \,,
\end{equation}
and since $r$ is an integer index and the cosines' arguments are multiples of $\pi/2$, we can reexpress it as 
\begin{equation}
\label{eq:check-CL4}
 \hat{C}\left(\frac{L}{4}\right) = C\left(0\right) + 2\sum_{r=1}^{L/2-1}C\left(r\right)\left[1+\left(-1\right)^r\right]\left(-1\right)^{r/2} + C\left(\frac{L}{2}\right)\cos{\left(\frac{\pi L}{4}\right)} \,. 
\end{equation}
These tests were performed both on the average and on the per-quantile correlation functions.

\chapter{Managing the errors\label{app:eah3d-errors}}\index{errors|(}
The observables $\obs$ measured in the numerical experiments shown in this dissertation suffer from two noises, one due to thermal fluctuations during a single 
run, and a second one deriving from the disorder. Since we perform measurements at equilibrium, 
we can treat these measurements as \ac{iid} random variables with two independent noises.

Given a a set of $\mathcal{N}$ measurements $\obs_i$, their expected value $E(\obs)$ can be evaluated through an estimator
\begin{equation}\label{eq:estimator}
\nomenclature[E...o]{$\Est(\obs)$}{estimator of the average}
 \Est(\obs)=\frac{1}{\mathcal{N}}\sum_{i=1}^\mathcal{N}{\obs}_i
\end{equation}
that for the central limit theorem is at a $o(\mathcal{N}^{-1/2})$ distance from $\E(\obs)$.

Nonlinear functions of the observables, $f(\obs)$, \footnote{For simplicity of notation we treat functions of a single observable, but our statements
are also valid for functions of many observables.} can be estimated by evaluating them over the estimator. This results in an estimator
$f(\Est{(\obs)})$ that reproduces the actual expected value $f\left(E(\obs)\right)$ with a bias of order $o(\mathcal{N}^{-1})$ (see section \ref{app:need4}). Since
this bias is smaller than the statistical error we can neglect it.

We present in this appendix the jackknife and the bootstrap method, that are the two resampling methods that were used to calculate error bars throughout
this dissertation. Since these techniques are treated extendedly in literature (see e.g. \cite{young:12}), 
we will limit ourselves to a description of the methodology, with no pretention of originality.
\clearpage

\section{The \acl{JK} method\label{app:JK}}\index{errors!jackknife|(}
Being the central value of the linear functions of the observables $f(\obs)$ estimated as $f\left(E(\obs)\right)$, the \acf{JK} method provides us
a way to compute an appropriate uncertainty on it.
The idea is to block the data in a way that suppresses fluctuations and time correlations. Given the full set $\mathcal{B}'$ \nomenclature[B...xx]{$\mathcal{B}'$}{full set of measurements}
of measurements $\obs_i (i=1,\ldots,\mathcal{N})$, 
we group them in $n$ blocks $b_j (j=0,\ldots,n-1)$ of size $\ell$, so $n\ell=\mathcal{N}$, getting $n$ per-block estimators
\begin{equation}
\nomenclature[E...jO]{$\Est_j\left(\obs\right)$}{estimator of $\obs$ for the $j^\mathrm{th}$ block}
 \Est_j\left(\obs\right) = \frac{1}{\ell}\sum_{i \in b_j}^\ell \obs_i
\end{equation}
of the expectation value $E(\obs)$.
From those we contruct \ac{JK} estimators by creating new \ac{JK} bins. Each \ac{JK} bin 
$b_j^{\mathrm{(JK)}}$ \nomenclature[b....jJK]{$b_j^{\mathrm{(JK)}}$}{$j^\mathrm{th}$ jackknife block}
contains the full data except that regarding precisely $b_j$,
so $b_j^{\mathrm{(JK)}} = \mathcal{B}'\backslash b_j$. The \ac{JK} estimators are 
\begin{equation}
\nomenclature[E...jO]{$\Est_j^\mathrm{(JK)}\left(\obs\right)$}{estimator of $\obs$ for the $j^\mathrm{th}$ jackknife block}
 \Est_j^\mathrm{(JK)}\left(\obs\right) = \frac{1}{\mathcal{N}-\ell}\sum_{i \notin b_j}^{\mathcal{N}-\ell} \obs_i = \frac{1}{\mathcal{N}-\ell}\sum_{i \in b_j^\mathrm{(JK)}}^{\mathcal{N}-\ell} \obs_i\,,
\end{equation}
and over each of them we evaluate the nonlinear function $f_j^\mathrm{(JK)}=f\left(\Est_j^\mathrm{(JK)}(\obs)\right)$. The \ac{JK} error estimate $\sigma_f$ is then
\begin{equation}
\nomenclature[sigma....f]{$\sigma_f$}{jackknife error}
\sigma_f = \sqrt{\left(n-1\right)\left[ \frac{1}{n}\sum_{j=0}^{n-1} {f_j^\mathrm{(JK)}}^2 - \left(\frac{1}{n}\sum_{j=0}^{n-1} f_j^\mathrm{(JK)}\right)^2\right]}\,. 
\end{equation}

From a programming point of view, it is often useful to define $n+1$ \ac{JK} blocks, using the the extra one, block $n$, to store the average, so in the following section we will 
use the notation $f_n^\mathrm{(JK)}=f\left(\Est_n^\mathrm{(JK)}(\obs)\right) = f\left(\Est(\obs)\right)$.
\nomenclature[f....nJK]{$f_n^\mathrm{(JK)}$}{$f\left(\Est(\obs)\right)$ (the last extra jackknife block stores the average)}

\subsection{Variations on the \acl{JK} blocks to reduce the numerical rounding errors}\index{errors!rounding}

Reducing the rounding errors often reveals fundamental in numerical analyses, since
computers only have a finite number of decimal digits to perform arithmetical operations (we always used double precision).

Had we an infinite precision, we would calculate the variance of an observable $\mathcal{O}$ as
\begin{equation}\label{eq:varo}
 \var{\mathcal{O}} = E(\mathcal{O}^2) - E(\mathcal{O})^2\,.
\end{equation}
Yet, this approach is not always numerically stable. If the relative fluctuations are very small there is a very large amount of significant
digits between the most significant digit of the averages and the most significant digit of the deviations. This gap may be larger than the 
numerical precision, and could imply, for instance, that positive-definite quantities such as (\ref{eq:varo}) assume negative values.
To suppress these rounding errors we exploit the translational invariance of the variance
\begin{equation}\label{eq:trasl-a}
 E(\mathcal{O}^2) - E(\mathcal{O})^2 = E\left((\mathcal{O}-c)^2\right) - E(\mathcal{O}-c)^2~~,~~~~\forall c\in\mathbb{R}
\end{equation}
to enhance numerical stability with the convenient choice $c=\Est(\mathcal{O})$. By measuring quantities with this offset we contain
the gap that causes the rounding errors.

Consequently, when we construct the \ac{JK} blocks we do it in two steps. First we calculate $\Est(\mathcal{O})$, and only later the variance (or higher moments).
With this election equation (\ref{eq:trasl-a}) becomes 
\begin{equation}
 \var{\mathcal{O}} = E\bigg(\left(\mathcal{O}-\Est(\mathcal{O})\right)^2\bigg)-E\bigg(\mathcal{O}-\Est(\mathcal{O})\bigg)^2\,.
\end{equation}
This translates in a correction that we have to apply to every \ac{JK} block but the $n^\mathrm{th}$ one, the one that stores the average,
because in that case the second term is zero.

One can extend this reasonment to the $r^\mathrm{th}$ moment of the observable. We show it for the quantile-dependent moments of $q$, since 
they were widely used in our programs. Let us use the contracted notations $\Est_j\equiv\Est_j(q|\hq)$ \nomenclature[E...j]{$\Est_j$}{$\Est_j(q|\hq)$}
when the estimator is not followed by parantheses,
and expand the polinomial
\begin{align}
  &\Est_j \left(\left[q-\Est_j(q|\hq)\right]^r\right) = \\[1ex]
 =&\Est_j \left(\left[(q-\Est_n)-(\Est_j-\Est_n)\right]^r\right) = \\[1ex]
 =&\sum_{s=0}^r \binom{r}{s}\Est_j\left((q-\Est_n)^{r-s}\right)(\Est_j-\Est_n)^s\,.
\end{align}
The first moments $r=2,3,4$ are 
\begin{align}
 \Est_j \left(\left[q-\Est_j(q|\hq)\right]^2\right)  =& \Est_j \left(\left[q-\Est_n(q|\hq)\right]^2\right) - \left(\Est_j(q|\hq)-\Est_n(q|\hq)\right)^2\,,\\[5ex]
 \Est_j \left(\left[q-\Est_j(q|\hq)\right]^3\right)  =& \Est_j \left(\left[q-\Est_n(q|\hq)\right]^3\right) -2\left(\Est_j(q|\hq)-\Est_n(q|\hq)\right)^3 +\nonumber\\[1ex]
						      &- 3\Est_j\left((\left[q-\Est_n(q|\hq)\right]^2\right)\left(\Est_j(q|\hq)-\Est_n(q|\hq)\right)\,,\\[5ex]
 \Est_j \left(\left[q-\Est_j(q|\hq)\right]^4\right)  =& \Est_j \left(\left[q-\Est_n(q|\hq)\right]^4\right) -3\left(\Est_j(q|\hq)-\Est_n(q|\hq)\right)^4 +\nonumber\\[1ex]
						      &- 4\Est_j\left((\left[q-\Est_n(q|\hq)\right]^3\right)\left(\Est_j(q|\hq)-\Est_n(q|\hq)\right) +\\[1ex]
						      &+ 6\Est_j\left((\left[q-\Est_n(q|\hq)\right]^2\right)\left(\Est_j(q|\hq)-\Est_n(q|\hq)\right)^2\,,\\\nonumber
\end{align}
where it is clear that in the $n^\mathrm{th}$ block all the terms of the right hand sides disappear except the first.
\index{errors!jackknife|)}

\section{The bootstrap method\label{app:bootstrap}}\index{errors!bootstrap}
The bootstrap method is a valuable tool to calculate mean and variance of an estimator, as well as other moments (see \cite{efron:94} for a detailed treatise). It comes in a wide variety of
variants, and we will give the procedure for a very simple one, that we have used in the work here described.
Similarly to the \ac{JK} method, the estimator of the central value is the one described in equation (\ref{eq:estimator}), and the procedure concerns the determination 
of its uncertainty.

Given a population $\mathcal{X}_0$ of $\mathcal{N}$ measurements we resample it $N_\mathrm{b}$ times. Each resampling consists in recreating a population of $\mathcal{N}$ elements,
by picking them at random from the initial population. This means that each element of $\mathcal{X}_0$
can appear several times or not appear at all in the generic resampled population $\mathcal{X}_i (i=1,\ldots,N_\mathrm{b})$.

From each of the $N_\mathrm{b}$ populations we extract quantities $x$ such as the average or the median, and calculate their simple and quadratic averages
\begin{align*}
 x^{(1)}_i &= \frac{1}{\mathcal{N}}\sum_{j\in\mathcal{X}_i}^{\mathcal{N}} x_j\,,\\
 x^{(2)}_i &= \frac{1}{\mathcal{N}}\sum_{j\in\mathcal{X}_i}^{\mathcal{N}} x_j^2\,.
\end{align*}
The bootstrap error is then
\begin{equation}
 \sigma_\mathrm{b} = \sqrt{\left(\frac{N_\mathrm{b}}{N_\mathrm{b}-1}\right)\left[\frac{1}{N_\mathrm{b}}\sum_{i=1}^{N_\mathrm{b}} x^{(2)}-\left(\frac{1}{N_\mathrm{b}}\sum_{i=1}^{N_\mathrm{b}} x^{(1)}\right)^2\right]}\,.
\end{equation}
The magnitude of $\sigma_\mathrm{b}$ does not depend on the number of resamplings $N_\mathrm{b}$, but to take best advantage out of the method it is good that each data point be represented 
in the resampling, so as a general rule we adopted $N_\mathrm{b}=10\mathcal{N}$ to be able to make a proper resampling of the data set.

\index{errors|)}

%FINITO
 \chapter{The \aclp{IS} \label{chap:hsgrf-is-algorithm}}\index{inherent structure|(}\index{numerical simulations|(}
This appendix referers mainly to chapter \ref{chap:hsgrf} (section \ref{app:gs} refers to chapter \ref{chap:hsgm}), and it is dedicated to show how we
found the \acfp{IS} (section \ref{app:minimizing}), to the comparison between \acp{IS} reached with different protocols (section \ref{app:sor-lambda-T}),
and to the derivation of the Hessian matrix at the local minimum of the energy.

An \ac{IS} is the configuration to which the system converges when we decide to relax it. When we talk about relaxing,
we mean to give the best satisfaction to all the local constraints, that is moving towards the nearest energy minimum.
Although this concept seems well-defined, there is an ambiguity related to what one means by \emph{nearest}.

One could in principle define a distance, find all the minima of the energy, and see which of those minimizes this distance. 
Yet, different definitions of a distance can give different results, and especially in discrete models degeneracies are not excluded by this definition.
\footnote{Two minima can be equivalent candidates for being the \ac{IS} of an excited configuration.} 
Moreover, we do not have a way to measure all the 
local minima of the energy, and even if we had, it is not granted that the physical evolution converge to a minimum defined this way.

More in general, since when we minimize the energy we are following a non-equilibrium procedure, there is a component of
arbitrariety on the protocol we use. The mostly used way to minimize the energy in spin systems is through a quench, i.e.
with the Gauss-Seidel algorithm (section \ref{app:gs}), that is local and minimizes maximally the energy in each update, 
and can be seen as a zero-temperature \ac{MC}. \index{Monte Carlo!zero-temperature}
Nonetheless, there is no solid reason to state that \acp{IS} found with one algorithm are more representative than others, but
there also is none to say that all the inherent structures are equivalent. It has been shown in \cite{baityjesi:11} that 
the algorithm choice does imply some differences on the average properties of the \acp{IS}, but we show in this appendix that they 
are small enough to be neglected.

\clearpage

\section{Minimizing the energy \label{app:minimizing}}
We discuss two very simple algorithms of energy minimization that were used in this thesis.

\subsection{Gauss-Seidel \label{app:gs}} \index{Gauss-Seidel}\index{quench@quench$ $|seealso{Gauss-Seidel}}
The most commonly used way to minimize the energy of a \ac{SG} is the Gauss-Seidel algorithm, that
consists in successive local rearrangements of the spins that decrease maximally the local energy.\index{energy!local}
The spin update with Gauss-Seidel consists in aligning each spin to its local field
\begin{equation}
  \vec{s}_\bx^{\,\mathrm{Q}} = \frac{\vec{h}_\bx}{|\vec{h}_\bx|}~~~,~~~\vec{h}_\bx=\sum_{\|\bx - \by\|}J_{\bx\by} \vec{s}_{\bx}
\end{equation}
Energy minimizations with the Gauss-Seidel algorithm are often called quenches, since they consist 
in lowering abruptly the energy (temperature) of the system.
Since sometimes in literature also variants of Gauss-Seidel have also been called quenches, one also
refers to Gauss-Seidel as a greedy quench.

The problem with Gauss-Seidel is that despite a very fast initial decrease of the energy, after few steps its convergence
to a local minimum becomes so slow that the algorithm is not usable to obtain \acp{IS} on large lattices (see e.g. \cite{sokal:92},
where it is explained that in systems with continuous degrees of freedom convergence problems arise).

\subsection{Successive Overrelaxation \label{app:sor}}\index{successive overrelaxation}
To overcome the convergence trouble of the quenches, we recur to the \acf{SOR}, that consists in an interpolation, 
through a parameter $\Lambda$, between a greedy quench with the Gauss-Seidel algorithm,
and the microcanonical \ac{OR}\index{overrelaxation} \index{Monte Carlo!overrelaxation}
update shown in appendix \ref{sec-app:algorithms}.

We propose sequential single-flip updates with the rule
\begin{equation}
 \vec{s}_\bx^{\,\mathrm{SOR}} = \frac{\vec{h}_\bx + \Lambda \vec{s}_\bx^\mathrm{\,OR}}{||\vec{h}_\bx + \Lambda \vec{s}_\bx^\mathrm{\,OR} ||}\,.
\end{equation}
The limit $\Lambda=0$ corresponds to a direct quench that notoriously presents convergence problems. On
the other side, with $\Lambda=\infty$ the energy does not decrease.

It is shown in \cite{baityjesi:11} that the optimal value of $\Lambda$ in terms of
convergence speed is $\Lambda\approx300$.

\section[Testing the dependency on \texorpdfstring{$T$}{T} and \texorpdfstring{$\Lambda$}{L}]
{Testing the dependency on \texorpdfstring{$\boldsymbol{T}$}{T} and \texorpdfstring{$\boldsymbol{\Lambda}$}{L}\label{app:sor-lambda-T}}
In chapter \ref{chap:hsgrf} we used \ac{SOR} with $\Lambda=300$ because the Gauss-Seidel algorithm, that is recovered by 
setting $\Lambda=0$, has strong convergence problems and it was not possible to reach the \acp{IS} for the system sizes we needed. 
To validate the generality of our results we compared the \acp{IS} reached with $\Lambda=300$ and $\Lambda=1$, at $H_\amp=0$ over a wide range of temperatures.
We took advantage, for this comparison, of the $L=48$ configurations that were thermalized in \cite{fernandez:09b}, that go from
$T_\SG$ to $\frac{5}{3}T_\SG$.

\begin{figure}[!b]
 \includegraphics[width=\columnwidth]{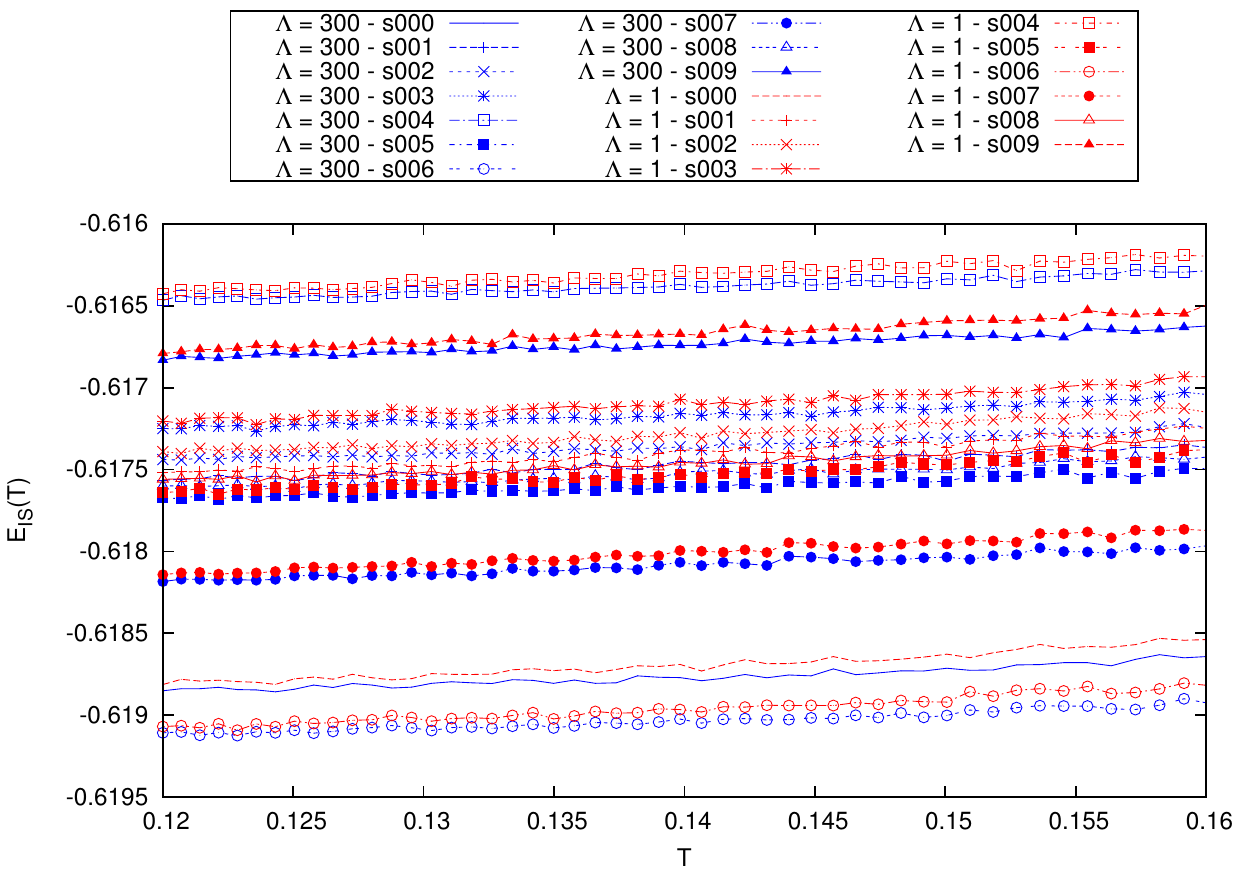}
  \caption[Comparison of the \acs{IS}'s energy for different $\Lambda$ and $T$.]{
  Energy of the inherent structure as a function of temperature for 10 samples chosen at random, for $H_\amp=0$.
  We use the same symbol for the same sample. \acp{IS} obtained with $\Lambda=300$ are in blue. Red represents $\Lambda=1$.
  Sample-to-sample fluctuations are the largest source of dispersion, compared with $\Lambda$ and $T$.}  
  \label{fig:ecompare-lambda}
\end{figure}

In figure \ref{fig:ecompare-lambda} we plot the energy $e_\IS$ of the reached \acp{IS}, as a function of the temperature $T$.
We show ten random samples, each minimized with $\Lambda=1, 300$. 
Increasing $\Lambda$ the energy of the inherent structures decreases but this variation is smaller than the 
dispersion between different samples. The energy of the \acp{IS} also decreases with $T$, but this decrease too is smaller than the
fluctuation between samples.
Since the dispersion on the energy is dominated by the disorder, rather than by $\Lambda$ or $T$, we can think of putting ourselves in the most
convenient situation: $T=\infty$, that does not require thermalization and $\Lambda=300$, that yields the fastest minimization.

\begin{figure}[!t]
 \includegraphics[width=0.49\columnwidth]{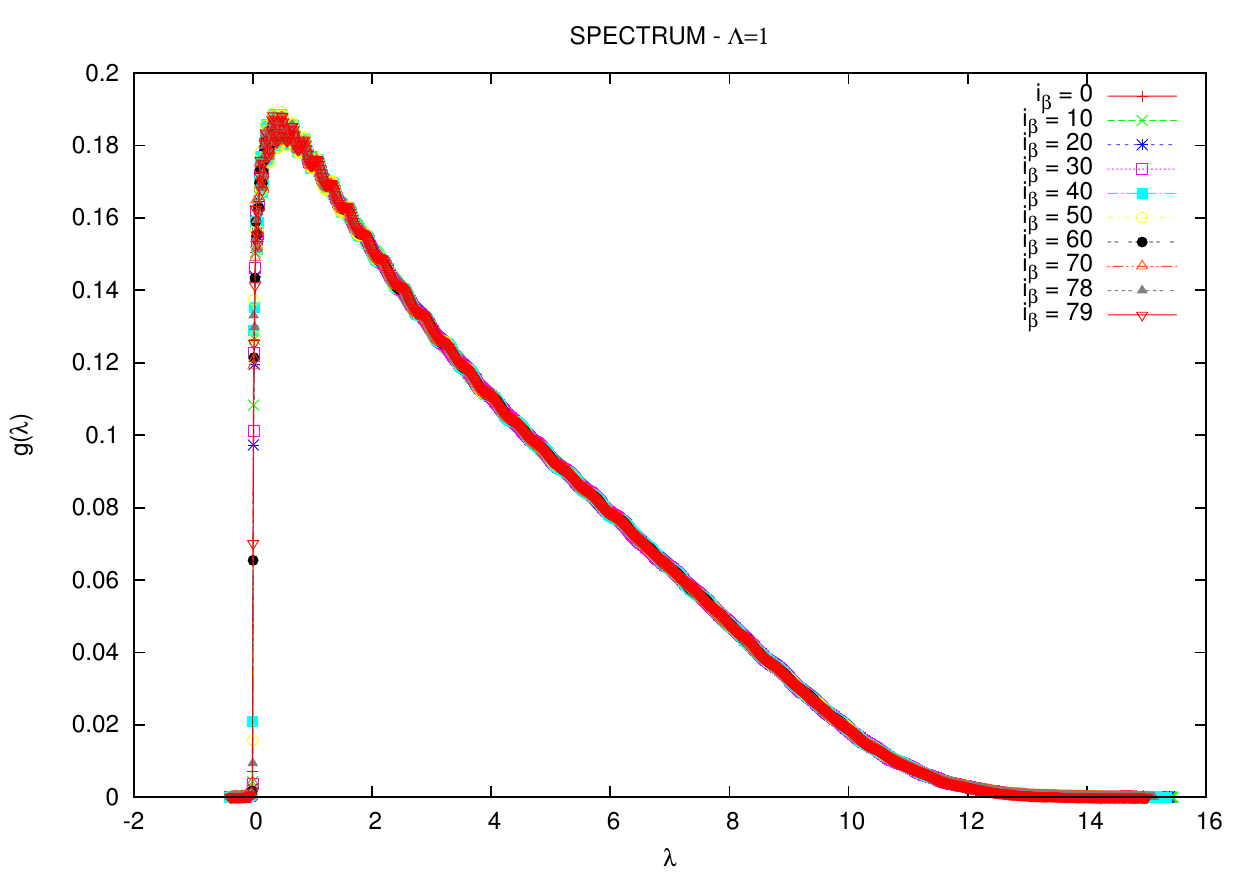}
 \includegraphics[width=0.49\columnwidth]{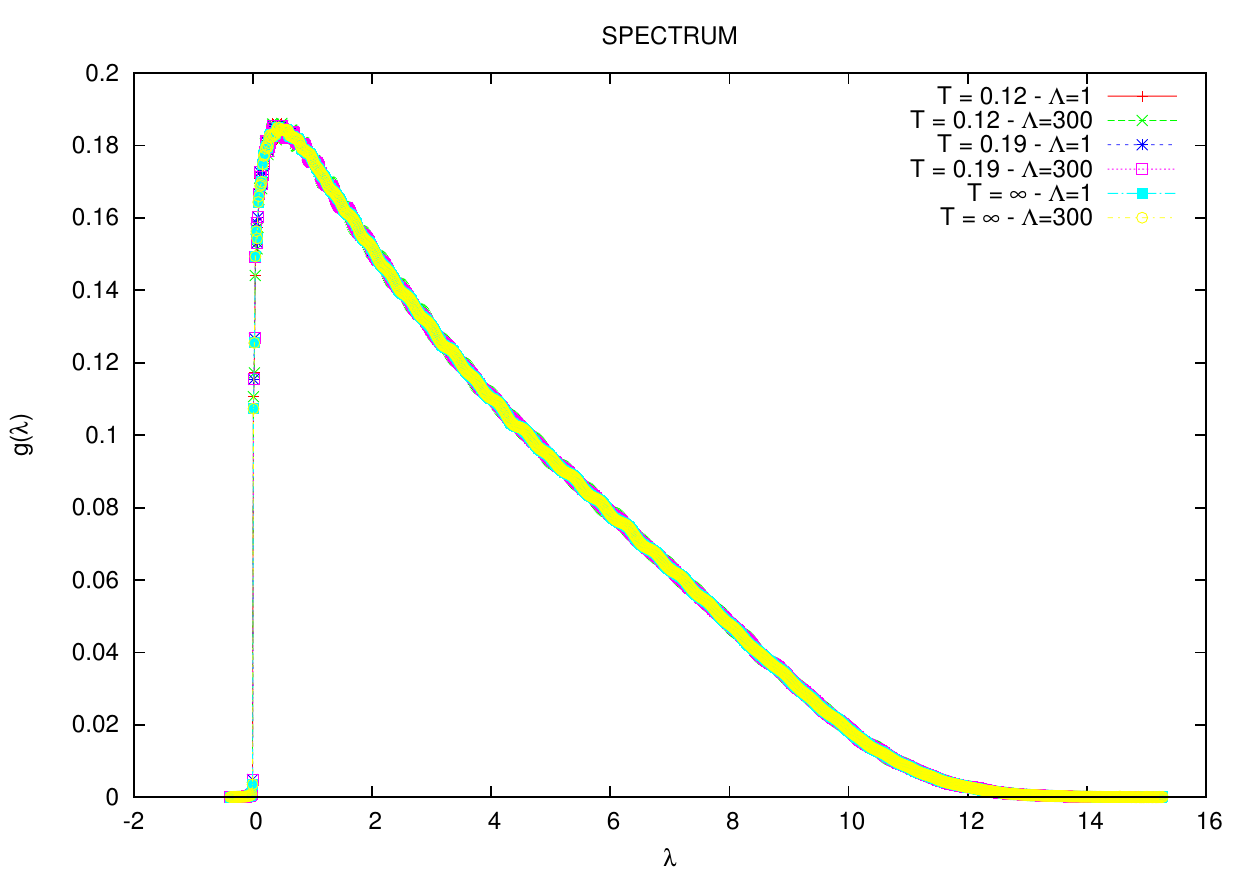}
 \caption[Dependency of the spectrum on $T$ and $\Lambda$.]{\index{Hessian matrix!spectrum}
  Spectrum $\rho(\lambda)$ of the Hessian matrix calculated at the inherent structure for $H_\amp=0$.
 \textbf{Left}: $\rho(\lambda)$ for different temperatures from $T=0.12$ to $T=\infty$. 
 \textbf{Right}: comparison of the spectrum between $\Lambda=1$ and $\lambda=300$ at $T=0.12, 0.19, \infty$ vary $\Lambda$.
 %Note: $i_\beta=0$ corresponds to $T=0.12$, $i_\beta=78$ to $T=0.19$ and $i_\beta=79$ to $T=\infty$.
 }
  \label{fig:spectrum-dep}
\end{figure}

Also the spectrum of the dynamical matrix, to which a great attention is dedicated in 
the whole chapter \ref{chap:hsgrf}, does not show relevant signs of dependency on either $T$ of $\Lambda$, as shown in figure \ref{fig:spectrum-dep}.
\index{numerical simulations|)}
\index{inherent structure}

\section[Derivation of \texorpdfstring{$\mathcal{M}$}{M}]{Derivation of \texorpdfstring{$\bm{\mathcal{M}}$}{M} \label{app:hsgrf-hessian}}
In this section we derive the expression of the Hessian matrix $\M$ of the Hamiltonian $\mathcal{H}_\RF$ \eqref{eq:HRF} that we implemented in our programs.
In terms of pionic perturbations, recall \eqref{eq:pions}, $\M$ would be defined as \index{pion|(}
$\M_{\bx\by}^{\alpha\beta}=\frac{\partial^2\mathcal{H}_\RF}{\partial\pi_{\bx,\alpha}\pi_{\by,\beta}}$.
An easy way to extract the Hessian is to write $\mathcal{H}_\RF$ as perturbations around the \ac{IS} and to pick only the second-order terms.

To rewrite $\mathcal{H}_\RF$ as a function of the pionic perturbations, it is simpler to compute separately the dot products $\left(\vec{s}_\bx\cdot \vec{s}_\by\right)$
and $\vec{h}_{\bx} \cdot \vec{s}_{\bx}$.
Including the $\epsilon$ factors into the perturbation $\pi_x$, the generic spin near the \ac{IS} is expressed as 
$\vec{s}_\bx = \vec{s}^{\,(\IS)}_\bx\sqrt{1-\vec{\pi}_\bx^2} + \vec{\pi}_\bx$.
We can make a second-order expansion of the non-diagonal part of the Hamiltonian by taking
the first-order expansion of the square root $\sqrt{1-\vec{\pi}_\bx^2}\simeq1-\vec{\pi}_\bx^2/2$,
\begin{align}\label{eq:term-coupling}
 &\left(\vec{s}_\bx\cdot \vec{s}_\by\right) =\\[3ex]
=&   \left(\vec{s}^{\,(\IS)}_\bx\sqrt{1-\vec{\pi}_\bx^2} + \vec{\pi}_\bx\right)
\cdot\left(\vec{s}^{(\,\IS)}_\by\sqrt{1-\vec{\pi}_\by^2} + \vec{\pi}_\by\right) =\nonumber\\[3ex]
=& \sqrt{1-\vec{\pi}_\bx^2}\sqrt{1-\vec{\pi}_\by^2}\left(\vec{s}^{\,(\IS)}_\bx\cdot\vec{s}^{\,(\IS)}_\by\right)
  +\sqrt{1-\vec{\pi}_\bx^2}\left(\vec{\pi}_\by\cdot\vec{s}^{\,(\IS)}_\bx\right)+\nonumber\\[2ex]
  &~~~~~~~~~~~~~~~~~~~~~~~~~~~~~~~~~~~~~~~~~~~~~~~~~~~~~~~~~~~~~~~~
  +\sqrt{1-\vec{\pi}_\by^2}\left(\vec{\pi}_\bx\cdot\vec{s}^{\,(\IS)}_\by\right)+\bigg(\vec{\pi}_\bx\cdot\vec{\pi}_\by\bigg) = \nonumber\\[3ex]
=& \left(1-\frac{\vec{\pi}_\bx^2}{2}\right)\left(1-\frac{\vec{\pi}_\by^2}{2}\right)
   \left(\vec{s}^{\,(\IS)}_\bx\cdot\vec{s}^{\,(\IS)}_\by\right)
  +\left(1-\frac{\vec{\pi}_\bx^2}{2}\right)\left(\vec{\pi}_\by\cdot\vec{s}^{\,(\IS)}_\by\right) +\nonumber\\[2ex]
 & ~~~~~~~~~~~~~~~~~~~~~~~~~~~~~~~~~~~~~~~~~~~~~~~~
 +\left(1-\frac{\vec{\pi}_\by^2}{2}\right)\left(\vec{\pi}_\by\cdot\vec{s}^{\,(\IS)}_\bx\right)
  +\left(\vec{\pi}_\bx\cdot\vec{\pi}_\by\right) + o(|\vec{\pi}|^3)\simeq\nonumber\\[3ex]
\simeq&   \left(\vec{s}^{\,(\IS)}_\bx\cdot\vec{s}^{\,(\IS)}_\by\right)+\left(\vec{s}^{\,(\IS)}_\bx\cdot\vec{\pi}_\by\right)+
\left(\vec{s}^{\,(\IS)}_\by\cdot\vec{\pi}_\bx\right)+\\[2ex]
 &~~~~~~~~~~~~~~~~~~~~~~~~~~~~~~~~~~~~~~~~~~~~~~~~~~~~~~~
 +\frac{1}{2}\left[\left(-\vec{\pi}_\bx^2-\vec{\pi}_\by^2\right)\left(\vec{s}^{\,(\IS)}_\bx\cdot\vec{s}^{\,(\IS)}_\by\right)+2\vec{\pi}_\bx\cdot\vec{\pi}_\by\right]
  \nonumber.\\\nonumber
 \end{align}
On the other hand the random-field term is
\begin{align}\label{eq:term-rf}
  \left(\vec{h}_\bx\cdot \vec{s}_\bx\right) 
  =&\vec{h}_\bx\cdot\left(\vec{s}^{\,(\IS)}_\bx\sqrt{1-\vec{\pi}_\bx^2} + \vec{\pi}_\bx\right)&\simeq&\\[1ex]
  \simeq&\vec{h}_\bx\cdot \left[\vec{s}^{\,(\IS)}_\bx\left(1-\frac{\vec{\pi}_\bx^2}{2}\right) + \vec{\pi}_\bx\right]&=&
  \left(\vec{h}_\bx\cdot\vec{s}^{\,(\IS)}_\bx\right) + \left(\vec{h}_\bx\cdot\vec{\pi}_\bx\right)
  -\frac{\vec{\pi}_\bx^2}{2}\left(\vec{h}_\bx\cdot \vec{s}^{\,(\IS)}_\bx\right) \,.\nonumber
 \end{align}
 By inserting eqs.(\ref{eq:term-coupling},\ref{eq:term-rf}) and taking only the second-order terms we obtain how the 
Hessian matrix acts on the fields $\ket{\pi}$
\begin{align}
 &\frac{1}{2}\bra{\vec\pi_\bx}\M\ket{\vec\pi_\by} =\\[2ex]
 =&-\frac{1}{2}\sum_{<\bx,\by>} J_{\bx,\by}\left[\left(-\vec{\pi}_\bx^2-\vec{\pi}_\by^2\right)
 \left(\vec{s}^{\,(\IS)}_\bx\cdot\vec{s}^{\,(\IS)}_\by\right)+2\vec{\pi}_\bx\cdot\vec{\pi}_\by\right]+
 \sum_\bx^N \frac{\vec{\pi}_\bx^2}{2}\left(\vec{h}_\bx\cdot \vec{s}^{\,(\IS)}_\bx\right) =\nonumber\\[2ex]
 =&\frac{1}{2}\sum_\bx^N\vec{\pi}_\bx^2\left[\vec{s}^{\,(\IS)}_\bx\cdot\left(\vec{h}^{\,(\IS)}_\bx+\vec{h}_\bx\right)\right]
 +\frac{1}{2}\sum_\bx\vec{\pi}_\bx\cdot\sum_{\by:\norm{\bx-\by}=1}J_{\bx\by}\vec{\pi}_\by\nonumber\,,
\end{align}
where we called $\vec{h}^{\,(\IS)}_\bx$ the local field of the \ac{IS}. 
The just-obtained expression represents a sparse matrix with a matrix element $\M_{\bx\by}$ that comfortably splits as 
$\M_{\bx\by}=\mathcal{D}_{\bx\by}+\mathcal{N}_{\bx\by}$ into a diagonal
term $\mathcal{D}_{\bx\by}$ and a nearest-neighbor one $\mathcal{N}_{\bx\by}$, with
\begin{align}
 \mathcal{D}_{\bx\by} &= \delta_{\bx\by} \left[\vec{s}^{\,(\IS)}_\bx\cdot\left(\vec{h}^{\,(\IS)}_\bx+\vec{h}_\bx\right)\right]\,,\\[1ex]
 \mathcal{N}_{\bx\by} &= -\sum_{{\mu}=-d}^d J_{\bx\by} \delta_{\bx+\hat{e}_\mu,\by}\,,
\end{align}
where $\hat{e}_\mu$ is the unit vector towards one of the 2$d$ neighbors.

\paragraph{$\M$ in the local reference frame} 
The last step is to get an expression of the Hessian matrix in the local reference frame, that includes the spin normalization constraint.

In the local reference frame the pions are written like $\vec{\pi}=a_1\hat{e}_{1,\bx}+ a_2\hat{e}_{2,\bx}$ because they are perpendicular to the first
vector of the basis, $\vec{s}^{\,(\IS)}_\bx$, and that is why we write them in a two-dimensional representation as 
$\tilde{\pi}=(a_1,a_2)$ (see section \ref{sec:hsgrf-localref}).

In this local basis, the matrix element acting on the pions is written as
\begin{align}
 \vec{\pi}_\bx\M_{\bx\by}\vec{\pi}_\by 
 =& (a_{1,\bx}, ~a_{2,\bx})\begin{pmatrix}
                                                         \M_{\bx\by}(\hat{e}_{1,\bx}\cdot\hat{e}_{1,\by})&\M_{\bx\by}(\hat{e}_{2,\bx}\cdot\hat{e}_{1,\by})\\
                                                         \M_{\bx\by}(\hat{e}_{1,\bx}\cdot\hat{e}_{2,\by})&\M_{\bx\by}(\hat{e}_{2,\bx}\cdot\hat{e}_{2,\by})
                                                        \end{pmatrix}
							\left(\begin{array}{c}
								a_{1,\by}\\a_{2,\by}
                                                              \end{array}\right)\,,
\end{align}
so in the 2$N$-dimensional reference $\M$ is expressed as
\begin{equation}
\M_{\bx\by}^{\alpha\beta}=\M_{\bx\by}\left(\hat{e}_{\alpha,\bx}\cdot\hat{e}_{\beta,\by}\right)\,, 
\end{equation}
and the elements of the diagonal and nearest-neighbor operators $\mathcal{D}$ and $\mathcal{N}$ become
\begin{align}
\nomenclature[D..xyab]{$\mathcal{D}_{\bx\by}^{\alpha\beta}$}{diagonal part of $\M_{\bx\by}^{\alpha\beta}$}
\nomenclature[N..xyab]{$\mathcal{N}_{\bx\by}^{\alpha\beta}$}{non-diagonal part of $\M_{\bx\by}^{\alpha\beta}$}
 \mathcal{D}_{\bx\by}^{\alpha\beta} &= \delta_{\bx\by}\,\delta^{\alpha\beta} \left[\vec{s}^{\,(\IS)}_\bx\cdot\left(\vec{h}^{\,(\IS)}_\bx+\vec{h}_\bx\right)\right]\,,\\[1ex]
 \mathcal{N}_{\bx\by}^{\alpha\beta} &= -\sum_{{\mu}=-d}^d J_{\bx\by} \delta_{\bx+\hat{e}_\mu,\by}\left(\hat{e}_{\alpha,\bx}\cdot\hat{e}_{\beta,\by}\right)\,.
\end{align}

\paragraph{A consistency check}
A consistency and debugging check we could run with the Hessian matrix is to control that the configurations were actually
inherent structures, by verifying that for small perturbations of order $\epsilon$ the energy variations were quadratic in $\epsilon$
\begin{equation}
 \mathcal{H}-\mathcal{H}(\epsilon) = \frac{\epsilon^2}{2}\bra{\pi}\M\ket{\pi} + o(\epsilon^3)\,.
\end{equation}

\index{pion|)}
 
% \section[Calculating the spectrum of \texorpdfstring{$\mathcal{M}$}{M}]{Calculating the spectrum of \texorpdfstring{$\bm{\mathcal{M}}$}{M}}
% To find the spectrum of $\M$ we use two similar algorithms, based on the decomposition
% of the matrix in Krilov's space. The first one, generically known as the method of the moments,
% yields the full $g(\omega)$ but is inexact at the tails of the distribution (App.~\ref{app:moments}). The second one is
% Arnoldi's method, and we use it to compute explicitly the smallest few eigenvalues of $\M$ (App.~\ref{app:arnoldi}).
% \subsection{Method of the moments \label{app:moments}}
% \subsection{Arnoldi's Algorithm \label{app:arnoldi}}

%FINITO

%  ************************************************************
%  BIBLIOGRAFÍA 
%  ************************************************************

\cleardoublepage
\phantomsection
\nolinenumbers
\addcontentsline{toc}{chapter}{\protect\numberline{}Bibliography}
%{\small\bibliography{/homenfs/rg/biblio}}
\begingroup
% \raggedright
% \sloppy
% \printbibliography
{\small\bibliography{biblio}}
\bibliographystyle{tesisalphnum}
\endgroup

%  ************************************************************
%  ÍNDICES VARIOS
%  ************************************************************

 \clearpage\phantomsection
 \chapter*{Acronyms}
 \chaptermark{Acronyms}
\addcontentsline{toc}{chapter}{\protect\numberline{}Acronyms}
\begin{multicols}{2}
\small
\begin{acronym}
\acro{1RSB}[1-RSB]{one-step replica symmetry breaking}
\acro{A}[A]{random (aleatory) dynamics}
\acro{CG}[CG]{chiral glass}
\acro{CPU}[CPU]{central processing unit}
\acro{Cu}[Cu]{copper}
\acro{CV}[CV]{conditioning variate}
\acro{dat}[dAT]{de Almeida-Thouless}
\acro{dof}{degrees of freedom}
\acro{dos}[DOS]{density of states}
\acro{DM}[DM]{Dzyaloshinskii-Moriya}
\acro{EA}[EA]{Edwards-Anderson}
\acro{EMCS}[EMCS]{elementary Monte Carlo step}
\acro{fp}[FP]{fixed point}
\acro{FP7}[FP7]{Seventh Framework Programme}
\acro{FPGA}[FPGA]{field programmable gate array}
\acro{FSS}[FSS]{finite-size scaling}
\acro{G}[G]{greedy dynamics}
\acro{GPU}[GPU]{graphics processing unit}
\acro{HB}[HB]{heat bath}
\acro{HPC}[HPC]{high performance computing}
\acro{IEA}[IEA]{Ising-Edwards-Anderson}
\acro{iid}[i.i.d.]{independent identically distributed}
\acro{IS}[IS]{inherent structure}
\acro{JK}[JK]{jackknife}
\acro{lhs}[l.h.s.]{left hand side}
\acro{MC}[MC]{Monte Carlo}
\acro{MF}[MF]{mean field}
\acro{Mn}[Mn]{manganese}
\acro{MPI}[MPI]{message passing interface}
\acro{MSC}[MSC]{multi-spin coding}
\acro{OR}[OR]{overrelaxation}
\acro{pdf}[pdf]{probability distribution function}
\acro{PRNG}[PRNG]{pseudo-random number generator}
\acro{PT}[PT]{parallel tempering}
\acro{R}[R]{reluctant dynamics}
\acro{REM}[REM]{random energy model}
\acro{RF}[RF]{random magnetic field}
\acro{RG}[RG]{renormalization group}
\acro{rhs}[r.h.s.]{right hand side}
\acro{RKKY}[RKKY]{Ruderman-Kittel-Kasuya-Yosida}
\acro{RS}[RS]{replica symmetric}
\acro{RSB}[RSB]{replica symmetry breaking}
\acro{RW}[RW]{random walk}
\acro{SG}[SG]{spin glass}
\acro{SK}[SK]{Sherrington-Kirkpatrick}
\acro{SOC}[SOC]{self-organized criticality}
\acro{SOR}[SOR]{successive overrelaxation}
\acro{TAP}[TAP]{Thouless-Anderson-Palmer}
\end{acronym}
\end{multicols}

\clearpage \phantomsection
 \addcontentsline{toc}{chapter}{\protect\numberline{}List of Figures}
 \listoffigures

 \clearpage \phantomsection
\addcontentsline{toc}{chapter}{\protect\numberline{}List of Tables}
 \listoftables

 \cleardoublepage\phantomsection
\renewcommand{\nompreamble}{\markboth{Notation}{Notation}}
\addcontentsline{toc}{chapter}{\protect\numberline{}Notation}
\printnomenclature[3.3cm]

%  ************************************************************
%  ÍNDICE ALFABÉTICO
%  ************************************************************

{
\scriptsize
 \cleardoublepage
 \phantomsection
 \addcontentsline{toc}{chapter}{\protect\numberline{}Alphabetic index}
 %\printindex[default][{\small Boldface page numbers refer to mentions in a figure or table.}]
 \printindex[default][{}]
}\normalsize

\begingroup
\thispagestyle{plain}
% pag                    Avant Garde
% fvs                    Bitstream Vera Sans
% pbk                    Bookman
% bch                    Charter
% ccr                    Computer Concrete
% cmr                    Computer Modern
% pcr                    Courier
% mdugm                  Garamond
% phv                    Helvetica
% fi4                    Inconsolata
% lmr                    Latin Modern
% lmss                   Latin Modern Sans
% lmtt                   Latin Modern Typewriter
% LinuxBiolinumT-OsF     Linux Biolinum (formerly 'fxb' in older package versions)
% LinuxLibertineT-OsF    Linux Libertine (formerly 'fxl' in older package versions)
% pnc                    New Century Schoolbook
% ppl                    Palatino
% ptm                    Times
% uncl                   Uncial
% put                    Utopia
% pzc                    Zapf Chancery
\fontfamily{pag}\selectfont

\chapter*{Retrospección}\thispagestyle{plain}\footnotesize
Ahora que esto acaba, quiero usar la excusa para mirarme un poco atr\'as, y ver con satisfacci\'on que mi doctorado en Madrid,
adem\'as de darme la posibilidad de trabajar cuatro a\~{n}os en algo que me gusta y me compensa, me ha regalado muchas satisfacciones también fuera
del ambiente acad\'emico, y me ha permitido de cultivar muchas inquietudes personales. Siento que he vivido.

He dejado estas consideraciones para el final, esperando el momento cuando finalmente me sintiera inspirado. 
El tiempo ha pasado y la inspiración se preocupó bien de no venir a visitarme, o de presentarse exactamente cuando era imposible tan solo escribir un mínimo apunte.
Ahora que tengo que imprimir la versión final del manuscrito, no me queda que hacer como siempre todo deprisa, al último momento.

Me gustaría hacer un resumen más o menos ordenado de mis años de tesis en Madrid, 
pero voy a terminar volcando un flujo de consciencia que podría acabar siendo injusto, y por esto pido disculpas. El tiempo es enemigo,
aunque si no lo fuese no nos moveríamos.
Así me esforzaré en escribir de cuantas más experiencias y personas pueda, recordando nombres y caras casi olvidadas, a la vez que presencias imperecederas,
que hayan marcado de una forma u otra mi transcurrido aquí.

Ante todo fue una tesis doctoral, así que empezaré por el ambiente académico, tan agradable, que me acogió cálidamente.
Le debo mucho a Víctor, que se me dedic\'o a pleno con un afecto casi paterno, permitiendo que este doctorado saliese tan bien,
a Luis Antonio, David y Bea, que siempre se mostraron muy disponibles para guiarme y asistirme con cualquier cosa,
y José Manuel, con quién escribí mi primer artículo.

Fuera de mi pequeño grupo de investigación también encontré todo el calor que podía necesitar. Ante todo en el \emph{Domilab}, la insigne
institución con sede en la mesa del Ilustrísimo y Excelentísimo Dr. \emph{Dominicus I}, que hospedó en los años a 
científicos y futbolistas de mundial renombre, entre los cuales el mismísimo Dominicus, Bea, Vivy, Santos, Édgar, Joserra, Rafa y el recién
llegado Ismael, que a lo mejor ni sabe de Domilab.

Saliendo del Domilab, en Teórica estuve rodeado de personas que es una suerte conocer, como Markus, Jenifer, David, Óscar, Alex, Laura,
José Antonio, Arkaitz, José Alberto, Nykos, Davide, Giovanni, Álvaro, Prado, Santiago, Andrés y Juan Miguel. Con ellos pude disfrutar de salidas de fiesta a 
las que Santos cumplía con la presencia, Vivy con la borrachera, y Domi cerrando las discotecas.
Muchos de ellos tuvieron la paciencia de soportar un año de \emph{Seminario Aperitivo},
el ciclo de seminario que organicé donde llevaba siempre picoteo y bebida para acompañar las charlas,
hasta el día que me petó el estómago y desde entonces hubo solo comida.
A la hora de hacer los certificados de partecipación tuve que cambiar el nombre a \emph{Transversal Seminars} porque Seminario Aperitivo
parecía demasiado de cachondeo.

Fuera de Teórica - no los voy a culpar por hacer una investigación menos interesante - hay que nombrar a los nucleares, y a los que no lo son pero
en mi mente es como si lo fueran, como Richi, Armando y Samu (que lo es, pero es como si no lo fuera). Vadym, sin embargo, al parecer es físico nuclear - dentro de poco va a sacar una aburridísima tesis
sobre núcleos pesados - y lo son también Viky, Paula, Esther, Edu, Jacobo, Maylin, Marie. El círculo puede seguir expandiéndose, de Izarra,
Belén y Samuel hasta el Pibe, Mariano y David.

En escuelas de verano y congresos he conocido a muchos cuyas caras ya son borrosas, pero no me olvido de Ludovica y Gino, con quienes conecté
especialmente en Hillerød, y claramente toda la pandilla de doctorandos y postdocs italianos, con quienes en principio debería haber compartido mucho tiempo en Roma, pero
que en la práctica solo encontraba en los congresos, ya que en Roma estuve muy poco. Son Jacopo, Paolo, Carlo, Aurelien, Bea (la mismísima), 
Matteo, Corrado, Pierfrancesco. Seguro que me olvido de alguien, pero esto está escrito en español, así que no se van a enterar. También quiero nombrar a Jacopo (otro!),
Mario y Manuel, los astrofísicos con quienes compartía despacho en Roma, y con quienes pasé mucho tiempo las veces que estaba ahí.

Abarcado vagamente el contexto académico, quiero recordar lo que fue este período además de una tesis doctoral en física.\\

\thispagestyle{plain}

Fue mi iniciación al tango. Llevaba mucho tiempo pensándolo, pero quitando una vez en París con María, nunca me había animado a aprenderlo, hasta que justo en mis primeros
meses Vivy y Domi me propusieron de ir al Patio Maravillas. Era gratis, lo que permitió que me animase a ir a pesar de un solapamiento parcial con mis entrenamientos de taekwondo.
Al cabo de un tiempo casi todos los físicos del departamento bailaban en el Patio.

Al principio me costaba más que lo normal conseguir concadenar dos pasos seguidos, pero me fascinaba ver a Carlos, Andrea y Amelia, y quería ser un día como ellos. Recuerdo
el momento que me cambió el chip y entendí cómo funciona eso. Fue cuando Jorge me cogió y me llevó por una tanda. Entendí el tango, fue mi contacto con el cambio
de roles, y especialmente, fue el principio de una amistad y la creación de un grande grupo de amigos que adoro, con quienes compartí tardes en el parque,
botellones, viajes y noches insomnes. Además de los ya citados comprende, Francesco y Rumi, y se extendió rapidamente en mil direcciones, contando Ginevra,
Itziar, Clara y nuestros entrenamientos en el salón, Roc\'{i}o, Andrea, Juanjo, Giuliana, Fernando, Marlena, Paloma, Maria, María, Julio, Luis, Santiago, 
Jorge, Carole, Maria Chiara, y podría seguir gastando kilos de tinta mencionando a
todas las personas con quienes tuve un contacto relevante a través del tango.

De ahí fue una pasión que no paró más. Siguió, a través de Amelia, con Fer y Vale. Me llevó a bailar durante las permanencias en el Patio, en el mercado de san Fernando, 
en Seúl con sus malas experiencias, y en las milongas en el Pier 45 del Hudson River. Ya no pienso
parar, ahora van a ser milongas en Philadelphia y clases con Claudia in París.

\thispagestyle{plain}

Fue mucho fútbol. Renormalotes, con Jacobo, Christian, Domi, Iribar. La mítica Gran Potra y sus terceros tiempos, con Armando, Vadym, 
Alberto, Richi, Samu, Keller, Rafón, Ricardo y también nuestro fichaje vallisoletano, el Nano, y sus goles a pesar de 40 kilos de sobrepeso
y una incipiente hepatitis; 
partidos de mercenario con Samu, en los Zores y con las chicas y chicos del Zulo, Cristina, Rober. Llegar a playoffs de coña con Personal, Carlos, Vadym, Jose, David.
Los Ángeles, Gonzalo, Vadym, Chema. Los futbol, birras y pibitas, entre la ESA y Canal, con Rithy, Vadym, Jose, Armando, David.

\thispagestyle{plain}

Fue el contacto con Dani, Janise, Tony, Dylan, Emma, Molly y esa mentalidad retorcida y sin responsabilidades con la cual he aprendido y disfrutado mucho.

\thispagestyle{plain}

Fue ligar sin parar, antes guiado por Gabba, con su extremo carisma, y luego con Jorge y Francesco, con esa inexplicable
ley del triángulo orientado, que siempre ha valido. Sin ninguna presión por realmente conseguir algo, dejando que las cosas viniesen solas. 
A veces. Otras estábamos de la olla, pero siempre lo principal era jugar disfrutando.

\thispagestyle{plain}

Fue MH41, con una pared de latas, Catán, partidas de PES a las cinco de la mañana, un solo rey del campo de juego, y fiestones mensuales con media Madrid, empezando por
Richi, Samu, Juli, Adri, Ceci, Oskar.

Fue reflexiones a la hora del café, y nachdenken um kaffeezeit.

Los fiestones continuaron, aunque con menos fuerza debido al tipo de piso, en Mesón de Paredes, donde Valentino traía a toda su peña Erasmus, entre los
cuales Livia y Shalini, María a los muertos vivos, Francesco a tangueros y folkeros, Dani a su pandilla \emph{ratchet}, y Noelia trajo a Marcel,
que fue una alegría tener. Baobab, zapatillas, piso status: ratchet, forza Roma.

\thispagestyle{plain}

Fue una introducción parcial, pero lo suficiente profunda, en el mundo de la moda, que bajo la mayoría de sus aspectos no me atrae mucho,
y que sin embargo me ha regalado experiencias muy fuertes y la pasión por las azoteas abandonadas.

\thispagestyle{plain}

Fue un menjunje de amistades y relaciones que se fueron construyendo y deshaciendo a lo largo de cuatro años. Jorge, Rumi, Amelia, Francesco, Gabba, Dani, Janise, Richi, Samu, 
Juli, Luciano, María, Danilo, Esther, Ginevra, Vivy, Bea, Javirulo, Pablo, Alex, Ana, María O, Dani, Irene, Clara, Inés,
Kike, Michelle, Rocío, Alejandro, Diego, Valentino, Claudia, Francesca, Marion, Alba y Miguel.

\thispagestyle{plain}

Fue voleiból en la Complu, con Héctor, Pablo, Samu, Xexu, Javi Perez, Estrellita, Piru, Sandra y todo el resto del equipo.

\thispagestyle{plain}

Fue una estancia en Nueva York, con Antoine y los franchutes de l'École Polytéchnique, Edan, tango, Andy, Saugerties, Kyle, Samantha, 
los WASPS que hay que eliminar de la Tierra, y Cheryl, que me acogió como un rey con Connor y Ryan.

\thispagestyle{plain}

Fue circo y acrobacia, comprenderlo realmente, saltos en la cama elástica, palomas que nunca salían, Pablo, Álvaro, Sonia,
Juán, Elena, Rafa, Alicia, tapas al salir, antes con Rumi, luego con María.

\thispagestyle{plain}

Fue taekwondo, Manolo que me guío paternamente hasta llegar a mi cinturón negro con la bandera de España, María, 
Irene, Manolo, Dani, Juanan, Miguel, Inés, cañas después de entrenar, Víctor, Marcos, Alex, Laura, David, Carlos, y un campeonato que me costó dos ligamentos y me dió
la oportunidad de renovarme con paracaidismo, relatos y folk.

\thispagestyle{plain}

Fue Nano en Valladolid, escapadas en ambas direcciones, un Viña Rock con Francesco y Bea.

\thispagestyle{plain}

Fue mucha bicicleta, romper en dos el cuadro de la del padre de Aurora, llegar ya cansado a los entrenamientos, y moverme en sincronía con Francesco y Amelia.

\thispagestyle{plain}

Fue paracaidismo, siguiendo el consejo de Davide de ahorrarme la pasta del tandem. Fue la lindísima experiencia con Dani Baelo, el mejor instructor de los cielos, 
Vicente, Lucía, saltar sólo, y la promesa de volver a hacerlo cuando las circunstancias sean favorables.

\thispagestyle{plain}

Fue el taller de relatos en el Patio, escuchar mucho y escribir poco, Kike, Michelle, Rocío, y por fin presentar mi cuento en castellano el 
día antes de irme a Nueva York. Me ayudó Aurora, ahora lo admito, yo no escribo tan bien.

\thispagestyle{plain}

Fue compartir momentos en Madrid con algunos amigos del erasmus, y volverme loco de la alegría cuando los volví a ver todos en París, Aurora, María, Claudia, piña, Jelen y Chris, 
Uri, Diego, parrapportausport, Alejandro, Antuán, Gonzalo, Irene, parrapportroyal, todo seguía tal como 6 años antes.

\thispagestyle{plain}

Fue mucha música. Clases de Juli en MH41 y una banda sonora muy bonita grabada con Rubén, soñando el Patio, tango bango, un corto nunca hecho, tango bukkango, tango obsession, 
banda sonora reciclada para Andromeda's Son, figuración de baile con Amelia. Other Names, donde Jesús, Abdel y Pancho, más que de música me enseñarón lo que es vivir realmente al día.
De lo mejor que me ha pasado en Madrid es que Francesco me liase con las jams de folk, con lo cual casi tengo que agradecer haber perdido los ligamentos. Retomé de una vez el clarinete,
y me apunté a las quedadas semanales, donde me divertía muchísimo con 
Pablo, Roberto, Francesco, Alex, Lucía, Marcel, Adara, Juan Sarrion, Fabio. La cosa mejoró aun más gracias a Domi, que nos vino a ver y nos propuso de tocar en las fiestas de Malasaña.
Fuimos solo Francesco, Alex, Rober y yo, con Clara que animaba a la gente a danzar. Una experiencia de cuento de hadas, con decenas de personas bailando divertida, como en los festejos
del medievo (los que vienen en las pelis de Hollywood). Ese concierto desencadenó muchos más, un sin parar de bolos para el grupo de la jam, sin capacidad de encontrarse un nombre. Podía agregarse quién quisiese,
pero al final éramos siempre los mismos, los cuatro de antes, Pablo, Juán, y Adara. Jam Folk Madrid.

\thispagestyle{plain}

Claramente fue Patio Maravillas, enamoramiento y vida en espacios sociales, entender cómo una mentalidad abierta y disponible se puede llegar a un mundo mejor, y 
en esto mi ejemplo siempre ha sido Domingo. Fue participación con ganas y alegría, con turnos de barra, domingos rojos y, al llegar de la orden de desahucio, 
permanencias alegres y pijama parties tangueros.
El Patio cambió rádicalmente mi manera de pensar, 
haciéndome entender cómo hay formas de pensamiento ortogonales a la dominante, llevándome a la convicción de que el anarquismo es la manera
en la que se gestiona una sociedad feliz. 
Todo esto empezó con la Milonga Antifascista del Hondo Bajo Fondo.

\thispagestyle{plain}

Fue ver que cuando volvía a casa nada había cambiado, que podía volver y Valerio, Dimitri, Nano, Guldo y Andrea eran los mismos de siempre. Tener la seguridad de que Madrid es mi casa y siempre lo será,
pero que también Roma nunca dejará de serlo. Y en esto es fundamental la estabilidad y el afecto de mi familia. Cada día que pasa me doy más cuenta del privilegio de tener a Tato, Maio,
Anna, Silvia, Sergino, Giorgia, Cristiano, Riccardo, Stefano, Valeria, Azzurra, Marzia, Fiammy, Nano,
Trilly, Flamy, Malena, Annablu, la Nonna, y Mamá. Son mi más grande seguridad, y nunca renunciaré a ustedes.

\thispagestyle{plain}

Fue mucha ayuda por parte de amigos y familia. Solo para mencionar unos ejemplos, Santos y Davide me cubrían las clases, 
Bea me siguió los papeleos de la FPU,  mamá me hizo muchos los papeleos desde Roma coordinándose con Valerio y Andrea, Claudia
me aguantó y me ofreció todo su apoyo con mudanza e impresión de tesis, Vadym me ayudó a preparar ciertas clases,
Pablo, Sergino y mi padre Keith revisaron varios de mis escritos, y Dimitri me dejó su moto en un momento fundamental.
Siempre viví en el umbral del fracaso, y siempre hubo un \emph{deus ex machina} que me salvó.
Me gusta ser ayudado. No solo por el favor recibido, sino también porque me siento querido, y hay poco mejor que eso.

\thispagestyle{plain}

Termina con mi decisión de intentar dar un paso adelante, de salir de mi mundo de juego, y tomarme la vida en serio, con resoluciones, haciendo que la buena suerte que me 
persigue sea también debida a mis decisiones. Por esto ahora me voy a París con quién más he querido en estos tiempos, con el valor para marcharnos, sin miedo al llegar,
con la ilusión de hacer juntos la próxima etapa.

\thispagestyle{plain}

Esto lo que me ha salido, más o menos una lista de nombres. No es lo que quería, pero es lo que salió, y el tiempo se me escurre entre los dedos.

\endgroup

\begin{flushright}
\textsc{Marco Baity Jesi}\\
\itshape
Madrid, septiembre de 2015
\end{flushright}

% 
% Recuerdo todavia mi primer dia de doctorado. Volvia de m\'as de seis meses de mochilero por Suram\'erica y no sab\'ia siquiera si tendr\'ia acceso a una fuente de financiaci\'on.
% Llevaba una pintas peculiares, con unos pantalones muy anchos y una camisa de cuadros que bajaba por fuera. Muy jipi, y aunque la gente en la universidad lo asoci\'o con el
% rollo mochilero - y yo se lo dej\'e creer - ese pantal\'on me lo hab\'ia comprado mi madre en la playa, y la camisa ven\'ia de una bolsa de ropa que mi primo ya no usaba. Tambi\'en
% llevaba una barba descuidada m\'as larga que mi pelo ahora. Esa s\'i era un vestigio del viaje.
% 
% Venía motivado con terminar el trabajo de mi tesis de máster, apuntarme a acrobacia y aprender a hacer los mortales, jugar al voleiból, retomar el taekwondo, volver a tocar 
% el clarinete y a escribir cuentos, volver a trabajar de modelo, y jugar al fútbol hasta no poder más. Mi tesis de máster sigue sin publicar, pero 
% Aun estaba con Aurora, 

\end{document}